\documentclass[twoside,a4paper,10pt,x11names,svgnames]{memoir}
\pdfoutput=1
	\usepackage{textcomp}
	\usepackage{amsmath}
	\usepackage{amsfonts}
	\usepackage{amssymb}
	\usepackage{graphicx}
	\usepackage{color}
	\usepackage{geometry}
	\usepackage{wrapfig}  
        \usepackage[utf8]{inputenc}
	\usepackage[spanish,english]{babel}
	\usepackage{booktabs}
    \usepackage{makeidx}
    \usepackage[square]{natbib}
    \usepackage{multicol}
\usepackage{fix-cm}
\usepackage{lmodern}
    
\usepackage{color}
\usepackage{hyperref}
\usepackage{multirow}
\usepackage{colortbl}
\usepackage[dvipsnames]{xcolor}
\usepackage{float}

\usepackage{wallpaper}
\usepackage{tikz}
\usetikzlibrary{shapes,positioning}

\usepackage{lipsum}

\newcommand{\ra}[1]{\renewcommand{\arraystretch}{#1}}

\definecolor{darkred}{rgb}{0.0,0.3,0.0}
\definecolor{darkverde}{rgb}{0.0,0.2,0.0}

\hypersetup{colorlinks,breaklinks,linkcolor=darkred,urlcolor=darkred, anchorcolor=darkred,citecolor=darkred}
\makeindex
\decimalpoint

      \pretitle{\begin{center}\sffamily\huge\MakeUppercase}
      \posttitle{\par\end{center}\vskip 0.5em}
	\maxtocdepth{subsection} 

	\maxsecnumdepth{subsubsection} 
	\setsecheadstyle{\Large\bfseries\raggedright} 
	\setsubsecheadstyle{\large\bfseries\raggedright} 
	\setsubsubsecheadstyle{\normalsize\bfseries\raggedright} 

    \setsubsubsecindent{30pt}
    \setsubsecindent{20pt}

    \setaftersecskip{1.8\baselineskip}
    \setaftersubsecskip{1.5\baselineskip}   
	\newcommand{\degree}{\ensuremath{^\circ}}
\usepackage[font=small,indention=-1.0em,labelfont=bf,hang]{caption}

\selectfont

\captiontitlefont{\small\sffamily}
\captionstyle{\centerlastline}
	\author{Ramiro Checa-Garcia}
\addto\captionsspanish{%
}

 \setlrmarginsandblock{33mm}{28mm}{*}
  \setulmarginsandblock{33mm}{33mm}{*}

\setheadfoot{\baselineskip}{10mm}
\checkandfixthelayout

\newcommand\chapterillustration{}

\makechapterstyle{FancyChap}{
\setlength\beforechapskip{0pt}
\setlength\midchapskip{0pt}
\setlength\afterchapskip{130mm}

\def\printchaptertitle##1{
\ThisULCornerWallPaper{1}{\chapterillustration}
\tikz[overlay,remember picture]
  \fill[fill=PaleGreen1,opacity=.7]
  (current page.north west) rectangle 
  ([yshift=-3cm] current page.north east);
  \strictpagecheck\checkoddpage
  \ifoddpage{
    \begin{tikzpicture}[overlay,remember picture]
    \node[anchor=south west,
      xshift=20mm,yshift=-30mm,
      font=\sffamily\bfseries\Huge] 
      at (current page.north west) 
      {\chaptername\chapternamenum\thechapter};
    \node[fill=DarkGreen!30!black,opacity=.9,text=white,
      font=\LARGE\bfseries, 
      inner ysep=12pt, inner xsep=20pt,
       rectangle, rounded corners=2mm,anchor=east,text width=115mm,
      xshift=-10mm,yshift=-30mm] 
      at (current page.north east) {##1};
    \end{tikzpicture}
  }
  \else {
    \begin{tikzpicture}[overlay,remember picture]
    \node[anchor=south east,
      xshift=-20mm,yshift=-30mm,
      font=\sffamily\bfseries\Huge] 
      at (current page.north east)
      {\chaptername\chapternamenum\thechapter};
    \node[fill=DarkGreen!30!black,opacity=.9,text=white,
      font=\LARGE\bfseries,
      inner sep=12pt, inner xsep=20pt,
       rectangle,rounded corners=2mm,anchor=west,text width=115mm,
      xshift=10mm,yshift=-30mm] 
      at ( current page.north west) {##1};
    \end{tikzpicture}
  }
  \fi
}
}

\makechapterstyle{FancyUnnumberedChap}{
\setlength\beforechapskip{0pt}
\setlength\midchapskip{0pt}
\setlength\afterchapskip{55mm}

\def\printchaptertitle##1{
\tikz[overlay,remember picture]
  \fill[fill=PaleGreen1,opacity=.5]
  (current page.north west) rectangle 
  ([yshift=-3cm] current page.north east);
  \strictpagecheck\checkoddpage
  \ifoddpage{
    \begin{tikzpicture}[remember picture, overlay]
    \node[fill=DarkGreen!30!black,text=white,
      font=\LARGE\bfseries, 
      inner ysep=12pt, inner xsep=20pt,
      rectangle, rounded corners=2mm,anchor=east, 
      xshift=-20mm,yshift=-30mm] 
      at (current page.north east) {##1};
    \end{tikzpicture}
  }
  \else {
    \begin{tikzpicture}[remember picture, overlay]
    \node[fill=DarkGreen!30!black,text=white,
      font=\LARGE\bfseries,
      inner sep=12pt, inner xsep=20pt,
      rectangle, rounded corners=2mm,anchor=west,
      xshift=20mm,yshift=-30mm] 
      at ( current page.north west) {##1};
    \end{tikzpicture}
  }
  \fi
}
}

\newlength\pagenumwidth
\tikzset{pagefooter/.style={
anchor=base,font=\sffamily\bfseries\small,
text=white,fill=DarkGreen!60!black,text centered,
text depth=30mm,text width=\pagenumwidth}}

\definecolor[named]{GreenTea}{HTML}{CAE8A2}
\definecolor[named]{MilkTea}{HTML}{C5A16F}

\nouppercaseheads

\makeoddhead{headings}
{\begin{tikzpicture}[remember picture,overlay]
\fill[MilkTea!00!white] (current page.north east) 
	rectangle (current page.south west);
\fill[white, rounded corners] 
	([xshift=-10mm,yshift=-20mm]current page.north east) rectangle 	
	([xshift=15mm,yshift=17mm]current page.south west);
\end{tikzpicture}}%
{}%
{\begin{tikzpicture}[xshift=-.75\baselineskip,yshift=.25\baselineskip,remember picture, overlay,fill=DarkGreen,draw=DarkGreen]\fill circle(3pt);\draw[semithick](0,0) -- (current page.west |- 0,0);\end{tikzpicture}\sffamily\itshape\small\rightmark}

\makeoddfoot{headings}{}{}{%
\tikz[baseline]\node[pagefooter]{\thepage};}
\makeoddfoot{plain}{}{}{\tikz[baseline]\node[pagefooter]{\thepage};}

\makeevenhead{headings}
{{\begin{tikzpicture}[remember picture,overlay]
\fill[MilkTea!00!white] (current page.north east) rectangle (current page.south west);
\fill[white, rounded corners] ([xshift=-15mm,yshift=-20mm]current page.north east) rectangle ([xshift=10mm,yshift=17mm]current page.south west);
\end{tikzpicture}}%
\sffamily\itshape\small\leftmark\ 
\begin{tikzpicture}[xshift=.5\baselineskip,yshift=.25\baselineskip,remember picture, overlay,fill=DarkGreen,draw=DarkGreen]\fill (0,0) circle (3pt); \draw[semithick](0,0) -- (current page.east |- 0,0 );\end{tikzpicture}}{}{}
\makeevenfoot{headings}{\tikz[baseline]\node[pagefooter]{\thepage};}{}{}
\makeevenfoot{plain}{\tikz[baseline]\node[pagefooter]{\thepage};}
{}{}

\begin{document}

\selectlanguage{english}

	\renewcommand{\chaptername}{Chapter}  
	\renewcommand{\contentsname}{Index}
	\renewcommand{\appendixname}{ }
	\renewcommand{\partname}{Part}
	\renewcommand{\figurename}{Figure}

 \thispagestyle{empty}
\ThisLLCornerWallPaper{1}{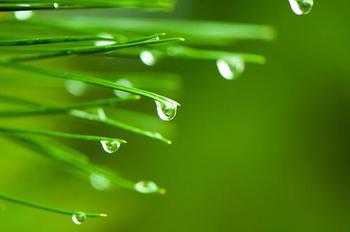}
 \tikz[remember picture,overlay]%
 \node[fill=DarkGreen,text=white,font=\Huge\bfseries,text=white,%
 minimum width=\paperwidth,minimum height=6em,anchor=north]%
 at (current page.north){PhD Thesis Dissertation};
 \vspace*{1\baselineskip}
\tikz[remember picture,overlay]%
\node[fill=white,opacity=.85,text=DarkGreen!10!Black,font=\Huge\bfseries,%
minimum width=0.90\paperwidth,minimum height=6em,anchor=north,text width=180mm]%
at (6.8,-1){First measurement of the small-scale spatial \\ variability of the rain drop size distribution};
\vspace*{2\baselineskip}

\tikz[remember picture,overlay]%
 \node[opacity=.8,text=DarkGreen!10!Black,font=\LARGE\bfseries,%
 minimum width=0.90\paperwidth,minimum height=4em,anchor=north,text width=180mm]%
 at (7.0,-1.3){Results from a crucial experiment and maximum entropy modeling
};

\tikz[remember picture,overlay]%
 \node[opacity=.7,text=DarkSlateGray,font=\LARGE\bfseries,%
 minimum width=0.90\paperwidth,minimum height=3.45em,anchor=north,text width=180mm]%
 at (7.0,-4.5){Primera medida de la variabilidad espacial\\ a pequeña escala de la distribución de tamaño de gota};

\tikz[remember picture,overlay]%
 \node[opacity=.7,text=DarkSlateGray,font=\Large\bfseries,%
 minimum width=0.90\paperwidth,minimum height=3em,anchor=north,text width=180mm]%
 at (7.0,-5.3){Resultados de un experimento crucial y modelización mediante máxima entropía};

\vspace*{1.1\baselineskip}


\tikz[remember picture,overlay]%
\node[fill=DarkGreen,opacity=.99,font=\Huge\bfseries,text=White,%
minimum width=\paperwidth,minimum height=5em,anchor=south]%
at (current page.south) {Ramiro Checa-García};

\begin{center}
\LARGE\bfseries\color{SaddleBrown!30!black}

\end{center}

\cleartorecto

\thispagestyle{empty}

\begin{figure}
  \flushleft \includegraphics[width=1.2\textwidth]{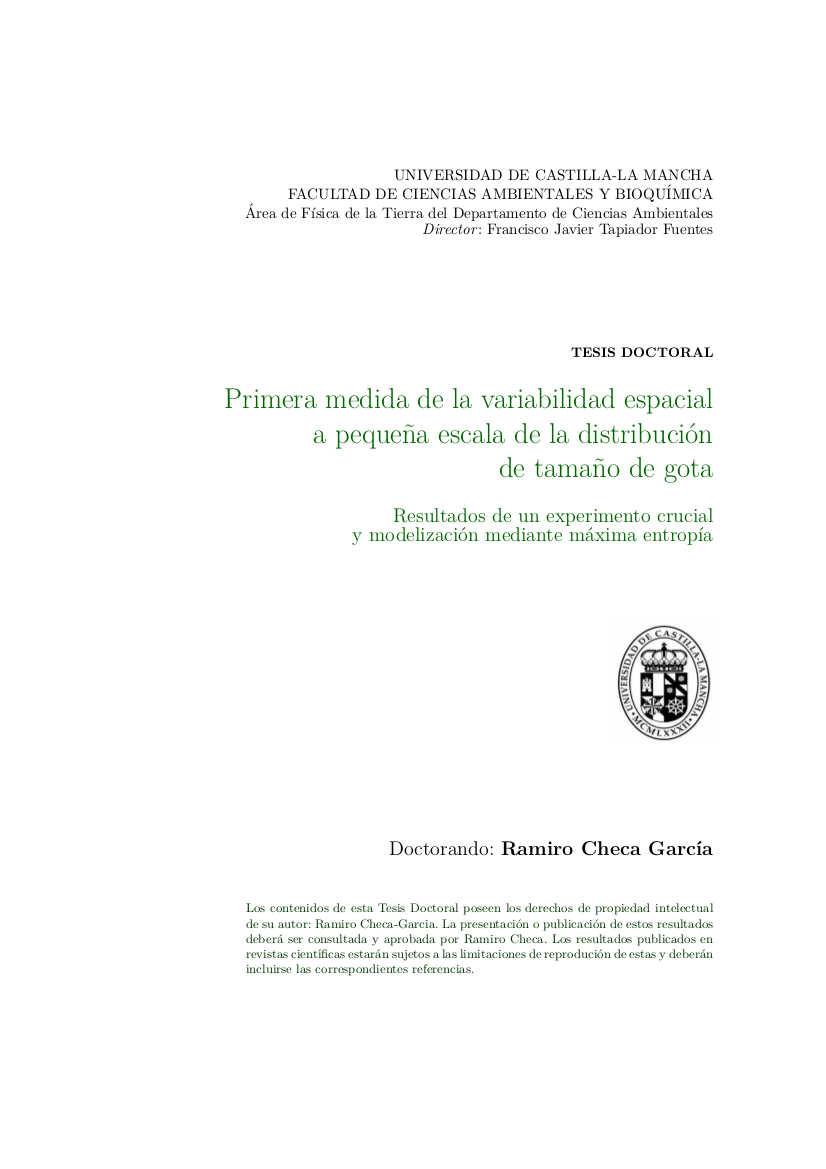}
\end{figure}

\frontmatter

\cleartorecto

\vspace*{4cm}

\begin{flushright}
 \large \emph{A mis padres \\y a mi hermana.}
\end{flushright}

\cleartorecto

\chapterstyle{FancyUnnumberedChap}
\tableofcontents

\mainmatter

\selectlanguage{spanish}

\chapterstyle{FancyChap}
\renewcommand\chapterillustration{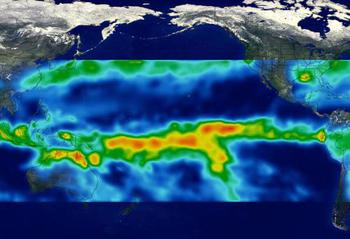}

\chapter{Introducción}

Esta memoria presenta una investigación novedosa y original sobre la distribución a nivel del suelo de los tamaños de gota en la precipitación  (DSD en sus siglas en inglés). La tesis avanza el estado del arte en dos direcciones: caracterización de la DSD y análisis de su variabilidad espacial.\\

Por primera vez, se ofrece evidencia empírica de que la DSD varía de manera sustancial a escala kilométrica. Este resultado es relevante en la práctica y cuenta con implicaciones para la medida remota de la precipitación, ya que los radares se calibran en función de la DSD a través de la relación reflectividad/precipitación (ó Z/R). Caracterizar la variabilidad espacial de la relación Z/R es fundamental para obtener medidas remotas más precisas de la precipitación, así como para estimar las incertidumbres en la medida.\\

También por primera vez, se ofrecen pruebas empíricas de que la modelización de la DSD utilizando el método de máxima entropía mejora las parametrizaciones previas basadas en la función exponencial y en la función gamma de distribución de probabilidad, cuando la resolución temporal de la DSD es alta, y cuando las características físicas de la precipitación implican una distribución de tamaños de gota multimodal. La relevancia de este resultado es que la nueva parametrización mejora la caracterización de la DSD, y por tanto, mejora el conocimiento de la Z/R con que se calibran los radares, tanto terrestres como a bordo de satélites.\\

La tesis cuenta con una base empírica construida \emph{ex profeso} para contrastar las hipótesis de los experimentos, y recoge los datos obtenidos con la red de disdrómetros de la UCLM en sendas campañas en los años 2010 y 2011. Se utilizan también otras bases de datos con objeto de contrastar los resultados experimentales, en la forma de simulaciones controladas de distribuciones de tamaños de gota generadas computacionalmente mediante métodos estadísticos contrastados. \\

La tesis se encuadra dentro de la misión \emph{Global Precipitation Measurement} (GPM) de la NASA, en cuyo \emph{science team} se encuentra el director de este trabajo, y en cuyos \emph{science meetings} se han presentado los resultados contenidos en esta tesis.

\section{Interés científico del tema de investigación}

El tema de esta tesis es la caracterización de la DSD y el estudio de su variabilidad espacial. Se trata de un tema que ha suscitado interés investigador desde los años sesenta del siglo pasado, y que continua generando literatura dada su relevancia para la estimación remota de precipitación. \\  

La medida de la DSD es imprescindible para calibrar las medidas remotas de precipitación con radar meteorológico. Conocer la variabilidad espacial de la DSD es también muy importante para los radares orbitales, puesto que su resolución espacial es del orden de kilómetros.\\

La medida de la DSD es importante para los radares de tierra, puesto que podría darse el caso de que en áreas amplias coexistieran diferentes relaciones Z/R. En este escenario, el uso de una misma relación para un mismo radar y una misma área introduciría sesgos en las estimaciones.\\

El estudio posee aplicaciones más allá de los radares meteorológicos, ya que los estudios de atenuación de señales incorporan hipótesis implícitas sobre la DSD, tanto en su caracterización funcional, como en su variabilidad espacial. Esto amplia el impacto de los resultados de esta tesis más allá de su campo de origen.\\


En relación al entronque de esta tesis con los estudios del ciclo hidrológico, en los últimos informes de la UNESCO \textemdash tanto sobre medio ambiente en general como sobre el problema del agua en particular \citep{loucks2005water}\textemdash \, se incide en la importancia del correcto entendimiento del ciclo hidrológico a nivel global, y de la importancia de la precipitación en dicho ciclo, máxime teniendo en cuenta su gran variabilidad espacial y temporal. En este sentido, las medidas de precipitación mediante sensores remotos localizados en radares orbitales son claves al ser la única fuente de medida sobre la mayor parte del planeta, ya que los océanos, las zonas montañosas o bosques tropicales no cuentan con instrumentación suficiente.\\


El contenido del trabajo también es relevante a nivel nacional y regional, dado que no existen otros grupos de investigación a nivel regional (y solo unos pocos a nivel nacional) que estudien medidas a nivel del suelo de distribuciones de tamaños de gota mediante disdrómetros.

\section{Novedad y originalidad}

Esta tesis representa una novedad absoluta en su campo, puesto que nadie había realizado antes una medida de la variabilidad espacial de la DSD a escalas kilométricas. Previamente, se habían realizado estudios de la variabilidad temporal, o de la variabilidad espacial a escalas de decenas,  centenas de kilómetros, o continentales, pero nunca a las escalas relevantes para los radares a bordo de satélites.\\

Uno de los rasgos originales de la investigación reside en el uso de una red de disdrómetros diseñada y construida para el experimento. Como se describe más adelante, los disdrómetros son los únicos instrumentos que permiten medir la DSD. Otros aparatos se limitan a inferirla. Contar con un número grande de disdrómetros (18) gracias a un proyecto FEDER ha permitido diseñar y llevar a cabo un \emph{experimentum crucis} para validar la hipótesis, algo que no siempre es posible, pero que sin duda resulta deseable para cualquier investigación.\\

La tesis también avanza otra línea de investigación original en forma de una nueva pa\-ra\-me\-tri\-za\-ci\-ón de la DSD que se puede enlazar con razonamientos físicos, no solo de ajuste estadístico, y que mejora sustancialmente las parametrizaciones previas.\\

Los resultados parciales de la investigación han sido publicados en dos revistas listadas en el ISI-JCR: \emph{Geophysical Research Letters}, y \emph{Entropy}. 

\section{Hipótesis del trabajo de investigación}

La primera hipótesis del trabajo de investigación es que la variabilidad espacial de la DSD es relevante a escala kilométrica. Es decir, que existen diferencias sustanciales en la distribución de tamaño de gota en un mismo episodio de precipitación entre puntos separados por menos de cinco kilómetros. Esta escala se corresponde con la mejor resolución espacial del sensor DPR del satélite GPM-core trabajando en los 85 GHz.\\

La segunda hipótesis de trabajo es que la parametrización de la DSD utilizando el método de entropía máxima mejora la caracterización de la DSD con respecto a los métodos utilizados anteriormente. 

\section{Tesis del trabajo de investigación}

La tesis del trabajo de investigación consiste en la validación de las hipótesis previas.\\

Para validar la primera hipótesis se diseñó un experimento consistente en la media simultánea de la DSD en un área limitada. Como se describe en el capítulo \S\ref{sec:chapVARSPACIAL1}, los resultados del experimento validan la hipótesis inicial.\\

Para validar la segunda hipótesis se realizó una modelización teórica que después se contrastó con medidas de campo, comparándola para su verificación con métodos preexistentes. Los resultados consiguieron validar esta la hipótesis.\\

Como resultado adicional se analizó la influencia de los métodos de cuantización de tamaños de gota en las medidas disdrométricas, evaluando la hipótesis de su  relevancia en la DSD y los parámetros integrales de la precipitación. El resultado fue la validación de la hipótesis en el curso del estudio realizado.

\part{Base teórica: Microfísica de precipitaciones} 
\renewcommand\chapterillustration{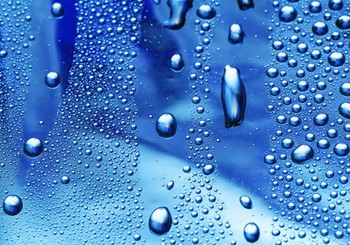}

\chapter{Distribución de tamaños de gota (DSD)}
\label{cap:DSD}
En este capítulo se introducen las definiciones tanto de la distribución de tamaños de gota (Drop Size Distribución, DSD) como de las magnitudes físicas que se obtienen de esta. También se analizan las relaciones más utilizadas entre parámetros de la DSD.

\section{Definición de la \textit{distribución de tamaños de gota} (DSD)}
\label{sec:defDSD}
A continuación se definen el concepto de distribución de tamaños de gota, junto con una serie de conceptos relacionados con este, fijando la notación utilizada a lo largo del trabajo.

\begin{itemize}
\item N(D) es la concentración de hidrometeoros con diámetro comprendido entre D y D+dD por unidad de volumen y unidad de intervalo de diámetros. Lo usual es tratar con una versión discreta de diámetros $N(D_{i})$, i=1,...,n que se denomina DSD\footnote{Acrónimo de \emph{Drop Size Distribution}. Dado que puede haber distribuciones de tamaños de gota de nubes o de agregados presentes en la atmósfera como aerosoles, se suele hablar de la RDSD, \emph{RainDrop Size Distribution}. En otros contextos se habla de PSD (\emph{Particle Size Distribution}). Algunos autores utilizan DSD para referirse a datos experimentales y N(D) para modelos de estos histogramas experimentales.}. Sus dimensiones son $[L^{-4}]$, y las unidades son $m^{-3}mm^{-1}$.

\item $f(D)$ se define como la probabilidad de encontrar partículas de tamaño D. Está normalizada a la unidad. Por definición, $N(D)=N_{t}f(D)$ donde $N_{t}$ es el número total de gotas (momento cero de la DSD).
\item n(D) es el número de partículas de tamaño D registradas en un evento determinado. Es dependiente, por ejemplo, del tiempo de recolección de partículas. Por ello se suele preferir N(D) para la caracterización de los fenómenos de precipitación.
\end{itemize}

\begin{figure}[h] 
\begin{center}
   \includegraphics[width=0.52\textwidth]{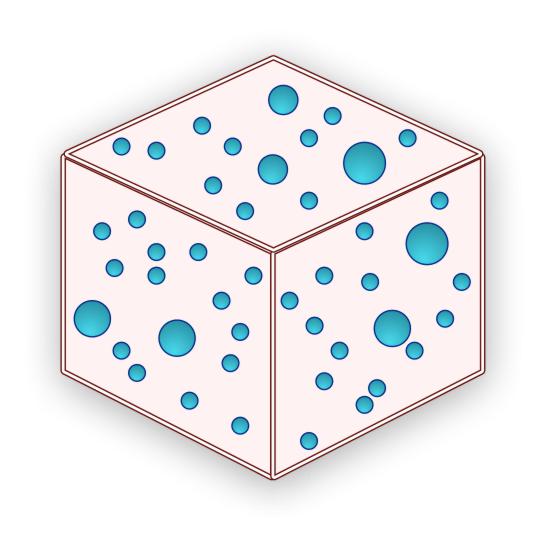} 
\end{center}
   \caption[Esquema del muestreo sobre un volumen de la precipitación]{\textbf{Esquema de la precipitación muestreada en un volumen}. Es el tipo de muestreo realizado mediante medidas de teledetección.}
\label{fig:CuboLluvia1}
\end{figure}

\begin{figure}[h] 
\begin{center}
   \includegraphics[width=0.52\textwidth]{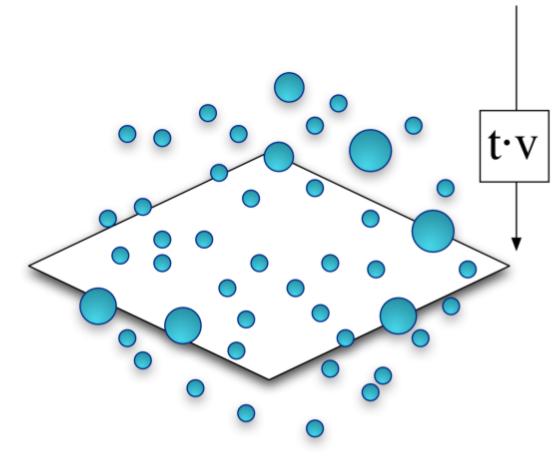} 
\end{center}
   \caption[Esquema del muestreo a través de una superficie de la precipitación]{\textbf{Esquema de la precipitación que atraviesa superficie (S) en un tiempo (t)}. El volumen muestreado depende del diámetro efectivo de cada gota, como consecuencia de que la velocidad vertical de caída típica está condicionada por el tamaño de gota. Es el tipo de muestreo realizado en medidas a nivel del suelo mediante disdrómetros.}
\label{fig:CuboLluvia2}
\end{figure}

Aparte del tamaño, la otra magnitud esencial para caracterizar un conjunto de gotas es la velocidad de caída. Generalmente se supone que, a nivel del suelo, es la terminal del hidrometeoro. El hecho de que no exista una relación biunívoca entre diámetro y velocidad del hidrometeoro motiva una definición más amplia:

\begin{itemize}
\item Denominamos $n(D,v)$ al número de gotas con diámetro entre D y D+dD y velocidad entre $v$ y $v+dv$. La correspondiente distribución de probabilidad bidimensional será $f(D,v)$.
\end{itemize}

La cantidad $n(D,v)$ se puede medir experimentalmente. De hecho, la estimación de N(D) se basa en datos experimentales para $n(D,v)$. Por tanto, la física del proceso recomienda el uso de las distribuciones bidimensionales $f(D,v)$. Sin embargo, no todos los aparatos de medidas microfísicas en superficie determinan la velocidad y el diámetro. En ocasiones se asume una relación entre diámetros y velocidades para dar lugar después a una medida de la DSD. Ello supone que existe una relación funcional $v(D)$ independiente de los parámetros concretos de la masa de aire bajo estudio, y única dentro de volumen muestreado. Por tanto, implica realizar una cierta aproximación, dado que, primero, la relación entre velocidad y diámetro presenta un carácter estocástico dentro del volumen de muestreo; y segundo, presenta una dependencia con, por ejemplo, la densidad de la masa de aire donde se registra la precipitación.\\

La relación más utilizada \citep{1973Atlas} se basa en un experimento realizado por \citep{GunnKinzer1948}. Así, \citep{1973Atlas}, modelizaron la $v(D)$ mediante la expresión:

\begin{equation}
v(D)=9.65-10.3e^{-0.6D}
\label{eqn:AtlasVDequation}
\end{equation}

donde D se mide en $[mm]$ y la velocidad en $[m/s]$. Desde el punto de vista de las aplicaciones de la DSD, resulta más útil una relación de potencias como $v(D)=\gamma D^{\delta}$. Basándose en el mismo conjunto de datos experimentales surge la relación:

\begin{equation} 
v(D)=3.78D^{0.67}
\label{eqn:AtlasVDequationPOWERLAW}
\end{equation}

\citep{AtlasUlbrich200microDSD}, aunque algunos autores (véase, por ejemplo, \citep{Beard1976vD}, y para un análisis más reciente \citep{ChinosPARSIVELestudio}) han propuesto otros valores para los parámetros $\gamma$ y $\delta$, basados en estudios experimentales, y han planteado\footnote{Es posible realizar también estudios basados en dinámica de fluidos, en los que aparecen \citep{RogersBOOK} para $\delta$ valores entre 0.5 y 1 . Por otra parte, las relaciones de potencias entre magnitudes relacionadas con el diámetro o la velocidad son de gran utilidad práctica, lo que ha llevado a ampliar este tipo de ajuste $v(D)$ a agregados de nieve o hielo, o a relacionar en forma de potencias la masa de estos agregados con el diámetro medio. Véase \citep{RogersBOOK}, \citep{StrakaBOOK} y \citep{brandes_ikeda_etal_2007_aa}} posibles dependencias en el valor del parámetro $\delta$ con la densidad del aire y turbulencias locales.\\

Si analizamos los métodos de medida de la DSD observamos que se ha de realizar una distinción entre dos modos esenciales \citep{uijlenhoet_pomeroy_2001_aa}. En primer lugar, la distribución de tamaños de gota, tal y como ha sido definida arriba, es el número de gotas en un intervalo de diámetros por \emph{unidad de volumen de aire}, $N_{v}(D)$. En segundo lugar, en la práctica, usando instrumentos a nivel del suelo, la medida es el número de gotas en un intervalo de diámetros que llegan a la superficie colectora por unidad de área y unidad de tiempo, y que podemos denotar por $N_{a}(D)$. Luego, en términos del número de gotas por diámetro D, se tiene que:

\begin{equation}
n_{a}(D)=n_{v}(D)v(D)
\label{eqn:tiposlluvia}
\end{equation}

Como se verá en los capítulos posteriores hay una amplia gama de instrumentos de medida de N(D) que, como los disdrómetros JWD y POSS, se basan en la existencia de una forma funcional de $v(D)$ a partir de la cual puede estimarse la cantidad $n_{v}(D)$.\\

En este punto, es necesario diferenciar conceptualmente estimaciones sobre un volumen, como en el caso de una medida por teledetección, figura (\ref{fig:CuboLluvia1}), de medidas basadas en el flujo de precipitación a través de una superficie (flujo de partículas), figura (\ref{fig:CuboLluvia2}). Si observamos la relación (\ref{eqn:tiposlluvia}) vemos que el volumen de muestreo para gotas de diferente diámetro varía para cada caso. Es necesario también notar que uno de los problemas que a menudo se plantea en la modelización de la DSD es la asignación de errores a una medida determinada, ya que muestreamos un colectivo de gotas y esto induce errores debido a la medida parcial sobre el colectivo total de gotas.\\

Por otra parte, la propia DSD puede variar por razones físicas tanto espacial como temporalmente. Las medidas de un flujo de precipitación a nivel del suelo intentan pues un compromiso entre volúmenes de muestreo pequeños, para poder describir estas variaciones físicas, y volúmenes de masa de aire lo suficientemente amplios para que el problema de muestreo no condicione en exceso las estimaciones de N(D). Esto se consigue diseñando instrumentos con una superficie colectora pequeña que acumulan las medidas de n(D,v) durante un intervalo de tiempo suficiente\footnote{Los dispositivos más recientes permiten variar el tiempo de acumulación. Tiempos pequeños permiten estudiar fluctuaciones en las cantidades $n(D,v)$; tiempos mayores permiten evitar la presencia de dichas fluctuaciones.}.\\

Respecto de los posibles errores de muestreo, se ha propuesto \citep{NormalizadaTestud2001} una me\-to\-do\-lo\-gía basada en la hipótesis de que la detección de una gota individual perteneciente a un colectivo de estas puede ser aproximado por un proceso de Poisson homogéneo. En este caso, la desviación estándar de $\hat{N}(D_{i})$ en un intervalo de longitud $\Delta D_{i}$ centrado en $D_{i}$ viene determinado por $\sqrt{N(D_{i})/u_{s}\Delta D_{i}}$, donde $u_{s}$ es el volumen de muestreo para el diámetro concreto $D_{i}$.\\



\section{Principales parámetros físicos de la DSD}
\label{sec:parametrosintegralesDSD}
La mayoría de los parámetros físicos relacionados con la precipitación se expresan en función de la distribución de tamaños de gota, a menudo mediante integrales de esta con determinadas funciones peso $\phi(D)$,
\vspace{2mm}
\begin{equation}
\Phi=\int_{D_{min}}^{D_{max}}\phi(D) N(D)dD
\end{equation} 
\vspace{2mm}

En general, estas cantidades pueden escribirse como momentos de la DSD aunque la terminología usada para referirse al conjunto de estas cantidades es la de \emph{parámetros integrales de la precipitación}, lo que incluye cantidades como la reflectividad de un conjunto de gotas, o el espectro de velocidades a una determinada frecuencia que, en principio, no tienen por qué ser momentos de la distribución.\\


Dados un diámetro mínimo $D_{min}$ y un diámetro máximo $D_{max}$ para el espectro de la precipitación, el momento de orden k-ésimo queda expresado como:
\vspace{2mm}
\begin{equation}
M_{k}=\int_{D_{min}}^{D_{max}}D^{k}N(D)dD
\end{equation}
\vspace{2mm}

donde las unidades usuales de D son $[mm]$ y de N(D) $[m^{-3}mm^{-1}]$. De cara a obtener las propiedades físicas en las unidades adecuadas, suelen expresarse como $C_{k}M_{k}$, donde $C_{k}$ es un factor de conversión apropiado. En el caso de no indicarse los límites $D_{min}$ y $D_{max}$, los naturales son [0, $\infty$).

\subsection{Número total de gotas}

El número total de gotas, llamado también \emph{concentración total}, se define como:

\vspace{2mm}
\begin{equation}
N_{t}=\int_{D_{min}}^{D_{max}}N(D)dD
\end{equation}
\vspace{2mm}

que expresa el número de gotas por unidad de volumen de aire. En general se asumen como límites: $[D_{min},D_{max})=[0,\infty)$ aunque para análisis concretos de la variación microfísica de la DSD puede ser conveniente estudiar variaciones del número total de gotas en intervalos concretos del espectro de tamaños. Sus unidades típicas son $[m^{-3}]$.\\

$N_{t}$ es un parámetro que se estima directamente en los modelos de nubes, mientras que en las parametrizaciones incluidas en los modelos de predicción numérica suele estimarse a partir de otras cantidades, y asumiendo una determina forma funcional parametrizada para la N(D).

\subsection{Diámetros característicos}

Existen una serie de parámetros que intentan expresar el hecho de que diferentes DSD pueden poseer diferentes tamaños típicos de gota. De este modo una DSD originada en un \emph{Nimbostratus} poseería un diámetro característico diferente de una originada en un \emph{Cumulonimbus}, más allá de la variabilidad estocástica que las DSDs posean. Se definen a continuación los diámetros típicos más utilizados:\\

El \emph{diámetro medio ponderado sobre la concentración total} se define como:
\begin{equation}
\overline{D}=\frac{1}{N_{t}}\int_{D_{min}}^{D_{max}}DN(D)dD=\frac{M_{1}}{M_{0}}
\end{equation}

Sus unidades son $[mm]$. Complementariamente se define su desviación estándar como:

\begin{equation}
s^{2}=\frac{1}{N_{t}}\int_{D_{min}}^{D_{max}}(D^{2}-\overline{D}^{2})N(D)dD=\frac{M_{2}M_{0}-M_{1}^{2}}{M_{0}^{2}}
\end{equation}

El \emph{diámetro medio ponderado sobre la masa} se escribe:

\begin{equation}
D_{m}=\frac{\int_{D_{min}}^{D_{max}}D^{4}N(D)dD}{\int_{D_{min}}^{D_{max}}D^{3}N(D)dD}=\frac{M_{4}}{M_{3}}
\end{equation}

que proviene de la definición:

\begin{equation}
D_{m}=\frac{\int_{D_{min}}^{D_{max}}D m(D) N(D)dD}{\int_{D_{min}}^{D_{max}}m(D)N(D)dD}
\end{equation}

donde $m(D)$ es la masa del hidrometeoro, dado que $m(D)\simeq \rho D^{3}$. Durante esta memoria se notará, bien como $D_{m}$, bien como $D_{mass}$. Su desviación estándar correspondiente es:

\begin{equation}
s_{m}^{2}=\frac{\int_{0}^{\infty}N(D)D^{3}(D-D_{m})^{2}dD}{\int_{0}^{\infty}N(D)D^{3}dD}
\end{equation}

En general, se puede definir un diámetro \emph{típico} como el cociente $M_{k+1}/M_{k}$. Así, el \emph{diámetro efectivo} es $D_{eff}=M_{3}/M_{2}$.

\begin{figure}[H] 
\begin{center}
   \includegraphics[width=0.70\textwidth]{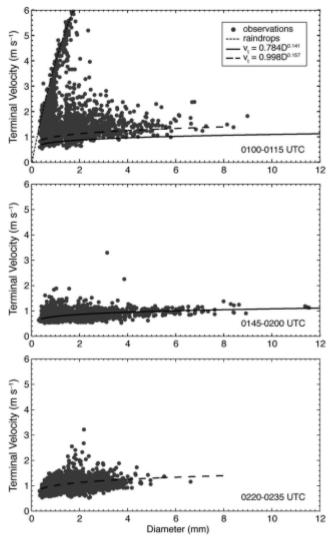}
\end{center}
   \caption[Diagramas experimentales de la $v(D)$ para tormentas de nieve. Fuente: \citep{brandes_ikeda_etal_2007_aa}]{\textbf{Diagramas experimentales de $v(D)$ para tormentas de nieve. Fuente: \citep{brandes_ikeda_etal_2007_aa}}. La figura recoge las velocidades terminales experimentales en función del diámetro bajo la hipótesis de copos de forma esférica para un mismo evento: tormenta del 5 de marzo del 2004 en Front-Range (Colorado) en tres momentos diferentes de su evolución. Se aprecia arriba un caso mixto mientras que las siguientes figuras representan únicamente copos de nieve. El caso mixto posee una rama con velocidades terminales en torno a la relación $v(D)=3.78\cdot D^{0.67}$, lo que permite diferenciar gotas de lluvia de copos de nieve en este periodo de la tormenta. Los casos puros de copos de nieve presentan una relación $v(D)$ diferente, caracterizada por un incremento mucho menor de las velocidades terminales con el diámetro.}
\label{fig:VDcolorado}
\vspace{1.5cm}
\end{figure}

También es posible usar la \emph{mediana}, es decir, definir $D_{0}$ mediante la ecuación implícita:

\begin{equation}
\int_{D_{min}}^{D_{0}}DN(D)dD=\int_{D_{0}}^{D_{max}}DN(D)dD
\end{equation}

Se usan unas u otras definiciones dependiendo tanto de la metodología utilizada para llegar a una modelización determinada de la DSD, como del campo concreto de aplicación. Todos estos parámetros están relacionados asumida una cierta forma funcional para N(D), y algunos de ellos poseen valores medios y distribuciones de valores similares para DSD experimentales, como por ejemplo $D_{0}$ y $D_{m}$.

\subsection{Contenido de agua líquida}

El contenido en agua líquida se suele notar como W o LWC (\emph{Liquid Water Content}), y se define como:

\begin{equation}
W=\frac{\pi}{6}\rho_{w}\int_{D_{min}}^{D_{max}}D^{3}N(D)dD=C_{w}M_{3}
\end{equation}

donde podemos aproximar la densidad del agua, $\rho_{w}$, como $10^{3}\,kg/m^{3}$, y tener $C_{w}=\frac{\pi}{6}10^{-3}$ expresando $W[g/m^{3}]$. El contenido de agua líquida posee relevancia práctica en física de nubes, donde suele estudiarse su variación con la altura. Los valores característicos oscilan entre $0.1\,g/m^{3}$ para una nube tipo \emph{stratus} hasta $1.1\,g/m^{3}$ para un \emph{cumulonimbus}. Es una magnitud importante de cara a la interpretación de medidas mediante sensores en satélites, ya que la atenuación de la señal recibida se relaciona con el contenido de agua líquida de la columna de la masa de aire observada desde el satélite (magnitud denotada como LWP -\emph{Liquid Water Path}- $[g/m^{2}]$, y que puede estimarse mediante sensores en el rango de las microondas).

\subsection{Intensidad de precipitación}
\label{sec:refINTENSIDADR}
Se define como\footnote{La medida de la intensidad de precipitación a nivel del suelo se basa en instrumentos que estiman la cantidad de agua acumulada durante un intervalo de tiempo. El método más sencillo, dada una unidad de superficie, consiste en medir la altura que adquiere en un pluviómetro, de ahí que las unidades clásicas sean mm/h. Las unidades usuales son $D$ en $[mm]$, $v_{t}(D)$ en $[m/s]$ y N(D) en $[mm^{-1}m^{-3}]$. Como el volumen de cada gota es $\frac{\pi}{6}D^{3}$, expresando R en $[mm/h]$ requiere utilizar un factor de unidades $\frac{6\pi}{10^{4}}$.}:
\vspace{2mm}

\begin{equation}
R[mm/h]=\frac{6\pi}{10^{4}}\int_{D_{min}}^{D_{max}}D^{3}N(D)v_{t}(D)dD
\label{eqn:definicionR}
\end{equation}
\vspace{2mm}

La relación $v(D)$ utilizada suele ser\footnote{Algunos autores expresan D en cm. En tal caso $v(D)=17.67\cdot D^{0.67}$.} $v(D)=3.78\cdot D^{0.67}$, con lo que R es proporcional a $M_{3.67}$. En este caso, $C_{R}=6\pi 3.78\cdot 10^{-4}$. A menudo en estudios climáticos o hidrológicos se detallan valores de intensidad de lluvia en \emph{mm/mes} o \emph{mm/a\~no}. 

En el caso de un episodio de precipitación de nieve es posible determinar el equivalente en precipitación líquida de los valores experimentales\footnote{La ecuación (\ref{eqn:definicionR}) no es la utilizada típicamente para calcular R, sino que se parte de la información dada por $n(D,v)$ sin hipótesis sobre $v(D)$. En consecuencia, es posible calcular R dada $n(D,v)$ sin importar la fase real (líquida o no) del hidrometeoro, aunque la interpretación del valor de la intensidad de precipitación sí que dependerá del tipo de precipitación. Lo más usual para realizar comparaciones es, dado el volumen estimado del agregado sólido, transformarlo en su equivalente agua líquida tras un proceso de fusión.} para R. Para ello se usa la relación \citep{brandes_ikeda_etal_2007_aa},

\begin{equation}
\rho_{s}(D)=aD^{b}
\end{equation}

con D en \emph{[mm]} y $\rho_{s}$ en $g/cm^{3}$. Relación complementaria de la que expresa la masa del hidrometeoro en gramos como $m(D)=\hat{a}D^{\hat{b}}$, típicamente $m(D)=8.90\cdot 10^{-5}D^{2.1}$. En la tabla (\ref{tablaDensidadCopoNieve}) se han recogido las relaciones más importantes utilizadas en la literatura y basadas en estudios experimentales. Su representación gráfica puede verse en la figura (\ref{fig:VDcolorado}). Para los cálculos realizados en este tesis se ha tomado la más reciente de las relaciones indicadas debida a \citep{brandes_ikeda_etal_2007_aa}.

\begin{table}[h]
\caption[Relaciones entre la densidad de copos de nieve y su tamaño, caracterizado por el diámetro efectivo. Fuente: \citep{brandes_ikeda_etal_2007_aa}]{\textbf{Relaciones entre la densidad de copos de nieve y su tamaño caracterizado por un diámetro efectivo. Fuente: \citep{brandes_ikeda_etal_2007_aa}}. La tabla indica relaciones experimentales entre la densidad $\rho_{s}$ en $g/cm^{3}$ y el diámetro efectivo en [mm]. Su representación gráfica se puede ver en la figura (\ref{fig:DensidadNieveFIGURA}).}
\vspace{0.5cm}
\begin{center}
\begin{tabular}{lll}
\toprule
& \textbf{Estudio experimental} & \textbf{Relación propuesta} \\
\midrule
A & Manogo \& Nakamura (1965) &  $\rho_{s}=2D^{-2}$   \\
B & Holroyd (1971)    & $\rho_{s}=0.17D^{-1}$  \\
C & Marumoto (1995) & $\rho_{s}=0.048D^{-0.406}$\\
D & Fabry \& Szymer (1999)  & $\rho_{s}=0.15D^{-1}$\\
E & Heymsfield (2004)           & $\rho_{s}=0.104D^{-0.95}$\\
F & Brandes (2007)  & $\rho_{s}=0.178D^{-0.922}$ \\
\bottomrule
\end{tabular}
\end{center}
\label{tablaDensidadCopoNieve}
\end{table}%

\begin{figure}[h] 
\begin{center}
   \includegraphics[width=0.60\textwidth]{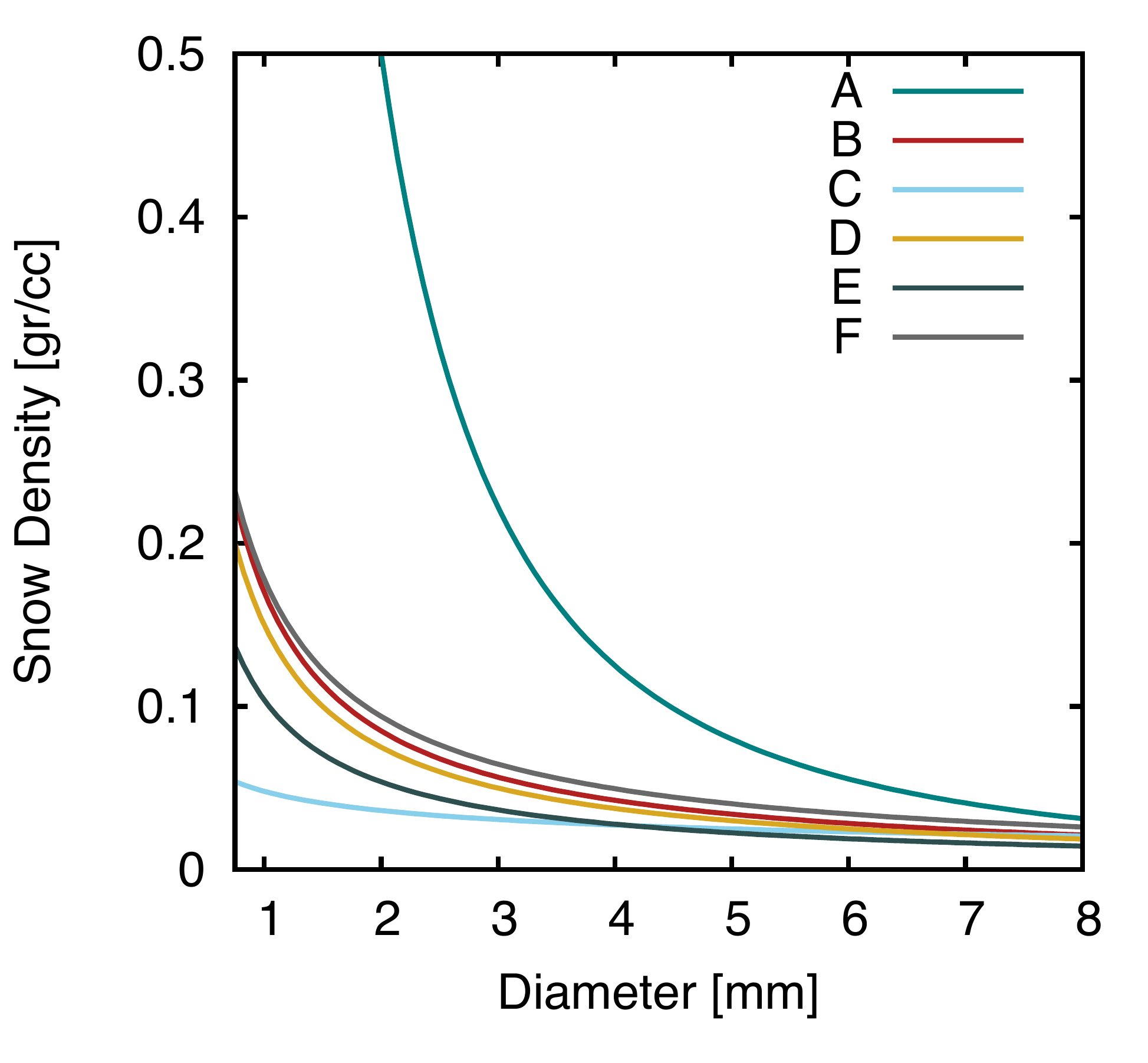}
\end{center}
   \caption[Densidad de los copos de nieve en función de diámetro.]{\textbf{Densidad de los copos de nieve en función de diámetro}. Las diferentes propuestas comparten el comportamiento asintótico de la densidad para valores altos del diámetro. Las diferencias se aprecian más claramente en el caso de copos de nieve de diámetros inferiores a 3 mm.}
\label{fig:DensidadNieveFIGURA}
\end{figure}

\subsection{Extinción óptica y energía cinética}

La \emph{extinción óptica} dada una distribución de tamaños de gotas se define como:

\begin{equation}
\Sigma= C_{\Sigma}\int_{D_{min}}^{D_{max}}D^{3}N(D)dD
\end{equation}

que posee unidades típicas de $[km^{-1}]$, y $C_{\Sigma}=1.57\cdot10^{9}\,[km^{-1} mm]$ para las unidades de N(D) de $[m^{-3}mm^{-1}]$.\\

El \emph{flujo de energía cinética} de la precipitación viene dado por,

\begin{equation}
K=C_{K}\int_{D_{min}}^{D_{max}}D^{3}v_{t}^{3}(D)N(D)dD
\end{equation}

donde si K está expresada en $[J m^{-2} h^{-1}]$, se tiene que $C_{K}=\frac{3\pi}{10^{4}}$. Esta magnitud ha recibido cierta atención debido a su relevancia en problemas de erosión, habiéndose estudiado la posibilidad de determinarla a partir de otras cantidades más directamente medibles, como la intensidad de precipitación o la reflectividad \citep{SteinerKINETIC2000}.

\subsection{Reflectividad}

La reflectividad es uno de los parámetros integrales de la precipitación de mayor interés en teledetección ya que es el resultado de las medidas mediante radar. Esencialmente la señal recibida por un radar meteorológico sobre un volumen específico de aire viene dada por la ecuación del radar:

\begin{equation}
\mathcal{P}_{r}=C\eta\frac{1}{r^{2}}
\end{equation}

donde la constante C aglutina propiedades específicas del radar pero no del volumen de aire estudiado, mientras que $\eta$ es la reflectividad del radar, definida como:

\begin{equation}
\eta=\int_{0}^{\infty}N(D)\sigma_{bs}(D)dD
\label{eqn:eta_reflectividad}
\end{equation}

N(D) es la distribución de tamaños de gota del volumen de aire estudiado, y $\sigma_{bs}$ es la sección eficaz de retrodispersión (\emph{backscattering}) de las gotas. Típicamente, $\sigma_{bs}$ depende tanto del material (forma, temperatura) como de la longitud de onda $\lambda$ de la señal. En el caso en que $D<0.1\lambda$ estamos en el régimen de Rayleigh, y $\sigma_{bs}$ puede ser aproximada por:

\begin{equation}
\sigma_{bs}=\frac{\pi^{5}|k|^{2}D^{6}}{\lambda^{4}}
\end{equation}

Donde k depende del índice de refracción. Fijado este y la longitud de onda se puede expresar la señal recibida por el radar en función de la DSD como:

\begin{equation}
Z=\int_{D_{min}}^{D_{max}}D^{6}N(D)dD
\end{equation}

Las unidades de Z son $[mm^{6}m^{-3}]$, aunque es preferible visualizarla en escala de decibelios. $ZdBZ=10log_{10}(Z)$. El rango típico de valores comprende desde 10dBZ para lluvias ligeras hasta 40-45 dBZ para lluvias intensas\footnote{Los valores umbrales típicos de sensibilidad para Z dependen del tipo de radar. Los primeros radares meteorológicos no eran capaces de discriminar señales correspondientes a lluvias del orden de 0.1 $mm/h$.}.\\ 

Otra magnitud relevante en las medidas remotas es la \emph{atenuación}. Interesa en el rango de microondas para estimaciones mediante medidas de satélite, así como en el rango de longitudes de onda del radar (centimétrico o milimétrico)\citep{AttenuationRelevance}. De manera similar a (\ref{eqn:eta_reflectividad}) se define como:
 
\begin{equation}
A=C_{A}\int_{D_{min}}^{D_{max}}\sigma_{ex}(D)N(D)dD
\end{equation}

donde $\sigma_{ex}(D)$ es la sección eficaz de extinción a la longitud de onda utilizada. El valor usual de la constante para expresar A en $[dB km^{-1}]$ viene dado por $C_{A}=4343$. Su estimación es importante para los métodos de calibración de los radares, máxime para aquellos de frecuencias mayores donde los procesos de atenuación con la distancia debido a la presencia de precipitación pueden llegar a ser más importantes; como en el caso de los radares en banda X. También es el caso de radares en satélites, donde un tratamiento no adecuado de la atenuación puede inducir errores relevantes (véanse las referencias al respecto en \citep{Tokay2010smallscaleDSD}).

\vspace*{1.15cm}
\begin{table}[H]
\caption[Parámetros integrales de la precipitación]{\textbf{Parámetros integrales de la precipitación}. Se indican las unidades y símbolos típicos de los diferentes parámetros integrales de la precipitación. También se muestran las constantes que se utilizan para obtener las unidades usuales.} 
\vspace{0.5cm}
\begin{center}
\ra{1.45}
\begin{tabular}{lccrr}
\toprule
\ra{1.75}
\textbf{Parámetro integral} & \textbf{Símbolo}  & \textbf{Orden}(*) & \textbf{Unidades típicas} & \textbf{Factor} $C_{k}$ \\
\midrule
Concentración total    & $ N_{T}$  &0         & $[m^{-3}]       $ & 1    \\
Extinción óptica    & $ \Sigma$    &2         & $[km^{-1}]      $ & $1.57\cdot10^{9}$ \\
Contenido agua líquida  & W / LWC  &3         & $[gr m^{-3}]    $ & $\rho_{w}\pi/6$ \\
Atenuación  & A         & $\sigma_{ex}$       &  $[dB km^{-1}]$                & 4343 \\
Intensidad precipitación             & R         &3+$\delta$ & $[mm/h]         $ & $6\pi 10^{-4}\gamma$ \\
Energía cinética       & K         &3+$\delta^{3}$ &  $[J m^{-2} h^{-1}]$ & $\frac{3\pi}{10^{4}}\gamma^{3}$       \\
Reflectividad          & $Z$       &6         & $[m^{-3}mm^{6}] $ & 1 \\
\midrule
Diám. medio     & $\overline{D}$ &1/0       & $[mm]           $ & 1 \\
Diám. medio masa líquida  & $D_{m}$   &4/3       & $[mm]           $ & 1 \\
\bottomrule
\end{tabular}
\end{center}
\label{tablaPARAMETROSintegrales}
 \small (*) En el caso de cociente entre dos momentos se indica con los órdenes como k/l. El parámetro $\delta$ es el exponente de la relación $v(D)=\gamma D^{\delta}$. En el caso de la atenuación es necesario estimar la sección eficaz de extinción.
\normalsize
\end{table}%
\vspace*{1.45cm}

\section{Relaciones entre parámetros integrales de la DSD: la relación Z-R}

Uno de los objetivos más importantes del estudio de la DSD es la posibilidad de establecer relaciones cuantitativas entre parámetros integrales clave. El parámetro independiente de estas relaciones dependerá de la metodología de la medida. En el caso de métodos de teledetección, la magnitud medida es la reflectividad, y suele intentar estimarse la intensidad de precipitación (así como la atenuación) en función de la reflectividad\footnote{Esta relación es importante en los algoritmos que estiman la atenuación de la señal devuelta por la masa de aire que contiene la precipitación al atravesar el volumen que caracteriza al muestreo realizado por el radar.}. A parte de ello, en caso de estudiar pro\-ble\-mas de erosión, se suele intentar determinar la energía cinética de la distribución de tamaños de gota (y la variable medida será la reflectividad en teledetección o la intensidad de lluvia si se utilizan medidas a nivel del suelo). Todas estas relaciones adquieren la forma de relaciones de potencias, es decir:

\begin{equation}
\phi=a \psi^{b}
\end{equation}
la relaciones que más relevancia poseen para nuestro estudio son:
\begin{equation}
Z=a_{R} R^{b_{R}}
\end{equation}
así como la recíproca\footnote{En el caso de $Z=a_{R} R^{b_{R}}$ se simplificará a $Z=a R^{b}$ siempre que el contexto no implique ambigüedad con respecto a la relación R(Z).},
\begin{equation}
R=a_{Z} Z^{b_{Z}}
\end{equation}

\begin{figure}[h] 
\begin{center}
   \includegraphics[width=0.85\textwidth]{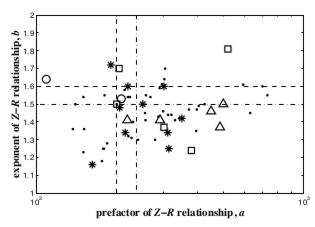}
\vspace{1.55cm}
   \caption[Coeficientes de la relación Z-R para varios eventos y experimentos diferentes. Fuente: \citep{uijlenhoet_pomeroy_2001_aa}]{\textbf{Coeficientes a y b de la relacion $Z=aR^{b}$ para eventos diferentes. Fuente: \citep{uijlenhoet_pomeroy_2001_aa}} Los eventos son clasificados según el tipo de precipitación que origina la DSD. \emph{Círculos}: Precipitación orográfica, \emph{Triángulos}: Tormentas, \emph{Estrellas}: Precipitación estratiforme (\emph{nimbostratus}), \emph{Cuadrados}: Chubascos, \emph{Puntos}: Precipitación no clasificada. La \emph{línea discontinua} representa la relación $Z=200 R^{1.6}$. La \emph{línea discontinua-punteada} representa la relación $Z=237 R^{1.5}$.}
\label{fig:ZR69relations1}
\end{center}
\end{figure}

La primera de ellas es una de las relaciones más estudiadas y de mayor interés. Se conoce como la relación Z-R. Una de las principales cuestiones de interés es la implicación que tiene la variabilidad de la DSD en la relación Z-R, y la relevancia de los dos aspectos de dicha variabilidad: el problema de muestreo estadístico y las propias variaciones naturales de la DSD. El nivel de incertidumbre en la estimación de R desde Z puede alcanzar el 40\%. Además, incluso conocida la relación Z-R para una localización concreta, e incluida una corrección debido al tipo de nubosidad que origina la precipitación la incertidumbre puede ser del 20 al 30\% \citep{SteinerKINETIC2000}. \\

La derivación de las relaciones R(Z) posee varias dificultades: cuantificar la variación natural de la DSD que puede condicionar la relación Z-R; la existencia de una variación de Z con la altura (este hecho aconseja el uso de medidas a nivel del suelo de la DSD ya que permiten estimar simultáneamente R y Z sobre la misma muestra de gotas); la estimación de la atenuación de la señal original del radar en el volumen muestreado, y la falta de una calibración apropiada de los componentes del radar, así como una cuantificación del ruido intrínseco de este. A este respecto se ha propuesto \citep{Ciach1999585} una metodología basada en diferenciar los errores propios del radar con los de muestreo a nivel del suelo. Para estos últimos serán relevantes los resultados que se han obtenido en esta tesis.\\
\label{sec:relacionesZR}


\begin{figure}[h] 
\begin{center}
   \includegraphics[width=0.80\textwidth]{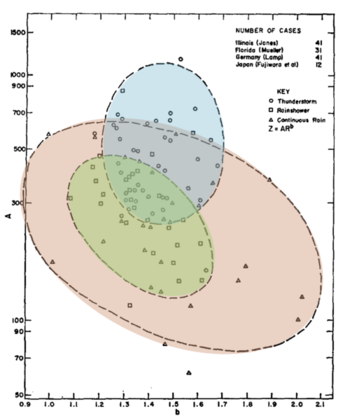}
\vspace{1.55cm}
   \caption[Coeficientes de la relación Z-R para varios eventos y experimentos diferentes. Fuente: \citep{Fujiwara1965ZRclasification}]{\textbf{Coeficientes a y b de la relación $Z=aR^{b}$ para eventos diferentes. Fuente: \citep{Fujiwara1965ZRclasification}. } Los eventos se clasifican según el tipo de precipitación que origina la DSD. \emph{Círculos}: Tormentas, \emph{Cuadrados}: Chubascos, \emph{Triángulos}: Precipitación continua. Se puede observar como la clasificación resulta difícil al haber zonas amplias de intersección para eventos de tipo diferente. Fuente: \citep{Fujiwara1965ZRclasification}.}
\label{fig:ZR69relations2}
\end{center}
\end{figure}

En el caso de los estudios microfísicos, estos indican que si se alcanza una DSD de equilibrio entre los procesos físicos que originan la distribución de tamaños de gota, esta dará lugar a relaciones lineales entre los diferentes momentos de la distribución \citep{1988List}. Esto equivale a afirmar que $b_{R}\simeq 1$, aunque $a_{R}$ puede variar notablemente. Sin embargo, para episodios típicos de pre\-ci\-pi\-ta\-ci\-ón, los valores de $b_{R}$ reales suelen ser manifiestamente distintos de la unidad, lo que implica la necesidad de determinar experimentalmente los parámetros $a_{R}$ y $b_{R}$ para diferentes eventos. Al respecto de la cuestión de qué relación Z-R elegir para un evento particular de características meteorológicas conocidas, hay registradas en la literatura multitud de propuestas. En la figura (\ref{fig:ZR69relations1}) se comparan 69 relaciones Z-R de distinta procedencia. Mientras que en la figura (\ref{fig:ZR69relations2}) se intenta una clasificación de la relaciones Z-R. Como consecuencia de la variedad, muchos autores han usado la media aritmética de los exponentes $b_{R}$ y la medida geométrica de los pre-factores $a_{R}$ de este conjunto amplio de valores, dando lugar a la relación:

\begin{equation}
Z=238\,R^{1.5}
\end{equation}

Similar a la predicción de \citep{MarshallPalmer1948}, que indica valores $a_{R}=237$ y $b_{R}=1.5$. Con todo, la relación más usada es $Z=200 R^{1.8}$, aunque en aplicaciones de predicción cuantitativa de la intensidad de precipitación se necesita calibrar cada radar, y los valores concretos de la relación utilizados en la práctica suelen referirse a estudios climáticos en cada localización geográfica, junto con clasificaciones entre el tipo de precipitación producida; principalmente distinciones entre precipitación estratiforme o convectiva.\\

En los últimos años el uso de radares polarimétricos permite medir no solo la reflectividad, sino además la reflectividad diferencial y la fase específica diferencial, lo que ha abierto nuevas pers\-pec\-ti\-vas en la medida remota de la precipitación. Uno de los objetivos es la estimación de la DSD a partir de las magnitudes anteriores, cuestión que en la mayoría de las metodologías requiere una propuesta funcional concreta para la DSD, usualmente caracterizada por la distribución gamma \citep{bringi_huang_etal_2002_aa}.\\

\section{Sumario/Summary}

This chapter summarizes the main facts and properties of the drop size distribution.

\begin{itemize}
\item The N(D) represents the concentration of drops within the volume unit and diameters between D and D+dD. The usual units are $[m^{-3}mm^{-1}]$.
\item There are two main kinds of measurements: those estimated from a volume of drops, and those estimated from the flow of drops across a specific surface. Knowledge of velocity of drops is needed to convert one measurement method into another. This is usually conceptualized in a power-law: $v(D)=\gamma D^{\delta}$.
\item The main physical information of the DSD is contained in the integral rainfall parameters. The more widely used are the liquid water content, the characteristic diameter or the reflectivity. The last one in the Rayleigh regime is written also as a moment of the DSD.
\item The rainfall expressed as a moment of the DSD implies the assumption of a functional relationship $v(D)$, usually written as a power-law.
\item One of the main objectives of the DSD studies is to build relationships between the integral rainfall parameters. The most widely analyzed $Z=a_{R}R^{b_{R}}$, is relevant for remote sensing estimation of rainfall using weather radars.
\end{itemize}

\renewcommand\chapterillustration{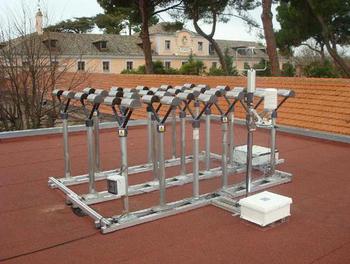}
\chapter{Instrumentos de medida de la DSD}
\label{sec:Instrumentacion}
De modo genérico, los instrumentos que miden la distribución de tamaños de lluvia se denominan \emph{disdrómetros}, acrónimo de DIStributiton DROps METERing (en ocasiones también se denominan espectro-pluviómetros). La base física para la medida de la DSD puede variar desde las estimaciones basadas en principios mecánicos hasta la medidas basadas en propiedades ópticas.\\

También es posible inferir la DSD mediante métodos de teledetección y en consecuencia, los aparatos diseñados específicamente con este fin igualmente se denominan disdrómetros. Sin embargo es importante tener en mente la diferencia entre estimar la DSD desde la medida de propiedades físicas de cada gota de un conjunto de ellas, y calcular la DSD \emph{infiriéndola} desde propiedades glo\-ba\-les de ese conjunto de gotas.\\

Los métodos pioneros \citep{MarshallPalmer1948, LawsParsons1943} de estimación de distribuciones de tamaños de gota permitieron establecer las primeras parametrizaciones de la DSD, aunque carecían de las posibilidades de automatización y transportabilidad de los instrumentos actuales. Los últimos modelos de disdrómetros están siendo diseñados principalmente para realizar estimaciones de la DSD basándose en propiedades ópticas de la precipitación.\\

En nuestro caso las medidas experimentales proceden de un red de disdrómetros en que los todos los instrumentos son Parsivel  (acrónimo de PARticula SIze VELocity) idénticos\footnote{Una análisis de la variabilidad a pequeña escala fue realizado por \citep{miriovsky_bradley_etal_2004_aa} utilizando una red heterogénea de instrumentos. Como resultado de dicha heterogeneidad fue necesaria una evaluación de las medidas de la DSD respecto de un instrumento de referencia, para intentar así hacer que las estimaciones de la distribución de tamaños de gota fueran lo menos sesgadas (\emph{biased}) posible. Este diseño enmascara la variación real de los parámetros integrales de la precipitación analizada con la distancia. Es pues preferible el uso de una red homogénea de instrumentos.}, manufacturados por OTT y calibrados de fábrica. Estos instrumentos han sido utilizados en diversos estudios previos, en particular en estudios comparativos en estaciones experimentales con diferentes tipos de instrumentos, comprobándose su fiabilidad adecuada \citep{Chapon200852,RainWetSnow2006parsivel,ChinosPARSIVELestudio,krajewski_kruger_etal_2006_aa}.\\ 

En este capítulo se describe el disdrómetro utilizado en los experimentos comparando sus propiedades con otros disdrómetros, con el fin de adquirir una perspectiva más completa. Pers\-pec\-ti\-va que se complementa indicando las diferencias generales con métodos de teledetección.\\

\section{Disdrómetros Ópticos}
\subsection{Principio físico}
\label{sec:opticalDISDRO}
\begin{figure}[h] 
\begin{center}
   \includegraphics[width=0.85\textwidth]{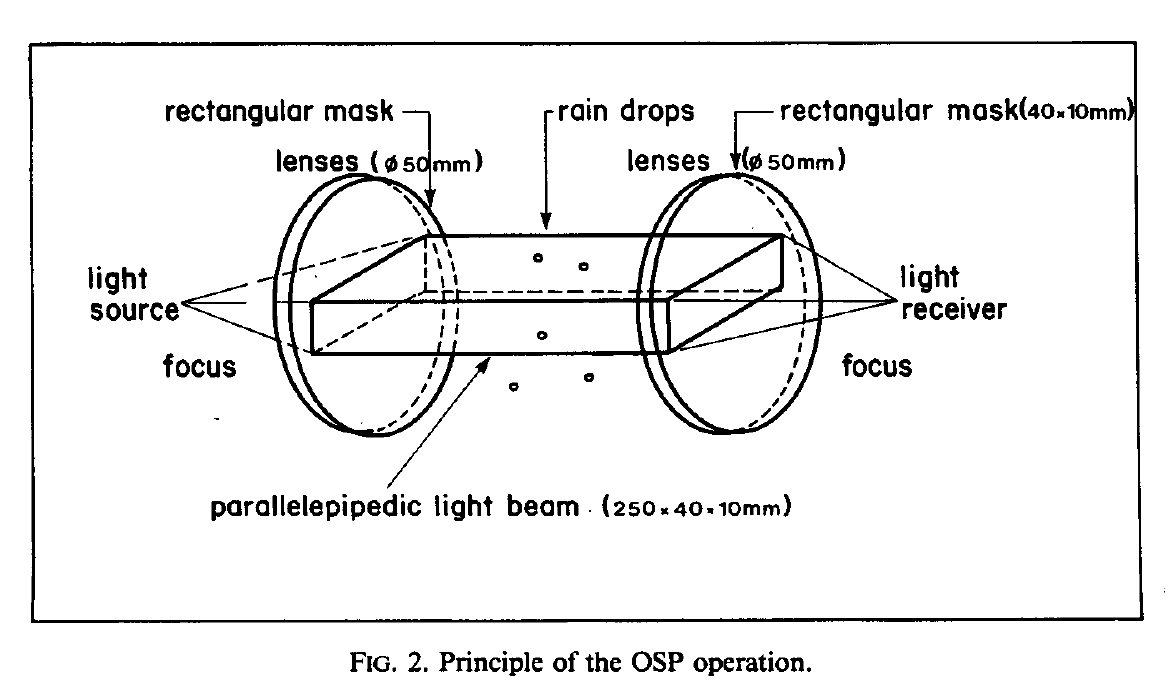} 
\vspace{0.75cm}
   \caption[Esquema de un disdrómetro óptico basado en la extinción óptica de un haz de luz monocromática.  Fuente: \citep{hauser_amayenc_etal_1984_aa}]{\textbf{Esquema de un disdrómetro óptico basado en la extinción óptica de un haz de luz monocromática.  Fuente: \citep{hauser_amayenc_etal_1984_aa} } Las dimensiones específicas en la figura se corresponden con el disdrómetro óptico tipo OSP o DBS. }
\label{figOptico1}
\end{center}
\vspace{1.25cm}
\end{figure}

El proceso físico en el que se fundamenta este instrumento es la extinción óptica que se produce en un haz de luz al ser atravesado por gotas de lluvia. El aparato posee una fuente y un receptor de luz monocromática. Las diferencias entre la señal emitida y la recibida son asociadas a propiedades de la precipitación. En la figura  (\ref{figOptico1}) se proporciona un esquema del sistema de medida.\\

En el caso de espectro-pluviómetros ópticos el haz de luz monocromática se localiza en el infrarrojo, usualmente en los 780 $nm$ o los 650 $nm$ y se produce en forma de un haz continuo de dimensiones fijas dx-dy-dz. En el caso del Parsivel, las dimensiones son dx=180 $mm$, dy=30 $mm$ y dz=1 $mm$, lo que define una superficie colectora de 54 $cm^{2}$ similar a la que poseen disdrómetros basados en otros principios de medición.\\ 

Este haz es recibido por un fotodiodo y produce una señal continua de 5 voltios si ningún objeto se interpone entre el emisor y el receptor. La homogeneidad de la señal depende, en parte, de la calidad con que es emitida la señal y de la eficiencia en la detección de los propios fotodiodos existiendo, por tanto, un umbral de ruido $\delta V$ por debajo del cual no es posible distinguir si las variaciones del voltaje se deben a la presencia de una gota o al propio ruido intrínseco a la señal.\\

Las variaciones con suficiente intensidad de esta señal se relacionan con la presencia de gotas. La señal puede variar en amplitud y hacerlo durante un intervalo determinado de tiempo, lo que permite inferir tanto el tamaño de gota como su velocidad (véase la figura \ref{figOptico2}). El instrumento está diseñado principalmente para la medida de gotas de lluvia, y no específicamente para la medida de granizo o nieve; por lo tanto los valores de diámetro y velocidad de la precipitación están sujetos a las siguientes hipótesis \citep{ParsivelSNOW}:

\begin{figure}[h] 
\begin{center}
   \includegraphics[width=0.90\textwidth]{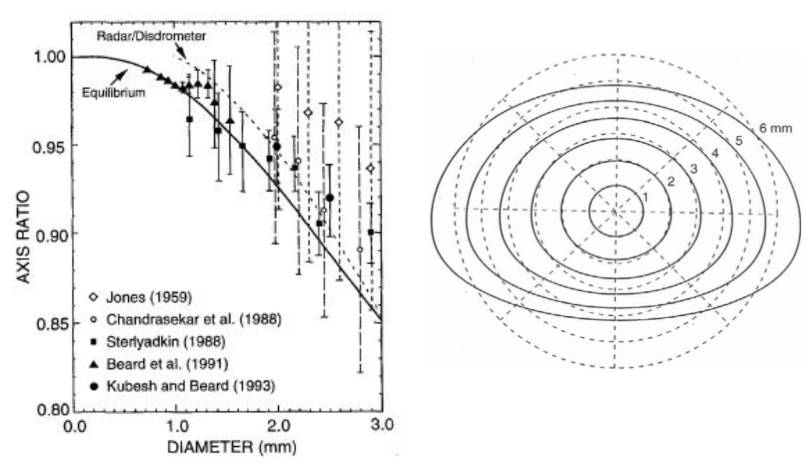}
\vspace{1cm}
   \caption[Esfericidad de las gotas en función del diámetro. Fuente: \citep{TokayBeard1996drops} y \citep{2008BAMSpolarizationRADARforGPM}]{\textbf{Esfericidad de las gotas en función del diámetro. Fuente: \citep{TokayBeard1996drops} y \citep{2008BAMSpolarizationRADARforGPM}} \textit{Izquierda}: Desviación de la hipótesis de esfericidad en función del diámetro. \textit{Derecha}: Se aprecia la desviación respecto de la esfericidad de modo visual, así como el crecimiento de la a-esfericidad con el diámetro. Es importante notar que esta figura muestra como una visualización lateral y una visión vertical del proceso de caída pueden ser diferentes, cuestión a tener en cuenta en radares terrestres \emph{vs.} radares en satélites.}
\label{fig:Esfericidad}
\end{center}
\end{figure}

\begin{itemize}
\item Las partículas tienen una forma aproximadamente esférica. Es posible que haya cierta asimetría entre los ejes vertical y horizontal. Véase al figura (\ref{fig:Esfericidad}). 
\item Las partículas caen con su eje de simetría alineado verticalmente (el eje mayor se supone está alineado horizontalmente).
\item Las esquinas del haz pueden tener problemas de medida (razón por la cual el Parsivel OTT incorpora dos fotodiodos extra).
\item Las partículas por medir deben poseer propiedades de extinción óptica respecto de la luz monocromática del haz similares a las de las gotas de lluvia.
\item La posible componente horizontal de la velocidad de caída no es tenida en cuenta. Como consecuencia, fuertes vientos locales o fenómenos de turbulencia pueden introducir errores en la estimación de la velocidad. La duración de una variación en la señal se relaciona con la velocidad terminal vertical de caída.
\item Únicamente se detecta una partícula a la vez, lo que tiene consecuencias en las medidas durante episodios de lluvia intensa.
\end{itemize}

%

\begin{figure}[h] 
\begin{center}
   \includegraphics[width=0.75\textwidth]{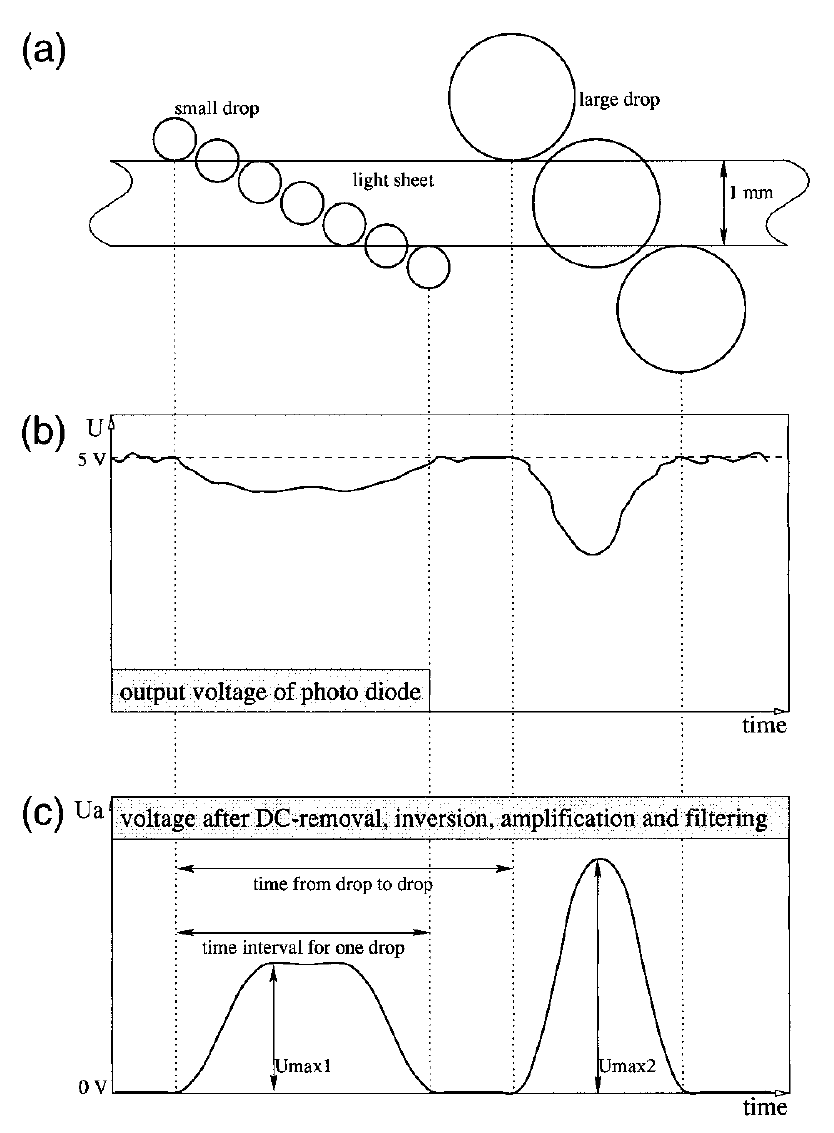}
\vspace{1.5cm}
   \caption[Metodología de medida de los disdrómetros ópticos. Fuente: \citep{loffler-mang_joss_2000_aa}]{\textbf{Metodología de medida de los disdrómetros ópticos. Fuente: \citep{loffler-mang_joss_2000_aa}.} La figura muestra la medida de gotas perfectamente esféricas aunque los algoritmos pueden ser adaptados a la situación más realista de gotas con dos ejes de simetría ligeramente diferentes. En el panel (b) se puede apreciar la presencia de ruido de fondo en la señal que limita el tamaño mínimo que es posible medir.}
\label{figOptico2}
\end{center}

\end{figure}

La metodología para obtener el valor del diámetro en el caso del Parsivel está basada en el área efectiva en vertical que intercepta el haz. Si la gota posee valores para los ejes de simetría a (eje mayor) y b (eje menor), el área interceptada viene dada por:

\begin{eqnarray}
S=\pi a b\qquad B<0.5 \, mm\\
S=2 a b \left [arcsin(c)-c\sqrt{1-c^{2}}\right] \qquad B<0.5\, mm
\end{eqnarray}
donde c=1/2b.\\

 Se asume un esferoide oblato, con lo que su volumen es $(4/3)\pi a^{2}b$ y el área interceptada se asocia con el volumen. La cantidad suministrada por el disdrómetro es el diámetro equivalente o efectivo, $D_{eff}$, que poseería una esfera de idéntico volumen. El tiempo de duración de la señal está relacionado con la velocidad de la gota, y se define como el tiempo entre dos valores sucesivos\footnote{En la figura (\ref{figOptico2}) aparece señalada la duración entre el inicio y el final del pulso pero suele ser más adecuado medirlo entre valores sucesivos de $U_{max}/2$ \citep{ParsivelSNOW}.} de $U_{max}/2$ en la señal (en la referencia \citep{hauser_amayenc_etal_1984_aa} se propone asociar la velocidad al área de la señal de voltaje recibida, que debe ser proporcional a $D^{2}/v$, con lo que dada la estimación previa del diámetro, es posible determinar la componente vertical de la velocidad a partir del cálculo del área encerrada por la señal).\\

Uno de los problemas que poseen los disdrómetros ópticos es la medida de gotas que atraviesan simultáneamente el haz. Los estudios realizados \citep{hauser_amayenc_etal_1984_aa} indican que el efecto principal es sobrevalorar el número de gotas grandes en lluvia intensa, e infravalorar el número de gotas pequeñas ($<1\,mm$). Un problema adicional que puede presentarse, es el choque y ruptura de gotas que colisionan en la estructura que sustenta el disdrómetro; estas gotas poseen velocidades terminales diferentes de las usuales (en principio notablemente menores) con lo que en teoría es posible eliminar parte de ellas asumiendo la forma funcional $v(D)$ junto con una tolerancia estadística sobre esta relación, e incorporar esta hipótesis en forma de filtro sobre el histograma-matriz que se obtiene directamente del instrumento \citep{hauser_amayenc_etal_1984_aa}. \\ 

\begin{figure}[h] 
\begin{center}
   \includegraphics[width=0.80\textwidth]{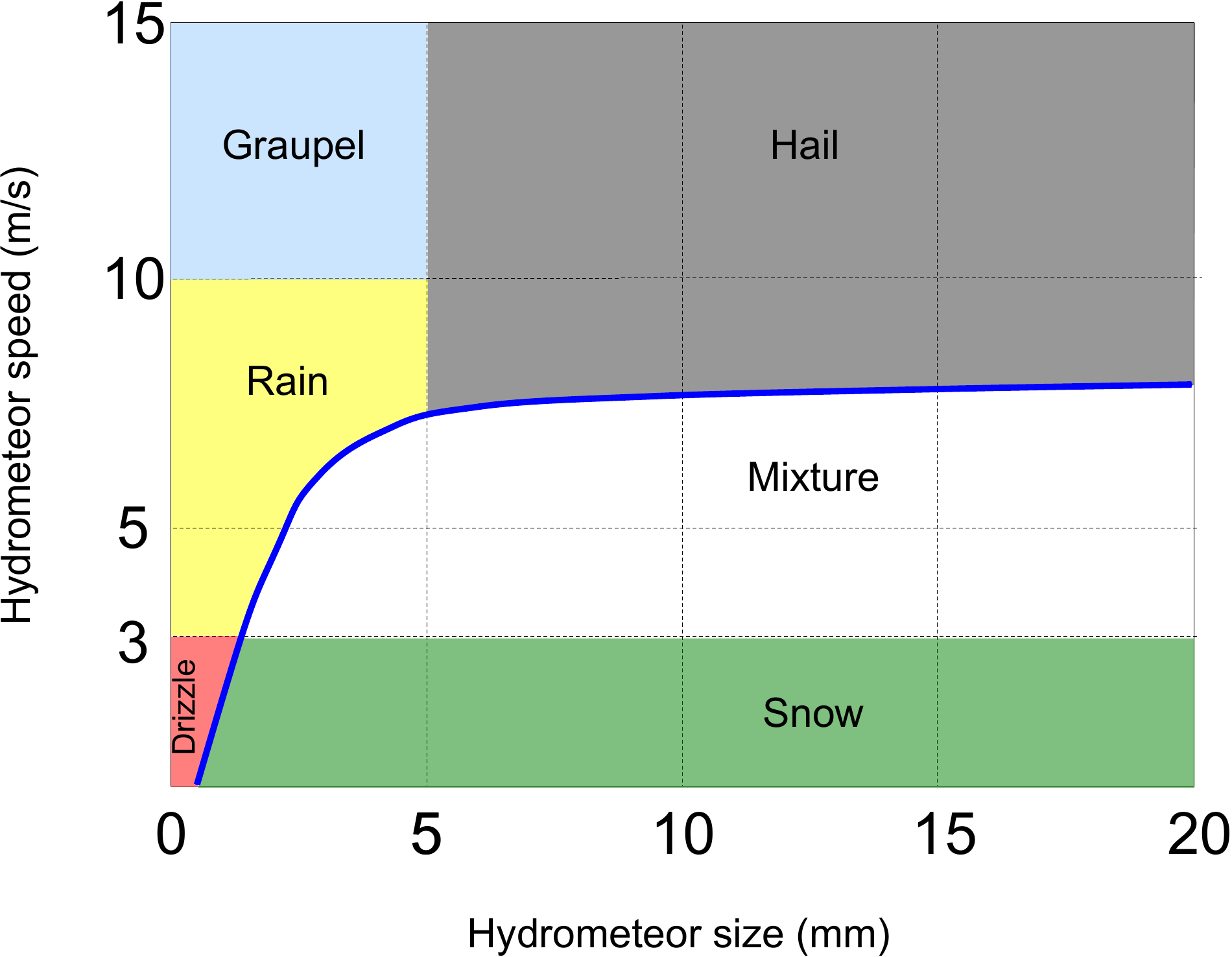}
\vspace{0.3cm}
   \caption[Diferenciación del tipo precipitación mediante diagrama $v-D$. Fuente: \citep{loffler-mang_joss_2000_aa}. ]{\textbf{Diferenciación del tipo de precipitación mediante la curva $v(D)$. Fuente: \citep{loffler-mang_joss_2000_aa}}. Se puede apreciar la viabilidad para estimar si la precipitación corresponde a diferentes agregados, además de poder ser clasificada según el tamaño (llovizna o lluvia).}
\label{figOptico3}
\end{center}
\end{figure}

El instrumento acumula la medida del número de gotas por intervalo de diámetro efectivo y velocidad durante un intervalo de tiempo $\delta t$, que en el caso del Parsivel OTT es de 30 o 60 $s$, devolviendo una matriz bidimensional $n(D,v)$ en forma de histograma para unos intervalos de clase fijos (véase tabla \ref{TablaBINSparsivel}). La construcción del diagrama $v(D)$ permite discriminar \texttwelveudash véase la figura (\ref{figOptico3})\texttwelveudash\, el tipo de agregado: \emph{graupel}, granizo, lluvia, llovizna, nieve o mezcla. Siendo por ello un instrumento interesante desde el punto de vista del registro del tipo de precipitación aunque, desde el punto de vista de la estimación de la distribución de tamaños de gota, sea más fiable para gotas de lluvia líquida.\\


\begin{table}[h]
\ra{1.0}
\caption[Comparativa entre diferentes disdrómetros ópticos]{\textbf{Comparativa de diferentes disdrómetros ópticos basados en la extinción de un haz de luz monocromático.} En el caso del Parsivel se indica entre paréntesis los intervalos de clase relevantes para la medida de precipitación líquida. El disdrómetro OSP está descrito en detalle en la referencia \citep{hauser_amayenc_etal_1984_aa}. En el caso del disdrómetro Thies, el área del sensor viene dada por: (228 $mm$ x 20 $mm$) x 0.75 $mm$. }
\vspace{0.45cm}
\begin{center}
\normalsize
\ra{1.00}
\begin{tabular}{lccc}
\toprule
& \textbf{Parsivel OTT} &  \textbf{Thies} &   \textbf{OSP} \\
\midrule
Área del sensor   & 180 x 30 $mm^{2}$ & 228 x 20 $mm^{2}$ & 250 x 40 $mm^{2}$ \\
Altura de haz     & 1.00 $mm$   & 0.75 $mm$     & 1.00 $mm$ \\ 
Haz (long. onda)  & 650 $nm$    & 780 $nm$      & 780 $nm$ \\
Intervalos (D)    & 32 (23)     & 22            & 16\\
Intervalos ($v$)  & 32 (23)     & 20            & 16\\
Tiempo muestreo   & 30 s / 60 s   & 60 s          & 30 s\\ 

\bottomrule
\end{tabular}
\end{center}
\label{TableComparativaOPTICOS}
\end{table}

\section{JWD - POSS - 2DVD - DBS}

Joss and Waldovogel (1967)  fueron los primeros en introducir un instrumento capaz de medir la distribución de tamaños de gota mediante principios mecánicos. Este instrumento, hoy día, se ha convertido en una referencia tras llevar operativo más de tres décadas. Multitud de experimentos y parametrizaciones de la DSD se han basado en medidas tomadas mediante este dispositivo \citep{TokayShort1996,Tokay2003disdrometerKAMP,Tokay2010smallscaleDSD,chandrasekar_gori_1991_aa}.\\

El fundamento físico por el que se determinan las características físicas de la gota es la transferencia de momento en la dirección vertical entre la gota y la superficie sobre la que impacta. El movimiento generado por el impacto es transformado en una corriente eléctrica susceptible de ser medida, mientras que la superficie está diseñada para recuperar rápidamente su posición original y detectar el siguiente impacto. La capa externa del detector es convexa e hidrófoba con objeto de facilitar el secado inmediato de la superficie de detección.\\

La determinación del tamaño de gota se basa en asumir una relación funcional entre el diámetro efectivo de la gota y la velocidad terminal que posee. En consecuencia el instrumento, al contrario que el disdrómetro óptico, no es capaz de medir simultáneamente velocidad y tamaño de gota\footnote{La relación $v(D)$ se considera que posee un carácter estocástico, es decir, la relación entre la velocidad terminal y el diámetro sigue aproximadamente una relación funcional, pero se esperan desviaciones estadísticas. Además en un estudio reciente \citep{Montero-Martinez:2009kx} se ha mostrado que las gotas de menor tamaño pueden caer con más velocidad de la esperada por sus diámetros, lo que implica que hay que realizar correcciones sobre las medidas aportadas por los disdrómetros JWD \citep{Leijnse2010}.}.

\begin{table}[H]
\ra{1.0}
\caption[Intervalos de clase (\textit{bins}) del Parsivel OTT]{\textbf{Intervalos de clase (\textit{bins}) del Parsivel OTT}. \textit{Rojo Oscuro}: Valores desestimados debido al ruido de fondo. \textit{Rojo}: diámetros para los cuales puede haber efectos de infra-valoración. \textit{Gris}: Valores de diámetros fuera del rango de gotas líquidas, usados principalmente para medidas de nieve/granizo.}
\vspace{0.05cm}
\begin{center}
\normalsize
\ra{1.05}
\begin{tabular}{lcccc|lcccc}
\toprule
\textbf{Clase} & \textbf{D} & $\mathbf{\Delta D}$ &  \textbf{v} & $\mathbf{\Delta v}$ & \textbf{Clase} &  \textbf{D} &  $\mathbf{\Delta D}$ & \textbf{v} & $\mathbf{\Delta v}$\\
\midrule
1   & \textcolor{darkred}{0.062}   & 0.125   & 0.05 & 0.10 & 17  & 3.250   & 0.50    & 2.60 & 0.40\\
2   & \textcolor{darkred}{0.187}   & 0.125   & 0.15 & 0.10 & 18  & 3.750   & 0.50    & 3.00 & 0.40\\
3   & \textcolor{red}{0.312}   & 0.125   & 0.25 & 0.10                  & 19  & 4.250   & 0.50    & 3.40 & 0.40\\
4   & \textcolor{red}{0.437}   & 0.125   & 0.35 & 0.10                  & 20  & 4.750   & 0.50    & 3.80 & 0.40\\
5   & \textcolor{red}{0.562}   & 0.125   & 0.45 & 0.10                  & 21  & 5.500   & 1.00    & 4.40 & 0.60\\
6   & \textcolor{red}{0.687}   & 0.125   & 0.55 & 0.10                  & 22  & 6.500   & 1.00    & 5.20 & 0.60\\
7   & 0.812   & 0.125   & 0.65 & 0.10                  & 23  & 7.500   & 1.00    & 6.00 & 0.80\\
8   & 0.937   & 0.125   & 0.75 & 0.10                  & 24  & \textcolor{gray}{8.500}   & 1.00    & 6.80 & 0.80\\
9   & 1.062   & 0.125   & 0.85 & 0.10                  & 25  & \textcolor{gray}{9.500}   & 1.00    & 7.60 & 0.80\\
10  & 1.187   & 0.125   & 0.95 & 0.10                  & 26  & \textcolor{gray}{11.00}   & 2.00    & 8.80 & 1.20\\
11  & 1.375   & 0.250   & 1.10 & 0.15                  & 27  & \textcolor{gray}{13.00}   & 2.00    & 10.4 & 1.60\\
12  & 1.625   & 0.250   & 1.30 & 0.20                  & 28  & \textcolor{gray}{15.00}   & 2.00    & 12.0 & 1.60\\
13  & 1.875   & 0.250   & 1.50 & 0.20                  & 29  & \textcolor{gray}{17.00}   & 2.00    & 13.6 & 1.60\\
14  & 2.125   & 0.250   & 1.70 & 0.20                  & 30  & \textcolor{gray}{19.00}   & 2.00    & 15.2 & 1.60\\
15  & 2.375   & 0.250   & 1.90 & 0.20                  & 31  & \textcolor{gray}{21.50}   & 3.00    & 17.6 & 2.40\\
16  & 2.750   & 0.50    & 2.20 & 0.30                  & 32  & \textcolor{gray}{24.50}   & 3.00    & 20.8 & 3.2\\
\bottomrule
\end{tabular}
\end{center}
\label{TablaBINSparsivel}
\end{table}

 \begin{figure}[H] 
 \begin{center}
    \includegraphics[width=0.95\textwidth]{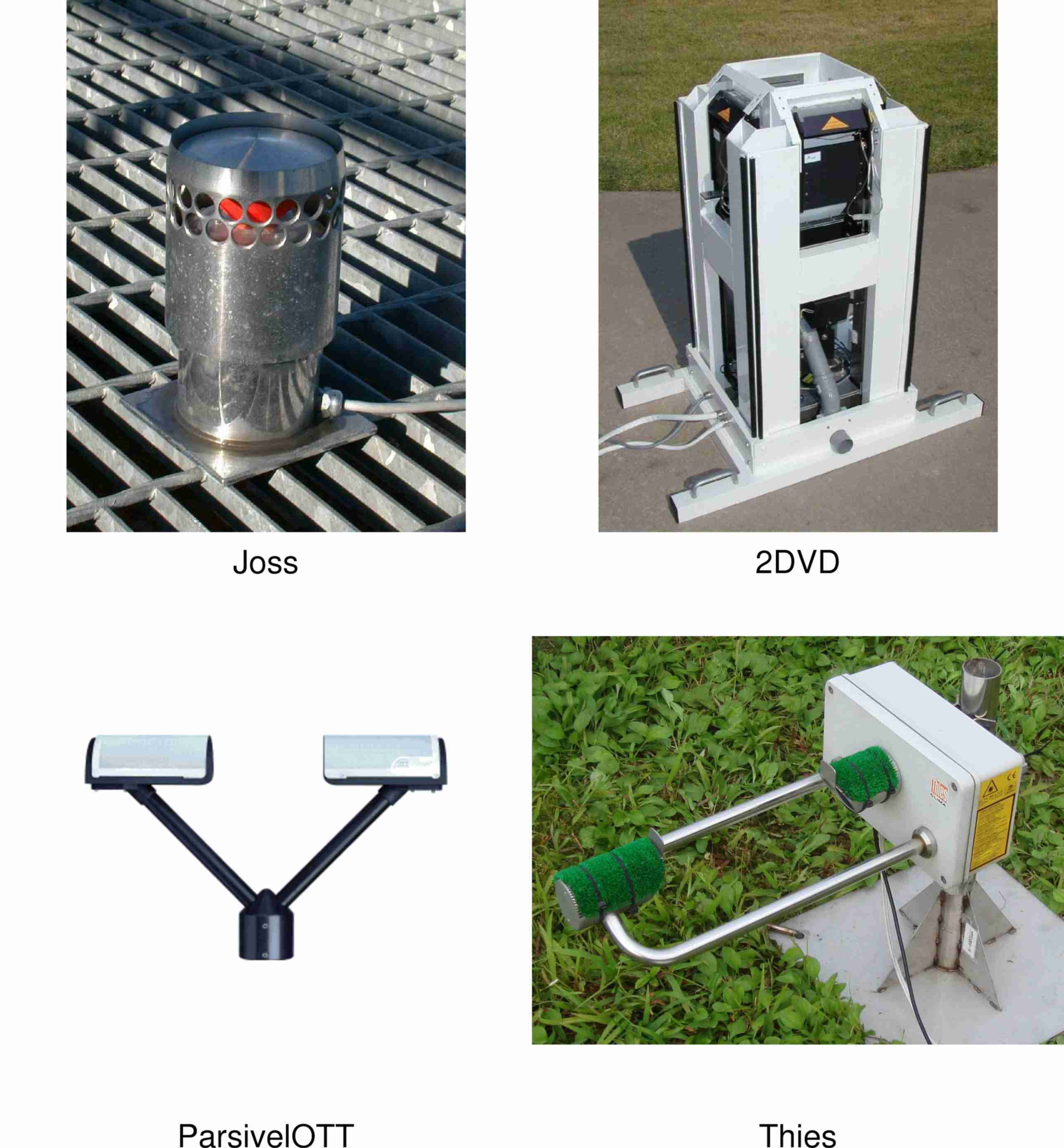}
\vspace{1.95cm}
    \caption[Fotos del disdrómetro Joss-Waldovogel y los disdrómetros ópticos Parsivel OTT, Thies y 2DVD.]{\textbf{Fotos del disdrómetro Joss-Waldovogel y los disdrómetros ópticos Parsivel OTT, Thies y 2DVD.} \textit{Arriba-Izquierda:} Disdrómetro de impacto JWD. \textit{Arriba-Derecha:} Disdrómetro óptico 2DVD. \textit{Abajo-Izquierda:} Disdrómetro Parsivel OTT. \textit{Abajo-Derecha:} Disdrómetro Thies. Se aprecian las diferentes características de cada instrumento. El más avanzado, el 2DVD, tiene el problema de su alto precio, mientras que el disdrómetro JWD, más portable, no permite medir simultáneamente D y $v$. Los disdrómetros ópticos Parsivel OTT y Thies, que se basan en el mismo fundamento físico, poseen un diseño diferente. En el caso del disdrómetro Thies, su mayor asimetría puede jugar un papel en casos de viento intenso.} 
 \end{center}
 \label{fig:JWDfoto}
\vspace{1.75cm}
 \end{figure}

Por otra parte existe un tiempo de retardo en la respuesta del aparato (\textit{Dead Time Effect}) para ser capaz de medir una gota tras un impacto previo. Este tiempo de respuesta depende de los tamaños respectivos entre dos gotas consecutivas. Para gotas de tamaño similar el tiempo es del orden de 1 $ms$, mientras que, si la segunda gota es menor que la primera, este tiempo de respuesta se incrementa. Como consecuencia el instrumento infravalora los casos de lluvia intensa. Para solucionar este problema el fabricante incorpora una matriz de corrección \citep{sheppard_joe_1994_aa}; sin embargo, otros autores prefieren no utilizarla debido a que su metodología es introducir un factor de corrección multiplicativo sobre el número de gotas registrado en cada intervalo de diámetros y esto implica que la corrección no es eficaz para valores del diámetro en que no haya un registro previo de gotas, lo que lleva a dificultades en la estimación de la DSD para estos casos \citep{TokayShort1996}.\\


Además, el ruido acústico interfiere en las medidas del JWD por lo que necesita ser situado en zonas carentes de fuentes de ruido ambiental intenso. En las situaciones típicas el ruido de fondo enmascara el impacto de gotas menores de 0.3 $mm$ de diámetro, mientras que la forma funcional típica de la $v(D)$ implica que las medidas del diámetro de gotas mayores posean mayor incertidumbre, por lo que el JWD acumula todas las medidas de diámetros mayores de 5 $mm$ en el mismo intervalo de clase de 5.1 $mm$. Desde el punto de vista práctico las gotas de mayores tamaños son escasas, aunque eventos en que su presencia pueda ser relevante darán lugar a estimaciones no muy precisas de la reflectividad. Por último, el propio principio físico en que se basa el instrumento imposibilita la medida de precipitación no líquida.\\

Otros dos disdrómetros ópticos usados en estudios comparativos con Parsivel son el 2DVD (\emph{two-dimensional video disdrometer}) y el DBS/OSP. El primero es un disdrómetro óptico avanzado que permite medir no solo el tamaño y la velocidad terminal de las gotas, sino también su forma, lo que es relevante en teledetección.\\

El OSP es otro dispositivo basado en la ocultación de un haz de luz infrarroja desarrollado en Francia \citep{hauser_amayenc_etal_1984_aa}. Para una comparativa de diversos instrumentos basados en el mismo principio véase la tabla (\ref{TableComparativaOPTICOS}). En la tabla (\ref{TableComparativaDISDROS}) se recoge una comparación del Parsivel con instrumentos basados en otras formas de medida.\\

El disdrómetro POSS, acrónimo de \emph{Precipitation Occurence Sensor System}, se denomina disdrómetro por estar destinado a medir la DSD, pero es un radar que opera en la banda X (9.4 $GHz$), y que por tanto posee diferencias notables respecto de los disdrómetros anteriores, ya que parte de hipótesis previas sobre las propiedades de las gotas para poder inferir la DSD\footnote{El instrumento se orienta hacia arriba y detecta la señal devuelta por hidrometeoros presentes en el volumen de muestreo. La medida es esencialmente el espectro de densidad Doppler, que a frecuencia f está relacionado con la N(D) por $S(f)=\int N(D)v(D)s(f,D)dD$. Invirtiendo la expresión anterior, es posible obtener N(D), que es estimada en 34 intervalos de diámetros diferentes \citep{sheppard_joe_1994_aa}. Al igual que el caso del JWD la expresión anterior implica asumir una relación $v(D)$ que suele ser también la dada por (\ref{eqn:AtlasVDequation}). }.

\begin{table}[h!]
\ra{1.5}
\caption[Comparativa entre disdrómetros basados en diferentes principios físicos]{\textbf{Comparativa entre disdrómetros basados en diferentes procesos físicos}. El volumen de muestreo depende del diámetro concreto de la partícula detectada, ya que gotas mayores se espera posean velocidades mayores. El valor indicado es para un diámetro de referencia de $D\simeq 1\, mm$. En el caso del POSS, el volumen muestreado varía más con el diámetro, de manera que indicamos su estimación entre $D\simeq 1\, mm$ y $D\simeq 3\, mm$.}
\vspace{0.5cm}
\begin{center}
\small
\ra{1.50}
\begin{tabular}{lcccc}
\toprule
&  \textbf{JWD} &  \textbf{Parsivel OTT} &  \textbf{POSS} & \textbf{2DVD} \\
\midrule
Superficie del sensor   & 50 $cm^{2}$              & 54 $cm^{2}$                & -                               & 100 $cm^{2}$\\
Tipo            & Impacto             & Extinción óptica     & Radar banda X                    & Haz de luz dual \\
Volumen muestreado (*)   & $10^{4} cm^{3}$   & $10^{4} cm^{3}$     & $10^{6}$ a $10^{8} cm^{3}$   & $10^{4} cm^{3}$\\
N. intervalos de clase (D)        & 20                  & 32 (23 para gotas) & 34                              & 40 \\
$D_{min}$ [mm]       & 0.353               & 0.321                 & 0.34 &  0.100 \\
$D_{max}$ [mm]      & 5.140               & 7.725                 & 5.32 &  8.100 \\

\bottomrule

\end{tabular}
\end{center}
\begin{footnotesize}  (*) Volumen muestreado por segundo en orden de magnitud. Respecto del valor concreto del volumen de muestreo para el disdrómetro 2DVD es el doble que en el caso de JWD.     \end{footnotesize}
\label{TableComparativaDISDROS}
\end{table}

\section{Cálculo de la DSD}

En el caso del disdrómetro óptico Parsivel, el cálculo de la DSD se realiza a partir de las medidas tanto del diámetro como de la velocidad\footnote{Un ejemplo de análisis de las relaciones $v(D)$ sobre la base de las medidas realizadas con el disdrómetro óptico Parsivel puede encontrarse en \citep{ChinosPARSIVELestudio}.}. Por tanto, recurriendo a la definición de la DSD dada en el capítulo anterior tenemos que:

\begin{equation}
N(D_{j})=\sum_{i}\frac{n(D_{j},v_{i})}{S\, \delta t\, v_{i}\, \delta D_{j}}
\end{equation}
donde $N(D_{j})$ es el valor experimental de la DSD para $D_{j}$, mientras que la suma se hace en el número de gotas que atraviesan el área de medida S en un tiempo $\delta t$ con velocidad $v_{i}$.\\

Internamente, el instrumento Parsivel proporciona la salida de la DSD (véase \citep{krajewski_kruger_etal_2006_aa} para un ejemplo de uso directo de la N(D) proporcionada por el dispositivo) así como de algunos parámetros integrales, como la intensidad de lluvia y la reflectividad. Sin embargo es importante hacer notar que todos los resultados expuestos en esta tesis parten, bien del cálculo de la DSD indicado arriba, bien del cálculo de la intensidad de precipitación a partir de la suma del volumen de las gotas como:

\begin{equation}
R=\sum_{i}\frac{4\pi (D_{i}/2)^{3}}{3 S }n(D_{i})
\label{eqn:calculoRexperimental}
\end{equation}

donde S es el área de medida del disdrómetro, y la suma se realiza en los valores de los intervalos de clase. En el caso estimar la intensidad de precipitación en mm/h tendremos que $n(D_{i})$ es el número de gotas de diámetro $D_{i}$ detectadas en 1 hora. Sin embargo lo usual es determinar la intensidad de precipitación a resoluciones temporales mayores manteniendo las unidades mm/h. Por tanto habremos de introducir un factor corrector tal que si $n(D_{i})$ representa el número de gotas detectadas en $\delta t$ segundos, habremos de multiplicar $n(D_{i})$ por $3600/\delta t$. Una medida alternativa se basa en utilizar la relación $v(D)$ y la expresión dada en el capítulo anterior (\ref{eqn:AtlasVDequationPOWERLAW}). Sin embargo, es preferible utilizar la ecuación (\ref{eqn:calculoRexperimental}) que no necesita hipótesis extra. A lo largo de los experimento llevados a cabo en esta tesis el valor de $\delta t$ es de 60 segundos.\\

En el caso de la reflectividad la expresión directa en $[m^{-3}mm^{-1}]$ es:

\begin{equation}
Z=\sum_{j}\sum_{i}\frac{n(D_{j},v_{i})D_{j}^{6}}{S\, \delta t \, v_{i}}
\end{equation}

Se han propuesto algunas correcciones para el cálculo de la N(D). En la tesis se evalúan los resultados tanto introduciendo estas posibles correcciones como con los datos directos de la matriz $n(D,v)$ sin preprocesado\footnote{En el capítulo \S\ref{chap:preprocesadoTOKAY} se detalla el método concreto de preprocesado de los datos. Desde el punto de vista del contraste de hipótesis sobre la variabilidad espacial, la corrección de posibles errores sistemáticos a lo largo de la red no es crucial, mientras que para obtener datos numéricos acerca de dicha variabilidad para incorporarla a otros estudios puede ser conveniente disponer de datos con y sin preprocesado.}.\\

En el caso de medidas utilizando JWD, no disponemos de medidas de la velocidad, por lo que la expresión anterior debe hacer uso de $v(D)$. Sobre la metodología concreta con este tipo de disdrómetro puede consultarse la referencia \citep{tokay_kruger_etal_2001_aa}, mientras que detalles concretos sobre las medidas con DBS/OSP pueden encontrarse en \citep{krajewski_kruger_etal_2006_aa}.

\begin{table}[h!]
\ra{1.5}
\vspace{0.55cm}
\caption[Comparativa entre diferentes sistemas de medida]{\textbf{Comparativa entre diferentes sistemas de medida}. Se comparan las diferentes ca\-rac\-te\-rís\-ti\-cas de los sistemas de medida de la precipitación. Se observa como las propiedades disdrométricas permiten investigar soluciones para las limitaciones de las medidas de teledetección.}
\vspace{0.15cm}
\begin{center}
\small
\ra{1.50}
\begin{tabular}{lcccc}
\toprule
                       & \textbf{Radar terrestre} &  \textbf{Satélite} &  \textbf{Disdrómetro} & \textbf{Pluviómetro} \\
 \midrule
 Resolución temporal    & Tiempo de escaneo       & Periodo de revisitado         & 1 min                      &  5 min\\
 Resolución espacial    & Pixel (1-5 $km^{2}$)    & IFOV (5-10 $km^{2}$)    & Puntual                  & Puntual \\
 Cobertura              & Terrestre y costa       & Global                       & Terrestre                       & Terrestre\\
 Limitaciones           & Variabilidad            & Variabilidad           & Muestreo                        & Muestreo/Viento \\
\bottomrule
\end{tabular}
\end{center}
\label{tabla:sistemasMedida}
\end{table}

\section{Comparación con otros métodos}

Como ya se indicó en \S\ref{sec:defDSD}, existen diferentes métodos de medida de la DSD. Las medidas disdrométricas se clasifican como medidas sobre un flujo de precipitación sobre superficies colectoras pequeñas, lo que da lugar a volúmenes de muestreo que a escala terrestre suelen considerarse puntuales. Los pluviómetros comparten esta propiedad, pero no realizan mediciones de la DSD sino solo de la intensidad de precipitación. Las medidas de teledetección, en cambio, se caracterizan por establecer medidas de la precipitación sobre masas de aire amplias, lo que hace que su principal reto no sean problemas de muestreo, sino de variabilidad natural de la DSD dentro del volumen de medida. En esta situación las medidas mediante instrumentos puntuales son complementarias con las anteriores ya que permiten estudiar la variabilidad natural de la DSD que escapa a las medidas de teledetección. Un resumen de las características de cada sistema de medida se da en la tabla (\ref{tabla:sistemasMedida}).

\section{Sumario/Summary}

The main properties of disdrometrics measurements explained in this chapter are:

\begin{itemize}

\item There are two kinds of measurements of precipitation: pointwise and areal. The first one has the problem of sampling, the second one has the issue of spatial and temporal variability of the rain fields.
\item The pointwise instruments that measure the Drop Size Distribution are called Disdrometers or Spectro-Pluviometers. There are several kinds of them, depending on the kind of physical principle in which the measuring process is based on.
\item The most widely used is the Joss-Waldovogel disdrometer (JWD) based on impact transference of moments from drops to the device. However it has a limitation on assuming an specific $v(D)$ relation to retrieve the DSD.
\item The optical disdrometers have been developed in the last years, and they provide a simple and effective method to measure $n(D,v)$ that relies in the optical extinction of drops crossing a continuous beam of monochromatic light.
\item While the optical disdrometers, like every device, have some shortcomings, they provide the $n(D,v)$ with good accuracy, which allows the development of advanced networks of measurement stations.  
\item The different kinds of instruments, disdrometers and radars, are complementary. The first can be used to estimate the natural variations of the DSD inside a radar pixel. Then the correct characterization of DSD variability allow us to improve the algorithms used for quantitative estimations of rainfall parameters from radars measurements.
\end{itemize}







\renewcommand\chapterillustration{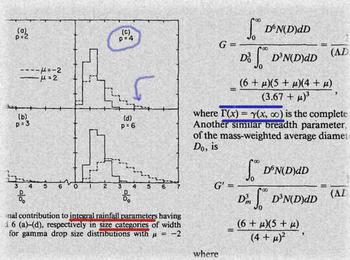}
\chapter{Modelizaciones de la DSD}
\label{sec:modelizacionesDSDchap}

La primera cuestión que emerge en la modelización de la DSD es si las distribuciones de tamaños de gota obedecen a formas funcionales concretas y, si tal es el caso, qué relación puede existir entre los parámetros libres de dichas formas funcionales y las propiedades concretas de los diferentes episodios de precipitación. Esta es de hecho una de las cuestiones principales que ha sido afrontada en la literatura. Disponer de modelizaciones de la DSD es importante, no solo en la construcción de algoritmos que obtengan esta desde parámetros integrales medidos experimentalmente sino, además, para su inclusión en las parametrizaciones de la precipitación y de la microfísica presentes en los modelos de predicción numérica de tiempo (NWPM) y en modelos regionales de clima (RCM). Además, las modelizaciones poseen relevancia en los modelos físicos de nubes cuando no se afronta el problema completo desde la dinámica de fluidos, sino desde una parametrización de los procesos microfísicos ya sea de manera total o parcial \citep{StrakaBOOK}.  

\section{Modelizaciones de la DSD}

Las modelizaciones de la DSD intentan tanto describir y conceptualizar los resultados de las medidas disdrométricas como realizar predicciones a partir de los modelos. De modo complementario, los procesos microfísicos pueden ser visualizados como transformaciones de dichos modelos. A continuación se describen los modelos de DSD más utilizados.

\subsection{Distribución exponencial}
Los resultados aportados por las teorías microfísicas\footnote{Entendiendo por teorías microfísicas aquellas que indagan en el origen de la precipitación desde la formación de las gotas que conforman las nubes.} indican que las gotas de mayor diámetro son menos frecuentes que las gotas de menor diámetro. Una representación paramétrica que tenga en cuenta este hecho, y que constituya una representación razonable de las medidas del espectro a nivel del suelo, es la forma funcional de la distribución exponencial:
\begin{equation}
N(D)=N_{0}e^{-\Lambda D}
\end{equation}

donde los valores de $N_{0}$ $[m^{-3}mm^{-1}]$  y $\Lambda\,[mm^{-1}]$ tienen que ser ajustados desde los datos experimentales. La propuesta más utilizada, conocida como Marshall-Palmer, asume la existencia de una dependencia funcional de estos parámetros con la intensidad de la precipitación. Es decir, que la intensidad de lluvia del evento condiciona el espectro de tamaños de gota. Por tanto, el proceso de análisis implica la clasificación en intervalos de acuerdo con los valores de R, y ajustes funcionales que dan lugar a los dos parámetros de la DSD. Los resultados obtenidos por \citep{MarshallPalmer1948} muestran pocas dependencias:  $\Lambda=41\,R^{-0.21}$, $N_{0}\simeq8000\,m^{-3}cm^{-1}$. Aunque, \citep{LawsParsons1943} obtuvieron resultados compatibles con $\Lambda=38\,R^{-0.20}$, $N_{0}\simeq5100\, R^{-0.03}\,m^{-3}cm^{-1}$. Más adelante, \citep{SekhonSrivastava1971} propusieron $N_{0}=0.07\,R^{0.37}\,cm^{-4}$ y $\Lambda=38\,R^{-0.14}\,cm^{-1}$.\\

\begin{figure}[h] 
\begin{center}
   \includegraphics[width=1.05\textwidth]{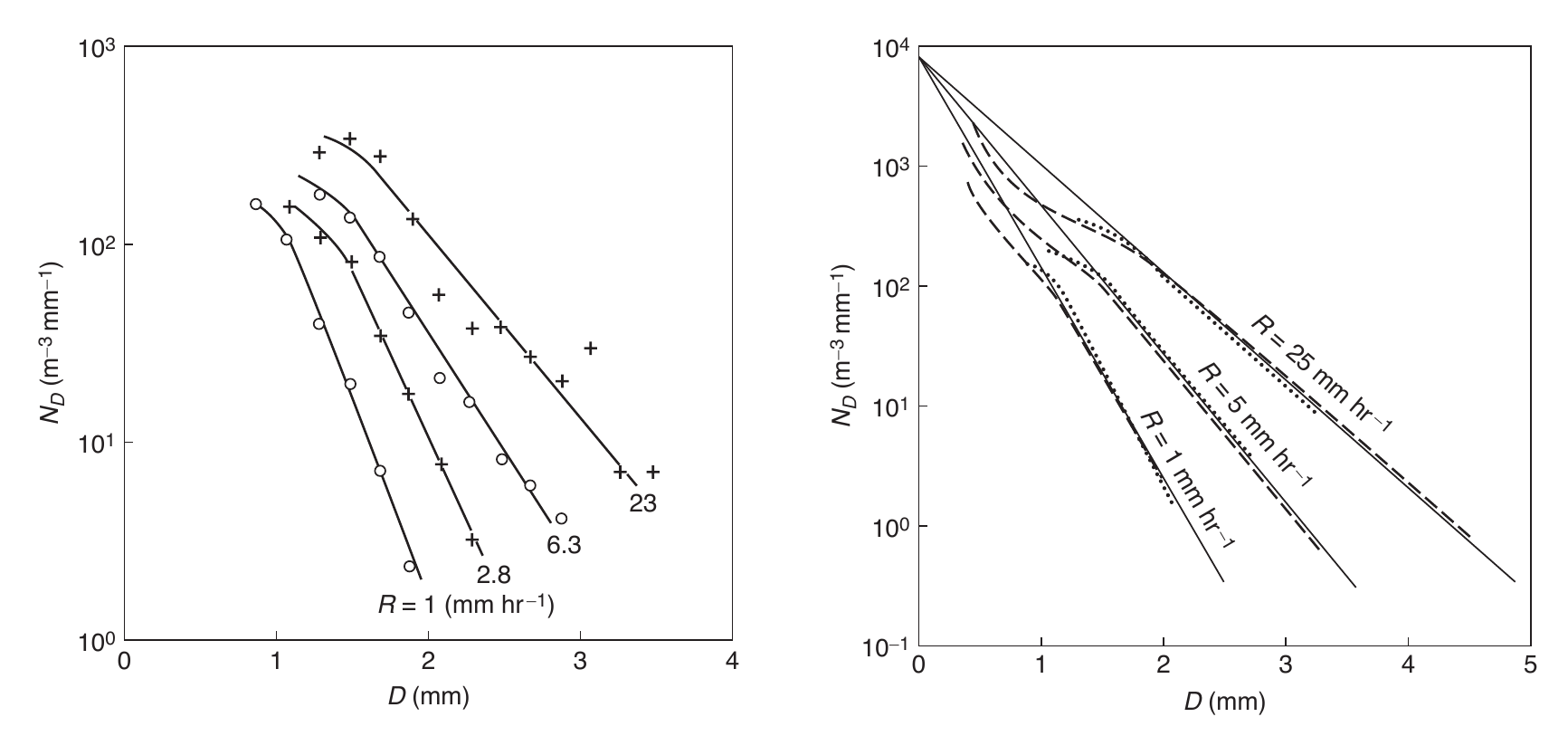}
\end{center}
\vspace{0.25cm}
   \caption[Modelización de la  DSD usando una distribución exponencial. Fuente: \citep{StrakaBOOK}.]{\textbf{Modelizacion de la DSD usando una distribución exponencial. Fuente: \citep{StrakaBOOK}}. En la parte \emph{izquierda} se observa la idea de Marshall-Palmer (1948): las DSD se clasifican en intervalos de intensidad de precipitación (R) y posteriormente se ajustan a una distribución exponencial. Los resultados de \citep{MarshallPalmer1948} son compatibles con $N_{0}$ constante y $\Lambda$ una función de R. \emph{Derecha}: Propuesta de \citep{LawsParsons1943}, obteniendo dependencias con R tanto para $\Lambda$ como para $N_{0}$. Los puntos en ambas figuras corresponden a datos experimentales obtenidos en Ottawa.}
\label{fig:MarshallPalmer}
\vspace{0.75cm}

\end{figure}



A partir de esta modelización se asume que las gotas de tamaños mayores son más improbables en su detección que las gotas de tamaños menores. Eso implica, para la medición a nivel del suelo, que los errores debidos a un muestreo insuficiente afectan en mayor medida a las gotas de mayor tamaño y en consecuencia, los errores de muestreo asociados a la reflectividad serán de mayor importancia con respecto de otros parámetros integrales.\\

Esta modelización ha sido ampliamente utilizada en esquemas de parametrización de la pre\-ci\-pi\-ta\-ción de modelos de numéricos de predicción. Sin embargo, está siendo poco a poco sustituida por otras modelizaciones debido a que, si bien la variabilidad está más o menos bien descrita para intervalos largos de tiempo (donde las fluctuaciones en el registro del número de gotas son promediadas), existen diferencias notables entre los registros experimentales y una función exponencial pura. Con el propósito de mejorar la forma exponencial surgen las propuestas de modelizar la DSD como una distribución gamma y como una distribución log-normal.

%
%
%
%

\subsection{Distribución gamma}
\label{sec:GAMMAdistrib}
Se ha propuesto \citep{ulbrich_1983_aa} que la DSD puede ser modelizada mediante una función de distribución gamma:

\begin{equation}
N(D)=N_{0}D^{\mu}e^{-\Lambda D}
\label{eqn:modelizacionGAMMA}
\end{equation}

que mediante los parámetros $N_{0}$, que ahora posee unidades $[m^{-3}mm^{-1-\mu}]$, $\mu$, un factor de forma adimensional, y $\Lambda$, un factor de escala con dimensiones $[mm^{-1}]$, permite en teoría capturar gran parte de la variabilidad que no es posible describir mediante la distribución exponencial.\\

Desde el punto de vista estadístico existen varios métodos para estimar los parámetros libres de esta distribución, siendo los más comunes el \emph{método de los momentos} y el \emph{método de máxima verosimilutud}. El primero ha sido el más usado por ser más intuitivo y por estar relacionado de forma más directa con los parámetros integrales de la precipitación. Parte de su estudio inicial ha sido realizado en  \citep{ulbrich_1983_aa,UlbrichAtlas1998}. Conviene notar (1) que el método de los momentos no es único, ya que del conjunto de todos los posibles momentos se pueden seleccionar tres para determinar los parámetros libres de la ecuación anterior (2), problemas y limitaciones en la medida de las DSD han sugerido que se pueden combinar los métodos tradicionales con los métodos para una distribución gamma truncada bien en $D_{min}$ bien en $D_{max}$. Se analizan seguidamente los métodos de estimación más utilizados. 


\subsubsection{Momentos de la distribución gamma}

Dada la distribución gamma, los momentos $M_{k}$ pueden ser resueltos analíticamente\footnote{No se exige que k sea un numéro natural o cero, ya que la función gamma es una generalización para cualquier numero real del factorial.} como para $(D_{min},D_{max})=[0,\infty)$:

\begin{equation}
M_{k}=N_{0}\frac{\Gamma(k+\mu+1)}{\Lambda^{(k+\mu+1)}}
\label{eqn:momentoORDENkGAMMA}
\end{equation}

Para una descripción de la función gamma, $\Gamma(x)$, véase el recuadro tras la sección (\ref{sec:ampliGAMMA}).

\subsubsection{Método de los momentos}
\label{sec:metodoMomentos}
La expresión de la distribución gamma posee tres parámetros libres. El método de los momentos aplicado a la estimación de dichos parámetros implica el uso de tres momentos de la distribución\footnote{Utilizaremos la notación MM-lkn para el método que se base en los momentos de orden l, orden k y orden n}, por tanto, según la elección de los momentos utilizados, es posible realizar diferentes estimaciones. Las más utilizadas incorporan momentos desde el orden 2 hasta el orden 6 debido principalmente a que los momentos de orden 0 y 1 están afectados por las limitaciones de las medidas disdrométricas. En este sentido, el hecho de que los momentos de orden mayor posean, en principio, más problemas de muestreo, hace al método que use los momentos 2, 3 y 4 un candidato ideal para la estimación adecuada de $N_{0},\Lambda,\mu$.\\

La capacidad real de predicción depende no solo de posibles problemas de muestreo, sino del hecho de si para una determinada muestra experimental la distribución elegida, en este caso la gamma, es adecuada o no \citep{cao_zhang_2009_aa}. Introducimos aquí los métodos de estimación que se han aplicado a medidas disdrométricas, siendo los métodos MM234 y MM346 los más usados.\\

Todos los métodos comparten un procedimiento genérico. Se parte de la definición de un parámetro G como el cociente de los momentos utilizados:
\begin{equation} 
G=\frac{M_{l}^{a}}{M^{b}_{k}M^{c}_{n}}
\end{equation}
de manera que se cumple que:
\begin{equation}
   a\cdot l=b\cdot k+c\cdot n      
\end{equation}
lo que permite tener una expresión adimensional para G. Si añadimos la condición:
\begin{equation}
a=b+c
\end{equation}
 La magnitud  $N_{0}$ no aparece en la expresión de G. Además permite eliminar los factores $\Lambda$, llegando a una expresión $G(\mu)$ que solo depende de $\mu$. Por tanto, G puede ser determinado a partir de valores experimentales (sea $G_{exp}$ su valor); también es posible estimar $\mu_{MM}=\mu(G_{exp})$ invirtiendo la expresión $G_{exp}=G(\mu)$. El siguiente paso es tomar los dos momentos menores del conjunto (l,k,n). Supongamos que son (l,k). Dado que:
\begin{equation}
N_{0}=M_{k}\frac{\Lambda^{(k+\mu+1)}}{\Gamma(k+\mu+1)}
\label{eqn:ConcentracionN0}
\end{equation}
es posible escribir la relación como:
\begin{equation}
\Lambda^{l-k}=\frac{M_{l}}{M_{k}}\frac{\Gamma(k+\mu_{MM}+1)}{\Gamma(l+\mu_{MM}+1)}
\end{equation}
Como último paso se estima $N_{0}$ usando la ecuación (\ref{eqn:ConcentracionN0}) para el menor de los valores de (l,k,n) y los valores estimados $\mu_{MM}$ y $\Lambda_{MM}$. Lo usual es tomar los valores menores del conjunto (l,k,n) dado que la estimación de $M_{k}$ desde medidas disdrométricas es menos sesgada para momentos de orden menor \citep{smith_kliche_2005_aa}. Los resultados aparecen en la tabla (\ref{tablaMMmetodosGAMMA}). A continuación se comenta el desarrollo típico, que puede ser entendido como una aplicación del método general introducido.\\

\begin{table}[h]
\caption[Estimación por el método de los momentos de los parámetros de una distribución gamma]{\textbf{Estimación por el método de los momentos de los parámetros de una distribución gamma}. Se introducen cuatro métodos diferentes más utilizados en la bibliografía. Respecto de la metodología de obtención de las expresiones ha sido introducida de modo general en el texto principal.}
\vspace{0.55cm}
\ra{1.50}
\begin{center}
\begin{tabular}{lcccc}
\toprule

\textbf{Método}          &  \textbf{Función G}                       &  $\mathbf{\mu(G)}$ & $\mathbf{\Lambda(\mu)}$ & $\mathbf{N_{0}(\Lambda,\mu)}$ \\

\midrule
\ra{1.75} 

MM012             &  $\displaystyle\frac{M_{1}^{2}}{M_{0}M_{2}}$   &$\displaystyle\frac{1}{1-G}-2$   & $\displaystyle \frac{(2+\mu)M_{1}}{M_{2}}$ & $M_{2}\frac{\Lambda^{(2+\mu+1)}}{\Gamma(2+\mu+1)}  $                    \\

MM246             &  $\displaystyle\frac{M_{4}^{2}}{M_{2}M_{6}}$   & $\displaystyle\frac{7-11G-\sqrt{14G^{2}+G+1}}{2(G-1)}$ & $\displaystyle \sqrt{\frac{(3+\mu)(4+\mu)M_{2}}{M_{4}}}$ & $M_{2}\frac{\Lambda^{(2+\mu+1)}}{\Gamma(2+\mu+1)}  $                    \\

MM346             &  $\displaystyle\frac{M_{4}^{3}}{M^{2}_{3}M_{6}}$            &   $\displaystyle\frac{-8+11G+\sqrt{G^{2}+8G}}{2(1-G)}$   &   $\displaystyle\frac{(4+\mu)M_{3}}{M_{4}}$             &     $M_{3}\frac{\Lambda^{(3+\mu+1)}}{\Gamma(3+\mu+1)}  $                 \\

MM234             &  $\displaystyle\frac{M_{3}^{2}}{M_{2}M_{4}}$            &  $\displaystyle\frac{1}{1-G}-4$           &  $\displaystyle\frac{(3+\mu)M_{2}}{M_{3}}$              &  $M_{2}\frac{\Lambda^{(2+\mu+1)}}{\Gamma(2+\mu+1)}  $                     \\

MM456             &  $\displaystyle\frac{M_{5}^{2}}{M_{4}M_{6}}$            &  $\displaystyle\frac{1}{1-G}-6$                 &   $\displaystyle\sqrt{\frac{(5+\mu)(4+\mu)M_{4}}{M_{5}}}$             &    $M_{4}\frac{\Lambda^{(4+\mu+1)}}{\Gamma(4+\mu+1)}  $                    
\\

\bottomrule
\end{tabular}
\end{center}
\label{tablaMMmetodosGAMMA}
\vspace{1.0cm}
\end{table}

Conviene notar que, en el caso de restringir las medidas con $D_{max}$ o $D_{min}$, es posible calcular los momentos de la distribución gamma haciendo uso de la función gamma incompleta inferior, véase la definición (\ref{eqn:gammaInferior}). Únicamente basta tener en cuenta que (véase la referencia \citep{UlbrichAtlas1998}):

\begin{equation}
M_{k}=N_{0}\frac{\gamma(k+\mu+1,D_{max})}{\Lambda^{(k+\mu+1)}}
\end{equation}

Obtenemos expresiones similares en el caso de fijar $D_{min}$ usando la función gamma incompleta superior. En caso de restringir  $D_{min}$ y $D_{max}$ se tiene que\footnote{Aspectos relacionados con el uso de estas relaciones que implican el uso de diámetros máximo y mínimo serán tratados en el capítulo \S\ref{chap:BINNING} donde se retomará la discusión dada en esta sección, aunque se ha dotado al capítulo \S\ref{chap:BINNING} de una redacción autónoma.}:

\vspace{2mm}
 \begin{equation}
M_{k}=N_{0}\frac{\gamma(k+\mu+1,D_{max})-\gamma(k+\mu+1,D_{min})}{\Lambda^{(k+\mu+1)}}
\end{equation}
\vspace{2mm}

\paragraph{MM246}

Dados $M_{k}$ para k=2,4,6 se define un parámetro $\eta$ como:

\vspace{2mm}

\begin{equation}
\eta=\frac{M_{4}^{2}}{M_{2}M_{6}}=\frac{(4+\mu)(5+\mu)}{(5+\mu)(6+\mu)}
\end{equation}

\vspace{2mm}

que es una función monótona para $\mu \geq -3.63$. La segunda igualdad se construye haciendo uso de la relación de recurrencia de la función gamma. Es posible resolver la ecuación anterior para $\mu$, teniendo en cuenta que se espera que la función DSD sea convexa ($\mu>0$), por lo que la solución se puede expresar como:
\begin{equation}
\mu=\frac{(7-11\eta)-\sqrt{\eta+14\eta^{2}+1}}{2(\eta-1)}
\end{equation}
Dado $\mu$ es posible obtener $\Lambda$ mediante:
\begin{equation}
\Lambda=\sqrt{\frac{(4+\mu)(3+\mu)M_{2}}{M_{4}}}
\end{equation}
y $N_{0}$ mediante\footnote{Puede usarse $M_{2}$ o también $M_{4}$ o $M_{6}$ vía expresión (\ref{eqn:momentoORDENkGAMMA}).}:
\begin{equation}
N_{0}=M_{2}\frac{\Lambda^{(2+\mu+1)}}{\Gamma(2+\mu+1)}
\end{equation}

\paragraph{MM346}

Dados $M_{k}$ para k=3,4,6 se define un parámetro $G$ como:
\vspace{2mm}

\begin{equation}
G=\frac{M_{4}^{3}}{M_{3}^{2}M_{6}}
\end{equation}
\vspace{2mm}

Es posible calcular $\mu$ mediante:

\begin{equation}
\mu=\frac{(8-11G)-\sqrt{G^{2}+8G}}{2(1-G)}
\end{equation}
\vspace{2mm}

Dado $\mu$ es posible obtener $\Lambda$ mediante:
\begin{equation}
\Lambda=\frac{(4+\mu)M_{3}}{M_{4}}
\end{equation}
y $N_{0}$ como,
\begin{equation}
N_{0}=M_{3}\frac{\Lambda^{(3+\mu+1)}}{\Gamma(3+\mu+1)}
\end{equation}

\paragraph{MM234}

Dados $M_{k}$ para k=2,3,4 se define un parámetro $G$ como:
\vspace{2mm}

\begin{equation}
G=\frac{M_{3}^{2}}{M_{2}M_{4}}
\end{equation}
\vspace{2mm}

Es posible calcular $\mu$ mediante:

\begin{equation}
\mu=\frac{1}{1-G}-4
\end{equation}
Dado $\mu$ es posible obtener $\Lambda$:
\begin{equation}
\Lambda=\frac{(3+\mu)M_{2}}{M_{3}}
\end{equation}
y $N_{0}$ utilizando:
\begin{equation}
N_{0}=M_{2}\frac{\Lambda^{(2+\mu+1)}}{\Gamma(2+\mu+1)}
\end{equation}
\vspace{2mm}

\paragraph{MM456}

Dados $M_{k}$ para k=4,5,6 se define un parámetro $G$ como:
\vspace{2mm}

\begin{equation} 
G=\frac{M_{5}^{2}}{M_{4}M_{6}}
\end{equation}
\vspace{2mm}

Es posible calcular $\mu$ mediante:

\begin{equation}
\mu=\frac{1}{1-G}-6
\end{equation}
Dado $\mu$ es posible obtener $\Lambda$:
\begin{equation}
\Lambda=\frac{(5+\mu)M_{4}}{M_{5}}
\end{equation}
y $N_{0}$ usando que:
\begin{equation}
N_{0}=M_{4}\frac{\Lambda^{(4+\mu+1)}}{\Gamma(4+\mu+1)}
\end{equation}
\vspace{2mm}
\subsubsection{Estimación mediante el método de máxima verosimilitud}
\label{sec:MLEexplicado}
Partimos de la expresión normalizada a la unidad de la distribución gamma:

\begin{equation}
f(D;\nu,\lambda)=\frac{\lambda^{\mu+1}}{\Gamma(\mu+1)}D^{\mu}exp\left[-\lambda D\right]
\end{equation}

La metodología consiste en, dado un conjunto S de N gotas de diámetros $D_{i}, i=1,...,N$ maximizar la función de máxima verosimilitud:

\begin{equation}
L(S;\mu,\lambda)=\prod_{i}^{N} f(D_{i};\mu,\lambda)
\end{equation}

Suele ser más cómodo maximizar el logaritmo de la función anterior, ya que se expresa como suma de los datos experimentales. En tal caso se obtiene un conjunto de dos ecuaciones relacionadas con las derivadas parciales respecto de $\mu$ y $\lambda$ que se condensan en:

\begin{equation}
ln(\mu_{MLE}+1)=\psi(\mu_{MLE}+1)+ln\left[ \frac{\bar{D}}{\left(\prod_{i=1}^{N}D^{i}\right)^{1/N}} \right]
\end{equation}
donde $\psi(x)$ es la función digamma dada por $\psi(x)=\Gamma'(x)/\Gamma(x)$.\\

La ecuación obtenida puede ser resuelta iterativamente dada una condición inicial adecuada, como puede ser la estimación mediante uno de los métodos por momentos. En nuestro caso se ha seguido la sugerencia de \citep{kliche_smith_etal_2008_aa}: dado el valor experimental de $D_{m}$, y definiendo $y_{i}=D_{i}/D_{m}$, se tiene la siguiente serie de valores:
\begin{equation}
\alpha_{k+1}=\alpha_{k}\frac{ln(\alpha_{k})-\psi(\alpha_{k})}{ln\left[ \frac{\bar{y}}{\left(\prod_{i=1}^{N}y^{i}\right)^{1/N}} \right]}
\end{equation}
cuando se obtenga que $\alpha_{k+1}-\alpha_{k}=\delta$, con $\delta$ una cota suficientemente pequeña se considera convergida la serie a $\alpha_{k+1}=\alpha$ y $\mu_{MLE}=\alpha-1$. Para la definición de la función digamma véase el siguiente recuadro.  
\newpage




\rule{1.00\linewidth}{0.75pt}
\small
\begin{center}
\textbf{Ampliación de la distribución gamma y de las funciones gamma}
\end{center}
\begin{multicols}{2}
\label{sec:ampliGAMMA}
La función gamma se define por la expresión:
\begin{equation}
\Gamma(x)=\int_{0}^{\infty}t^{x-1}e^{-t}dt
\end{equation}
que posee la siguiente propiedad de recurrencia:
\begin{equation}
\Gamma(x+1)=x\Gamma(x)
\end{equation}
Esta propiedad implica que si n es un número na\-tu\-ral, $\Gamma(n+1)=n!$ La propiedad de recurrencia es utilizada en la deducción algebraica de los diferentes métodos de momentos desarrollados para estimar los parámetros de la distribución gamma. Complementarias de la expresión anterior son las funciones gamma incompletas inferior y superior:
\begin{equation}
\gamma(a,x)=\int_{0}^{a}t^{x-1}e^{-t}dt
\label{eqn:gammaInferior}
\end{equation}

\begin{equation}
\Gamma(a,x)=\int_{a}^{\infty}t^{x-1}e^{-t}dt
\end{equation}
que cumplen que $\gamma(s,x)+\Gamma(s,x)=\Gamma(x)$ sea cual sea el valor de s. Estas funciones han sido utilizadas para algunas expresiones basadas en la distribución gamma truncada \citep{UlbrichAtlas1998}. Es útil, además, la definición de la función digamma $\psi$:
\begin{equation}
\psi(x)=\frac{\Gamma'(x)}{\Gamma(x)}
\end{equation}
donde para la función $\psi(x+1)$ es útil el desarrollo en serie de potencias, que viene dado por:
\begin{equation}
\psi(x+1)\simeq \frac{x}{x+1}-\gamma+\frac{x^{7}}{2}+\sum_{i=1}^{6}c_{i}(x^{i}-x^{7})
\end{equation}
La constante $\gamma$ es la constante de Euler y los parámetros $c_{i}$ están dados en la tabla adjunta (\ref{TablaDesarrolloDigamma}). La aproximación dada por estos coeficientes es adecuada para valores de la variable independiente del orden de la unidad. Luego en la práctica este desarrollo es adecuado para valores del diámetro que sean del mismo orden de magnitud que $D_{m}$. En ocasiones se denota a la función digamma como $\psi_{0}(x)$ y a su derivada n-esima por $\psi_{n}(x)$.\\

Una de las distribuciones más versátiles para la parametrización de la microfísica de la precipitación es la distribución gamma modificada, dada por:
\begin{equation}
f(D;\nu,\lambda,c)=\frac{c\lambda}{\Gamma(\nu)}(\lambda D)^{c\nu-1}e^{\left[-(\lambda D)^{c}\right]}
\end{equation}
ya que esta expresión está normalizada. Teniendo en cuenta que $N(D)=N_{t}f(D)$ es posible escribir:
\begin{equation}
N(D;\nu,\lambda,c)=N_{t}\frac{c\lambda}{\Gamma(\nu)}(\lambda D)^{c\nu-1}e^{\left[-(\lambda D)^{c}\right]}
\end{equation}
Lo más común es restringir esta función, que además de los parámetros indicados depende de $N_{t}$. De esta manera con $c=1$ se obtiene la distribución gamma:
\begin{equation}
N(D;\nu,\lambda)=N_{t}\frac{\lambda}{\Gamma(\nu)}(\lambda D)^{\nu-1}exp\left[-\lambda D\right]
\end{equation}
que se expresa usualmente como:
\begin{equation}
N(D)=N_{0}D^{\mu}exp\left[-\lambda D\right]
\end{equation}
Si $\mu=0$ obtenemos la distribución exponencial.\\

Es habitual parametrizar la distribución gamma del siguiente modo:
\begin{equation}
f(x;k,\theta)=x^{k-1}\frac{e^{-x/\theta}}{\theta^{k}\Gamma(k)}\quad x>0\quad k,\theta>0
\end{equation}
El parámetro k es llamado factor de forma y el parámetro $\theta$ factor de escala. La función de distribución acumulada (cdf), que se obtiene desde esta función de distribución de probabilidad (pdf), es la función gamma incompleta inferior, que por el hecho de no poseer una forma analítica y tener que ser expresada en forma integral hace que metodologías basadas en la pdf sean más usuales que las basadas en la cdf. Otra parametrización posible es,
\begin{equation}
g(x;\alpha,\beta)=\beta^{\alpha}x^{\alpha-1}\frac{e^{-\beta x}}{\Gamma(\alpha)}\quad x>0
\end{equation}

Las últimas dos expresiones son las más usuales para la distribución gamma, y como tales son las que incorporan las librerías de \emph{software} estándar de análisis estadístico y numérico.

\begin{table}[H]
\caption[Coeficientes desarrollo en serie de la función digamma]{\small\textbf{Coeficientes del desarrollo en serie de la función digamma.}}
\begin{center}
\small
\begin{tabular}{lr}
\toprule
 Coeficiente &  Valor \\
\midrule
$\gamma$   & 0.57721566   \\
$c_{1}$    & 0.65593313   \\
$c_{2}$    & -0.20203181  \\
$c_{3}$    & 0.08209433   \\
$c_{4}$    & -0.03591665  \\
$c_{5}$    & 0.01485925   \\
$c_{6}$    & -0.00472050  \\

\bottomrule
\end{tabular}
\end{center}
\label{TablaDesarrolloDigamma}
\end{table}

\end{multicols}
\normalsize
\rule{1.00\linewidth}{0.75pt}
%

\subsubsection{Relaciones $\mu-\lambda$}
\label{teo:relacionmulb}
Una de las cuestiones más ampliamente discutidas en la literatura reciente es la posibilidad de establecer relaciones empíricas entre los parámetros de la distribución gamma, y por tanto lograr la posible estimación de $N_{0},\Lambda,\mu$ a partir de tan solo dos cantidades experimentales \citep{ulbrich_1983_aa}.\\

Este aspecto ha tenido su relevancia tras el desarrollo de medidas de teledetección ya que, si tales relaciones tuvieran una base física, sería técnicamente posible determinar la parametrización de la DSD usando solo dos medidas remotas. Una parte importante de la discusión se ha centrado en si las relaciones entre el factor de escala ($\Lambda$) y el factor de forma ($\mu$) son viables, o si son únicamente una relación espuria debida a los problemas de muestreo antes comentados \citep{moisseev_chandrasekar_2007_aa}. Los estudios parecen indicar que ambas variabilidades, la real y la estadística, se entrelazan en las relaciones entre $\Lambda=\Lambda(\mu)$. Si tal relación existe físicamente sería por tanto posible determinar la parametrizacion de la DSD desde por ejemplo la medida de Z y de $Z_{DR}$ \citep{zhang_vivekanandan_etal_2003_aa}. Al respecto, véase también el apéndice \S\ref{sec:newBINS}
.

\begin{figure}[h] 
\begin{center}
   \includegraphics[width=0.90\textwidth]{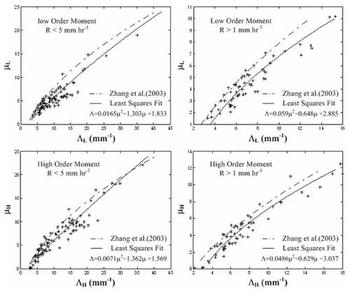}
   \caption[Relaciones $\mu-\lambda$ experimentales propuestas por \citep{2008ChuandSu}.]{\textbf{Relaciones $\mu-\lambda$ experimentales propuestas por \citep{2008ChuandSu}.} Se muestran relaciones $\mu-\lambda$ donde los parámetros de la DSD son estimados mediante momentos de orden bajo y momentos de orden alto. Se incluyen diferentes acotaciones en la intensidad de pre\-ci\-pi\-ta\-ci\-ón. Los autores compararon sus resultados con el estudio previo \citep{zhang_vivekanandan_etal_2003_aa}. Los resultados han sido cuestionados por varios autores debido a la posible relación entre problemas de estimación de los parámetros de la DSD y relaciones artificiales entre estos.}
\label{fig:relacionesMULB_chu}
\end{center}
\vspace{0.5cm}
\end{figure}

\begin{figure}[h] 
\begin{center}
   \includegraphics[width=0.90\textwidth]{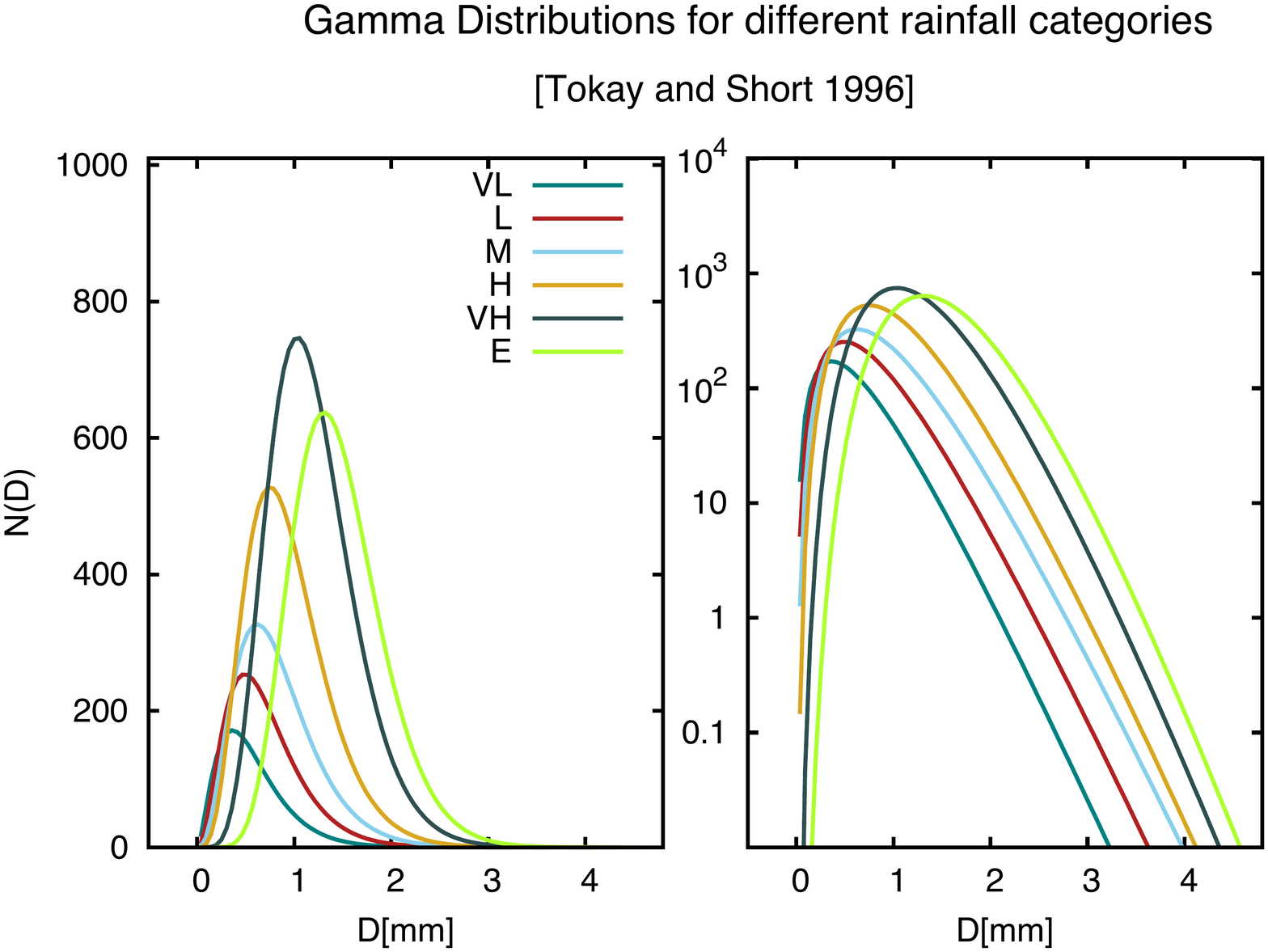}
   \caption[Forma típicas de la DSD mediante la función de distribución gamma para diferentes intensidades de lluvia. Categorías basadas en estudio de \citep{TokayShort1996}.]{\textbf{Forma típicas de la DSD mediante la función de distribución gamma para diferentes intensidades de lluvia. Categorías basadas en estudio de \citep{TokayShort1996}.} Se muestra la modelización mediante la ecuación (\ref{eqn:modelizacionGAMMA}) para intensidades de precipitación propuestas por \citep{TokayShort1996} que van desde lluvia muy débil (VL) hasta lluvia extrema (E), pasando por lluvia moderada (M) y lluvia intensa (H). En estas modelizaciones los casos más intensos se desplazan hacia tamaños mayores de gota, tal y como sucedía en la modelización dada por la distribución de Marshall-Palmer.}
\label{fig:gammaDISTRIBUTIONclass}
\end{center}
\vspace{0.5cm}
\end{figure}

\subsection{Distribución log-normal}
Fue propuesta por \citep{FeingoldLevin1986} y ha sido usada en varios estudios de precipitación, aunque no tan extensivamente como la distribución gamma. Se modeliza mediante:
\vspace{2mm}
\begin{equation}
N(D)=N_{T}\frac{1}{\sqrt{2\pi}Dln\sigma}e^{-\frac{ln^{2}(D/D_{g})}{2ln^{2}\sigma}}
\end{equation}
donde $N_{T}$ representa el número total de gotas por unidad de volumen y los parámetros $D_{g}$ y $\sigma$ están relacionados con la media y la dispersión cuadrática media, es decir, $ln D_{g}=<ln D>$ y $ln^{2}\sigma=<(lnD-lnD_{g})^{2}>$. Para la determinación de los parámetros\footnote{Además propusieron: $N_{t}=172R^{0.22}$, $D_{g}=0.72R^{0.23}$ y $\sigma=1.43$.} se pueden utilizar diferentes métodos. El más directo es mediante momentos usando los de orden 0, 1 y 2. Así el estimador para $ln^{2}\sigma$ es:
\vspace{2mm}
\begin{equation}
\nu^{2}=e^{-ln^{2}\sigma}=\frac{M_{1}^{2}}{M_{0}M_{2}}
\end{equation}
y
\begin{equation}
D_{g}=\frac{M_{1}}{M_{0}}\sqrt{\nu}
\end{equation}
\vspace{2mm}
Como en el caso de la distribución gamma tenemos un parámetro relacionado con la forma, $\sigma$, y otro con el tamaño característico de las gotas, $D_{g}$, mientras que la cantidad de gotas viene determinada por $N_{T}$. \\

Es más cómodo expresar los momentos usando el parámetro $\eta=ln\sigma$:
\vspace{2mm}
\begin{equation}
M_{k}=N_{t}D_{g}^{k}e^{0.5k^{2}\eta^{2}}
\end{equation}
De modo similar al caso de la distribución gamma, es posible utilizar no solo el método MM012 comentado anteriormente, sino otros métodos definiendo un parámetro $G$ dado por:
\vspace{2mm}
\begin{equation}
G_{a,b,c}=\frac{M_{a}^{l}}{M^{k}_{b}M^{m}_{c}}
\end{equation}
\vspace{2mm}
de manera que $a\cdot l=b \cdot k+c \cdot n$ y $l=k+m$. De este modo:
\vspace{2mm}
\begin{equation}
G_{a,b,c}=e^{(l\cdot a^{2}-k\cdot b^{2}-m \cdot c^{2})0.5\eta^{2}}
\end{equation}
con lo que es posible despejar $\eta$. Del cociente de dos de estos momentos es posible determinar $D_{g}$, mientras que $N_{t}$ se puede obtener de uno de los momentos (como en el caso de la distribución gamma, se suele elegir el menor si todos son de orden mayor o igual que dos) dados los valores de $D_{g}$ y $\eta$ obtenidos previamente. Así pues el método es, en su procedimiento general, similar para la estimación de los parámetros de la distribución log-normal y gamma.

\begin{table}[h]
\caption[Estimación por el método de los momentos de los parámetros de una distribución log-normal]{\textbf{Estimación por el método de los momentos de los parámetros de una distribución log-normal}. Se introducen cuatro métodos, siendo el primero de ellos el más utilizado en la bibliografía.}
\vspace{0.35cm}
\ra{1.75}
\begin{center}
\begin{tabular}{lcccc}
\toprule

\textbf{Método}&  \textbf{Función G}                    &  $\mathbf{\eta(G)}$ & $\mathbf{D_{g}(\eta)}$ & $\mathbf{N_{T}(D_{g},\eta)}$ (*)\\
\midrule
 
MM012          &$\displaystyle\frac{M_{1}^{2}}{M_{0}M_{2}}$ & $\displaystyle\sqrt{-ln G}$ & $\displaystyle e^{-\frac{1}{2}\eta^{2}}\frac{M_{1}}{M_{0}}$ & $M_{2}D_{g}^{-2}e^{-2\eta^{2}}$   \\

MM123          &$\displaystyle\frac{M_{2}^{2}}{M_{1}M_{3}}$ & $\displaystyle\sqrt{-ln G}$ & $\displaystyle e^{-\frac{3}{2}\eta^{2}}\frac{M_{2}}{M_{1}}$             & $M_{3}D_{g}^{-3}e^{-3\eta^{2}}$ \\

MM234          &$\displaystyle\frac{M_{3}^{2}}{M_{2}M_{4}}$ &  $\displaystyle\sqrt{-ln G}$ &  $\displaystyle e^{-\frac{5}{2}\eta^{2}}\frac{M_{3}}{M_{2}}$  &  $M_{4}D_{g}^{-4}e^{-8\eta^{2}} $                     \\

MM246             &  $\displaystyle\frac{M_{4}^{2}}{M_{2}M_{6}}$            &  $\displaystyle\sqrt{-\frac{1}{4}ln G}$                 &   $\displaystyle \sqrt{e^{-6\eta^{2}}\frac{M_{4}}{M^{2}_{2}}}$             &    $M_{2}D_{g}^{-2}e^{-2\eta^{2}}   $                    
\\

\bottomrule
\end{tabular}
\end{center}
\small (*) No existe un único modo para determinar $N_{t}$, y de hecho puede calcularse eligiendo un parámetro integral (momento) u otro dentro de los que definen cada método.
\label{tablaMMmetodosLOGNORMAL}
\vspace{1.5cm}
\end{table}

\section{Método de escalado usando un momento}
\label{sec:Scaling1moment}
En la propuesta de una exponencial, los parámetros se hacen depender de una magnitud relacionada con los momentos de la distribución, en concreto la intensidad de precipitación R (como sucedía en la propuesta de Marshall-Palmer), mientras que en los estudios en los que se utiliza la distribución gamma se proponen toda una serie de relaciones entre los diferentes momentos de la distribución que aparecen relacionados con las propiedades físicas de interés. La siguiente cuestión es, por tanto, qué significado poseen estos hechos y, llegado el caso, si existe alguna magnitud que, escalando respecto de ella, permita eliminar gran parte de la variabilidad observada. En tal caso se obtendría una posible forma intrínseca para la distribución de tamaños de gota. Esto sugiere proponer:
\begin{equation}
N(D,R)=R^{\alpha}g(D R^{-\beta})
\end{equation}
junto con relaciones en forma de ley de potencias entre los parámetros integrales\footnote{El típico ejemplo de relación entre momentos es $Z=a_{R}R^{b_{R}}$.}. La idea subyacente es que la nueva función g(x) pase a tener una variabilidad mucho menor que la original $N(D)$ y se puedan relacionar cambios en la DSD con cambios en la intensidad de la precipitación, mientras que son los procesos microfísicos que producen la DSD los que fundamentan, desde el punto de vista experimental, la presencia de relaciones de potencias entre los parámetros integrales de la precipitación. Este proceso de análisis fue realizado por \citep{semperetorres_porra_ea_1994} y permite enfocar el punto de partida del problema en las relaciones entre los momentos, más que en la forma funcional específica para la DSD. Veámoslo en una forma general dado un parámetro integral $\phi$,
\begin{equation}
N(D,\phi)=\phi^{\alpha}g(D \phi^{-\beta})
\label{eqn:sempere-torres-1moment}
\end{equation}
donde $\phi$ es un cierto parámetro integral dado por:
\begin{equation}
\phi=C_{\phi}\int N(D)D^{k}dD
\label{eqn:phirefscaling}
\end{equation}
La \emph{hipótesis de partida} son las relaciones entre potencias para el resto de parámetros integrales $\psi$ mediante:
\begin{equation}
\psi=a_{\psi}\phi^{b_{\psi}}
\label{eqn:powerlaw-1moment}
\end{equation}
Los valores de $a_{\psi}$ y $b_{\psi}$ han de ser calculados a partir de datos experimentales. Imaginemos que la función $\psi$ es un parámetro integral de orden n, dado por:
\begin{equation}
\psi=C_{\psi}\int N(D)D^{n}dD
\end{equation}
Sustituyendo la definición dada por (\ref{eqn:sempere-torres-1moment}),  obtenemos que:
\begin{equation}
\psi=C_{\psi}M_{n}^{(g)}\phi^{\alpha+\beta(n+1)}
\label{eqn:powerlaw-1moment_cong}
\end{equation} 
para $M_{n}^{(g)}$ el momento de orden n de la función g. Es decir,
\begin{equation}
M_{n}^{(g)}=\int_{0}^{\infty} g(x)x^{n}dD
\end{equation}
Comparando con (\ref{eqn:powerlaw-1moment}) observamos como:
\begin{equation}
b_{\psi}=\alpha+\beta(n+1)
\end{equation}
y estudiando esta relación para diferentes funciones $\psi$ podemos obtener un ajuste lineal para ha\-llar $\alpha$ y $\beta$. Mientras que $a_{\phi}=C_{\phi}M_{n}^{(g)}$.\\

Una vez obtenidos los valores $\alpha$ y $\beta$ es posible representar la nube de puntos que define a g(x), y estimar si sigue o no una cierta función parametrizada. Los resultados experimentales \citep{SempereTorres1998} parecen indicar que para un tipo fijo de precipitación  la variabilidad presente en la nube de puntos que define g(x) será menor.  

\begin{figure}[H] 
\begin{center}
   \includegraphics[width=0.98\textwidth]{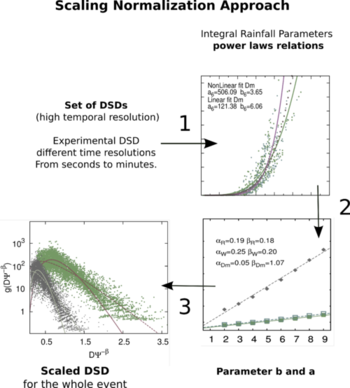}
\vspace{1.5cm}
   \caption[Esquema de aplicación del método de escalado usando un momento.]{\textbf{Esquema de aplicación del método de escalado usando un momento.} Se muestra esquematicamente el proceso de aplicación del método de escalado, incidiendo en una de las hipótesis subyacentes es que los momentos se relacionan mediante relaciones de potencias para cada episodio de precipitación (o cada muestra experimental considerada) analizado.}
\label{fig:esquema_Escalado1momento}
\end{center}
\vspace{0.5cm}
\end{figure}

\subsection{Relaciones de consistencia}

Es fácil comprobar que si la función $\psi$ es la propia función $\phi$ elegida para escalar la N(D), obtenemos dos relaciones que han de verificarse y que pueden ser entendidas como relaciones de consistencia dentro del método. Dado que hemos supuesto que $\phi$ es un parámetro de orden k se cumple:
\begin{equation}
1=\alpha+(k+1)\beta
\label{eqn:consistencia1-1moment}
\end{equation}
y
\begin{equation}
C_{\phi}\int_{0}^{\infty}x^{k}g(x)dx=1
\label{eqn:consistencia2-1moment}
\end{equation}

Como consecuencia, tenemos un grado de libertad en la pareja $\alpha,\beta$. Por otra parte la relación de consistencia impone una relación que ha satisfacer g(x), de manera que los modelos funcionales para esta poseerán un grado de libertad menos que los correspondientes para N(D).

\subsection{Aplicación a la distribución gamma}

Si proponemos para g(x) la forma de una función gamma dada por:
\begin{equation}
g(x)=\kappa x^{\mu}e^{-\lambda x}
\label{eqn:gammaScaling1moment}
\end{equation}

podemos, ajustar los tres parámetros libres mediante alguna técnica de ajuste no-lineal y comprobar la relación de consistencia (\ref{eqn:consistencia2-1moment}), o a través de la relación de consistencia determinar el parámetro $\kappa$ en función de $\mu,\lambda$ y el valor de $C_{\psi}$. Esto permite además mediante la ecuación (\ref{eqn:powerlaw-1moment_cong}) expresar, dada g(x), todos los momentos de orden k como $M_{k}=F(C_{\psi},\mu,\lambda,k)\psi^{\alpha+(n+1)\beta}$. Finalmente, comparando la expresión de g(x) con la función gamma es posible determinar los parámetros de la función para N(D) a partir de la información anterior obtenida.\\

Para $\phi$ de orden k tenemos por (\ref{eqn:consistencia2-1moment}) que:
\begin{equation}
C_{\phi}\int_{0}^{\infty}x^{k}g(x)dx=C_{\phi}\kappa I(\lambda,\mu+k)=1
\end{equation}
dado que para una función exponencial la integral:
\begin{equation}
I(\lambda,\mu+k)=\frac{\Gamma(\mu+k+1)}{\lambda^{(\mu+k+1)}}
\end{equation}
podemos escribir:
\begin{equation}
\kappa=\frac{1}{C_{\phi}}\frac{\lambda^{(\mu+k+1)}}{\Gamma(\mu+k+1)}
\label{eqn:consistenciaGammaK-1moment}
\end{equation}
Los valores de $C_{\phi}$ se indican en la tabla (\ref{tablaPARAMETROSintegrales}) para las magnitudes de interés en la medida de la precipitación.

\subsection{Aplicación a la distribución log-normal}

Si proponemos para g(x) la forma de una función log-normal dada por:
\begin{equation}
g(x)=\tau \frac{1}{\sqrt{2\pi} x s}e^{-\frac{ln^{2}(x/\theta)}{2s^{2}}}
\end{equation}
De modo análogo:
\begin{equation}
C_{\phi}\int_{0}^{\infty}x^{k}g(x)dx=C_{\phi}\tau J(\theta,k,s)=1
\end{equation}
y dado que $J(\theta,k,s)=\theta^{l}e^{0.5k^{2}s^{2}}$ se puede escribir:
\begin{equation}
\tau=\frac{1}{C_{\phi}}\theta^{-l}e^{-0.5k^{2}s^{2}}
\end{equation}

\begin{figure}[H] 
\begin{center}
   \includegraphics[width=0.58\textwidth]{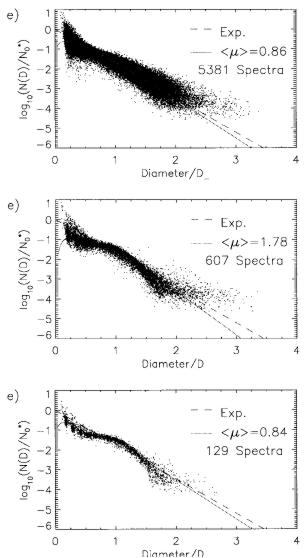}
\vspace{1.5cm}
   \caption[Ejemplos de aplicación del concepto de distribución normalizada \citep{NormalizadaTestud2001}]{\textbf{Ejemplos de aplicación del concepto de distribución normalizada \citep{NormalizadaTestud2001}.} Se muestran los tres ejemplos clásicos obtenidos por \citep{NormalizadaTestud2001} para el conjunto de datos TOGA-COARE en tres situaciones diferentes: Estratiforme (panel superior), Convectiva con intensidades menores de 30 mm/h (panel central) y Convectiva con intensidades mayores de 30 mm/h (panel inferior). El eje de abscisas corresponde al diámetro de las gotas escalado respecto del diámetro medio ponderado sobre la masa.}
\label{fig:esquema_Escalado1momento}
\end{center}
\vspace{0.5cm}
\end{figure}
\section{Distribución normalizada}
\label{sec:metodoNormalizada}
La idea central del método de escalado es normalizar la DSD usando un parámetro integral experimental, y explicar de este modo la variabilidad de la DSD. Sin embargo, hay que decir que los resultados descritos hasta ahora solo satisfacen parcialmente el objetivo de encontrar una forma intrínseca.\\

El trabajo de \citep{SempereTorres1998} parte de las relaciones entre los momentos de la distribución como hecho experimental y busca una caracterización acorde con las relaciones entre los parámetros integrales de la DSD.\\ 

Con la información dada por estas premisas, \citep{NormalizadaTestud2001} se propusieron encontrar dicha forma funcional intrínseca mediante una distribución normalizada que, no basándose directamente ni en la hipótesis de las relaciones de potencias entre los parámetros integrales ni en una forma funcional prescrita, pudiera sin embargo representar situaciones de precipitación muy diversas.\\

Para ello se sugiere que las principales cuestiones que plantean muchos análisis de la DSD son cuál es la \emph{intensidad} de lluvia medida mediante algún parámetro integral adecuado y cuál es el \emph{tamaño característico} de las gotas. Su idea es que si existe una forma intrínseca de la DSD esta no debe depender de dicha intensidad ni del diámetro medio. Por tanto, una forma de modelización apropiada a este razonamiento heurístico sería\footnote{\citep{SempereTorres1998}: $N(D)=\frac{N_{T}}{D_{c}}\hat{\rho}(D/D_{c})$ con $D_{c}$ un diámetro característico. Del mismo modo se puede generalizar a $N(D)=\frac{M_{k}}{D_{c}^{k+1}}\hat{\rho}(D/D_{c})$, que resulta ser equivalente conceptualmente a la propuesta contenida en \citep{NormalizadaTestud2001}.}:
\begin{equation}
N(D)=\hat{N}_{0}F(D/\hat{D}_{m})
\label{eqn:normalizadaTestud}
\end{equation}
La sugerencia de \citep{NormalizadaTestud2001}, es usar:
\begin{equation}
D_{m}=\frac{\int_{0}^{\infty}D^{4}N(D)dD}{\int_{0}^{\infty}D^{3}N(D)dD}
\label{eqn:normalizadaDM}
\end{equation}
y
\begin{equation}
N_{0}^{*}=\frac{4^{4}}{\pi \rho_{w}}\frac{LWC}{D^{4}_{m}}
\end{equation}
De este modo la interpretación directa de $N_{0}^{*}$ es el valor de $N_{0}$ de una distribución exponencial que posea el mismo valor de LWC y $D_{m}$ (en contraste con $N_{0}$ de la distribución gamma, que no posee una interpretación tan directa). Sin embargo, no es necesario restringirse ni a una distribución dada ni a estas definiciones concretas para $\hat{N}_{0}$ y $\hat{D}_{m}$. Es posible utilizar otro parámetro integral adecuado para $\hat{N}_{0}$, como puede ser la concentración total de gotas. También para $D_{m}$ podría ser otra definición con una interpretación en términos de diámetro característico (por ejemplo el cociente entre dos momentos consecutivos diferentes k+1 y k). Siguiendo \citep{Tokay2010smallscaleDSD} es posible comparar los resultados para $\hat{N}_{0}=N_{0}^{*}$ y $\hat{N}_{0}=N_{t}/D_{mass}$.\\

Así, es posible dada la función $F(D/D_{m})$ ajustar esta a formas funcionales dadas, observando si los valores de los parámetros libres de dichas propuestas funcionales poseen de hecho menor variabilidad que una modelización similar directamente realizada en N(D). Los ajustes más usuales son a la distribución gamma y a la distribución log-normal. Se detalla seguidamente la primera de ellas. 

\subsection{Aplicación a una distribución gamma}
\label{sec:AplicacionGammaNormalizada}
El procedimiento implica asumir que N(D) es parametrizable por una distribución gamma. En el caso en que se prefiera normalizar usando $N_{0}^{*}$, dadas las expresiones para el momento $M_{k}$ según una distribución gamma podemos escribir:
\begin{equation}
N_{0}=N_{0}^{*}D_{m}^{4}\frac{\Gamma(4)}{4^{4}}\frac{\Lambda^{(\mu+4)}}{\Gamma\mu+4}
\label{eqn:normalizadaTestudgammaN0}
\end{equation}
mientras que tomando $N_{t}$ tendríamos:
\begin{equation}
N_{0}=N_{t}\frac{\Lambda(\mu+1)}{\Gamma(\mu+1)}
\label{eqn:normalizadaTestudgammaN0Nt}
\end{equation}
En ambos casos se puede usar la relación $\Lambda=\frac{4+\mu}{D_{m}}$. Esto implica que la función F(X) queda expresada, haciendo uso de $X=D/D_{mass}$ como:
\begin{equation}
F(X; \mu)=\frac{\Gamma(4)(\mu+4)^{(\mu+4)}}{4^{4}\Gamma(\mu+4)}X^{\mu}exp[-(\mu+4)X]
\label{eqn:normalizadaTestudgammaNw}
\end{equation}
y
\begin{equation}
F(X; \mu)=\frac{(\mu+4)^{(\mu+1)}}{\Gamma(\mu+1)}X^{\mu}exp[-(\mu+4)X]
\label{eqn:normalizadaTestudgammaNt}
\end{equation}

Observamos como la función F(X) depende solo de $\mu$, para determinar $\mu$ son posibles diferentes métodos. Es posible bien minimizar las diferencias por un método de mínimos cuadrados respecto de una DSD experimental normalizada, o bien buscar el valor $\mu$ que hace menos sesgada la estimación de un determinado parámetro integral. Lo usual en este último caso es tomar la intensidad de la precipitación, aunque es posible el uso de otros parámetros integrales cuyas medidas se consideren fiables o con sentido físico.






\section{Sumario/Summary}

The main points covered in this section are:

\begin{itemize}
 \item The DSD modelings are introduced to explain the general properties of drop size distribution using parameters that allow us to reduce the sampling issues and to study the physical variations of DSD as expressed in the parameters of the DSD model.
\item There are a large amount of DSD models. The most widely used are: exponential, log-normal and gamma distributions. The exponential model is useful to analyze accumulated DSDs for long periods of time. The others allow to study the time variability of DSDs within a rain event.
\item There are several methods to get the parameters of each model, which sometimes means differences depending on the estimation method chosen to retrieve the parameters for a data set. 
\item The methods of one moment scaling and normalized DSD allow us to handle the high variability of the DSD. It is supposed that after the scaling/normalization process the DSD obtained has less variability than the original and it may be possible to study the general properties of DSD studying models of them. 
\end{itemize}

\renewcommand\chapterillustration{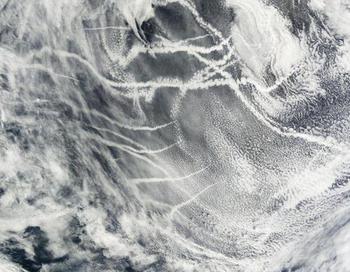}
\chapter{Procesos microfísicos y DSD}
\label{sec:MicroANDdsd}

La distribución de tamaños de gota de lluvia es el resultado de una serie de procesos que tienen lugar desde el momento en que empieza a condensarse el agua líquida en la atmósfera hasta el momento en que la DSD es medida al nivel del suelo. Estos procesos físicos están condicionados por las propiedades de la masa de aire en que se producen, desde la presencia de núcleos de condensación de diferentes tamaños y características físico-químicas hasta las propiedades sinópticas de la masa de aire.

\section{Procesos físicos y su relevancia en la DSD}
\label{sec:MicroDSDrelevancia}

Los procesos principales que dan lugar a una distribución de tamaños de gota desde las condiciones en que se crean los diferentes sistemas nubosos que dan lugar a la precipitación pertenece al campo de la física de nubes. Sin embargo el conocimiento de dichos procesos puede aportar información valiosa para entender propiedades de la DSD y sus variaciones físicas. Los procesos microfísicos principales son evaporación, agregación, colisión, coalescencia y ruptura\footnote{Estos procesos y su relevancia estan contrastados en múltiples estudios. Sin embargo, recientemente se ha propuesto una revisión de la relevancia del proceso de ruptura \citep*{Villermaux2009Nature}.}. Aquí resumimos su papel en la forma de la DSD así como la relevancia de factores dinámicos locales en este proceso, asumiendo que esta descrita por una función de distribución gamma.\\

\begin{description}
\item[Coalescencia] Consiste en la agregación de dos gotas para formar una de mayor tamaño\footnote{La unión de más de dos gotas en una sola posee una probabilidad de producirse mucho menor.}. Este proceso implica el decrecimiento del número total de gotas pequeñas y el aumento del número total de gotas grandes. Si se mantiene constante la cantidad total de agua líquida, una de las implicaciones es la disminución del número total de gotas. Por tanto, desde el punto de vista de los parámetros integrales, implica un aumento del diámetro característico y una disminución de $N_{t}$. En el caso de parametrizar mediante una distribución gamma el parámetro $N_{0}$ debe disminuir mientras que $\mu$ puede experimentar un leve aumento.
\item[Ruptura] Es el proceso por el cual una gota se divide en dos o más de menor tamaño\footnote{Al igual que en la coalescencia, la ruptura en más de dos gotas posee menor probabilidad que en dos.}. Al ser este proceso el opuesto del anterior, su implicación es el cambio contrario de los parámetros integrales y de los parámetros que definen la distribución gamma.
\item[Coalescencia y ruptura] Al actuar juntos, el proceso global viene condicionado por ser más relevante la coalescencia en gotas pequeñas y la ruptura en las grandes. La consecuencia desde el punto de vista de la DSD es un incremento del valor de $\mu$.
\item[Agregación] Sucede al incorporarse vapor de agua, presente en la masa de aire, a las gotas. Implica el aumento del tamaño de las gotas homogéneamente en todo el conjunto. Por tanto, si bien $N_{t}$ debe ser aproximadamente constante, el efecto es un aumento del diámetro característico.
\item[Evaporación] Este proceso implica sobre todo a gotas pequeñas, y por tanto $N_{t}$ debe disminuir y $\mu$ aumentar.
\end{description}
En cuanto a los procesos dinámicos locales que no son, estrictamente hablando, procesos microfísicos pero que pueden tener una relevancia en la DSD, tendremos las corrientes ascendentes y descendentes, así como una corriente de aire que dé lugar a una diferenciación (u ordenación) local por tamaños.
\begin{description}
\item[Corriente ascendente] Debe afectar principalmente a las gotas con menor masa y en consecuencia la DSD registrada a nivel del suelo posee un menor número de gotas pequeñas, sin alterar significativamente las grandes. Puede implicar un leve aumento del valor de $\mu$.
\item[Corriente descendente] El efecto opuesto de la corriente ascendente. 
\item[Ordenación por tamaños] Este proceso permite explicar intensos cambios en la DSD que o\-cu\-rren debido a procesos dinámicos \citep{AtlasUlbrich200microDSD}. La DSD se hace más estrecha al clasificarse las partículas por tamaños. El valor de $\mu$ se incrementa significativamente. 
\end{description}

Es posible además justificar las afirmaciones anteriores desde la base de una modelización concreta de la DSD. Partamos del modelo dado por la distribución gamma (que engloba a la exponencial) y supongamos un colectivo de gotas que caen en una masa de aire no saturado\footnote{Caso usual que además evita tener que incluir la condensación en la discusión. Para una explicación del significado de masa de aire saturado y no saturado, véase la referencia \citep{RogersBOOK}.}. La evaporación provocará la pérdida de agua líquida; por otra parte, esto es solo relevante para las gotas pequeñas de modo que la variación que implica en el colectivo total es pequeña. Esto permite suponer que el contenido de agua líquida total es constante. Por otra parte los procesos de ruptura y coalescencia actuando en conjunto producen un decrecimiento del número total de gotas. Dadas estas dos hipótesis y las expresiones introducidas en (\ref{sec:GAMMAdistrib}) se obtiene que una distribución exponencial evoluciona en una distribución gamma ($\mu>0$) al tiempo que $N_{0}$ y $\lambda$ aumentan \citep{ulbrich_1983_aa}. Esto justifica desde el punto de vista microfísico la preferencia de una distribución gamma sobre una exponencial al poder dar cuenta de variaciones físicas que realmente ocurren en las DSDs.\\

\begin{figure}[h] 
\begin{center}
   \includegraphics[width=0.5\textwidth]{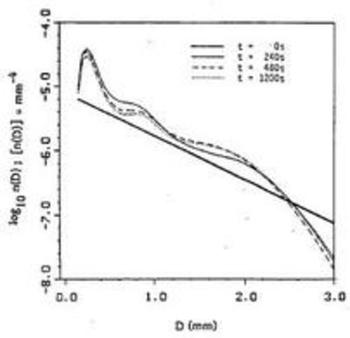}
\vspace{0.10cm}
   \caption[DSD de equilibro para los procesos de coalescencia y ruptura. Fuente: \citep*{MicroDSDBROWN}] {\textbf{DSD de equilibro para los procesos de coalescencia y ruptura. Fuente: \citep*{MicroDSDBROWN}} Se muestra la evolución de la DSD desde el caso inicial dado por una distribución exponencial con los valores propuestos por Marshall-Palmer, junto con la evolución para tres tiempos diferentes de 4 min, 8 min y 20 min.}
\label{fig:DSDequilibrium1}
\end{center}
\end{figure}

\begin{figure}[h] 
\begin{center}
   \includegraphics[width=0.5\textwidth]{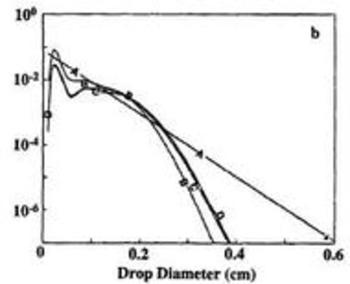}
\vspace{0.10cm}
   \caption[DSD de equilibro para los procesos de coalescencia, ruptura y evaporación. Fuente: \citep*{MicroDSDSrivastava1995}]{\textbf{DSD de equilibro para los procesos de coalescencia, ruptura y evaporación. Fuente: \citep*{MicroDSDSrivastava1995}}. Se muestra la evolución de la DSD desde el caso inicial A, dado por una distribución exponencial, para 10 min (línea B), 40 min (línea C) y 60 min (línea D). El caso D es la forma funcional de una distribución de tamaños de gota en equilibrio.}
\label{fig:DSDequilibrium2}
\end{center}
\end{figure}

El hecho de que la distribución gamma permita dar cuenta de procesos microfísicos de la DSD no implica que permita una descripción completa de toda la fenomenología presente. Las figuras (\ref{fig:DSDequilibrium1}) y (\ref{fig:DSDequilibrium2}) muestran el resultado que se obtiene mediante estudios de simulación 
 para la distribución de tamaños de gota. La idea general de estos estudios es, dada una distribución inicial de tamaños de gota (se suele tomar una distribución de Marshall-Palmer) determinar la forma que adquiere la DSD cuando los procesos de coalescencia, ruptura y evaporación están en \emph{equilibrio} y se alcanza una forma funcional estable. La conclusión general de estas simulaciones es la presencia de oscilaciones en la DSD, que en la mayoría de los estudios aparece en forma de tres picos. La posición donde están centrados es de modo aproximado 0.2-0.3 mm, 0.6-0.8 mm y 1.5-1.8 mm. En la práctica, esta DSD estable no se suele alcanzar completamente y es necesario sopesar qué procesos microfísicos tienen mayor relevancia en la forma funcional estimada para la distribución. Al tiempo, se ha observado cómo una distribución gamma no es una representación completa de la variabilidad real de la DSD, aunque posea gran utilidad en la modelización de medidas disdrométricas. \\

\section{Procesos físicos y su relevancia en la relación Z-R}

Los diferentes procesos físicos implican cambios en la DSD que se expresan también en la relación entre la reflectividad y la intensidad de la precipitación \citep{Steiner2004microphysisofZRgamma}.

\begin{table}[h]
\caption[Relevancia de los procesos microfísicos en la relación Z-R]{Relevancia de los procesos microfísicos en la relación Z-R, expresada como: 
$Z=a_{R}R^{b_{R}}$.}
\begin{center}
\begin{tabular}{llcc}
\toprule

\textbf{Tipo}   & \textbf{Proceso} & $\mathbf{a_{R}}$ & $\mathbf{b_{R}}$ \\
\midrule

Microfísico           &      &   &   \\

\midrule

&Evaporación           &  Descenso     &    Aumento \\
&Acrección             &  Aumento      &    Descenso   \\
&Coalescencia          &  Aumento      &    Descenso   \\
&Ruptura               &  Descenso     &    Descenso   \\

\midrule

Dinámico             &                  &            \\

\midrule

&Corriente ascendente  &  Aumento      &    Descenso   \\
&Corriente ascendente  &  Descenso     &    Descenso   \\
&Ordenación en tamaños &  Aumento      &    Descenso   \\

\bottomrule
\end{tabular}
\end{center}
\label{tablaZRmicrofisica}
\end{table}

Los procesos microfísicos y dinámicos, que se han comentado implican una variación de la DSD, tienen también su reflejo en las relaciones de potencias entre los diferentes parámetros integrales, en particular en la relación Z-R. Esta información es resumida en la tabla (\ref{tablaZRmicrofisica}).\\

En el caso de usar una modelización concreta de la DSD las relaciones entre los momentos aparecen expresadas en función de los parámetros libres de la distribución elegida. Un modo complementario de entender las relaciones entre los parámetros $(a_{R},b_{R})$ y los fenómenos microfísicos es expresar la relación Z-R para una parametrización dada. Estos estudios han sido realizados principalmente para la distribución exponencial, log-normal \citep{SmithKrajewski1993} y gamma \citep{Steiner2004microphysisofZRgamma}. En el caso de la distribución gamma, las expresiones para Z y R son:
\begin{equation}
Z=\frac{N_{0}}{\Lambda^{(7+\mu)}}\Gamma(7+\mu)
\end{equation}
y
\begin{equation}
R=C_{R}\frac{N_{0}}{\Lambda^{(4+p+\mu)}}\Gamma(4+p+\mu)
\end{equation}
Lo usual de cara a buscar una interpretación microfísica es sustituir los parámetros $N_{0}$ por $N_{t}=\frac{N_{0}}{\Lambda^{(1+\mu)}}\Gamma(1+\mu)$ e introducir $D_{m}=\frac{4+\mu}{\Lambda}$, que poseen una interpretación física más clara. Dado que existen multitud de fenómenos microfísicos interactuando que pueden dar lugar a diferentes relaciones Z-R, es útil diferenciar tres posible situaciones:
\begin{itemize}
\item Si igualamos $N_{t}$ en las expresiones podemos escribir la relación como:
\begin{equation} 
Z=A(D_{m},\mu) R^{1}
\end{equation}
De este modo en caso de no variar $D_{m}$ y $\mu$, tenemos una relación que se mantiene para cambios en el número total de gotas $N_{t}$.

\item Si despejamos $D_{m}$ en las expresiones podemos escribir la relación como:
\begin{equation} 
Z=A(N_{t},\mu) R^{6/(3+p)}
\end{equation}
De este modo en caso de no variar $N_{t}$ y $\mu$, tenemos una relación que se mantiene para cambios en $D_{m}$.
\item A partir de las expresiones anteriores es posible obtener un caso mixto:
\begin{equation} 
Z=A(D_{m}/N_{t},\mu) R^{(7+\mu)/(4+p+\mu)}
\end{equation}
Donde el exponente está condicionado por el factor de forma de la distribución gamma. Este caso es útil cuando $D_{m}/N_{t}\simeq cte$.
\end{itemize}
Los tres casos anteriores reflejan tres situaciones microfísicas diferentes. En el primero las variaciones naturales de la DSD están condicionadas solo por el número total de gotas, con lo que desde el punto de vista de la forma funcional se ha alcanzado un equilibrio entre los procesos de coalescencia, colisiones y ruptura entre gotas. En tal situación todos los momentos de la distribución están relacionados de forma lineal \citep{1988List}\footnote{También ha sido interpretada por \citep{jameson_kostinski_2002_aa} como precipitación en condiciones estadísticamente homogéneas.}. Cabe notar que no existe una única relación Z-R para esta situación física ya que dicho equilibrio podría ser alcanzado por diferentes parejas de valores $(D_{m},\mu)$.\\

La segunda situación física se daría en el caso de no producirse los procesos de coalescencia o ruptura, de tal modo que $N_{t}$ sea aproximadamente constante y las gotas aumenten su masa principalmente por agregación. Este tipo de situación ha sido propuesta para la microfísica de lluvias estratiformes estacionarias poco intensas, células convectivas en proceso de disipación y también la presencia de diferenciación por tamaños debido a fenómenos de turbulencia.\\

El último caso ha sido típicamente analizado para una exponencial donde $\mu=0$. En esta situación el exponente resulta $\beta=1.5$, donde el factor A crece con $D_{m}$ y decrece con $N_{t}$. Este es el caso obtenido por Marshall y Palmer (1948) ya que se supone constante el cociente $D_{m}^{1+\mu}/N_{t}$, que se puede interpretar para $\mu=0$ como constancia del parámetro $N_{0}$.\\

\section{Microfísica del método de escalado en R de la DSD}
\label{sec:microESCALADOdsd}
Respecto de la interpretación en términos microfísicos de la modelización mediante el método de escalado respecto de R, se han venido relacionando los procesos microfísicos descritos con el parámetro $\beta$ obtenido del ajuste de las relaciones de potencias entre los parámetros integrales de la precipitación. De esta manera los modelos de crecimiento de la DSD cuando la intensidad de precipitación aumenta son asociados a $\beta$ y se clasifican como:

\begin{description}
 \item[$\beta = 0$] Implica relaciones lineales entre los momentos de la DSD, lo que, como se ha comentado previamente se relaciona con el equilibrio entre los procesos de coalescencia y ruptura. Se considera que esta situación puede producirse en ciertos procesos convectivos.
\item[$0 < \beta < 0.214$] Valores menores se suelen relacionar con convección y valores mayores con procesos típicos de la precipitación estratiforme. En este último caso, a menudo se registra también un incremento del diámetro medio. El número de gotas crece con R en todo el espectro de tamaños.
\item[$\beta = 0.214$]  Las diferencias en la distribución de tamaños de gota al incrementarse R se acentúan para gotas grandes, pero se mantiene el valor de $N_{0}$ constante tal y como describe el modelo de Marshall-Palmer.
\item[$\beta > 0.214$] Las diferencias en la DSD al incrementarse R se acentúan para gotas grandes mientras que la concentración de gotas pequeñas disminuye. Los fenómenos de agregación son los dominantes.
 \end{description}

\begin{figure}[h] 
   \centering
   \includegraphics[width=0.85\textwidth]{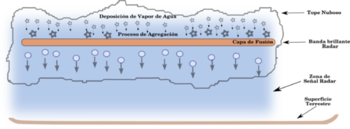} 
\vspace*{1.0cm}

   \includegraphics[width=0.80\textwidth]{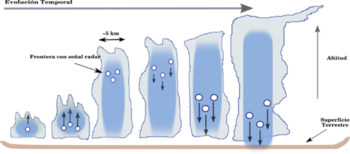} 
\vspace{0.3cm}
   \caption[Precipitación Estrafitorme \textit{vs.} Convectiva.]{\textbf{Precipitación Estratiforme \textit{vs.} Convectiva.} Se aprecia la diferencia entre un sistema estratiforme y un sistema convectivo. Se ve como en el primero aparece una banda brillante en la imagen radar relacionada con la capa de fusión. En la parte de abajo se ve el sistema convectivo, más propio de sistemas mesoescalares y más rápido en la formación final de precipitación debido a las altas velocidades de ascensión de masas de aire que lo caracterizan.}
\label{fig:ConvectivoVSestratiforme}
\end{figure}

\section{Precipitación convectiva \emph{vs.} estratiforme}

De modo genérico, la precipitación se suele clasificar en estas dos categorías generales. Su utilidad viene determinada por el hecho de ser la DSD ligeramente diferentes en ambos casos, cuestión que se puede relacionar con los procesos microfísicos que predominan en cada caso. En la figura (\ref{fig:ConvectivoVSestratiforme}) se esquematizan las dos categorías. Como se ha comentado en el capítulo precedente, de cara a clasificar las relaciones Z-R también se alude a esta clasificación.

\section{Sumario/Summary}

A brief summary of the topics commented in this chapter is:

\begin{itemize}
\item The functional form of the DSD depends on the micro-physical processes. For this reason, different cloud systems mean differences on the DSD. The dynamical processes, like local turbulence or updrafts/downdrafts, have an extra influence on the functional form of the DSD.
\item The exponential distribution model can't explain the full correlation between micro-physics and observed DSD, while the gamma distribution allows us to build the connection between the funcional form of DSD and the different micro-physical processes. However the direct simulation of coalescence, breakup and evaporation shows an oscillating functional form which needs more elaborated models.
\item The Z-R relationship is also conditioned by the micro-physics of precipitation showing that it is important an analysis of the Z-R relationship for each kind of event and geographical localization.  
\item The relationships between the time series of $D_{m}$, $\mu$ y $N_{t}$ can be related with the values of Z-R relationships using the gamma model of the drop size distribution.
\end{itemize}


\part{Base empírica}
\renewcommand\chapterillustration{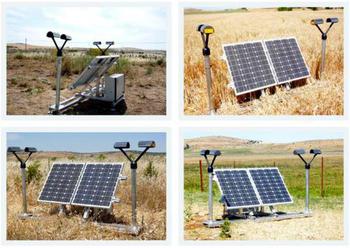}
\chapter{La red de disdrómetros de la UCLM}
\label{sec:baseempirica}

En este capítulo se describe la base empírica utilizada para el estudio de la variabilidad espacial de la precipitación. El primer paso es contextualizar las posibles escalas de estudio, para después describir el diseño del experimento que se ha llevado a cabo en el curso de esta tesis, y con él analizar una de estas escalas. Los datos generales de la base empírica obtenida en la primera campaña de recogida de datos se describen también en este capítulo. 

\section{Escalas de análisis experimental}

La variabilidad espacial y temporal de la precipitación se produce a diferentes escalas, que comprenden desde distancias del orden del tamaño de los propios hidrometeoros que constituyen la precipitación hasta distancias propias de la escala sinóptica. La metodología experimental para estudiar cada rango diferirá, condicionada tanto por sus propiedades físicas relevantes como por las limitaciones técnicas de cada método. A grandes rasgos se pueden diferenciar las siguientes escalas espaciales:

\begin{description}
\item [0.1-1 m] Esta escala se suele denominar micro-escala de la precipitación, ya que los estudios aluden a las propiedades de micro-estructura de esta. La variación espacial a la resolución comprendida entre centímetros y metros se denomina \emph{textura de la precipitación}. El principal problema de estudio será si, a esta escala, la distribución espacial de gotas es razonablemente homogénea o si, por contra, existen efectos de agrupamiento notables \citep{Kostinski200638, 2009NPGeo16287U}.\\

En el primer caso se trataría de una ausencia total de correlaciones entre las gotas a escalas del propio tamaño de estas; en el segundo caso se encontraría una correlación espacial (po\-si\-ti\-va o negativa) junto con la presencia de fluctuaciones locales.\\

Este segundo caso poseería una longitud de escala característica de estas fluctuaciones (dependiendo de las propiedades de las correlaciones con la distancia se puede hablar de comportamiento fractal si se aprecia cierta invariancia de escala). \\

\begin{figure}[h] 
\begin{center}
   \includegraphics[width=1.00\textwidth]{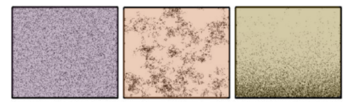}
   \caption[Micro-escala de la precipitación. Fuente:  \citep{Kostinski200638}.]{\textbf{Ejemplos de posibles micro-estructuras de precipitación.  Fuente:  \citep{Kostinski200638}.} \emph{Izquierda}: Caso homogéneo e isótropo. \emph{Centro}: caso no homogéneo con la presencia de agregados y fluctuaciones del tamaño típico de estos agregados, pero con isotropía global. \emph{Derecha}: caso no homogéneo y además no isótropo en dirección vertical.}
\end{center}
\label{fig:Var1a}
\end{figure}

La hipótesis homogénea inicial ha sido poco a poco cuestionada por evidencias de posible agrupamiento (\emph{clustering}). En todo caso, en este ámbito no se han realizado hasta el momento experimentos concluyentes. Algunos trabajos \citep{1990JApMe291167L} proponen una interpretación en que se considera la naturaleza de la precipitación como fractal, que ha sido cuestionada por otros autores \citep{2009NPGeo16287U}. Una de las consecuencias prácticas del problema es que la textura de la precipitación tendría que ver con la simulación de distribuciones realistas de tamaños de gota y si es viable generar dichas distribuciones mediante un proceso de Poisson, un proceso no homogéneo de Poisson u otro método.\\

Los experimentos a esta escala han sido realizados mediante el análisis de depósito de gotas en materiales específicos con objeto de estudiar la distribución espacial de estas gotas, aunque en los últimos años se han diseñado experimentos con videocámaras para estudiar la distribución espacial tridimensional \citep{2006JHydropExperiment} y no solo su proyección bidimensional.

\item [1-10 metros] A esta escala, una de las principales cuestiones es la relevancia de las turbulencias locales en la distribución de tamaños de gota. Variaciones locales, a distancias de metros, en el campo de viento, pueden condicionar las medidas nominales de la DSD si se utiliza un único disdrómetro. Para paliar este efecto es conveniente utilizar varios disdrómetros situados en el mismo lugar. Esta es una de las razones por las que se aconseja usar disdrómetros duales en experimentos de variabilidad. Algunos estudios han sido realizados en este sentido \citep{TokayCOLLOCATEDjwd2005}; sin embargo, hasta el momento no se ha realizado ningún experimento de campo con un número suficiente de disdrómetros que permita relacionar el campo de velocidades del viento con variaciones a esta escala de la DSD.

\item [100-1000 m] El principal objetivo es caracterizar la variabilidad espacial de la DSD dentro de un pixel tanto de un radar meteorológico terrestre como de los últimos radares incorporados a satélites (TRMM) o los que van a ser incorporados (GPM)\footnote{La fecha de lanzamiento del GPM-core está prevista para el 14 de febrero del 2014.}. Su estudio es importante, tanto para la validación de productos relacionados con medidas de teledetección\footnote{En \citep{Ciach1999585} se define el \textit{problema de validación} como la cuantificación y caracterización estadística del error en la predicción mediante teledetección de la intensidad de precipitación. El \textit{problema de la estimación} sería el de construir algoritmos que impongan las propiedades caracterizadas sobre el error.} como para la realización de análisis de variabilidad que cubran escalas geográficas mayores. Esta será la escala analizada más adelante, siendo su estudio uno de los proyectos que se están realizando con la red de disdrómetros utilizada en esta tesis. \\

Es importante notar que algunos modelos numéricos de predicción (NWP) están llegando a resoluciones del orden de 1-5 $km$, con lo que estos estudios de variabilidad son relevantes para las parametrizaciones de precipitación y microfísica incluidas en estos modelos. Además son útiles en combinación con estudios de simulación de la física de nubes. Los estudios rea\-li\-za\-dos en distancias en este rango se denominan de \emph{pequeña escala}\footnote{En la referencia \citep{Tokay2010smallscaleDSD} se precisa que pequeña escala se define, para un sensor remoto determinado, como aquella escala en el estudio de variabilidad que necesita de otros instrumentos.}.\\

Al respecto de estudios previos estos se han realizado principalmente mediante radar \citep{2009ChoongLEE} lo que, como se comentó en \S\ref{sec:relacionesZR}, presenta ciertas dificultades, y solo es posible estudiar de este modo la variabilidad cuando todos los otros posibles factores diferentes de las variaciones naturales de la DSD han sido eliminados o controlados. Igualmente se han utilizado redes de pluviómetros que permiten estudiar la variabilidad en esta escala de la intensidad de la precipitación. Por otra parte no hay experimentos fiables que evalúen la variabilidad de la DSD y en consecuencia un análisis directo de su implicación en las relaciones Z-R.\\

Existe además otra aplicación importante ya que, caracterizada la variabilidad de la precipitación en la escala de resolución espacial de los satélite actuales, es posible determinar el número de pasos que ha de realizar un satélite polar por una zona determinada para identificar los errores en los algoritmos de predicción de la intensidad de lluvia basados en imágenes de satélite \citep{1999HaNorth}. Esto implica estudiar la variabilidad en diferentes latitudes con redes densas de instrumentos puntuales.

\item[$\mathbf{>10}$ km ] Estas escalas son importantes en estudios que van desde la mesoescala (análisis de eventos extremos \citep{Berne2009flood}, por ejemplo) hasta la climatológica, donde la variación temporal de la distribución espacial es relevante. Establecer conclusiones requiere conocer la incertidumbre en las medidas aportadas en escalas menores.
\end{description}


\begin{figure}[h] 
\begin{center}
   \includegraphics[width=0.65\textwidth]{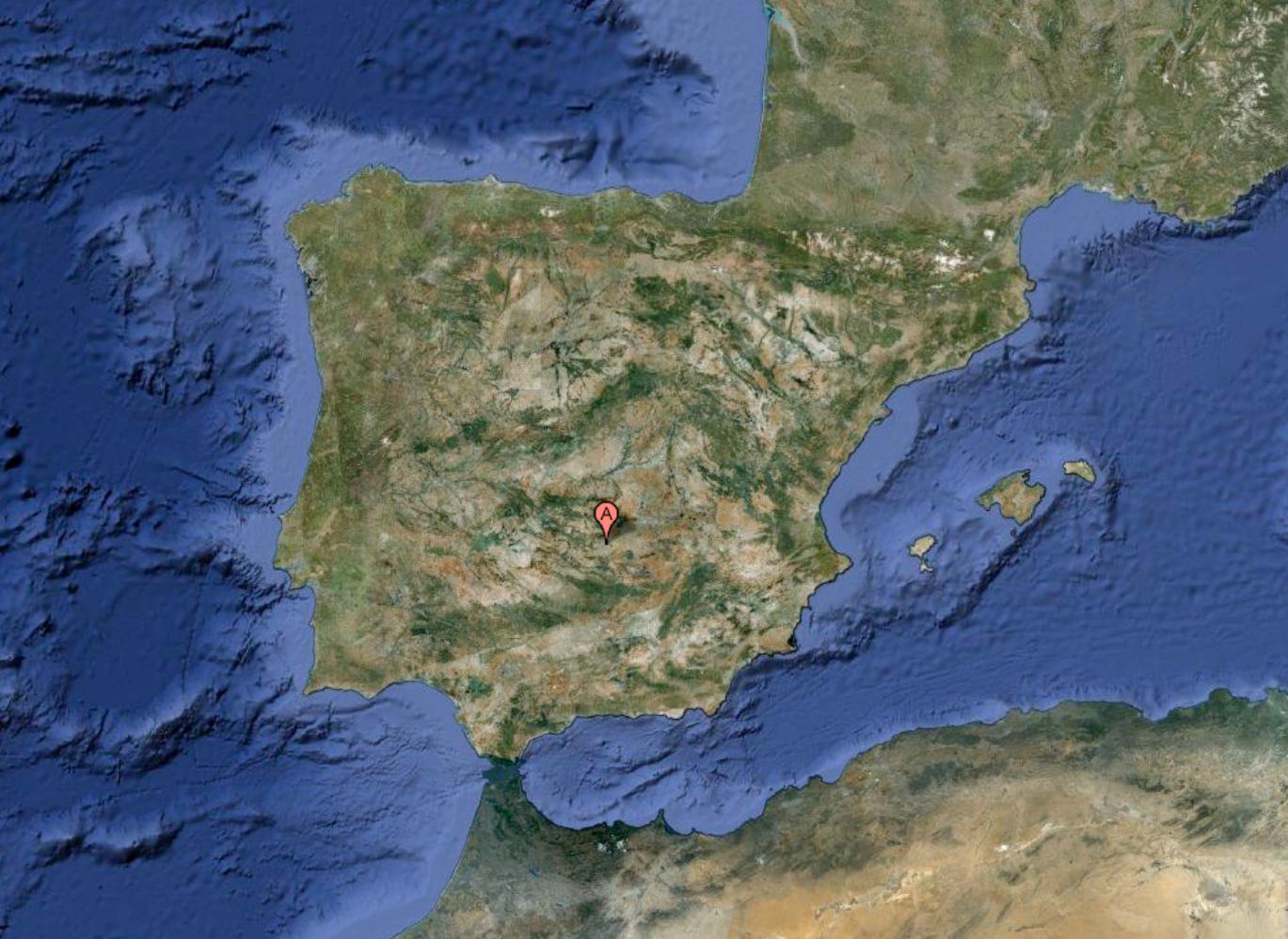}
\vspace{0.45cm}
   \caption[Localización geográfica del experimento]{\textbf{Localización geográfica del experimento.} La red de disdrómetros se encuentra situada en la Meseta, en la provincia de Ciudad Real, a una altitud promedio de 600 $m$, en una zona con orografía aproximadamente plana, que presenta solo leves variaciones del 5\% entre diferentes estaciones.}
\label{fig:Var1:LOC}
\vspace{0.25cm}
\end{center}
\end{figure}

\section{Diseño experimental}

\begin{figure}[h] 
\begin{center}
   \includegraphics[width=0.87\textwidth]{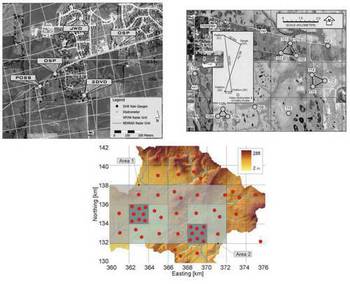}
\vspace{0.25cm}
   \caption[Redes de instrumentos desplegadas en otros experimentos]{\textbf{Redes de instrumentos desplegadas en otros experimentos.} Arriba a la izquierda se aprecia la red estudiada en \citep{miriovsky_bradley_etal_2004_aa} mediante un sistema heterogéneo de instrumentos. Arriba a la derecha: experimento de validación del proyecto TRMM mediante la campaña TEFLUN-B \citep{2002TeflunBcampaingHabib} mediante pluviómetros. Abajo: un experimento a varias escalas usando pluviómetros \citep{villarini_mandapaka_etal_2008_aa}. Se aprecia como en cada sub-celda densamente poblada de instrumentos se desecha la disposición regular en forma de cuadrados.}
\label{fig:Var1redOTROS}
\end{center}
\vspace{0.25cm}
\end{figure}

La escala espacial que se ha estudiado en esta tesis comprende distancias de hasta 2.5 $km$, por tanto se enmarca en estudios de \emph{pequeña escala} de la DSD correspondientes al programa de validación sobre el terreno del proyecto GPM\footnote{La misión de medida de precipitación global (GPM) es una red internacional de satélites que constituirá la base de las medidas futuras de lluvia y nieve a escala global. Surgió por el éxito obtenido en la misión \textit{Tropical Rainfall Measuring Mission} (TRMM), que combinaba diferentes sensores de precipitación en un mismo satélite. Entre los objetivos científicos se espera que el proyecto GPM contribuya al mejor entendimiento del ciclo hidrológico a nivel global, mejorar la predicción de eventos extremos, así como una mayor comprensión de la microfísica de la precipitación y su interrelación con los procesos de dinámica atmosférica.  El principal componente de la red de satélites que constituirán el GPM se compone de un sensor de microondas y dos radares operando en las bandas Ku y Ka. Además se espera que otros satélites y sensores relacionados con el proyectos GPM se beneficien de los nuevos sensores. Algunos ejemplos son: MADRAS, SAPHIR, SSMIS, el WindSat, el satélite Aqua y el sensor AMSU.} (véase \citep[chap. 6]{BookPrecipitation}). En la figura (\ref{fig:Var1redOTROS}) se comparan diferentes redes de instrumentos diseñadas también para investigar la variabilidad de la precipitación a  escala kilométrica. Tal y como se indica en \citep{villarini_mandapaka_etal_2008_aa}, con\-fi\-gu\-ra\-cio\-nes diferentes de la regular-cuadrangular son preferibles para estudios de hidro-meteorología.\\

\begin{figure}[h] 
\begin{center}
   \includegraphics[width=0.45\textwidth]{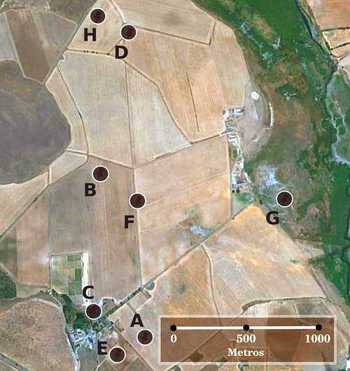}
\vspace{0.55cm}
   \caption[Red de disdrometros Galiana 2009-2010]{\textbf{Esquema de la distribución espacial de los disdrómetros en la red.} La disposición es irregular para cubrir el máximo número de distancias posibles al tiempo que en cada sub-escala se disponen varias subredes. Obsérvese que las parejas H-D, B-F y C-A están a distancias similares pero con leves diferencias. Las distancias G-D, G-B, G-E y G-C son del mismo orden pero también con pequeñas diferencias. Contrastar con figura (\ref{fig:Var1redOTROS}).}
\label{fig:Var1red}
\end{center}
\vspace{0.05cm}
\end{figure}

El experimento está constituido por 8 parejas de disdrómetros Parsivel calibrados de fábrica que han sido colocados en una red homogénea (aparatos idénticos) y no regular de 6 $km^{2}$. Uno de los aspectos más importantes del diseño experimental ha sido el disponer de instrumentos capaces de ser autónomos, tanto en el suministro eléctrico como en la emisión y registro de los datos experimentales. Con este fin los instrumentos poseen:
\begin{itemize}
\item Un panel solar por cada pareja de disdrómetros que, junto con una batería, permite una autonomía suficiente tanto en horas/días despejados como en horas nocturnas/días nubosos. Los experimentos previos realizados han mostrado que el voltaje suministrado por la batería es el apropiado para obtener un sistema de medición continuo. Uno de los datos registrados de modo continuo por los instrumentos es la carga de dicha batería. Los paneles solares están orientados al sur con una inclinación dependiente de la latitud para maximizar la irradiancia solar.
\item Cada pareja de disdrómetros posee una CPU integrada usando un entorno Linux que, junto con un sistema GPRS, permite enviar los datos a un servidor apropiado donde se almacenan sincronizadamente minuto a minuto.
\item Para este experimento cada pareja de disdrómetros X1-X2, ha sido colocada en orientaciones Norte-Sur (disdrómetro X1) y Este-Oeste (disdrómetro X2) para estimar las posibles di\-fe\-ren\-cias debidas al viento local en algunos eventos y poder estimar cuando un disdrómetro presenta problemas operativos. Esto puede ser útil además ya que la orografía no es completamente plana y en algunos eventos específicos son posibles leves turbulencias de origen local. 
\item Cada pareja es georeferenciada mediante GPS (los datos se pueden encontrar en la tabla (\ref{tabla:GPSred}) junto con la altitud de cada estación sobre el nivel del mar) mientras que desde estos datos GPS es posible calcular con una resolución adecuada los valores de las distancias entre parejas de disdrómetros de la red. Véase tabla (\ref{TablaDistances}).
\item Los elementos no han sido dispuestos en una red regular, ya que esto permite cubrir un conjunto de distancias entre parejas mayor y de modo que cada sub-escala aparezca representada más de una vez. Véanse las figuras (\ref{fig:Var1red}) y (\ref{fig:Var1redOTROS}).
\item Las características de cada disdrómetro han sido dadas en el capítulo \textsection\ref{sec:Instrumentacion}.
\end{itemize}


\begin{figure}[h] 
\begin{center}
   \includegraphics[width=0.55\textwidth]{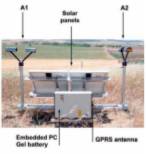}
\vspace{0.35cm}
   \caption[Dual Parsivel OTT]{\textbf{Detalle de los instrumentos duales utilizados en el experimento.} Se compone de dos disdrómetros Parsivel OTT situados perpendicularmente a una distancia de 1.2 metros. El suministro eléctrico se consigue mediante la combinación de paneles solares y una batería. La transmisión de datos se realiza sincronizadamente entre todos los disdrómetros de la red mediante una antena GPRS.}
\end{center}
\label{fig:Var1Dual}
\vspace{0.25cm}
\end{figure}

\begin{table}[htdp]
\ra{1.15}
\caption[Coordenadas GPS disdrómetros duales Galiana 2009-2010.]{\textbf{Coordenadas GPS disdrómetros duales Galiana 2009-2010.} Para apreciar su lo\-ca\-li\-za\-ción véase la figura (\ref{fig:Var1red}) a una escala kilométrica y la figura (\ref{fig:Var1:LOC}) a una escala sinóptica. Se ha incluido también la altitud a la que está cada uno de los instrumentos. No hay desniveles significativos relativos a la distancia horizontal entre parejas excepto entre las parejas B-F, que estando próximas tienen un desnivel de 8 metros.}
\vspace{0.15cm}
\begin{center}
\begin{tabular}{crrr}
\toprule
\textbf{Disdrómetro Dual} & \textbf{Latitud}  & \textbf{Longitud} & \textbf{Altitud} \\
\midrule
A   & 38.9788 Norte  &  4.0396 Oeste & 595 m\\ 
B   & 38.9888 Norte  &  4.0431 Oeste & 593 m\\ 
C   & 38.9804 Norte  &  4.0436 Oeste & 594 m\\ 
D   & 38.9973 Norte  &  4.0408 Oeste & 593 m\\ 
E   & 38.9778 Norte  &  4.0417 Oeste & 594 m\\ 
F   & 38.9871 Norte  &  4.0401 Oeste & 585 m\\ 
G   & 38.9872 Norte  &  4.0286 Oeste & 576 m\\ 
H   & 38.9983 Norte  &  4.0433 Oeste & 595 m\\ 

\bottomrule
\end{tabular}
\end{center}
\label{tabla:GPSred}
\end{table}%

\begin{table}[t]
\ra{1.15}
\caption[Matriz Distancias Red Galiana 2009-2010]{\textbf{Matriz Distancias Red Galiana 2009-2010.} Las distancias están medidas en metros para cada instrumento dual. La distancia entre los elementos de cada instrumento dual es de 1.2 metros. Cabe resaltar que un histograma de distancias construido a partir de esta tabla posee siempre varios disdrómetros en las escalas 0-500 metros, 500-1000 metros, 1000-1500 metros, 1500-2000 metros, 2000-3500 metros.}
\vspace{0.15cm}
\begin{center}
\begin{tabular}{rrrrrrrrr} 
\toprule
       & \textbf{A1,A2} & \textbf{B1,B2}& \textbf{C1,C2} &  \textbf{D1,D2}& \textbf{E1,E2}& \textbf{F1,F2}  &  \textbf{G1,G2} & \textbf{H1,H2} \\
\midrule
\textbf{A1,A2}  &   0  &1\,159 &   397 &  1\,960 & 232 &    923 &  1\,333  & 2\,199 \\
\textbf{B1,B2}  &      &   0 &   938 &   961 & 1\,127 &    326 &  1\,276  & 1\,057 \\
\textbf{C1,C2}  &      &     &     0 &  1\,893 & 320 &    801 &  1\,506  & 1\,994 \\
\textbf{D1,D2}  &      &     &       &     0 & 2\,171 &   1\,138 &  1\,548  &  244 \\
\textbf{E1,E2}  &      &     &       &       &    0 &   1\,047 &  1\,558  & 2\,282 \\
\textbf{F1,F2}  &      &     &       &       &      &      0 &  1\,003  & 1\,283 \\
\textbf{G1,G2}  &      &     &       &       &      &        &     0  & 1\,782 \\
\textbf{H1,H2}  &      &     &       &       &      &        &        &    0 \\
\bottomrule
\end{tabular}
\end{center}
\label{TablaDistances}
\end{table}

\section{Base empírica}

La localización geográfica del experimento puede verse en la figura (\ref{fig:Var1:LOC}). Está situado en latitudes medias: el centro de la red tiene coordenadas GPS (WGS-84)  38.98\degree Norte 4.04 \degree Oeste. Está situada en la Meseta Ibérica a una altitud media de 600 metros sobre el nivel del mar.\\

\subsection{Control de calidad}

El experimento consta de una fase previa a la toma sistemática de datos en que se realiza un análisis comparando datos obtenidos por los aparatos desplegados en Galiana y un instrumento situado en Toledo en la Universidad de Castilla-La Mancha. La fase previa se realizó entre los meses de septiembre y noviembre del 2009 y en ella se constató el comportamiento idéntico en términos de fiabilidad entre los instrumentos situados en diferentes localizaciones.\\

Durante la fase de toma sistemática de datos se realizaron comprobaciones continuadas tanto de la transmisión como de la correlación de los instrumentos duales, realizando un mantenimiento de los aparatos y verificando la influencia mínima de factores no meteorológicos en las medidas (vegetación y fauna).\\

La elección final del tiempo de acumulación de datos de 60 segundos mostró ser, en la fase de control inicial, la más adecuada para que posibles problemas de sincronización en los relojes internos de las CPU-integradas de cada estación (del orden de 1 segundo) no tengan relevancia estadística. Resoluciones temporales mayores incrementarían las posibilidades de pérdida de datos en la transmisión por GPRS.

\subsection{Características climáticas y meteorológicas}
\label{sec:basemepiricaCLIMA}

Los datos climáticos (fuente: AEMET) revelan que el número típico de días con lluvia al año es de 62. En los últimos 30 años (periodo 1971-2000) el mayor registro de precipitación ha sido de 620 mm/año y el mínimo de 235 mm/año. El promedio de precipitación acumulada es de 396 mm/año. Los datos climáticos por meses relevantes para la campaña de datos analizada en esta tesis pueden verse en la tabla (\ref{TablaCLIMATICdata}), donde se aprecia que los meses analizados (diciembre, enero, febrero y marzo) poseen mayor precipitación.\\

\begin{figure}[h] 
\begin{center}
   \includegraphics[width=0.75\textwidth]{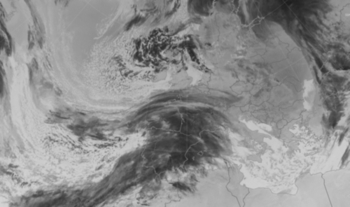}
\vspace{0.15cm}
   \caption[Imagen METEOSAT 9. Sensor SEVIRI. IR 10.8$\mu$m. Evento: 20-21 de diciembre del 2009. Hora: 02:00 UTC.]{\textbf{Imagen METEOSAT 9. Sensor SEVIRI. IR 10.8$\mu$m. Evento: 20-21 de diciembre del 2009. Hora: 02:00 UTC.} Imagen suministrada por SATREP online/EUMETSAT. La imagen ha sido invertida para apreciar mejor la menor temperatura de los topes nubosos.}
\label{fig:IMAGENevento21DicIR}
\end{center}
\end{figure}

En general, los episodios de precipitación proceden de frentes que atraviesan la Península Ibérica, siendo muy poco probable la presencia de tormentas de carácter convectivo que, en el caso de la meseta en la Península Ibérica, se concentran en los meses del verano (mayo hasta agosto). Un ejemplo del tipo de nubosidad que genera la precipitación estudiada se aprecia en la figura (\ref{fig:IMAGENevento21DicIR}) donde se muestra una imagen infrarroja (IR 10.8) centrada sobre la Meseta para uno de los episodios que se analizan en esta tesis.\\

\begin{figure}[H] 
\begin{center}
\vspace{0.65cm}
    \includegraphics[width=1.03\textwidth]{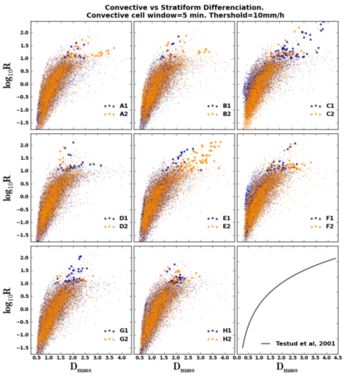}
\vspace{0.75cm}
   \caption[Intensidad de precipitación \emph{vs.} $D_{mass}$ a lo largo de la red de disdrómetros para toda la base empírica]{\textbf{Intensidad de precipitación R \emph{vs.} $\mathbf{D_{m}}$ a lo largo de la red de disdrómetros para toda la base empírica} Se representa $log_{10}R$ vs $D_{mass}$ para todos los disdrómetros. Los puntos representan minutos clasificados como estratiformes por el criterio de \citep{NormalizadaTestud2001}. Los círculos de mayor tamaño representan minutos clasificados como convectivos. La línea representa $R=1.64\,D^{4.65}$, relación introducida en \citep{NormalizadaTestud2001} para diferenciar precipitación estratiforme y convectiva.}
\label{fig:Var1RvsDm}
\end{center}
\vspace{0.75cm}
\end{figure}

Por tanto, desde el punto de vista de la distinción entre tipos de precipitación, la base empírica utilizada en este trabajo se compone de eventos que dan lugar a precipitación estratiforme\footnote{Véase la figura (\ref{fig:DSDequilibrium1}) para una explicación esquemática de la distinción entre ambos tipos de precipitación.}. Dados los resultados experimentales de la base empírica completa es posible contrastarlo, para ello se pueden analizar las relaciones típicas entre parámetros integrales \citep{NormalizadaTestud2001} y de una distribución gamma \citep{TokayShort1996}\footnote{Otra metodología propuesta por \citep{NormalizadaTestud2001} consiste en clasificar una DSD como estratiforme si y solo si la intensidad de precipitación es menor de 10 $mm/h$, y en los 10 minutos precedentes y posteriores también. La idea de este procedimiento es evaluar la posible influencia de una célula convectiva en las DSD. Véase también \citep{moumouni_gosset_etal_2009_aa}.}. El análisis en el primero de los casos comprende la comparación de diagramas de intensidad de precipitación frente al diámetro característico (dia\-gra\-ma $R/D_{m}$).\\


\begin{table}[h]
\ra{1.10}
\caption[Datos climáticos de la estación meteorológica de la AEMET más cercana al experimento. Fuente: AEMET.]{\textbf{Datos climáticos de la estación experimental situada en Galiana. Fuente: AEMET.} Datos climáticos indicados por meses, promediados sobre el periodo 1971-2000 (30 años). Los datos se refieren a la estación meteorológica más cercana del experimento a 10 km de distancia. Altitud (m): 628. Latitud: 38\degree 59' 22'' N -  Longitud: 3\degree 55' 11'' O.}
\vspace{0.50cm}
\begin{center}
\begin{tabular}{lcccc}
\toprule
\textbf{Mes} & \textbf{R total (mm/mes)}  & \textbf{Tormentas}  & \textbf{Nieve} & \textbf{Días} $R>1mm$ \\
\midrule
Sep    & 22             &   2  & 0   & 6\\ 
Oct    & 47             &   1  & 0   & 6\\
\midrule
Nov    & 42             &   0  & 0   & 6\\
Dic    & 55             &   0  & 0   & 8 \\
Ene    & 36             &   0  & 1   & 6 \\
Feb    & 34             &   0  & 1   & 6\\
Mar    & 28             &   1  & 0   & 5 \\
Abr    & 44             &   1  & 0   & 8 \\
\midrule
May    & 43             &   3  & 0   & 7\\
Jun    & 29             &   4  & 0   & 4\\
Jul    & 9             &   2  & 0   & 1\\
Ago    & 7             &   2  & 0   & 1\\
\bottomrule
\end{tabular}
\end{center}
\label{TablaCLIMATICdata}
\vspace{0.70cm}

\end{table}

La figura (\ref{fig:Var1RvsDm}) representa la base empírica completa junto con la línea de separación entre precipitación estratiforme y convectiva propuesta por \citep{NormalizadaTestud2001} que aparece en el último panel. En la figura se han incluido todos los minutos de la base empírica sobre un conjunto consistente de datos en toda la red y se ha clasificado cada distribución de tamaños de gota como convectiva o estratiforme siguiendo el siguiente criterio: si en una ventana que abarca 5 minutos previos y cinco posteriores existe al menos un minuto con intensidad de precipitación mayor de 10 $mm/h$ y el mínimo en todo el intervalo es de 9.5 $mm/h$ se clasifica como convectivo. En caso contrario, el espectro en ese minuto es clasificado como estratiforme. Se han probado ventanas de tiempo mayores lo que disminuye el número de minutos convectivos, aunque no de modo notable. Se observan algunas cuestiones significativas: los disdrómetros C1 y E2 registraron un mayor número de minutos convectivos, mientras que todos los minutos convectivos registrados aparecen concentrados en un mismo evento el 23 de diciembre del 2009. Con respecto a la propuesta de \citep{NormalizadaTestud2001} para la relación $R/D_{m}$ límite entre ambos tipos de precipitación, se cumple de modo aproximado, dado que el conjunto de datos utilizados para su elaboración (TOGA-COARE data-set) procedía de instrumentos diferentes en otra latitud y altitud. Véase también \citep{moumouni_gosset_etal_2009_aa}.\\

\subsection{Propiedades generales}

La base empírica analizada se extiende desde 01/12/2009 hasta 18/03/2010. El número total de minutos por disdrómetro varía entre cada instrumento dual debido a problemas lógicos de transmisión GPRS. En consecuencia un cierto número de minutos no es posible registrarlos en su totalidad. De cara al estudio de las correlaciones entre series temporales es necesario realizar un filtrado correspondiente a la coherencia o congruencia de minutos entre toda la red, lo que lleva a eliminar todos aquellos minutos en que cualquiera de los disdrómetros no aporte información, de manera que el conjunto total de datos analizado congruente corresponde a una base empírica de 14\,227 minutos.\\

Adicionalmente, es posible aplicar acotaciones diversas a los valores de la intensidad R para los episodios con precipitación líquida, además de incluir solo minutos con un número mínimo de gotas, lo que corresponde con una distinción entre lluvia y no lluvia (cuyo valor se ha establecido en 10 gotas). Es además un umbral mínimo bajo el que puede considerarse posible realizar mo\-de\-li\-za\-ci\-o\-nes de la DSD a una forma funcional fija mediante los métodos tradicionales. La tabla (\ref{TablaMINUTOSdata}) muestra el número total de minutos que persisten bajo este filtrado en toda la red de disdrómetros a un tiempo (el caso de $R > 0.1\,mm/h$ es el considerado estándar en la separación entre minutos de lluvia y de no lluvia).\\

\begin{table}[h]
\vspace{0.25cm}

\ra{1.15}
\caption[Número de minutos de la base empírica bajo diferentes acotaciones.]{\textbf{Número de minutos de la base empírica bajo diferentes acotaciones.} Se ha aplicado un filtrado en que solo se seleccionan minutos con precipitación líquida y con al menos 10 gotas \textit{en todos los disdrómetros}. Se procede después a aplicar acotaciones en R, a un tiempo, \textit{en toda la red de disdrómetros}. El número de minutos resultantes con datos de precipitación se muestra en esta tabla.}
\vspace{0.45cm}
\begin{center}
\begin{tabular}{llrrrrr}
\toprule
  \textbf{Acotaciones} & & \multicolumn{5}{c}{$R_{min}$}   \\
\cmidrule(l{.5em}){3-7}
 R [mm/h] & &  0.0 &  0.05 &  0.1 &  0.2 &  0.5 \\
\midrule
        & 10 &  7\,551 &  6\,103 &  5\,167 &  4\,111 & 2\,796 \\
$R_{max}$ & 20 &  8\,078 &  6\,628 &  5\,688 &  4\,626 & 3\,283 \\
        & $\infty$ &  8\,254 &  6\,802 &  5\,861 &  4\,794 & 3\,450 \\

\bottomrule
\end{tabular}
\end{center}
\label{TablaMINUTOSdata}
\vspace{0.75cm}
\end{table}%

\clearpage
Sobre esta base de minutos totales se ha procedido a estudiar y comparar el total de lluvia acumulada en cada instrumento. La tabla (\ref{tablaBIASwhole}) muestra el sesgo en porcentaje según la expresión:\footnote{Otras definiciones usuales que provienen de los estudios pluviométricos son (tomamos como referencia la media (\textit{ave}) pero puede ser un instrumento específico que se considere fiable):

\begin{equation}
Bias(X_{d})=\frac{1}{N_{min}}\sum_{i=1}^{N_{min}}x_{i}^{d}-x_{i}^{ave}
\label{eqn:BIAS}
\end{equation}

\begin{equation}
Absolute\, Bias(X_{d})=\frac{1}{N_{min}}\sum_{i=1}^{N_{min}}|x_{i}^{d}-x_{i}^{ave}|
\label{eqn:BIAS}
\end{equation}

\begin{equation}
Weighted\,Bias(X_{d})=\frac{1}{N_{min}}\sum_{i=1}^{N_{min}}\omega_{i}(x_{i}^{d}-x_{i}^{ave})
\label{eqn:BIAS}
\end{equation}

\begin{equation}
Weighted\, Absolute\, Bias(X_{d})=\frac{1}{N_{min}}\sum_{i=1}^{N_{min}}\omega_{i}|x_{i}^{d}-x_{i}^{ave}|
\label{eqn:BIAS}
\end{equation}

La función peso $\omega_{i}$ se suele definir como la media de instrumento bajo estudio y el de referencia para ese minuto entre la media sobre todos los minutos analizados.
},

\begin{equation}
Percent\, Bias(X_{d})=\frac{\sum_{i=1}^{N_{min}}(x_{i}^{d}-x_{i}^{ave})}{\sum_{i=1}^{N_{min}}(x_{i}^{ave})}
\label{eqn:percentBIAS}
\end{equation}
y en la tabla (\ref{tablaBIASwhole}) también aparecen:
\begin{equation}
Percent\, Absolute\, Bias(X_{d})=\frac{\sum_{i=1}^{N_{min}}|x_{i}^{d}-x_{i}^{ave}|}{\sum_{i=1}^{N_{min}}(x_{i}^{ave})}
\label{eqn:abspercentBIAS}
\end{equation}

La magnitud \textit{Percent Bias} la denominaremos sesgo porcentual en la magnitud x (de cada uno de los disdrómetros (d-ésimo) respecto del valor medio) y donde la suma se realiza en todos los minutos considerados de la base empírica bajo cada una de las acotaciones. En ambas tablas se aprecia que el acuerdo es bueno aunque algunos disdrómetros como el E2 y el C1 muestran una desviación respecto de la media de entre\footnote{Una interpretación completa se puede realizar diferenciando cada episodio para relacionar las desviaciones globales con episodios concretos. En el caso de diferencias globales del 10\% análogas en toda la red serían similares a las encontradas en otros experimentos, no siendo necesario determinar si un sesgo anómalo procede, o no, de un momento particular. En nuestro caso, viendo que las diferencias aparecen en tres instrumentos específicos, conviene determinar si es una propiedad de estos instrumentos o procede de un evento concreto de gran variación en la red; para este fin el análisis de la serie temporal de precipitación acumulada nos será de ayuda.} el 15\% y el 20\%. En el caso del disdrómetro B1 tendríamos un sesgo negativo del 18\%. Estos valores están en el mismo rango que en el estudio realizado por \citep{Tokay2003disdrometerKAMP} donde las diferencias en valores de precipitación acumulada dada por pluviómetros pueden ser de este orden para instrumentos anexos. Comparando con el experimento \citep{2009ChoongLEE} las desviaciones que encontramos aquí son algo menores.

\begin{table}[h]
\ra{1.02}
\caption[Sesgo porcentual y absoluto porcentual en la red para diferentes acotaciones en la intensidad de precipitación.]{\textbf{Sesgo porcentual y absoluto porcentual en la red para diferentes acotaciones en la intensidad de precipitación.} Se muestran los valores del sesgo porcentual respecto de la media para el conjunto de toda la base empírica dado por la ecuación (\ref{eqn:percentBIAS}) junto con los valores del sesgo absoluto porcentual respecto de la media para el conjunto de toda la base empírica dado por la ecuación (\ref{eqn:abspercentBIAS}).}
\vspace{0.15cm}
\begin{center}
\begin{tabular}{crrrr}
\toprule
 & \multicolumn{2}{r}{Sesgo Porcentual}  & \multicolumn{2}{r}{Sesgo Abs. Porcentual}   \\
\cmidrule(l{.5em}){2-3} \cmidrule(l{.5em}){4-5}
   \textbf{Disdrómetro}                      & $ 0.0<R$ &$0.5<R$  & $ 0.0<R$ &$0.5<R$  \\
\midrule

  A2 &   -0.060    &   -0.005  &   0.340    &   0.305    \\
  B2 &   -0.021    &   -0.029  &   0.298    &   0.265    \\
  C2 &   -0.036    &   -0.025  &   0.301    &   0.268    \\
  D2 &   -0.053    &   -0.054  &   0.352    &   0.316    \\
  E2 &   +0.199    &   +0.215  &   0.423    &   0.388    \\ 
  F2 &   +0.057    &   +0.073  &   0.266    &   0.246    \\
  G2 &   -0.063    &   -0.049  &   0.297    &   0.268    \\
  H2 &   -0.039    &   -0.054  &   0.352    &   0.305   \\
\midrule
  A1 &   -0.070    &   -0.075  &   0.320    &   0.282   \\
  B1 &   -0.187    &   -0.178  &   0.286    &   0.259   \\
  C1 &   +0.230    &   +0.210  &   0.387    &   0.347   \\
  D1 &   -0.088    &   -0.089  &   0.361    &   0.321   \\
  E1 &   -0.045    &   -0.046  &   0.312    &   0.272   \\
  F1 &   +0.053    &   +0.036  &   0.259    &   0.225   \\
  G1 &   +0.074    &   +0.086  &   0.322    &   0.291   \\
  H1 &   -0.002    &   -0.014  &   0.345    &   0.299   \\
\bottomrule
\end{tabular}
\end{center}
\label{tablaBIASwhole}
\end{table}

%

\subsection{Distribuciones de tamaños de gota compuestas}

\begin{figure}[H] 
\vspace{0.75cm}
\begin{center}
  
\includegraphics[width=1.08\textwidth]{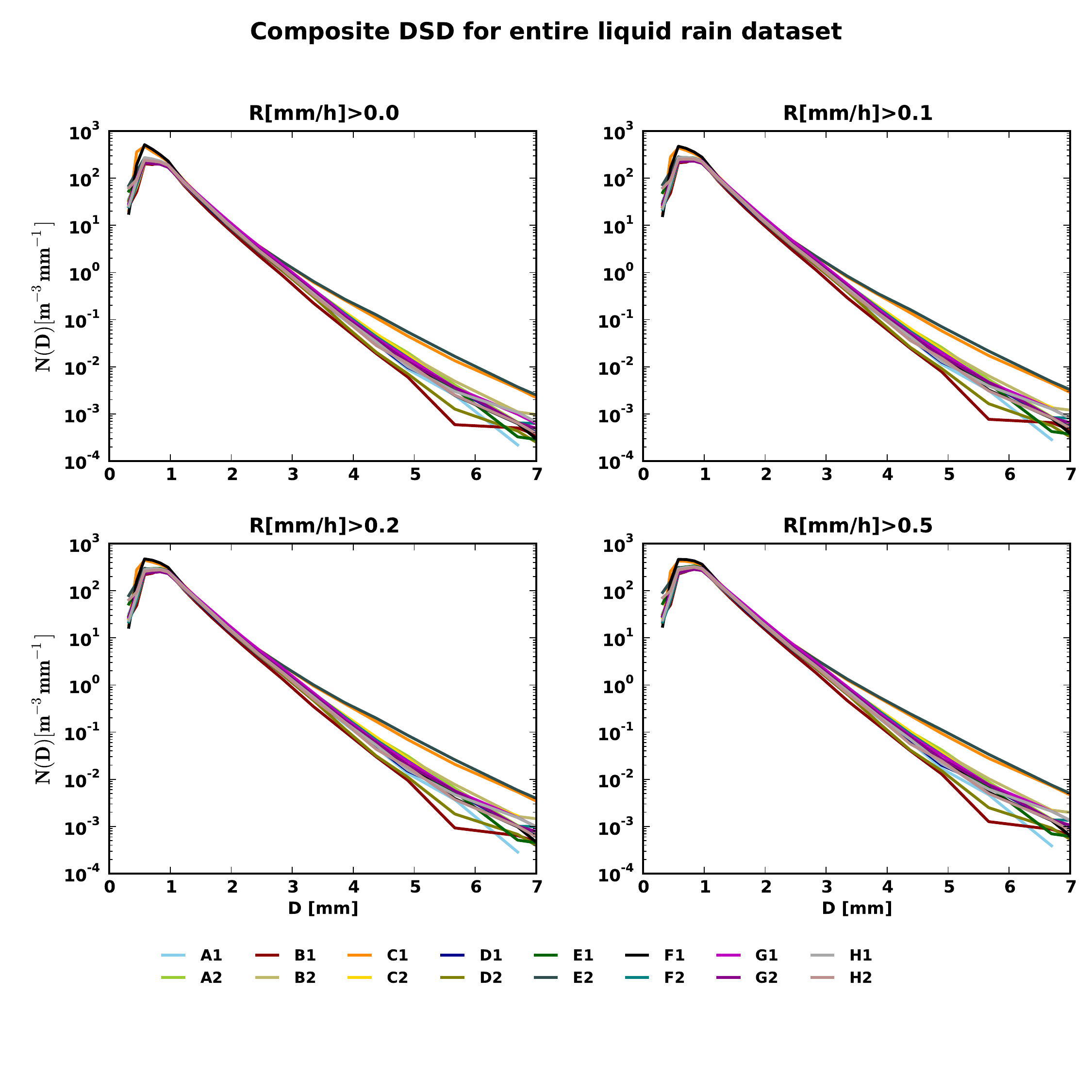}
\vspace{0.40cm}
   \caption[DSD compuestas para toda la base empírica sobre un conjunto consistente de datos]{\textbf{DSD compuestas para toda la base empírica sobre un conjunto congruente de datos.} Se han incluido acotaciones que se especifican en la figura: $R>0.1\,mm/h$, $R>\,0.1mm/h$ y $R>\,0.5mm/h$, que se corresponde con umbrales de detección de lluvia de diferentes sensores. R = 0.1 $mm/h$ es la cota de muchos radares terrestres tradicionales de Institutos Nacionales de Meteorología. R = $0.2\, mm/h$ es la sensibilidad estimada del futuro sensor radar de precipitación que incorporara el satélite GPM. R = 0.5 $mm/h$ es el umbral de detección del radar que incorpora el satélite TRMM. Los diferentes umbrales son necesarios para establecer inferencias sobre los modelos de DSD que se incorporan en los diferentes algoritmos de cada radar. Observamos como los disdrómetros C1 y E2 poseen una mayor cantidad de gotas de tamaños grandes. El sensor B1 posee menor número de estas. Estos hechos explican los resultados de la tabla (\ref{tablaBIASwhole}). Los disdrómetros C1 y F1 detectan menor cantidad de gotas pequeñas aunque las diferencias son leves y están en una parte del espectro donde todos los disdrómetros presentan resultados diferentes.}
\label{fig:Var2WholeComposite}
\end{center}
\vspace{0.95cm}
\end{figure}

Los resultados para las DSD compuestas obtenidas en cada uno de los disdrómetros para la base empírica completa se muestran en la figura (\ref{fig:Var2WholeComposite}) calculados mediante:
\begin{equation}
N_{c}(D)=\frac{1}{n_{t}}\sum_{k=1}^{n_{t}} N_{k}(D)
\label{eqn:DSDcomposite}
\end{equation}
donde se consideran $n_{t}$ distribuciones de tamaños de gota diferentes para componer la DSD global\footnote{Los valores de $n_{t}$ se corresponden con los minutos dados en la tabla (\ref{TablaMINUTOSdata}).}. \\

En la figura se muestran las DSD compuestas para cuatro acotaciones en el valor de la intensidad de precipitación (minuto a minuto y en todos los disdrómetros de la red) diferentes. Además se comprueba como la DSD compuesta se ajusta de manera aproximada a una distribución exponencial, en que el valor de $N_{0}$ no se ve condicionado por las acotaciones en la intensidad de precipitación (pero sí condicionaría el valor de $\Lambda$, como consecuencia del filtrado progresivo de gotas de mayor tamaño que se realiza de modo efectivo en caso de clasificar las DSD en función de intervalos para la intensidad de lluvia). Este hecho se corresponde con el modelo de Marshall-Palmer.\\


Una interpretación de carácter microfísico de la DSD compuesta similar a la DSD de equilibrio, introducida en el capítulo \textsection \ref{sec:MicroDSDrelevancia}, y que presenta oscilaciones en diámetros en torno a $D\simeq 1.4\, mm$, es difícil, por el carácter de promedio sobre diferentes episodios de precipitación en los que se pueden haber alcanzado de manera parcial, y en distinto grado, situaciones de equilibro entre los diversos procesos microfísicos\footnote{En otros estudios experimentales con disdrómetros se han encontrado DSD multimodales. En \citep{PeterHartmann2007} se relacionan con propiedades sinópticas en que se genera la DSD, esencialmente diferencias entre masas de aire de origen continental y de origen oceánico.}.\\

Las DSD compuestas permiten explicar el origen de los sesgos que se apreciaban en la precipitación acumulada de tres de los disdrómetros de la red. Los C1 y E2 han registrado un número mayor de gotas con diámetros superiores a 2.5 mm, siendo este el origen de las diferencias (además en los diferentes umbrales de precipitación mínima el tamaño de gotas en que ambos disdrómetros se diferencian del resto aparece fijo no siendo por tanto el sesgo una acumulación anormal de gotas de menores diámetros).\\

En el caso del disdrómetro B1, apreciamos una diferencia sistemática en todo el espectro de tamaños, no solo en gotas de diámetros mayores de 2.5 mm, aunque estas son las que tienen un efecto más nítido en los sesgos de intensidad de lluvia. Cabe notar que esta desviación en la DSD compuesta es menor que la desviación entre instrumentos diferentes que se encuentra en el experimento \citep{krajewski_kruger_etal_2006_aa} y está en el rango del encontrado entre instrumentos similares en otros estudios \citep{PeterHartmann2007}; con todo, este sesgo ha de ser tenido en cuenta en las cantidades acumuladas (no tanto en los procesos de correlación entre estaciones si es un sesgo sistemático).



\subsection{Series de precipitación acumulada}

Es posible comparar también la serie temporal de datos obtenidos en los diferentes disdrómetros. Se ha realizado para la intensidad de precipitación mostrando su acumulación a lo largo de la base empírica\footnote{Excluyendo los casos de precipitación en forma de nieve que tal y como se introdujo en \S\ref{sec:refINTENSIDADR} requieren un análisis específico.}. Además se calcula la precipitación acumulada media junto con la desviación estándar sobre la red mediante (d indica el disdrómetro y $n_{d}$ el número de disdrómetros de la red):

\begin{equation}
\bar{R}_{acc}=\frac{1}{n_{d}}\sum_{d=1}^{n_{d}}{R^{(d)}_{acc}}
\label{eqn:Racc_red}
\end{equation}
y
\begin{equation}
s^{2}(R_{acc})=\frac{1}{n_{d}-1}\sqrt{\sum_{d=1}^{n_{d}}{(R^{(d)}_{acc}-\bar{R}_{acc})^{2}}}
\label{eqn:Racc_red_std}
\end{equation}

\begin{figure}[H] 
\begin{center}
   \includegraphics[width=0.99\textwidth]{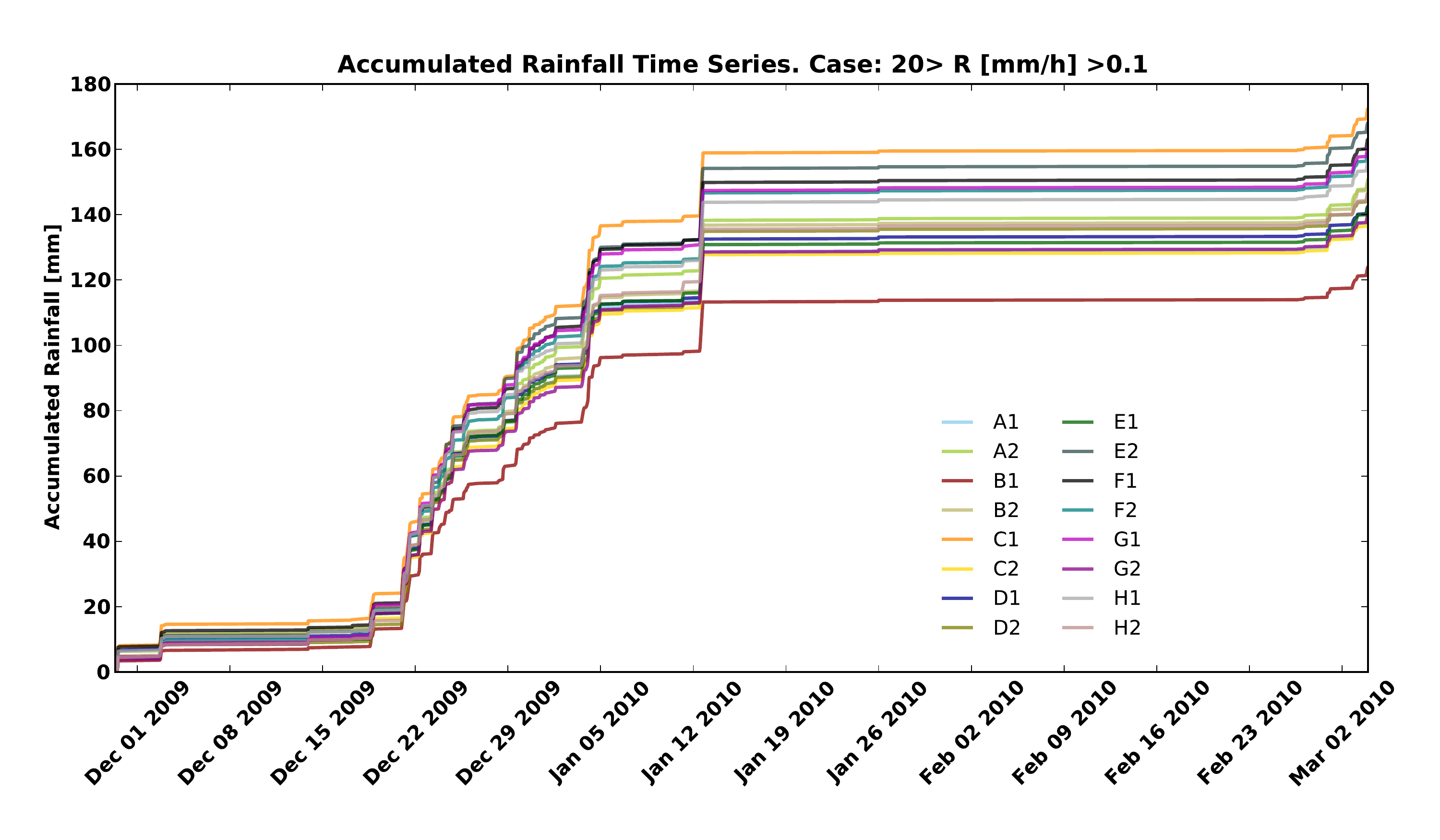}
   \caption[Acumulación de lluvia durante la base empírica. Precipitación líquida.]{\textbf{Acumulación de lluvia durante la base empírica. Precipitación líquida.} Cálculo desde conjunto consistente para minutos con al menos 10 gotas registradas en cada instrumento. Se aprecian las diferencias en los instrumentos C1, E2 y B1 respecto del resto de la red; una parte importante de la diferencia en la precipitación acumulada procede del evento registrado el 23 de diciembre, sobre todo en lo que se refiere al instrumento E2. Observamos que el sesgo negativo del disdrómetro B1 aparece de modo sistemático en toda la base empírica. A la vista de la figura (\ref{fig:Var2WholeComposite}) proviene de un déficit igualmente sistemático.}
\label{fig:RaccTimeSeries}
\includegraphics[width=0.99\textwidth]{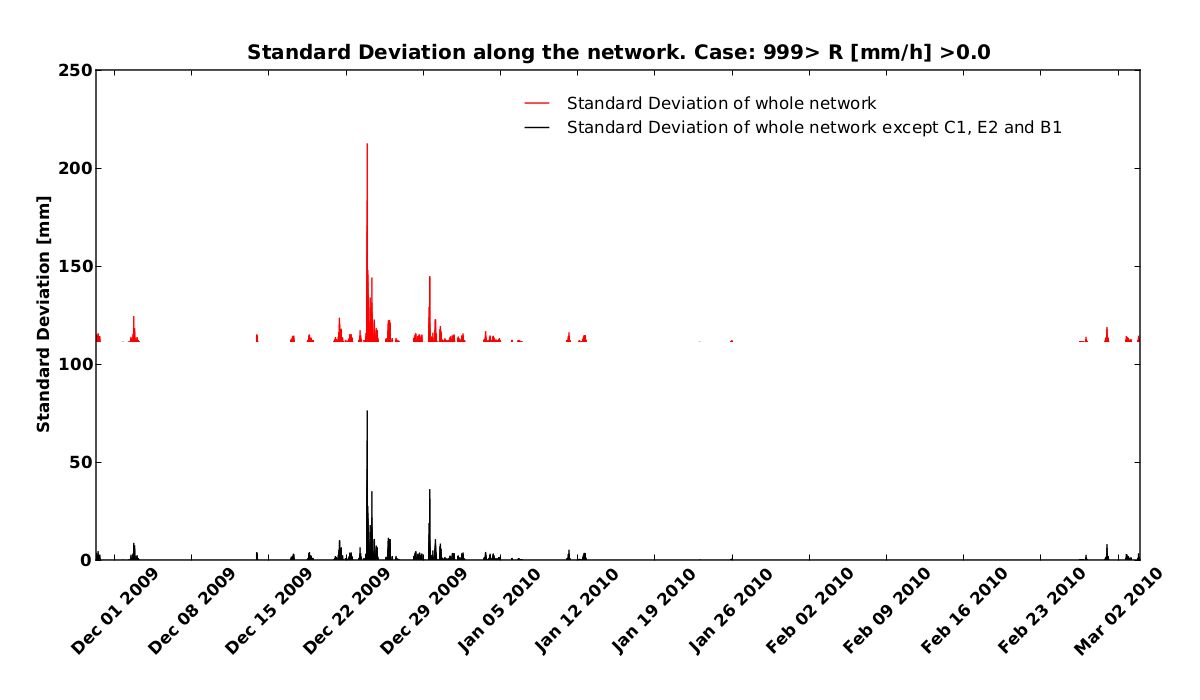}
   \caption[Desviación estándar de la serie temporal de precipitacion calculada a lo largo de la red cada minuto.]{\textbf{Desviación estándar de la serie temporal de precipitación calculada a lo largo de la red cada minuto.} Cálculo realizado desde un conjunto consistente (entre todos los disdrómetros de la red) para minutos con al menos 10 gotas registradas en cada instrumento. Se indica la desviación estándar calculada en toda la red, sustrayendo del cálculo los disdrómetros C1, E2 y B1. Observamos un episodio de gran variación el 23 de diciembre. Los episodios con mayor intensidad de precipitación son los que implican mayor desviación estándar; sin embargo, este episodio parece aportar mayores diferencias al sesgo de los disdrómetros C1 y principalmente E2 que el resto de episodios, no solo por su intensidad sino por una alta variabilidad específica.}
\label{fig:Std4RaccTimeSeries}
\end{center}
\end{figure}

Es interesante además incorporar acotaciones diferentes para observar su relevancia en el total de la base empírica. En la figura (\ref{fig:RaccTimeSeries}) se muestra el caso sin acotaciones. Se comprueba cómo los principales eventos están concentrados en el intervalo de 19 de diciembre al 25 de enero junto con el intervalo entre los últimos días de febrero y primeros de marzo, cuestión compatible por tanto con los valores climáticos medios en la localización del experimento, al tiempo que la localización de los episodios principales de precipitación no depende de los umbrales sobre la intensidad de precipitación que se añadan.\\

La desviación estándar calculada mediante (\ref{eqn:Racc_red_std}) se puede ver en la figura (\ref{fig:Std4RaccTimeSeries}), tanto para el total de disdrómetros como para el subconjunto que no incluye los de mayor sesgo (C1, E2 y B1) para comprobar si existe algún episodio con mayor peso en los sesgos encontrados\footnote{Las figuras mostradas en esta sección y en la sección \S\ref{sec:liquidACCpreprocess} incluyen un escalado adimensional de la precipitación acumulada para llevar todas las figuras a un rango común. El factor de escalado se basa en el cociente entre el número de minutos de la acotación $0.5<R$ y el número de minutos real bajo cada acotación tal y como se describe en la tabla (\ref{TablaMINUTOSdata}).}.

\section{Relaciones Z-R para la base empírica completa}

Se han determinado las relaciones Z-R a lo largo de toda la red de instrumentos para el caso de un filtrado que incluye tanto restricciones a las posibles velocidades terminales\footnote{Se hará un análisis más exhaustivo de las implicaciones del pre-procesado de datos disdrométricos en \S\ref{chap:preprocesadoTOKAY}.} como varias intensidades mínimas de precipitación.

\begin{figure}[H] 
\begin{center}
   \includegraphics[width=0.95\textwidth]{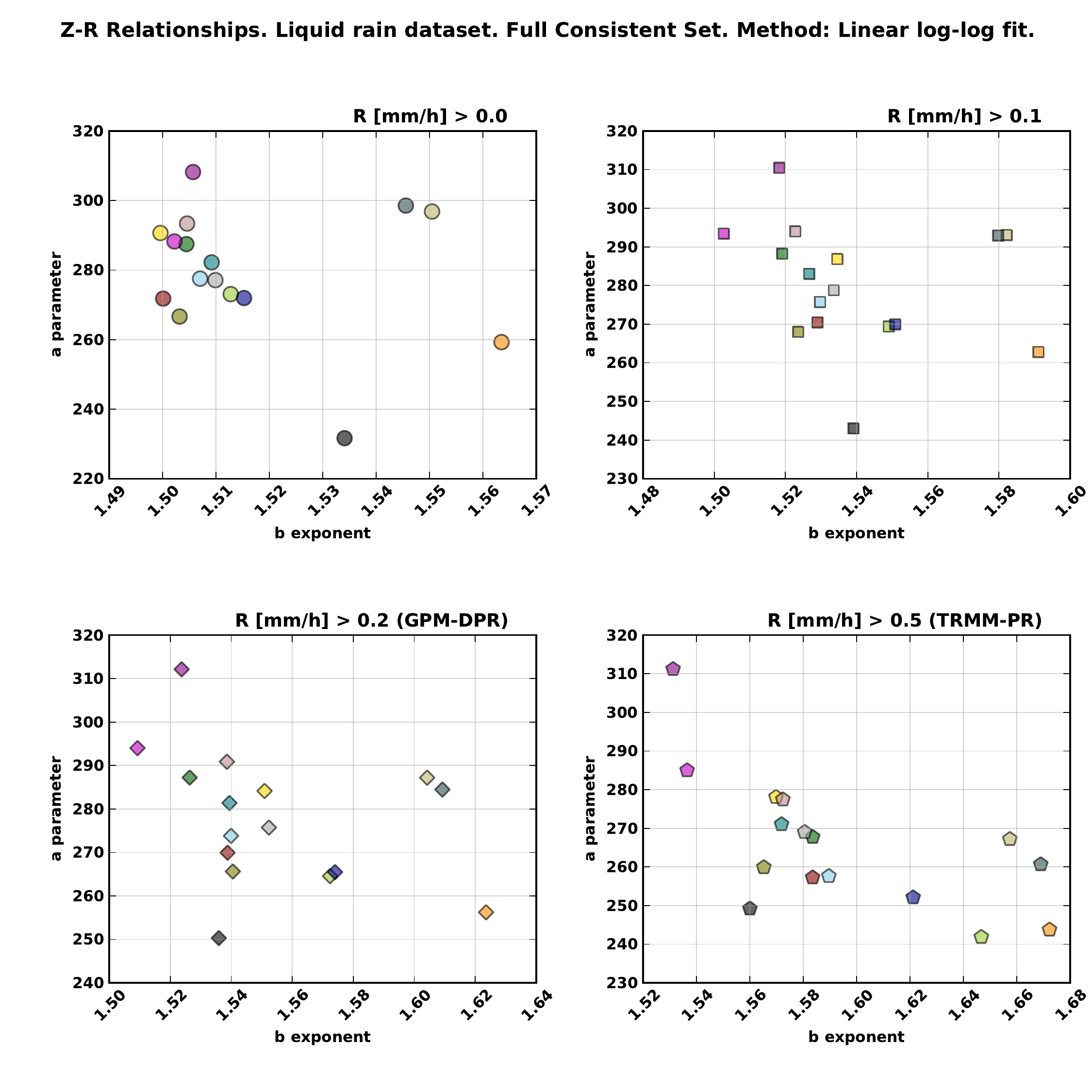}
   \caption[Relaciones Z-R para la base empírica completa y diferentes intensidades de lluvia mínimas.]{\textbf{Relaciones Z-R para la base empírica completa y diferentes intensidades de lluvia mínimas.} Se muestran las relaciones Z-R para la base empírica completa (restringida a todos los episodios de precipitación líquida con intensidades de precipitación máximas de 20 mm/h ) calculada para todos los disdrómetros de la red y bajo cuatro intensidades de precipitación mínima en todos los disdrómetros a un tiempo. Simula las sensibilidades estimadas para varios sensores radar en términos de intensidad de precipitación mínima detectable. Los coeficientes de la relación Z-R han sido calculados mediante regresión lineal simple en escala logarítmica.}
\label{fig:ZRwhole}
\end{center}
\vspace{0.5cm}
\end{figure}

Se han utilizado dos metodologías diferentes:
\begin{itemize}
   \item Ajuste de regresión lineal tras una transformación logarítmica de Z y R. Se observa un desplazamiento leve a valores de $a_{R}$ menores y $b_{R}$ mayores conforme las acotaciones de intensidad mínima aumentan.
   \item Ajuste no-lineal de la relación Z-R: se observan menos diferencias debido los umbrales de intensidad mínima de precipitación utilizados. Llama la atención la presencia de una aparente relación lineal entre los parámetros $a_{R}$ y $b_{R}$. La correlación que se observa implica que valores altos de $a_{R}$ se corresponden con valores bajos de $b_{R}$.
\end{itemize}

\begin{figure}[H] 
\begin{center}
   \includegraphics[width=0.95\textwidth]{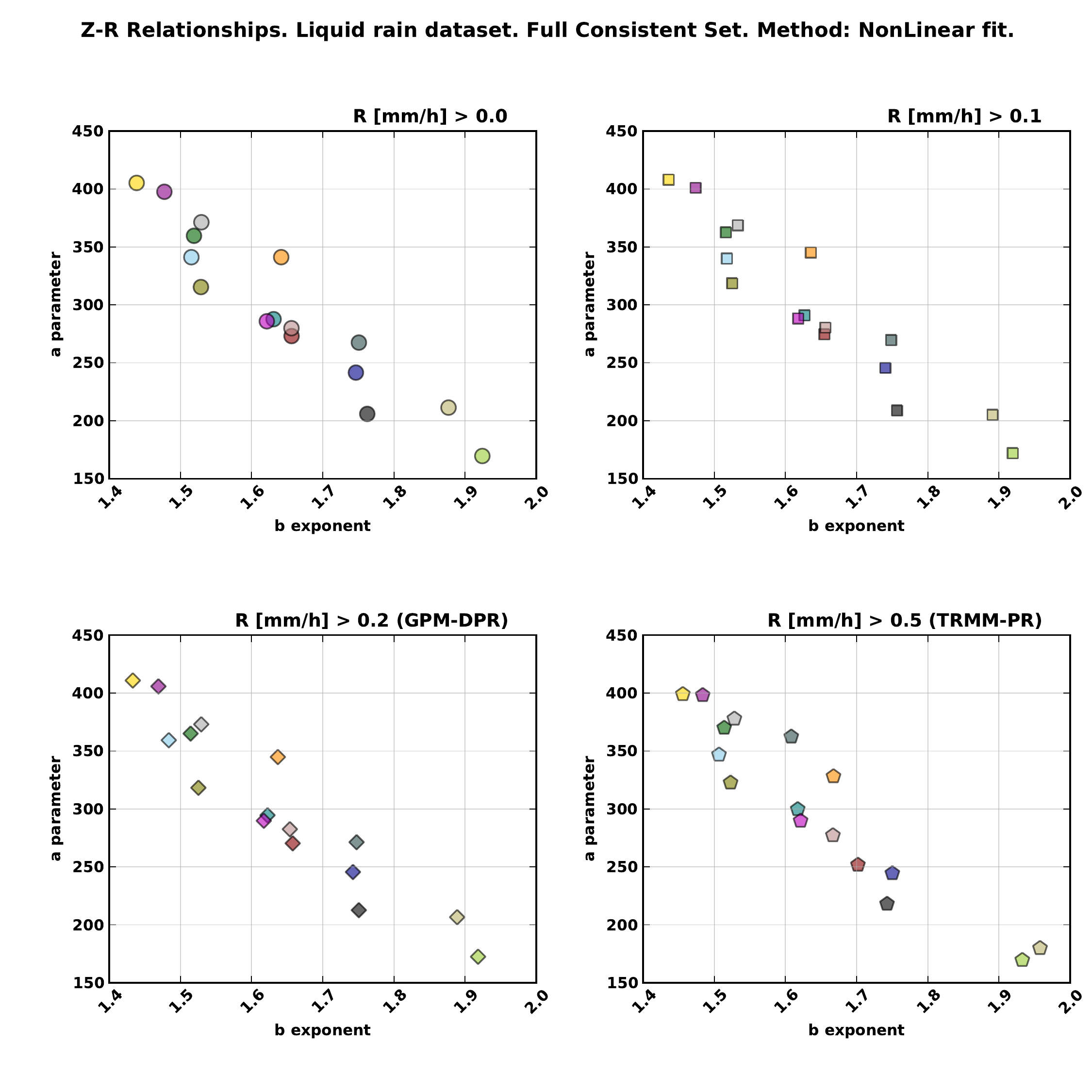}
   \caption[Relaciones Z-R para la base empírica completa y diferentes intensidades de lluvia mínimas.]{\textbf{Relaciones Z-R para la base empírica completa y diferentes intensidades de lluvia mínimas.} Se muestran las relaciones Z-R para la base empírica completa (restringida a todos los episodios de precipitación líquida con intensidades de precipitación máximas de 20 mm/h) calculada para todos los disdrómetros de la red y bajo cuatro intensidades de precipitación mínima en todos los disdrómetros a un tiempo. Simula las sensibilidades estimadas para varios sensores radar en términos de intensidad de precipitación mínima detectable. Los coeficientes de la relación Z-R han sido calculados mediante un ajuste no-lineal de mínimos cuadrados.}
\label{fig:ZRwhole}
\end{center}
\end{figure}

\section{Histogramas de los parámetros integrales de la DSD}

\begin{figure}[H] 
\begin{center}
   \includegraphics[height=0.80\textheight]{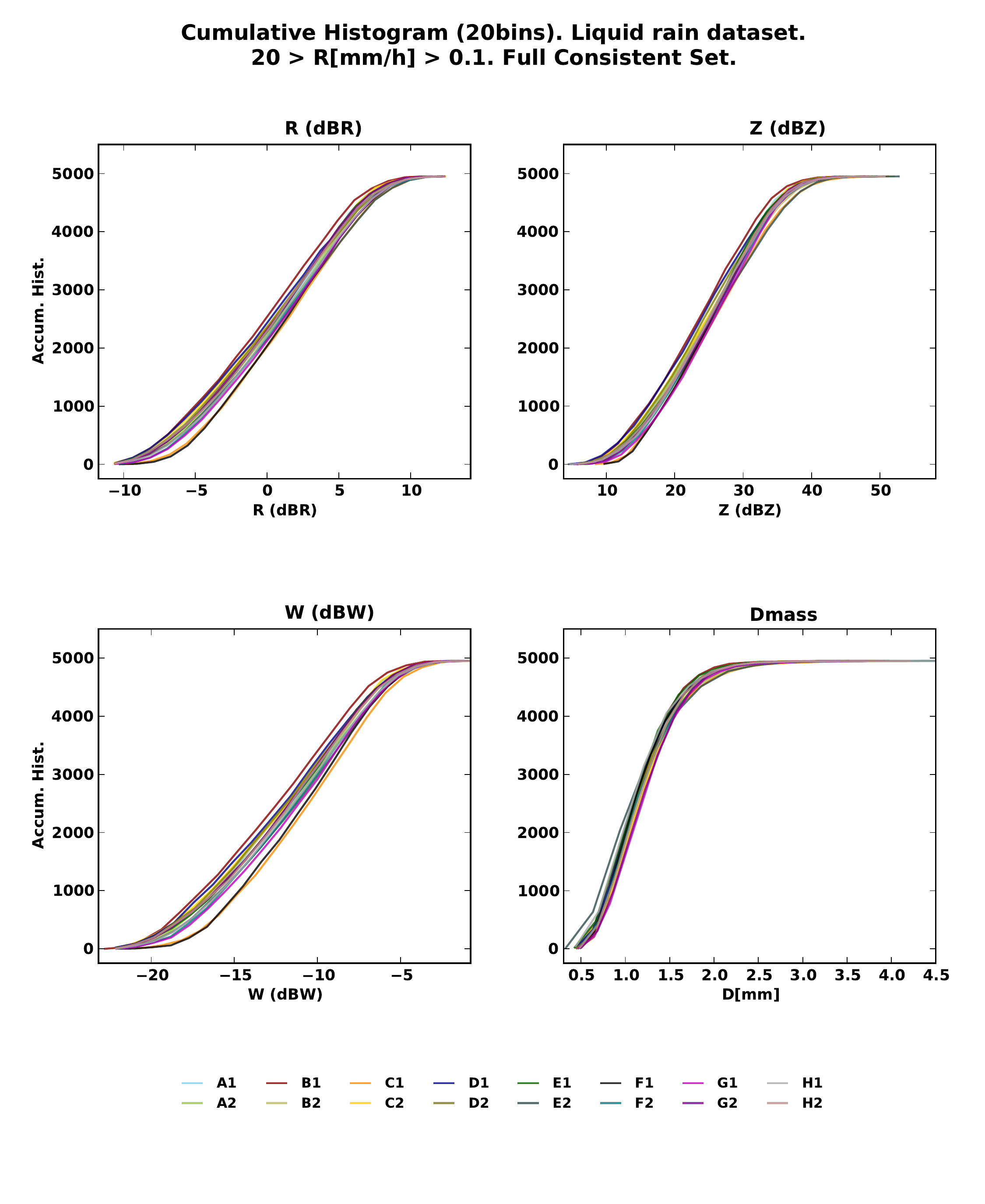}
   \caption[Diagramas de frecuencia acumulada para toda la base empírica sobre un conjunto congruente de datos. R , Z , W y $D_{mass}$. $20>R>0.1\,mm/h$.]{\textbf{Diagramas de frecuencia acumulada para toda la base empírica sobre un conjunto congruente de datos. R , Z , W y $\mathbf{D_{mass}}$. $\mathbf{20>R>0.1\,mm/h}$.} Primera fila: intensidad de precipitación y reflectividad, ambos en escala de decibelios. Segunda fila: Contenido en agua líquida en la escala de decibelios y diámetro medio ponderado sobre la masa. El diagrama no aparece normalizado para apreciar visualmente el número total de distribuciones de tamaño de gota analizadas.}
\label{fig:cumfreqRZCW}
\end{center}
\end{figure}

\begin{figure}[H] 
\begin{center}
   \includegraphics[height=0.85\textheight]{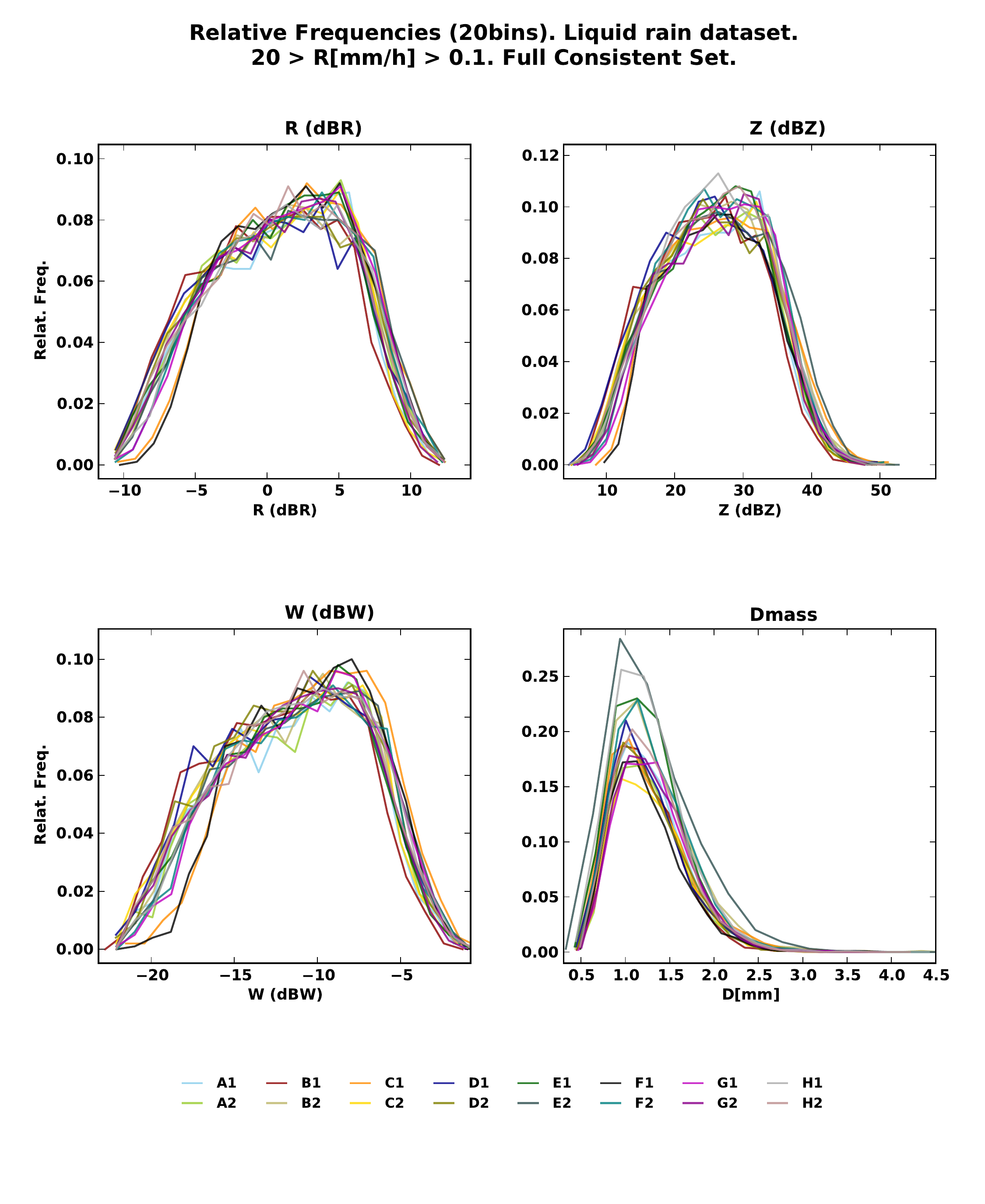}
 \caption[Histogramas de frecuencias para toda la base empírica sobre un conjunto congruente de datos. R , Z , W y $D_{mass}$. $20>R>0.1\,mm/h$ ]{\textbf{Histogramas de frecuencias para toda la base empírica sobre un conjunto congruente de datos. R , Z , W y $\mathbf{D_{mass}}$. $\mathbf{20>R>0.1\,mm/h}$} Primera fila: intensidad de precipitación y reflectividad, ambos en escala de decibelios. Segunda fila: Contenido en agua líquida en la escala de decibelios y diámetro medio ponderado sobre la masa. }
\label{fig:relfreqRZCW}
\end{center}
\end{figure}

Se han comparado además los histogramas tanto de los parámetros integrales de la precipitación (contenido en agua líquida, intensidad de precipitación, reflectividad y diámetro característico) como del diámetro máximo detectado en los disdrómetros a lo largo de la red. En las figuras (\ref{fig:cumfreqRZCW}) y (\ref{fig:relfreqRZCW}) se muestran los histogramas para todos los disdrómetros, tanto con orientación NS como EO. Los histogramas presentan una similitud notable consecuencia de que la DSD promediada sobre largos periodos de tiempo disminuye notablemente su variabilidad natural. Los instrumentos C y F presentan menor número de minutos con valores pequeños de los parámetros integrales de precipitación: se comprueba en W y en R (y levemente en Z).\\

Los histogramas de $D_{mass}$ muestran cómo las estaciones C y F registran un menor número de gotas pequeñas que otros disdrómetros aunque entra dentro de un margen razonable de variabilidad. También estos disdrómetros poseen un pico menos pronunciado entorno a $D\, =\,1\,mm$ aunque su posición sea la misma. El disdrómetro E2 presenta un histograma normalizado cuyas diferencias pueden ser debidas a que el filtrado de minutos con alta intensidad de lluvia ha implicado descartar una parte importante de gotas de tamaños mayores de 3 mm que poseen relevancia en el calculo de $D_{mass}$. Las gotas pequeñas son más afectadas por fenómenos de turbulencia (y corrientes de aire que den lugar a una diferenciación por tamaños) que pueden ser más acusados en determinadas localizaciones. Así el disdrómetro F presenta mayor número de gotas pequeñas que B ya que a pesar de distar solo 300 metros de este, la diferencia de altitud puede ser relevante en este aspecto. En el caso de la intensidad de precipitación, la acotación utilizada se aprecia en el histograma, no así en los otros parámetros integrales. \\

\section{Histogramas de los parámetros de la distribución gamma normalizada}

\begin{figure}[H] 
\begin{center}
   \includegraphics[height=0.85\textheight]{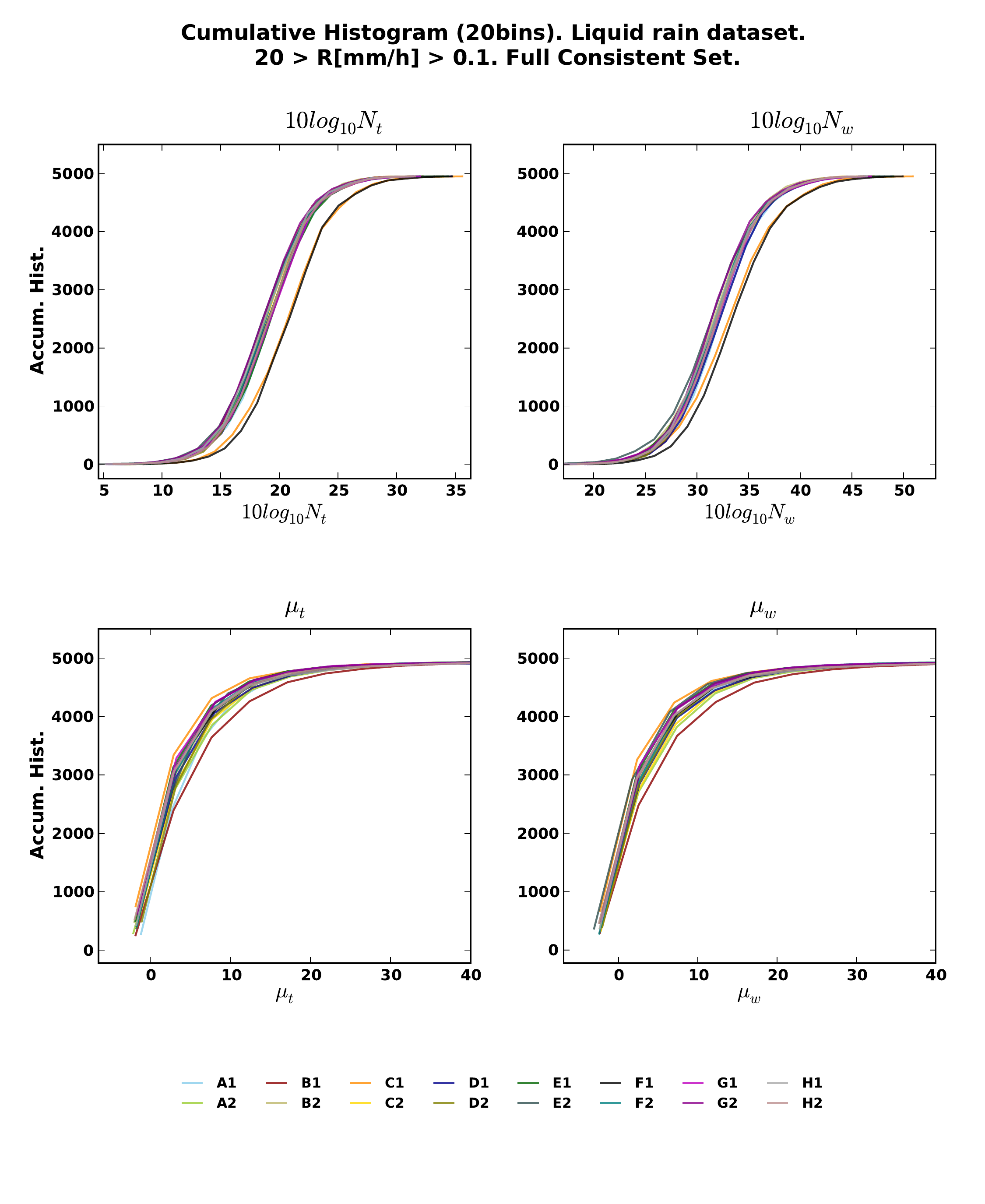}
 \caption[Diagramas de frecuencia acumulada para toda la base empírica sobre un conjunto congruente de datos. Parámetros de la modelización de la DSD como una distribución gamma normalizada. $20>R>0.1\,mm/h$ ]{\textbf{Diagramas de frecuencia acumulada para toda la base empírica sobre un conjunto congruente de datos. Parámetros de la modelización de la DSD como una distribución gamma normalizada. $\mathbf{20>R>0.1\,mm/h}$ } Primera fila: Valores de la concentración para las ecuaciones (\ref{eqn:normalizadaTestudgammaN0}) y (\ref{eqn:normalizadaTestudgammaN0Nt}). Segunda fila: Factores de forma incluidos en las expresiones (\ref{eqn:normalizadaTestudgammaNw}) y (\ref{eqn:normalizadaTestudgammaNt}). El diagrama no aparece normalizado para apreciar visualmente el número total de distribuciones de tamaño de gota analizadas.}
\label{fig:cumfreqDSDpar}
\end{center}
\end{figure}

\begin{figure}[H] 
\begin{center}
   \includegraphics[height=0.85\textheight]{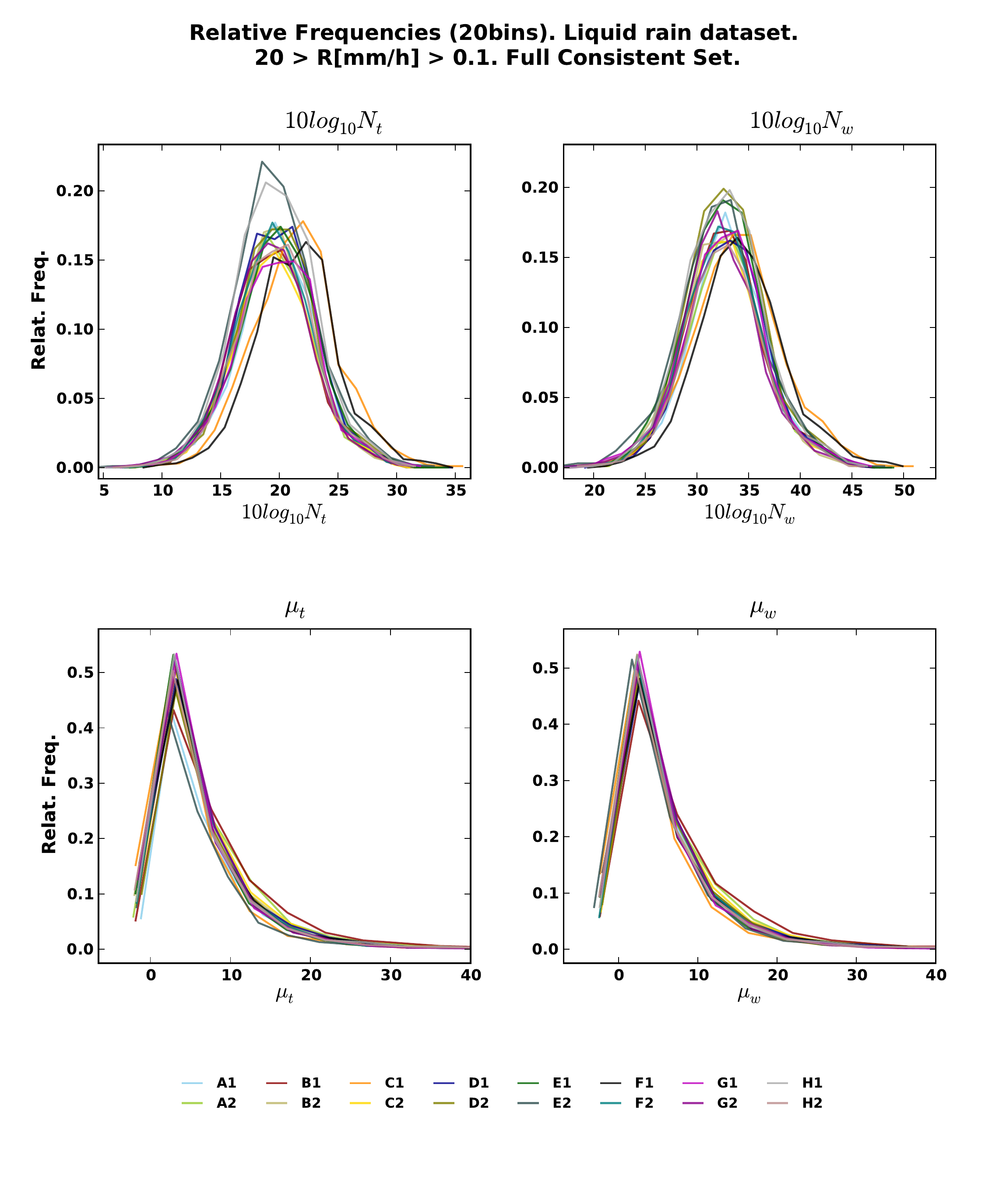}
 \caption[Histogramas de frecuencias para toda la base empírica sobre un conjunto congruente de datos. Parámetros de la modelización de la DSD como una distribución gamma normalizada. $20>R>0.1\,mm/h$ ]{\textbf{Histogramas de frecuencias para toda la base empírica sobre un conjunto congruente de datos. Parámetros de la modelización de la DSD como una distribución gamma normalizada. $\mathbf{20>R>0.1\,mm/h}$ } Primera fila: Valores de la concentración para las ecuaciones (\ref{eqn:normalizadaTestudgammaN0}) y (\ref{eqn:normalizadaTestudgammaN0Nt}). Segunda fila: Factores de forma incluidos en las expresiones (\ref{eqn:normalizadaTestudgammaNw}) y (\ref{eqn:normalizadaTestudgammaNt}).}
\label{fig:relfreqDSDpar}
\end{center}
\vspace{1cm}
\end{figure}

En el caso de la modelización de la DSD se han representado histogramas para los parámetros de una distribución gamma normalizada, figuras (\ref{fig:cumfreqDSDpar}) y (\ref{fig:relfreqDSDpar}), introducida en \S\ref{sec:AplicacionGammaNormalizada}. Se han tomado dos esquemas diferentes: ambos utilizan como diámetro característico el valor de $D_{m}$ dado por (\ref{eqn:normalizadaDM}), mientras que para la concentración se han usado (\ref{eqn:normalizadaTestudgammaN0}) y (\ref{eqn:normalizadaTestudgammaN0Nt}). Como consecuencia se determina el parámetro libre de forma $\mu$ correspondiente a las expresiones (\ref{eqn:normalizadaTestudgammaNw}) y (\ref{eqn:normalizadaTestudgammaNt}), para el primer caso se escribe $\mu_{w}$, para el segundo $\mu_{t}$.\\

Los parámetros $N_{w}$ y $N_{t}$ en escala logarítmica muestran diferencias para los disdrómetros C2 y F1 reflejo de que, tanto la concentración total de gotas como el $D_{mass}$ son los que poseen mayores diferencias entre disdrómetros. En ambos casos los histogramas poseen una forma aproximadamente simétrica entorno al valor máximo, pudiéndose sugerir que las magnitudes originales $N_{w}$ y $N_{t}$ poseen una asimetría similar a una distribución log-normal.\\

Los valores de los parámetros $\mu_{w}$ y $\mu_{t}$ han sido obtenidos mediante el procedimiento indicado por \citep{Testud2001}, y que esta basado en minimizar la suma de cuadrados de la diferencia entre los puntos generados por la función y los valores suministrados por los datos. En este caso la minimización se ha realizado entre $log_{10}F_{\mu}(D/D_{mass})$ y $log_{10}F_{i}(D/D_{mass})$. Los histogramas muestran una consistencia notable, siendo únicamente el disdrómetro B1 el que posee mayor diferencia en el caso del histograma acumulado. Este hecho se relaciona con el mayor registro de gotas pequeñas de este disdrómetro relativo a las gotas de mayor tamaño, ya que descarta distribuciones de probabilidad con $\mu_{t}$ y $\mu_{w}$ pequeño. La forma general del histograma es similar a otros trabajos recientes \citep{Tokay2010smallscaleDSD} aunque en dicho trabajo la metodología de estimación de $\mu$ es distinta (véase \S\ref{sec:metodoNormalizada}).\\ 

\vspace{1cm}

\section{Método de escalado basado en un parámetro integral de la precipitación}
\label{sec:basempirica_escalado}

En la sección \S\ref{sec:Scaling1moment} se describió la metodología desarrollada por \citep{semperetorres_porra_ea_1994} para caracterizar las distribuciones de tamaños de gota por episodios o por estaciones. La base central de la metodología es la existencia de relaciones de potencias entre los diferentes parámetros integrales de la precipitación, lo que permite estimar los valores $\alpha$ y $\beta$ que posibilitan escalar las DSDs de un episodio basándose en un parámetro integral de referencia\footnote{Obteniendo, por tanto, los parámetros necesarios de la ecuación (\ref{eqn:sempere-torres-1moment}).}. Una de las propiedades que se deducen de esta metodología es la existencia de una relación implícita entre $\alpha$ y $\beta$  mediada por el valor del orden del momento que caracteriza el parámetro de referencia, relación que ya deducíamos en el capítulo \S\ref{sec:modelizacionesDSDchap} y expresabamos en la ecuación (\ref{eqn:consistencia1-1moment}).\\

En este capítulo de análisis global de la base empírica nos interesa comprobar, para diferentes parámetros integrales de referencia, los resultados para $\alpha$ y $\beta$, y verificar la consistencia general de estos con las relaciones analíticas que obtuvimos previamente. Dado que la relación de consistencia se basa en la hipótesis inicial de leyes de potencias entre los momentos de la DSD, conviene comparar los resultados para diferentes subconjuntos de estos y estimar así, las dificultades prácticas de estimación que implica la hipótesis inicial de este método. Para ello hemos seleccionado tres conjuntos de momentos:
\begin{enumerate}
 \item $(M_{0}, M_{1},\ldots,M_{7})$ que incluye tanto, momentos de orden alto más afectados por problemas de muestreo como momentos de orden bajo, afectados por problemas de estimación de gotas pequeñas.
 \item $(M_{1}, M_{2},\ldots,M_{6})$, se excluyen los momento de orden cero y siete para estimar su relevancia en la relación de consistencia al compararlo con el procedimiento (1).
 \item $(M_{2}, M_{3},\ldots,M_{6})$, se excluye el momento de orden 1 respecto del caso (2) pero se mantiene la reflectividad por su importante significado físico.
\end{enumerate}

\begin{figure}[H] 
\begin{center}
   \includegraphics[height=0.70\textheight]{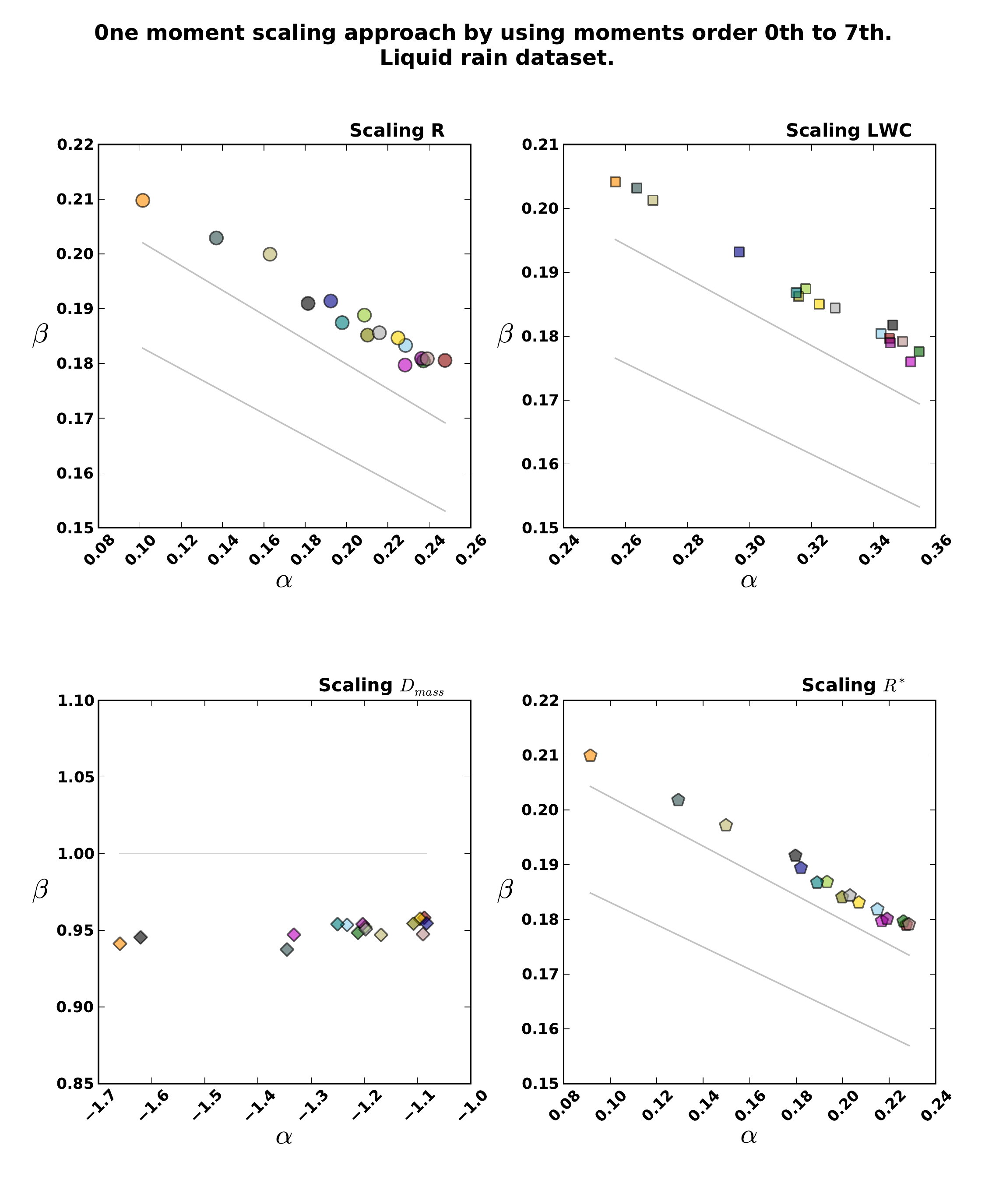}
 \caption[Relación $\alpha$ vs. $\beta$ para el método de escalado usando $(M_{1}, M_{2},\ldots,M_{6})$. $20>R>0.1\,mm/h$.]{\textbf{Relación $\mathbf{\alpha}$ vs. $\mathbf{\beta}$ para el método de escalado usando $\mathbf{(M_{1}, M_{2},\ldots,M_{6})}$.  $\mathbf{20>R>0.1\,mm/h}$.} Se presentan los valores de $\alpha$ y $\beta$ obtenidos para cada disdrómetro de la red y para todos los episodios de precipitación líquida. Las variables de referencia utilizadas en la aplicación del método descrito en \S\ref{sec:Scaling1moment} son la intensidad de precipitación, el contenido de agua líquida, el diámetro medio ponderado sobre la masa y la intensidad de precipitación estimada desde la DSD como el momento de orden 3.67. Las líneas en gris representan un intervalo de confianza del 5\% sobre la relación de consistencia (\ref{eqn:consistencia1-1moment}).}
\label{fig:Scaling1_0to7_season}
\end{center}
\vspace{1cm}
\end{figure}

\begin{figure}[H] 
\begin{center}
   \includegraphics[height=0.85\textheight]{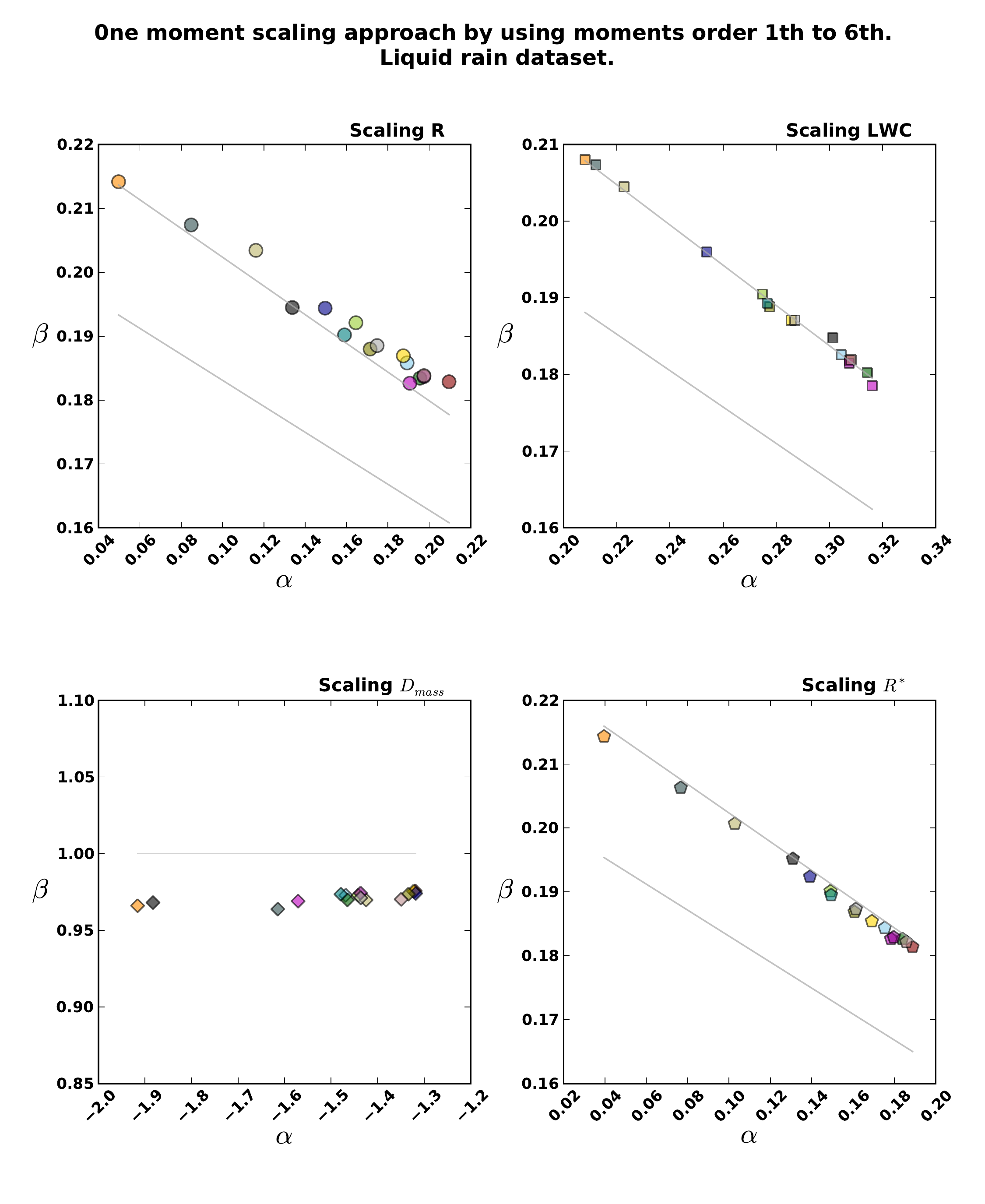}
\caption[Relación $\alpha$ vs. $\beta$ para el método de escalado usando $(M_{2}, M_{3},\ldots,M_{6})$. $20>R>0.1\,mm/h$.]{\textbf{Relación $\mathbf{\alpha}$ vs. $\mathbf{\beta}$ para el método de escalado usando $\mathbf{(M_{2}, M_{3},\ldots,M_{6})}$.  $\mathbf{20>R>0.1\,mm/h}$.} Se presentan los valores de $\alpha$ y $\beta$ obtenidos para cada disdrómetro de la red y para todos los episodios de precipitación líquida. Las variables de referencia utilizadas en la aplicación del método descrito en \S\ref{sec:Scaling1moment} son la intensidad de precipitación, el contenido de agua líquida, el diámetro medio ponderado sobre la masa y la intensidad de precipitación estimada desde la DSD como el momento de orden 3.67. Las líneas en gris representan un intervalo de confianza del 5\% sobre la relación de consistencia (\ref{eqn:consistencia1-1moment}).}
\label{fig:Scaling1_1to6_season}
\end{center}
\vspace{1cm}
\end{figure}

\begin{figure}[H] 
\begin{center}
   \includegraphics[height=0.85\textheight]{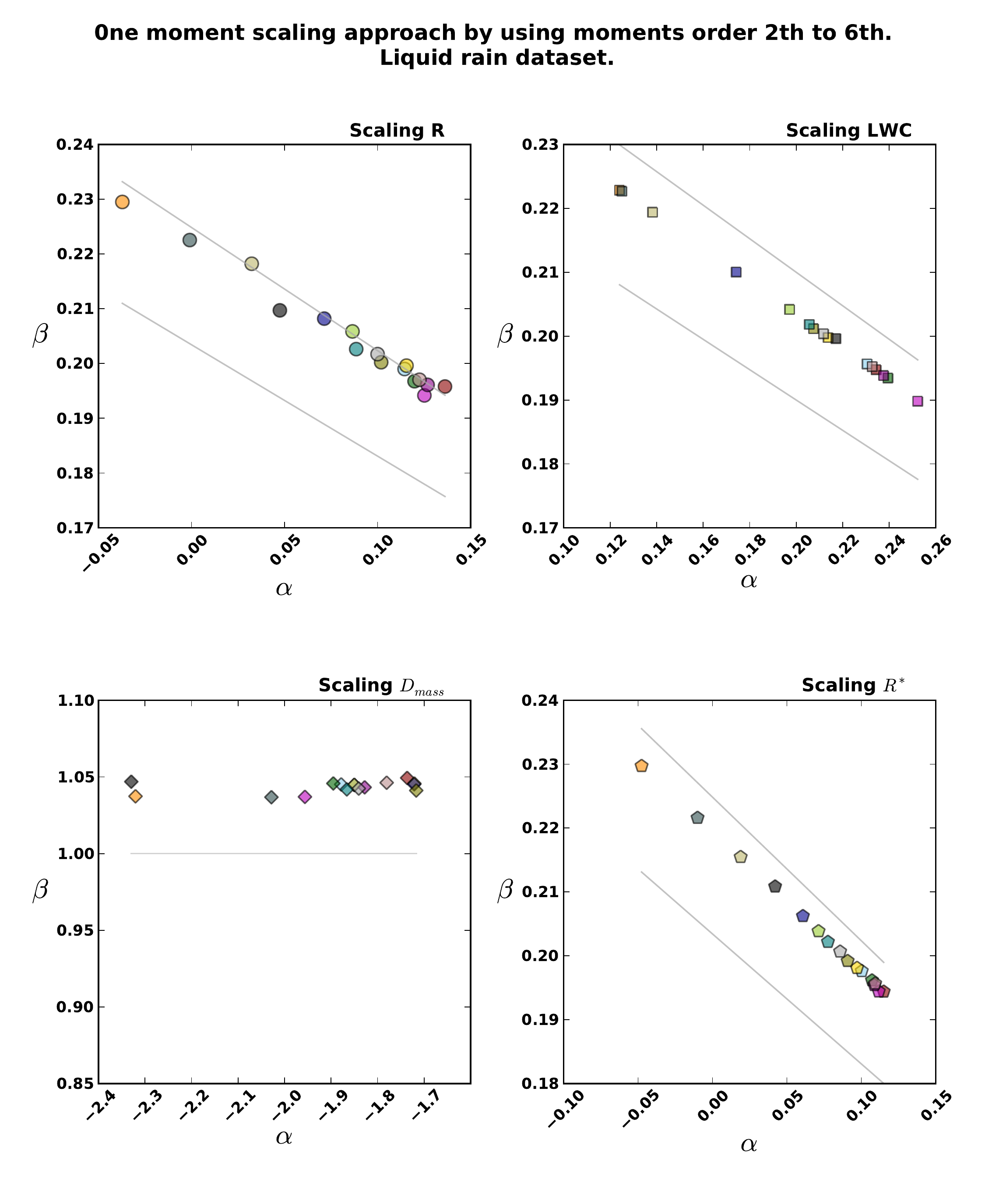}
\caption[Relación $\alpha$ vs. $\beta$ para el método de escalado usando $(M_{0}, M_{1},\ldots,M_{7})$. $20>R>0.1\,mm/h$.]{\textbf{Relación $\mathbf{\alpha}$ vs. $\mathbf{\beta}$ para el método de escalado usando $\mathbf{(M_{0}, M_{1},\ldots,M_{7})}$.  $\mathbf{20>R>0.1\,mm/h}$.} Se presentan los valores de $\alpha$ y $\beta$ obtenidos para cada disdrómetro de la red y para todos los episodios de precipitación líquida. Las variables de referencia utilizadas en la aplicación del método descrito en \S\ref{sec:Scaling1moment} son la intensidad de precipitación, el contenido de agua líquida, el diámetro medio ponderado sobre la masa y la intensidad de precipitación estimada desde la DSD como el momento de orden 3.67. Las líneas en gris representan un intervalo de confianza del 5\% sobre la relación de consistencia (\ref{eqn:consistencia1-1moment}).}
\label{fig:Scaling1_2to6_season}
\end{center}
\vspace{1cm}
\end{figure}

Los resultados aparecen en las figuras (\ref{fig:Scaling1_0to7_season}), (\ref{fig:Scaling1_1to6_season}) y (\ref{fig:Scaling1_2to6_season}). En ellas se muestran tres pa\-rá\-me\-tros de referencia: la intensidad de precipitación obtenida mediante la ecuación (\ref{eqn:calculoRexperimental}) y que por tanto es solo aproximadamente un momento de la DSD. El contenido en agua líquida, que es proporcional al momento de orden 3 de la DSD. El parámetro que hemos llamado $R^{*}$ que definimos como el momento de orden 3.67, y que es una aproximación adecuada para la intensidad de lluvia si se cumple la ecuación (\ref{eqn:AtlasVDequationPOWERLAW}). Como complemento hemos incluido como referencia $D_{mass}$, que es un cociente de dos momentos de la DSD, en este último caso la relación de consistencia (\ref{eqn:consistencia1-1moment}) toma la forma de $\beta \simeq 1$.\\

En todas las figuras se muestran dos líneas que representan un intervalo de confianza del 5\% en el valor de $\beta$ dado por la relación de consistencia. Observamos varios hechos:

\begin{itemize}
 \item Dado el conjunto total de datos de precipitación líquida y valores de intensidad de precipitación entre 0.1 y 20 mm/h. La red de disdrómetros presenta, en sus predicciones, una consistencia interna al obtener siempre relaciones entre $\alpha$ y $\beta$ similares a la esperada.
 \item La presencia de momentos afectados bien por problemas de muestreo, bien por problemas de estimación de gotas pequeñas implican la aparición de un sesgo positivo en el valor de $\beta$ supuesto el valor $\alpha$. 
 \item El conjunto de momentos $(M_{2}, M_{3},\ldots,M_{6})$ es el más adecuado para la estimación mediante el método de escalado usando un momento, ya que permite un procedimiento consistente cuando el parámetro de referencia es un momento exacto de la DSD.
 \item  En el caso de utilizar $D_{mass}$ como referencia se ha de tener en cuenta un posible sesgo sistemático, bien positivo, bien negativo dependiendo del conjunto de momentos considerado.
\end{itemize}

\vspace{1cm}

\vspace{1cm}

\vspace{1cm}

\section{Sumario/Summary} 

The main properties of the UCLM network of disdrometers and the first campaign of measurements are: 

\begin{itemize}
 \item A specific experiment has been designed to be able to study the spatial variability of drop size distribution. This experiment uses a dense network of 18 disdrometers. They are deployed in an irregular spatial configuration that allow us to have estimations of DSD in the whole range between 0 and 3 $km$. Also at the same time the number of disdrometers at each scale inside the network is constant (at intervals around 500 $m$).
\item The experiment uses autonomous disdrometers that use solar panels to have a continuous electrical power. The data transmission is done by a GPRS antenna.
\item The empirical database has enough DSD measurements to build estimations of the physical variables under study. The mean climatic values are similar to the general properties obtained by the network.
\item The consistence along the network has been done from different points of view:  
\begin{itemize}
 \item Raingauge simulation.
\item Composite DSD.
\item Time series of accumulated rainfall.
\item Histograms of Integral rainfall parameters.
\item Histograms of normalized DSD parametes.
\item Consistency within the one moment scaling approach.
\end{itemize}
showing that any of them has a global anomalous behavior during the whole season. We only found differences in the B1, E2 and C1 disdrometers. This differences  could be related with the non-homogeneous issues of the laser beam \citep{deMoraes2011}, and they are similar to other experiments.
\end{itemize}

\renewcommand\chapterillustration{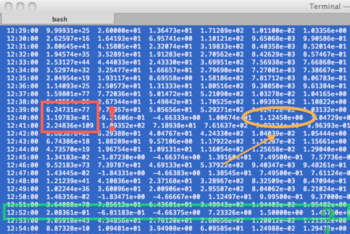}
\chapter{Preprocesado de datos disdrométricos}
\label{chap:preprocesadoTOKAY}
Los disdrómetros son los instrumentos más fiables para la medida de la DSD. Sin embargo, como se indicó en \S\ref{sec:Instrumentacion}, cada instrumento posee unas particularidades que reflejan sus ventajas y defectos en las estimaciones. Estas características específicas se deben tanto a aspectos técnicos (calidad de componentes) y ambientales (ruido ambiental) como a la propia capacidad de extraer información del principio físico en que se basa la medida. 

\section{Características propias del disdrómetro Parsivel}

En el caso de los disdrómetros Parsivel las principales cuestiones que deben tenerse en cuenta son: 

\begin{itemize}
\item Los intervalos de clase con valores desde 0.062 a 0.197 $mm$ están desactivados debido a las altas tasas de ruido. Se espera que futuros desarrollos de instrumento mejoren este aspecto.
\item El número de gotas con diámetros de hasta aproximadamente 0.8 $mm$ puede ser infravalorado (en comparación con disdrómetros JWD o POSS)\footnote{En el proceso de escritura de esta tesis se tuvo acceso a uno de los nuevos modelos de Parsivel constatando que las medidas son consistentes en todo el espectro de tamaños de gota excepto en gotas pequeñas donde en efecto la medida es infravalorada, y gotas grandes en intensidades de precipitación altas que no sobrevalora en exceso la tasa de gotas mayores de 2.5 $mm$. Esto no altera los resultados de esta tesis, ya que no se han basado en la estimación de esta parte del espectro. De cara a las correlaciones entre instrumentos del mismo tipo esta cuestión no tiene relevancia alguna.}.
\item Se ha comprobado la existencia de sobrestimación sistemática de gotas grandes, $D>2.5\,mm$, para eventos de precipitación intensos caracterizados por una intensidad de precipitación superior a $15\,mm/h$ (algunos autores establecen este límite entre 10 y 20 $mm/h$).
\item Es posible que, debido a choques con la estructura del disdrómetro, se registren gotas pequeñas procedentes no de la DSD original, sino de la ruptura de gotas mayores al interaccionar con el disdrómetro. En caso de producirse este fenómeno estas gotas poseen relaciones $v(D)$ notablemente diferentes de las usuales, con lo que es posible añadir como hipótesis una relación $v(D)$ junto con un intervalo de confianza sobre esta para discriminar gotas pequeñas que proceden de la DSD de las que pueden proceder de procesos de colisión con el aparato (véase \S\ref{sec:opticalDISDRO}).
\item En el reciente estudio de \citep{deMoraes2011} se muestra el papel jugado por la homogeneidad del haz láser en las medidas de instrumentos ópticos \emph{Thies} que tienen un principio de medida análogo al Parsivel.
\end{itemize}

\section{Estandarización de las medidas del disdrómetro Parsivel}

Las características del proceso de estimación mediante el disdrómetro Parsivel aconsejan estandarizar las medidas\footnote{En otros instrumentos se habría de realizar también un preprocesado de datos. En particular para el disdrómetro JWD sería necesario incorporar los puntos (1), (2) y (5), además de detalles específicos de este disdrómetro como el filtro para corregir el problema de \emph{Dead Time Effect}.}, en el caso de disponer de redes heterogéneas de instrumentos, mediante un pre-procesado que consta de los siguientes puntos:

\begin{figure}[h] 
\vspace{0.5cm}
\begin{center}
    \includegraphics[width=0.85\textwidth]{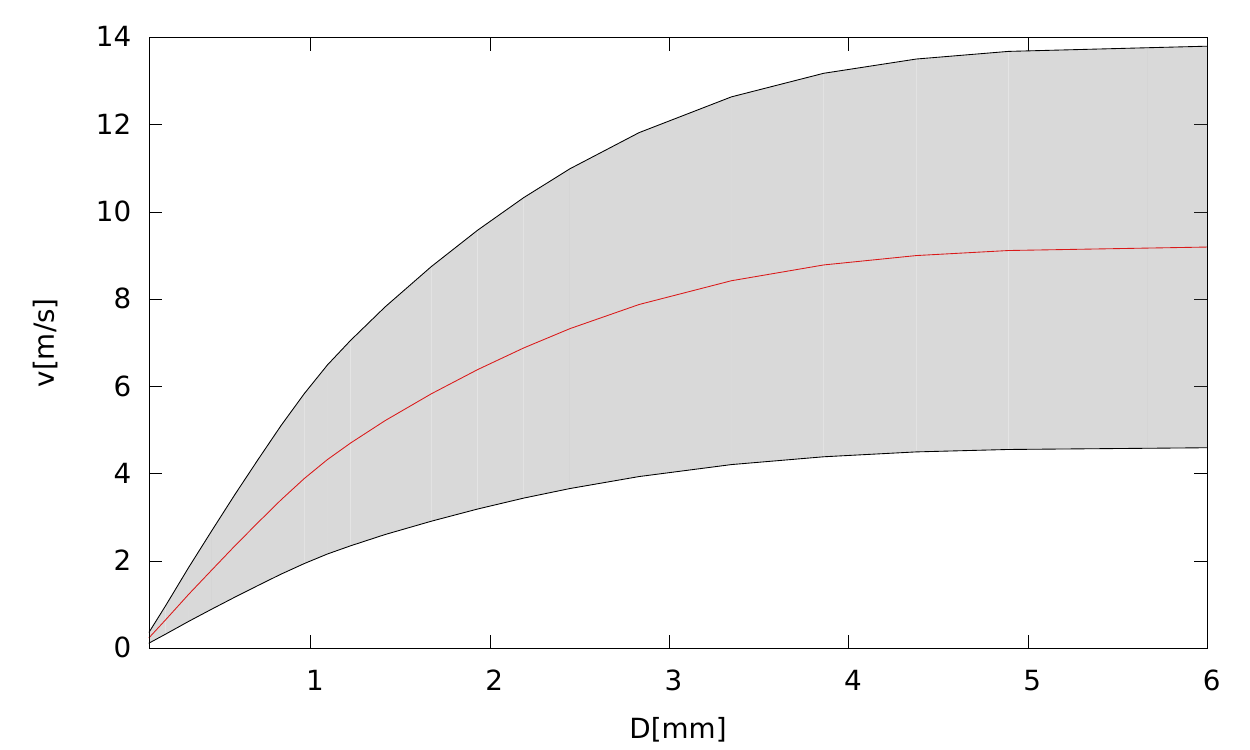}
\vspace{0.5cm}
    \caption[Filtro en velocidades basado en la relación dada por ecuación la (\ref{eqn:AtlasVDequation})]{\textbf{Filtro en velocidades basado en la relación dada por ecuación la (\ref{eqn:AtlasVDequation})}. El filtro se construye eliminando aquellas gotas cuyas estimaciones no pertenezcan al área dibujada en gris en la figura.}
 \label{fig:filtrovD}
 \end{center}
\vspace{0.5cm}
 \end{figure}

\begin{figure}[H] 
\vspace{0.5cm}
\begin{center}
    \includegraphics[width=1.00\textwidth]{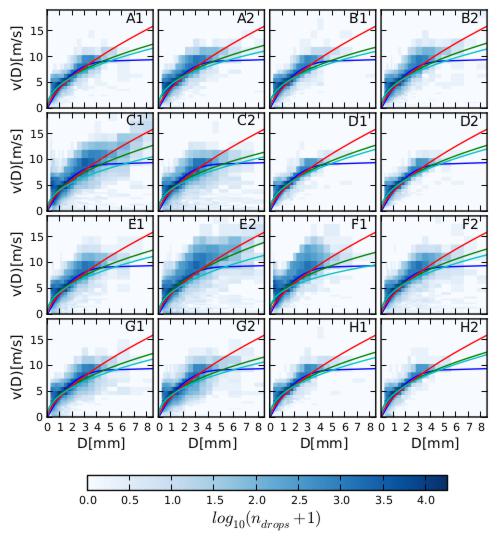}
\vspace{0.5cm}
    \caption[Matriz de valores $n(D,v)$ para episodio  23/12/2009 sin filtrado]{\textbf{Matriz de valores $n(D,v)$ para episodio 23/12/2009 sin filtrado.} Se muestran los resultados experimentales en todos los disdrómetros de la red para $n(D,v)$ en escala logarítmica. Se muestra la relación dada por ecuación (\ref{eqn:AtlasVDequation}) en rojo. Los ajustes para obtener $v(D)$ desde $n(D,v)$ tanto mediante un ajuste lineal de  $v(D)=\gamma D^{\delta}$ como mediante un ajuste no-lineal se muestran usando líneas verde y azul claro. La línea azul oscuro representa una modificación de la propuesta dada por (\ref{eqn:AtlasVDequation}) que incorpora una corrección por no esfericidad así como una saturación más rápida a partir de diámetros de 4 $mm$.}
 \label{fig:filtrovD_dia23}
 \end{center}
 \end{figure}

\begin{figure}[H] 
\vspace{0.5cm}
\begin{center}
    \includegraphics[width=1.00\textwidth]{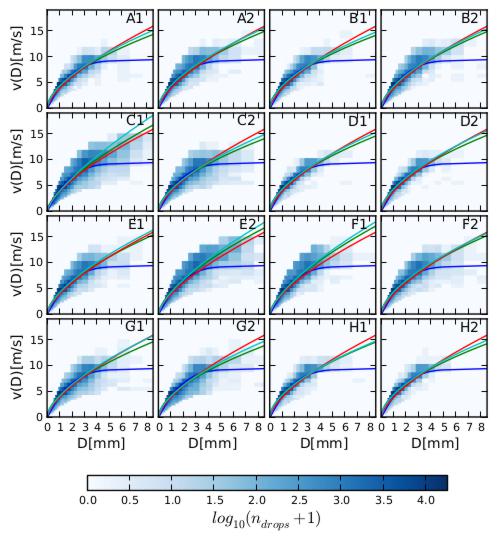}
\vspace{0.5cm}
    \caption[Matriz de valores $n(D,v)$ para episodio 23/12/2009 con filtrado]{\textbf{Matriz de valores $n(D,v)$ para episodio 23/12/2009 con filtrado}. Se muestran los resultados experimentales en todos los disdrómetros de la red para $n(D,v)$ en escala logarítmica. Se muestra la relación dada por ecuación (\ref{eqn:AtlasVDequation}) en rojo. Los ajustes para obtener $v(D)$ desde $n(D,v)$ tanto mediante un ajuste lineal de  $v(D)=\gamma D^{\delta}$ como mediante un ajuste no-lineal se muestran usando líneas verde y azul claro. La línea azul oscuro representa una modificación de la propuesta dada por (\ref{eqn:AtlasVDequation}) que incorpora una corrección por no esfericidad así como una saturación más rápida a partir de diámetros de 4 $mm$.}
 \label{fig:filtrovD_dia23filtrado}
 \end{center}
\vspace{1cm}
 \end{figure}

\begin{enumerate}
\item Espectros $n(D,v)$ con número total de gotas menor de 10 son considerados como minutos de no precipitación suficiente para inferir una DSD y se excluyen del análisis.

\item Se considera que la relación $v(D)$ de \citep{1973Atlas}, véase ecuación (\ref{eqn:AtlasVDequation}), se satisface razonablemente, por lo que se construye un filtro basado en aceptar gotas con velocidades terminales que difieran como máximo un 50\% por exceso o por defecto sobre esta relación\footnote{Es decir, es un filtro multiplicativo basado en la relación $v(D)$ que se aplica a $n(D,v)$.}. En la figura (\ref{fig:filtrovD}) se presenta el filtro comentado y en las figuras (\ref{fig:filtrovD_dia23}) y (\ref{fig:filtrovD_dia23filtrado}) se presenta su relevancia mediante figuras que contienen diagramas $v(D)$ para el episodio de precipitación del 23 de diciembre del 2010 sin el proceso de filtrado y tras el proceso de filtrado.

\item El número de gotas pequeñas puede estar infravalorado para $D<0.8\,mm$ con lo que el análisis se puede realizar con y sin los intervalos de clase hasta la cifra de 0.9 $mm$. Para estudios basados en la DSD compuesta que representen un evento completo se puede proceder a evaluar posibles correcciones a las medidas en estos intervalo de clase (bajo la hipótesis, por ejemplo, de estar bien modelizadas como una distribución exponencial).
\item Con objeto de evaluar la posible sobre-estimación de gotas grandes, se realizan estudios combinando acotaciones entre 10 y 20 $mm/h$.
\item Se incluyen acotaciones inferiores basadas en la intensidad de precipitación con valores 0.05 y 0.1 $mm/h$, para la distinción de eventos con y sin precipitación. El valor 0.1 $mm/h$ es un valor de referencia utilizado en experimentos con JWD; el valor 0.05 $mm/h$ se puede incluir para una comparación con el anterior. La tabla (\ref{TablaMINUTOSdata}) contiene la relevancia de este filtrado.

\item Se pueden incluir también dos acotaciones inferiores con objeto de simular las estimaciones mínimas de los radares incorporados en los satélites TRMM y el futuro GPM. Las acotaciones son respectivamente 0.5 y 0.2 $mm/h$. Consúltese la figura (\ref{fig:Var2WholeComposite}). 

\item Se corrige el área de medida del disdrómetro usando el hecho de que si hay una gota en el haz el área efectiva será la nominal menos la que posea la sección vertical de una gota del diámetro efectivo de la gota \citep{ParsivelSNOW,loffler-mang_joss_2000_aa}:

\vspace{2mm}
\begin{equation}
S_{eff}(D_{i})=L(W-D_{i}/2)
\label{eqn:AreaEfectiva}
\end{equation}
\vspace{2mm}

\end{enumerate}

\begin{table}[h]
\vspace{0.95cm}
\caption[Parámetros de la relación de potencias $v(D)=\gamma D^{\delta}$ para cada episodio de precipitación líquida tras el preprocesado]{\textbf{Parámetros de la relación de potencias $v(D)=\gamma D^{\delta}$ para cada episodio de precipitación líquida tras el preprocesado}. Comparativa de la estimación $v(D)=\gamma D^{\delta}$ mediante regresión lineal sobre la red de disdrómetros. El coeficiente de correlación medio es la media aritmética sobre los 16 disdrómetros del coeficiente de correlación obtenido mediante los datos de cada uno de ellos.}\label{tabla:preprocesadoVD}
\vspace{0.95cm}
\begin{center}
\small \ra{1.30}
\begin{tabular}{lcc>{\columncolor[gray]{0.95}}c>{\columncolor[gray]{0.95}}ccc>{\columncolor[gray]{0.95}}c}

\toprule
\textbf{Evento} & \multicolumn{2}{c}{\textbf{Ajuste no lineal}} & \multicolumn{2}{c}{\textbf{Ajuste lineal}} &\multicolumn{3}{c}{\textbf{Correlaciones lineales}}  \\

\cmidrule(r{.5em}){2-3} \cmidrule(l{.5em}){4-5} \cmidrule(l{.5em}){6-8}

       & $\bar{\delta}\pm\Delta \delta$ & $\bar{\gamma}\pm\Delta \gamma$ &  $\bar{\delta}\pm\Delta \delta$ & $\bar{\gamma}\pm\Delta \gamma$ & $\rho$ max.        &  $\rho$ min. & $\rho$ medio \\
\midrule
02-dic-2009   &$0.63\pm0.04$        & $4.05\pm0.16$         & $0.69\pm 0.05$&  $4.05\pm0.19$ & 0.89 & 0.81 & 0.87 \\
20-dic-2009   & $0.53\pm0.03$       & $4.26\pm0.12$         & $0.56\pm0.04$  & $4.24\pm0.12$  & 0.94 &0.88       &0.92\\
24-dic-2009   & $0.62\pm0.04$       & $4.14\pm0.13$         & $0.67\pm0.04$  & $4.09\pm0.14$  & 0.89 &0.86       &0.87\\
03-ene-2010     & $0.56\pm0.04$       & $4.25\pm0.11$         & $0.58\pm0.05$  & $4.23\pm0.10$  & 0.94 &0.88       &0.92\\
06-ene-2010        & $0.61\pm0.05$       & $4.20\pm0.12$         & $0.67\pm0.05$  & $4.21\pm0.13$  & 0.89 &0.83       &0.86\\
12-ene-2010       & $0.60\pm0.04$       & $4.13\pm0.14$         & $0.65\pm0.05$  & $4.07\pm0.14$  & 0.93 &0.86      & 0.89\\  
                                                                             
\bottomrule
\end{tabular}
\end{center}
\vspace{0.75cm}
\end{table}


Los resultados presentados en el capítulo anterior presentan este tipo de pre-procesado, específicamente se ha usado el punto (2) así como una intensidad mínima de 0.1 mm/h y máxima de 20 mm/h. En el caso de la publicación \citep{Tapiador2010} se optó por incluir el mínimo pre-procesado posible dado que el objetivo era realizar un contraste de hipótesis que no dependiera de la inclusión o no de hipótesis previas, con todo sí fueron evaluadas diferentes acotaciones que incluyen de modo efectivo tanto los puntos (1) como (5). El punto (2) ha sido evaluado mostrando los diagramas $v(D)$ de modo que las hipótesis de variabilidad espacial sobre los episodios analizados no están condicionadas por la inclusión o no de este punto. Esto se aprecia en el hecho de que los valores experimentales de la intensidad de precipitación no utilizan los valores de $v(D)$ y su correlograma es similar a la reflectividad. El punto (4) se satisface genéricamente para los eventos analizados.\\

Los resultados presentados en los capítulos sucesivos incluyen un pre-procesado de datos dis\-dro\-mé\-tri\-cos, con objeto de entender la relevancia de las diferentes acotaciones y su posible implicación en los algoritmos presentes en el futuro GPM-DPR y el actual TRMM-PR. En el mismo sentido son necesarios en caso de intentar estimaciones cuantitativas de la variabilidad espacial, tal y como se muestra en el apéndice \S\ref{sec:newVariability}. En último término las figuras mostradas en \S\ref{sec:newVariability} son conceptualmente análogas a las mostradas en \citep{Tapiador2010}. El realizar un preprocesado no varía el contraste de hipótesis realizado aunque, para estandarizar las medidas respecto de otras bases empíricas y desarrollar aplicaciones concretas, pueden ser aplicados los puntos anteriores.\\

La tabla (\ref{tabla:preprocesadoVD}) muestra los resultados tras el preproceso para la relación $v(D)=\gamma D^{\delta}$ tanto mediante un ajuste no lineal directo como mediante un ajuste lineal tras una transformación logarítmica. Los resultados muestran coeficientes de correlación altos, valores menores para el exponente $\delta$ y valores de $\gamma$ algo mayores respecto de la relación (\ref{eqn:AtlasVDequationPOWERLAW}). Los datos sin el proceso de filtrado aparecen en \S\ref{sec:relacionVD}.\\

Esta parte del preprocesado se basa en el filtrado directo en la matriz $n(D,v)$ para una $v(D)$ dada. Lo lógico es, por tanto, que sean los valores de la relación $v(D)=\gamma D^{\delta}$ los que presenten más diferencias. Es importante señalar que este proceso de filtrado solo puede ser realizado tras un análisis del tipo de hidrometeoro del episodio de precipitación bajo estudio. En un evento mixto nieve-lluvia podría dar lugar a inconsistencias con otros aparatos de medida al aparecer, tras el filtrado, como un episodio típico de lluvia sin serlo en realidad.
\newpage

\section{Relevancia del preprocesado}

En esta sección se recopilan algunas comparativas específicas con el fin de mostrar la relevancia del preprocesado en la estandarización de las medidas disdrómetricas, cuya utilidad es importante de cara a poder comparar los resultados de esta tesis, tanto con resultados previos como con resultados futuros.

\begin{figure}[H] 
\begin{center}
  
\includegraphics[width=1.00\textwidth]{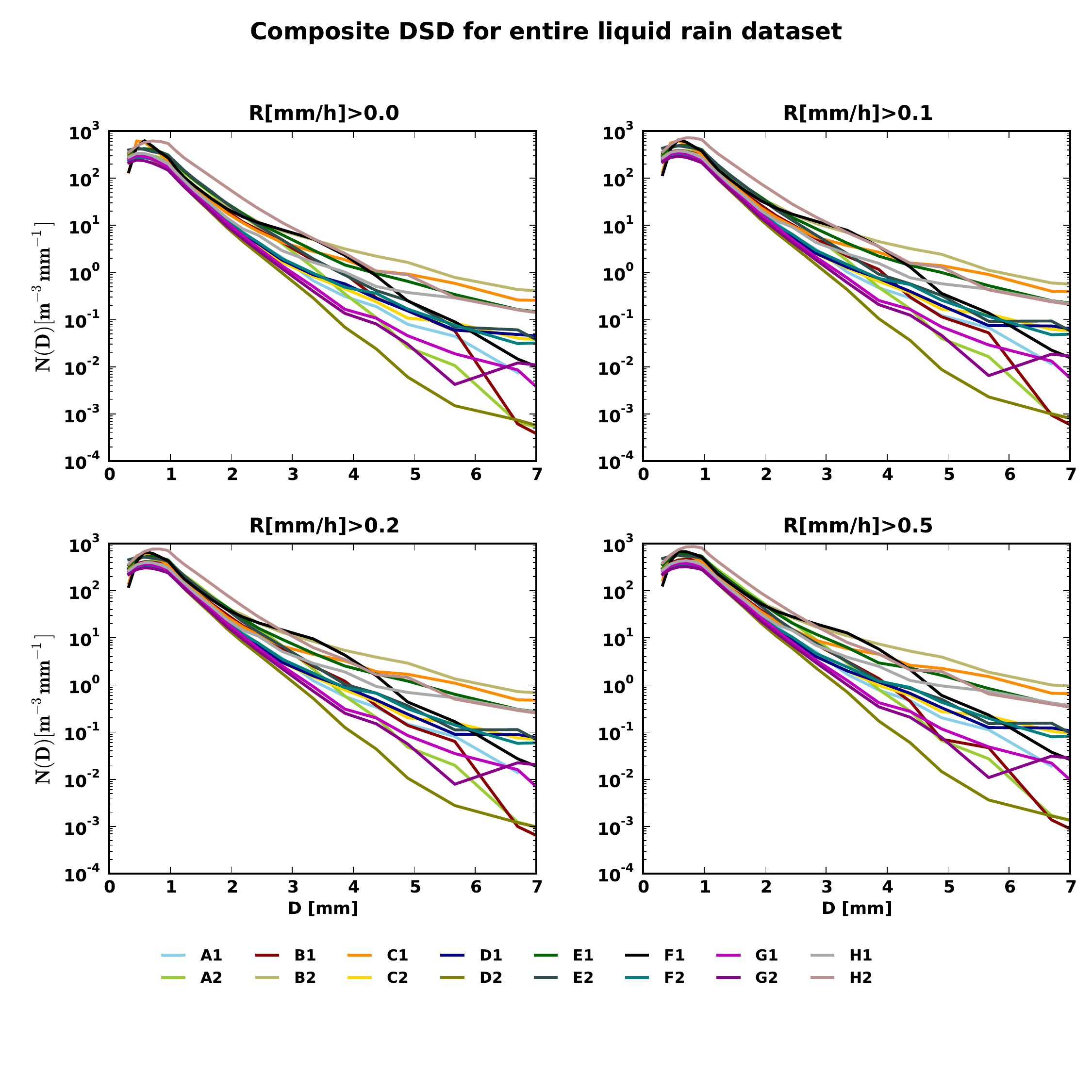}
\vspace{0.55cm}
   \caption[DSD compuestas para toda la base empírica sin el filtrado dado en Figura (\ref{fig:filtrovD})]{\textbf{DSD compuestas para toda la base empírica sin el filtrado dado en Figura (\ref{fig:filtrovD}).} Se han incluido acotaciones que se especifican en la figura: $R\,>\,0.1\,mm/h$, $R\,>\,0.2\,mm/h$ y $R\,>\,0.5\,mm/h$, que se corresponde con umbrales de detección de lluvia de diferentes sensores. El significado de los diferentes umbrales puede consultarse en la figura (\ref{fig:Var2WholeComposite}).}
\label{fig:Var2WholeComposite_nonf_nontv}
\end{center}
\vspace{0.95cm}
\end{figure}

\begin{figure}[H] 
\vspace{1.75cm}
\begin{center}
  
\includegraphics[width=1.05\textwidth]{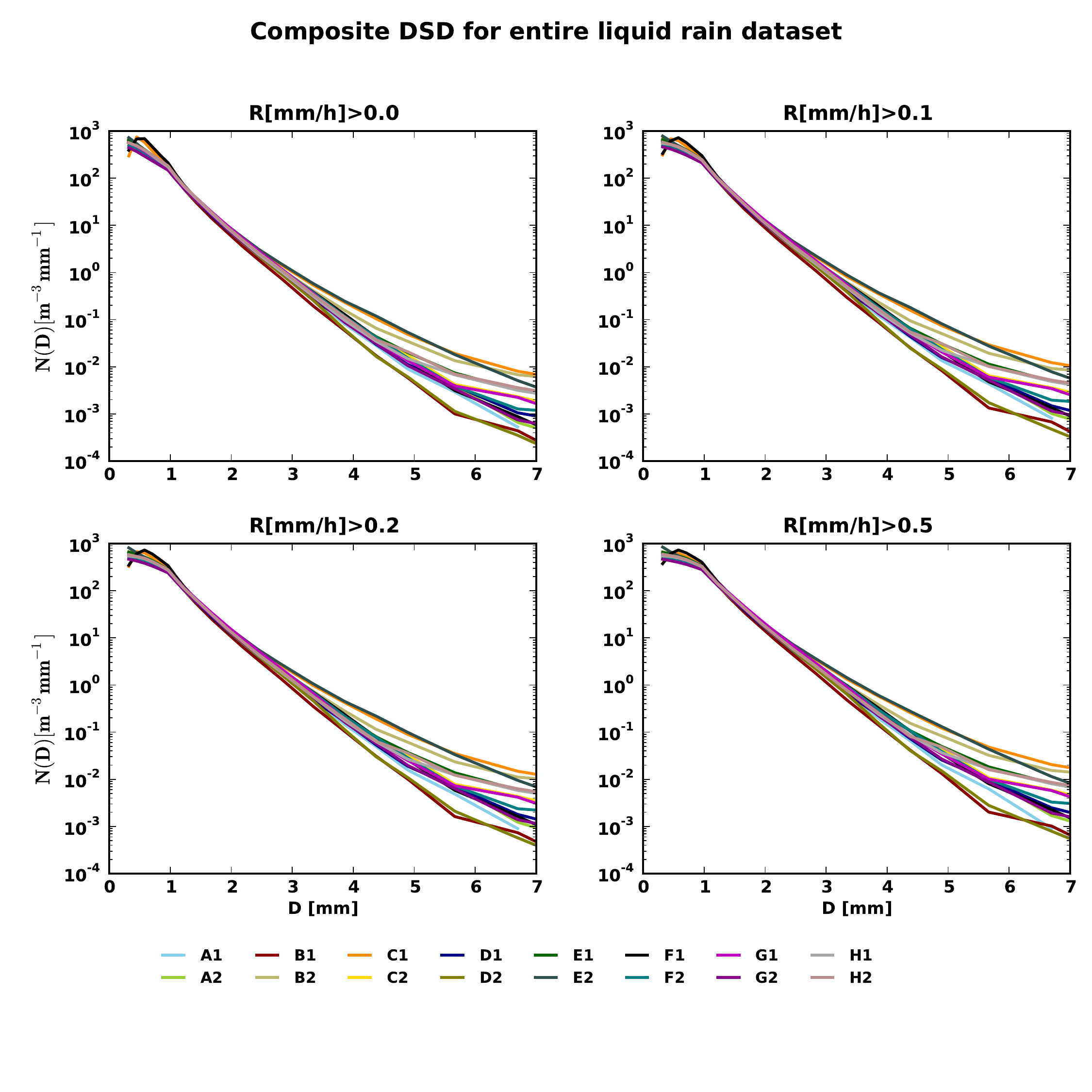}
\vspace{0.55cm}
   \caption[DSD compuestas para toda la base empírica sin el filtrado dado en Figura (\ref{fig:filtrovD}) pero con velocidades de caída obtenidas mediante la ecuación (\ref{eqn:AtlasVDequation})]{\textbf{DSD compuestas para toda la base empírica sin el filtrado dado en Figura (\ref{fig:filtrovD}) pero con velocidades de caída obtenidas mediante la ecuación (\ref{eqn:AtlasVDequation}).} Se han incluido acotaciones que se especifican en la figura: $R\,>\,0.1\,mm/h$, $R\,>\,0.2\,mm/h$ y $R\,>\,0.5\,mm/h$, que se corresponde con umbrales de detección de lluvia de diferentes sensores. El significado de los diferentes umbrales puede consultarse en la figura (\ref{fig:Var2WholeComposite}).}
\label{fig:Var2WholeComposite_nonf_yestv}
\end{center}
\vspace{0.95cm}
\end{figure}
\newpage

\subsection{DSD compuesta}

Observando las figuras (\ref{fig:Var2WholeComposite_nonf_nontv}), (\ref{fig:Var2WholeComposite_nonf_yestv}) y (\ref{fig:Var2WholeComposite}) es posible entender la relevancia en las DSD compuestas de los procesos de estandarización de medidas.\\

La figura (\ref{fig:Var2WholeComposite_nonf_nontv}) muestra el efecto de solo incluir los datos brutos obtenidos por los diferentes disdrómetros. En ella la presencia de un número pequeño de minutos en que las gotas de mayor tamaño posee velocidades anómalas tiene un impacto nítido en las DSD compuestas. La alta sensibilidad de estas a variaciones en las velocidades verticales se comprueba observando la figura (\ref{fig:Var2WholeComposite_nonf_yestv}) donde el asumir velocidades según ecuación (\ref{eqn:AtlasVDequation}) introduce una mayor consistencia entre los diferentes instrumentos. En consecuencia, la mayoría de los estudios realizados con instrumentos Parsivel incluyen filtros similares a la figura (\ref{fig:filtrovD}), de cara a comparar las distribuciones de tamaños de gota promedio.

\subsection{Precipitación líquida acumulada}
\label{sec:liquidACCpreprocess}
\begin{figure}[H] 
\begin{center}
  
\includegraphics[width=0.95\textwidth]{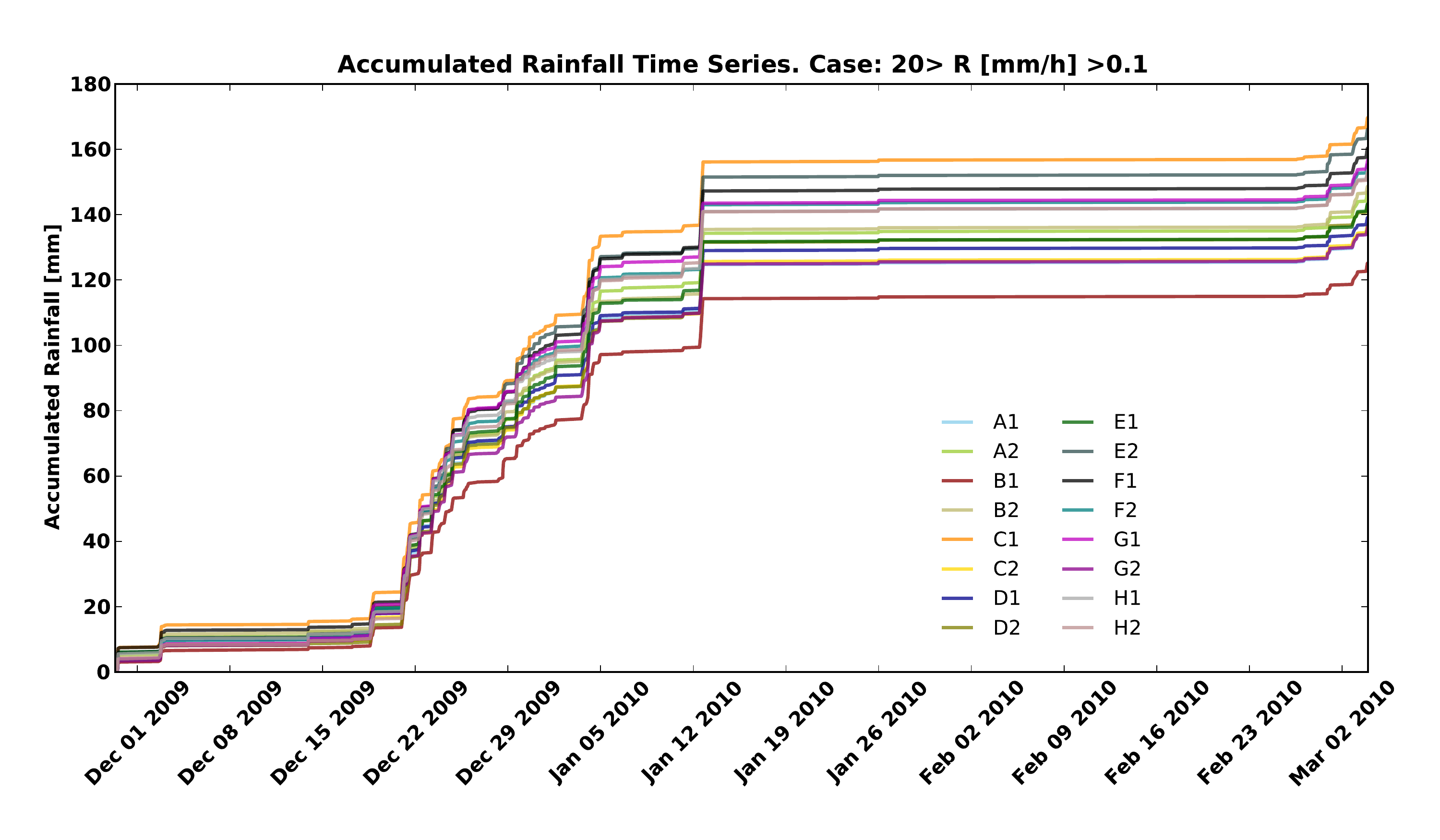}
   \caption[Acumulación de lluvia durante la base empírica. Precipitación líquida.Sin filtrado.]{\textbf{Acumulación de lluvia durante la base empírica. Precipitación líquida. $20\, >\,R\,[mm/h]\, >\, 0.1$. Sin filtrado.} Cálculo desde conjunto consistente para minutos pero sin filtrado basado en velocidades terminales. Se aprecian diferencias en instrumentos B1 respecto del resto de la red; A la vista de la figura (\ref{fig:Var2WholeComposite}) proviene de un déficit igualmente sistemático.}
\label{fig:RaccTimeSeries_nonf_nontv_20}
\end{center}

\end{figure}

\begin{figure}[H] 
\begin{center}
  
\includegraphics[width=0.99\textwidth]{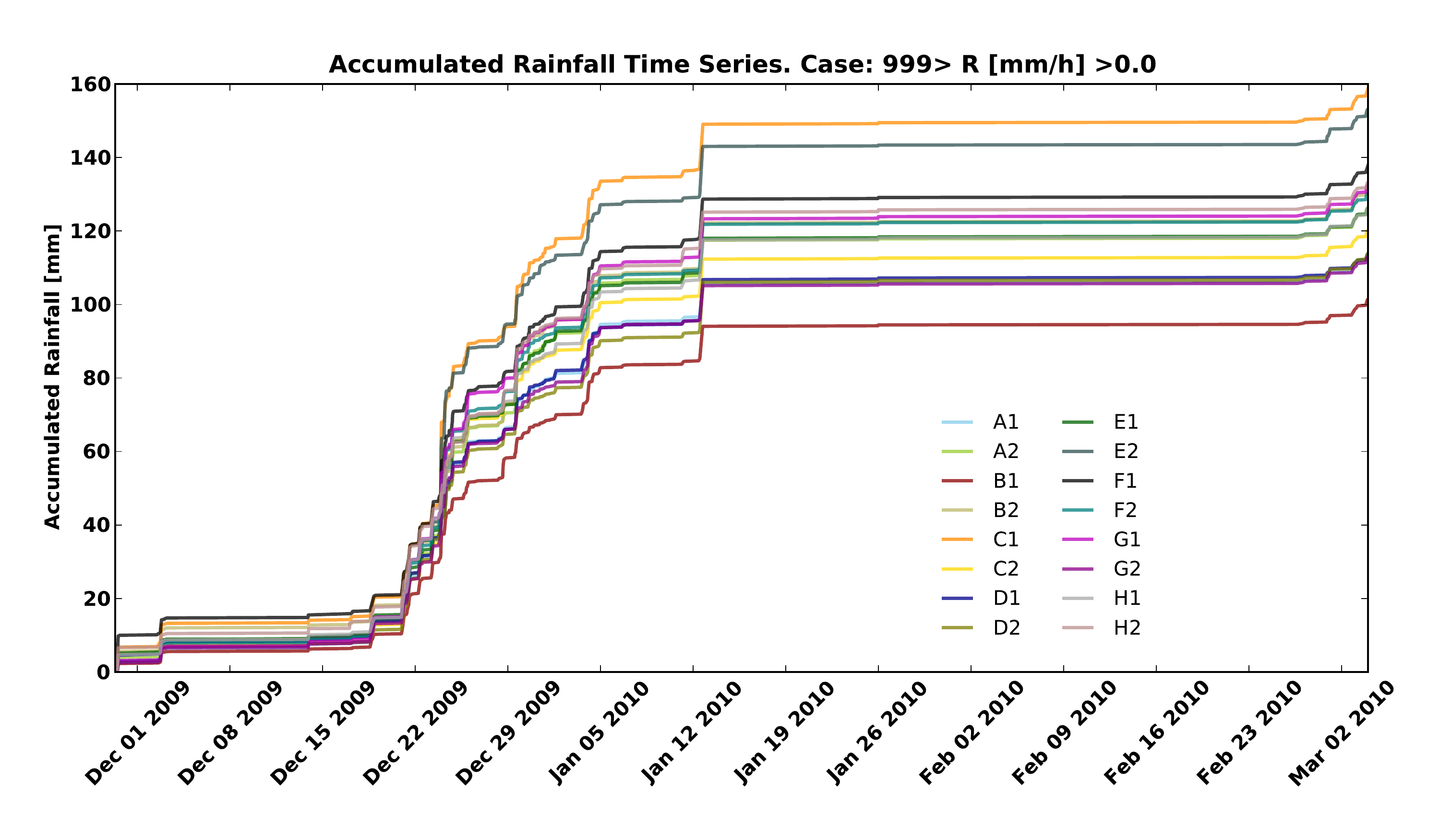}
   \caption[Acumulación de lluvia durante la base empírica. Precipitación líquida. Sin filtrado.]{\textbf{Acumulación de lluvia durante la base empírica. Precipitación líquida. $999\, >\, R\, [mm/h]\, >\, 0.0$. Sin filtrado. }. Se aprecian las diferencias en los instrumentos C1, E2 y B1 respecto del resto de la red; una parte importante de la diferencia en la precipitación acumulada procede del evento registrado el 25 de diciembre, sobre todo en lo que se refiere al instrumento E2.}\label{fig:RaccTimeSeries_nonf_nontv_0_999}

\includegraphics[width=0.99\textwidth]{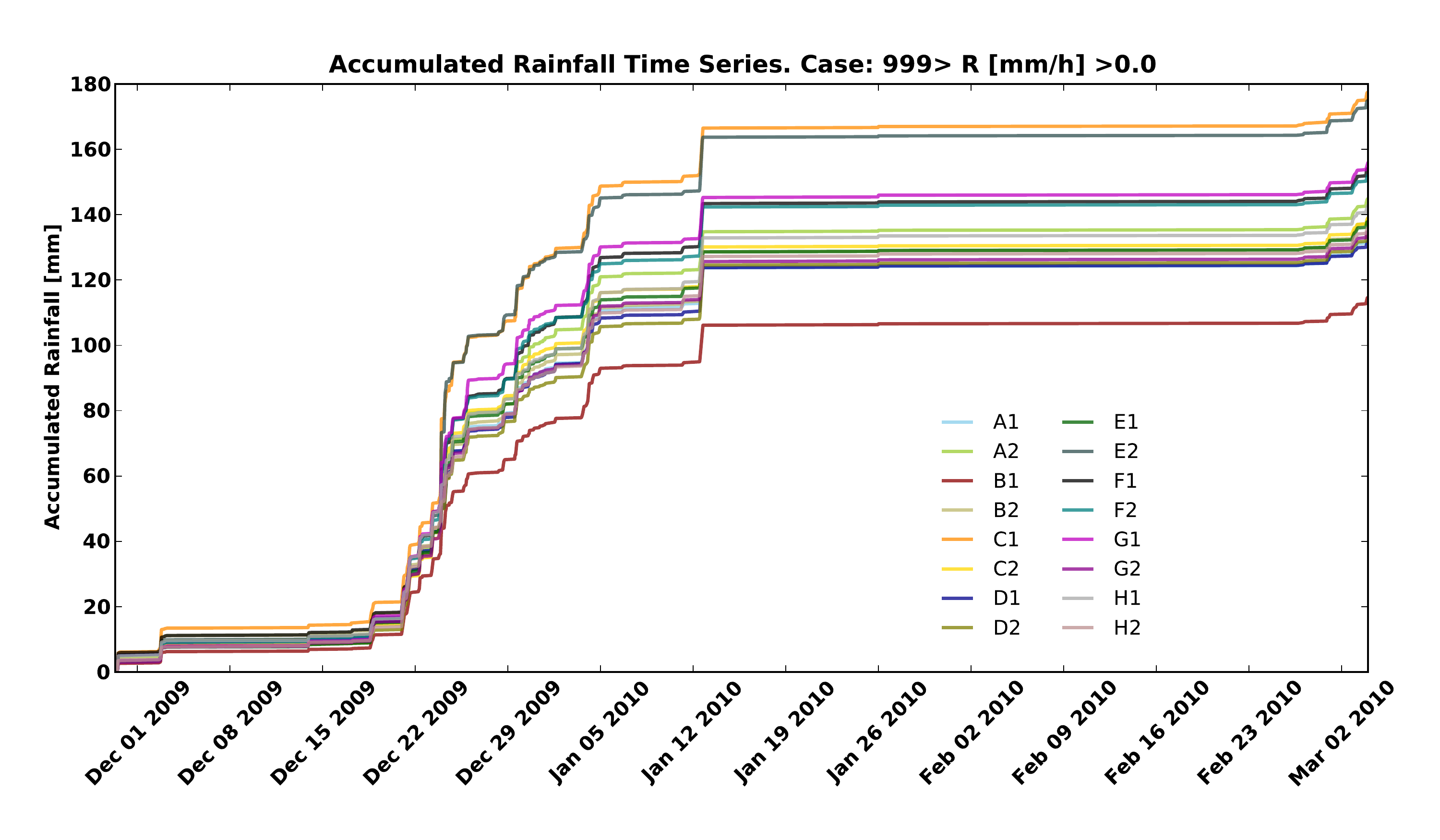}
   \caption[Acumulación de lluvia durante la base empírica. Precipitación líquida. Con filtrado.]{\textbf{Acumulación de lluvia durante la base empírica. Precipitación líquida.  $999\,>\, R\, [mm/h]\, >\, 0.0$. Con filtrado.} Cálculo desde conjunto consistente para minutos con al menos 10 gotas registradas en cada instrumento. También se aprecian las diferencias en los instrumentos C1, E2 y B1 respecto del resto de la red.}
\label{fig:RaccTimeSeries_yesf_nontv_0_999}
\end{center}
\end{figure}

Para entender el papel de los diferentes filtros y umbrales comparamos cuatro figuras de intensidad de precipitación acumulada sujetas, bien a diferencias en el proceso de filtrado, bien a diferencias en los umbrales de intensidad de precipitación introducidos. Como se aprecia en las figuras (\ref{fig:RaccTimeSeries_nonf_nontv_20}), (\ref{fig:RaccTimeSeries_nonf_nontv_0_999}) y (\ref{fig:RaccTimeSeries_yesf_nontv_0_999}), el factor clave desde el punto de vista de la precipitación acumulada no es la presencia de filtrado de gotas en función de sus velocidades terminales, sino que es la selección de umbrales de precipitación mínima y máxima. Es además este último el que más puede condicionar la comparación adecuada entre disdrómetros y pluviómetros que pudieran aparecer en un experimento multi-instrumental.  

\subsection{Relaciones Z-R}

\begin{figure}[H] 
\begin{center}
   \includegraphics[width=0.95\textwidth]{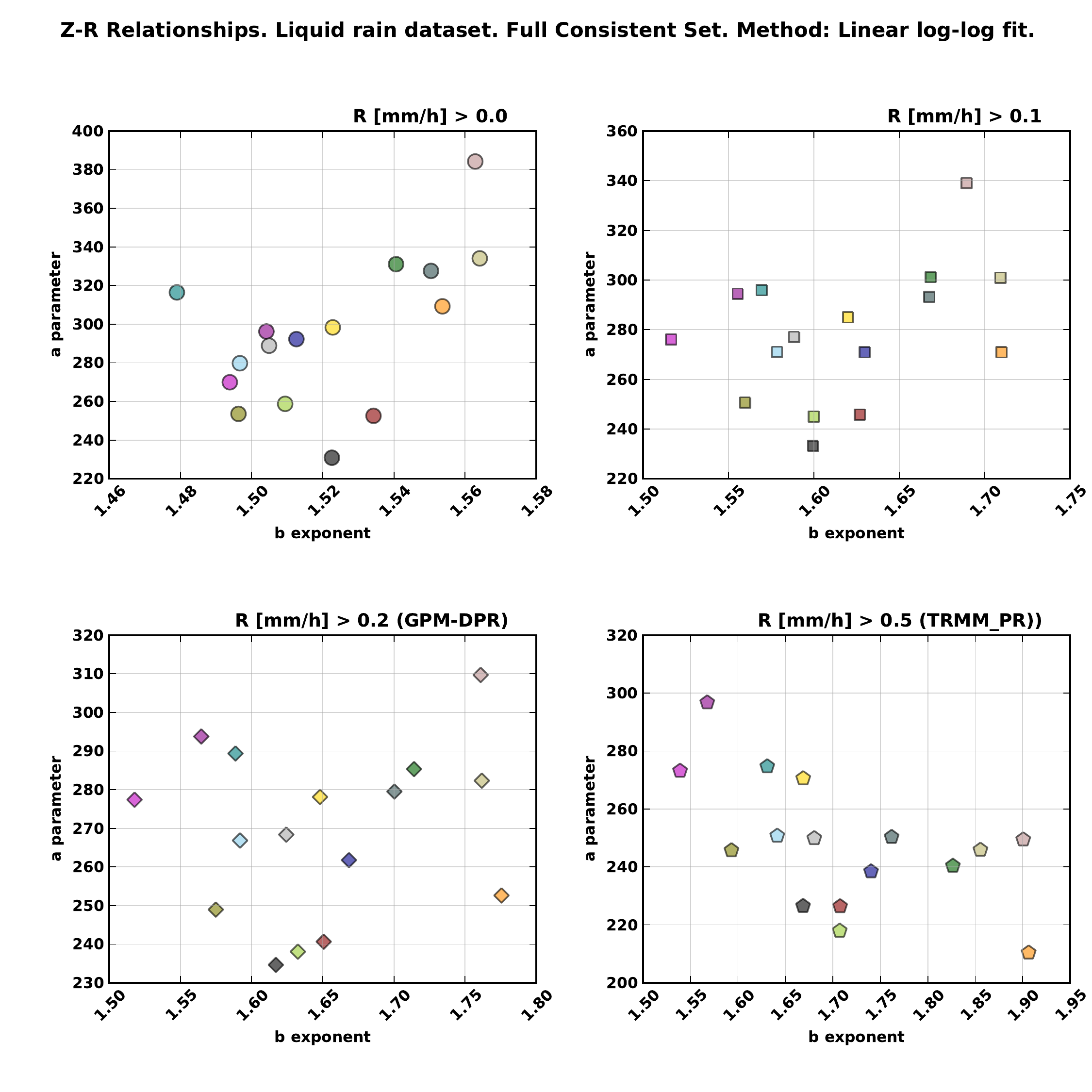}
   \caption[Relaciones Z-R para la base empírica completa y diferentes intensidades de lluvia mínimas.]{\textbf{Relaciones Z-R para la base empírica completa y diferentes intensidades de lluvia mínimas.} Se muestran las relaciones Z-R para la base empírica completa (restringida a todos los episodios de precipitación líquida con intensidades de precipitación máximas de 20 mm/h) calculada para todos los disdrómetros de la red y bajo cuatro intensidades de precipitación mínima en todos los disdrómetros a un tiempo. Simula las sensibilidades estimadas para varios sensores radar en términos de intensidad de precipitación mínima detectable. Los coeficientes de la relación Z-R han sido calculados mediante regresión lineal simple en escala logarítmica. Este caso se basa en un pre-procesado que no incluye filtro en velocidades.}
\label{fig:ZRwhole_nonf_nontv}
\end{center}
\vspace{0.5cm}
\end{figure}

\begin{figure}[H] 
\begin{center}
   \includegraphics[width=0.95\textwidth]{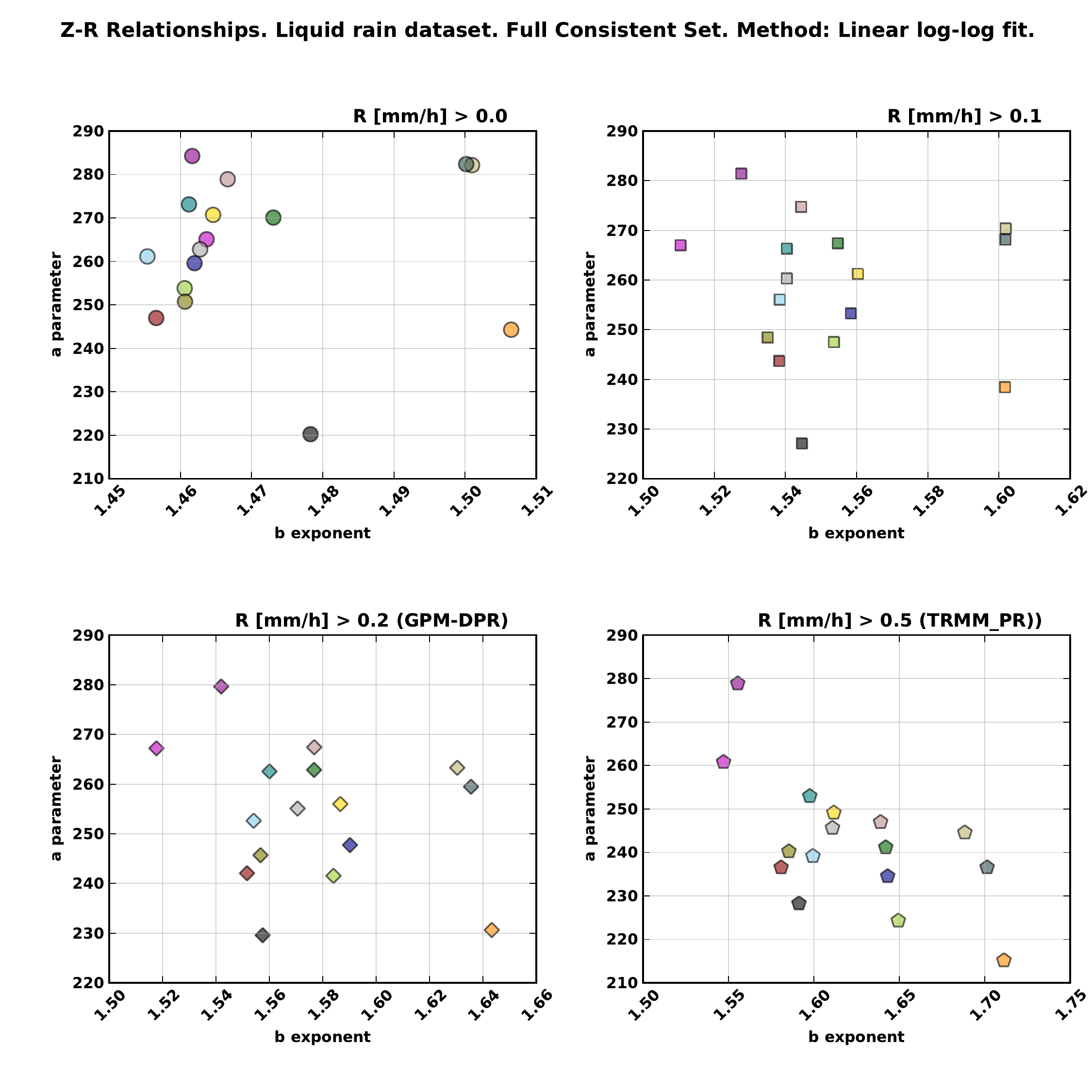}
   \caption[Relaciones Z-R para la base empírica completa y diferentes intensidades de lluvia mínimas.]{\textbf{Relaciones Z-R para la base empírica completa y diferentes intensidades de lluvia mínimas.} Se muestran las relaciones Z-R para la base empírica completa (restringida a toda la base empírica de precipitación líquida con intensidades de precipitación máximas de 20 mm/h) calculada para todos los disdrómetros de la red y bajo cuatro intensidades de precipitación mínima en todos los disdrómetros a un tiempo. Simula las sensibilidades estimadas para varios sensores radar en términos de intensidad de precipitación mínima detectable. Los coeficientes de la relación Z-R han sido calculados mediante regresión lineal simple en escala logarítmica. Este caso se basa en un pre-procesado que no incluye filtro en velocidades pero utiliza la relación (\ref{eqn:AtlasVDequation}) para estimar las velocidades de caída.}
\label{fig:ZRwhole_nonf_yestv}
\end{center}
\vspace{0.5cm}
\end{figure}

En el caso de las relaciones Z-R la presencia de umbrales, así como, las diferencias que las gotas con velocidades terminales más anómalas pueden condicionar las estimaciones de la relación Z-R. Comparamos dos casos con la figura (\ref{fig:ZRwhole}) mostrada en el capítulo anterior que se basaba en el pre-procesado mediante el filtrado en velocidades de caída y unos umbrales $20 > R[mm/h]$. En este capítulo mostramos, para esos mismos umbrales, pre-procesados sin el filtrado en velocidades y sin asumir la relación (\ref{eqn:AtlasVDequation}) \textemdash figura (\ref{fig:ZRwhole_nonf_nontv})\textemdash \, y asumiendo que las velocidades terminales siguen dicha relación \textemdash figura (\ref{fig:ZRwhole_nonf_yestv})\textemdash.

Se observa la similitud entre la figura (\ref{fig:ZRwhole}) y (\ref{fig:ZRwhole_nonf_yestv}), concluyendo que las velocidades terminales anómalas tiene un efecto relevante en los coeficientes de las relación Z-R. Más aun en caso de no incluir umbrales de intensidad de precipitación mínima que excluyan parte de los valores marginales, valores que pueden condicionar la estimación de $a_{R}$ y $b_{R}$. En el caso de la estimación de dichos coeficientes mediante un método de ajuste no-lineal, la presencia de estos valores marginales condiciona la estabilidad del algoritmo de estimación, pudiendo dar lugar a valores en $a_{R}$ y $b_{R}$ artificiales. Resulta por tanto adecuado, tanto utilizar el pre-procesado estándar como evaluar el papel de los diferentes umbrales en los análisis sucesivos.

\begin{figure}[H] 
\begin{center}
   \includegraphics[height=0.65\textheight]{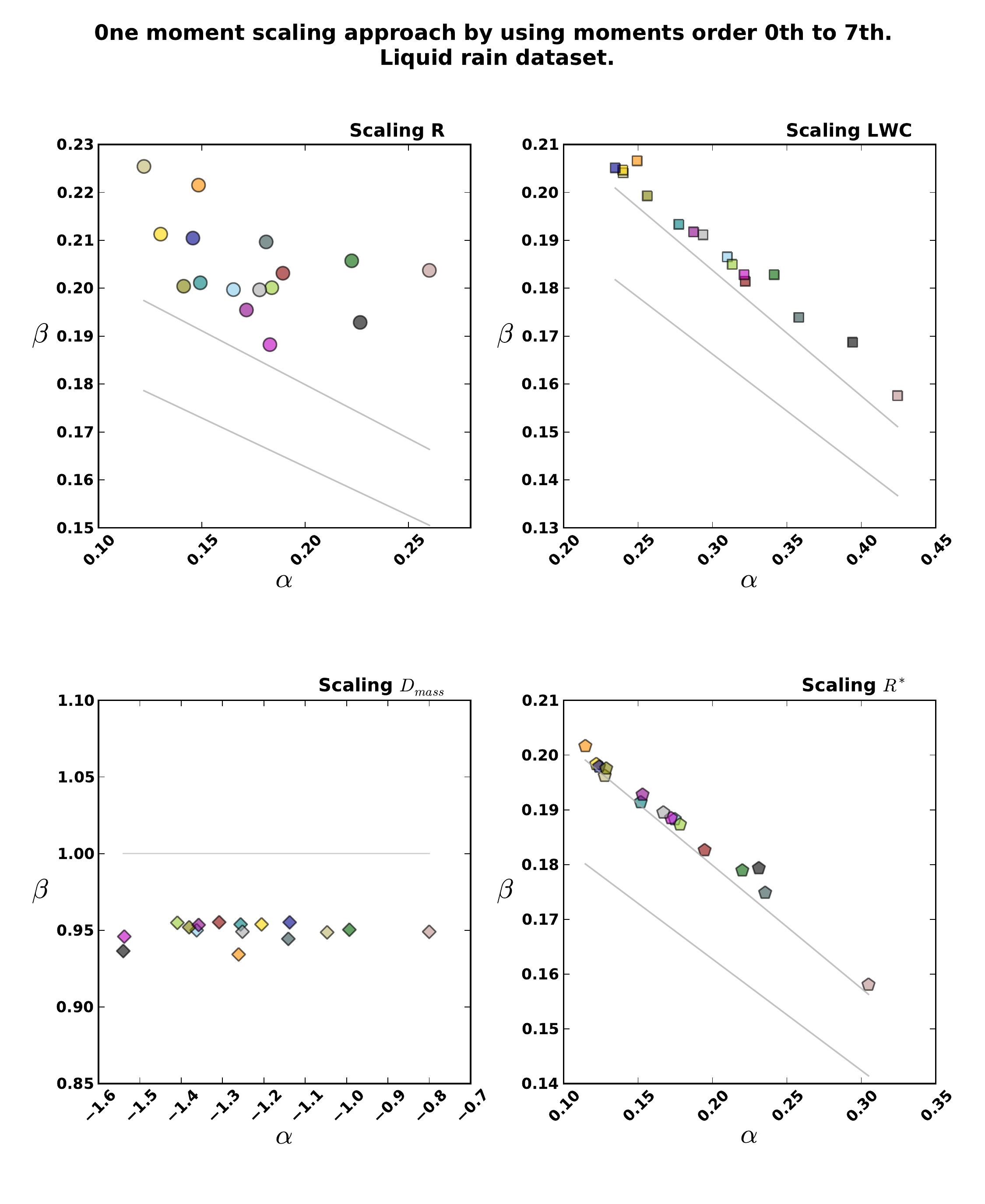}
 \caption[Relación $\alpha$ vs. $\beta$ para el método de escalado usando $(M_{0}, M_{1},\ldots,M_{7})$. $20>R>0.1\,mm/h$. Sin filtrado.]{\textbf{Relación $\alpha$ vs. $\beta$ para el método de escalado usando $(M_{1}, M_{2},\ldots,M_{6})$.  $20>R>0.1\,mm/h$. Sin filtrado.} Se presentan los valores de $\alpha$ y $\beta$ obtenidos para cada disdrómetro de la red y para toda la base empírica de precipitación líquida. Las variables de referencia utilizadas en la aplicación del método descrito en \S\ref{sec:Scaling1moment} son la intensidad de precipitación, el contenido de agua líquida, el diámetro medio ponderado sobre la masa y la intensidad de precipitación estimada desde la DSD como el momento de orden 3.67. Las lineas en gris representan un intervalo de confianza del 5\% sobre la relación de consistencia (\ref{eqn:consistencia1-1moment}). Este caso se basa en un pre-procesado que no incluye filtro en velocidades.}
\label{fig:Scaling1_0to7_season_nonf_nontv}
\end{center}
\vspace{1cm}
\end{figure}
\vspace{1cm}

\begin{figure}[H] 
\begin{center}
   \includegraphics[height=0.65\textheight]{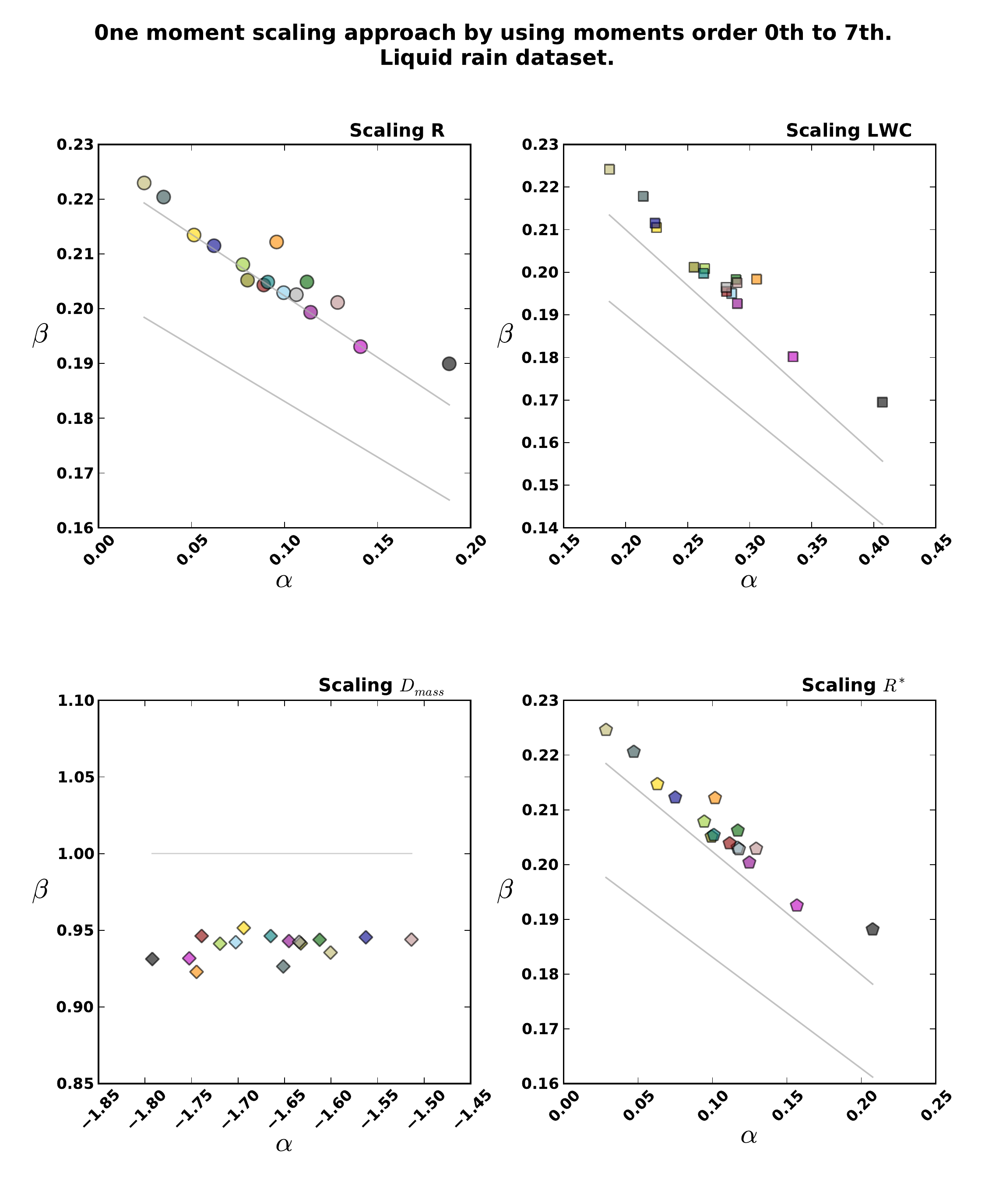}
\caption[Relación $\alpha$ vs. $\beta$ para el método de escalado usando $(M_{0}, M_{1},\ldots,M_{7})$. $20>R>0.1\,mm/h$. Sin filtrado. Con velocidades terminales dada por (\ref{eqn:AtlasVDequation}).]{\textbf{Relación $\alpha$ vs. $\beta$ para el método de escalado usando $(M_{1}, M_{2},\ldots,M_{6})$.  $20>R>0.1\,mm/h$.} Se presentan los valores de $\alpha$ y $\beta$ obtenidos para cada disdrómetro de la red y para toda la base empírica de precipitación líquida. Las variables de referencia utilizadas en la aplicación del método descrito en \S\ref{sec:Scaling1moment} son la intensidad de precipitación, el contenido de agua líquida, el diámetro medio ponderado sobre la masa y la intensidad de precipitación estimada desde la DSD como el momento de orden 3.67. Las lineas en gris representan un intervalo de confianza del 5\% sobre la relación de consistencia (\ref{eqn:consistencia1-1moment}). Este caso se basa en un pre-procesado que no incluye filtro en velocidades pero utiliza la relación (\ref{eqn:AtlasVDequation}) para estimar las velocidades de caída.}
\label{fig:Scaling1_0to7_season_nonf_yestv}
\end{center}
\vspace{1cm}
\end{figure}
\vspace{1cm}

\subsection{Método de escalado basado en un parámetro integral}

Las implicaciones del pre-procesado en el método de escalado usando un parámetro integral (introducido en \S\ref{sec:Scaling1moment}) se han evaluado desde la relación de consistencia entre $\alpha$ y $\beta$. Para ello es necesario comparar las figura (\ref{fig:Scaling1_0to7_season}) con los resultados análogos sin la introducción de filtrados. Estos resultados aparecen en las figuras (\ref{fig:Scaling1_0to7_season_nonf_nontv}) y (\ref{fig:Scaling1_0to7_season_nonf_yestv}). Las diferencias entre ellas permiten inferir las siguientes conclusiones:

\begin{itemize}
   \item La introducción de un filtrado en velocidades terminales es necesaria para que el escalado respecto de la intensidad de precipitación cumpla de modo razonable la relación de consistencia al comparar instrumentos similares.
   \item El no realizar un filtrado, pero si utilizar las velocidades dadas por (\ref{eqn:AtlasVDequation}) lleva a satisfacer la relación de consistencia más fielmente en el caso de escalar respecto de R.
   \item Los escalados respecto de otros parámetros integrales no resultan tan dependientes del pre-procesado pudiendo considerarse en este sentido más robustos. También podemos observar como el resultado el rango de valores de $\alpha$ es mayor cuando no se introduce un filtro en velocidades terminales, incluso si se adopta la ecuación (\ref{eqn:AtlasVDequation}). Por tanto, las gotas que no se eliminan tras el filtrado pueden tener relevancia en las relaciones entre parámetros integrales, principalmente en el conjunto $(M_{0}, M_{1},\ldots,M_{7})$.
\end{itemize}

\section{Sumario/Summary}

In this chapter a general preprocessing procedure was presented. The main characteristics are:

\begin{itemize}
 \item The preprocessing allows to build more standard data estimation of the integral rainfall parameters and the DSDs. This is necessary to reduce the bias when we are comparing results of several kinds of instruments.
 \item The condition on minimum number of drops and rainfall to define a rainy minute allows us to obtain a set of DSDs that are more easy to model. In the case of a network the application needs to check how this set changes along the network.
 \item The filter based on the relation (\ref{eqn:AtlasVDequation}) is designed to avoid spurious drops generated in the collision process with the disdrometer. After this filter we obtain minimal differences in the experimental $v(D)$ relations along the network (in an event by event analysis).
 \item The disdrometers are supposed to show an underestimation of the smallest drops. Another filter could be included to correct this issue in the case of composite DSD. In the case of high temporal resolution more insight is needed to build this kind of heavy preprocessing.
\end{itemize}

\part{Results}

\renewcommand\chapterillustration{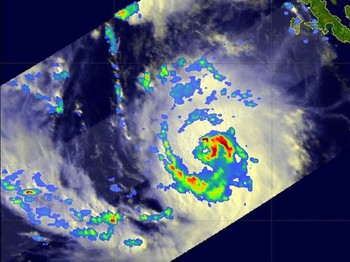}
\chapter{First measurement of small-scale DSD variability}

\label{sec:chapVARSPACIAL1}

The initial analysis of the entire empirical basis carried out in \textsection\ref{sec:baseempirica} shows coherent functioning of the experimental system. Additionally, the total minutes and available episodes allow a detailed analysis of the proposed objective in this thesis, which is to study the variations of precipitation in fields on the kilometer scale.\\

This chapter presents results for the \textit{verification of the hypothesis} on the kilometer scale. To this end, the procedure is first introduced, and  each one of the relevant analyses is subsequently developed, including episode-selection processes, studies of velocity-diameter relationships, time-series analysis of rain intensity (phase lag and correlation), stability of Z-R relations and DSD modeling throughout the network. Each of the sections introduces the methodologies employed, which complement those already introduced in \S\ref{cap:DSD}.\\

This chapter contains a figure (\ref{fig:Var1SerieRmean}) published in the journal \textit{Geophysical Research Letters}.

\section{Initial concepts and methodologies}
\label{sec:var1conceptos}
For the study of the precipitation texture, the use of concepts related to both the theory of stochastic processes (Poisson processes, Cox processes, scale analysis) and the theory of fluctuation statistics has been proposed \citep{Kostinski200638}. However, in light of the scale of the interest to us, the main tool utilized is statistical data analysis in addition to the use of geostatistical techniques (which provide the most useful approach at larger scales). This study should be carried out in two steps. First, the hypothesis concerning whether there is kilometer-scale spatial variation in precipitation should be tested, identifying the areas in which it is reflected; next, an attempt to perform a statistical characterization of the variation can be performed. This approach allows the \textit{problem associated with validation} to be addressed, as it has been defined by \citep{Ciach1999585}.\\

In this chapter, the hypothesis is validated; additionally, an analysis of the relevance of various aspects, ranging from the integral parameters of precipitation to DSD modeling, is presented through the role in Z-R relationships.\\

Next, we introduce several definitions of practical use throughout this chapter. Given a physical quantity related to precipitation, we characterize it as a random variable X with concrete values $x_{i}$ in accordance with a probability distribution that in general should be determined (modeled from the random variable X) by physical criteria or that also can be inferred based on a set of measurements.\\

The probability distribution can be dependent upon the location in which the measurement is made; in this case, we write $X(\vec{u})$ and speak of the \textit{random field}\footnote{Direct translation of \textit{random field}.} Additionally, it is possible that we have $X(\vec{u},t)$, in which case we speak of the spatiotemporal variability. In a case in which $X(\vec{u})$ does not depend on the location, $X(\vec{u})=X_{0}$, we speak of a homogeneous random field. In contrast, if it only depends on distance and not direction, $X(\vec{u})=X(|\vec{u}|)$, an isotropic random field is usually spoken of. It is also of interest to analyze the variability of $X(\vec{u})$ at different scales \footnote{Although here only the typical distances of the typical size of a ground-radar pixel and the satellite spatial resolution that make up a radar sensor are studied. It is interesting to compare the results of isotropy and homogeneity at various scales to a global compression of the factors that determine the DSD 
variability.}.\\

Given that, generically, we have a random field $X(\vec{u},t)$, the questions are presented in the following order:
\begin{enumerate}
\item The episodes of precipitation are studied from the point of view of the intensity of accumulated precipitation and the $v(D)$ relation, to test the data consistency by episode. To avoid excluding processes of precipitation intermittency in the network, the rain intensity data under study has not been filtered; in fact, the only requirement is that simultaneous reliable data must be in the entire network.
\item Various time series of rain intensity are compared $R(\vec{u},t)$ in different positions to test whether they exhibit a temporal deviation (phase lag), i.e., whether spatial variability studies comparing different instruments should be carried out via $R(\vec{u}_{k},t_{k})$ y $t_{k}$ depending on the distance between disdrometers. Conversely, the data are also tested for possible comparison in terms of the strictly simultaneous values.
\item After verifying the second issue, it is important to understand the correlation between the time series of two disdrometers for different positions\footnote{In addition to evaluating the possible dependencies in the methodology for estimating the correlation between time series, this issue requires investigation of whether, when facing a verification of hypothesis, the choice of a particular method is determining.}. Answering the questions of whether the random fields are (a) homogeneous and (b) isotropic is of particular interest.
\item One of the most important applications of the variability studies is determined by their application to remote sensing. It is of interest to understand the relevance of the previous section's conclusions in light of the stability of the relationships between rain intensity and reflectivity. That is to say, it is important to demonstrate whether the power law hypothesis between the Z-R variables is valid at all scales, and if, being so, it exhibits stable values of the parameters that define it, especially taking into account the sampling geometry of a radar system, as well as the increase in the sampled volume with distance in the case of ground radar \citep{MGossetNUBF2001}.
\item As a complement, the variation in the parameters contained in the scaling proposal has been studied using a variable (a methodology that was introduced in \S\ref{sec:Scaling1moment}) with the additional objective of investigating the stability of its estimation throughout the available network of disdrometers in addition to its potential impact on DSD modeling.
\end{enumerate}

\section{Analysis of precipitation episodes}

The issues put forth in the preceding points can be dependent, at least partly, on precipitation events. Therefore, in this chapter, a joint study of seven different time series is carried out to establish reasonably general answers to the above questions. In Table (\ref{TablaEventosVar1}), we can see the seven episodes analyzed\footnote{We have included a mixed snow-rain episode to complete the type of episodes that the empirical basis has at its disposal, although it will not be analyzed in detail. This analysis requires differentiating drops of snow and rain through filtrates that would necessitate additional new hypotheses.}, which have been selected for having the following:
\begin{itemize}
\item A sufficient amount of total accumulated precipitation (greater than 9 mm) \footnote{In principle, an accumulated rainfall of less than 9 mm was established as the requirement, but a low-intensity episode has been included to analyze whether the same conclusions are also possible in cases of accumulated rainfall under 3 mm.}.
\item A sufficient number of minutes (greater than 100 min).
\item Not having prolonged transmission losses during the event (greater than 30 consecutive min), once consistency across the network has been achieved. One should have data with sufficient continuity at one time in all experimental devices.
\item An attempt was made to represent episodes with different average intensity values and possible variabilities of the DSD.
\end{itemize}

The total number of minutes is therefore 2\,718 for liquid precipitation, distributed over six stratiform episodes. The selection that was made is similar to that studied in \citep{2009ChoongLEE}, in which the spatial variability of the DSD was analyzed using other instruments and a less dense network for four stratiform episodes totaling 1\,800 recorded minutes. In that study, differences were found in terms of the accumulated precipitation on the order of 22\% in the range of 1.3 $km$ in distance; these values were somewhat higher than those obtained in our experiment, as seen in Table (\ref{tablaBIASwhole}).\\

The other episode analyzed in this chapter relates to January 10, 2010, which is characterized as having precipitation in the form of snow. For this case, several aspects of its analysis are specific, as will be demonstrated in the following sections.\\

\subsection{Simulation of pluviometers}

In table (\ref{TablaEventosVar1}), the average accumulated precipitation can be observed in addition to the calculation of the standard deviation over the network calculated by the expressions (\ref{eqn:Racc_red}) and (\ref{eqn:Racc_red_std}). The minimum and maximum values in the network are also indicated in the table, presented with the experimental station in which they were recorded (each for a consistent set of minutes).\\

The network is designed with identical devices to avoid the bias that would otherwise be introduced when comparing measurements gleaned from instruments utilizing different measurement methods. In this sense, this thesis represents a complete novelty in the field because homogeneous instrumentation has never before been used to perform this measurement.\\

\begin{table}[h]
\ra{1.20}
\caption[Precipitation episodes analyzed]{\textbf{Precipitation episodes analyzed for verification of the hypothesis about small-scale spatial variability}. The total number of minutes recorded in each episode is indicated after the consistency-ensuring process of the DSD over the network. The total accumulated precipitation values represent the precipitation obtained by the expression (\ref{eqn:calculoRexperimental}) averaged over the network in addition to its standard deviation in the disdrometer network. Also shown are the maximum and minimum accumulated precipitation values, described with the relevant disdrometer giving the entry.}
\vspace{0.55cm}
\begin{center}
\begin{tabular}{lcccrrr}
\toprule
\small \textbf{Event} &  \textbf{Julian Day} & \textbf{Type}& \textbf{Min.}  & $\mathbf{\overline{R}_{acc}\pm s^{2}(R_{acc})}$ & \textbf{Max. (dis.)} & \textbf{Min. (dis.)} \\
\midrule
\small 
02-dic-2009  & 336& Rain &  204   &   $9.40\pm2.10$ mm & 13.77 (C1)   & 6.8 (B1)\\
20-dic-2009  & 354& Rain &  743   &  $31.76\pm3.68$ mm & 38.27 (H2)   & 22.30 (B1)\\
24-dic-2009  & 358& Rain &  186   &  $10.20\pm2.01$ mm & 14.34 (C1)   & 7.31 (G2)\\
03-ene-2010  & 003& Rain &  725   &  $21.35\pm1.08$ mm & 24.59 (C1) & 18.41 (D1) \\
06-ene-2010  & 006& Rain &  351   &  $2.01 \pm0.27$ mm & 2.58 (F1)  & 1.70 (B1)\\
12-ene-2010  & 012& Rain &  509   &  $25.42\pm3.53$ mm & 32.10 (E2) & 20.74 (A1)\\
\midrule
25-dic-2010  & 359& Mixed  & 398 &  $22.397\pm3.90$ mm & 29.43 (E2) & 16.83 (E1)\\
\midrule
10-ene-2010 & 010&\emph{Snow}& 564   &  $193.9\pm23.7$ mm & 238.93 (E2) & 152.70 (B2)\\
\bottomrule
\end{tabular}
\label{TablaEventosVar1}
\end{center}
\vspace{0.70cm}
\end{table}%

\begin{center}
\begin{figure}[h]
\vspace{0.50cm}
   \includegraphics[width=1.00\textwidth]{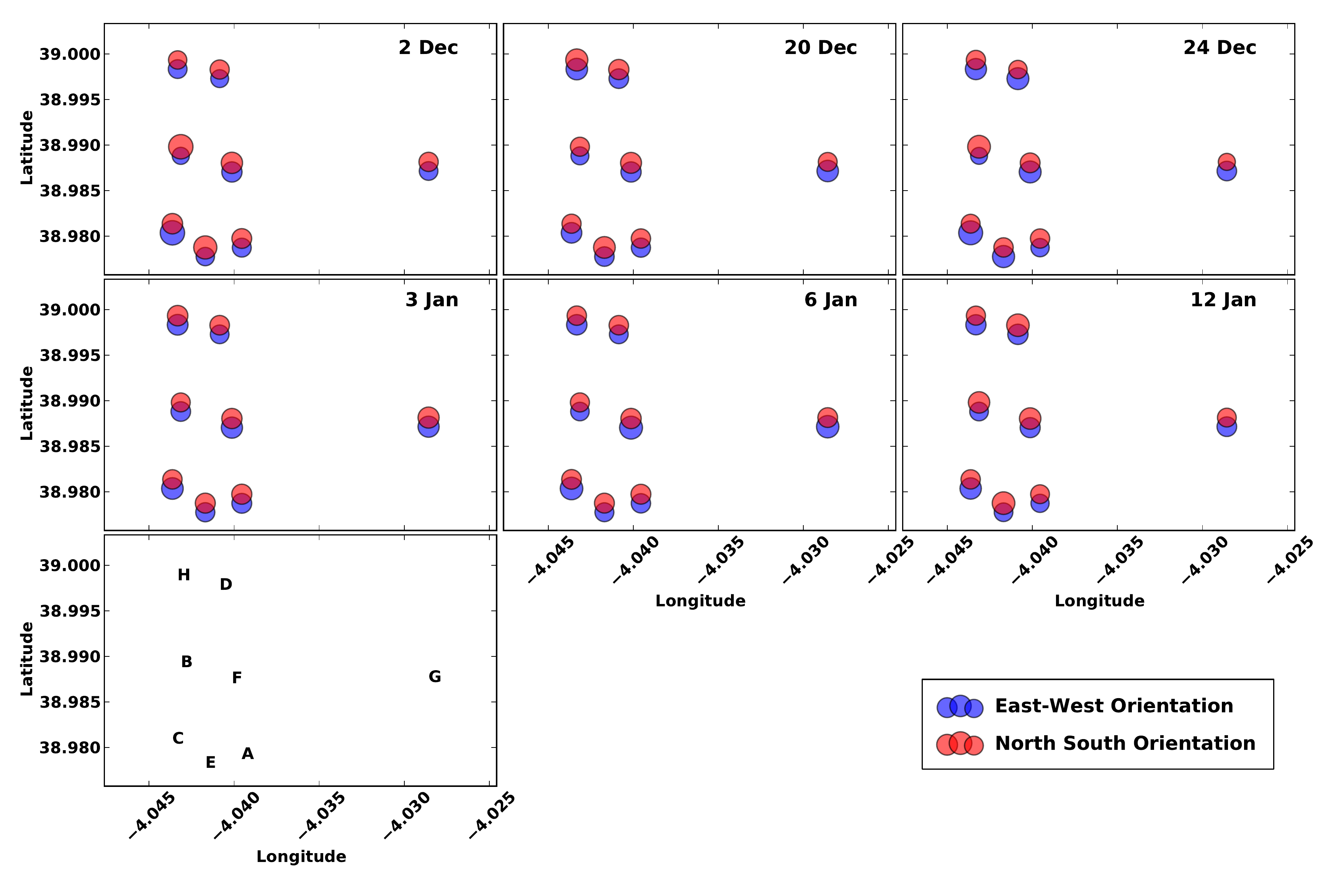}
\vspace{0.45cm}
   \caption[Precipitation accumulated in each disdrometer of the network for episodes analyzed between the average of the whole network. Spatial distribution in the network]{\textbf{Precipitation accumulated in each disdrometer of the network for episodes analyzed between the average of the entire network. Spatial distribution network}. The area of each circle is proportional to the accumulated precipitation for the corresponding episode. \emph{Red}, North-South orientation, \emph{Blue}, East-West orientation. The values that are shown are relative to the average value of the entire network. It should be noted that in the events of 20 December and 12 January, the H and D pairs exhibit different behaviors with no specific directionality in the accumulated precipitation.}
\label{fig:RaccREDVar1}
\vspace{0.30cm}
\end{figure}
\end{center}

From the point of view of validation, it is of interest to verify that no disdrometer records its values in a systematic and significant way that is different from those in any of the other episodes. We can observe this phenomenon in figure (\ref{fig:RaccREDVar1}), where circles proportional in size to the intensity of accumulated precipitation are shown for each disdrometer. The reference is given by the mean value in the network.\\

In the analysis of this figure, no disdrometer recording systematic differences in rainfall amounts (neither major nor minor) can be observed in comparison to the average values. The events on 6 January 2010 and 12 January 2010 appear to exhibit larger variations than the events on 20-21 December 2009 and 3-4 January 2010, indicating that there might be an intrinsic variation with dependence on the event. Thus, disdrometers that in one event record greater precipitation can record lower values in another event, which means that from the point of view of accumulated precipitation the network exhibits reasonable isotropy, with no apparent preferential directionality in terms of the pluviometer simulation \citep{TokayCOLLOCATEDjwd2005,tokay_kruger_etal_2001_aa}.\\

\subsection{Velocity-diameter relationships}
\label{sec:relacionVD}

It is of interest to verify that the relationships between the velocity and diameter remain stable throughout the disdrometer network, both in the sense of measurement dispersion for a given event throughout the network and in the recording of different velocity-diameter relationships for the snow event recorded on 10 January 2010.\\

\begin{center}
\begin{figure}[H] 
\vspace{0.25cm}
   \includegraphics[width=1.00\textwidth]{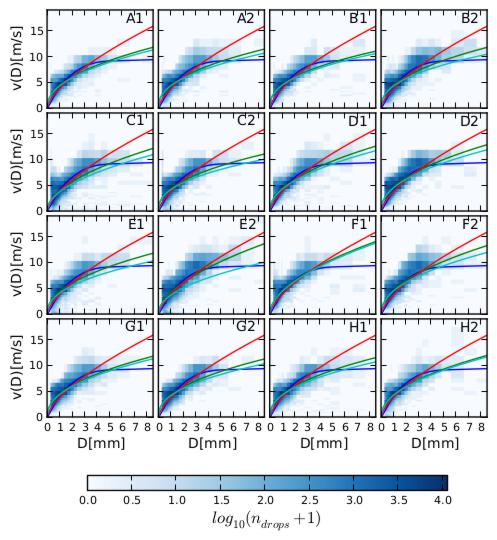}
\vspace{1cm}
   \caption[The experimental matrix $n(D,v)$ is represented in a $v(D)$ diagram. Episode on 12 January 2010]{\textbf{The experimental relationship between the velocity and diameter recorded by the 16 disdrometers in the network. Episode on 12 January 2010.} The curve in red represents $v(D)=\gamma D^{\delta}$ for Equation (\ref{eqn:AtlasVDequation}). The curve in blue represents the said law with a correction such that it saturates at diameters greater than 5 $mm$. The curves in green represent fits of the experimental data to $v(D)=\gamma v^{\delta}$.}
\label{fig:VDlluvia}
\end{figure}
\end{center}

\begin{center}
\begin{figure}[H] 
\vspace{0.55cm}
   \includegraphics[width=1.00\textwidth, angle=270]{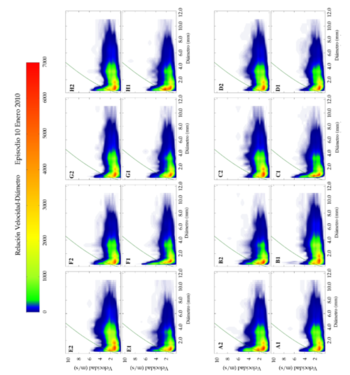}
\vspace{1cm}
   \caption[The experimental matrix $n(D,v)$ represented in a $v(D)$ diagram. Episode on 10 January 2010. Snow.]{\textbf{ The experimental relationship between the velocity and diameter recorded by the 16 disdrometers in the network. Episode on 10 January 2010.} This episode is a precipitation event in the form of snow. This figure can be compared to Figures (\ref{fig:VDcolorado}) and (\ref{figOptico2}). The theoretical model of liquid precipitation has been included according to Eq.(\ref{eqn:AtlasVDequation}) for comparison. The maximum frequency is located at approximately $v=1[m/s]$ and $1<D[mm]<2$, which is in accordance with the experiment carried out by \citep{ChinosPARSIVELestudio}. Therefore, in converting to the equivalent in liquid water, the F relationship from Table (\ref{tablaDensidadCopoNieve}) is also used, as illustrated in Figures (\ref{fig:RtimeseriesA1B1E2_Var1}) and (\ref{fig:Var1SerieRmean}).}
\label{fig:VDnieve}
\end{figure}
\end{center}

Figures (\ref{fig:VDlluvia}) and (\ref{fig:VDnieve}) present these relationships for the events of 12 January 2010 and 10 January 2010. The first is a rain event that according to Figure (\ref{fig:RaccREDVar1}), exhibits more potential variability, while the event on 10 January is a typical case of snow. If we compare these figures with the comparative study carried out in \citep{krajewski_kruger_etal_2006_aa}, we illustrate the importance of having a homogeneous network of instruments (See Figure 14 in the cited reference) because the variations in the velocity-diameter relationship would be more significant with a heterogeneous network, which would make it difficult to answer the issues raised in section \S\ref{sec:var1conceptos}.\\

\vspace{1.05cm}

\begin{table}[h]
\ra{1.05}

\caption[The parameters of the power-law relationship $v(D)=\gamma D^{\delta}$ for each episode of liquid precipitation.]{\textbf{Parameters of the power-law relationship $v(D)=\gamma D^{\delta}$ for each episode of liquid precipitation}. A comparison of the $v(D)=\gamma D^{\delta}$ estimate via linear regression over the disdrometer network for the different events analyzed, obtaining a value for the linear correlation coefficient $\rho$. The minimum correlation coefficient ($\rho$) possesses the minimum value in parentheses, excluding the absolute minimum. The average correlation coefficient is the arithmetic mean of the correlation coefficient obtained over 16 disdrometers via data from each one. This table was built without preprocessing of the disdrometric data. In chapter \S\ref{chap:preprocesadoTOKAY}, the standard preprocessing values are shown. It should be noted that the main variation resides in the correlations and not in the values of the coefficients because a limited number of drops with 
anomalous terminal velocities significantly decreases the linear correlation coefficient.}

\vspace{0.01cm}
\ra{1.05}
\begin{center}

\begin{tabular}{lccccc}
\toprule
\textbf{Event}       & $\mathbf{\bar{\delta}\pm\Delta \delta}$ & $\mathbf{\bar{\gamma}\pm\Delta \gamma}$ & $\mathbf{\rho}$ \textbf{max.}        &  $\mathbf{\rho}$ \textbf{min.} & $\mathbf{\rho}$ \textbf{avg.} \\
\midrule
02-dic-2009  & $0.37\pm0.07$       & $4.00\pm0.23$         & 0.68                     & 0.10 (0.26)       &0.50\\
20-dic-2009  & $0.44\pm0.05$       & $4.17\pm0.18$         & 0.82                     & 0.23 (0.46)       &0.62\\
24-dic-2009  & $0.39\pm0.05$       & $4.17\pm0.20$         & 0.72                     & 0.28 (0.29)       &0.52\\
03-ene-2010  & $0.47\pm0.06$       & $4.28\pm0.11$         & 0.91                     & 0.38 (0.61)       &0.87\\
06-ene-2010  & $0.44\pm0.05$       & $4.16\pm0.15$         & 0.77                     & 0.52 (0.62)       &0.69\\
12-ene-2010  & $0.45\pm0.04$       & $4.20\pm0.22$         & 0.80                     & 0.38 (0.42)       &0.66\\
\bottomrule
\end{tabular}
\label{TableRelacionesVD}
\end{center}
\vspace{0.01cm}
\end{table}

Table (\ref{TableRelacionesVD}) shows the arithmetic mean of the coefficients of the $v(D)=\gamma D^{\delta}$ relationship obtained via linear regression for each of the analyzed events \footnote{It should be noted that the $\gamma$ values are similar to those shown in chapter \S\ref{chap:preprocesadoTOKAY}, which were calculated utilizing a filter in the experimental data preprocessing step.}. Qualitatively, the results coincide with those found by \citep{ChinosPARSIVELestudio}, in which the velocities tend to be somewhat greater than in the relationship given by Equation (\ref{eqn:AtlasVDequation}) for drop diameters of up to 2 $mm$. In this same study, two possible fundamental factors are indicated:

\begin{itemize}
 \item The experimental data concerning where the relationship (\ref{eqn:AtlasVDequation}) is based were obtained at an atmospheric pressure of 1\,013 hPa, a temperature of 20\degree C and a relative humidity of 50\%. Lesser values of air density imply slightly greater terminal velocities for the same diameter, which explains the general trend that also was found in this study for drops with diameters less than 2 mm (i.e., drops that are affected by thermodynamic parameters).
 \item Another factor unites more stochastic constraints, such as the local turbulence, organized air movements or asystematic errors in measurement, and in principle is capable of explaining the differences between the values obtained by the instruments and the reference given by (\ref{eqn:AtlasVDequation}).
\end{itemize}

The fact that higher terminal velocities were generally found throughout the entire network indicates that velocity components due to other factors can be relevant locally and in particular episodes. Generally, the deviation appears to be explained by the difference in the physical magnitudes that control air density\footnote{The following correction is often used: \begin{equation}v(D)=\left [9.65-10-3exp(-0.6D)\right]\left (\frac{\rho}{\rho_{0}}\right )^{-0.4} \end{equation}See, for example, \citep{MicrofisicaNOAAprofiler}.}, especially when taking into account that the location of the experiment is a plateau of 600 meters in average altitude.\\

\begin{figure}[h] 
\begin{center}
\vspace{0.75cm}
   \includegraphics[width=1.00\textwidth]{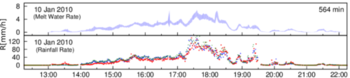}\\
\vspace{0.90cm}
   \includegraphics[width=1.00\textwidth]{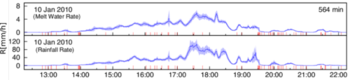}
\vspace{0.75cm}
   \caption[The time series of the average precipitation intensity throughout the network for the episodes analyzed. Episode on 10 January 2010. Snow.]{\textbf{The time series of the average precipitation intensity throughout the network for the episodes analyzed. Episode on 10 January 2010. Snow.} The standard deviation throughout the network also is represented per minute. In the case of snow, it is important to equivalently represent the data as liquid precipitation by transforming the time series of the direct experimental values. For the conversion to an equivalent in liquid water, one starts from the F relation in Table (\ref{tablaDensidadCopoNieve}). The difference among the time series of both rain intensity and liquid water content are not only reflected in the general values but also in the positions of the relative maxima and minima, as in the interval between 18:00 and 18:30. The different relationships in Table (\ref{tablaDensidadCopoNieve}) do not significantly affect the results.}
\label{fig:Var1SerieRmean}
\end{center}
\vspace{0.85cm}
\end{figure}

\subsection{Rainfall-intensity time series}

The precipitation-intensity time series were compared for the various episodes and different disdrometers. Figure (\ref{fig:RtimeseriesA1B1E2_Var1}) shows these series for three different disdrometers, A1, B1 and E2 (see the map). The disdrometers were chosen to represent different estimates of the total accumulated rainfall, both for episodes and for percentage bias, as shown in the previous chapter (Table \ref{tablaBIASwhole}). The following should be noted:

\begin{itemize}
\item The differences among the values in the time series cannot be directly associated with the distances between disdrometers (in the sense that $R_{A1}(t)\simeq R_{E2}(t+\delta t)$ for a certain $\delta t$ associated with the distance between stations A1 and E2) such that in the following, the time series in UTC same time values are directly compared .
\item The differences increase with increasing rainfall intensity, suggesting the presence of multiplicative differences, as typically found in other experimental studies \citep{krajewski_kruger_etal_2006_aa}. 

\begin{center}
\begin{figure}[H] 
\vspace{1.00cm}
   \includegraphics[width=1.05\textwidth]{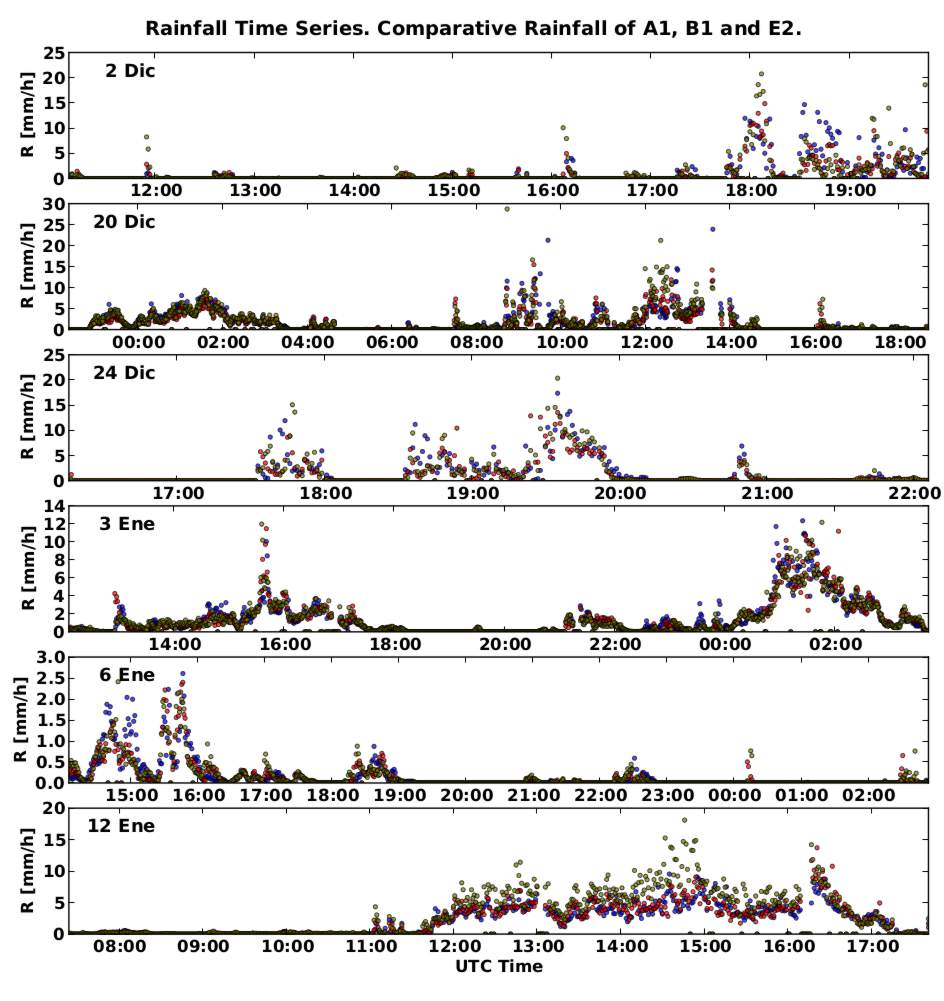}
\vspace{0.75cm}
   \caption[The precipitation-intensity time series for disdrometers A1, B1 and E2 and for the six episodes analyzed]{\textbf{ The precipitation-intensity time series for disdrometers A1, B1 and E2 and for the six episodes analyzed }. The rainfall-intensity time series are compared for all episodes. \textit{Blue}: disdrometerA1, \textit{Red}: disdrometer: B1, \textit{Green}: disdrometerE2. It should be noted that the episodes on 20 December and 3 January extend into the following day.} 
\label{fig:RtimeseriesA1B1E2_Var1}
\end{figure}
\end{center}

\begin{figure}[H] 
\begin{center}
\vspace{1.00cm}
   \includegraphics[width=1.05\textwidth]{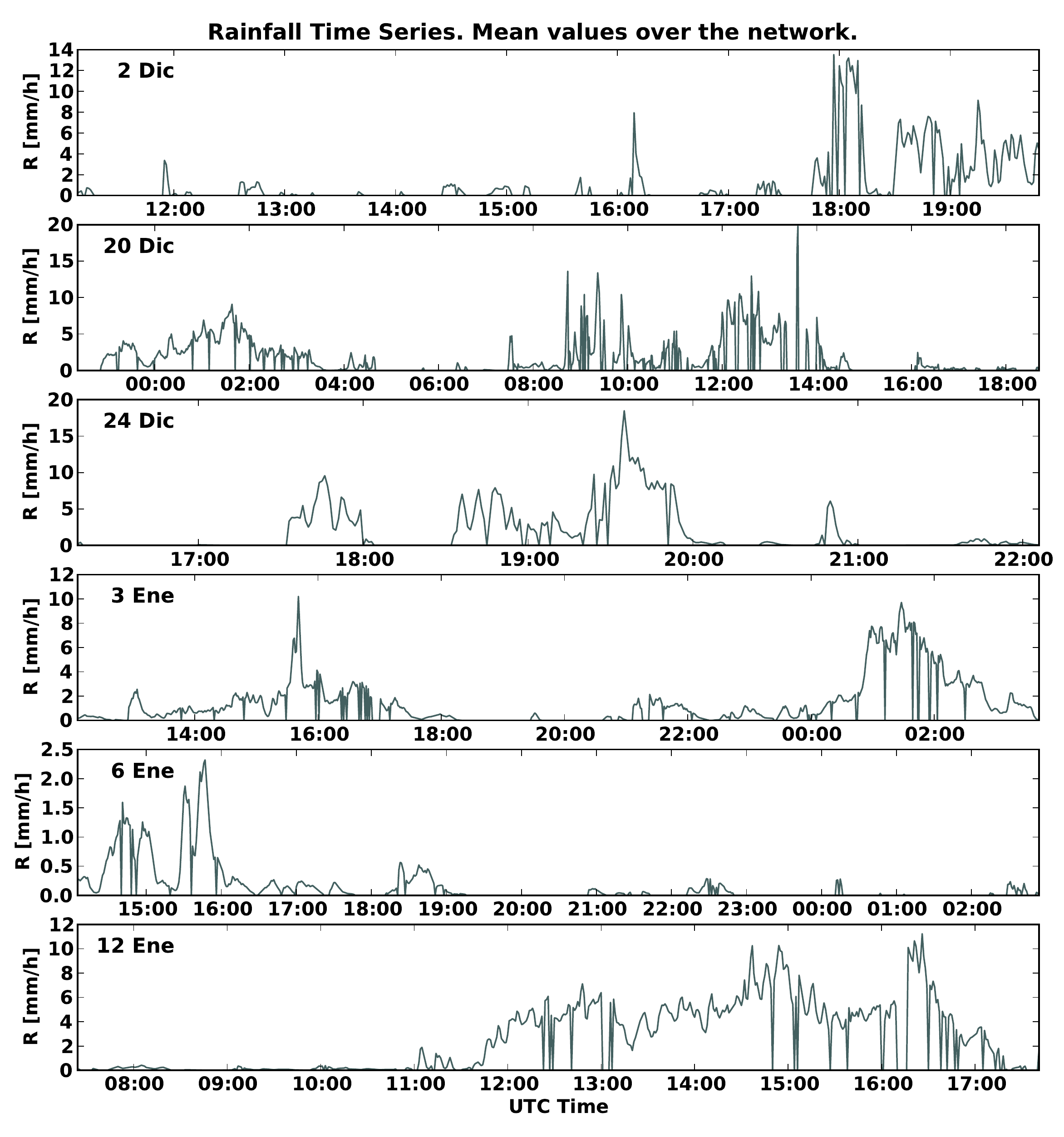}
\vspace{0.75cm}
   \caption[The time series of the average precipitation intensity throughout the network for the episodes analyzed.]{\textbf{The time series of the average precipitation intensity throughout the network for the episodes analyzed}. The apparent spikes in precipitation intensity are due to minutes that were deleted, lost or filtered after applying the consistency process over the entire set of disdrometers. If the curves of this figure are compared with the values shown in Figure (\ref{fig:RtimeseriesA1B1E2_Var1}), a reasonable correlation is observed between the average values and the values given by the subset of instruments A1, B1 and E2.}
\label{fig:Var1SerieRmean}
\end{center}
\vspace{1.2cm}
\end{figure}

\item In the 12 January event, we can observe the significant differences among E2 and either A1 or B1. This event exhibits lower values in the correlation of time series. Generally, this finding could be due to operating errors (a loss of key minutes in the transmission) or peculiar characteristics of the specific event (intense wind located at one station, factors external to the measurement). In this particular case, the second factor is the likely contributor. It should be noted that one of the advantages of using a dense network of disdrometers is the ability to record these differences both spatially and temporally.
\end{itemize}

To quantify these differences in a practical way, we calculated the average value of R in addition to the standard deviation throughout the network. This is equivalent to assuming that we can estimate the average R value ($\overline{R}$) for the area covered by the network using the simple arithmetic average without incorporating the spatial-distribution information of the measurements; similarly, a similar process is applied to calculate the standard deviation ($\sigma_{R}$). If the random field $R(\vec{x})$ were homogeneous, this approximation would be valid insofar as measurements of the different disdrometers represent measurements of a DSD with the same value of R throughout the whole network. Thus, the results in Figure (\ref{fig:Var1SerieRmean}) represent the values of $\overline{R}\pm\sigma_{R}$ under the hypothesis of spatial homogeneity, and the results can give us an approximate idea of the errors made in estimating the precipitation using a single disdrometer that would represent the entire 
disdrometer network \footnote{As discussed in the next section, a rigorous estimate of the error would require calculating the spatial correlations and their use in determining the \textit{sampling uncertainty}.}.

\section{Spatial correlation of precipitation}

Most previous studies on the spatial variability of rainfall were carried out with data from pluviometers. This entails an analysis of the random field of the precipitation intensity only.\\

These studies used the time series of data obtained at each station. To determine their variation throughout a set series deployed at the location of the study, concrete statistical parameters are analyzed. The most widely used parameter is the correlation between time series, from which data regarding the spatial and temporal dependencies is of interest (the latter obtained from the time resolution of the experimental data series)\footnote{The process of validation over a field of remote radar sensors is based on separating two errors from different sources. One error is derived from the radar sensor itself and from errors in determining the actual reflectivity. The other error originates in the sampling error, which arises when identifying the precipitation intensity throughout the sampling volume by a single value. More concretely, the \textit{sampling error} that is made upon identifying the precipitation value at a point with precipitation in an area is given by $\epsilon=\hat{R}-R_{A}$, while the 
standard deviation of $\epsilon$ is called the \textit{sampling uncertainty} is denoted by $\sigma_{\epsilon}$ and defined as
\begin{equation}
 \sigma_{\epsilon}=\sigma_{R}\phi\left[\rho(\vec{r}),R\right]
\end{equation} 
where $\sigma_{R}$ is the variance of a point[Ed.3]-measurement instrument and $\phi$ is the \textit{variance reduction factor} that functionally depends on the correlogram, i.e., on the correlation based on the position defined as $\rho(r)$, for the precipitation field under study, normally the rainfall intensity.}.\\

Therefore, the main estimator used for the spatial correlations between the two series is the Pearson correlation coefficient. Given two time series $X_{a}$ and $X_{b}$ of magnitude X, obtained from two instruments located at positions $\vec{y}_{a}$ and $\vec{y}_{b}$, this coefficient is expressed as (see for example \citep{Brommundt2007spatialcorre})

\begin{equation}
\rho_{X}(\vec{y}_{a},\vec{y}_{b})=\frac{Cov(X_{a},X_{b})}{\sqrt{Var(X_{a})Var(X_{b}})}
\label{eqn:PearsonTradicional}
\end{equation} 
where the covariance between $X_{a}$ and $X_{b}$ is defined by:
\begin{equation}
Cov(X_{a},X_{b})=E(X_{a}X_{b})-E(X_{a})E(X_{b})
\end{equation}

and the denominator\footnote{In other studies, see \citep{2009ChoongLEE}, defining a correlation that does not subtract the average values has been proposed, but its use is restricted to the comparison of radars at larger scales and does not constitute an extended methodology.} is
\begin{equation}
Var(X_{a})Var(X_{b})=\left[E(X_{a}^{2})-E(X_{a})^{2}\right]\left[E(X_{b}^{2})-E(X_{b})^{2}\right]
\end{equation} 
where $E(X)$ is the expected value of the random variable X. Often, $X(y_{a}, y_ {b})$ is conceived as a random field in which most studies suggest a variation with distance but not with direction (an isotropic but not homogeneous field), which is reasonable if the study area does not have a complex topography.\\

In addition to the hypothesis of isotropy, one of the possible problems that can exist is the non-normality of the processed data, which may imply a biased estimate. Several authors recommend using a logarithmic transformation of the time series to reduce the skewness of the integral parameters of the DSD, although this transformation can only be performed for the data set with precipitation values greater than zero in both time series \citep{2001HabibKrajewski}.\\

A methodology based on determining this same Pearson coefficient has been proposed but under a \textit{bivariate mixed lognormal distribution} \citep{1993Shimizu,2004Gebremichael,2001HabibKrajewski}, which we denote as BMLN and which permits analysis in joint fashion of cases in which precipitation is not recorded and the hypothesis of log-normality exists in the time series. Other studies appear to indicate that if the tail in the precipitation-intensity distribution is less than a lognormal, it is preferable to utilize the traditional Pearson coefficient. This finding makes the methodology dependent on the empirical basis that is being analyzed.\\

In our case, we verified the results by utilizing all previous methodologies. That is, we compared the correlation between the precipitation time series assuming normality in the series, including both cases in which no precipitation is recorded at any station as well as the case in which we exclude this fact. In this last case, we carried out the logarithmic transformation in both time series before applying the definition of the traditional Pearson coefficient. Finally, for the complete data series (including minutes with and without precipitation), we applied the BMLN method, which performs a differentiation of four different cases of the initial, bidimensional, random variable $(X_{a},X_{b})$ and later proceeds to the logarithmic transformation of the data once the cases have been differentiated. These four cases are described in Table (\ref{TablaSHIMIZU}).\\

With the quantities defined in Table (\ref{TablaSHIMIZU}), the calculations used for the magnitudes that appear in the Pearson-coefficient equation for the bivariate mixed lognormal distributions are given as
\begin{equation}
E(X_{a})=\delta_{1}e^{\left[\mu_{1}^{*}+(\sigma_{1}^{*2}/2)\right]}+\delta_{3}e^{\left[\mu_{1}+(\sigma_{1}^{2}/2)\right]}
\end{equation}
\begin{equation}
E(X_{b})=\delta_{2}e^{\left[\mu_{2}^{*}+(\sigma_{2}^{*2}/2)\right]}+\delta_{3}e^{\left[\mu_{2}+(\sigma_{2}^{2}/2)\right]}
\end{equation}
\begin{equation}
E(X_{a},X_{b})=\delta_{3}e^{\left[\mu_{1}+\mu_{2}+(\sigma_{1}^{2}+\sigma_{2}^{2}+2\sigma_{1}\sigma_{2}\rho_{N})/2\right]}
\end{equation}
\begin{equation}
Var(X_{a})=\delta_{1}e^{[2\mu_{1}^{*}+2\sigma_{1}^{*2}]}+\delta_{3}e^{[2\mu_{1}+2\sigma_{1}^{2}]}-E(X_{a})^{2}
\end{equation}
\begin{equation}
Var(X_{b})=\delta_{2}e^{[2\mu_{2}^{*}+2\sigma_{2}^{*2}]}+\delta_{3}e^{[2\mu_{2}+2\sigma_{2}^{2}]}-E(X_{b})^{2}
\end{equation}

 \begin{figure}[H] 
 \begin{center}
 \vspace{0.05cm}
    \includegraphics[width=0.90\textwidth]{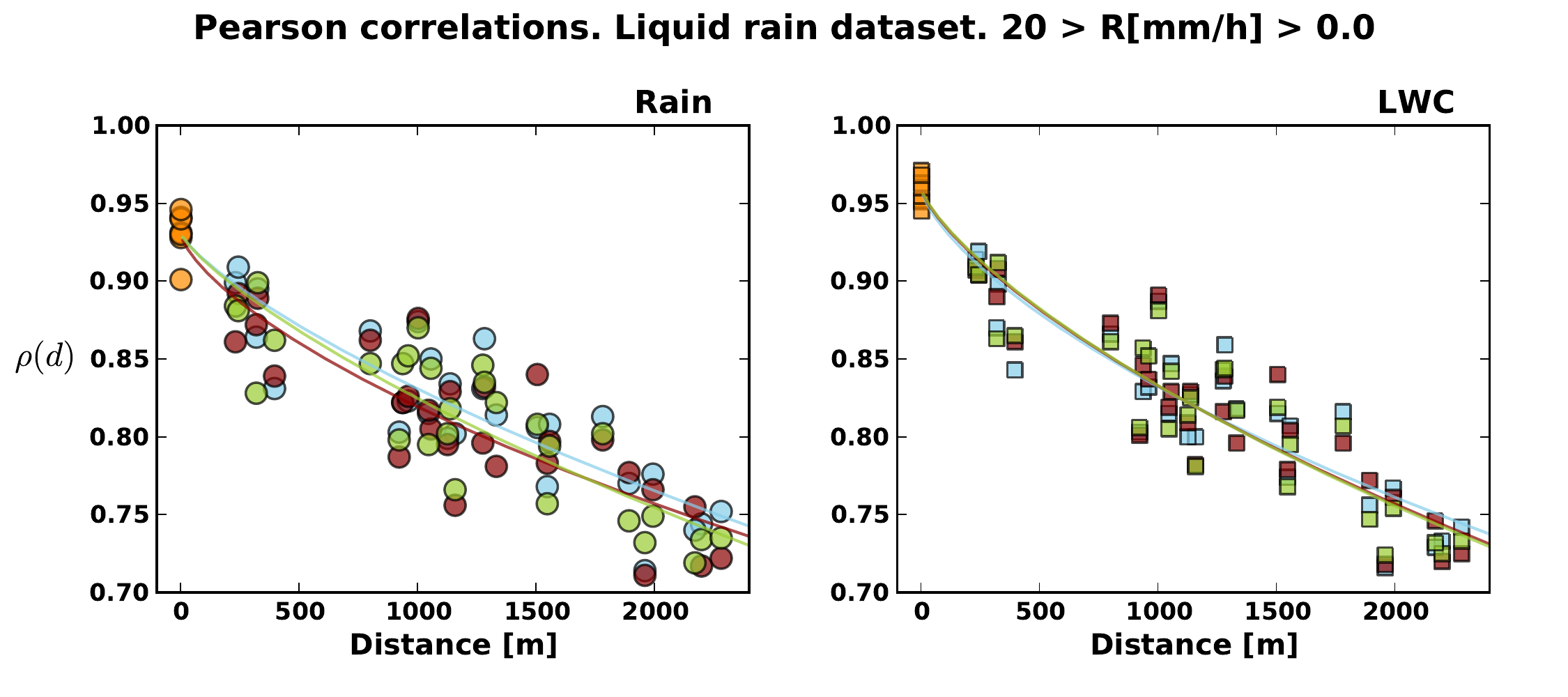}
    \includegraphics[width=0.90\textwidth]{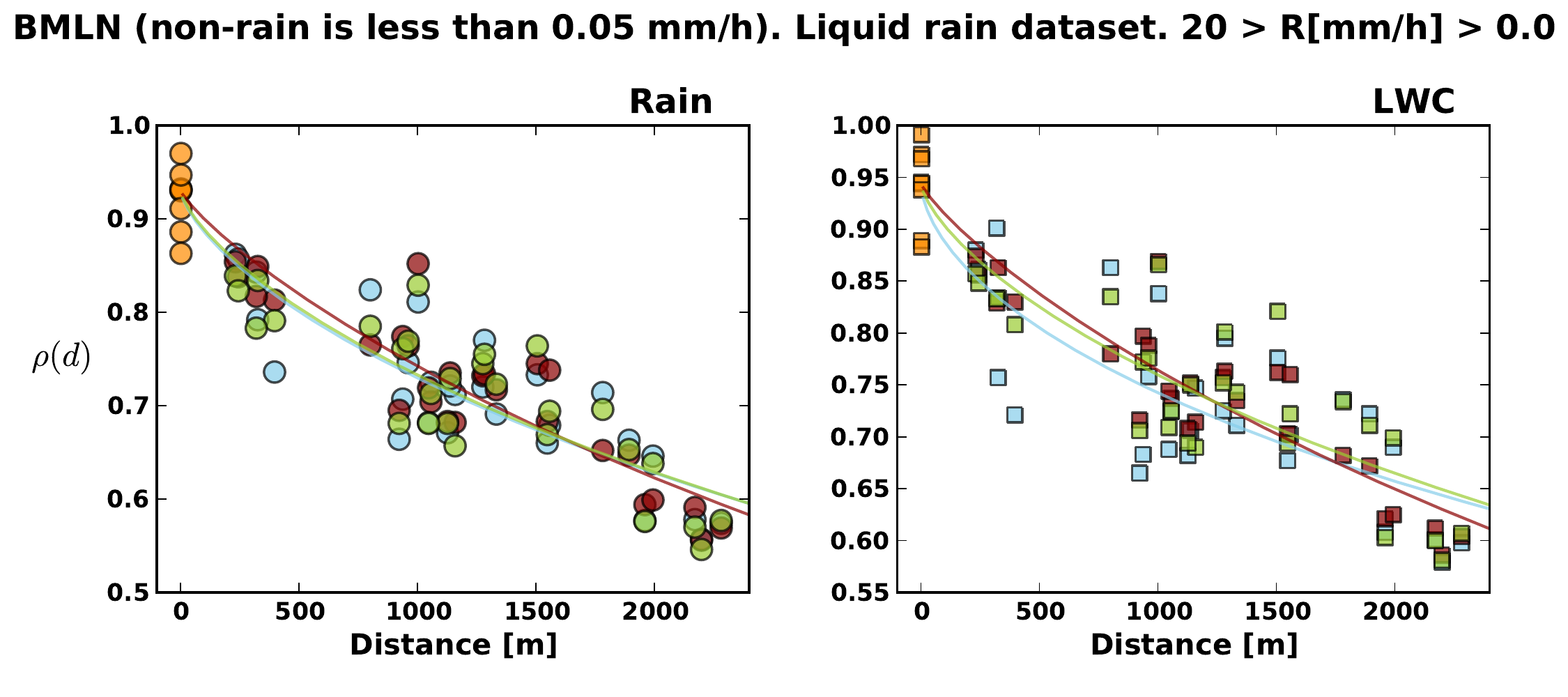}
    \includegraphics[width=0.90\textwidth]{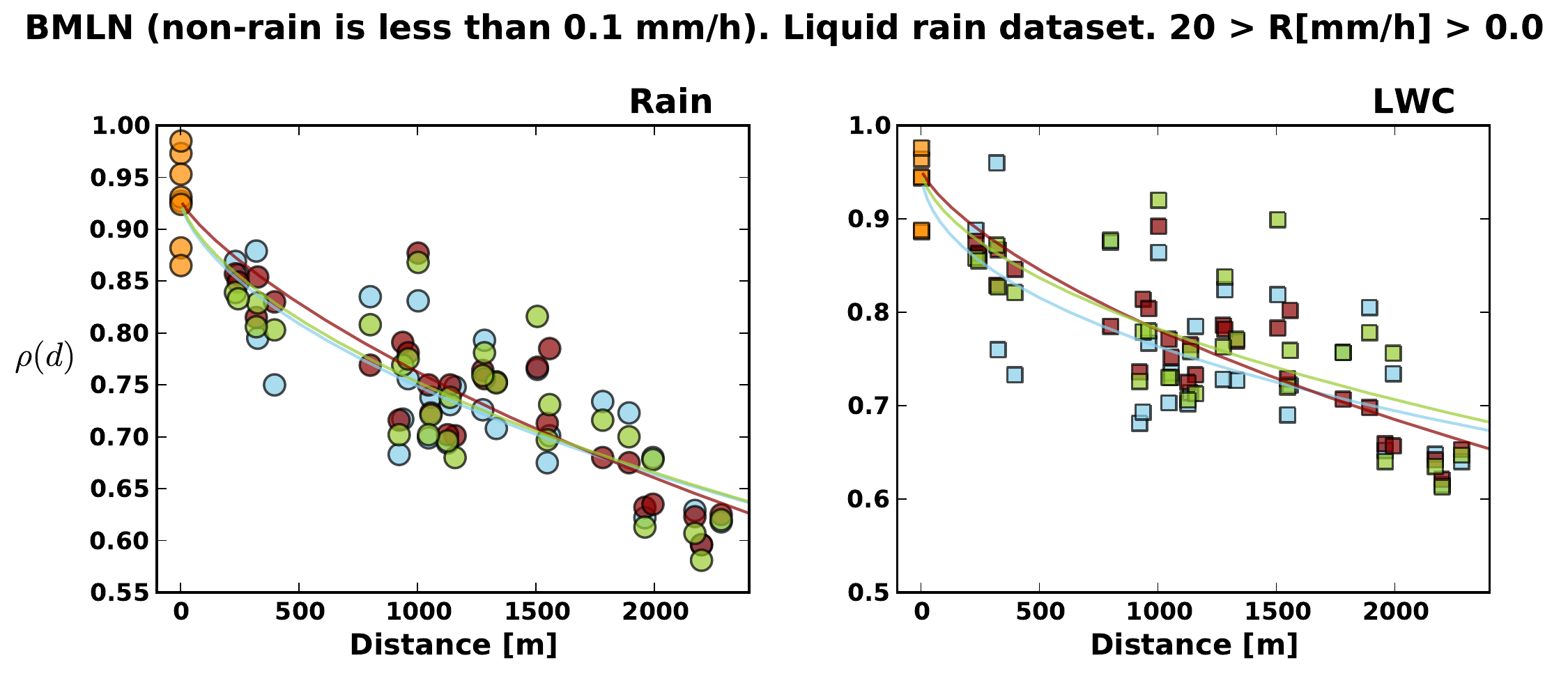}
    \includegraphics[width=0.60\textwidth]{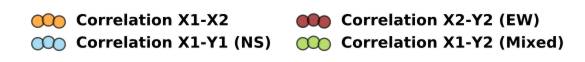}
 \vspace{0.05cm}
    \caption[A comparison of the estimated correlograms for the entire empirical basis of liquid precipitation. A direct method based on the traditional Pearson coefficient, and a method based on the bivariate mixed log-normal distribution.]{\textbf{ A comparison of the correlograms for the whole empirical basis of liquid precipitation. A direct method based on the traditional Pearson coefficient, and a method based on the bivariate mixed log-normal distribution. The comparison is made for the R and LWC time series. A comparison for the entire empirical basis of liquid precipitation.} \textit{Blue} is the correlation between disdrometers from different stations but with the same North-South orientation, and \textit{red}, for East-West. The correlations between stations where disdrometers are taken with different orientations appear in \textit{green}. The correlations for dual instruments (connected) appear in \textit{orange}. The lines represent \emph{trends} based on a model that is introduced in \S\ref{sec:newVariability}. In all cases of greater spatial variability, the linear models do not connect with the values of the attached stations; for this reason, the decay model was proposed, which is discussed in Section \S\ref{sec:newVariability}.}
 \vspace{0.05cm}
 \label{fig:Var1metodosCORRELACION_BMLN}
 \end{center}
 \end{figure}

 \begin{figure}[H] 
 \begin{center}
 \vspace{0.35cm}
    \includegraphics[width=0.95\textwidth]{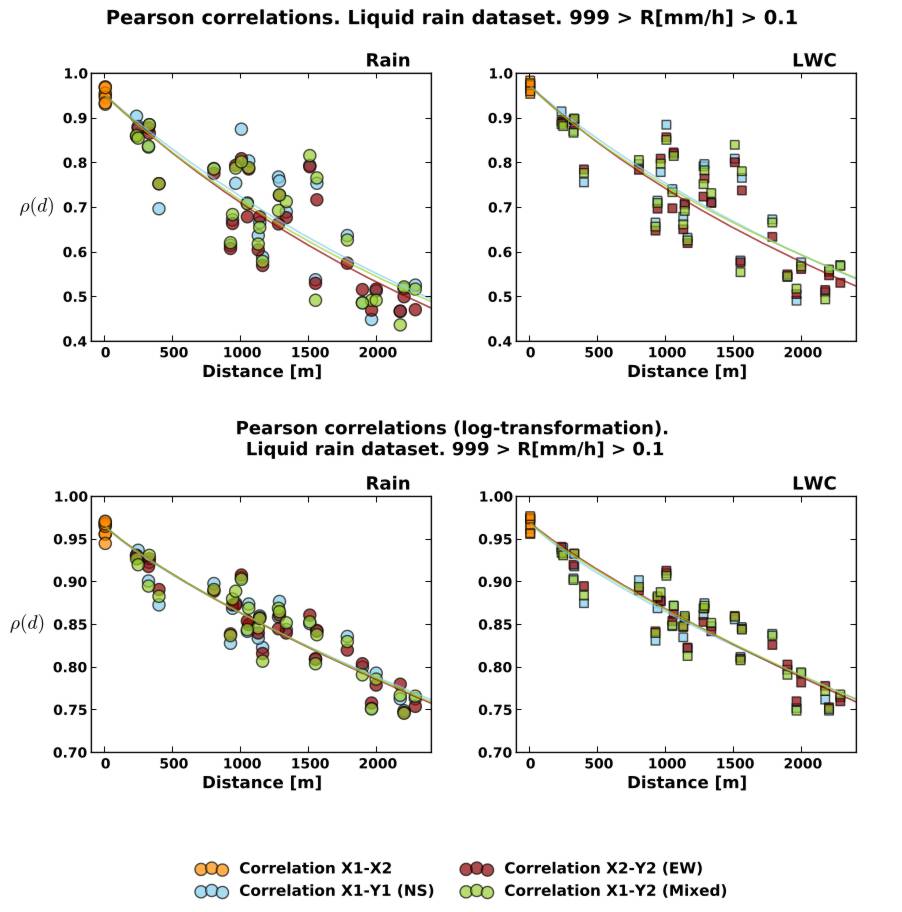}
 \vspace{0.65cm}
    \caption[A comparison of the correlograms for the entire empirical basis of liquid precipitation. A direct method based on the traditional Pearson coefficient and under a logarithmic transformation.]{\textbf{ A comparison of the correlograms for the entire empirical basis of liquid precipitation. A direct method based on the traditional Pearson coefficient and with the performance of a prior logarithmic transformation. The comparison is made for the R and LWC time series. A comparison for the entire empirical basis of liquid precipitation.} \textit{Blue} is the correlation between disdrometers from different stations but with the same North-South orientation, and \textit{red}, for East-West. Correlations between stations where disdrometers with different orientations are taken appear in \textit{green}. The correlations for dual instruments (connected) appear in \textit{orange}. The lines represent \emph{trends} based on a model that is introduced in \S\ref{sec:newVariability}. In all cases of greater 
spatial variability, the linear models do not connect with the values of the attached stations; for this reason, the decay model was proposed, which is discussed in Section \S\ref{sec:newVariability}.}
 \vspace{1.45cm}
 \label{fig:Var1metodosCORRELACION_LOGN}
 \end{center}
 \end{figure}

\begin{table}[H]
\caption[The statistical parameters utilized in the bivariate mixed log-normal methods (BMLN)]{\textbf{ The statistical parameters utilized in the bivariate mixed lognormal methods (BMLN).} The statistical parameters for the calculation of correlations between time series for a bivariate mixed lognormal distribution (BMLN). This process entails the separation of the original data $(X_{a},X_{b})$ in 4 different cases, permitting the separation of situations in which one of the two instruments (or both) does not record precipitation. The probabilities $\delta$ are conditional probabilities, as are the averages and standard deviations. The method also makes use of the Pearson correlation for the case $(x_{a},x_{b})$ given by Equation (\ref{eqn:PearsonTradicional}), which we call $\rho_{N}$. Along this table \textit{Cond.} means conditional.}
\vspace{0.0cm}
\begin{center}
\ra{1.00}
\begin{tabular}{lccrr}
\toprule
\small \textbf{Case}            & \small \textbf{Sample size} & \small \textbf{Cond. Prob.}  & \small \textbf{Cond. Avg.} & \small \textbf{Cond. Standard Deviation.}  \\

\midrule
$(0,0)$          & $n_{0}$        &  $\delta_{0}$    &  -  & -\\
$(x_{a},0)$      & $n_{1}$        &  $\delta_{1}$    & $\mu_{a}^{*}$ & $\sigma_{a}^{*}$ \\
$(0,x_{b})$      & $n_{2}$        &  $\delta_{2}$    & $\mu_{b}^{*}$  & $\sigma_{b}^{*}$\\
$(x_{a},x_{b})$  & $n_{3}$        &  $\delta_{3}$    & $\mu_{a}$ , $\mu_{b}$ & $\sigma_{a}$ , $\sigma_{b}$\\
Todos            & N              &  - & - & -\\
\bottomrule
\end{tabular}
\end{center}
\label{TablaSHIMIZU}
\end{table}

The results are shown in Figures (\ref{fig:Var1metodosCORRELACION_BMLN}) and (\ref{fig:Var1metodosCORRELACION_LOGN}) for the entire empirical basis with liquid precipitation. We observe how
\begin{itemize}
\item All methods show a decrease in the correlation of both the rain intensity and the liquid water content with increasing distance.
\item In all cases, there also appears to be an isotropy in the distance dependence; however, for the distance of 1\,000 to 1\,500 meters, a greater variation appears, indicating the possible relevance of topography. This effect is greater in the case of the BMLN method and lower in the case of an applied logarithmic transformation of the data, but all methods imply a significant and clear decrease in the correlation with increasing distance.
\item Whether minutes in which one of the disdrometers records rainfall and the other does not should be included because it appears not to change the value of the correlations significantly\footnote{Generally, the number of minutes of the form $(non-rain, non-rain )$ or $(non-rain, rain)$ is less than 8\%, with little relevance to the methods shown. In Figure (\ref{fig:Var1metodosCORRELACION_BMLN}), by \textit{non-rain} we denote two situations $R<0.05 mm/h$ and $R<0.1 mm/h$. These values were selected for their usefulness with respect to its relevance to the actual sensitivities of ground radar systems.}.
\item The BMLN method has a lower correlation value and varies more significantly with distance, while the lognormal transformation of the time series gives several of the highest correlation values calculated; this point has been corroborated by subsequent studies\citep{Tokay2010smallscaleDSD}.
\end{itemize} 

Generally, from the point of view of verification of the hypothesis of kilometer-scale spatial variability for integral parameters of precipitation, the methodology adopted to estimate the correlations is non-determinant. In the remainder of the chapter, the traditional method of Pearson is adopted, which is also the most widely used method in works by other authors.\\

\clearpage

\subsection{Analysis by episodes}

\begin{figure}[H]
 \begin{center}
 \vspace{0.15cm}
    \includegraphics[width=0.95\textwidth]{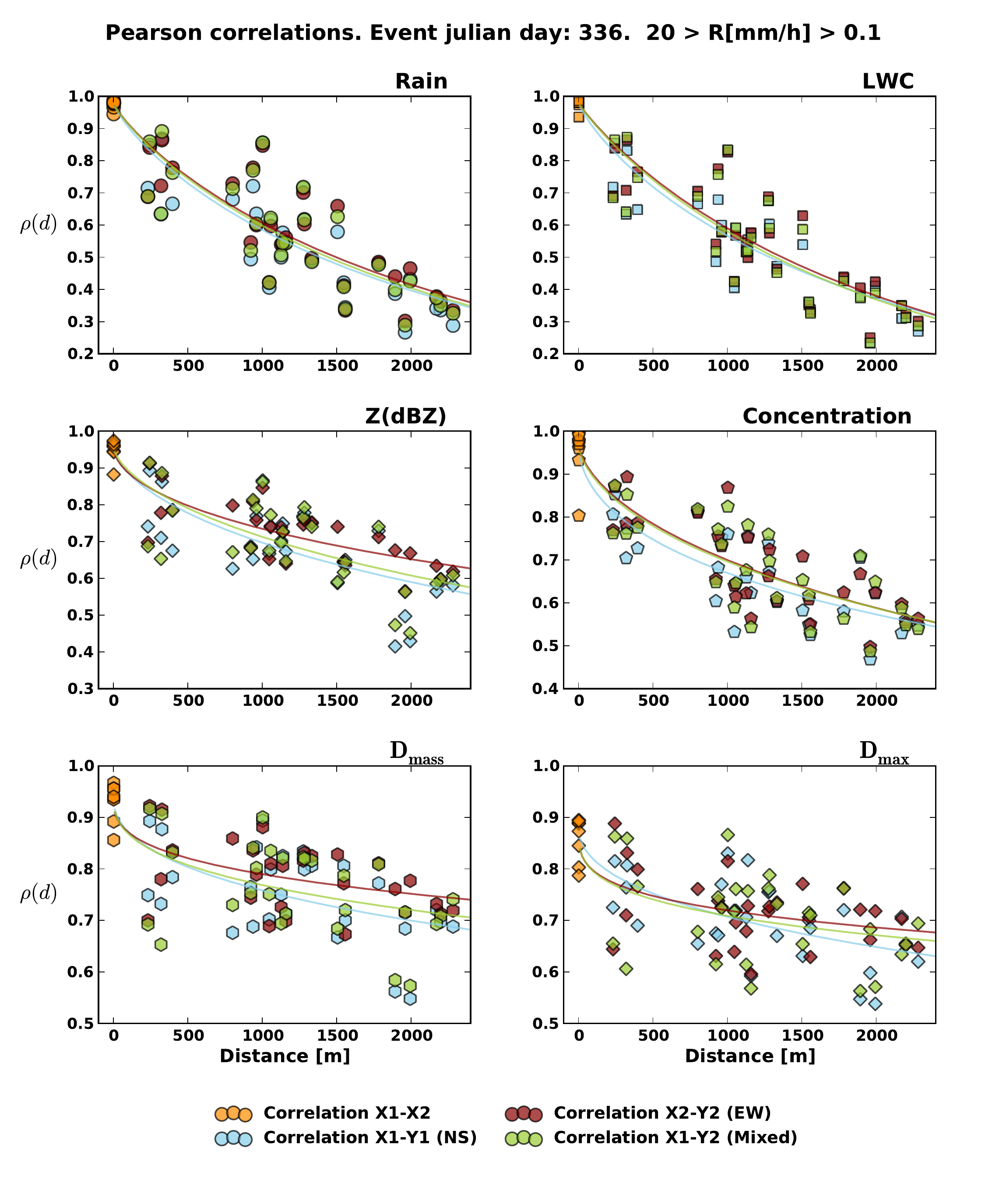}
 \vspace{0.05cm}
    \caption[The correlogram based on the Pearson coefficient for various integral parameters. Event from 2 December 2009.]{\textbf{The correlogram based on the Pearson coefficient for various integral parameters. Event from 2 December 2009.} $\rho(d)$ was determined via Equation (\ref{eqn:PearsonTradicional}). \textit{Blue} is the correlation between disdrometers from different stations but with the same North-South orientation, and \textit{red}, for East-West. The correlations between stations where disdrometers with different orientation are taken appear in \textit{green}. The correlations for dual instruments (connected) appear in \textit{orange}. The lines represent \emph{trends} based on a model that is introduced in \S\ref{sec:newVariability}. In the cases of greatest spatial variability, the linear models do not connect with the values of the attached stations; for this reason, the decay model was proposed, which is discussed in Section \S\ref{sec:newVariability}.}
 \vspace{0.05cm}
 \label{fig:Var1CorreEpisodio2DIC}
 \end{center}
 \end{figure}

\begin{figure}[H]
 \begin{center}
 \vspace{0.15cm}
    \includegraphics[width=0.95\textwidth]{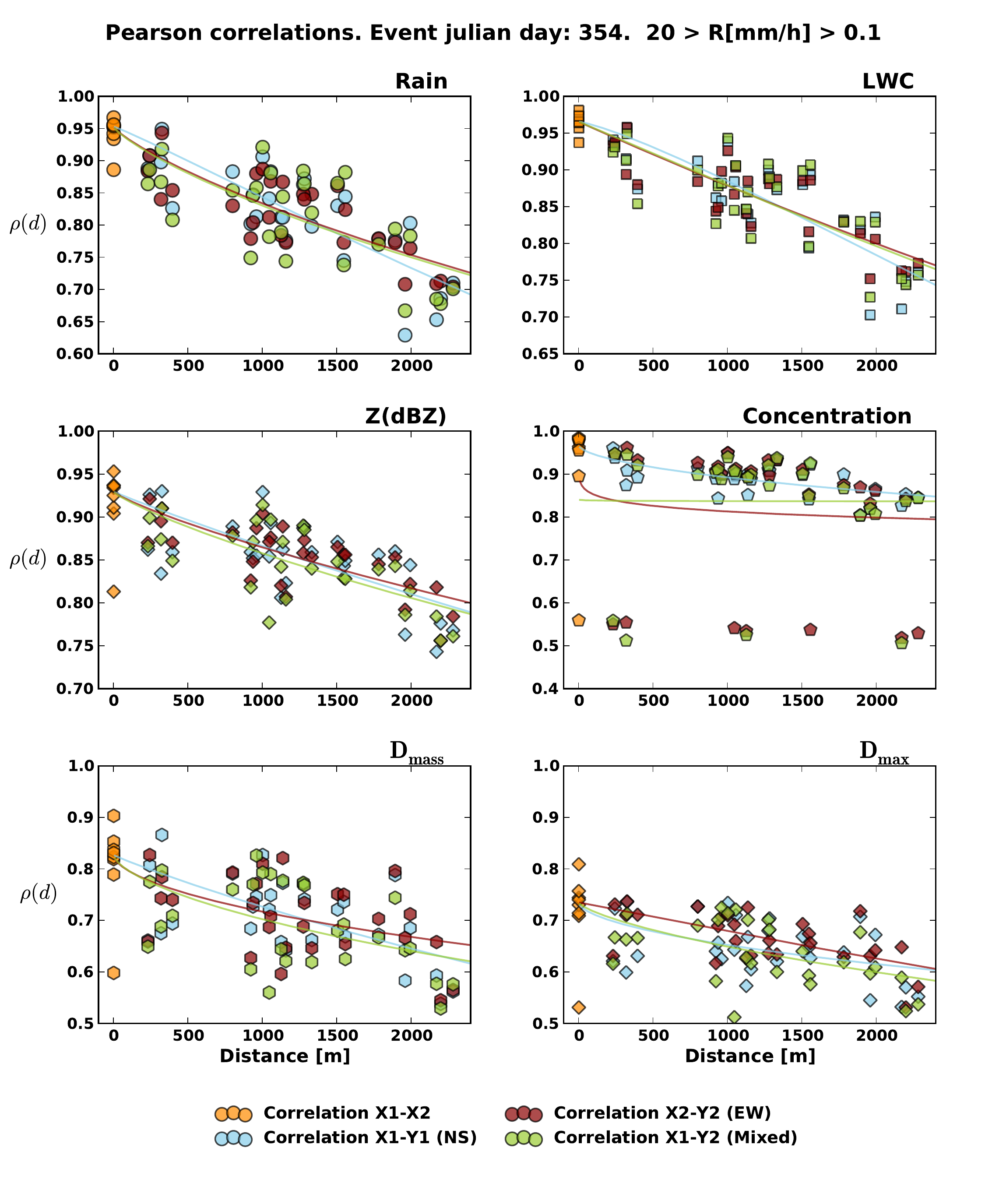}
 \vspace{0.05cm}
    \caption[The correlogram based on the Pearson coefficient for various integral parameters. Event from 20 December 2009.]{\textbf{ The correlogram based on the Pearson coefficient for various integral parameters. Event from 20 December 2009.} \textit{Blue} is the correlation between disdrometers from different stations but with the same North-South orientation, and \textit{red}, for East-West. The correlations between stations where disdrometers with different orientation are taken appear in \textit{green}. The correlations for dual instruments (connected) appear in \textit{orange}. The lines represent \emph{trends} based on a model that is introduced in \S\ref{sec:newVariability}. In the cases of greatest spatial variability, the linear models do not connect with the values of the attached stations; for this reason, the decay model was proposed, which is discussed in Section \S\ref{sec:newVariability}.}

  \vspace{0.05cm}
 \label{fig:Var1CorreEpisodio20DIC}
 \end{center}
 \end{figure}

\begin{figure}[H]
 \begin{center}
 \vspace{0.15cm}
    \includegraphics[width=0.95\textwidth]{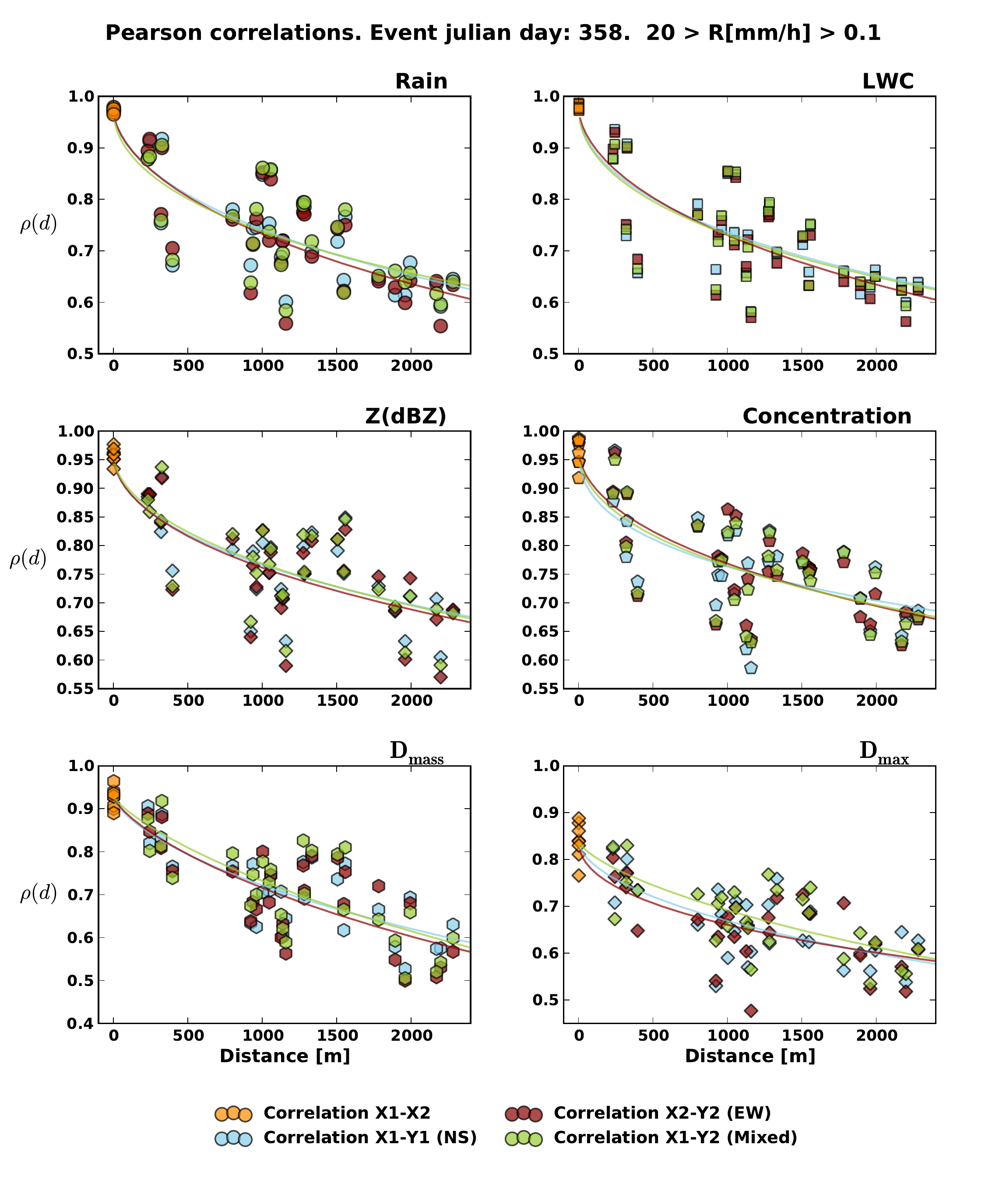}
 \vspace{0.05cm}  
    
    \caption[The correlogram based on the Pearson coefficient for various integral parameters. Event from 24 December 2009.]{\textbf{ The correlogram based on the Pearson coefficient for various integral parameters. Event from 24 December 2009.} \textit{Blue} is the correlation between disdrometers from different stations but with the same North-South orientation, and \textit{red}, for East-West. The correlations between stations where disdrometers with different orientation are taken appear in \textit{green}. The correlations for dual instruments (connected) appear in \textit{orange}. The lines represent \emph{trends} based on a model that is introduced in \S\ref{sec:newVariability}. In the cases of greatest spatial variability, the linear models do not connect with the values of the attached stations; for this reason, the decay model was proposed, which is discussed in Section \S\ref{sec:newVariability}.}

  \vspace{0.05cm}
 \label{fig:Var1CorreEpisodio24DIC}
 \end{center}
 \end{figure}

\begin{figure}[H] 
 \begin{center}
 \vspace{0.15cm}
    \includegraphics[width=0.95\textwidth]{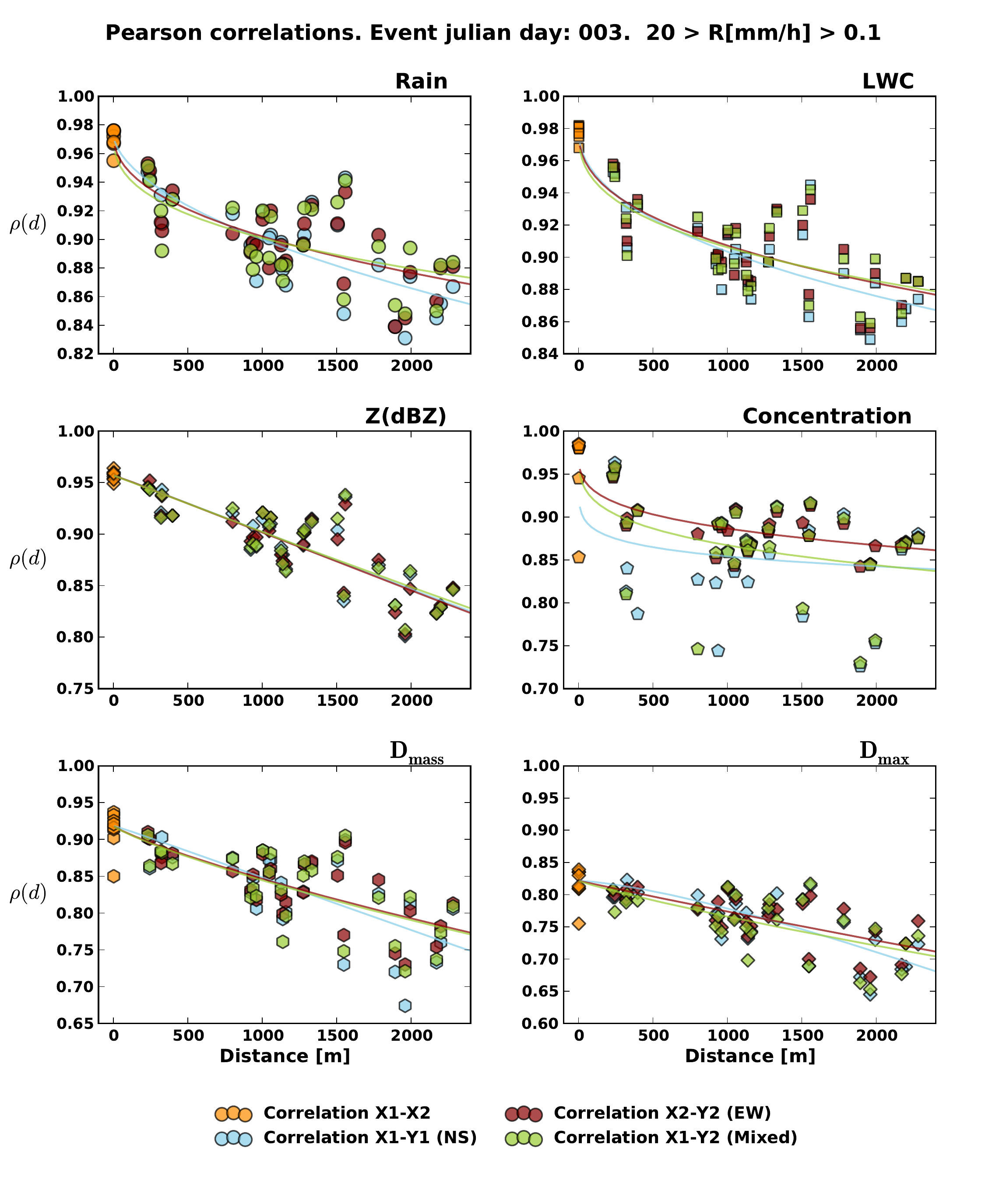}
 \vspace{0.05cm}   
    
    \caption[The correlogram based on the Pearson coefficient for various integral parameters. Event from 3 January 2009.]{\textbf{ The correlogram based on the Pearson coefficient for various integral parameters. Event from 3 January 2009.} \textit{Blue} is the correlation between disdrometers from different stations but with the same North-South orientation, and \textit{red}, for East-West. The correlations between stations where disdrometers with different orientation are taken appear in \textit{green}. The correlations for dual instruments (connected) appear in \textit{orange}. The lines represent \emph{trends} based on a model that is introduced in \S\ref{sec:newVariability}. In the cases of greatest spatial variability, the linear models do not connect with the values of attached stations; for this reason, the decay model was proposed, which is discussed in Section \S\ref{sec:newVariability}.}

    \vspace{0.05cm}
 \label{fig:Var1CorreEpisodio3_4ENE}
 \end{center}
 \end{figure}

\begin{figure}[H] 
 \begin{center}
 \vspace{0.15cm}
    \includegraphics[width=0.95\textwidth]{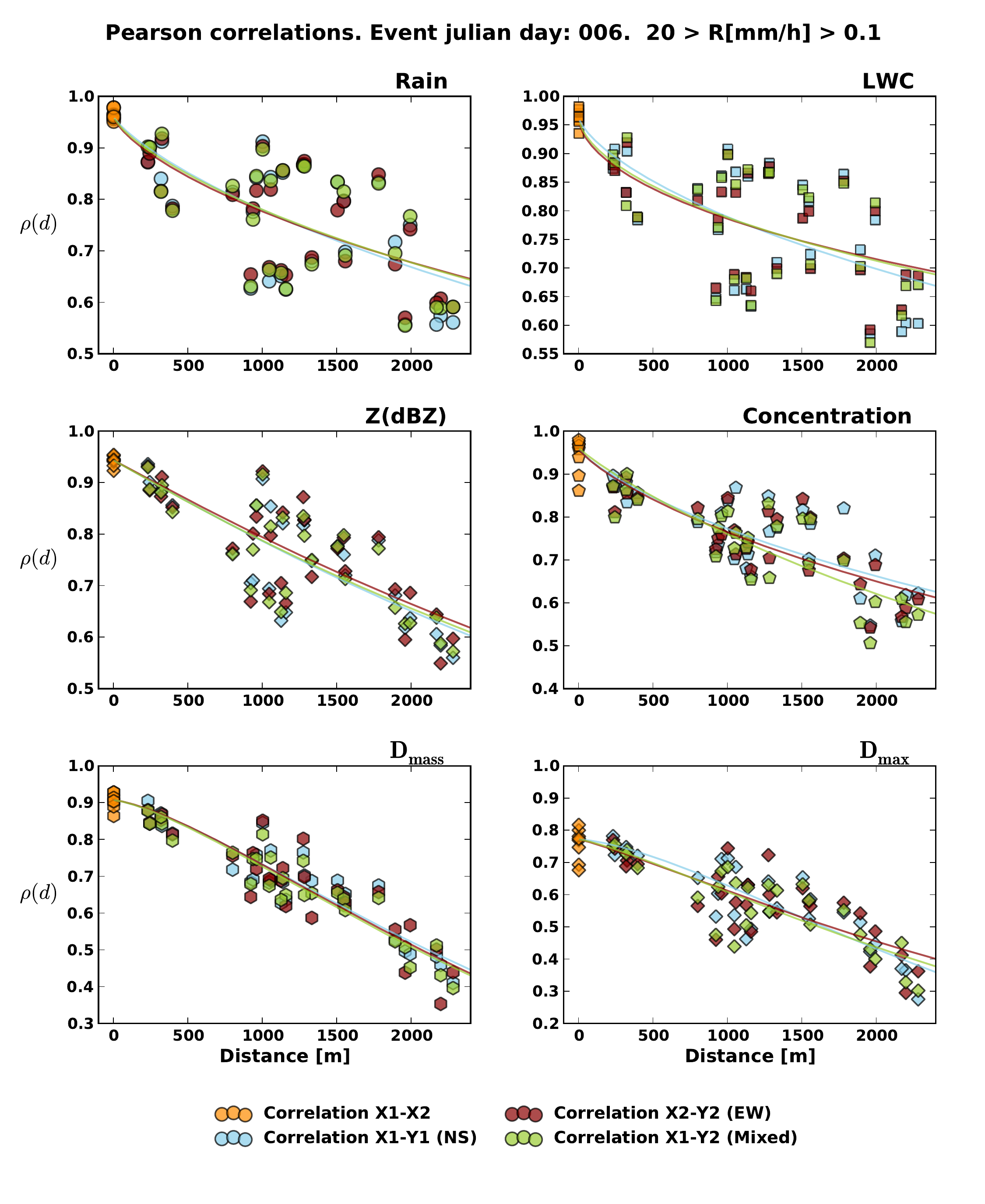}
 \vspace{0.05cm}
    
    \caption[The correlogram based on the Pearson coefficient for various integral parameters. Event from 6 January 2009.]{\textbf{ The correlogram based on the Pearson coefficient for various integral parameters. Event from 6 January 2009.} \textit{Blue} is the correlation between disdrometers from different stations but with the same North-South orientation, and \textit{red}, for East-West. The correlations between stations where disdrometers with different orientation are taken appear in \textit{green}. The correlations for dual instruments (connected) appear in \textit{orange}. The lines represent \emph{trends} based on a model that is introduced in \S\ref{sec:newVariability}. In the cases of greatest spatial variability, the linear models do not connect with the values of attached stations; for this reason, the decay model was proposed, which is discussed in Section \S\ref{sec:newVariability}.}
 \vspace{0.05cm}
 \label{fig:Var1CorreEpisodio6ENE}
 \end{center}
 \end{figure}

\begin{figure}[H] 
 \begin{center}
 \vspace{0.15cm}
    \includegraphics[width=0.95\textwidth]{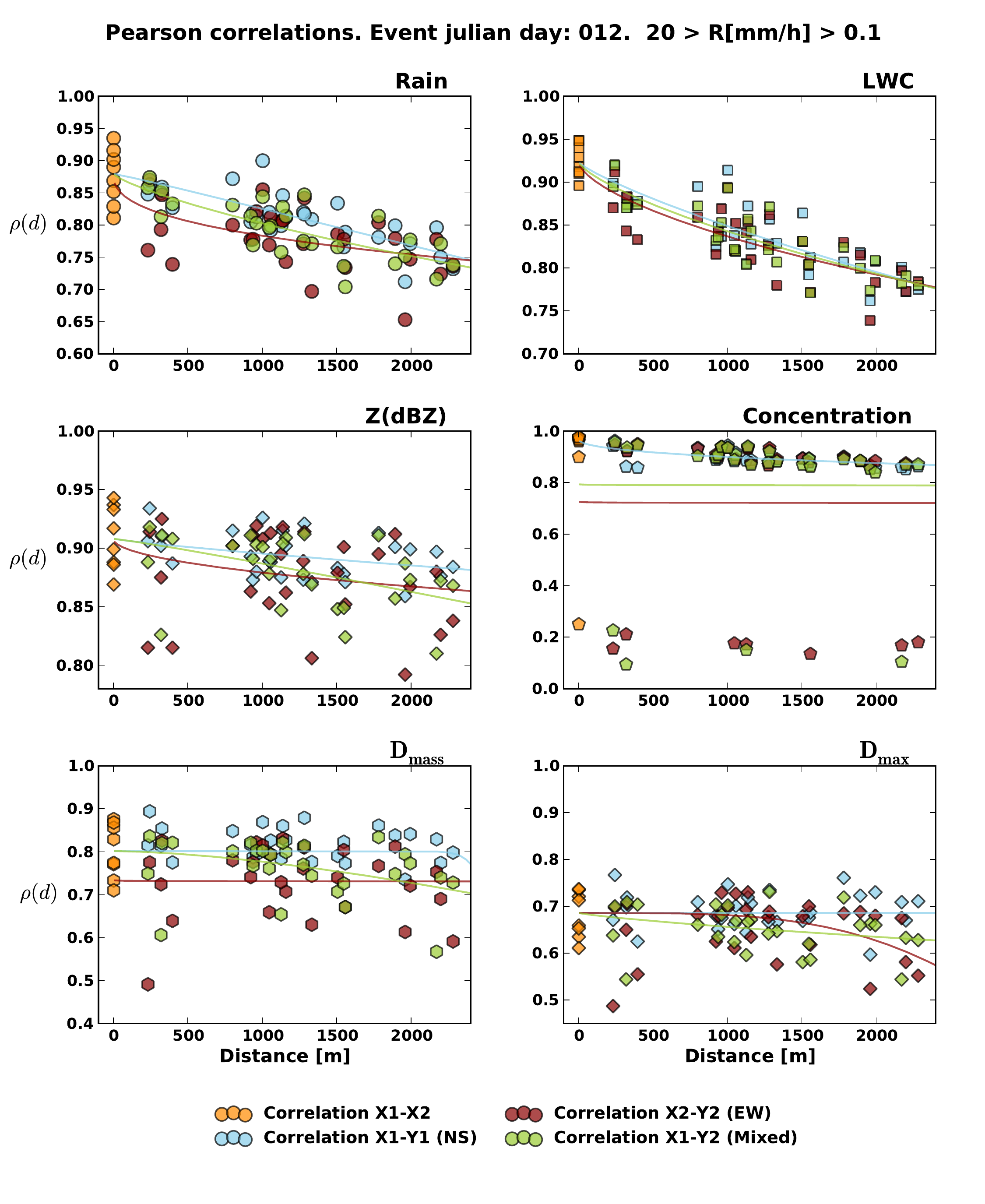}
 \vspace{0.05cm}
    \caption[The correlogram based on the Pearson coefficient for various integral parameters. Event from 12 January 2009.]{\textbf{ The correlogram based on the Pearson coefficient for various integral parameters. Event from 12 January 2009.} \textit{Blue} is the correlation between disdrometers from different stations but with the same North-South orientation, and \textit{red}, for East-West. The correlations between stations where disdrometers with different orientation are taken appear in \textit{green}. The correlations for dual instruments (connected) appear in \textit{orange}. The lines represent \emph{trends} based on a model that is introduced in \S\ref{sec:newVariability}. In the cases of greatest spatial variability, the linear models do not connect with the values of attached stations; for this reason, the decay model was proposed, which is discussed in Section \S\ref{sec:newVariability}.}
 \vspace{0.05cm}
 \label{fig:Var1CorreEpisodio12ENE}
 \end{center}
 \end{figure}

The availability of disdrometric measurements permits analysis of the time series correlations of other parameters in addition to the precipitation intensity. In this section, various integral parameters are analyzed, including the reflectivity and precipitation intensity, which are necessary to radar estimation, as well as other integral parameters, such as the concentration or the characteristic diameters (which are of relevance to the DSD models, such as determining the gamma distribution by means of the concept of normalized distribution\footnote{In section \S\ref{sec:newVariability}, the models are detailed specifically from the normalized gamma distribution parameters for the entire empirical basis of liquid precipitation.}). The analysis is conducted for the 6 episodes of liquid precipitation that include different situations of possible variability. The main observations derived from figures (\ref{fig:Var1CorreEpisodio20DIC}), (\ref{fig:Var1CorreEpisodio2DIC}), (\ref{fig:Var1CorreEpisodio24DIC}), (\ref{fig:Var1CorreEpisodio3_4ENE}), (\ref{fig:Var1CorreEpisodio6ENE}) and (\ref{fig:Var1CorreEpisodio12ENE}) are summarized.  

\begin{itemize}

\item Small-scale variation is present in all episodes in the different integral parameters, although the variation is more distinct in the cases of precipitation intensity, reflectivity, and the liquid-water content.
\item In the case of concentration, the episodes from 20 December 2009, 3 January 2010 and 12 January 2010, Figures (\ref{fig:Var1CorreEpisodio20DIC}), (\ref{fig:Var1CorreEpisodio3_4ENE}) and (\ref{fig:Var1CorreEpisodio12ENE}) also exhibit a decay; however, the presence of an anomalous time series, which is clearly identifiable in the episodes from 20 December and 12 January, corresponding to the E2 instrument, implies the existence of anomalous correlogram values when correlations with this instrument are involved \footnote{We understand that the cause of this observation arises from the presence of a wind power generator located near the instrument at a measuring station that is approximately 10 meters from the instrument, contributing intense local and directional turbulence to certain episodes, which implies that the E2 disdrometer records a different number of drops than its dual instrument E1.}. Aside from the set of measurements of the correlations that implicate this station, the results for all 
episodes show a decrease in correlation with distance as well as for the concentration of drops.
\item For the maximum diameter and for $D_{mass}$, the decorrelation with distance is present though its modeling in the case of 12 January 20120 is uncertain, with a possible decorrelation appearing between sub-networks N-S and E-W. It is worth noting that the analysis of $D_{max}$ is more dependent on specific sampling problems than the integral precipitation parameters because it depends exclusively on the drops of greatest size (which are statistically less probable). We conclude that the greatest difficulty found in modeling the decay of these parameters is derived from this fact.
\item Some episodes, such as those shown in Figures (\ref{fig:Var1CorreEpisodio20DIC}) and (\ref{fig:Var1CorreEpisodio12ENE}), possess greater intrinsic variability (mainly in the case of 12 January), which is an issue that can be identified via dispersion in the correlation of connected instruments. The use of dual instruments allows these circumstances to be identified in the analysis by episode.
\item The variability characterized by the correlogram was verified and depends on the concrete episode of analysis. Events, such as that recorded on 2 December 2009, possess a decorrelation in the range of 2\,500 meters of approximately 80\% in terms of precipitation intensity, while the episode on 3 January 2010 exhibits a value of 20\%. Generally, we can establish that for the rainfall intensity in stratiform episodes, an average value of the decorrelation at a distance of 2 km oscillates between 25\% and 40\%, exhibiting similar results both for the reflectivity as well as for the liquid water content.
\end{itemize}

\clearpage

\clearpage

\clearpage

\clearpage

\clearpage
\clearpage

\section{Stability of the Z-R relationships}
\label{sec:ZRrelations}
One of the most relevant implications of the spatial variability of precipitation is determined by the variability of the Z-R relationship in the typical size covered by a radar pixel (either with a terrestrial base or from satellite measurements), as in the case of the TRMM satellite or the GPM project, which incorporate radar systems for the study of precipitation.\\

\begin{figure}[h]
\begin{center}
\vspace{0.05cm}
   \includegraphics[width=0.75\textwidth]{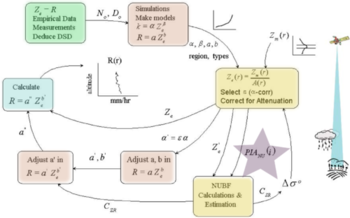}
\vspace{0.01cm}
   \caption[The scheme used for precipitation estimation via radar incorporated in to the TRMM satellite. The algorithm is based on the Z-R relationships. Source: \citep{2008BAMSpolarizationRADARforGPM}.]{\textbf{The scheme used for precipitation estimation via radar incorporated in to the TRMM satellite. The algorithm is based on the Z-R relationships. Source: \citep{2008BAMSpolarizationRADARforGPM}.} The appearance of the Z-R relationships at various points in the TRMM algorithm can be proven. This fact is related to different aspects of the estimate, which have been discussed throughout the test, such as evaluating the attenuation suffered by the signal itself, the use of parameterized DSD models or the problem of non-uniform filling of precipitation within the volume sampled by the radar.}
\vspace{0.05cm}
\label{fig:TRMMPRmethodVar1}
\end{center}
\end{figure}

A practical example of the use of Z-R relationships is outlined in figure (\ref{fig:TRMMPRmethodVar1}), which clearly highlights how the estimates of R as a function of Z via a power-law relationship appear in several steps of the algorithm. The DSD variability itself appears in the attenuation estimates within the volume of the radar beam, which is not uniformly filled with precipitation \citep{Tokay2010smallscaleDSD,MGossetNUBF2001}. This makes the variability in Z and R have with respect to distance one of the most significant issues. As seen in Figures  (\ref{fig:Var1CorreEpisodio2DIC}), (\ref{fig:Var1CorreEpisodio20DIC}), (\ref{fig:Var1CorreEpisodio24DIC}), (\ref{fig:Var1CorreEpisodio3_4ENE}), (\ref{fig:Var1CorreEpisodio6ENE}) and (\ref{fig:Var1CorreEpisodio12ENE}), the $\rho_{Z}(h)$ and $\rho_{R}(h)$ correlations are similar but not identical (for a relatively homogeneous case, as well), which is a fact that should be reflected in the Z-R relationships.

\subsection{Methodologies and limits}

As introduced in \S\ref{cap:DSD}, the Z-R relationships are based on the estimation of the parameters $a_{R}$ and $b_{R}$ present in the equation:

\begin{equation}
Z=a_{R}R^{b_{R}}
\end{equation}

Different methods exist for estimation of the values of $a_{R}$ and $b_{R}$. The most common is to determine, given the above equation, the parameters, either by a linear method or by the method of a nonlinear fit to the indicated expression\footnote{An extensive comparison of the differences between both relationships appears in \citep{Steiner2004microphysisofZRgamma}. Although for a certain event, the method can lead to a lack of consistency between the values of the $(a_{R},b_{R})$ pair and the $(a_{Z},b_{Z})$ pair, nonetheless, from the point of view of the dispersion of parameter values within the extensive analysis of 102 cases, the methods are assessed as equivalent such that for the purposes of the study carried out in this thesis, the method is sufficient for the presentation of results for $a_{R}$ and $b_{R}$.}.\\

 The fact that the results are not always consistent among the different methods, due either to physical variability or statistical sampling problems, has led to the complementation of these previous methods with other methodologies that attempt, given the initial dataset, to perform processing that presumably allows elimination of the possible variability caused by sampling problems. Hence, only the variability arising from physical processes remains \footnote{There is a method called PMM (\textit{Probability Matching Method}) that we have not treated in this section because the previous study carried out by \citep{PeterHartmann2007} revealed that the possible differences present in $a_{R}$ and $b_{R}$ based on the other methods mask those that we would see if we were also to apply PMM, i.e., that from the point of view of variations in the network, applying another method does not contribute new information. Additionally, from a more formal point of view, the PMM method has been questioned by several 
authors \citep{ciach_krajewski_ea_1997}.}.\\

 In this sense, one of most utilized methods is called SIFT, \emph{Sequential Intensity Filtering Technique} \citep{LeeSIFTmethod2005}. According to the authors, SIFT permits the elimination of the unnatural variability that can exist in time series of data, further reflected in the Z-R relationships. The steps that comprise this procedure are\footnote{In this section, to establish comparative conclusions based on the SIFT method without the use of prior hypotheses, the use of pre-processing has been minimized. In this way, a consistent set of minutes is required in addition to the condition that each minute must record at least 5 drops. Based on this empirical requirement for each episodes, different limits in the minimum intensity are compared.}:

\begin{itemize}
\item A reference variable that we shall call P is selected. Generally, P is an integral parameter of precipitation, where Z and R are the most used. 
\item Based on the parameter P, the DSD time series is arranged in increasing order. We go from N(D;t) to a series in increasing order, N(D;P), based on the values generated for P. 
\item Given the N(D,P) series, \emph{moving averages} of \emph{m} elements of the series are carried out\footnote{The value of DSD in the position given by the P value is obtained from the average in a window of a size determined by m. In our case, as we approach the edges of the series N (D, P), we progressively reduce the value of m such that the series of DSDs after the process of \emph{moving averages} has the same number of DSDs as the original series.} to build $N^{(m)}(D, P)$, which consequently has the same number of DSDs as the original series.
\item With the filtered set of DSD $N^{(m)}(D;P)$, the R and Z time series are determined \footnote{In our case for the calculation of the average R over the m distributions of drop sizes, we have directly taken the R values obtained from $n(D,v)$, which is more precise than performing an R estimate using the DSD and assuming a $v(D)$.}, over which the a and b values are calculated using either the linear fitting methods or the nonlinear methods.
\end{itemize}

The main characteristic of this procedure is that after averaging the original time series in a window of size \emph{m}, the possible stochastic variation due to sampling problems is mitigated by taking the average, although the implicit hypothesis is that averaging DSD with similar values of P makes sense (i.e., this implies that P allows for classifying the DSD, and this has physical sense).\\

Previous studies seem to indicate that after this processing, the consistency of the obtained values of a and b is greater, and the variability they represent is essentially due to natural processes, if R or Z are taken as P reference parameters.\\

In this study, the values of parameters $a_{R}$ and $b_{R}$ have been determined, both with and without application of the SIFT method: 
\begin{itemize}
 \item The typical method (NO-SIFT) is to utilize the original time series at a 1-min resolution. 
\item The SIFT method:
\begin{itemize}
 \item Reference Z with windows of sizes m $\in \left\lbrace 5, 7, 9, 13 \right\rbrace$.
 \item Reference R with windows of sizes m $\in \left\lbrace 5, 7, 9, 13 \right\rbrace$.
\end{itemize}
\end{itemize}

 This approach allows a verification, within the philosophy of the method, of whether the variability we found without applying SIFT might simply be due to DSD sampling problems for large drops.\\

\begin{figure}[H] 
\vspace{0.7cm}
\begin{center}
   \includegraphics[width=1.10\textwidth]{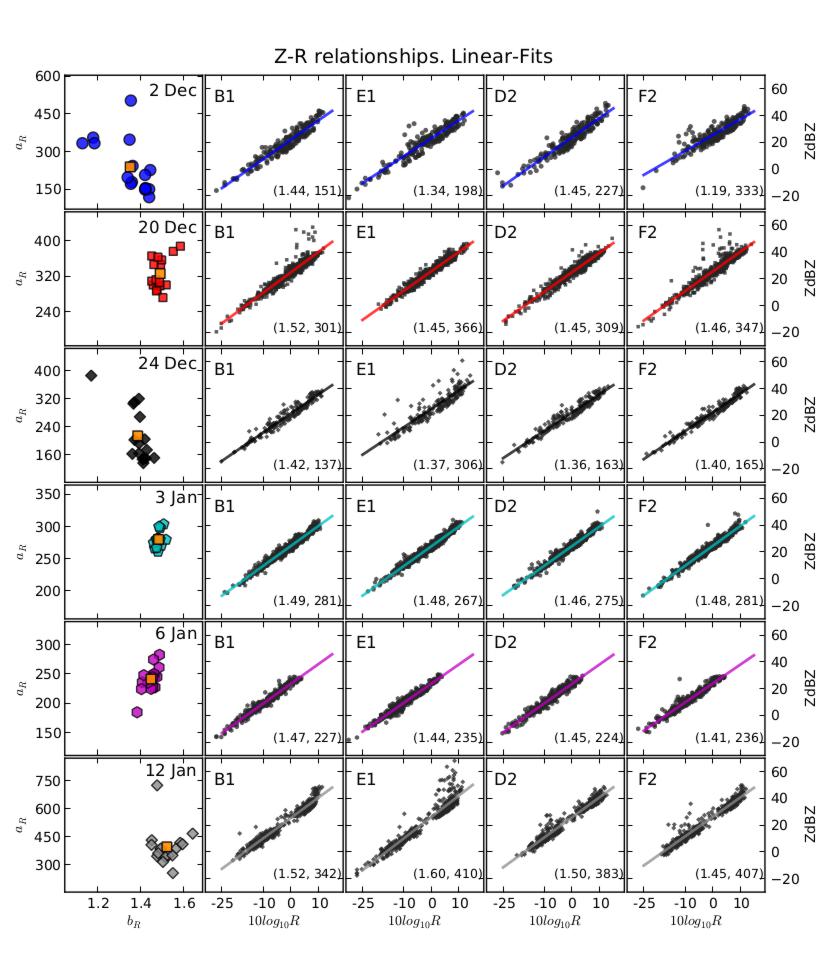}
   \caption[The Z-R relationships for the episodes under analysis. Linear fitting method without SIFT filtering.]{\textbf{ The Z-R relationships for the episodes under analysis. Original time series without SIFT filtering. Linear fitting method. The minimum number of drops is 5.} \textit{Left}: The (a, b) values are given for the set of 16 disdrometers in addition to the value utilizing the time series of average Z and R values in the network. Each graph represents one of the analyzed precipitation episodes. \textit{Right}: The Z-R scattering relationships are shown for four disdrometers in the network along with the fits to the Z-R relationship. The (a, b) values, given in parentheses, represent the value by means of a nonlinear fit with the goal of showing that a similar dispersion in the Z-R relationship is obtained with another fitting method.}

 \label{fig:Var1_ZR_lineal}
\end{center}
\vspace{1.5cm}
\end{figure}

\begin{figure}[H] 
\vspace{1.5cm}
\begin{center}

   \includegraphics[width=1.10\textwidth]{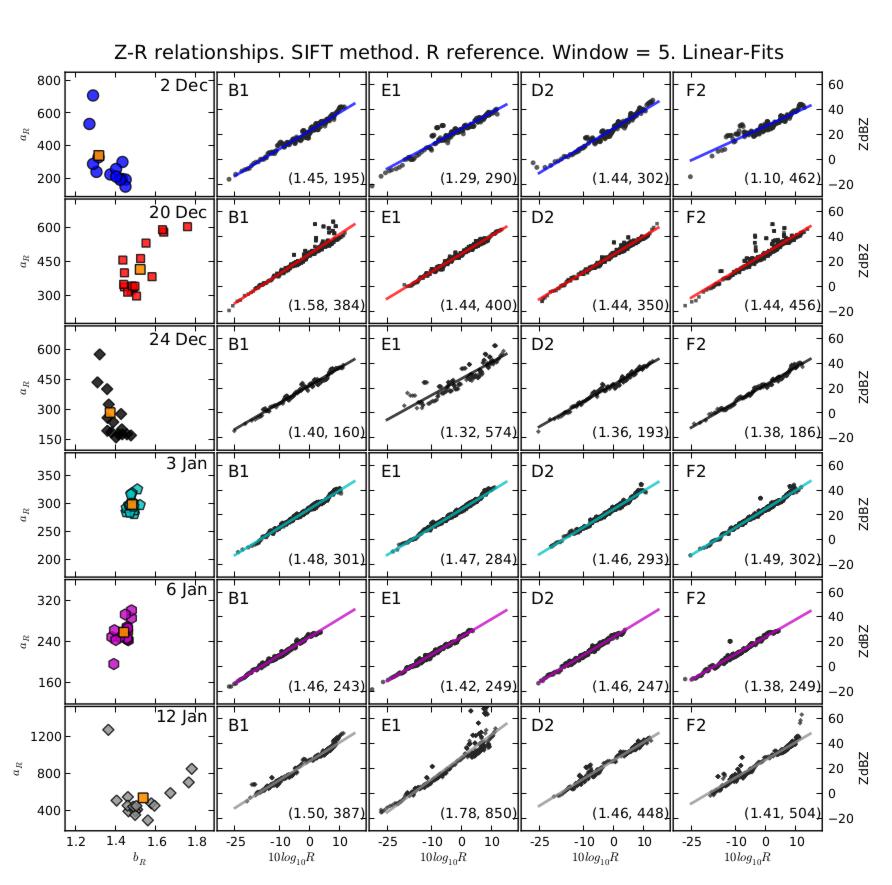}
   \caption[The Z-R relationships for the episodes under analysis. Linear fitting method with SIFT filtering (window m=5 and reference R).]{\textbf{The Z-R relationships for the episodes under analysis. Linear fitting method with SIFT filtering (window m=5 and reference R). The minimum number of drops is 5.} \textit{Left}: The (a, b) values are given for the set of 16 disdrometers in addition to the value utilizing the time series of average Z and R values in the network. Each graph represents one of the analyzed precipitation episodes. \textit{Right}: The Z-R scattering relationships are shown for four disdrometers in the network along with the fits to the Z-R relationship. The (a, b) values, given in parentheses, represent the value by means of a nonlinear fit with the goal of showing that a similar dispersion is obtained in the Z-R relationship with another fitting method.}

\label{fig:Var1_ZR_Z5_lineal}
\vspace{1.5cm}
\end{center}
\end{figure}

As in studies conducted with individual instruments, both methods give different values of $a_ {R}$ and $b_ {R}$ (although their differences are small if we use the rainfall intensity as variable P); however, from the point of view of the variability within the network, we found that both methods give similar results, see Figures (\ref{fig:Var1_ZR_lineal}), (\ref{fig:Var1_ZR_Z5_lineal}), y (\ref{fig:Var1_ZR_Z7_lineal}).\\

The results in the three preceding figures show how
\begin{itemize}
\item The values of $a_{R}$ and $b_{R}$ slightly depend on the chosen method, i.e., SIFT vs. NO-SIFT, but the general dispersion in the network is not determined as a result of adopting one method or another. This suggests that the dispersion is a consequence of the natural variation in DSD.
\item The window sizes of the SIFT method imply a change in the $a_{R}$ and $b_{R}$ values, but again from the network point of view, this change is not significant. 
\item In addition to taking R as a reference, Z was taken as a reference. The results indicate that the $a_{R}$ and $b_{R}$ values are somewhat different (less dispersion). However, of more relevance is the particular minimum precipitation intensity that we choose. 
\end{itemize}

The results indicate that, out of the methods designed to eliminate sampling problems, similar results for the Z-R relationship variation are obtained throughout the network. The relevance of including limits on the rainfall intensity also was shown, i.e., including a previous filtering of minutes below different R thresholds. The results are shown in Figure (\ref{fig:Var1_ZR_red_metodos}). The inclusion of thresholds of $R\,[mm/h] > 0.2$ does indeed eliminate part of the dispersion \textit{between episodes} such that the cluster of dots is displaced to higher $b_{R}$ and lower $a_{R}$ parameters (independent of the method used). The implication is that the difference within each episode is similar but the differences between episodes are less\footnote{The case of the episode with the greatest variability does not follow this pattern, and in the application of limits and the SIFT methods, the episode exhibits a more erratic behavior.}. As a result, the spatial variability of each episode expressed by means 
of the Z-R relationships for each episode does not depend on the R thresholds (although these determine in part the concrete values of $a_{R}$ and $b_{R}$).

\vspace{2.05cm}

\begin{figure}[H] 
\vspace{1.5cm}
\begin{center}
   \includegraphics[width=1.10\textwidth]{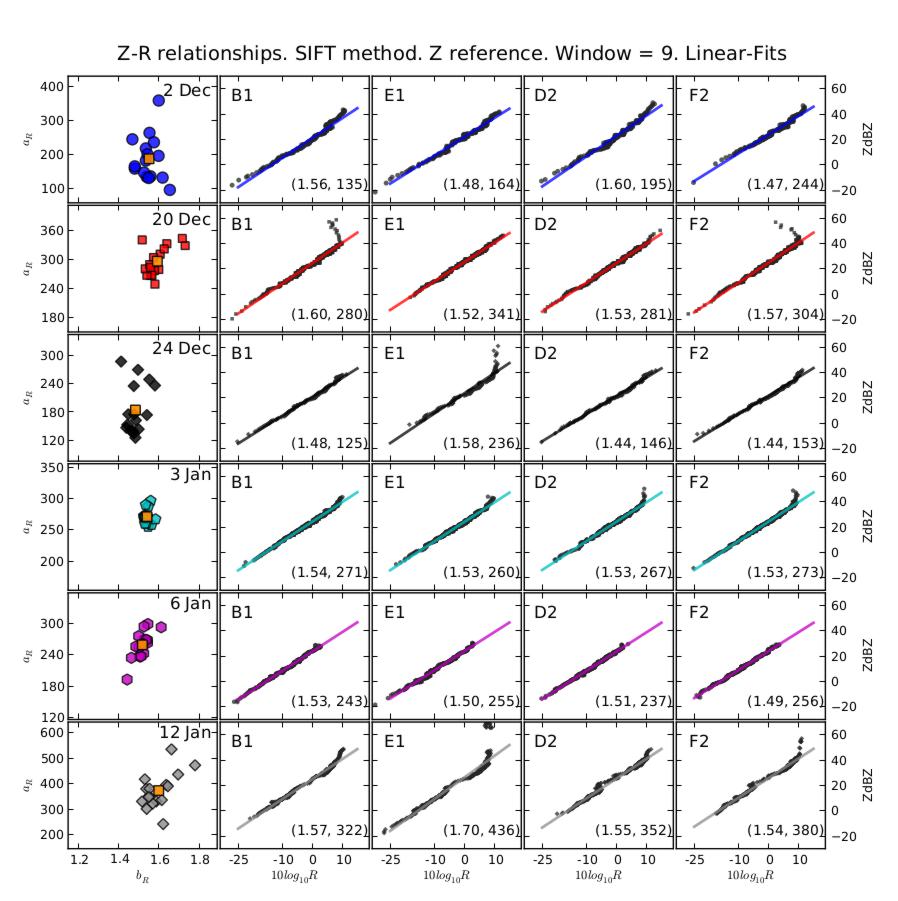}
      \caption[The Z-R relationships for the episodes under analysis. Linear fitting method with SIFT filtering (window m=9 and reference R).]{\textbf{The Z-R relationships for the episodes under analysis. Linear fitting method with SIFT filtering (window m=9 and reference R). The minimum number of drops is 5.} \textit{Left}: The (a, b) values are given for the set of 16 disdrometers in addition to the value utilizing the time series of average Z and R values in the network. Each graph represents one of the analyzed precipitation episodes. \textit{Right}: The Z-R scattering relationships are shown for four disdrometers in the network along with the fits to the Z-R relationship. The (a, b) values, given in parentheses, represent the value by means of a nonlinear fit with the goal of showing that a similar dispersion is obtained in the Z-R relationship with another fitting method.}
\label{fig:Var1_ZR_Z7_lineal}
\end{center}
\vspace{1.5cm}
\end{figure}

\begin{figure}[H] 
\vspace{1.5cm}
\begin{center}
   \includegraphics[width=1.10\textwidth]{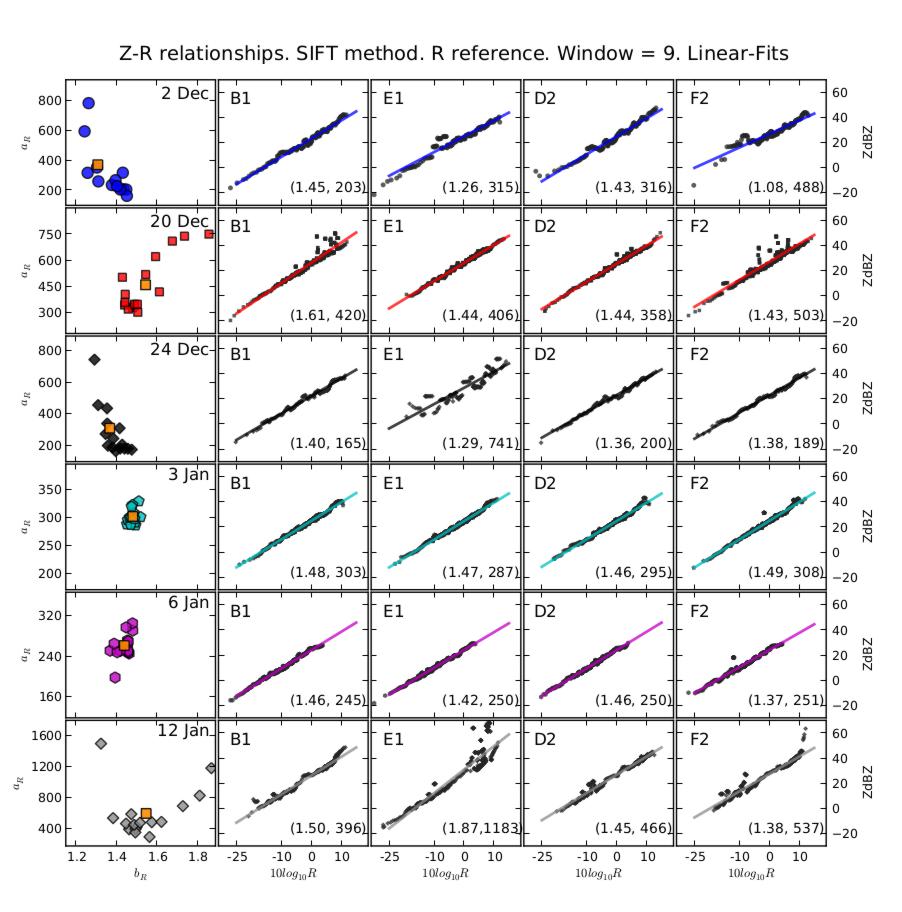}
      \caption[The Z-R relationships for the episodes under analysis. Linear fitting method with SIFT filtering (window m=9 and reference Z).]{\textbf{The Z-R relationships for the episodes under analysis. Linear fitting method with SIFT filtering (window m=9 and reference Z). The minimum number of drops is 5.} \textit{Left}: The (a, b) values are given for the set of 16 disdrometers in addition to the value utilizing the time series of average Z and R values in the network. Each graph represents one of the analyzed precipitation episodes. \textit{Right}: The Z-R scattering relationships are shown for four disdrometers in the network along with the fits to the Z-R relationship. The (a, b) values, given in parentheses, represent the value by means of a nonlinear fit with the goal of showing that a similar dispersion is obtained in the Z-R relationship with another fitting method.}
\label{fig:Var1_ZR_Z7_lineal}
\end{center}
\vspace{1.5cm}
\end{figure}


\begin{figure}[H]
\begin{center}
\vspace{1.05cm}
   \includegraphics[width=1.05\textwidth]{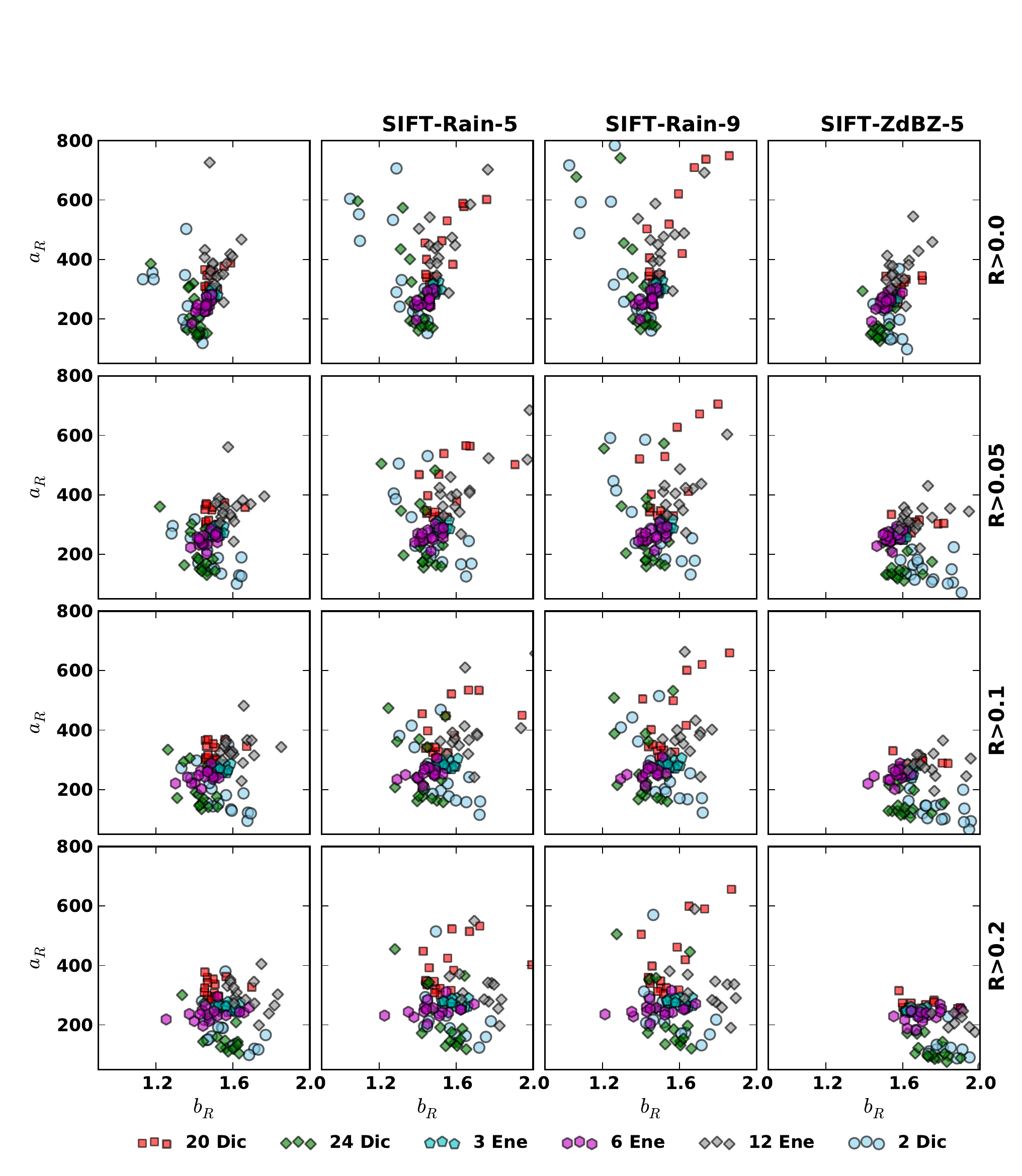}
\vspace{0.85cm}
   \caption[A comparison of the Z-R relationships over the network for NON-SIFT and SIFT methods. Various rainfall-intensity thresholds.]{\textbf{A comparison of the Z-R relationships over the network for NON-SIFT and SIFT methods (using Z and m=5). Various rainfall-intensity thresholds.} Each circle represents the resulting relationship for one of the disdrometers by episode. The following thresholds have been included in the precipitation intensity, the first row for $R [mm/h]>0.0$, the second row $R [mm/h]>0.05$, the third row $R [mm/h] > 0.1$ (most common limit) and the last row $R > 0.2$ (The estimated sensitivity for the future GPM-core Radar).}
\label{fig:Var1_ZR_red_metodos}
\end{center}
\vspace{1.05cm}
\end{figure}

\section{DSD scaling method}
\label{sec:escaladado_variab_esp}

The scaling method of the DSD that is based on one moment was introduced in Section \textsection\ref{sec:Scaling1moment} and was applied to the set of all disdrometers in the episodes under analysis. The integral parameters R, W and $D_{mass}$ were taken as a reference, in addition to utilizing the moment of order 3.67 that is usually considered the DSD moment closest to the rainfall intensity. Regarding the use of $D_{mass}$ because it is not the moment of the distribution, it implies that the consistency analysis possible with R and W under which, for example, equation (\ref{eqn:consistencia1-1moment}) is satisfied, ceases to be direct. However, it is interesting to compare the results using $D_{m}$ with respect to R and W, thereby gaining insight concerning the general properties of this modeling approach\footnote{The case of snow also was analyzed, although it is shown in Appendix \S\ref{sec:newVariability} because it entails making a series of fine tunings. That is, the method has traditionally been 
used for the N(D) of liquid rain, though formally, it can be utilized more widely. Additionally, the quantities defined as R, W and $D_{m}$ are calculated from either n(D,v) or N(D) as they were defined in \S\ref{cap:DSD}, which can be applied formally to data from n(D,v) without conversions, although they correspond to non-liquid aggregates. As a result, scaling with respect to R in the case of snow involves formally following the same procedure, even though the physical significance of R is different (now, scaling must be applied with respect to the moment of the distribution). The same approach is applied to W and $D_{m}$.}.\\

As in the case of the Z-R relationships, it is of interest to investigate whether the variability within a given episode is of similar magnitude to the variability among several episodes. Similarly, the method is based on power-law relationships, applied between the moments of distribution, which allows a verification of the relevance of different subsets of moments in the possible variability. That is, because the sampling problems are more significantly felt at high moments, while underestimation occurs in the disdrometer measurements of small drops, affecting the smaller moments more; hence, we have included the three analyses found in \S\ref{sec:basempirica_escalado}. One includes moments from 0 to 7, and another, from 2 to 6, stressing that\footnote{Some authors include noninteger moments, which lead to more data for successive fits imposed by the methodology; however, including noninteger moments does not appear to alter the general results \citep{uijlenhoet_steiner_etal_2003_aa}.}:

\begin{itemize}
\item The problems associated with small drops do not affect the moments in a relevant way beyond the second moment. Hereafter, the second-order moments are usually representative of the moments of the underlying distribution (excluding statistical sampling problems).
\item Because the analyzed episodes are not highly intense in terms of the precipitation intensity ($R \leq 20 $[mm/h]), the possible over-estimates of the Parsivel OTT in the case of intense precipitation are not critical, given that they are confined to a small set of minutes.
\end{itemize}

The above two points raise the possibility that the results obtained in the hypothesis test conducted using this methodology would be similarly obtained with other disdrometers and are not specific to the Parsivel OTT \footnote{In Chapter \S\ref{chap:BINNING}, a detailed analysis of the differences is made in terms of estimating the moments of different instruments with respect to the construction of drop-size histograms and sampling.}.\\

\begin{figure}[H]
\begin{center}
   \includegraphics[width=1.00\textwidth]{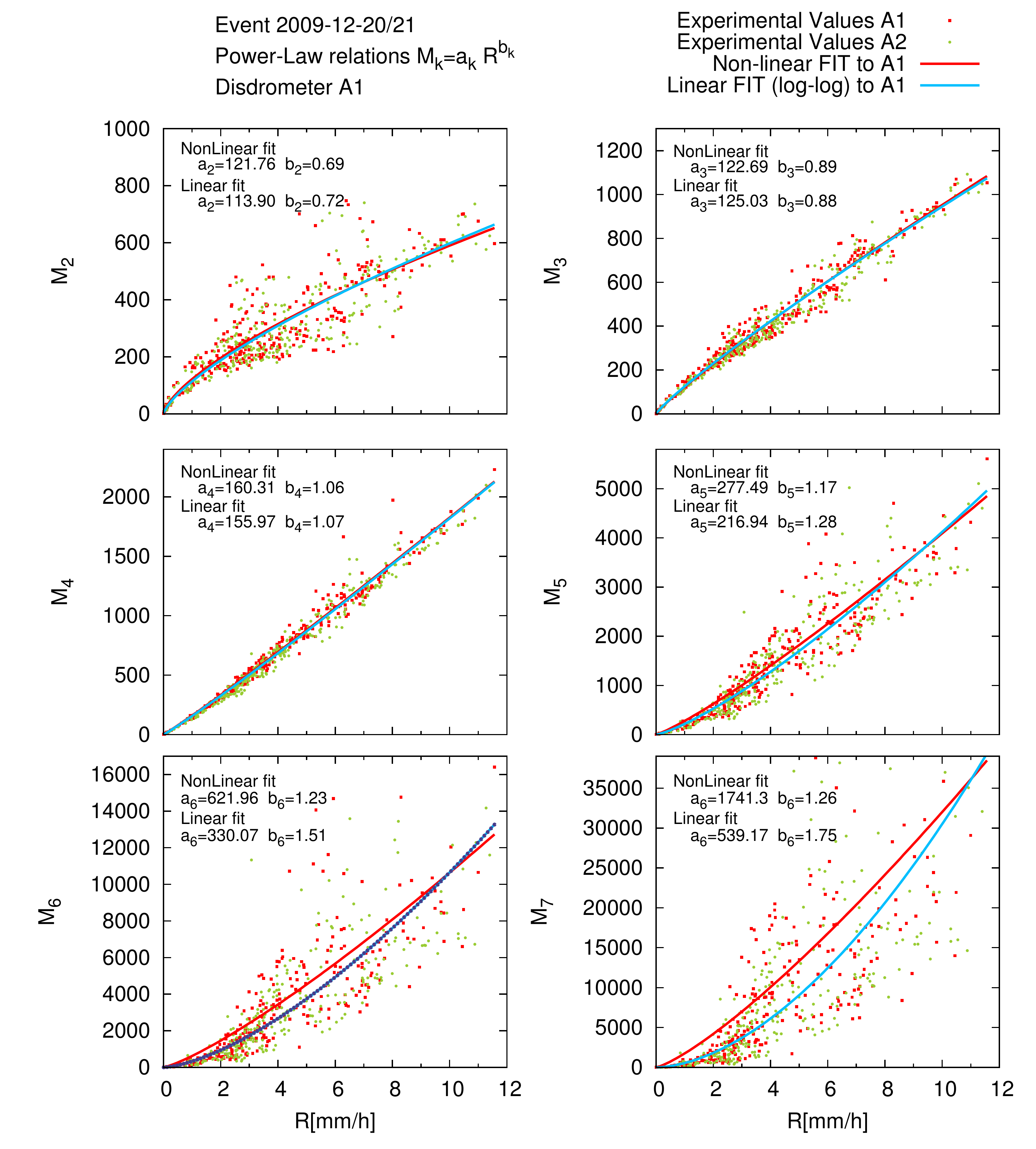}
\vspace{1cm}
   \caption[The power law of the n-th moment with respect to R. Linear and nonlinear fits. Event: 20-21 December 2009.]{\textbf{The power law of the n-th moment with respect to R. Linear and nonlinear fits. Event: 20-21 December 2009.} The experimental results are shown for moments with order 2 up to moments with order 7 and the \textit{scatterplots} of R. In red, disdrometer A1; in green, disdrometer A2. The solid blue lines represent the linear fits a in log-log scale, and the red lines represent nonlinear fits that use the Levenberg\textendash Marquardt algorithm. The case of the dark blue dotted line, shown in $M_{6}$, represents the same nonlinear algorithm applied in a log-log scale. As the order of the moment increases, it is apparent that the nonlinear Levenberg\textendash Marquardt algorithm becomes more sensitive to dispersed and high-value points.}

\label{fig:Var1ScalingMomentsvsR}
\end{center}
\vspace{1cm}
\end{figure}

\begin{figure}[H] 
\begin{center}
   \includegraphics[width=0.95\textwidth]{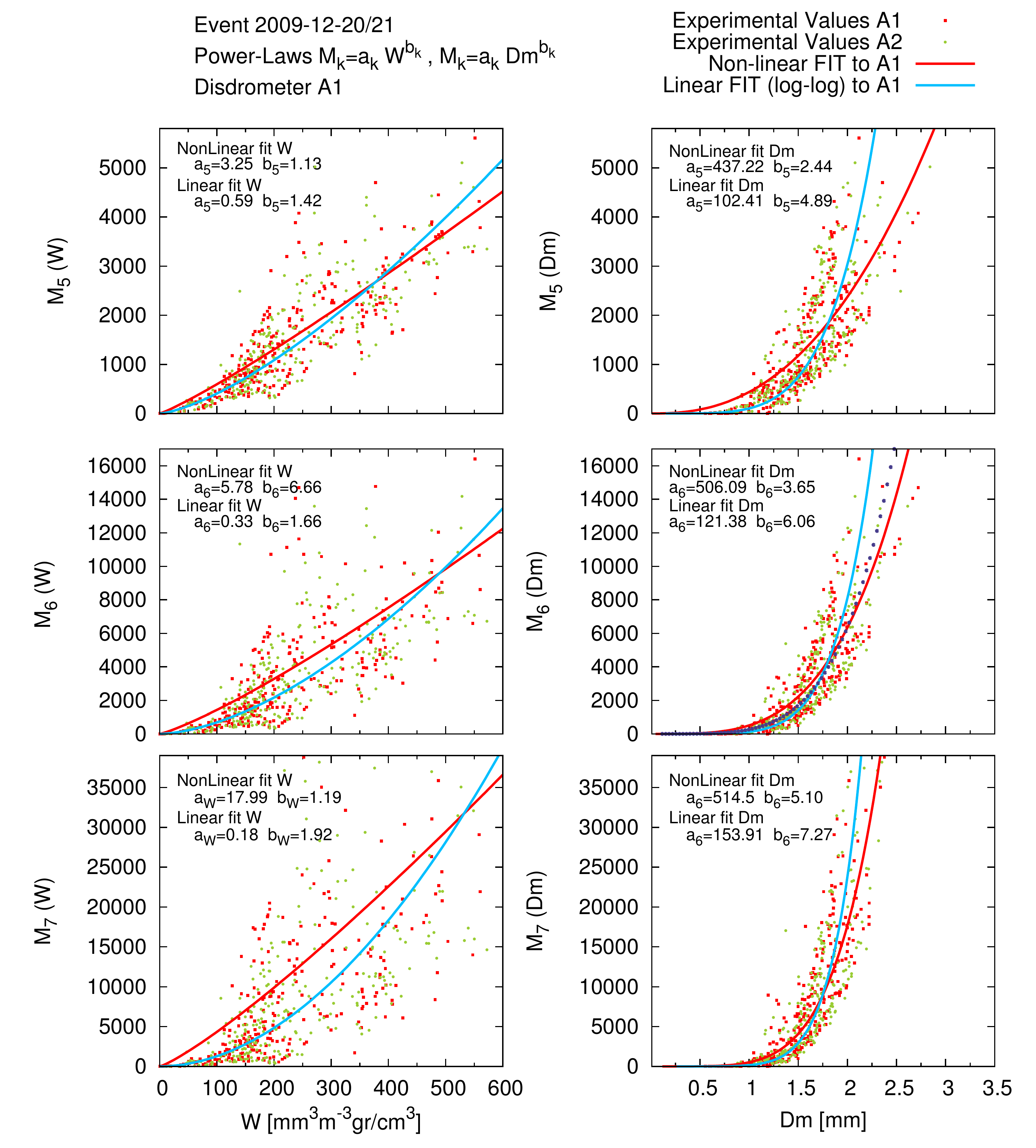}
\vspace{0.8cm}
   \caption[The power law of the n-th moment with respect to W and $D_{m}$. Linear and nonlinear fits. Event: 20-21 December 2009.]{\textbf{The power law of the n-th moment with respect to W and $D_{m}$. Linear and nonlinear fits. Event: 20-21 December 2009.} The experimental results are shown for the moments with order 5 up to moments with order 7 and the \textit{scatterplots} of W and $D_{m}$. In red, disdrometer A1; in green, disdrometer A2. The solid blue lines represent linear fits in a log-log scale, and the red lines represent nonlinear fits that use the Levenberg\textendash Marquardt algorithm. The case of the dark blue dotted line, shown in $M_{6}$, represents the same nonlinear algorithm applied in a log-log scale. As the order of the moment increases, it is apparent that the nonlinear Levenberg\textendash Marquardt algorithm becomes more sensitive to dispersed and high-value points (outliers). In the case of W, the results are very similar to scaling using R. In the case of $D_{m}$ for moments 
of order 1 to 3, the dispersion is greater, and the fits, more uncertain. Here, orders 5 to 7 are represented.}
\label{fig:Var1ScalingMomentsvsWDm}
\end{center}
\vspace{1.8cm}
\end{figure}

Additionally:
\begin{itemize}
\item Given the function g(x), it is possible to model it according to a gamma or log-normal distribution and to prove the variation in the fitting parameters throughout the network in similar fashion to the parameters of the Z-R power-law relationship.\\

\item In the process of constructing g(x), it is possible to relate the obtained data via a similar comparison to that performed for the Z-R relationships. That is, the differences in parameters $\alpha$ and $\beta$ contained in Equation (\ref{eqn:sempere-torres-1moment}) can be compared throughout the network and by episodes.
\end{itemize}

We have tried to include limits on the precipitation intensity in the scaling using the liquid water content (W or LWC). It has been proven how, when using Equation (\ref{eqn:consistencia1-1moment}), the values of $\alpha+\beta(4+1)\simeq 1.04$ are obtained, independent of the R thresholds. This finding implies maximum deviations on the order of 4\% in the consistency relation. We understand that the subsequent variability study is not determined in terms of its general properties by the precipitation-intensity limits, even though from the point of view of the adequate estimation of comparative parameters with other disdrometers, the indicated limitations can be introduced\footnote{These results are accordingly based on disdrometric data in the interval of rainfall intensity from 0.0 mm/h to 20 mm/h.}.

\subsection{Methods for estimating the $M_{k}=a\Psi^{b}$ relationships}

During the calculation process, it is necessary to obtain the (a, b) parameters of the relation:
\begin{equation}
M_{k}=a\Psi^{b}
\label{eqn:MomentosVar1}
\end{equation}
Systematically, we have proceeded to calculate the values of the parameters via two different techniques, one corresponding to the fit of the classic linear regression, and the other corresponding to a nonlinear fit.

\begin{itemize}
\item \textit{Nonlinear method}: The essential idea is to compare the values of the data series $M_{k}$ with the values given by the model $a\Psi^{b}$ such that the residuals $\delta=M_{k}-a\Psi^{b}$ have the property that their sum squared throughout the series of experimental values is minimal with respect to the parameters (a, b) and that they turn out to be the parameters obtained via the nonlinear fit. The algorithm used is called Levenberg\textendash Marquardt and is available in a number of visualization libraries and programs.
 
\item \textit{Linear method}: The usual fit achieved by linear regression via least squares, applied to the logarithmic transformation of expression (\ref{eqn:MomentosVar1}).
\end{itemize}

The results are shown in Figures (\ref{fig:Var1ScalingMomentsvsR}) and (\ref{fig:Var1ScalingMomentsvsWDm}) for one episode and one disdrometer. The dispersion diagrams are fitted to the power-law relationships in a similar fashion to the episodes analyzed by \citep{SempereTorres1998}. The linear and nonlinear methods give similar values of a and b for the low-order moments, while the nonlinear methods have proven to be exceedingly sensitive for marginal values of the power-law relationship\footnote{In the case of $M_{6}$ in Figure (\ref{fig:Var1ScalingMomentsvsR}), it is shown that the main difference is the use of a prior logarithmic transformation that eliminates part of the difference among methods.}. In the face of subsequent analyses, a and b estimated values have been taken via linear fits.\\

\begin{figure}[H] 
\begin{center}
   \includegraphics[width=1.05\textwidth]{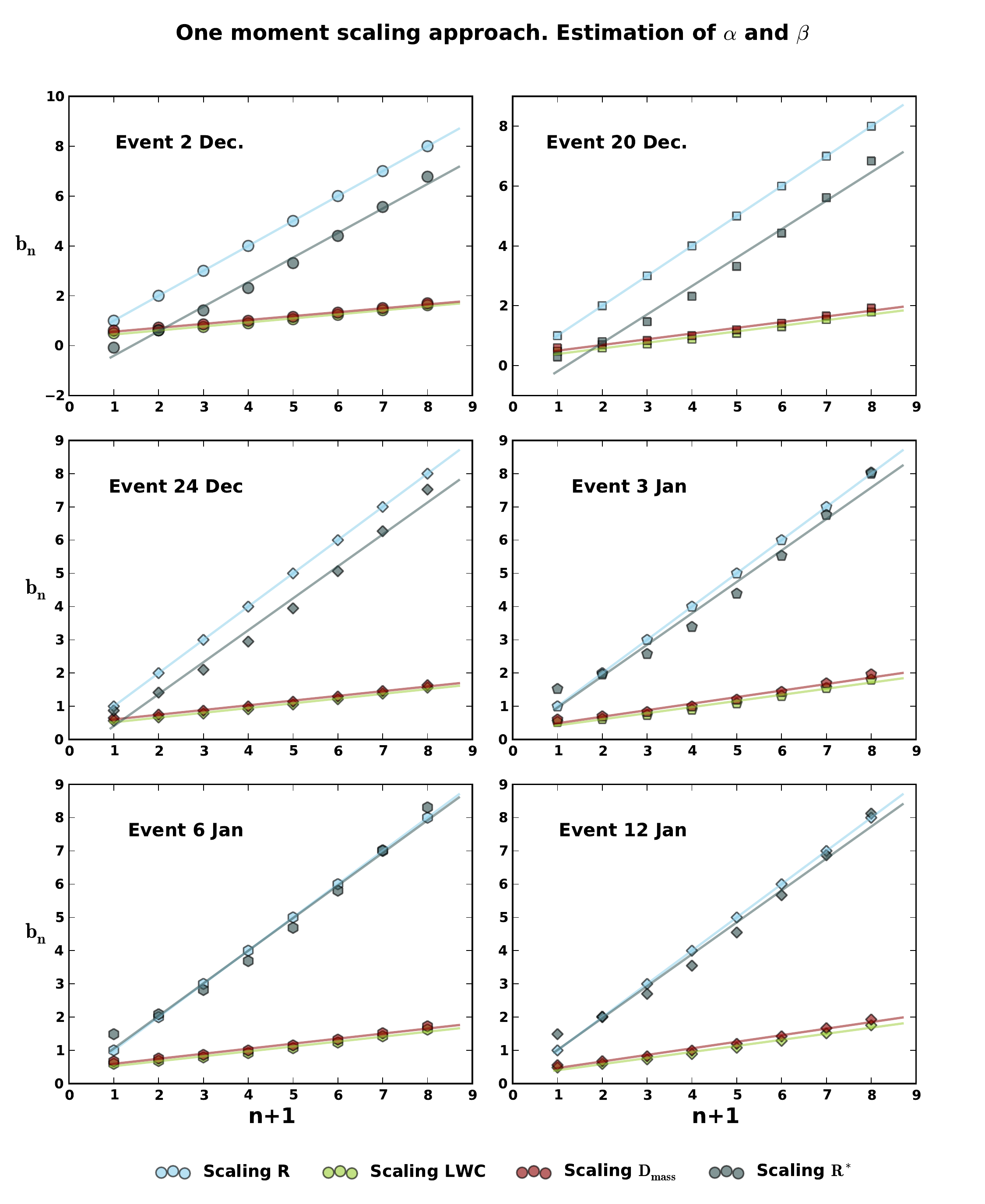}
\vspace{1.0cm}
   \caption[The fit of the $\beta$ and $\alpha$ \textit{scaling} parameters from Equation (\ref{eqn:sempere-torres-1moment}) corresponding to the scaling method of a moment over R, W, $D_{m}$ and $R^{*}$. The case of disdrometer A1 for the analyzed episodes is shown in the illustration.]{\textbf{ The fit of the $\beta$ and $\alpha$ \textit{scaling} parameters from Equation (\ref{eqn:sempere-torres-1moment}) corresponding to the scaling method of a moment over R, W, $D_{m}$ and $R^{*}$. The case of disdrometer A1 for the analyzed episodes is shown in the illustration.} The values of parameters $a_{n}$ and $b_{n}$ are shown for n=1 up to 8 and for the scaling with respect to R (sky-blue), W (green), $D_{m}$ (red) and $R^{*}$ (gray). The process was carried out over the whole network; the result for disdrometer A1 is illustrated in the figure.}
\label{fig:Var1scalingBeta1}
\end{center}
\end{figure}

\begin{figure}[H]
\begin{center}
 
\includegraphics[width=1.02\textwidth]{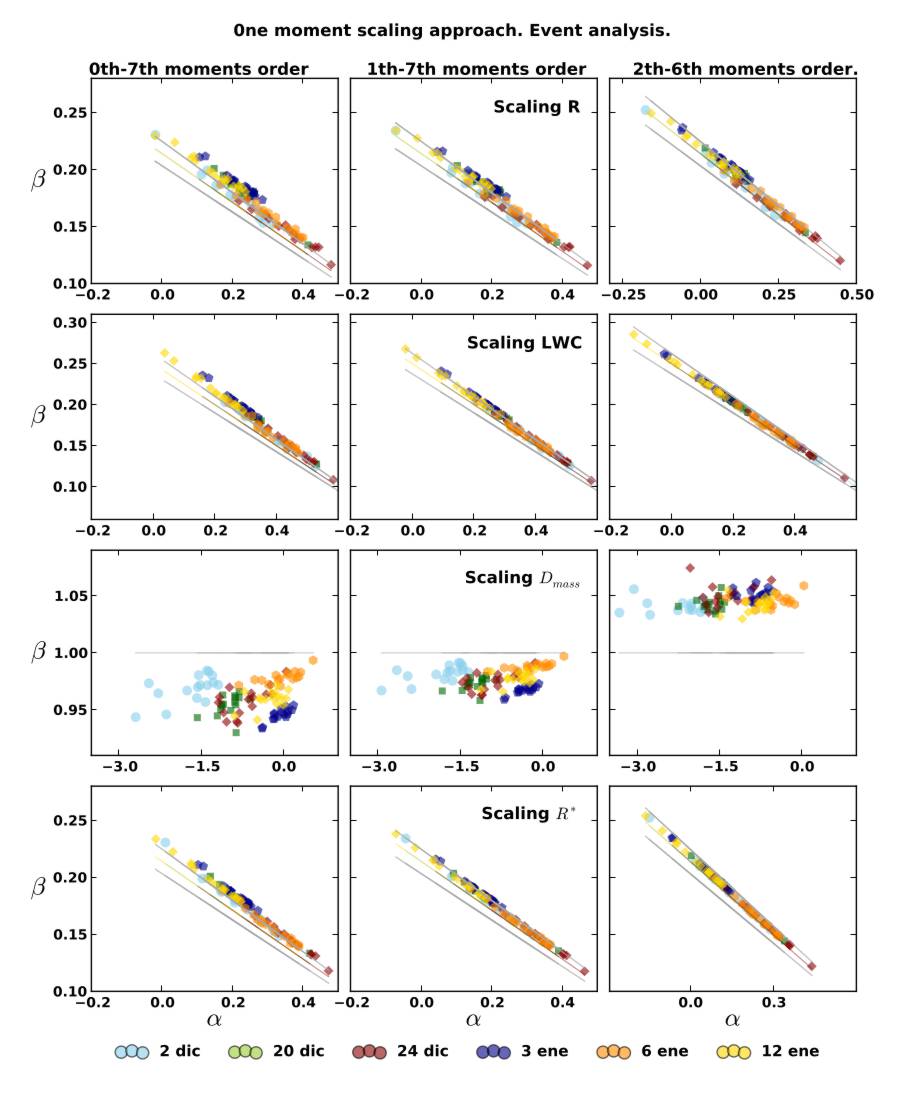}
\vspace{0.05cm}

   \caption[A representation of the $\alpha$ and $\beta$ estimates for each event throughout the disdrometer network.]{\textbf{ A representation of the $\alpha$ and $\beta$ estimates for each event throughout the disdrometer network.}. Each point represents modeling by means of the scaling of one reference parameter for each event and each disdrometer. Event on 2 December 2009 in blue. Event on 20 December 2009 in green. Event on 24 December 2009 in dark red. Event on 3 January 2010 with dark blue triangles. Event on 6 January 2010 with orange circles. Event on 12 January 2010 with yellow squares.  The results are similar to those obtained from the Z-R relationships, with respect to dispersion of each events. The gray lines represent an error interval of 5\% in the $\beta$ estimate assuming a correct $\alpha$ given the consistency relationship between both. The center lines represent the values of $\beta$ given for the consistency relationship for the $\alpha$ interval of each episode.}

\label{fig:Var1ScalingAlfa1}
\end{center}
\end{figure}
\vspace{1.0cm}

The information gleaned from the relationships between powers for all episodes is shown in Figure (\ref{fig:Var1scalingBeta1}) for disdrometer A1, in which the $\alpha$ and $\beta$ parameters are estimated (in this case, using moments 0 to 7). This process is carried out for each disdrometer and episode for each scaling with respect to R, W and $D_{m}$ and for n from 2 to 6. The results appear in Figure (\ref{fig:Var1ScalingAlfa1}). In this figure, the consistency relationships imply that for R and W the values are distributed in a line (in the case of T, this fact is approximate, given that R is only approximately a moment of fixed order for the distribution\footnote{The point spread along the consistency line is greatest in those episodes that least obey a $v(D)=\gamma D^{\delta}$ relationship with the $\delta$ parameter constant in space and time. This fact already appears in an original study by \citep{semperetorres_porra_ea_1994}, where the arbitrary weighting functions with an expression of the form (\
ref{eqn:phirefscaling}) break the self-consistency relationship. This finding explains why the snow episode yields similar consistency results for W but not for R.}, while W is a constant by the moment of order 4), while in the case of $D_{m}$, the consistency implies that $\beta\simeq 1$, the relative error of which has been greater than 6\%.\\

The results are compatible with the data in the analysis of the Z-R relationships. There, the variation within an episode can be greater than the variation between two episodes, which is a matter that is also reflected in the values of $\beta$. If the $v(D)=\gamma D^{\delta}$ relationship is verified, it is possible to obtain estimates of the exponent $b_{R}$ of the Z-R relationships from the $\beta$ values in the scaling using R R\footnote{The Z-R relationship is written as $Z=C_{R}R^{\alpha+\beta(6+1)}$, while $\alpha=1-(4+\delta)\beta$. In the case of $\delta \simeq 0.67$, $b_{R}=1+2.33\beta$ is obtained.}. The consistency between the direct Z-R relationships and those derived from the $\beta$ values is maintained (and by introducing the values from Table (\ref{TableRelacionesVD}), closer values are obtained than by utilizing $\delta \simeq 0.67$).\\

It is possible to understand the correlations found between disdrometers, given the DSD modeling study. Thus, in the episode on 12 January 2010 in which the E2 disdrometer has a lower correlation in the time series than the remaining disdrometers, it exhibits several $(\alpha,\beta)$ values that, although obeying the consistency relationships, nonetheless stray from the cluster of values for this episode. In the same way, the F1 disdrometer has significantly different $(\alpha,\beta)$ values for the episode on 6 January 2010, with its correlation not being different in terms of the rainfall-intensity time series. However, the F1 disdrometer has a more stochastic $v(D)$ relationship than the rest of the disdrometers due to the lower number of drops encountered in addition to the greater number of small drops at higher velocities. The scaling method allows analysis of the specific disdrometers within the network, which are subjected to more intense local variations. 

\subsection{Microphysics and $\alpha$ and $\beta$ relationships}
\label{sec:microfisica_escalado_esp}

The obtained $\beta$ and $\alpha$ parameters have been associated with the physical processes that occur in the DSD formation \citep{GyuWonLee2004}, such as that introduced in Section \S\ref{sec:microESCALADOdsd}. The $\beta$ parameter is related to the scaling diameter, while the $\alpha$ parameter is related to the scaled concentration. Therefore, the processes in which the concentration remains constant and that register an increase in the characteristic diameter of the distribution imply a variation in $\beta$, holding $\alpha$ constant. If we had an equilibrium DSD among the processes of coalescence, rupture and evaporation, we would obtain $\alpha=1$ and $\beta=0$; the episodes from 3 and 6 January 2010 would be the closest to this microphysical situation. This result does not imply greater spatial homogeneity, as was found in \S\ref{sec:ZRrelations}.

\begin{figure}[H]
\begin{center}
   \includegraphics[width=1.00\textwidth]{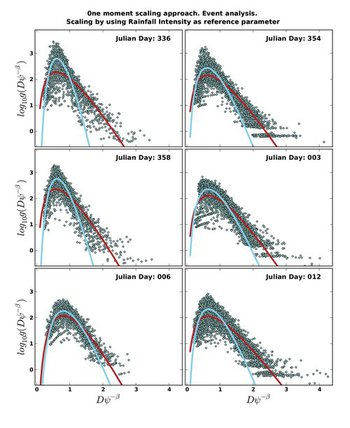}
   \caption[A representation of the N(D), scaled using the rainfall intensity R as a parameter in addition to modeling that utilizes the gamma distribution for the 6 episodes studied and the A1 disdrometer.]{\textbf{A representation of the N(D), scaled using the rainfall intensity R as a parameter in addition to modeling that utilizes the gamma distribution for the 6 episodes studied and the A1 disdrometer.} The symbols in gray constitute data obtained from the experiment for $g(D\,R^{-\beta})$. The red lines represent the two fitting methodologies, which are based on nonlinear Levenberg\textendash Marquardt fits; the blue lines are direct fits; and the solid lines have undergone a logarithmic transformation. This transformation permits both a better fit of the whole size spectrum and adherence to the consistency condition (\ref{eqn:consistenciaGammaK-1moment}).}
\label{fig:Var1ScalingFUNCIONgDISDROA1_rain}
\end{center}
\end{figure}

\begin{figure}[H] 
\begin{center}

\includegraphics[width=1.00\textwidth]{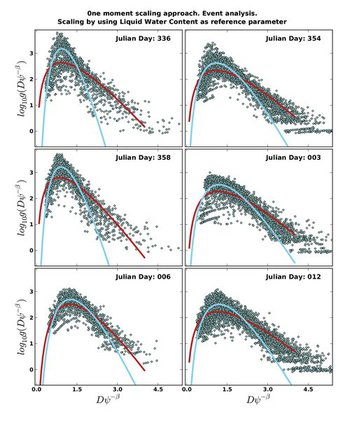}
   \caption[A representation of the N(D), scaled using the Liquid Water Content W as a parameter in addition to modeling that utilizes the gamma distribution for the 6 episodes studied and the A1 disdrometer.]{\textbf{A representation of the N(D), scaled using the Liquid Water Content W as a parameter in addition to modeling that utilizes the gamma distribution for the 6 episodes studied and the A1 disdrometer.} The symbols in gray constitute data obtained from the experiment for $g(D\,W^{-\beta})$. The red lines represent the two fitting methodologies, which are based on nonlinear Levenberg\textendash Marquardt fits; the blue lines are direct fits; and the solid lines have undergone a logarithmic transformation. This transformation permits both a better fit of the whole size spectrum and adherence to the consistency condition (\ref{eqn:consistenciaGammaK-1moment}).}
\label{fig:Var1ScalingFUNCIONgDISDROA1_lwc}
\end{center}
\end{figure}

\begin{figure}[H] 
\begin{center}
   \includegraphics[width=1.00\textwidth]{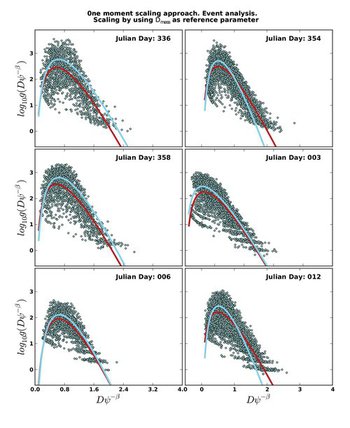}
   \caption[A representation of the N(D), scaled using the characteristic diameter $D_{mass}$ as a parameter in addition to modeling that utilizes the gamma distribution for the 6 episodes studied and the A1 disdrometer.]{\textbf{A representation of the N(D), scaled using the characteristic diameter $D_{mass}$ as a parameter in addition to modeling that utilizes the gamma distribution for the 6 episodes studied and the A1 disdrometer.} The symbols in gray constitute data obtained from the experiment for $g(D\,D^{-\beta}_{mass})$. The red lines represent the two fitting methodologies, which are based on nonlinear Levenberg\textendash Marquardt fits; the blue lines are direct fits; and the solid lines have undergone a logarithmic transformation. This transformation permits both a better fit of the whole size spectrum and adherence to the consistency condition (\ref{eqn:consistenciaGammaK-1moment}).}
\label{fig:Var1ScalingFUNCIONgDISDROA1_dmss}
\end{center}
\end{figure}

\begin{figure}[H]
\begin{center}
   \includegraphics[width=1.00\textwidth]{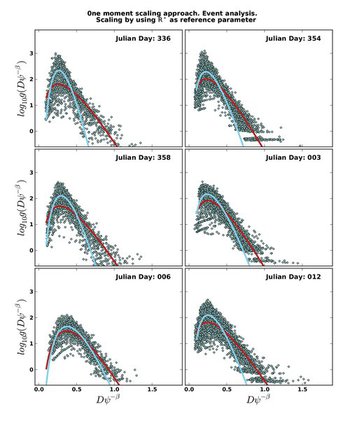}
   \caption[A representation of the N(D), scaled using the moment of order 3.67 ($R^{*}$) as a parameter in addition to modeling that utilizes the gamma distribution for the 6 episodes studied and the A1 disdrometer.]{\textbf{A representation of the N(D), scaled using the the moment of order 3.67 ($R^{*}$) as a parameter in addition to modeling that utilizes the gamma distribution for the 6 episodes studied and the A1 disdrometer.} The symbols in gray constitute data obtained from the experiment for $g(D\,(R^{*})^{-\beta})$. The red lines represent the two fitting methodologies, which are based on nonlinear Levenberg\textendash Marquardt fits; the blue lines are direct fits; and the solid lines have undergone a logarithmic transformation. This transformation permits both a better fit of the whole size spectrum and adherence to the consistency condition (\ref{eqn:consistenciaGammaK-1moment}).}
\label{fig:Var1ScalingFUNCIONgDISDROA1_R367}
\end{center}
\end{figure}

\begin{table}[H]
\caption[The gamma distribution fitting parameter $\kappa$ for DSD scaled with R. See Eq. (\ref{eqn:gammaScaling1moment}). Episode: 12 January 2010]{\textbf{ The gamma distribution fitting parameter $\kappa$ for DSD scaled with precipitation intensity. See Eq. (\ref{eqn:gammaScaling1moment}). Episode: 12 January 2010.} The values are compared via the fit after logarithmic transformation (columns 2, 3 and 4), with $\kappa$ values for $\lambda$ and $\mu$ of the fit calculated via the consistency relationship, using both the theoretical $v(D)$, as well as the experimental $v(D)$ given by Table (\ref{tabla:preprocesadoVD}). Columns 7 and 8 show the estimated $\mu$ and $\lambda$ results for the fit without a logarithmic transformation, which does not correctly fit the tail in the distribution; furthermore, this finding implies values of $\kappa$ higher than those given in this Table by one or two orders of magnitude.}
\vspace{0.70cm}
\begin{center}
\ra{1.50}
\begin{tabular}{c>{\columncolor[gray]{0.95}}r>{\columncolor[gray]{0.95}}r>{\columncolor[gray]{0.95}}rrr>{\columncolor[gray]{0.95}}r>{\columncolor[gray]{0.95}}r}

\toprule
                    & \textbf{Fit}   &  \textbf{transf.}  & \textbf{logarithmic}   & $\kappa(\lambda,\mu)\,\,\,$      &  $\kappa(\lambda,\mu)\,\,\,$  & \textbf{Fit}  & \textbf{Direct fit.}  \\
 \cmidrule(r{.5em}){2-4}  \cmidrule{4-6}  \cmidrule(l{.5em}){7-8}
\textbf{Disdr.}             & $\mu\,\,\,$ &  $\lambda\,\,\,$  & $\kappa\,\,\,$ & $\,\,v=3.78D^{0.67}$  & $\,v(D)\,\,$ exp.  & $\mu\,\,\,$  &  $\lambda\,\,\,$  \\

\midrule
A1         &   3.066    &   4.807     &  9\,571.632   &  8\,893.309 &  8\,327.362   & 6.675      &      10.194    \\     
A2         &   2.452    &   3.990     &  3\,713.798   &  2\,952.655 &  2\,770.165   & 6.283      &       9.609    \\    
B1         &   2.569    &   4.120     &  4\,254.735   &  3\,508.171 &  3\,290.381   & 7.030      &      10.499    \\
B2         &   1.715    &   3.122     &  1\,432.698   &   859.627   &     808.552   & 4.595      &       6.854    \\    
C1         &   1.668    &   3.313     &  2\,211.427   &  1\,287.881 &  1\,209.738   & 4.016      &       6.404    \\
C2         &   2.592    &   4.136     &  4\,582.594   &  3\,567.716 &  3\,346.202   & 4.942      &       7.361    \\
D1         &   1.792    &   3.373     &  2\,017.043   &  1\,346.984 &  1\,265.328   & 5.720      &       9.067    \\
D2         &   1.892    &   3.291     &  1\,875.679   &  1\,081.293 &  1\,016.583   & 4.514      &       6.607    \\
E1         &   2.192    &   3.959     &  3\,571.580   &  3\,179.498 &  2\,981.057   & 5.915      &       9.234    \\
E2         &   1.224    &   2.686     &     877.568   &     473.834 &     446.246   & 2.520      &       4.768    \\
F1         &   2.033    &   3.926     &  4\,441.846   &  3\,230.928 &  3\,028.269   & 5.227      &       8.436    \\
F2         &   2.320    &   3.667     &  2\,633.309   &  1\,748.652 &  1\,642.685   & 5.600      &       7.938    \\
G1         &   2.614    &   4.196     &  5\,278.633   &  3\,918.598 &  3\,674.477   & 5.899      &       9.027    \\
G2         &   2.199    &   3.743     &  2\,847.763   &  2\,153.814 &  2\,021.707   & 5.066      &       7.857    \\
H1         &   2.209    &   3.758     &  3\,241.848   &  2\,201.886 &  2\,066.729   & 6.180      &       9.593    \\
H2         &   2.240    &   3.832     &  3\,356.330   &  2\,478.867 &  2\,326.027   & 6.225      &       9.705    \\

\bottomrule
\end{tabular}
\label{TablaKAPPA}
\end{center}
\end{table}%
\vspace{3.00cm}

\begin{figure}[H] 
\begin{center}
   \includegraphics[width=1.00\textwidth]{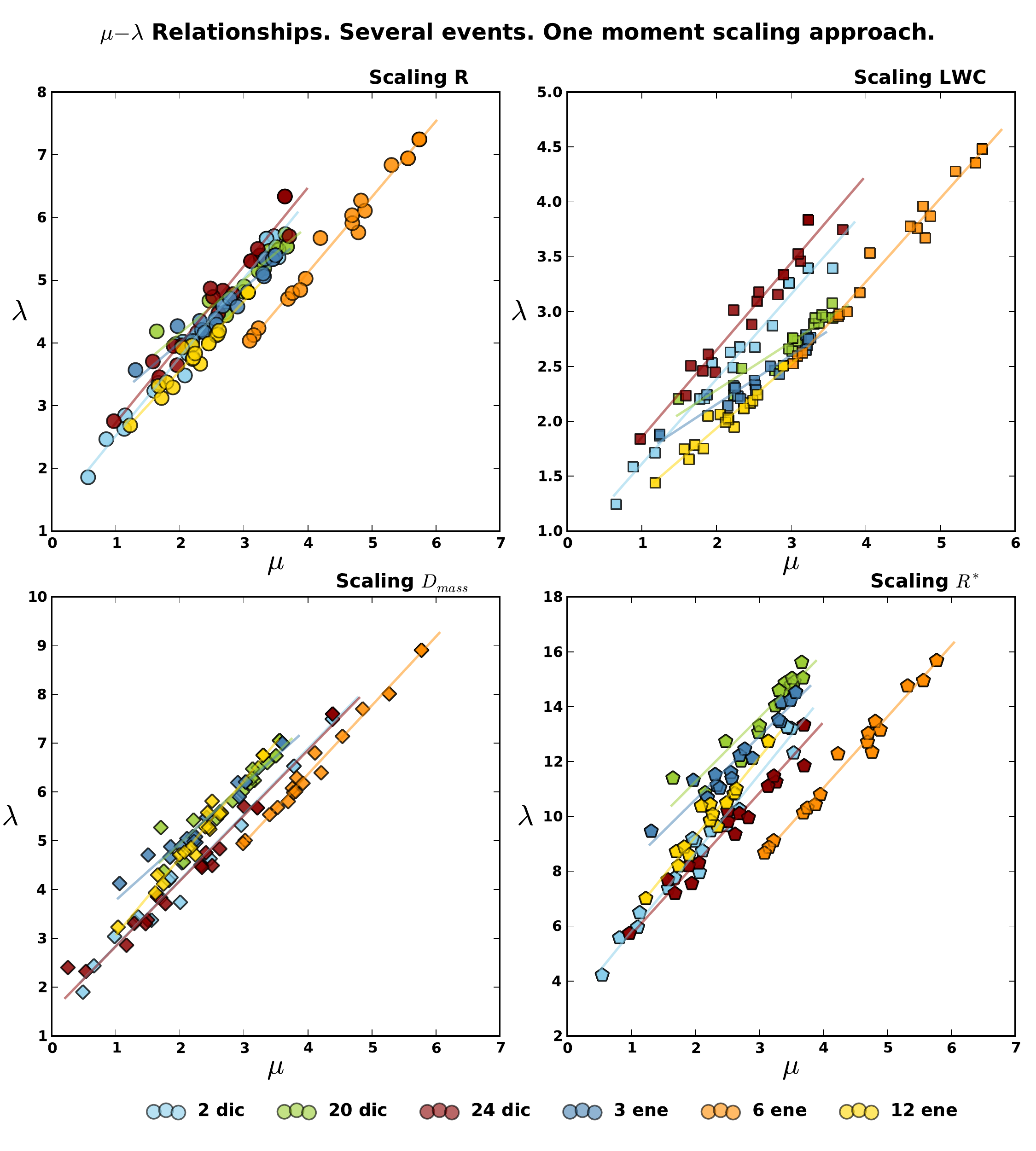}
\vspace{0.5cm}
   \caption[A representation of the $\mu$ and $\lambda$ estimates corresponding to the gamma distribution, given by Eq. (\ref{eqn:gammaScaling1moment}) under the R, W, $D_{mass}$ and $R^{*}$ scalings.]{\textbf{ A representation of the $\mu$ and $\lambda$ estimates corresponding to the gamma distribution, given by Eq. (\ref{eqn:gammaScaling1moment}) under the R, W, $D_{mass}$ and $R^{*}$ scalings.} Each point represents the modeling for each disdrometer in the network. The values come from nonlinear fits, such as those given in Figure (\ref{fig:Var1ScalingFUNCIONgDISDROA1_rain}).}
\label{fig:Var1Scalingmulambda}
\end{center}
\end{figure}
\vspace{1.0cm}

\begin{table}[H]
\caption[The $\mu$ and $\lambda$ parameters from the gamma-distribution fit for the scaled DSD. See Eq. (\ref{eqn:gammaScaling1moment}). Six episodes of liquid precipitation.]{\textbf{The $\mathbf{\mu}$ and $\mathbf{\lambda}$ parameters from the gamma-distribution fit for the scaled DSD. See Eq. (\ref{eqn:gammaScaling1moment}). Six episodes of liquid precipitation.} The estimated $\mu$ and $\lambda$ parameters are shown for a fit after a logarithmic transformation; the indicated values are average values, obtained throughout the network of instruments. The estimated standard deviation of the values obtained from 16 disdrometers is shown.}
\vspace{0.70cm}
\begin{center}
\ra{1.20}
\begin{tabular}{c>{\columncolor[gray]{0.95}}crr>{\columncolor[gray]{0.95}}r>{\columncolor[gray]{0.95}}r}

\toprule
\textbf{Scaling Parameter}  & Julian Day         & $\mu\,\,\,$ &  $\Delta \mu\,\,\,$& $\lambda\,\,\,$  &  $\Delta \lambda\,\,\,$  \\

\midrule
Rainfall Intensity & & & & &\\
\midrule
&   336   & 2.0630    & 0.8768    & 3.8319    & 1.0849 \\
&   354   & 3.0884    & 0.5569    & 5.0763    & 0.5033 \\
&   358   & 2.5097    & 0.7544    & 4.5871    & 0.9229 \\
&   003   & 2.6809    & 0.5876    & 4.5867    & 0.5148 \\
&   006   & 4.3424    & 0.8182    & 5.5345    & 0.9953 \\
&   012   & 2.1736    & 0.4359    & 3.7449    & 0.4885 \\
\midrule
Liquid Water Content & & & & &\\
\midrule
&   336   & 2.0919    & 0.7975    & 2.4426    & 0.6116 \\
&   354   & 3.0064    & 0.5436    & 2.7249    & 0.2537 \\
&   358   & 2.4116    & 0.7014    & 2.9576    & 0.5493 \\
&   003   & 2.5839    & 0.5536    & 2.4108    & 0.2519 \\
&   006   & 4.2617    & 0.8131    & 3.4704    & 0.6259 \\
&   012   & 2.0795    & 0.4195    & 1.9808    & 0.2503 \\
\midrule
$D_{mass}$ & & & & &\\
\midrule
&   336   & 1.9799    & 1.0153    & 4.1616    & 1.3697 \\ 
&   354   & 2.8327    & 0.5773    & 5.9502    & 0.7307 \\
&   358   & 2.0279    & 1.0100    & 4.1964    & 1.3260 \\
&   003   & 2.3026    & 0.6030    & 5.3010    & 0.7000 \\
&   006   & 4.0303    & 0.7362    & 6.4060    & 1.0446 \\
&   012   & 2.2109    & 0.5113    & 4.9811    & 0.8176\\ 
\midrule
$R^{*}$ & & & & &\\
\midrule
&   336   & 2.0562    & 0.8936    & 8.7844    & 2.5473 \\
&   354   & 3.0983    & 0.5602   & 13.8081    & 1.3826 \\
&   358   & 2.5207    & 0.7616    & 9.5692    & 1.9449 \\
&   003   & 2.6905    & 0.5983   & 12.1769    & 1.4011 \\
&   006   & 4.3426    & 0.8232   & 11.9166    & 2.1570 \\
&   012   & 2.2083    & 0.4452    & 9.8435    & 1.3067 \\

\bottomrule
\end{tabular}
\label{Tablamulb}
\end{center}
\end{table}%
\vspace{3.00cm}

\vspace{0.5cm}
\subsection{Modeling under a gamma distribution}
\vspace{0.5cm}

The last variability analysis studied was modeling the DSD scaled by means of a gamma distribution. For that purpose, the three parameters contained in (\ref{eqn:gammaScaling1moment}) were estimated via the nonlinear Levenberg\textendash Marquart fitting methods, in an analogous methodology as that utilized in \citep{NormalizadaTestud2001} and \citep{Chapon200852}. Additionally, following the procedure for comparing different techniques to interpret the relevance of the statistical methods, the fit was carried out by transforming the data and the expression logarithmically.\\

This scaling methodology was applied to the study by episode. The results appear in Figures (\ref{fig:Var1ScalingFUNCIONgDISDROA1_rain}). In these figures, the scaled N(D) are represented along with the estimates for the gamma distribution. Both in the case of scaling via W and in the case of scaling via R, it can be observed that the log scale fits permit adequate representation of the tail in the distribution. If this transformation is not used, then the tail, having a comparatively lower concentration of points and lower weight in the residuals to be minimized, is not conveniently modeled by the Levenberg-Marquart method without a scale transformation. This fact is implicit in the selection of the method utilized by \citep{NormalizadaTestud2001}, even though in that case, the method used was that introduced in \S\ref{sec:metodoNormalizada}.\\

The correct fit of the tail in the distribution is not a sufficient reason to prefer a specific method, given that additionally the consistency relationship given by Equation (\ref{eqn:consistenciaGammaK-1moment}) should be verified. This verification is shown in Table (\ref{TablaKAPPA}) for one episode, with the results found to be analogous for the remaining episodes.\\

In the face of variability problems, it is highly relevant to analyze the differences in the $\lambda$ and $\mu$ values, both throughout the network and by episode. From Figure (\ref{fig:Var1Scalingmulambda}), three main conclusions can be drawn: 

\begin{enumerate}
\item The dispersion in one episode with respect to the dispersion among episodes is similar to that which we found for the Z-R relationship. 
\item The errors in $\lambda$ and $\mu$ can be considered relatively small with respect to the variability that we find within each episode (from the point of view of obeying the consistency relationship and obtaining an adequate fit to the distribution tail).
\item Effective relationships between $\lambda$ and $\mu$ appear. In the case of scaling based on the precipitation intensity, all episodes appear to obey the same relationship, except the case of 6 January, which exhibits a light precipitation intensity and a total accumulation of liquid precipitation of approximately 2 mm/h. Therefore, it appears that the $\mu-\lambda$ relationship given in the figure is generally satisfied for stratiform precipitation episodes with sufficient precipitation intensity. In the case of scaling with respect to $D_{mass}$, the results also suggest a possible relationship common to all episodes. In the remaining cases, a linear relationship can be asserted by episode with an approximately common slope\footnote{There are two episodes, 20 December and 3 January, exhibiting a different slope. A more significant difference can be observed in the scaling with respect to LWC. It has not been possible for us to discern whether this implies the existence of two different classes of 
episodes (classifiable via some criterion external to the methodology) or that the presence of a $(\mu,\lambda)$ pair for one disdrometer is sufficiently different from the rest, making it difficult in some episodes to obtain a similar slope as that for the rest of the episodes. It has been considered that generally, the slopes are relatively similar such that we can consider all episodes to share an increase in $\lambda$ based on the analogous increase in $\mu$. A future investigation to clarify this issue could employ a broader study with more precipitation episodes that would include clear convective cases.}.\\
\end{enumerate}

Points (1) and (2) are interesting from the perspective of understanding the variability in the DSD modeling by episode, and they prove to be compatible with the results of the Z-R relationships and the $\alpha$ and $\beta$ parameters. The episodic variations observed in the fitting parameters appear to be due to natural variations in the DSD at the kilometer scale.\\

The results for (3) point to the possible existence of a relationship of physical origin between the parameters $\mu$ and $\lambda$ in the sense that the natural DSD variations are reflected in the dispersion values in Figure (\ ref {fig: Var1Scalingmulambda}), first row. However, this dispersion is subject to the constraint that a relationship between $\mu$ and $\lambda$ exists. This fact has been discussed in the literature from various points of view, including its origin in sampling problems and a physical explanation of such variation (see the comments at the end of Section \S\ref{sec:GAMMAdistrib}). The results might indicate that the natural variability of the DSD is subject to changes synchronized between $\mu$ and $\lambda$; even so, without a complementary analysis at other scales, it is not possible to conclude whether a portion of these relationships is due to uncertainties arising from an insufficient or biased sampling of the DSD. In any case, Table (\ref{Tablamulb}) indicates for each of the 
episodes analyzed the average values of $\mu$ and $\lambda$ and the expected uncertainty based on the standard deviation of the set of values obtained from the network. These values can be interpreted as the expected errors in $\mu$ and $\lambda$ at the spatial scale analyzed when we estimate $\mu$ and $\lambda$ using a single instrument. 

\vspace{0.95cm}
\section{Conclusions}

\vspace{0.5cm}

The main results obtained, related with the small-scale spatial variability of the DSD, and discussed in this chapter are:\\
\begin{itemize}
\item The relationships $v(D)$ for the different events are similar to previous studies with only one instrument. The estimation of the coefficients of the power-law relation $v(D)=\gamma D^{\delta}$ shows consistency along the network in terms of the values of the retrieved parameters. However without any kind of filtering technique the correlations are low, mainly due to few drops with large differences in vertical velocities. If we introduce a filtering pre-processing technique we will obtain higher correlations. Also the value of $\delta$ is closer to the traditional law (\ref{eqn:AtlasVDequation}), while the coefficient $\gamma$ is not very sensible to the filtering pre-processing and shows larger values. This means larger vertical velocities for smaller drops. This last conclusion was obtained with one disdrometer previously but here we have shown that is not an spurious result of one instrument or a particular experiment because the differences with the equation (\ref{eqn:AtlasVDequation}) are the 
same for the full network.

\item Studying time series of rainfall we have seen that the typical errors have a multiplicative bias that grows with the rainfall intensity. This is obtained calculating the standard deviation in the network using the differences respect to the mean rainfall (calculated minute by minute values between all instruments). But this simple characterization of the error is only enough in the case of uniform rainfall in space and time. In the real cases we should make deeper studies of the correlations between instruments.

\item The correlogram for the integral rainfall parameters R and Z shows a decreasing value of the Pearson correlation coefficient (calculated between time series across the network) when the distance between instruments grows.

\item The previous result is not dependent of the method used to calculate the correlogram, in the sense that, if we compare different methodologies we will always obtain a decreasing behavior. The decorrelation distance may slightly depend on the methodology, and is lower for the BMLN method.

\item We have reported differences in the estimation of Z-R relationships along the network. The conclusion is that the variation inside an event can be higher than the variation between events. This fact has been shown using several methods to calculate the Z-R relationship. In particular we have compared the traditional method with the SIFT method. Also we have reported that the values of the Z-R relationships would change if we introduced thresholds in the minimum rainfall allowed in any disdrometer. At the same time the fact of large variability inside an event remains true.

\item The modeling process using the one moment scaling procedure showa us a similar conclusion to the Z-R relationship study. The variability, in terms of the coefficients retrieved by the one moment scaling theory, can be larger inside an event that between events.

\item Also we have detected possible relationships between the  $\mu$ and $\lambda$ parameters of the scaled DSD when they are modeled by a gamma DSD. 

\item But this last result demands more insight to check if, after a scaling process, the sampling issues and their relevance for artificial $\mu-\lambda$ relationships are still relevant.

\end{itemize}
\vspace{0.5cm}

So the main conclusion of the chapter is that the variability of DSD at the kilometer scale is relevant and demands a physical and statistical characterization. Our research shows that this variability is relevant for the Z-R relationships, for the ground validation campaigns of the orbital based radars and for the modeling methods of the DSD.

\renewcommand\chapterillustration{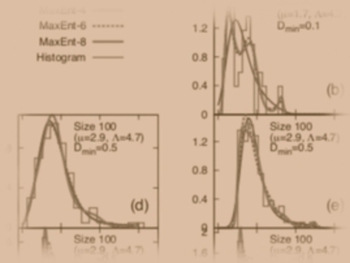}

\chapter{New DSD parametrization by using MaxEnt}

\selectlanguage{english}
\label{sec:chapMAXENT}

\vspace{0.5cm}

In the previous chapter the hypothesis of spatial variability of precipitation in the range
of a pixel radar was tested. It was shown that the spatial variability is enough to decrease the correlation
in the rainfall field in more than 25\% at a distance of three kilometers. This variability is relevant for Z-R relationship and is related to natural variations of the DSD.\\

As a consequence the modelization of DSD should take into account these facts. Also, because one of the main objectives of modelization methods is to prune the non-natural variability due to sampling issues and to differentiate it from physical variability (spatial and temporal).\\

In this situation the second objective of this thesis is to develop a DSD modelling to study the DSD faithfully to represent the actual variability that the DSD has\footnote{Este capítulo es una elaboración de los contenidos publicados en la revista \textit{Entropy}, véase \citep{checa_tapiador_2011_aa}.}. Also, as it has been explained in the introduction (see \textsection\ref{sec:MicroANDdsd}) the equilibrium DSD between the microphysical processes of coalescence, breakup and evaporation doesn't have an exponential form neither a functional form exactly described by gamma or log-normal distributions. But also the experimental DSDs doesn't follow those functional prescribed forms.\\

It is proposed a new method that allows:
 
\begin{itemize}
 \item Model multimodal distributions and oscillating DSDs.
 \item To incorporate experimental information in a progressive scheme to the modelling process.
 \item To build a link between the modelling process and the underlying microphysical process.
\end{itemize} 

The first two points are developed in this chapter, the last one is part of the future applications of the developed methodology, and it will be commented in the appropriate part of the thesis.\\

In the following sections the context which motivate the application of Maximum Entropy, the description the logic of the method, and a comparison with other modelling methodologies used by other authors for the same problem are explained.

\section{Context}
The actual functional form of the RDSD is the result of
microphysical processes transforming the water vapour into
hydrometeors such as rain. Thus, multiple factors determine the
RDSD, including small-scale turbulence, the interplay between
several variables such as relative humidity, and differences in the
cloud systems in which the RDSD emerges. Therein, one of the main
characteristics and challenges in RDSD research is the high spatial
and temporal variability in the spectrum of drops sizes. This is a
major drive to replace the original model based on exponential
distribution \citep{MarshallPalmer1948}, $N(D)=N_{0}e^{-\Lambda D}$
with a log-normal distribution \citep{FeingoldLevin1986,
FeingoldLevin1987} or the most frequently employed gamma equation,
as follows,
\vspace{2mm}
\begin{equation}
N(D)=N_{0}D^{\mu}e^{-\Lambda D} \label{eqn:gammaDSD}
\end{equation}
\vspace{2mm}
The gamma distribution has three free parameters: $N_{0}$ indicates
raindrop concentration (with typical units [$mm^{-1-\mu}m^{-3}$]), $\mu$ is a shape
parameter and $\Lambda$ is a scale parameter [$mm^{-1}$], see \S\ref{sec:GAMMAdistrib}. In
principle, three free parameters ensure high representability of
the experimental measurements obtained from different sources
\citep{ulbrich_1983_aa}, although experimental measurements have shown
that it is not enough to represent all observed RDSD accurately \citep{Multimodal1,Multimodal2}: if RDSDs are assumed to
follow a gamma distribution, it may be necessary to know the model
error \citep{cao_zhang_2009_aa}.\\

Alternative to these models, some studies do not assume a fixed
functional form. Rather, the RDSD is obtained as a consequence of
the relations between integral parameters of N(D)
\citep{SempereTorres1998,GyuWonLee2004}. This allows to address at
least a part of the high variability of the N(D) through a scaling
technique \citep{NormalizadaTestud2001}. In all these methods, the integral
parameters of the sample are key experimental parameters. That
includes the radar reflectivity, the rainfall rate, the liquid water
content or the total number concentration. On such situation, the
method of the moments is widely used, even though the method is well
known to be biased \citep{smith_kliche_2005_aa}.\\

Another classical
method that could outperform the method of moments includes the
maximum likelihood estimation (MLE), which directly uses the values
from the sample and the hypothetical functional form to estimate the
free parameters. However, typical disdrometric measurements cannot
obtain values across all spectra of sizes because the noise is too
high for the smallest drops; then as showed in \citep{kliche_smith_etal_2008_aa}
truncating the low end of the RDSD deteriorates the performance of
MLE more that it affects to moments method. As a result the typical
MLE could be even more biased than the method of moments for typical
experimental measurements \citep{cao_zhang_2009_aa}.\\

Other possibilities for modelling can be derived by drawing on
\emph{Information Theory} whereby the physical properties of the RDSD
simply constrain the possible functional form of RDSD. As such, it
is reasonable to use the maximum entropy method (MaxEnt) to model
the RDSD. Following the seminal work of Jaynes \citep{Jaynes1957a}
the MaxEnt method should have theoretical advantages over other
methods. Namely, it retrieves the least biased probability density
function (pdf) for a given amount of information, making also
possible to update a prior model if additional empirical information
becomes available. A major additional bonus is that thought MaxEnt,
physical interpretations of the RDSD can be extracted. The
application of MaxEnt to atmospheric sciences presents an
opportunity to develop new conceptual models for physical process in
the atmosphere with a stochastic component, such as droplet or
aerosol size distributions.\\

To the best of our knowledge, this approach has not yet been applied
to RDSD modelling.  Indeed, the MaxEnt has already been  fruitfully
applied to model drop-size distribution in sprays
\citep{Dumouchel2009, Babinsky2002}, or in the remote sensing of
precipitation \citep{Tapiador2007,Tapiador2004,Tapiador2002}. In the
field of cloud microphysics, some models have been developed using
Jaynes entropy functional aiming to explore the idea that total
water content available during the microphysical process of
coalescence and break-up, together with the total concentration of
drops may be the main constrains on the final cloud drop size
distribution \citep{LiuYangang1995, LiuYangang1998,LiuYangang2002}.\\

In the next section is introduced an inductive inference method for the DSD relying in the maximum entropy principle. This method obtains the less biased distribution for an experimental set of data that is introduced in the method in the form of constrains on the entropy. Then this method relies in the existence of a functional of the distribution function (in our case N(D)), that it is called $\mathcal{S}[N(D)]$ or S(N). Some important properties of $\mathcal{S}[N(D)]$ together with the sense of the application of this method are described:

\begin{itemize}
\item There is a huge amount of possible drop size distributions fixed, for example, the rain intensity, o any other physical condition  $\Phi$ that it is supposed relevant for the characterisation of a DSD, and it is not possible choose an fixed functional form. However, it would be convenient to have a measure which chooses between all possible distributions functions, the distribution functions that fulfill a given condition $\Phi_{1}$ those that also fulfill $\Phi_{2}$. That is, if a given DSD $N_{A}(D)$, that fulfill  $\Phi_{1}$ and $\Phi_{2}$, is chosen because it over $N_{B}(D)$, that only fulfills $\Phi_{1}$, the consequence is that $\mathcal{S}[N_{A}(D)]>\mathcal{S}[N_{B}(D)]$. 
\item Also the method should carry on the transitive property: if $N_{A}(D)$ is chosen over $N_{B}(D)$ and $N_{B}(D)$ over $N_{C}(D)$, then $\mathcal{S}[N_{A}(D)]>\mathcal{S}[N_{C}(D)]$.
\item If a condition is focused or localized on a subset of values of D, the maximum entropy method shouldn't change the DSD values in the region in which the condition is not relevant. In other words, if a constrain is important for a part of the spectrum of drops sizes, the inclusion of it in the maximum entropy method should not interfere with the conditions that only constrains the DSD in other part of the spectrum of sizes.
\end{itemize}

These general ideas has been formalized in axioms, from which it is possible to build \citep{CatichaEntropyFluctuations2001}  the functional form of the $\mathcal{S}[N(D)]$. However while the full logic funding of the inference process could have some theoretical difficulties \citep{Uffink1995223} , the methodology of the application, the relation with several fields of physics and statistics, together with the capacity of developing, physical interpretations have been widely completed. In the next section the application method of the maximum entropy principle to a DSD is explained.

\section{Methodology of the maximum entropy method}

Often, only some of the properties of a probability density function
are known, and indeed, not only the set of free parameters but even
the functional distribution shape itself may be unknown. In his
seminal work, Jaynes proposed the maximum entropy (MaxEnt) principle
as a method of inference to solve indeterminate problems with
several origins using the concept of the relative entropy function
as defined in information theory \citep{Shannon1948}:
\vspace{2mm}
\begin{equation}
S_{I}[N]=-\int N(D)log \frac{N(D)}{m(D)} dD \label{eqn:entropy}
\end{equation}

The goal of this inference process \citep{Jaynes1957a} is to update a
prior probability distribution, m(D) to a posterior distribution,
N(D), when new information about the probability density function
becomes available. On this situation the MaxEnt provides a
systematic and objective way to construct a distribution based on
information given in the form of constraints on the family of
possible distributions.\\

These constraints are defined as follows:
\vspace{2mm}
\begin{equation}
\Phi_{i}=\int N(D)\phi_{i}(D)dD\qquad i=0,...,L
\label{eqn:constrains}
\end{equation}
where the functions $\phi_{i}(D)$ are selected according to the
specific system being analysed \citep{POME1986} or the available
empirical information, and the values of $\Phi_{i}$ are supposed to
be known.\\

The formalism of MaxEnt relies on the maximisation of Equation
(\ref{eqn:entropy}) subject to the constraints given as
(\ref{eqn:constrains}), to yield the least biased probability
distribution under these constraints. As the result of the
maximisation, we formally obtain the following equation:
\vspace{2mm}
\begin{equation}
\hat{N}(D)=\frac{m(D)}{\mathbb{Z}}exp\left[-\sum_{i=0}\lambda_{i}\phi_{i}(D)\right],
\qquad \mathbb{Z}(\lambda_{1},...,\lambda_{k})=\int
m(D)e^{-\sum_{i=0}\lambda_{i}\phi_{i}(D)} \label{eqn:pdf}
\end{equation}
where the values $\lambda_{0},...,\lambda_{L}$ represent the values
of the free parameters that cause to N(D) to satisfy the Equation
(\ref{eqn:constrains}) reporting with a maximum value of the
entropy.\\

As a consequence, the method, with appropriately selected choices
for the functions $\phi_{i}(D)$ can be applied to estimate, for
example, the free parameters of a gamma distribution or a log-normal
distribution. In the case of the gamma distribution, the functions
$\phi_{i}$ of Equation (\ref{eqn:constrains}) that define the constrains
are, $\phi_{0}=1$, $\phi_{1}(D)=lnD$ and $\phi_{2}(D)=D$, and the
results for $\lambda_{i}$ are algebraically dependent on the values
of the parameters in (\ref{eqn:gamma}). Moreover, given an estimate
for RDSD, it is possible to evaluate how the functional form of the
distribution would change if new empirical information in the form
of $\Phi_{i}$ became available, simply by substituting the given
previous estimate as m(D) in the formalism.\\

In our case, m(D)=1 as there is no information initially available.
The constraints are related to the integral parameters of rainfall
\citep{ulbrich_1983_aa}. The general solution introduced in the
constraints (\ref{eqn:constrains}) provides a non-linear system of
equations that must be solved numerically to obtain the values of
$\lambda_{i}$, and the pdf (\ref{eqn:pdf}). The
mathematical details are explained in next section.\\

\section{Numerical method for maximizing the Entropy Functional}
\label{sec:metodoMaxEnt}
The analytical maximisation of Equation (\ref{eqn:entropy}) is only
possible for selected sets of constrains \citep{POME1986}, and a
computational framework is needed. Here is presented the application
of the classical iterative method of Newton-Raphson
\citep{Djafari1991} for the numerical solution of the
moment-constrained maximum entropy problem.

\subsection{Non-linear systems of equations}
The method of Lagrange multipliers provides a strategy for finding
the maxima of a function, such as Equation (\ref{eqn:entropy})
subject to a set of constraints, such as Equation
(\ref{eqn:constrains}). In this method a new function
$\mathcal{S}^{*}[N]$ is defined as follows:
\vspace{2mm}
\begin{equation}
\mathcal{S}^{*}[N]=-\int_{a}^{b} \left\lbrace
N(D)ln(N(D))+\sum_{i=1}^{L}\lambda_{i} \phi_{i}(D)N(D)\right\rbrace
dD
\end{equation}
which is the function to be maximised; then the formal solution
obtained for the probability density function is given by Equation
(\ref{eqn:pdf}). With this solution the constraints are now given
by:
\vspace{2mm}
\begin{equation}
\mathcal{F}_{k}(\lambda_{0},...,\lambda_{L})=\int
\phi_{k}(D)\frac{m(D)}{\mathbb{Z}}exp\left[-\sum_{i=0}\lambda_{i}\phi_{i}(D)\right]=\Phi_{k},\qquad
k=1,...,L \label{eqn:systemEQU}
\end{equation}
This is a set of non-linear equations in the unknowns
$\lambda_{0},...,\lambda_{L}$ which must now be solved.

\subsection{Numerical Solution by Newton-Raphson Method}
The numerical method consists of calculating the linear
approximations of $\mathcal{F}_{k}$ around trial values of
$\lambda_{0},...,\lambda_{L}$, and solving the resulting linear
system iteratively. We define:
\vspace{2mm}
\begin{equation}
\boldsymbol\Phi=(\Phi_{0},...,\Phi_{L})
\end{equation}
and
\begin{equation}
\boldsymbol\lambda=(\lambda_{0},\lambda_{1},...,\lambda_{L})
\end{equation}
and the problem Equation (\ref{eqn:systemEQU}) is then given by:
\vspace{2mm}
\begin{equation}
\mathbf{F}(\boldsymbol\lambda)-\boldsymbol\Phi=0
\label{eqn:nonlinearsystem}
\end{equation}
The Jacobian of the vector function $\mathbf{F}(\boldsymbol\lambda)$
is given by:
\vspace{2mm}
\begin{equation}
(\mathbf{J}_{F})_{i}^{j}=\frac{\partial \mathcal{F}_{j}}{\partial
\lambda_{i}}
\end{equation}
Then, given a initial trial,
$\boldsymbol\lambda^{0}=(\lambda_{0}^{0},\lambda_{1}^{0},...,\lambda_{L}^{0})$
it is possible to solve the Equation (\ref{eqn:nonlinearsystem}) by
the iterative method:
\vspace{2mm}
\begin{eqnarray}
\mathbf{J}_{F}(\boldsymbol\lambda^{(k)})\Delta\boldsymbol\lambda^{(k)}=-\mathbf{F}(\boldsymbol\lambda^{(k)}) \nonumber\\
\boldsymbol\lambda^{(k+1)}=\boldsymbol\lambda^{(k)}+\Delta\boldsymbol\lambda^{(k)}
\label{eqn:metodoNewton}
\end{eqnarray}
This system is solved for $\Delta\boldsymbol\lambda^{(k)}$ from
which we drive
$\boldsymbol\lambda^{(k+1)}=\boldsymbol\lambda^{(k)}+\Delta\boldsymbol\lambda^{(k)}$
, which becomes our new initial vector $\boldsymbol\lambda^{(k)}$
and the iterations continue until $\Delta\boldsymbol\lambda^{(k)}$
is sufficiently small, {\em i.e.},
$||\boldsymbol\lambda^{(k+1)}-\boldsymbol\lambda^{(k)}||<<\delta$,
where $\delta$ is the tolerance parameter of convergence. The system
is considered solved when the difference between k-th and (k+1)-th
steps in the iterative process is less than $\delta=10^{-9}$, and a
fixed limitation on the maximum number of iterations was also added
to the solution algorithm.\\

Given the values of $\boldsymbol\Phi$, it can be a challenge to
choose a convenient value for $\boldsymbol\lambda^{0}$ to ascertain
that the iterative process converges to the fixed point representing
the maximum of the entropy functional. To avoid spurious cyclic
points, small perturbations $\delta \lambda_{i}$  were cyclically
added to the value of $\lambda_{i}$ if the iterative process
oscillated periodically around a fixed value of $\lambda_{i}$.\\

The matrix equation in (\ref{eqn:metodoNewton}) is solved by the
method of Gauss. To update the $\boldsymbol\lambda^{(k+1)}$ values,
in each iteration, a scaling factor is introduced to stabilise the
algorithm. This scaling factor $\nu$ is selected based on the
convergence process and the number of constraints, in an interval
from $10^{-2}$ to $10^{-5}$, then the second equation in
(\ref{eqn:metodoNewton}) is written as
$\boldsymbol\lambda^{(k+1)}=\boldsymbol\lambda^{(k)}+\nu\Delta\boldsymbol\lambda^{(k)}$.

\section{Comparison with other methods}

\label{Methods} The main goal is to compare the models of RDSD
including the two most widely used methods, that is, the method of
moments and the maximum likelihood estimation, and the method of
maximum entropy principle under different sets of constrains.

\subsection{Method of moments}

Given a sample of a population of drops, we can estimate the value
of the moments, which are expected to be unbiased but skewed. The
question is how to estimate accurately the parameters of a
hypothetical distribution that describes the population using the
information provided by the moments of the sample.\\

In general, the modelling of the RDSD using the \emph{method of
moments} applied to a hypothetical distribution of m free parameters
requires the information on m moments $M_{i}$ to estimate the entire
set of parameters. This makes it possible to define different
methods of moments by choosing several subsets of m moments of a
given sample. For the gamma distribution, the free parameters that
must be calculated include $\{N_{0},\mu,\Lambda \}$, while the most
widely used subset of moments are $\{M_{2},M_{3},M_{4}\}$
\citep{Smith2003}, $\{M_{2},M_{4},M_{6}\}$ \citep{UlbrichAtlas1998}
and the frequently applied $\{M_{3},M_{4},M_{6}\}$
\citep{ulbrich_1983_aa, kozu_nakamura_1991_aa, TokayShort1996}. Applications
with seventh moment of beyond are generally not used. The method
$\{M_{2},M_{3},M_{4}\}$  is claimed to be the least biased
\citep{smith_kliche_2005_aa} while the $\{M_{3},M_{4},M_{6}\}$ method is more
widely used in the estimation of ZR relations. While all of these
methods are well known, there is no general agreement regarding
which method should be adopted, and it has been suggested that if
model errors are included the position of the least biased method
$\{M_{2},M_{3},M_{4}\}$ becomes less attractive because the overall
differences among all methods of moments are then non substantial
\citep{cao_zhang_2009_aa}.\\

Therefore, we used the method $\{M_{2},M_{3},M_{4}\}$ for the
synthetic data and the methods of moments $\{M_{2},M_{3},M_{4}\}$
and $\{M_{3},M_{4},M_{6}\}$ for the empirical data, in order to
ascertain the advantages or disadvantages of the MaxEnt. These
methods are hereafter called MM234 and MM346, respectively.

\subsection{Maximum likelihood estimation}

Much like the method of moments, the maximum likelihood (MLE) method
is used by statisticians to estimate the parameters of an assumed
parametric model. It is based on the existence of a likelihood
function that attempts to indicate how likely a particular
population is to produce an observed sample. To implement the MLE
method mathematically over a sample of size n, the MLE method
requires the minimization of the likelihood function given by,
\vspace{2mm}
\begin{equation}
\mathcal{L}(D_{i};\mu_{0},\Lambda_{0})=\prod_{i=1}^{n}
f(D_{i};\mu_{0},\Lambda_{0})
\end{equation}
for the two parameters $\mu_{0}$ and $\Lambda_{0}$ of the gamma
function, see Equation (\ref{eqn:gamma}), as applied in previous
studies \citep{kliche_smith_etal_2008_aa}.


\section{Application of different methods to syntetic and experimental data sets}

In the following sections the Maximum Entropy Method is applied to the modelization of the DSD and the results are explained.
This is done from two different points of view. In the first case the drops are artificially generated,
in the other the experimental data explained in the previous chapters are modelled.\\

Also, the generation processes to produce synthetic samples are showed, together with the different measure of performance  to compare
the results of different modelization methods.\\

The analysis systematically covers three different sets of
constraints where $\phi_{i}(D)=D^{i}$, where i can be an integer
value where 0 until $i_{max}$ that is set to 4, 6 and 8. These
models will be known as \emph{MaxEnt-4}, \emph{MaxEnt-6} and
\emph{MaxEnt-8}, respectively. The first configuration is designed
to reproduce the general properties of the histograms while the last
case may reproduce full detailed multimodal cases. MaxEnt-6 shows an
intermediate situation. It is designed to provide the empirical
integral rainfall parameters and as results it retrieves the
histograms with enough detail to represent typical multimodal cases.\\

The value of each MaxEnt model depends on the specific study carried
out. The MaxEnt-8 model is introduced in order to test and
demonstrate how the MaxEnt method is able to reproduce the details
of a given sample and introduce information progressively. However,
from the point of view of physical applications the use of integral
parameters beyond the reflectivity makes its use more restricted. On
the other hand, for the MaxEnt-8 model, the values of the
$\lambda_{i}$ may show histograms with larger standard deviations
because disdrometric measurements include larger sampling errors for
larger drops, as shown in the integral rainfall parameter ahead of
the reflectivity. The MaxEnt-4 is a reliable model together with the
least bias property provided by the methodology.\\

A microphysical description of a precipitation process is typically
based on the integral rainfall parameters from the total number of
drops to the reflectivity. As such, to differentiate between the
convective core, stratiform rain or the region ahead the convective
core,  the usual arguments are based on the mean diameter,
reflectivity, rainfall rate and total number of drops together. The
MaxEnt-6 model provides the least biased functional from these given
values.\\

\section{Data}

The data used to test our proposal has been obtained using two
complementary approaches. The first is a synthetic method, in which
is theoretically possible to distinguish the various sampling
problems and natural physical variations of the RDSD, and the second
approach involves experimental data, which also presents both
sources of the aforementioned potential variability. The goals with
each of these datasets are different. The synthetic data may allow
us to test the representability of a histogram with different
methods, by allowing us to change conditions in a controlled way;
thus, it demonstrates the capacities of MaxEnt before applying the
method to experimental measurements, which is our main goal.

\subsection{Synthetic Data}
\label{DataSynthetic} The prototypical method to generate artificial
raindrop size distributions assumes a functional form that serves as
the underlaying distribution function. It represents the population
of raindrops from which the samples of data are obtained. In our
case, to generate synthetic samples the selected distribution is the
normalised gamma distribution f(D):
\vspace{2mm}
\begin{equation}
f(D;\mu,\Lambda)=\frac{\Lambda^{\mu+1}}{\Gamma(\mu+1)}D^{\mu}e^{-\Lambda
D} \label{eqn:gamma}
\end{equation}
To provide a comparison between N(D) and f(D), is defined a
concentration of drops $N_{d}$, by 
$N(D)=N_{d}f(D)=N_{0}D^{\mu}e^{-\Lambda D}$ and
$N_{d}=N_{0}\Gamma(\mu+1)/\Lambda^{\mu+1}$. Other authors \citep{
kliche_smith_etal_2008_aa} apply $D_{m}=(\mu+4)/\Lambda$ as free parameter for f(D)
instead of $\Lambda$ but the method is equivalent.\\

For Equation (\ref{eqn:gammaDSD}), a large amount of previous
experimental studies have reported different estimates of the free
parameters, thus showing that they can cover a wide range of values.
Following \citep{TokayShort1996,brawn_upton_2008_aa} the selection of
typical values for the free parameters, $\mu$ and $\Lambda$,
represents events with different rainfall intensity categories as
shown in Table \ref{TableGAMMAparameters}.
\vskip9mm
\begin{table}[h!]
\ra{1.3} \caption[Propiedades de los escenarios utilizados para la generación de DSD sintéticas \citep{TokayShort1996,
brawn_upton_2008_aa}]{\textbf{Properties of rainfall \textit{scenarios} used for
synthetic generation of RDSD} \citep{TokayShort1996,
brawn_upton_2008_aa}.\vspace{0.4cm}}
\begin{center}
\begin{tabular}{lcclr}
\toprule
\textbf{Category} & $\mathbf{\mu}$ & $\mathbf{\Lambda[mm^{-1}]}$  & $\mathbf{D_{min}[mm]}$ & \textbf{Size (Number drops)}  \\

\midrule

Very Light & 1.7 & 4.7  & 0.0 , 0.1 , 0.5 & 50, 100, 200, 500\\
Moderate   & 2.9 & 4.7 & 0.0 , 0.1 , 0.3,  0.5& 50, 100, 200, 500\\
Heavy   & 3.9 & 5.2  &  0.0 , 0.1 , 0.3, 0.5& 50, 100, 200, 500\\
Very Heavy  & 6.1 & 6.3 & 0.0 , 0.1 , 0.5& 50, 100, 200, 500\\

\bottomrule
\end{tabular}
\end{center}
\label{TableGAMMAparameters} \vspace{0.4cm}
\end{table}

To generate these samples, the Mersenne Twister pseudo-random number
generator \citep{MersenneTwister} is used, which has been widely
implemented in statistical packages such as MATLAB or R. To
ascertain the differences in the models a histogram comparison is
performed for different, fixed sizes of the samples. This allows us
to also address challenging problems present in the case with a low
total number of drops. To mimic the experimental samples without
values in lower diameters, several thresholds, $D_{min}$, are
considered as the minimum allowable size of drops in the samples,
see Table \ref{TableGAMMAparameters}. A maximum limit for $D_{min}$
was selected according to \citep{bringi_huang_etal_2002_aa}. This value was used to
evaluate whether the $\mu-\Lambda$ relations under gamma estimations
using the method of moments are due to a sampling problem. In
addition, the same value is considered a confident threshold to
discriminate among faithful measurements of smaller and medium drops
from noisier measurements characteristic of the smallest drops
\citep{mallet_barthes_2009_aa,kliche_smith_etal_2008_aa}.\\

Along the paper, the word \emph{category} denotes the pair of values
$\mu$ and $\Lambda$. Each category represent a functional form as
given by the Equation (\ref{eqn:gamma}). Also each simulated
situation is called scenario, and it is defined by: $\mu$,
$\Lambda$, $D_{min}$ and Size of the sample. A sample is a
particular realization of a given scenario, while the size of the
sample is defined by the number of drops (or number of elements
taken from the population defined by the category). For each
scenario 50 samples were generated all with the same number of
drops.\\

Simulating the statistical properties of the underlying measurement
methods several studies have generated a RDSD by choosing sample
sizes according to a Poisson distribution with a given 
mean \citep{bringi_huang_etal_2002_aa}. In addition, a second step to simulate
observational errors has been suggested to add to previous Poisson
ones \citep{cao_zhang_2009_aa}. Such methods are built to evaluate the bias and
errors of the method of moments or the maximum likelihood method of
estimation for a gamma RDSD (or theoretically another distribution
function) as near as possible to the supposed experimental
situation. In the context of the present research, this would
partially mask our main objective, which is to evaluate the capacity of
different methods to represent a given sample. Thus, in our study,
we used fixed values of $\Lambda$ and $\mu$ for several sample sizes
ranging from 50 to 500 drops; nevertheless, in each case, 50 samples
with the same characteristics are generated.\\

\subsection{Experimental}

The empirical data set corresponds to the first Spanish-GPM
Observation Program (SGPM/OP1) carried out from 15 December 2009, to
15 January 2010, as part of the Spanish contribution to the Ground
Validation segment of the NASA/JAXA Global Precipitation Measuring
(GPM) mission. The general characteristics of the experiment are
explained in \citep{Tapiador2010} where information on the instrument
is also supplied. This paper also contains a description of the four
rainfall events under analysis, called hereafter 21-Dec-2009,
3-Jan-2010, 6-Jan-2010 and 12-Jan-2010.\\

\section{Performance Measures}

The differences between the experimental histogram, $h_{exp}$, and
the models represented by $f_{0}$ (given by Equation
(\ref{eqn:gamma}) for MLE and method of moments, and by Equation
(\ref{eqn:pdf}) in our application of MaxEnt), are defined as
follows:
\begin{equation}
d_{a}[k]=\frac{1}{N}\sum_{i=1}^{N}|D^{k}_{i}h_{exp}(D_{i})-D^{k}_{i}f_{0}(D_{i})|
\label{eqn:dk}
\end{equation}
where the sub-index a indicates the absolute value of the difference
and the sum is over the bins of the experimental or synthetic
histogram. The goal of this general measure is to quantify the
deviation of the quantity $f_{0}$ from the histogram weighted to
analyse the relevance in the integral parameters. To give an account
of the accumulation of the differences in the entire rain event, the
following definition is used:
\begin{equation}
D_{a}[k]=\sum_{j=1}^{N_{e}}d_{a}^{(j)}[k]
\end{equation}
where $N_{e}$ is the total number of histograms evaluated in the
event, and $d_{a}^{(j)}[k]$ is the reported difference in the j-th
histogram.\\

From an experimental point of view, a histogram-based comparison is
possible if disdrometric measurements are available. The main
measure for this comparison is $d_{a}[0]$. Depending on the details
of the specific study it may be interesting to compare, for example,
$d_{a}[3]$, as this may more directly demonstrate the different
behavior between a rain process observed in a convective core region
and the one obtained in stratiform rain. For these such studies, the
general measure $d_{a}[k]$ is a valuable piece of information.\\

The two main measures of performance used in this paper are
$d_{a}[0]$ and $D_{a}[0]$. It is also possible to sum over the bins
without using any absolute values in Equation (\ref{eqn:dk}) to
analyse if the differences along the histogram are set off. This is
denoted as $d[0]$.\\

As a complement another measure of comparison has been introduced
based on the relative moment errors for the simulated RDSD
\citep{cao_zhang_2009_aa}, where the errors that the different
methods produce to estimate using the model, (1) the values of the
empiric integral rainfall parameters, (2) the values for the
integral parameters of the hypothetical distribution (based directly
in the values of $\mu$ and $\lambda$) are compared. Then, for each scenario:
\vspace{2mm}
\begin{equation}
F^{(1)}_{X}=\Bigg < \left
|\frac{X^{i}_{s}-\hat{X}^{i}}{X^{i}_{s}}\right | \Bigg >
\end{equation}
The values of $X^{i}_{s}$ represents the interested integral
variable in the histogram i-th of the scenario given. $\hat{X}^{i}$
represent the value in each method of modelling (for the sample
i-th) of the integral parameter X. The $< \cdot >$ represents the
mean value over the scenario, that is the mean value of the 50
samples.
\vspace{2mm}
\begin{equation}
F^{(2)}_{X}=\Bigg <\left |\frac{X_{d}-\hat{X^{i}}}{X_{d}}\right |
\Bigg >
\end{equation}
the $X_{d}$ represents the value of the integral parameter obtained
using the underlying gamma distribution function, then is a constant
value for each scenario. For scenarios defined by a total number of
drops higher than 100 drops and $D_{min}<0.3$  the values of
$X^{i}_{s}$ are similar to the values of $X_{d}$.\\

\section{Analysis}

\subsection{Analysis of Synthetic Data}
\label{sec:dataanalysisSynthetic}
\begin{figure}[h]
    \includegraphics[width=1.00\textwidth]{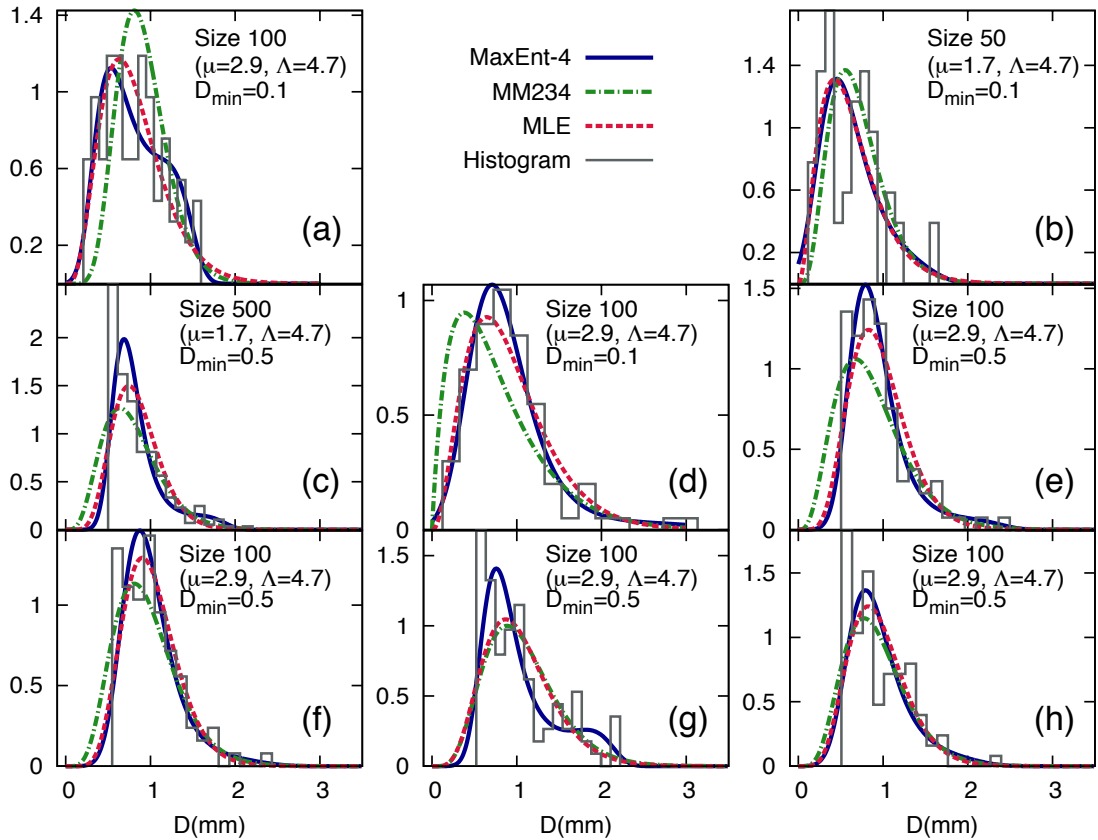}
\vspace{0.5cm}
    \caption[Histogramas sintéticos y sus modelizaciones utilizando MLE, MM234 y MaxEnt-4]{\textbf{Eight syntetic histograms and models using MLE, MM234 and MaxEnt-4.} Eight synthetic histograms together with the models obtained using the methods MLE, MM234 and MaxEnt-4 are shown. Cases with several values of $D_{min}$ and sample size are shown.}
    \label{fig1:Syntetic1}
\end{figure}

With the synthetic samples obtained as explained in Section
\ref{DataSynthetic} a binning procedure was carried out to construct
histograms of 15 bins \footnote{In generating the histograms, histograms of 20 bins were also computed, and, in the case of the smallest sample sizes, histograms of 10 bins were produced. Analogous general behaviour was reported
in both of these cases.}. The histograms generated $\left(
H_{1},...,H_{15}\right)$ were normalised to $\left(
h_{1},...,h_{15}\right)$ to satisfy the constraint:
$\sum_{i}h_{i}\delta D_{i}=1$. Note that $\delta D_{i}$ is the
distance between consecutive bins which in our case is a constant
value. The histograms produced by typical disdrometers are designed
to follow a log-scale; the motivation to choose a constant
linear-scale value is explore the robustness of MaxEnt to different
binning procedures. Now the values $\left( h_{1},...,h_{15}\right)$
can be compared directly with the methods explained in Section
\ref{Methods}. An example of the histograms generated is presented
in Figure \ref{fig1:Syntetic1}.\\

\vskip9mm

For all histograms of each scenario, the measure $d_{a}[0]$ was
calculated for each of the methods shown in the previous section.
The measure $d_{a}[0]$ has numerical values that depend on the
histogram, especially for samples with small number of drops. For
this reason the direct comparison of the $d_{a}[0]$ values for two
different histograms is a less valuable tool than the comparison of
$d_{a}[0]$ for different methods but the same histogram \footnote{It is also possible to compare the mean value of $d_{a}[0]$ over a
large set of histograms, or to analyse the values of $D_{a}[0]$ for
an entire precipitation event.} For these reasons, we study the following measures:
(1) the number of times each method reports the lowest values of
$d_{a}[0]$ of all methods, and (2) the relative values
$d_{a}[0]/d^{MLE}_{a}[0]$, taking MLE as a reference. By studying a
significant number of histograms for each scenario we will be able
to show the systematic behaviours of different methods. For the 50
different scenarios studied, 50 histograms were enough to highlight
comparative properties between methods\footnote{This is the same number of histograms per scenario that were used in the study of \citep{mallet_barthes_2009_aa}.}.\\

\begin{figure}[h]
    \includegraphics[width=1.00\textwidth]{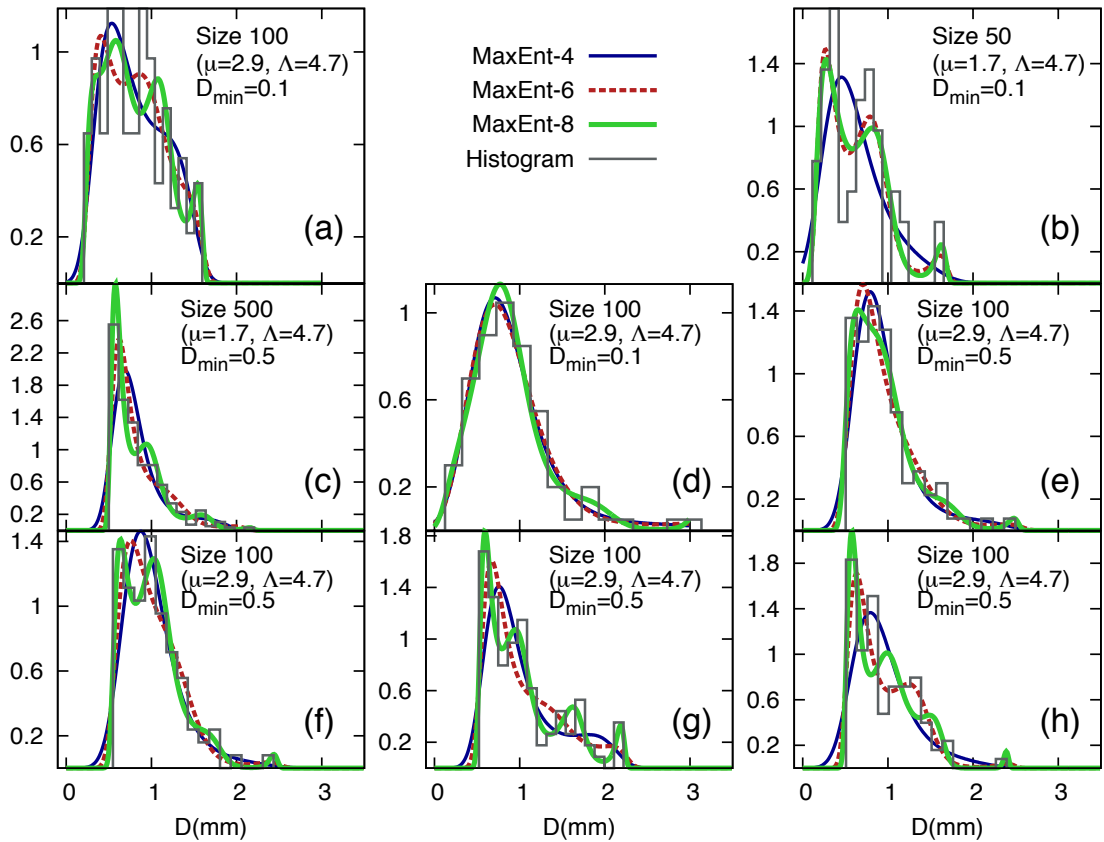}
\vspace{0.5cm}
    \caption[Histogramas sintéticos y sus modelizaciones utilizando MaxEnt-4, MaxEnt-6 y MaxEnt-8]{\textbf{Eight syntetic histograms and models using MaxEnt-4, MaxEnt-6 and MaxEnt-8.} Eight synthetic histograms
together with the models obtained using methods MaxEnt-4, MaxEnt-6,
MaxEnt-8 are shown. Cases with several values of $D_{min}$ and
sample size are shown.}
    \label{fig1:Syntetic2}
\end{figure}

\begin{figure}[h!]
    \includegraphics[width=0.99\textwidth]{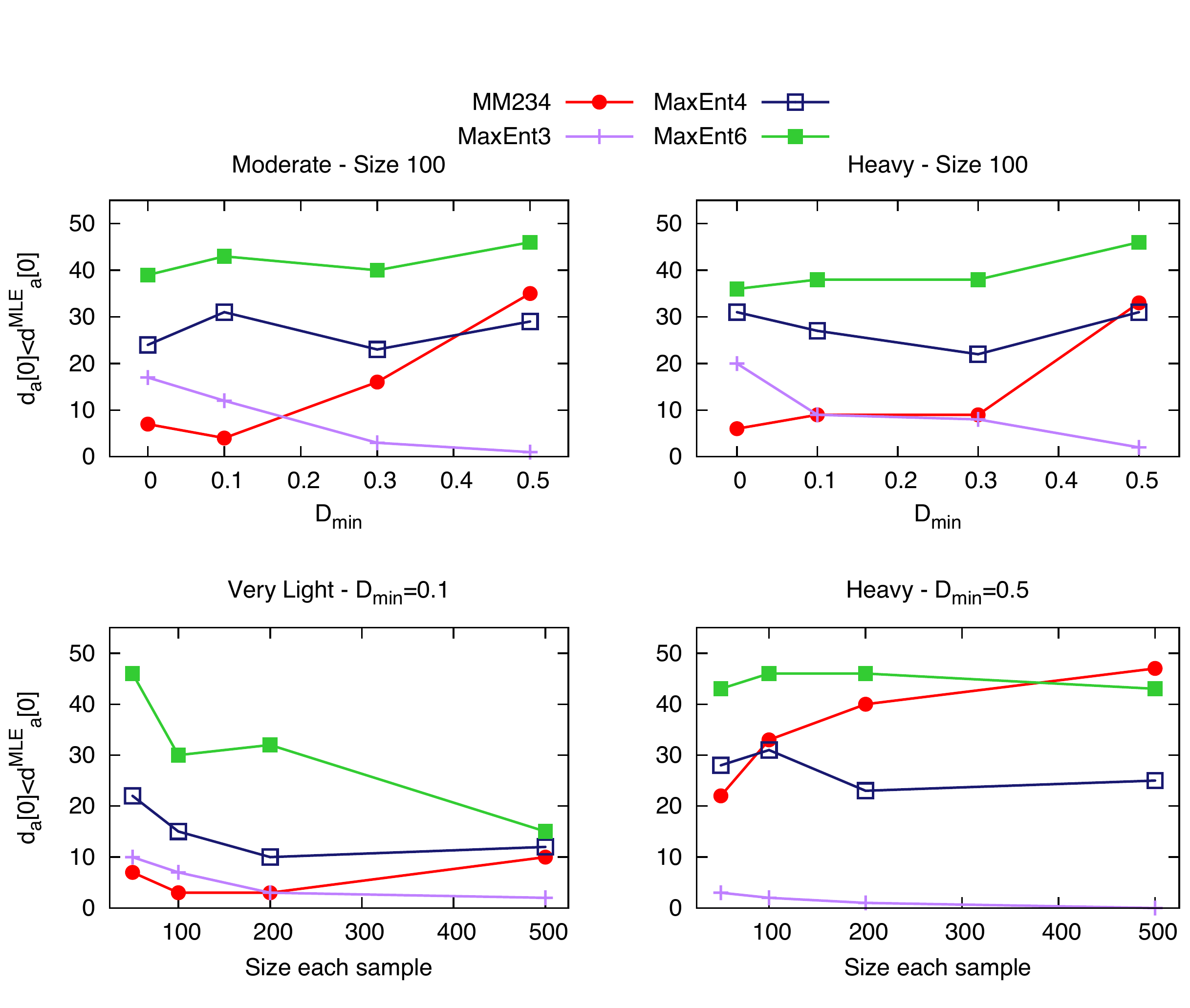}
\vspace{0.75cm}
    \caption[Número de casos en los que cada modelo mejora a MLE.]{\textbf{Comparative between different models using $d_{a}[0]$.} Number of cases in which each model reports lower value of $d_{a}[0]$ than MLE. Different scenarios are shown.}
    \label{fig:expHISTO}
\end{figure}

\begin{figure}[h!]
    \centering
    \includegraphics[width=0.97\textwidth]{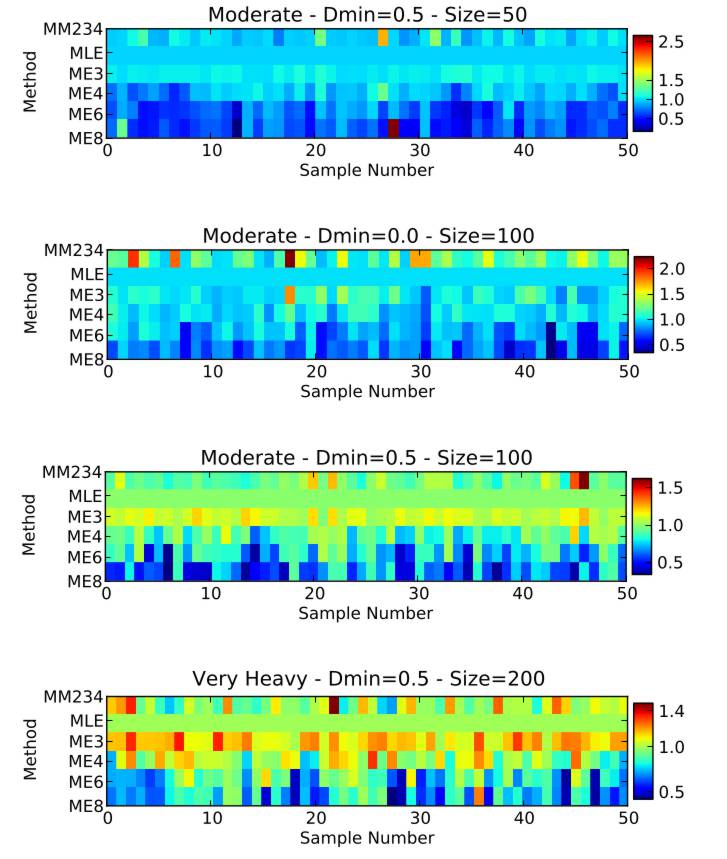}
\vspace{0.5cm}
    \caption[Comparación entre métodos relativa a MLE.]{\textbf{Comparison between models using relative values with respect to MLE.} Values of $d_{a}[0]$ relative to the value
for the MLE method, $d_{a}[0]/d^{MLE}_{a}[0]$. Cases with several
values of $D_{min}$ and sample size are shown. In first subplot, the
sample 28 is a case of non-convergence.}
    \label{fig2:Sintetic}
\end{figure}

Table \ref{TableALLcases} illustrates the results obtained for all
scenarios studied with a fixed category. The number of cases in
which each method produces the best value of $d_{a}[0]$ is shown
together with the number of times in which each method improves the
MLE. Illustrative
examples of the histograms are present in Figure
\ref{fig1:Syntetic1} comparing the methods MLE, MM234 and MaxEnt-4.
Figure \ref{fig1:Syntetic2} compares  MaxEnt-4, MaxEnt-6 and
MaxEnt-8 for the same histograms. Figure \ref{fig:expHISTO}
shows significance of $D_{min}$ and the number of drops, again using
MLE as a reference. Finally, the relatives values,  $d_{a}[0]/d^{MLE}_{a}[0]$, are shown in
Figure \ref{fig2:Sintetic} for 4 different scenarios.

\vskip9mm
\begin{table}[h]
\caption[Resultados utilizando diferentes métodos de modelización para la categoría \textit{Moderate} (véase también la Tabla
(\ref{TableGAMMAparameters}) utilizada para la generación sintética)]{\textbf{Results for the category \emph{Moderate}} For the category \emph{Moderate} (see Table
\ref{TableGAMMAparameters}) used for synthetic generation: The
number of times that each method produces lower levels of difference
with respect to the histogram, as measured with $d_{a}[0]$. Between
parentheses is the number of cases in which a method has a value of
$d_{a}[0]$ lower than that value derived under MLE. The symbol (*)
indicates the presence of a sample without convergence (see the main
text). As explained in the main text, size represents the number of
drops of each histogram, for each scenario  ($\mu$, $\lambda$,
$D_{min}$ and Size) 50 histograms were generated.}
\begin{center}
\small \ra{1.20}
\begin{tabular}{llrc>{\columncolor[gray]{0.95}}r>{\columncolor[gray]{0.95}}rrr>{\columncolor[gray]{0.95}}r>{\columncolor[gray]{0.95}}rrrr}
\toprule
\multicolumn{3}{c}{\textbf{Scenario}} &\multicolumn{10}{c}{\textbf{Methods of Modelling}}\\
\cmidrule(r{.5em}){1-3} \cmidrule(l{.5em}){4-13}

\textit{Rain Category} & $D_{min}$ & \textit{Size} & \textit{MLE} \qquad &\multicolumn{2}{c}{\textit{MM234}}  & \multicolumn{2}{c}{\textit{MaxEnt-3}} & \multicolumn{2}{c}{\textit{MaxEnt-4}} & \multicolumn{2}{c}{\textit{MaxEnt-6}} & \textit{MaxEnt-8} \\
\midrule

\multirow{12}{*}{Moderate} & \multirow{4}{*}{0.0}
              & 50  & 0    & 0 &(10)  & 0 & (22) & 1 & (35) & 6  &(45) & (*)43 (50) \\
            & & 100 & 1    & 0 &(7)   & 0 & (17) & 1 & (24) & 9  &(39)  & 39 (50)\\
            & & 200 & 3    & 0 &(5)   & 1 & (11) & 4 & (21) & 11 &(32) & 31 (50)  \\
            & & 500 & 3    & 0 &(6)   & 0 & (4)  & 0 & (11) & 11 &(36)  & 36  (50)  \\
\cmidrule{2-13}
            &  \multirow{4}{*}{0.1}

               & 50 & 1&0 &(5)  & 1 &(17) & 1 &(24) & 8 & (41) & 39   (50) \\
            &  & 100& 1&0 &(4)  & 1 &(12) & 3 &(31) & 12 &(43) & 33   (50) \\
            &  & 200 & 3&0 &(3)  & 0 &(6) & 1 &(19)  & 7 &(33) & 39   (50) \\
            &  & 500 & 7&0 &(1)   & 0 &(4) & 4 &(15) & 2 &(25) & 37   (50) \\
\cmidrule{2-13}
            &  \multirow{1}{*}{0.3}

              & 100& 1& 0 &(16)  & 0 &(3) & 1 &(23) & 4 &(40) & 44  (50)  \\
\cmidrule{2-13}
            &  \multirow{4}{*}{0.5}
               & 50  & 0 & 1 &(25)  & 0 &(0) & 1 &(31) & 14 &(48) & (*)34 (50)  \\
            &  & 100 & 0 & 0 &(35)  & 0 &(1) & 0 &(36) & 7  &(46) & (*)42 (50) \\
            &  & 200 & 0 & 0 &(37)  & 0 &(0) & 0 &(29) & 3  &(46) & 47  (50) \\
            &  & 500 & 0 & 0 &(45)  & 0 &(0) & 0 &(32) & 0  &(48) & 50  (50) \\

\bottomrule
\end{tabular}
\end{center}
\normalsize \label{TableALLcases}
\end{table}

The drawback of MaxEnt respect to other methods in the synthetic
cases relates to the numerical solution of the problem. With our
convergence parameters the system always solved for MaxEnt-3 and
MaxEnt-4. Meanwhile, only 3 cases out of 2500 histograms evaluated
failed to achieve convergence for MaxEnt-6. For MaxEnt-8 around 1\%
of the samples either required further iterations or simply did not
converge under the algorithm. These problems of convergence are
present in the more challenging samples involving few drops and
higher $D_{min}$. It important to note that while MM234 and MLE
always provide solutions to the problem, these solutions may not
faithfully represent the underlying histogram. On these cases
MaxEnt-4 and MaxEnt-6 provide better approximations to the problem.\\

From 80\% of the entire set of 2500 histograms MaxEnt-8 produces
lower values of $d_{a}[0]$. This fact makes logical sense, as a
larger set of free parameters allow a more accurate representation
of the histogram. For 68\% of the scenarios with $D_{min}=0$,
MaxEnt-8 produces lower values of $d_{a}[0]$, while for the
scenarios with $D_{min}=0.5$ the percentage is 85\%. This data shows
that for lower values of $D_{min}$ (with large values of the size)
the histograms are well described by MaxEnt-4 and MLE. If MaxEnt-8 is
excluded from the comparison, MaxEnt-6 produces the best performance
in the 63\% of the total number of histograms. Comparing only MLE,
MM234 and MaxEnt-4, the latter one produces lower values $d_{a}[0]$
for 40\% of the scenarios of the entire set. This shows how MaxEnt
is a progressive method of incorporating information, while the
results for MaxEnt-4 are interesting because a more realistic RDSD
using a fixed distribution function would need add model errors and
a direct consequence is a decrease in the performance of MLE and
MM234.\\

In the 15\% of the total histograms analysed MaxEnt-6 produces lower
values of $d_{a}[0]$ than any other method. This, together with the
generally similar values of $d_{a}[0]$ for MaxEnt-6 and MaxEnt-8
(see Figure \ref{fig2:Sintetic}) may be interpreted to mean that the
information contained in the integral rainfall parameters from the
total number of drops until to the reflectivity is enough to
retrieve an accurate representation of the samples. In the values of
$d_{a}[0]$ a small dependence in the histogram binning procedure is
possible, as shown in the panels (c) and (h) of Figure
\ref{fig1:Syntetic2}, while different binning procedures produce the
same global results when the total number of histograms generated
are compared.\\

\vskip9mm

MaxEnt-3 model of the RDSD, which includes the integral rainfall
parameters up to the liquid water content, is a good method for
small samples compared with MLE or MM234. Including just one more
constraint, MaxEnt-4 obtained better results than MLE in 51\% of the
histograms, with suitable performance for $D_{min}\neq0$ and
versatile behaviour for samples with different number of drops.\\

MM234 outperforms the MLE method when larger values $D_{min}$ are
introduced. These drawbacks of MLE in estimating the parameters of a
distribution are well known, but here the study focuses on analysing
the case with the measure $d_{a}[0]$. This is shown in the Figure
\ref{fig:expHISTO}.\\

\vskip13mm

In the case of MLE, the results show that MLE optimises the model
using values of the samples instead of the moments; however, the
capacity of this method is strongly conditioned by the
characteristics of the sample. In such situations for values of
$D_{min}=0.5$, MM234 provides a better representation of the
histogram; see Table \ref{TableALLcases}, particularly the
second column on MM234. This justifies the general preference for
the method of moments over MLE. The panels (d) and (e) Figure
\ref{fig1:Syntetic1} illustrate cases in which MLE outperforms MM234
for $D_{min}=0.5$ whereas for 33 cases out of 50, the result is the
opposite. This shows how necessary it is to consider a large number
of histograms. Finally, these results also suggest that methods of
MLE applied to a truncated gamma distribution can improve compared
to MM234 \citep{mallet_barthes_2009_aa}, although these methods  there are not
widely used \citep{cao_zhang_2009_aa,kliche_smith_etal_2008_aa}. However, from the point of
view of comparing with MaxEnt a completely realistic RDSD simulation
includes model errors that also decrease the capacities of MLE and
MM234 for specific truncated methods, and the differences with
MaxEnt may remain similar.\\

The roles of $D_{min}$ and the sample size are visually illustrated
for all methods in Figure \ref{fig2:Sintetic}. It shows the
progressive improvement of MaxEnt. The relative differences in
$d_{a}[0]$ between  MLE and MaxEnt-6 (or MaxEnt-8) can reach a
factor of two. As noted above, with $D_{min}=0.5$, MM234 generally
improves the estimate of MLE but lacks consistency. However, due to
sampling variability regarding the presence of larger drops in some
samples MM234 may lose part of the information present in the
histogram; see Figure \ref{fig1:Syntetic1} subplot (d).\\

A different estimate of the advantages of each method may be
obtained by comparing $F^{(1)}_{X}$ and $F^{(2)}_{X}$. In our cases,
it is interesting to compare the capacities of the different methods
to represent different integral rainfall parameters: the mean
diameter is $\Phi_{1}$, the mean mass diameter is the quotient
$\Phi_{4}/\Phi_{3}$, and the reflectivity is related to $\Phi_{6}$,
so we will compare all values from $\Phi_{1}$ until $\Phi_{6}$. We
ask two main questions, (1) What is the ability of each method to
reproduce given values of empiric integral rainfall parameters? For
this objective, it is useful to consider $F^{(1)}_{X}$. (2) What are
the differences when we compare  the integral rainfall parameters
obtained from a hypothetical gamma distribution function?\\

The results are shown in Tables \ref{TableF1} and \ref{TableF2}. The
first shows how MaxEnt-6 represents a model whose expected values
for the integral rainfall parameters are the empirical values. It
also shows how MaxEnt-4 gives excellent results for integral
rainfall parameters related to the kinetic energy of drops,
$\Phi_{5}$, and to the reflectivity $\Phi_{6}$, without constraints
on the probability density function as in MaxEnt-6. This shows how
MaxEnt-4 may be a valuable tool to study Z-R relations based on
empirical values of rainfall. It also shows another interesting
result: the information contained in the integral parameters up to
the rainfall fixes all of the remaining integral parameters.\\

\begin{table}[h]
\ra{1.3} \caption[Errores fraccionales $F^{(1)}$ para los parámetros
integrales de la precipitación.]{\textbf{Fractional Error $F^{(1)}$ for the Integral
Parameters.} Fractional Error $F^{(1)}$ for the Integral
Parameters. The field with the character "-" means that the error is
lower than $0.0001$, typically is $10^{-5}$ to $10^{-7}$.}
\begin{center}
\small \ra{1.30}
\begin{tabular}{llrcc>{\columncolor[gray]{0.95}}rr>{\columncolor[gray]{0.95}}rr>{\columncolor[gray]{0.95}}r}
\toprule
\multicolumn{3}{c}{\textbf{Scenario}} & \textbf{Method} &\multicolumn{6}{c}{$\mathbf{F^{(1)}}$ }\\
\cmidrule(r{.5em}){1-3} \cmidrule(l{.5em}){5-10}

\textit{Rain Category} & $D_{min}$ & \textit{Size} &\textbf{Modelling} &$\Phi_{1}$  & $\Phi_{2}$ & $\Phi_{3}$ & $\Phi_{4}$& $\Phi_{5}$ & $\Phi_{6}$ \\

\midrule

 \multirow{4}{*}{Very Heavy} &  \multirow{4}{*}{0.1} &  \multirow{4}{*}{100} & MLE & - & 0.033 & 0.122 & 0.260 & 0.418 & 0.571\\
 & & & MM234 & 0.088 & 0.105 & 0.109 & 0.117 & 0.125 & 0.123\\
 & & & MaxEnt-4 & - & - & - & - & 0.004 & 0.018\\
 & & & MaxEnt-6 &- & - & - & - & - & -\\

\midrule

 \multirow{4}{*}{Moderate} &  \multirow{4}{*}{0.0} &  \multirow{4}{*}{500} &MLE & - & 0.013 & 0.050 & 0.109 & 0.180 & 0.250\\
 & & & MM234 & 0.048 & 0.057 & 0.057 & 0.057 & 0.054 & 0.041\\
 & & & MaxEnt-4 & - & - & - & - & 0.011 & 0.045\\
 & & & MaxEnt-6 &- & - & - & - & - & -\\

\midrule

 \multirow{4}{*}{Moderate} &  \multirow{4}{*}{0.5} &  \multirow{4}{*}{50} & MLE & - & 0.037 & 0.133 & 0.271 & 0.415 & 0.539\\
 & & & MM234 & 0.052 & 0.055 & 0.055 & 0.055 & 0.034 & 0.035\\
 & & & MaxEnt-4 & - & - & - & - & 0.003 & 0.012\\
 & & & MaxEnt-6 &- & - & - & - & - & -\\

\midrule

 \multirow{4}{*}{Very Light} &  \multirow{4}{*}{0.5} &  \multirow{4}{*}{200} & MLE & - & 0.053 & 0.208 & 0.457 & 0.744 & 1.012\\
 & & & MM234 & 0.230 & 0.283 & 0.283 & 0.283 & 0.289 & 0.281\\
 & & & MaxEnt-4 & - & - & - & - & 0.008 & 0.035\\
 & & & MaxEnt-6 &- & - & - & - & - & -\\

\bottomrule
\end{tabular}
\end{center}
\label{TableF1}
\end{table}

\begin{table}[h]
\ra{1.3} \caption[Errores fraccionales $F^{(2)}$ para los parámetros
integrales de la precipitación.]{\textbf{Fractional Error $F^{(2)}$ for the Integral
Parameters.} Fractional Error $F^{(2)}$ for the Integral
Parameters.}
\vskip9mm
\begin{center}
\small \ra{1.30}
\begin{tabular}{llrcc>{\columncolor[gray]{0.95}}rr>{\columncolor[gray]{0.95}}rr>{\columncolor[gray]{0.95}}r}
\toprule
\multicolumn{3}{c}{\textbf{Scenario}} & \textbf{Method} &\multicolumn{6}{c}{$\mathbf{F^{(2)}}$ }\\
\cmidrule(r{.5em}){1-3} \cmidrule(l{.5em}){5-10}

\textit{Rain Category} & $D_{min}$ & \textit{Size} &\textbf{Modelling} &  $\Phi_{1}$  & $\Phi_{2}$ & $\Phi_{3}$ & $\Phi_{4}$& $\Phi_{5}$ & $\Phi_{6}$ \\

\midrule

\multirow{4}{*}{Very Heavy} &  \multirow{4}{*}{0.1} &  \multirow{4}{*}{100} &MLE & 0.023 & 0.048 & 0.076 & 0.105 & 0.135 & 0.166\\
 & & & MM234 & 0.065 & 0.027 & 0.091 & 0.282 & 0.544 & 0.880\\
 & & & MaxEnt-4 & 0.023 & 0.083 & 0.212 & 0.425 & 0.713 & 1.047\\
 & & & MaxEnt-6 &0.023 & 0.083 & 0.212 & 0.425 & 0.719 & 1.075\\

\midrule
 \multirow{4}{*}{Moderate} &  \multirow{4}{*}{0.0} &  \multirow{4}{*}{500} & MLE & 0.016 & 0.050 & 0.097 & 0.150 & 0.208 & 0.269\\
 & & & MM234 & 0.063 & 0.093 & 0.103 & 0.099 & 0.085 & 0.062\\
 & & & MaxEnt-4 & 0.016 & 0.038 & 0.048 & 0.044 & 0.041 & 0.066\\
 & & & MaxEnt-6 &0.016 & 0.038 & 0.048 & 0.044 & 0.031 & 0.022\\
 \midrule
 \multirow{4}{*}{Moderate} &  \multirow{4}{*}{0.5} &  \multirow{4}{*}{50} & MLE & 0.032 & 0.147 & 0.436 & 0.749 & 1.034 & 1.268\\
 & & & MM234 & 0.020 & 0.164 & 0.372 & 0.600 & 0.821 & 1.021\\
 & & & MaxEnt-4 & 0.032 & 0.113 & 0.326 & 0.562 & 0.804 & 1.041\\
 & & & MaxEnt-6 &0.032 & 0.113 & 0.326 & 0.562 & 0.802 & 1.036\\

\midrule

 \multirow{4}{*}{Very Light} &  \multirow{4}{*}{0.5} &  \multirow{4}{*}{200} & MLE & 0.937 & 1.514 & 1.655 & 1.432 & 0.978 & 0.428\\
 & & & MM234 & 0.599 & 1.098 & 1.502 & 1.818 & 2.056 & 2.225\\
 & & & MaxEnt-4 & 0.937 & 1.608 & 2.079 & 2.448 & 2.723 & 2.829\\
 & & & MaxEnt-6 &0.937 & 1.608 & 2.079 & 2.448 & 2.742 & 2.914\\

\bottomrule
\end{tabular}
\end{center}
\label{TableF2}
\end{table}
\vskip10mm

In the results for $F^{(2)}_{X}$, larger differences between MaxEnt
and MLE or MM234 were expected, as the functional form for N(D) is
different in our application of MaxEnt, however, the actual results
are similar to those of MM234 and, depending on the scenario, could
be better or worse than MLE, but always with reasonable values. This
means that even in the hypothetical cases in which the underlying
distribution is a gamma distribution function MaxEnt-4 is a useful
model. Obviously if the gamma distribution is considered to be the
real distribution function, then this hypothesis may be included in
the maximum entropy formalism as explained above.\\

\subsection{Analysis of Experimental Measurements}

\begin{figure}[h!]
    \includegraphics[width=1.00\textwidth]{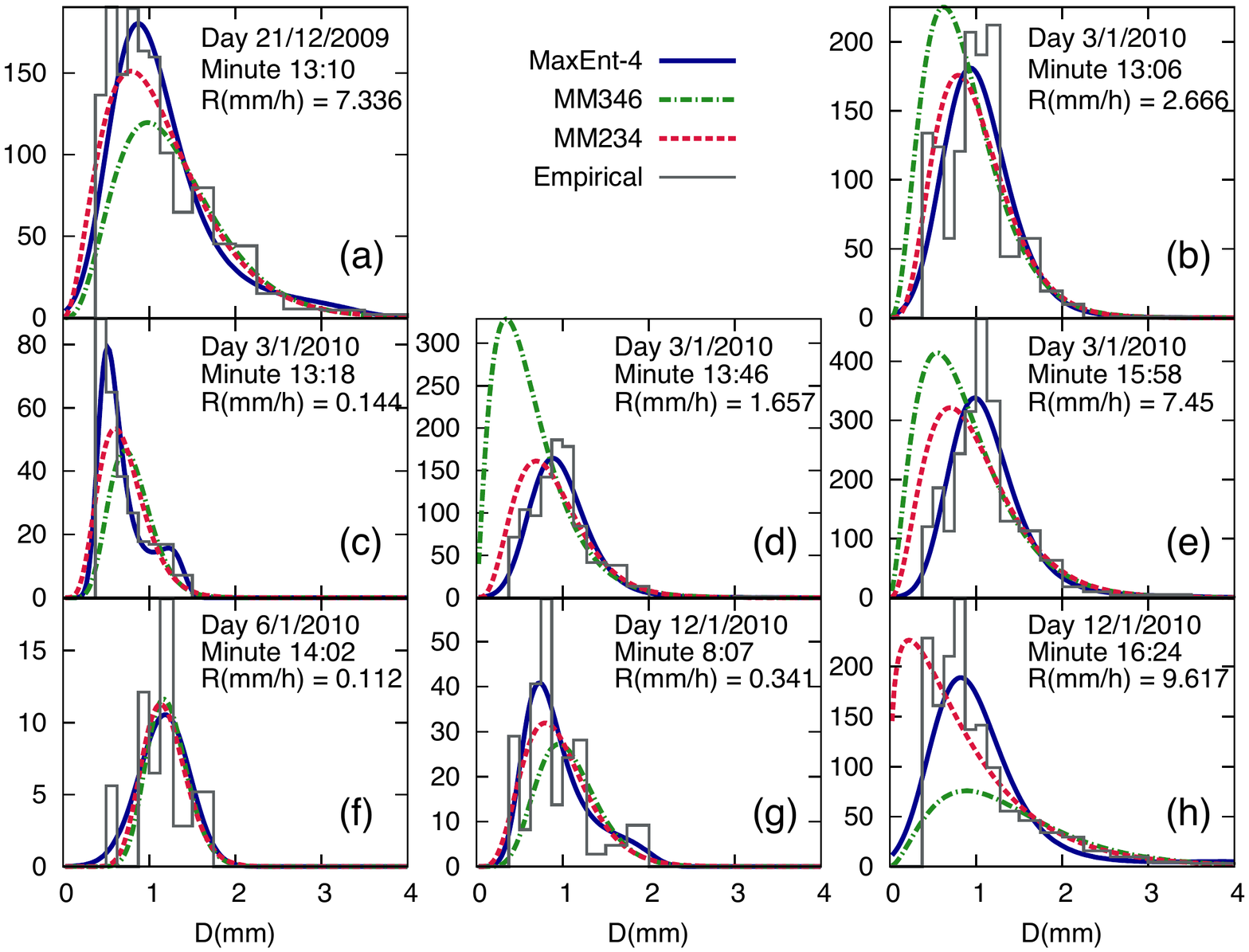}
\vspace{0.1cm}
    \caption[Histogramas experimentales y sus modelizaciones utilizando MM234, MM346 y MaxEnt-4]{\textbf{Eight syntetic histograms and models using MM234, MM346 and MaxEnt-4.} Eight Experimental histograms together with the models obtained using different methods are shown. The value of Rainfall as calculated from the sample is shown for comparative. MM234, MM346 and MaxEnt-4 are shown.}
    \label{fig:expHISTO1}
\vskip9mm
\end{figure}

\begin{figure}[h!]
    \includegraphics[width=1.00\textwidth]{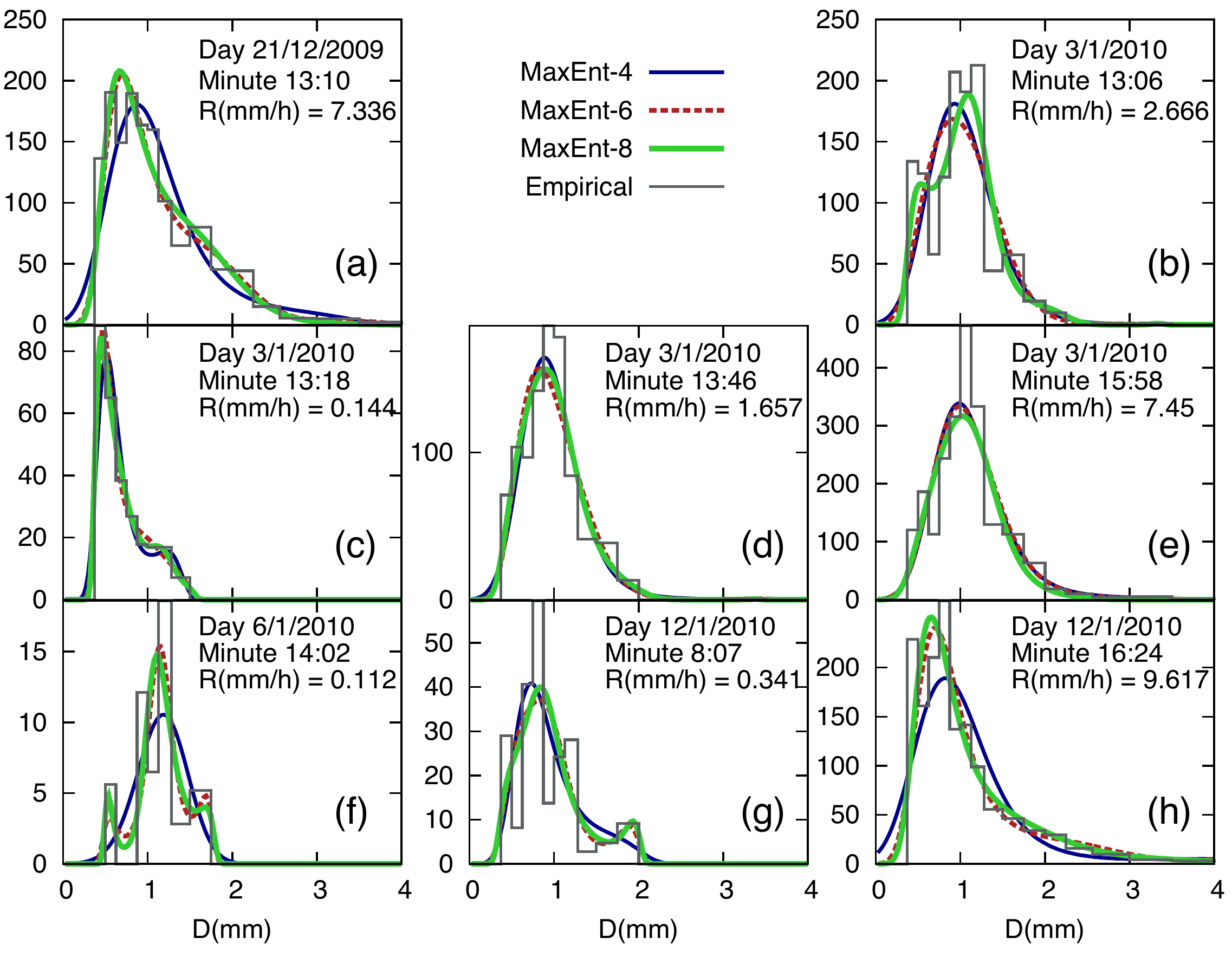}
\vspace{0.5cm}
    \caption[Histogramas experimentales y sus modelizaciones utilizando MaxEnt-4, MaxEnt-6 y MaxEnt-8]{\textbf{Eight syntetic histograms and models using MaxEnt-4, MaxEnt-6 and MaxEnt-8.} Eight Experimental histograms together with the models obtained using different methods are shown. The value of Rainfall as calculated from the sample is shown for comparative. Maximum Entropy models are shown.}
    \label{fig:expHISTO2}
\vskip9mm
\end{figure}

Experimental histograms are shown in Figures \ref{fig:expHISTO1} and \ref{fig:expHISTO2} for
the four different rain events. The general results are analogous to
the synthetic histograms. Subplots (f) and (g) of Figure \ref{fig:expHISTO2} show
that MaxEnt-6 and MaxEnt-8 can model adequately several peaks.
MaxEnt-6 is able to model cases with a second peak with larger drops
(typical multimodal cases), while MaxEnt-8 may model cases with
several peaks as well as small drops with low rainfall intensity.
The Figure \ref{fig:expHISTO1} illustrates how---for subplots (d) and (h)---MM346 fails to
provide a reasonable prediction for the smallest drops and,
consequently, underestimates the total number of drops. The method
is designed to represent experimental liquid water content $M_{3}$
and reflectivity $M_{6}$. However, when comparing the subplots (a)
and (e), both with similar amount of rainfall, the presence of a few
large drops results in an uncontrolled prediction for smaller drops.
MM234 shows more stability and results in a more systematic
representation of the histogram, while this representation is
improved using the MaxEnt methods. The comparison of subplots (a)
and (b) is illustrative of the role of larger drops in rainfall.\\

\begin{figure}[h!]
\vspace{0.75cm}
    \includegraphics[width=1.00\textwidth]{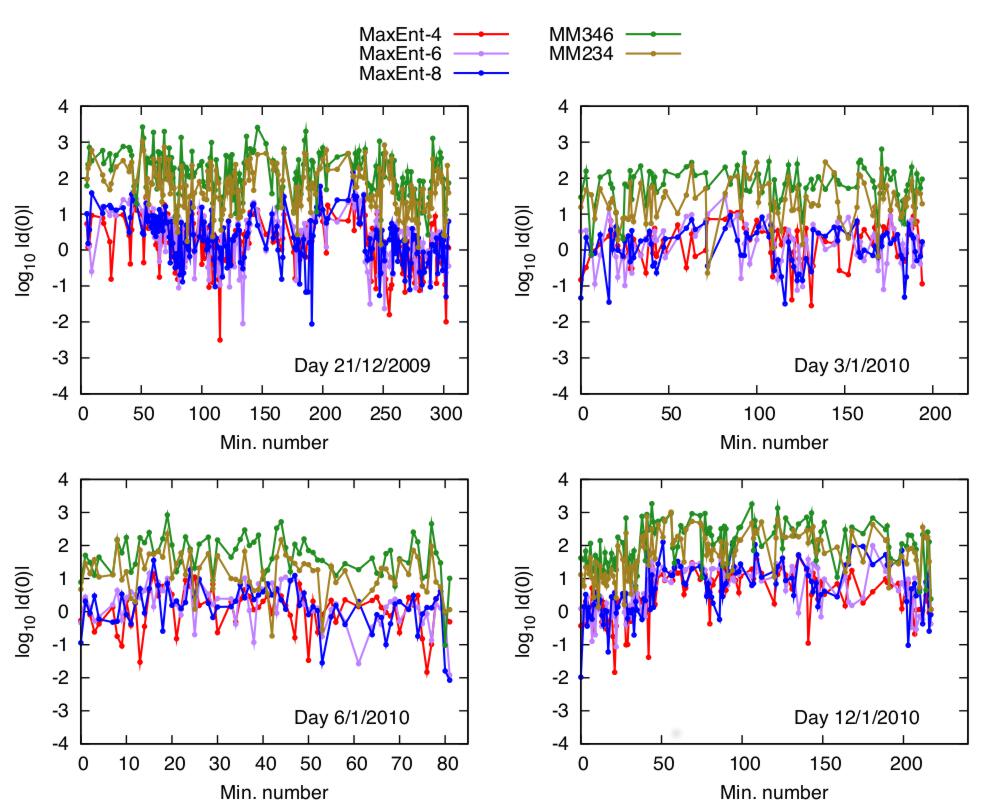}
\vspace{0.95cm}
    \caption[Comparación de diferencias entre varios modelos y los datos experimentales para cuatro episodios.]{\textbf{Comparison between different models in terms of $log_{10}|d[0]|$.} Comparison of $log_{10}|d[0]|$ for the four episodes of rain analyzed and for all methods used in Experimental Modeling.}
    \label{fig:sinABS}
\end{figure}
\vskip9mm

These facts can be appreciated in Figure \ref{fig:sinABS}, where
measure $d[0]$ is applied. The value of $log_{10}|d[0]|$ is shown
for the four different rain events. MM346 systematically results in
larger values, though similar in magnitude to MM234, while the
results of MaxEnt are always lower. In fact, MM346 in many cases
provides a value that is too high or too low with respect to the
total number of drops. This happen occurs for MM234, though less
frequently. On the other hand MaxEnt-4 without capturing all details
of the histogram is a good representation of it.\\

\begin{figure}[h!]
    \includegraphics[width=1.00\textwidth]{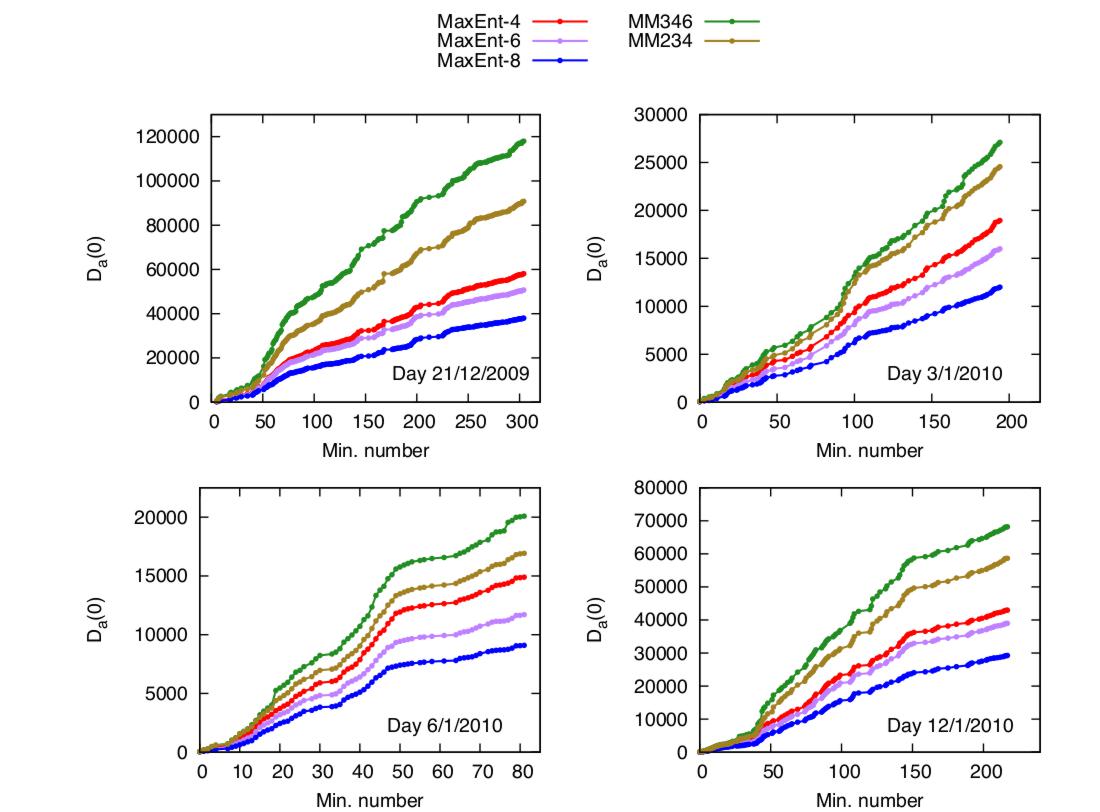}
    \caption[Comparación de diferencias acumuladas entre varios modelos y los datos experimentales para cuatro episodios.]{\textbf{Comparison between different models in terms of $D_{a}[0]$.} Comparison of $D_{a}[0]$ for the four episodes of rain analyzed and for all methods used in Experimental Modeling.}
    \label{fig:ACUM}
\end{figure}

As a result of the behaviour previously noted, the accumulation of
differences, $D_{a}[0]$, is an increasing monotonic function with
different rates. It is larger under the method of moments that under
the  MaxEnt methods. However, Figure \ref{fig:ACUM} shows a
correlation between all methods, all of which allow discriminate
periods of larger relative rates of increase, related presumably
with intrinsic properties of the rainfall process.\\

Looking to the experimental dataset, findings on convergence are
similar to the synthetic analysis. However, a careful comparison
allows us to discriminate two cases of convergence problems. Some
histograms with fewer that 10 drops required more effective
algorithms to solve the system and maximise of the entropy
functional. As noted before, problems related to a small number of
drops are present in all methods, and for practical purposes, some
authors preprocess data so that data with fewer than 10 drops are
classified as noise. For those analyses the algorithm applied to
retrieve the RDSD using MaxEnt is simple.\\

A previous study on sprays \citep{SpraysMaxEntMODAL2004} reported
situations in which this algorithm fails to convergence instead
reaches a narrow, non-physical fixed peak. Both situations explain
why in around a 5--7\% of the cases do not convergence for MaxEnt-8
(that is, given the reasonable difference within the data and a
limited number of iterations), though convergence is better under
MaxEnt-6. For the cases reported here, it may be possible improve
the Newton-Raphson algorithm \citep{Abramov2006} or avoid using it by
instead using more recently introduced algorithms
\citep{Abramov2009}.\\

In our comparative study with other methods, see Figures
\ref{fig:ACUM} and \ref{fig:sinABS}, the cases with fewest number of
drops that did not converge are not considered as in the previous
methods, because they cannot be used to provide a convenient
solution. Meanwhile the case of a narrow, large peak is easily
detected as indicating higher differences within histogram (or
integral parameters such as rainfall or reflectivity). Figures
\ref{fig:ACUM} and \ref{fig:sinABS} do not include histograms in
which any of the methods show a value of $d_{a}[0]$ higher than the
prescribed threshold. The criterion to determinate this threshold is
aimed at eliminating similar number of samples with high values of
$d_{a}[0]$ under method of moments as the number eliminated for
non-convergence under MaxEnt-8 method. The resulting, an improved
algorithm could provide higher differences under MaxEnt methods. In
other words, the threshold is selected to prune cases of
non-convergence under MaxEnt-8 as well as the cases of null
representability with respect to the histograms generated under the
method of moments.

\section{Discussion}

The maximum entropy modelling applied to RDSDs outperforms other
analysed methods for both synthetic and experimental datasets, in
terms of providing the empirical values of the integral rainfall
parameters and reproducing a given histogram.\\

In those cases where there is a sufficient amount of drops and
measurements of all sizes, the MLE will satisfactory represent a
sample. However, whenever these conditions are not true, the method
of moments can improve the predictions of MLE.\\

On the other hand, the MaxEnt method shows a systematic and
progressive approximation of the empirical information; while
MaxEnt-3 is a coarse representation of the histogram, MaxEnt-4 and
MaxEnt-6 achieve a gradual improvement by simply incorporating more
information. The only drawback of MaxEnt method are the occasional
difficulties related to convergence with the exposed simple
algorithm under the MaxEnt-8 model; therefore, MaxEnt-4 or MaxEnt-6
are reliable models for RDSD in the majority of the cases. The first
is a general valuable model, the latter is more useful to study
multimodal cases.\\

In the synthetic case, we tested the methods with an underlying
hypothetical distribution. However in the experimental case, model
errors appear, decreasing the performance when using traditional
methods. In contrast, MaxEnt provides less biased probability
density function which fulfills a given a set of constraints imposed
by empirical information. Thus, the study of the variability of RDSD
then does not rely on the capacity of a prefixed functional form to
represent the RDSD for all different cases. Rather, it focuses on
the physical and empirical meaningful selection of constraints.
Meanwhile, the analysis of the values of $\lambda$ retrieved by
MaxEnt provided the necessary information to develop a deeper
understanding of the questions related with the RDSD.\\

This opens the possibility of improving the predictions of
precipitation in two different aspects; the first requires the
incorporation of a more physical parametrisation of RDSD to the
numerical weather prediction models, while the second requires a new
method of analysis and prediction of ZR relations, which should
prove useful for ground and space-borne meteorological radars.

\section{Conclusions}

The formulation of statistical mechanics as proposed by Jaynes
\citep{Jaynes1957a} is conceptualised as a general inference method
resulting in the maximum entropy principle, which allows to
understand physical systems of stochastic origin and formulate more
objective and systematic models. This formulation of statistical
mechanics, and its underlying entropy concept have been applied to
many fields in earth sciences \citep{Tapiador2008}, from problems
similar to the presented in this paper to problems related to
reformulations of the MaxEnt as a maximum entropy production
postulate; for more details, see \citep{Niven2009} and the reference
there contained.\\

Under the maximum entropy principle the probability density function
represents the best guess according to available knowledge. The
method also has a physical significance when used to understand the
processes of the  earth's physics. This paper provides an
application to a real problem that is present in the microphysics of
precipitation resulting in a physically based method rather than
merely experimental fit.

\selectlanguage{spanish}
\renewcommand\chapterillustration{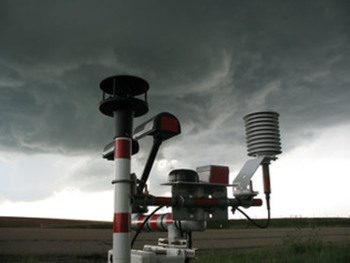}
\chapter{Discretization processes analysis of disdrometric measurements}

\label{chap:BINNING}

This chapter evaluates the binning effects on drop size distribution (DSD) measurements obtained by real disdrometers.  To achieve this goal, an evaluation using several simulated gamma DSDs was performed by considering the DSD moments and estimators of the DSD parameters. Those results were complemented by real measurements of Parsivel OTTs and 2DVD disdrometers. By comparing the measurements of several Parsivel OTTs a simple model of instrumental errors for the smallest drops was proposed and applied. Moreover, the 2DVD allowed us to study real cases and assess sampling and binning effects when drop-by-drop data were assumed to have been measured by other real instruments. It has been found that binning effects exist both in integral rainfall parameters and in the evaluation of DSD parameters of a gamma distribution. This study indicates an underestimation of low moments of the DSD and an overestimation of the large moments with a pattern, that differs between each instrument. Finally, the real 
measurements of 2DVD are affected by sampling uncertainty, but a sampling error estimation of instruments with smaller capture areas is possible. The results show that the differences due to sampling were a relevant uncertainty but that concentration, reflectivity and mass-weighted diameter were sensitive to binning. The main conclusion is that, while sampling errors and instrumental bias are the most acknowledged sources of differences between disdrometers, depending on the particular device used and the estimated parameters the binning effects can be as relevant as the other ones-

\section{Introduction}

Precipitation has a micro-structure that arises from the interaction between the hydrometeors that governs clouds and atmospheric dynamics. This micro-structure is conceptualised using the drop size distribution (DSD) which is the concentration of drops (per unit of volume and unit on diameter). The DSD has proved to be relevant for studies ranging from remote sensing to hydrology, climate and atmospheric prediction models \citep{michaelides_levizzani_etal_2010_aa,HydroPower2011,TapiadorGPM2011}.\\

 This has led to a significant amount of research based on in situ measurements of the DSD using disdrometers \citep{krajewski_smith_2002_aa}. However, the measurement of DSD remains a difficult challenge because it requires the discrimination between physical variability and a number of error sources, including the differences between several instruments whose measurements rely on different physical assumptions. In particular this requires the evaluation of the potential bias of every aspect of the associated measurement processes, particularly with respect to the binning process.\\

From a general point of view, a crucial problem concerning in situ measurements is the high temporal and spatial variability of the DSD. Because of this problem, disdrometers use short measurement times (typically 1 minute) and gather data from small areas (typically 50 to 100 $cm^{2}$). Although this approach partially solves the variability problem, it leads to insufficient sampling of the original population of drops. Furthermore, the errors in each type of instrument must be characterised to determine their optimum measuring conditions as well as their limitations \citep{deMoraes2011,cao_zhang_2009_aa}. As a consequence, numerous experimental studies have been based on disdrometric measurements \citep{tokay_short_etal_1999_aa,moumouni_gosset_etal_2009_aa,chandrasekar_gori_1991_aa,bringi_chandrasekar_etal_2003_aa,brandes_ikeda_etal_2007_aa,Jafrain_Berne_2011}, and a number of instruments have been designed based on mechanical properties \citep{tokay_kruger_etal_2001_aa} or on the optical properties of 
precipitation \citep{hauser_amayenc_etal_1984_aa,loffler-mang_joss_2000_aa,tokay_kruger_etal_2001_aa,sheppard_joe_1994_aa}.\\

Each instrument classifies data into intervals of size, and many studies have compared the results of different instruments to gain more information about a given DSD and its properties \citep{miriovsky_bradley_etal_2004_aa, krajewski_kruger_etal_2006_aa, thurai_petersen_ea_2011}. The two main issues addressed by these studies are the following: establishing the relevance of this sampling issue in inference and statistical analysis \citep{smith_kliche_2005_aa,joss_waldvogel_1969_aa,villarini_mandapaka_etal_2008_aa} and analysing the natural variability of a DSD \citep{ulbrich_1983_aa,uijlenhoet_steiner_etal_2003_aa,bringi_chandrasekar_etal_2003_aa}. The sampling issue is usually addressed by conducting a simulation of a DSD with known parameters, while the natural variability is assessed by examining precipitation experimentally, ideally using a network of instruments. In real cases, these two issues are related and coupled.\\

Aside from the sampling issue, the problem regarding the classification of continuous values of drop sizes into discrete size categories remains. This matter has been acknowledged by several authors \citep{krajewski_kruger_etal_2006_aa, Marzuki2010} but has not been addressed systematically when comparing the results obtained from different instruments. In this way, different disdrometric measurements present particular characteristics that are not always interpreted with the potential for discretisation bias in mind. The analysis of this bias was the main objective of this chapter.\\

 \begin{figure}[H]
 \begin{center}
\vspace{0.8cm}
 \includegraphics[width=0.90\textwidth]{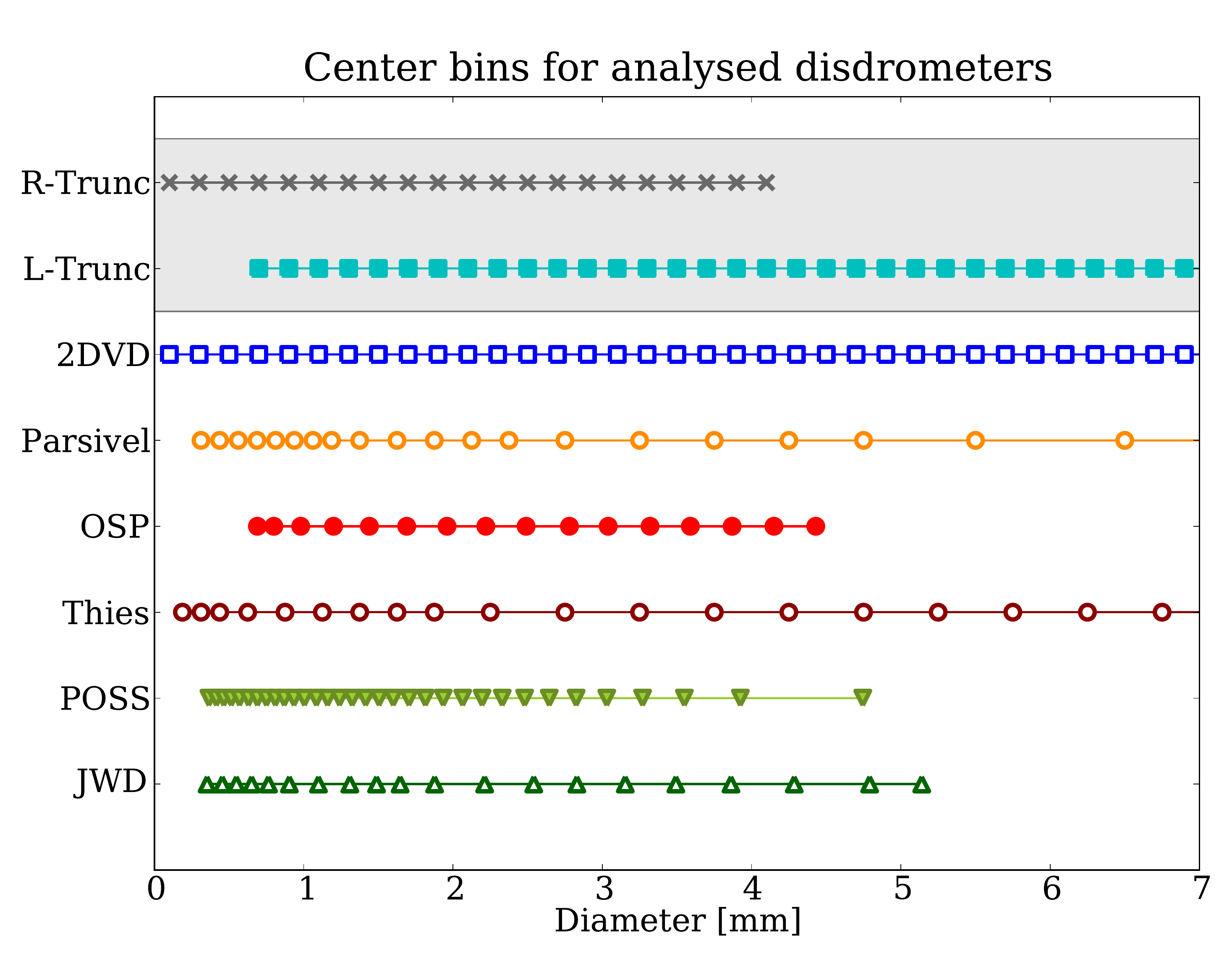}
 \end{center}
\vspace{0.9cm}
\caption[Esquema de los diferentes procesos de discretización analizados]{\textbf{Real instrumental bins analysed in this study.} Bins analysed in this study showing the central size classification values used by each instrument, as extracted from \citep{2000Campos,sheppard_joe_1994_aa,loffler-mang_joss_2000_aa}. The information of \textit{Thies} and Parsivel OTT disdrometers were provided by the manufacturer. In the case of Parsivel OTT the first and second bins were eliminated as the real instrument does not record information on these bins. }\label{fig1}
\vspace{1.3cm}
 \end{figure}

A pressing issue is that several sources of errors appear to be coupled in real DSD measurements. For this reason, studies should combine different sources of data, which also includes  simulated DSDs. Using a specific model distribution as a part of precipitation studies allows for the analysis of statistical inference problems with a known distribution. In sampling studies, the gamma distribution is most often used to represent the population of drop sizes.\\

 The gamma distribution allows for a reasonable representation of micro-physical variations that exist in real precipitation episodes \citep{kozu_nakamura_1991_aa,zhang_vivekanandan_etal_2003_aa,bringi_huang_etal_2002_aa,haddad_meagher_etal_2006_aa}. Thus, the first step in this study was to analyse binning effects on simulated DSD from several gamma distributions and estimate the relative relevance of the many sources of errors. However, studies on the estimation of DSD parameters have shown that each methodology used to estimate the DSD possesses virtues and weaknesses with respect to the sampling problem, an issue that must be evaluated jointly with the binning processes used by each instrument. All of these issues are addressed in the first part of the chapter (\S\ref{sec:estab_bin_effect}).\\

The second part of the chapter (\S\ref{sec:samp_bin}) discuss real DSD measurements of a 2DVD disdrometer, which allow us to build a drop-by-drop dataset. These DSD measurements are affected by sampling errors due to the limited size of the sensor. However, it is still possible to:
\begin{itemize}
 \item Estimate the increase in sampling errors obtained from instruments with a smaller sensing area than that of the 2DVD device.
 \item Compare binning effects for sensors with the same capture area as that of the 2DVD.
 \item Analyse the binning effects between sensors with smaller sensing areas.
\end{itemize}
These analyses were performed in the second part of this study.\\

 \begin{table*}[t]
\vspace{0.7cm}
\caption[Tabla de categorías de precipitación de \citep{TokayShort1996} aplicada al problema de discretización de la distribución de tamaños de gota]{\textbf{Precipitation categories from \citep{TokayShort1996}.} Precipitation categories from \citep{TokayShort1996} combined with Gaussian width values used as a complement in section (\S\ref{sec:defsigmas}). The minutes calculated under each category for real episodes 2009-12-21, 2010-03/04 and 2010-01-12 in \citep{Tapiador2010} are also indicated. This calculation allows for the establishment of the relevance of each class in stratiform precipitation cases. Episode 2011-06-07 was registered by the same group of disdrometers and represents a convective episode. 
}\label{t1}
\ra{1.44}
\vskip5mm
 \centering
 \begin{tabular}{ccccccccccc}
 \toprule
 12-21& 01-03 & 01-06 & 01-12 & 06-07 & Category & R[mm/h] & $N^{(g)}$ & $\lambda$ & $\mu$   & $\sigma(\mu)$  \\
 \midrule
364 & 617 & 242 & 168 &67 & very light (vl)   & $R<1$     &  5290  & 4.7 & 1.7 &  0.17 \\
145 & 246 & 19  & 33  &16 & light (l)         & $1<R<2$   & 13100  & 4.7 & 2.3 &  0.23 \\
238 & 232 & 5   & 224 &20 & moderate (m)      & $2<R<5$   & 24100  & 4.7 & 2.9 &  0.29 \\
76  & 81  & 0   & 74  &18 & heavy (h)         & $5<R<10$  & 80100  & 5.2 & 3.9 &  0.39 \\
9   & 6   & 0   & 4   &19 & very heavy (vh)   & $10<R<20$ & 332000 & 6.3 & 6.1 &  0.61 \\
0   & 0   & 0   & 0   &28 & exterme (e)       & $20<R$    & 426000 & 6.8 & 8.9 &  0.89 \\
 
 \bottomrule
 \end{tabular}
\vspace{1cm}
 \end{table*}

Previous studies \citep{Marzuki2010} have considered binning effects but without analysing the direct implications for a number of real instruments. The study by \citep{2000Campos} compared three types of disdrometers and concluded that discarding drops with diameters smaller than 0.7 mm did not affect the composite DSD but that different methods of analysis led to differences in the parameter estimates made by DSD models. More recently, \citep{brawn_upton_2008_aa} showed that incorporating bins of greater size affects the parameter estimation for the gamma distribution. Therefore, it is adequate to compare discretisation methods with differences in the minimum drop size considered and in bin sizes. This analysis reveals the relevance of features of the binning process, including the density of bins in different parts of the spectrum of drop sizes and the effect of ignoring certain sizes, such as small sizes or drops with diameters larger than 5 mm, as in the case of the JWD disdrometer.\\

 \begin{figure}[H]
 \vspace*{4mm}
 \begin{center}
 \includegraphics[width=0.54\textwidth]{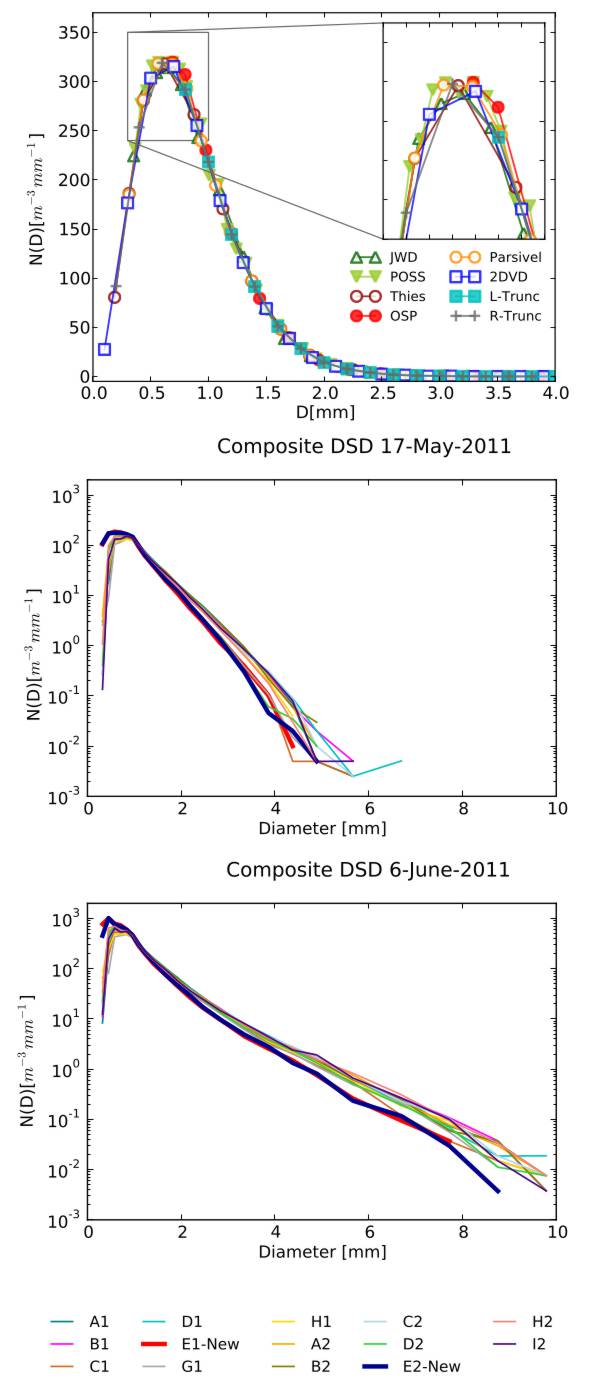}
 \end{center}
\caption[DSD compuesta para la categoría de precipitación moderada. DSD compuestas reales para episodios convectivos y estratiformes obtenida por instrumentos Parsivel OTT]{\textbf{Real and simulated composite DSDs.} \textit{Top panel}: The composite DSD resulting from the generation of 100 samples is shown for the category of moderate rainfall intensity. The relevance of the binning process is observed, even in smaller drops  where the density of bins is greater. \textit{Middle and bottom panels}: Two figures that correspond to two precipitation events are shown: stratiform (17 May 2011) and convective (6 Jun 2011). The DSDs shown were measured by 14 collocated Parsivel OTT disdrometers located in a rectangular net measuring 10 x 10 metres. E1 and E2 are advanced versions of the Parsivel OTT disdrometer, with more accurate drop size measurements.}\label{fig2}
 \end{figure}

This chapter is organised as follows. In the first part, the obtained simulated distributions are used to compare the different discretisation processes and their relevance. A subsection explains the methodology used to generate the simulated DSD and classify into size intervals, which is followed by details of the methods used to estimate the distribution function of drop sizes, N(D). These data are analysed by comparing the integral rainfall parameter values together with the moments and maximum likelihood estimators of the gamma distribution parameters. The second part analyses the 2DVD drop-by-drop dataset by comparing the results obtained with different instruments if this collection of drops were detected by those other devices. The last section concisely discusses the results and provides general conclusions. Further details about the physical assumptions made in generating the simulated DSDs are provided in the section \S\ref{app:DSDgenerationmethod}. \\

\section{Establishing the binning effects}
\label{sec:estab_bin_effect}

\subsection{Data}

The data used in this part are derived from two sources. The primary data source was provided by a computerised simulation of DSDs using a gamma distribution. The second data set was derived from disdrometric measurements using Parsivel OTT instruments from the first and second experimental ground validation campaigns from NASA-GPM in Spain. The experimental data were used to compare some of the hypotheses of this study with real precipitation events.\\

\subsubsection{Generation of artificial DSDs}
\label{sec:gen_gotas}

It is useful to know the original precipitation distribution when studying the bias and asymmetries in the integral rainfall parameters derived from the experimental drop size distribution, which is possible through computational DSD simulations. The same technique can be applied when analysing the relevance of class intervals in the experimental DSD estimates and their integral parameters. The procedure followed herein is similar to that performed in other studies \citep{smith_kliche_2005_aa, kliche_smith_etal_2008_aa, mallet_barthes_2009_aa,cao_zhang_2009_aa}. We begin with the following relationship which defines the gamma raindrop size distribution,
\begin{equation}
N(D)=N^{(g)}D^{\mu}e^{-\lambda D}=N^{(g)}\frac{\Gamma(\mu+1)}{\lambda^{\mu+1}}f(D)=N_{t}f(D)
\label{eqn:gamma}
\end{equation}

Once $N^{(g)}$, $\mu$, and $\lambda$ are set, we have a population with an average value of $N_{t}$ drops per volume unit. The values of the parameters of the gamma distribution are chosen following the classification given by \citep{TokayShort1996} in six different categories (Table \ref{t1}) and used by other authors \citep{brawn_upton_2008_aa,checa_tapiador_2011_aa}. A broad study \citep{2004Nzeukou} also showed a similar classification for rain with rainfall intensity lower than 20 mm/h and certain variations in the gamma distribution parameters depending on the experimental sample but with a similar range of values.\\

The sampling process used to select the set of measured drops is based on the initial selection of a category to define the average number of drops. This figure is derived using a Poisson distribution with an average of $N_{t}$ from which the effective number of drops of $n_{t}$ collected in the disdrometer is obtained. Then, in a second step, $n_{t}$ random drop sizes that correspond to the selected gamma distribution are generated.\\

\subsubsection{Variations in the distribution parameters}
\label{sec:defsigmas}

In addition to the previously simulated DSDs, we generated artificial DSDs that begin with the parameters that are defined in Table (\ref{t1}) but include uncertainties characterised by $\sigma_{\mu}$. This second process of DSD generation includes an extra step in which the nominal values are not taken for each category but are instead generated using the Gaussian distribution $\mathcal{N}(\mu,\sigma^{2}_{\mu})$, with an average of $\mu$ and a typical deviation of $\sigma_{\mu}$ , whose values for the case of relative errors of 10\% are indicated in Table (\ref{t1}). This analysis is designed to consider the impact of errors of the shape parameter $\mu$ , on the integral rainfall parameters.\\

\subsubsection{Classification of drops}

Eight classifications in different bins used by real instruments were systematically analysed with respet to both optical disdrometers and impact disdrometers. The procedure is as follows: each sample is classified into the bins shown in Figure (\ref{fig1}), which represent the centre of the class $D^{(d)}_{i}$ interval, while the class interval is given by, 

\begin{equation}
\Delta D^{(d)}_{i}=(D^{(d)}_{i+1}-D^{(d)}_{i})/2
\end{equation}
Frequency histograms are constructed for each sample $h^{(d)}_{i}$, leading to
\begin{equation}
N^{(d)}(D^{(d)}_{i})=h^{(d)}_{i}/\Delta D^{(d)}_{i}
\end{equation}

The histograms present jumps as a result of the different values of $\Delta D^{(d)}_{i}$, and these differences are reduced when the value of the class interval is divided by the value of the size of the class interval and when a magnitude is obtained per unit volume and distance.\\

It is important to note that the JWD disdrometer internally classifies the drops into 127 original bins that are later classified into 20 bins. The choice of these bins varies slightly between experiments. Here, the binning shown for JWD is similar to that reported by \citep{caracciolo_prodi_etal_2006_aa}.\\

Notably, for drops with diameters larger than 2.5 mm, the number of bins from the Parsivel disdrometer includes class intervals that are greater and smaller in number than what can be relevant for higher-order moments. The \emph{Thies} disdrometer \citep{deMoraes2011} possesses different bins even though it works according to the same physical basis as the Parsivel OTT. \emph{Thies} disdrometer presents class intervals that are somewhat greater than those for the Parsivel OTT ranging, from 0.5 mm to 2.5 mm, while for drops with diameters larger than 5.1 mm, the class interval is half that of the Parsivel. \\

The case of the 2DVD is different, as it provides drop-by-drop measurements, and the binning process is usually a user-made post process. However, the most widely used binning is uniform with a width of 0.2 mm. Additionally, to compare the results from the different disdrometers, we have also introduced artificial binning with the same bins width as the 2DVD instrument but with a maximum diameter of 4.3 mm (Right-Truncated) and minimum diameter of 0.7 mm (Left-Truncated). The binning process of the POSS disdrometer is included because, while it relies on remote-sensing measurement, the results also are classified into bins, as in other instruments that are also conditioned by binning effects. \\

\subsubsection{Information about the generation of artificial DSDs }
 \label{app:DSDgenerationmethod}

The proposed methodology is based on the modelling of precipitation as a homogeneous Poisson process which is the preferred method in the literature. The methodology is based on the assumption of stationary rain, a physical situation that arises in several types of real precipitation \citep{larsen_kostinski_etal_2005_aa,jameson_kostinski_2002_aa}. Additionally, the study \citep{uijlenhoet2006analytical} indicates that this approach allows for a lower level of statistical fluctuations than that observed in more general situations, and as consequence it may provide a lower threshold on the bias. In our study regarding the relevance of binning, the differentiation between homogeneously distributed rain and rain distributed in clusters is not necessary; in both cases, we expect the binning process to produce the same level of error relative to other error sources. \\

We could include an estimate of the sampled volume (given a collection area of S and a measurement time of T) for each diameter \citep{uijlenhoet_pomeroy_2001_aa,mallet_barthes_2009_aa} based on a value of $v(D)$ as $STv(D)$, which is calculated using $v(D)=\delta D^{\epsilon}$. Some authors \citep{moisseev_chandrasekar_2007_aa} do not consider this distinction relevant for the majority of analyses.
The procedure introduced by \citep{mallet_barthes_2009_aa} involves choosing a concrete relationship, $v(D)$, and is useful in the case of JWD-type disdrometers, which presume an specific $v(D)$ relationship in the measurement process. However, this approach is less practical for optical disdrometers that measure terminal drop velocity. These instruments usually include a tolerance interval of 50-60\% over a given $v(D)$ relationship, which in practice can eliminate the differences in sampling volumes between adjacent bins. Above all, this approach would make the analysis process dependent on the velocity distribution generation hypothesis for each diameter. We also observed that the sampled function, including $v(D)$, is analogous to the former function, $f(D)$, but with $\tilde{\mu}=\mu+\epsilon$ and $\tilde{N}_{t}=N_{t}\delta ST$. In our case, we chose a constant volume sampling solution (as we could attempt multiple combinations of $N_{t}$ and $\mu$) and we also introduced the possibility of moderated 
variations for $\mu$.\\

Other authors \citep{cao_zhang_2009_aa} introduce an observational error for each bin based on the comparison of two collocated disdrometers. In our case, it is inconvenient to include this type of error from the beginning. We compared the binning processes of disdrometers with different physical measurement processes that give rise to slightly different observational errors but do not alter the discretisation of the spectrum. Furthermore, we have included a generic model of observational error to evaluate its consequences. \\

Regarding with the values of $\sigma_{\mu}=10\%$ proposed. They are moderate in contrast to other references \citep{moumouni_gosset_etal_2009_aa} where they can reach 40-50\% of the average value. The main difference in our case is that we deal with errors within each category of rainfall intensity, while other studies assign variations for whole events. These typical moderated variations allow for the implicit inclusion of possible variations in sampling volumes, as well as variations over the intensity intervals of the studied precipitation. In this regard, with \citep{2004Nzeukou} as a reference, the average values for four different campaigns are similar to those included in Table (\ref{t1}), while the differences in the values of the $\mu$ and $\lambda$ parameters range from 20\% to 25\%. \\

\subsubsection{Experimental data: Parsivel OTT}

The Parsivel OTT is an optical disdrometer \citep{loffler-mang_joss_2000_aa} that measures both the equivalent volumetric diameter and fall velocity of the hydrometeors that cross a laser dimming the signal. We have included an effective measurement area in this calculation (as explained in chapter \S\ref{chap:preprocesadoTOKAY}), which is expressed as\citep{loffler-mang_joss_2000_aa}: 
\begin{equation}
 S_{eff}(D_{i})=L(W-D_{i}/2)
\end{equation}
The accumulation time was selected to be 1 minute. For the calculation, we proceeded according to the description provided by equation (8) of \citep{krajewski_kruger_etal_2006_aa}. We required a minimum rainfall intensity of 0.05 mm/h and at least 10 drops to process each minute, and a tolerance of 50\% was allowed over the usual relationship $v(D)=3.86D^{0.67}$. Real temporal series were considered and classified into the same intervals of rainfall intensity shown in Table (\ref{t1}) to verify the possible relevance of each category in real precipitation cases. \\

 \begin{figure}[H]
 \vspace*{8mm}
 \begin{center}
\vspace{1.3cm}

\includegraphics[width=1.05\textwidth]{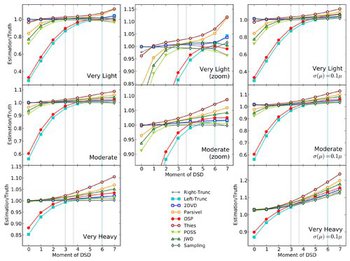}
 \end{center}
\vspace{1.1cm}
 \caption[Desviación relativa de los momentos de la DSD simulada para diferentes disdrómetros]{\textbf{Relative difference of the DSD moments for several real disdrometers.} The first column reports the estimates for each of the DSD moments ranging from 0 to 7 for three precipitation categories (very light, moderate and very heavy) based on 5000 samples. The second column shows the differences in the higher-order moments. The third column shows the results when a Gaussian noise is introduced in the $\mu$ variable for each sample. The true value is obtained from the given analytical values $\mu$, $\lambda$ and $N^{(g)}$ and the expression (\ref{eqn:Mk}). The sampling case is based on the sample moment estimates without carrying out a classification into bins. In the third column, the sampling represents the effective combination of the sampling case and the uncertainty in the moment estimates due to the Gaussian noise introduced in $\mu$. To illustrate the relationship with integral rainfall parameters the 
three vertical lines represent, from left to right, the position of the LWC ($=C_{LWC}M_{3}$), R ($\simeq C_{R}M_{3.67}$) and Z ($=M_{6}$), where the constants $C_{LWC}$ and $C_{R}$ allow for the retrieval of the usual units, which are presented in \citep{ulbrich_1983_aa}, and in the table (\ref{tablaPARAMETROSintegrales}).}\label{fig3}
\vspace{2.0cm}
 \end{figure}

\subsection{Methods}

The methodologies utilised to analyse the binning effects of the instruments are focused on comparing the integral rainfall parameters and the DSD parameters. For the integral rainfall parameters, the most practical method is to compare the moments of the DSD retrieved by each instrument after the binning process, while for the DSD parameters it is necessary to evaluate several approaches. For this reason, two different methodologies to estimate the DSD parameters were compared: one based on the distribution moments and the other on the maximum likelihood method. The first method included a second version that considered the absence of small drop measurements by some instruments and was therefore adapted to the specific case of disdrometric measurements.\\


 \begin{table*}[t]
\vspace{1.3cm}
\caption[Estimation of gamma distribution parameters using the moment method. ]{\textbf{Estimation of gamma distribution parameters using the moment method.} Five broadly used methods are shown. With regard to the methodology used to obtain the expressions, the generic method is introduced in the text. The moments used to calculate $\lambda$ are shown in parenthesis.}\label{t2}

 \vskip8mm
 \centering
\small
 \begin{tabular}{lcccc}
 \toprule
 Method          &  Function G                       &  $\widehat{\mu}(G_{exp})$ & $\widehat{\Lambda}(\widehat{\mu})$ & $\widehat{N}^{(g)}(\widehat{\Lambda},\widehat{\mu})$  \\
 \midrule MM012(01)             &  $\displaystyle\frac{M_{1}^{2}}{M_{0}M_{2}}$   &$\displaystyle\frac{1}{1-G}-2$   & $\displaystyle(1+\mu)\frac{M_{0}}{M_{1}}$  & $M_{0}\frac{\Lambda^{(\mu+1)}}{\Gamma(\mu+1)}  $                    \\[1.0 ex]
MM246(24)             &  $\displaystyle\frac{M_{4}^{2}}{M_{2}M_{6}}$   & $\displaystyle\frac{7-11G-\sqrt{14G^{2}+G+1}}{2(G-1)}$ & $\displaystyle \sqrt{(3+\mu)(4+\mu)\frac{M_{2}}{M_{4}}}$ & $M_{2}\frac{\Lambda^{(2+\mu+1)}}{\Gamma(2+\mu+1)}  $                    \\[2.0 ex]

MM346(34)             &  $\displaystyle\frac{M_{4}^{3}}{M^{2}_{3}M_{6}}$            &   $\displaystyle\frac{-8+11G+\sqrt{G^{2}+8G}}{2(1-G)}$   &   $\displaystyle(4+\mu)\frac{M_{3}}{M_{4}}$             &     $M_{3}\frac{\Lambda^{(3+\mu+1)}}{\Gamma(3+\mu+1)}  $  \\[2.0 ex]

MM234(23)             &  $\displaystyle\frac{M_{3}^{2}}{M_{2}M_{4}}$            &  $\displaystyle\frac{1}{1-G}-4$           &  $\displaystyle(3+\mu)\frac{M_{2}}{M_{3}}$              &  $M_{2}\frac{\Lambda^{(2+\mu+1)}}{\Gamma(2+\mu+1)}  $                 \\[2.0 ex]

MM456(45)             &  $\displaystyle\frac{M_{5}^{2}}{M_{4}M_{6}}$            &  $\displaystyle\frac{1}{1-G}-6$                 &   $\displaystyle(5+\mu)\frac{M_{4}}{M_{5}}$             &    $M_{4}\frac{\Lambda^{(4+\mu+1)}}{\Gamma(4+\mu+1)}  $   \\[1.0 ex]

 \bottomrule 
\end{tabular}
\vspace{1.8cm}
 \end{table*}

\subsubsection{Moment method}

The sampled and discretised gamma distribution can be estimated by different methods \citep{cao_zhang_2009_aa}. The most widely used technique is the moment method, in which three free DSD parameters are estimated from a subset of three integral rainfall parameters. The freedom in the choice of these integral parameters requires that estimates be compared from as many different subsets as possible (to achieve the best subset in each case). Given the distribution of drop size in equation (\ref{eqn:gamma}), the moment of order k is 
\begin{equation}
M_{k}=N^{(g)}\frac{\Gamma(\mu+k+1)}{\lambda^{\mu+k+1}}
\label{eqn:Mk}
\end{equation}

The methodology developed here to reach the estimate expressions is general and can in fact be applied to other distributions. We begin from the definition of a G parameter as follows:
\begin{equation}
G_{exp}=\frac{M^{a}_{l}}{M^{b}_{k}M^{c}_{m}}
\label{eqn:G}
\end{equation}
where l, k and m are the orders of the integral rainfall parameters used, and a, b and c are three real numbers. Then by using equation (\ref{eqn:Mk}):
\begin{equation}
G(\mu,\lambda,N^{(g)})=\left [N^{(g)} \right]^{a-b-c}\frac{\lambda^{(\mu+1)(b+c)+(k\cdot b+m\cdot c)}}{\lambda^{(\mu+1)a+l\cdot a}}g(\mu)
\label{eqn:Gmu}
\end{equation}
where $g(\mu)$ is an expression involving only $\Gamma$ functions. If the following is true, 
\begin{equation} 
l \cdot a=k \cdot b+m \cdot c
\label{eqn:MMrestric1}
\end{equation}
then G is a dimensionless quantity. If we also impose 
\begin{equation}
a=b+c
\label{eqn:MMrestric2}
\end{equation}
then eliminating the dependence of the G function on $N^{(g)}$ and eliminating the $\lambda$ factors are possible. We thus obtain an expression for $G$ that only depends on the value of $\mu$.
Therefore, given the experimental values of $M_{l},M_{k},M_{m}$, we can determine $G_{exp}$ and obtain an estimate $\hat{\mu}(G_{exp})$ by using the equation (\ref{eqn:Gmu}) with the restrictions (\ref{eqn:MMrestric1}) and (\ref{eqn:MMrestric2}).\\

Given $\hat{\mu}$ and the two moments (moments of a lower order usually have less severe sampling issues) from the set (k, l, m), we can determine $\lambda$ and immediately $N^{(g)}$. It is important to note that $\lambda$ can be calculated using any combination of two moments from the set (l, k and m). \\

The analytical expressions of the estimators are given in Table (\ref{t2}). For the remainder of this chapter, we will use the notation MMlkm to denote the method that uses the order l, k and m moments. This study systematically analysed the estimates using methods MM012, MM234 and MM456. The most frequently used methods in studies of disdrometers are MM234 and MM346. However, the behaviour of the last method (from the perspective of this study) can be understood from the study of the other moment methods. \\

\subsubsection{Truncated moment method}

Figure (\ref{fig3}) shows that the disdrometers have minimum and maximum diameters, which indicates that the moments estimated from the sample correspond to 

\begin{equation}
\widetilde{M}_{k}=\int_{D_{min}}^{D_{max}} D^{k}N(D)dD
\end{equation}
\begin{equation}
\widetilde{M}_{k}=N^{(g)}\frac{\gamma(\mu +k+1,D_{max}\lambda)-\gamma(\mu +k+1,D_{min}\lambda)}{\lambda^{\mu+k+1}}
\label{eqn:momentotrunkado}
\end{equation}
where $\gamma(a,l)$ is the incomplete gamma distribution that is given by

\begin{equation}
\gamma(\delta,l)=\int_{0}^{\delta}D^{l-1}e^{-D}dD
\end{equation}
Equation (\ref{eqn:momentotrunkado}) is based on the assumption that N(D) is a gamma distribution given by equation (\ref{eqn:gamma}). Given the expressions $\widetilde{M}_{k}$, it is not possible to write $G$ (equation \ref{eqn:Gmu}) as an uni-parametric function of $\mu$,l and a system of two joint equations has to be solved as
\begin{equation}
G_{exp}=G(\mu,\lambda)
\label{eqn:sistem1}
\end{equation}

\begin{equation}
\lambda^{k-m} =\frac{\widetilde{M}_{m}}{\widetilde{M}_{k}}
\label{eqn:sistem2}
\end{equation}
 
where the quotient $\widetilde{M}_{m}/\widetilde{M}_{k}$ is also a function of $\mu$ and $\lambda$.  The solutions of the non-linear system can be found numerically by the Newton-Rapshon algorithm starting from the initial values of the DSD parameters given by the previous procedure. 
The system of equations formed by (\ref{eqn:sistem1}) and (\ref{eqn:sistem2}) for specific moment subsets has been used in the past \citep{Vivekanandam2004} and more recently \citep{KumarIEEE2010,KumarIEEE2011}. In our case, we evaluated the relevance of $D_{min}$ (given that the relevance of $D_{max}$ requires that it should be compared at all times with the large drop sampling problems). The expression used for the moments that will be introduced in (\ref{eqn:sistem1}) and (\ref{eqn:sistem2}) is therefore,
\begin{equation}
\widetilde{M}_{k}=N^{(g)}\frac{\Gamma(\mu +k+1)-\gamma(\mu +k+1,D_{min}\lambda)}{\lambda^{\mu+k+1}}
\end{equation}

 \begin{figure}[H]
 \vspace*{2mm}
 \begin{center}
 \includegraphics[width=0.80\textwidth]{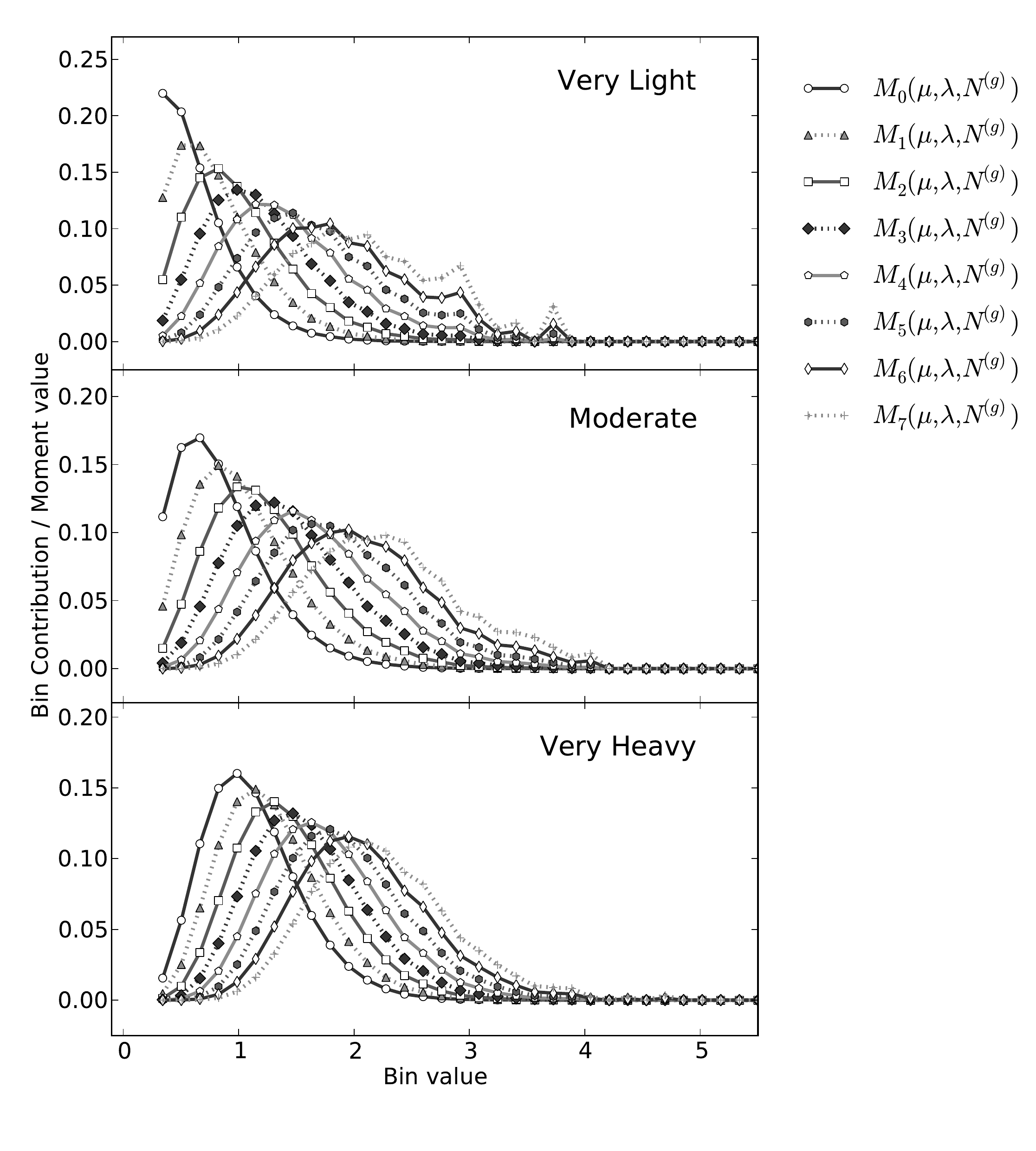}
 \end{center}
\vspace{0.75cm}
\caption[Relevancia de cada intervalo de clase en los parámetros integrales de la precipitación]{\textbf{The relative contribution of each bin to each DSD moment.} The relative contribution of each bin to the value of each moment was determined for each category and for the uniform bins similar to those of the 2DVD disdrometer. The curves are progressively displaced to greater diameters and approach functions that can be modelled by means of Gaussian distributions. This modelling allows for the interpretation of Figure (\ref{fig3}). These results are based on the simulation of 1000 samples for each category. }\label{fig4}
\vspace{0.75cm}
 \end{figure}

\subsubsection{Maximum Likelihood Estimation}

This method is based on the existence of a likelihood function (ML) that, with a given population (a distribution function). could generate the observed sample. The ML function is defined as follows: 
\begin{equation}
\mathcal{ML}(\{D_{i}\};\mu,\Lambda)=\prod_{i=1}^{n}
f(D_{i};\mu,\Lambda)
\end{equation}
for a sample of size n, where the two parameters $\mu$ and $\Lambda$ of the gamma function $f(D)$ are given by equation (\ref{eqn:gamma}). The mathematical procedure used to determine the estimators of both parameters requires maximising function ML \citep{kliche_smith_etal_2008_aa}.

\subsection{Results}

The results were structured as follows: a visual study of the artificial and experimental composite DSDs is shown. A detailed analysis of the results for the integral parameters of the precipitation in each type of disdrometer was presented, considering the relevance on the correlation between them and evaluating the role of modelled instrumental errors prior to performing the binning processes. Regarding the DSD parameters, different estimation methods were compared. The experimental data that originated in the Parsivel OTT disdrometers were also presented. 

\subsubsection{Overview of composite DSDs}

The generated DSDs are similar to the underlying gamma distribution functions if we analyse the average DSD for a sufficient number of cases (a stable form is usually reached after accumulating 50 DSDs). There is the possibility that slight instabilities may remain for drop diameters of $D<1$ mm after the binning processes (see Figure \ref{fig2}a), and depending on the rain intensity, variations may also persist for large drop sizes (of diameters $\gtrsim 4$ mm), similar to real cases.\\

For typical stratiform rain situations, the use of the classifications in Table (\ref{t1}) combined with the temporal series of precipitation intensity values produces monotonous composite DSDs similar to those of the experimental studies. To determine the relevance of each category of intensity in real cases, Table (\ref{t1}) indicates the minutes registered for each interval of intensity for stratiform and convective episodes using Parsivel OTT instruments. Figure (\ref{fig2})b shows two episodes, first stratiform and then convective, which are registered using a dense network of 14 Parsivel OTT instruments.

\subsubsection{Integral rainfall parameters}

The first issue is the relevance of the binning process to the estimation of the various integral parameters for the precipitation, which we write generically as
\begin{equation}
M_{k}=\int_{0}^{\infty} D^{k}N(D)dD\simeq \sum_{i=0}^{N_{bins}}N(D_{i})D_{i}^{k}\Delta D_{i}
\label{eqn:aprox_Mk}
\end{equation}

The usual approach is to approximate the integral using the sum over the disdrometer bins as indicated in (\ref{eqn:aprox_Mk}). The values of $D_{min}=D_{0}$ and $D_{max}=D_{N_{bins}}$, as well as the bin density in specific zones of the spectrum, led to systematic deviations in the estimates for the hypothetical underlying population values of $M_{k}$. This clarifies the results in Figure (\ref{fig3}) based on Figure (\ref{fig4}), where the relevance of each zone of the spectrum of sizes is observed in the DSD moments for each category of rain intensity (under the assumtion of a uniform binning process).\\

These results should be interpreted together with the general bias properties of the moment estimators \citep{smith_kliche_2005_aa}. It is acknowledged that due to sampling, the integral rainfall parameters of the gamma distribution are biased and the differences between the analytical value and sampled value increases with the order of the moment. The ratio between sampled and analytical values is shown in Figure (\ref{fig3}).\\

 \begin{table}[t]
\vspace{1cm}
\caption[Relevancia de los procesos de discretización en las correlaciones entre momentos de diferentes ordenes]{\textbf{Relevance of binning process on correlations between DSD moments.} The results of the correlations between moments of different orders for 5000 samples are compared after the binning process. The resulting theoretical equation (\ref{eqn:correlacionMomentos}) is shown, and the experimental results obtained from different instruments are also reported.}\label{t3}
\vspace{0.8cm}
 \centering
\small
\ra{1.15}
 \begin{tabular}{lcccccc}
 \textbf{Category}  &  \textemdash     & \textbf{Very}&\textbf{Light}& & & \\
\toprule
Disdr.   & $\rho_{0,3}$     &$\rho_{0,4}$  &$\rho_{0,6}$ &$\rho_{2,6}$  &$\rho_{3,4}$  &$\rho_{3,6}$ \\
\midrule
R-Trunc      & 0.39   & 0.25  & 0.10 & 0.50  & 0.95 & 0.73\\
L-Trunc      & 0.63   & 0.39  & 0.06 & 0.26  & 0.93 & 0.45\\
2DVD         & 0.39   & 0.23  & 0.06 & 0.28  & 0.93 & 0.47\\
Parsivel     & 0.41   & 0.23  & 0.06 & 0.21  & 0.91 & 0.41\\
OSP          & 0.61   & 0.41  & 0.05 & 0.50  & 0.95 & 0.71\\
Thies        & 0.39   & 0.23  & 0.16 & 0.26  & 0.93 & 0.46\\
POSS         & 0.45   & 0.28  & 0.11 & 0.48  & 0.95 & 0.70\\
JWD          & 0.43   & 0.26  & 0.10 & 0.49  & 0.95 & 0.71\\
Theoretical  & 0.40   & 0.25  & 0.09 & 0.45  & 0.95 & 0.66\\
\midrule
\\
\vspace{0.5cm}
\\
 \textbf{Category}&  \textemdash   &  \textbf{Moderate} & & & &  \\
\midrule
Disdr.   & $\rho_{0,3}$     &$\rho_{0,4}$  &$\rho_{0,6}$ &$\rho_{2,6}$  &$\rho_{3,4}$  &$\rho_{3,6}$ \\
\midrule
R-Trunc      & 0.49   & 0.33  & 0.14 & 0.53  & 0.96 & 0.73\\
L-Trunc      & 0.61   & 0.42  & 0.15 & 0.47  & 0.95 & 0.68\\
2DVD         & 0.48   & 0.32  & 0.11 & 0.47  & 0.95 & 0.68\\
Parsivel     & 0.49   & 0.32  & 0.11 & 0.47  & 0.95 & 0.68\\
OSP          & 0.60   & 0.41  & 0.16 & 0.50  & 0.95 & 0.72\\
Thies        & 0.48   & 0.32  & 0.11 & 0.47  & 0.95 & 0.69\\
POSS         & 0.51   & 0.34  & 0.13 & 0.50  & 0.95 & 0.71\\
JWD          & 0.50   & 0.33  & 0.12 & 0.47  & 0.95 & 0.68\\
Theoretical  & 0.48   & 0.32  & 0.13 & 0.49  & 0.95 & 0.69\\
 \bottomrule
 \end{tabular}
\vspace{1.8cm}
 \end{table}

The first implication observed in Figure (\ref{fig3}) is a bias at the moment $M_{k}$, which depends on the category but has systematic characteristics. Disdrometers that do not have bins with small diameters underestimate the first moments (most notably in cases of slight precipitation intensity in which the differences can be greater than 20\%), while the Parsivel OTT and Thies overestimate the greater moments (note that because of the sampling bias the effective deviation of Parsivel for higher-order moments due only to binning is slightly less than that shown in the figures). For those disdrometers, this is explained by the fact that they have fewer bins in the 2 to 4 mm interval. The effect of the difference on the size range of this bin quantity is also observed in POSS disdrometers for moderate to heavy intensities. In general, for the intense rain case, the differences in the smaller moments are smaller because the DSD has a less significant role for small drops. Only in the case of the OSP and Left-
Truncated do these differences persist and interfere with comparisons for smaller diameters. \\

When an uncertainty is introduced in $\mu$ (representing possible small fluctuations in the shape parameter of the gamma distribution) the results are analogous, but the sampling bias obtained is mainly increased for intense rainfall, while the binning effects seem additive regarding this kind of sampling issue.\\

\paragraph{Correlation between integral parameters:}
The essence of the scaling-law formalism \citep{semperetorres_porra_ea_1994} rely on relationships between integral rainfall parameters (see also, \S\ref{sec:Scaling1moment}). Therefore we evaluated the binning effects on the correlation between DSD moments. The pearson correlation estimate between the integral rainfall parameters can be compared with the expression from \citep{1987Chandra}:

\begin{equation}
\rho_{k,l}=\frac{\Gamma(\mu+k+l+1)}{\left[\Gamma(\mu+2k+1)\Gamma(\mu+2l+1)\right]^{1/2}}
\label{eqn:correlacionMomentos}
\end{equation}
In Table \ref{t3}, we compare the pearson correlation values for the following categories: very light and heavy for five different combinations of the k and l values. \\

In general, the absence of small drop measurements implies an overestimation of the correlations that involves lower order moments. This overestimation is systematic for low intensity episodes; however, according to the DSDs, it is shifted to greater diameters when the rain intensity increases. The overestimation may change to a underestimation by disdrometers with theoretically improved estimates of moments such as 2DVD if higher-order moments are involved.\\

\subsubsection{Inclusion of instrumental errors in the simulated DSDs }

As noted in section \S\ref{sec:gen_gotas}, aside from sampling problems and drop size spectrum discretisation, each instrument has its own characteristics that govern its precision (difference between measured values using similar devices) and accuracy (similarity between estimated and real values). Therefore, we performed a qualitative study using a network of Parsivel OTT disdrometers, as illustrated in Figure (\ref{fig2}).\\

In this study, two events with different characteristics were measured with 14 Parsivel OTT disdrometers. Two of these instruments were more advanced versions that are more capable of measuring small drops, leading us to believe that the original disdrometers exhibit a deficit in the measurement of drops with diameters of $D<0.8$ mm. Between 0.8 and 2.5 mm, the measurement is quite precise and, as inferred from comparisons with other instruments in several previous studies, is also accurate. Lower precision was observed for sizes of $D>2.5$ mm, a fact that arises from sampling issues. Regarding the accuracy in this range of sizes, a comparison with new devices indicates that traditional Parsivel OTT devices overestimate the presence of large drops in convective cases (greater rain intensity). For all other purposes the deviation was systematic. \\

 \begin{figure}[H]
 \vspace*{2mm}
 \begin{center}
 \includegraphics[width=0.70\textwidth]{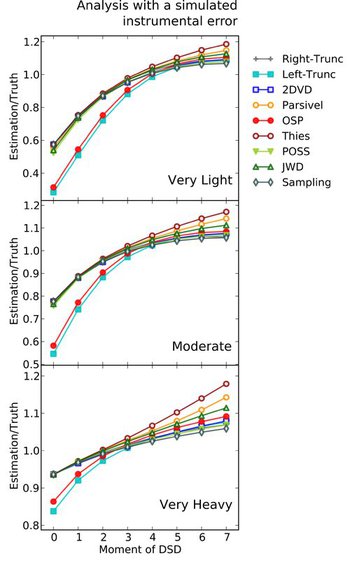}
 \end{center}
  \caption[Relevancia en los momentos de la DSD de la inclusión de un modelo de error instrumental en la estimación de gotas pequeñas]{\textbf{Relevance on DSD moments of the instrumental errors.} For the very light, moderate and very heavy precipitation categories, an instrumental error simulated using the common scheme explained in the main text has been included. As in Figure (\ref{fig3}), these results represent the evaluation of 5000 samples. In this case, the sampling represents the combination of the sampling problems and uncertainties caused by the simulated instrumental error. }\label{fig5} 
\vspace{0.5cm}
 \end{figure}

 \begin{figure}[H]
 \begin{center}
 \includegraphics[width=1.0\textwidth]{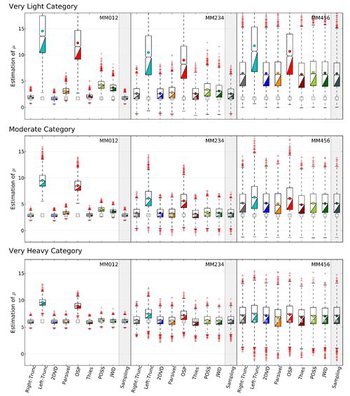}
 \end{center}
  \caption[Estimación de los parámetros de la distribución gamma por el método de los momentos]{\textbf{Estimates of DSD shape parameter compared for three different categories by three moment methods }The estimates of $\mu$ are compared for three different categories by three moment methods. The results are shown for the entire group of binnings analysed with experimental distributions of the estimated $\mu$ and the corresponding box-plot diagrams. The central line in the box-plot represents the median, while the two inferior and superior lines that define the box represent the first ($Q_{1}$) and third ($Q_{3}$) quartiles. The lines that define the box-plot extend to 1.5 times the value of $IQR=Q_{3}-Q_{1}$, which is further than the first and third quartiles. If the values are not within this range, they are considered outliers. The empty squares represent the real values (reported in Table \ref{t1}); the circles represent the average values of the distributions. The outliers are represented as red 
addition symbols. The box-plots are based on 5000 samples. From left to right each column reports the results obtained with the moment methods MM012, MM234 and MM456.}\label{fig6}
\vspace{0.5cm}
 \end{figure}

These qualitative characteristics of instrumental error were previously suggested \citep{hauser_amayenc_etal_1984_aa} to arise from the physical principle by which optical disdrometers measure with a \emph{non-perfect} laser beam (even though their quantification depends on each specific design). Here, we experimentally show these characteristics for many identical instruments. According to the original hypothesis, accuracy problems are due to the intrinsic noise of the instrument signal. This hypothesis is verified, given that the improvement in the signal has allowed for the differentiation of small drops from the background noise and made it possible to distinguish between simultaneous processes in drops in the laser beams, which implies an estimate of a greater number of small drops and fewer larger drops (differentiated now as several drops of a smaller size). \\

Similary, the errors of each type of instrument must be analysed using similar experimental designs \citep{tokay_bashor_etal_2005_aa}. The underestimation of the number of small drops appears to be a common characteristic for the majority of disdrometers, while the overestimation of large drops is characteristic of the first version of the Parsivel OTT. Given that comparing the different devices errors for each instrument with sampling and discretisation issues obscures the ability to identify the source of the error, an artificial filter was constructed to introduce some of the instrumental error characteristics that can be considered common to more than one type of instrument (for example, it is acknowledged that the JWD is also unable to measure the smallest drops due to background noise and the \textit{dead time} effect). The main question that we attempt to address with this filter is whether non-critical instrument uncertainties make the presence of binning irrelevant or whether the analysis of the 
binning process remains necessary despite the introduction of these instrumental errors. \\

The instrumental error was simulated under the following assumptions: the error in the interval of sizes $\leq 0.7$ incorporates a linear progressive filter that discards 85\% of the drops that are $< 0.2$ mm and 10\% of the 0.8 mm drops. Furthermore, for diameters between 0.8 and 3, a Gaussian error of 5\% in the real drop width was incorporated. For larger diameters, we suppose that the sampling was the greater problem and we did not introduce errors in the real drop size estimates, although different disdrometers can undervalue or overvalue the number of drops in these sizes depending on the precipitation conditions. The results are shown in Figure (\ref{fig5}). \\

Under these conditions, we observed how the differences between the different disdrometers decrease for lower-order moments, although we visualised differences between the OSP disdrometer and Left-Truncated binning and the other devices. All of the instruments undervalue the first moments to a similar degree. However, the differences in higher-order moments follow the same trend, and additives may be considered in sampling problems featuring large drops. Finally, in the case of the first Parsivel OTT model for cases of intense rain, the global deviations in the measurements of this instrument imply a positive bias for reflectivity of approximately 5-10\% due to binning (over the error due to sampling) and another overestimation due to instrumental error (at least in the case of first Parsivel OTT generation). This situation will also be true, although more moderately, for the precipitation intensity. In the case of the Thies, the impact of lower density of bins in the range from 1 to 3 mm is shown. \\

 \begin{figure}[H]
 \begin{center}
 \includegraphics[width=1.20\textwidth,angle=90]{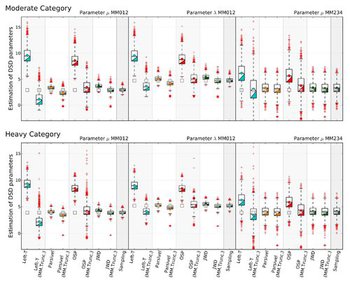}
 \end{center}
   \caption[Comparación entre las estimaciones los parámetros de la distribución gamma por el método de los momentos usual y truncado para diferentes instrumentos]{\textbf{The gamma distribution estimates are compared using the moment and truncated moment methods for two different categories and for a subset of disdrometers analysed that are more sensitive to small drops.} The experimental distributions of the estimated $\mu$ and $\lambda$ parameters were constructed as box-plot type diagrams. The empty squares represent the real values, and the circles represent the average values of the samples. The line that divides the box-plot is the median, and the boxplot shows different quartiles. Compare the presence of outliers in this Figure with those shown in Figure (\ref{fig6}). The box-plots are based on 5000 samples.}\label{fig7}
 \end{figure}

\subsection{DSD parameter estimates}

Comparing the different estimation methods for DSDs implies deciding that uncertainties in the estimation can have a greater effect in practice, which can depend on the specific use of the DSD measurements. One of the most commonly used methods is the mean squared error (MSE), defined for the case of the $\mu$ parameter as $MSE_{\mu}(\hat{\mu})=\left<(\hat{\mu}-\mu)^{2}\right>=Var(\hat{\mu})_{\mu}-bias(\hat{\mu})_{\mu}$, where the bias is the deviation from the average: $Bias(\hat{\mu})_{\mu}=<\hat{\mu}>-\mu$ which is another statistic used to determine the method of estimation. Each estimator $\hat{\mu}$ would have an average quadratic error and a bias that would depend (or not) on the value of $\mu$. Worse difficulties exist, such as having to characterise the estimator more broadly using other statistics (if the distribution of values of $\hat{\mu}$ presents peculiar properties) or including more robust estimators than usual. One practical way of comparing the different estimators based on our objective 
is to use box-plot diagrams that show in a compact manner many properties for the distribution of values found using each methodology. \\

\paragraph{Moment method:}

For the case in which the N(D) is estimated, understanding the relevance of binning for each of the existing methodologies is significant. The different method estimates for a broad sample of DSDs and the corresponding statistical properties were studied according to the categories of very light rain, moderate rain and very heavy rain and were compared to an estimate that directly uses the sample unclassified in bins (whose error originates only from the sampling), rather than performing discretisation. \\

The statistical properties of the estimator $\hat{\mu}$ are shown in the figure (\ref{fig6}). To build the box-plots, 5000 different samples were considered (more than $5\cdot 10^{5}$ drops were analysed in each case). This allowed us to assert which moment method is preferred according to the rain intensity and the several binning processes. \\

As shown in the figure (\ref{fig6}), for the MM456 case, the binning is less relevant than in other cases, as the sampling process masked the discretisation process, although major errors exist in the accuracy of the estimates. Cases MM234 and MM012 are more sensitive to the concrete characteristics of the disdrometer, implying that the bin selection of, for example, the JWD, POSS or Parsivel OTT disdrometers produces sensible deviations. The MM346 (not shown) exhibits intermediate behaviour between the behaviour of the MM234 and MM456 cases. \\

\paragraph{Truncated moment method:}

The truncated moment method, which incorporates a hypothesis regarding the size interval in the DSD estimation process, is used when DSD parameter prediction problems arise for the traditional moment method in which the bins fail to measure or undervalue small drops. We have restricted these analyses to the MM012 and MM234 methods, which exhibit sensitivity to the smaller diameters, and we report a comparison of the JWD, OSP and Parsivel disdrometers. \\

 The distribution of the resulting parameters has an average value that is similar to the real value and a distribution that is similar to that derived from the sampling process. The estimates change from overestimates to underestimates with the significant caveat that the distribution of values in the case of parameter $\lambda$ is notably biased. Apparently, the median is preferred over the average for this estimate. This caveat is explained by a significant growth in the marginal distribution values under a calculation that progressively involves up to 5000 DSDs in each of the categories. The averages in the heavy and very heavy cases are notably displaced, an aspect that is not observed in the remaining categories. These observations indicate that, the use of the median appears to be more robust than the use of the average, and the robust alternative is to use a trimmed mean or a Winsorised mean. \\

 \begin{figure}[H]
 \begin{center}
 \includegraphics[width=0.60\textwidth]{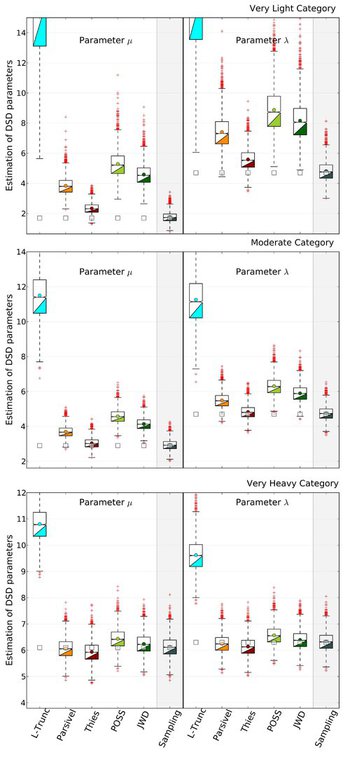}
 \end{center}
  \caption[Estimación de los parámetros de la distribución gamma por el método de de máxima verosimilitud]{\textbf{The predictions are compared by means of the MLE method for several binning methods.} The predictions are compared by means of the MLE method for several binning methods. The Left-Truncated and OSP provide results similar to those of MM012, thus the OSP is omitted in this figure. Additionally, the Left-Truncated is not entirely visualised to allow for a better visualisation of the detailed differences between the disdrometers shown here. Similarly, the 2DVD case is located between that of the Thies device and sampling, and it is omitted. The box-plots are based on 5000 samples.} \label{fig8}
 \end{figure}

\paragraph{Maximum likelihood estimation method:}
The problem for small drops persists in the maximum likelihood estimation (MLE) method, as reported in other studies \citep{mallet_barthes_2009_aa,cao_zhang_2009_aa}. Here, the objective of applying the MLE method is mainly to observe if the distributions of estimator parameters of the DSD are similar to those obtained with the moment method. The distribution of 2DVD sizes was sufficient to continue with the sampling process; verifying the DSD differences at this level is interesting. The MLE results are very similar to those of the MM012 method, implying that the measurement of small drops in the spectrum is highly sensitive. Figure (\ref{fig8}) includes a comparison of three disdrometers with uniform cases and distribution due to the sampling. \\

\section{Sampling vs. binning effects on experimental DSDs}
\label{sec:samp_bin}

Real measurements must deal with both, sampling issues and binning processes. The measurement of 2DVD disdrometers offers us the possibility of addressing both issues. In the following sections are explained the properties of the data-set and the methods used in the analysis are explained.\\

\subsection{Experimental data: 2DVD disdrometer}

The dataset was measured by a 2DVD video disdrometer from the MC3E (Mid-Latitude Continental Convective Clouds Experiment) in Central Oklahoma during the   spring and summer campaigns of 2011. The 2DVD disdrometer is an advanced optical instrument that measures three properties (drop size, vertical velocity and shape) of the collection of drops that cross the sampling area. \\

 One primary advantage of the 2DVD instrument is the possibility of recording a drop-by-drop database. This property was used to analyse different binning processes with real data. With the goal of obtaining a consistent dataset, a filtering technique was applied to filter spurious drops whose terminal velocities differ by more than 50\% from the usual relationship, which is the same scheme previously used to pre-process Parsivel OTT measurements. \\

 \begin{table}[H]
\caption[Episodios de precipitación obtenidos de la base empírica medida por un disdrómetro 2DVD del experimento MC3E]{\textbf{Precipitation events from 2DVD data-set.} Precipitation events from 2DVD data-set. (*) Number of minutes with more than 100 drops after appling a typical filter for terminal velocities. Accumulated rainfall is measured in [mm] and maximum rainfall in [mm/h].}\label{t4}
\vskip8mm
 \centering
\small
 \begin{tabular}{cccrrrrrr}
 \toprule
 Event & Minutes(*) & Date          & $R^{2DVD}_{acc}$ & $R^{2DVD}_{max}$ & $R^{OTT}_{acc}$ & $R^{OTT}_{max}$ & $R^{OSP}_{acc}$ & $R^{OSP}_{max}$  \\
      
 \midrule
A & 54  & 24 April 09:40 to 12:15   &  2.20         & 14.63           & 2.19          & 14.87    &  2.15  & 14.37 \\
B & 78  & 24 April 17:38 to 20:41   &  0.69         & 1.85            & 0.66          & 1.88     &  0.12  & 1.57 \\
C & 184 & 25 April 09:06 to 16:26   & 19.46         & 56.4            & 19.26         & 55.39    &  16.98  & 47.40 \\
D & 240 & 27 April 05:36 to 13:58   &  6.83         & 9.19            & 6.78          & 9.09     &  6.60  & 9.01  \\
E & 109 & 01 May   16:05 to 20:28   &  3.14         & 8.73            & 3.10          & 8.67     &  2.90 & 8.54 \\
F & 220 & 11 May   18:08 to 23:01   &  7.13         & 11.97           & 7.04          & 11.95    &  6.89 & 11.26  \\
All & 885 & -                       &  39.45        & 56.4            & 39.04         & 55.39    & 35.64 & 47.40 \\
 \bottomrule 
\vspace{1cm}
\end{tabular}
 \end{table}

 \begin{figure}[H]
 \begin{center}
 \includegraphics[width=1.00\textwidth]{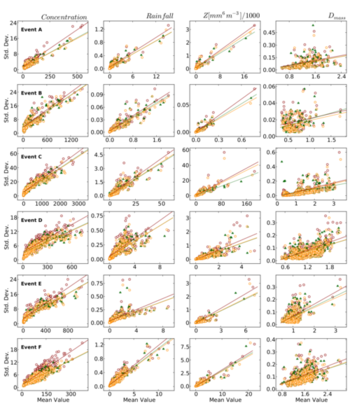}
 \end{center}
 \caption[Estimación de la desviación estandar de cada estimador de los momentos de la DSD para cada episodio de precipitación contenido en la Tabla (\ref{t4})]{\textbf{Estimation of standard deviation of each DSD moment estimator by a re-sampling technique.} Estimation of standard deviation of each DSD moment estimator by a re-sampling technique. For each experimental DSD, the standard deviation over the built sub-samples is calculated. Then, the value of standard deviation is interpreted as an estimator of the sampling error of the mean value. The meanings of colours are the same as those in Figure (\ref{fig1}). The values of reflectivity are scaled by a factor 1000. Linear regressions were included to indicate the general increasing tendency in the estimation of sampling error with the mean value. Each point represents the experimental DSD over a time resolution of 1 minute with at least 20 drops. }\label{fig9}
 \end{figure}

 \begin{figure}[t]
 \vspace*{8mm}
 \begin{center}
 \includegraphics[width=1.05\textwidth]{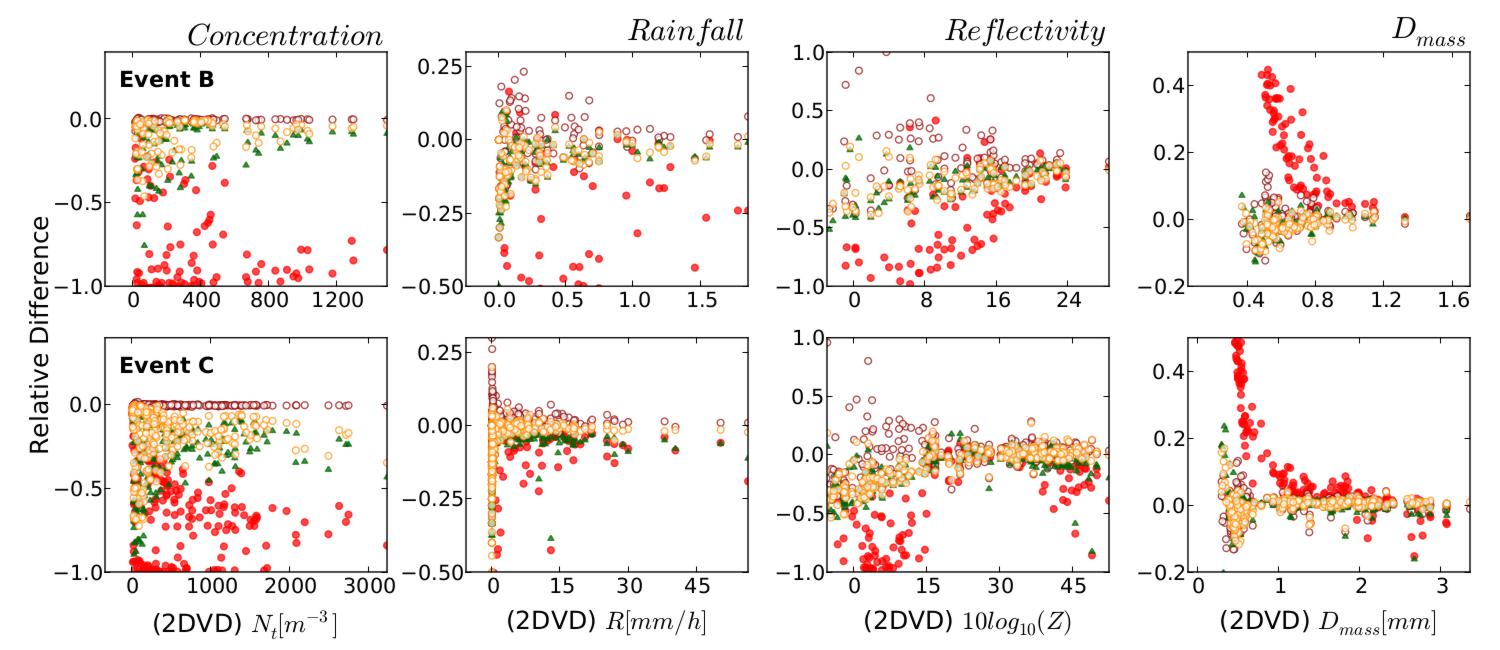}
 \end{center}
\vspace{0.5cm}
 \caption[Diferencia relativa entre la estimación de cada disdrómetro respecto de la medida del disdrómetro 2DVD para varios parámetros integrales]{\textbf{Relative difference $(\overline{X}_{D}-X_{2DVD})/X_{2DVD}$}. Relative difference $(\overline{X}_{D}-X_{2DVD})/X_{2DVD}$ where the disdrometer D was successively OSP, Thies, Parsivel OTT and JWD, and X is an integral rainfall parameter. The difference is calculated between the estimation of mean values for each disdrometer by a re-sampling technique and the original value of the 2DVD disdrometer.  The meanings of the colours are the same as those in Figure (\ref{fig1}). Each point represents the experimental DSD over a time resolution of 1 minute with at least 100 drops. Events B and C were the events with the fewest and greatest values of total accumulated rainfall, respectively.}\label{fig10}
\vspace{0.5cm}
 \end{figure}

\subsection{Generating DSDs detected by different instruments}

To be able to faithfully simulate the binning process of different disdrometers, we need to include information about the sensing areas, such as that shown in Table (\ref{t5}). For this reason, the collection of drops detected by different instruments is estimated by a two-step method: 
\begin{itemize}
 \item (a) using the drop-by-drop dataset a random subset with a number of drops proportional to the sampling area is selected \textemdash see Table (\ref{t5})\textemdash \,
 \item (b) classification into bins according to the disdrometers is performed.
\end{itemize}

In the case in which the sensing area is smaller than that of the 2DVD, it was necessary to perform an estimation of the sampling error. This was performed by a standard re-sampling bootstrap technique \citep{Efron1979}. The idea is to perform the steps (a) and (b) M times to be able to calculate the reliable estimator characteristics of each instrument for the underlying population of drops. The number of random subsets (DSDs) M of the original 2DVD measurement was chosen to be 50 samples for the 100 drops cases and 100 samples for the 1000 drops cases (with a linear increase of M with the number of drops).  This allowed us to estimate both the average value measured by M identical instruments with smaller sampling areas and estimate the standard deviation of the under-sampling. An analysis of 6 events was performed; the details of those events are provided in Table (\ref{t4}).\\

 \begin{table}[t]
\caption[Áreas de medida de cada disdrómetro utilizadas para el análisis comparativo de los problemas de discrectización y muestreo.]{\textbf{Sampling area of analysed disdrometers.} Sampling area of analysed disdrometers. OSP has a second version with a smaller sampling area but the widely used features a sampling area of 100 $cm^{2}$.  POSS has a much larger sampling volume because it relies on a remote-sensing measurement method.}\label{t5}
 \vskip4mm
 \centering
\small
\ra{1.10}
 \begin{tabular}{lll}

\toprule
 \textbf{Disdrometer}  &  \textbf{Sampling Area} &  \textbf{Measurement Method}   \\
\midrule
Parsivel OTT      & 54 $cm^{2}$  & Optical \\
2DVD              & 100 $cm^{2}$ & Optical (two beams)  \\
Thies             & 45.6 $cm^{2}$ & Optical   \\
JWD               & 50 $cm^{2}$  & Impact \\
OSP               & 100 $cm^{2}$ & Optical\\
POSS              & $>>100\,cm^{2}$ & Radar X-band\\
 \bottomrule
\vspace{0.3cm}
 \end{tabular}
 \end{table}

\subsection{Integral rainfall values for 2DVD measurements}

It is interesting to compare several integral rainfall parameters typically used in DSD studies. To achieve this objective, the total concentration of drops, rainfall intensity, reflectivity and mass-weighted diameter ($M_{4}/M_{3}$) were compared.\\

The first step is to understand the role of the sensing area. The challenge in determining the sampling error characteristics of a 2DVD sensing area is usually met by comparing identical collocated instruments. In our case, given an isolated instrument it is still possible to appreciate the role played by the sampling errors in devices with smaller sensing areas. To better understand these sampling issues, a relationship between the mean values and the standard deviation obtained by the re-sampling technique is shown in Figure (\ref{fig9}). The results show similar patterns for the Parsivel OTT, JWD and Thies instruments; however they also show slight differences. In the case of the Thies larger sampling errors (more obvious in concentration) are observed due to the smaller sensing area of this disdrometer. A roughly multiplicative bias appears for the concentration, rainfall and reflectivity, while in the case of $D_{mass}$, which is the quotient of two consecutive DSD moments, it would be difficult to 
model the relationship between mean values and standard deviation. \\

The second step is to evaluate the binning effects. We study the mean values of the integral rainfall parameters after the re-sampling process because they are supposed to be less dependent on sample-by-sample deviations. Therefore, they should be more efficient in reveling the real differences due to binning. To address those binning effects we used the relative difference with respect to the value of 2DVD, $(\overline{X}_{D}-X_{2DVD})/X_{2DVD}$ where the disdrometer D was successively OSP, Thies, Parsivel OTT and JWD, and X is an integral rainfall parameter . The collection of results is shown figure (\ref{fig10}), where the deviations between relative differences are mainly due to binning effects (an analogous result for simulated DSDs is shown in the figure (\ref{fig4}). \\

\begin{figure}[H]
 \begin{center}
 \includegraphics[width=1.05\textwidth]{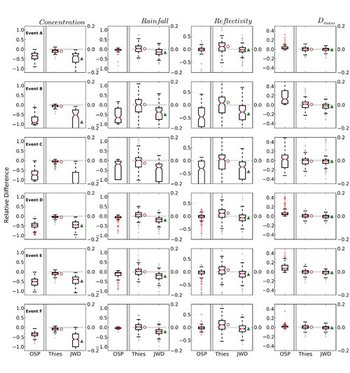}
 \end{center}
\vspace{0.5cm}
 \caption[Diagramas de cajas de la desviación relativa entre cada disdrómetro y el disdrómetro 2DVD para los episidos de precipitación contenidos en la Tabla (\ref{t4})]{\textbf{Box-plots of the relative difference $(\overline{X}_{D}-X_{2DVD})/X_{2DVD}$.} Box-plots of the relative difference $(\overline{X}_{D}-X_{2DVD})/X_{2DVD}$ where the disdrometer D was successively OSP, Thies and JWD, and X is an integral rainfall parameter. The difference is between the estimation of the mean values for each disdrometer, as determined by a re-sampling technique, and the original value of the 2DVD disdrometer.  The meanings of colours are the same as those in Figure (\ref{fig1}). The box-plots are calculated from the experimental DSD over a time resolution of 1 minute with at least 100 drops. The adjacent symbols are the mean values of the relative difference for each event. The results for the Parsivel OTT were intermediate between those of the Thies and JWD disdrometers.}\label{fig11}
\vspace{0.5cm}
 \end{figure}

 The most obvious effect was that of OSP instrument showing that discarded drops with diameters of 0.6 mm indicate relevant differences, as expected from the previous analysis with simulated DSDs. The Thies presents a faithful correspondence with the 2DVD with respect to concentration, in contrast with the JWD and Parsivel OTT. However, the Figure (\ref{fig10}) also shows that Thies presents a tendency for positive bias with respect to Rainfall and Reflectivity, as observed for simulated DSD. These facts are more obvious when histograms of the relative difference or box-plots are compared. The Figure (\ref{fig11}) supports the notion that the deviations present in the simulated gamma DSD persist in real DSD measurements.\\

However, it is important to note that while two different collocated disdrometers should exhibit binning effects, these effects should be considered an asymptotic statistical property. As a result, two disdrometers may have differences due to the sampling masking the binning effects but data accumulated over large periods or statistical analyses performed on an entire dataset show binning effects. This is illustrated in Figure (\ref{fig11}), where the deviations between mean values demonstrate the role of binning on statistical analysis. \\

\section{Conclusions}
 
The simulation of drop size distributions according to the size classifications performed by real instruments determined the significance of the binning process. The sensitivity of each moment and different region of the drop size spectrum explains systematic deviations in the estimation of moments. A smaller density of bins for drop diameters of $D>3$ mm implies a systematic reflectivity overestimation of approximately 5\%, which is additive with respect to other sources of error, such as sampling, and the uncertainties that arise due to errors in the parameter estimates that define the DSD. 
Deviations in the moments depend on both the intensity of the precipitation (through the category classifications used in this study) and on the order of the analysed moment, both of which will be considered in the error evaluations in the moment estimations from DSD modelling. The relevance of the DSD parameter estimates of the binning process has also been evaluated, demostrating that measurement problems for small drops are the most relevant, as they affect both the estimated means of both the moment method and the method based on maximum verisimilitude. \\

Estimates can be improved with the truncated moment method (and MLE analogue), but this method requires robust estimators for the distribution of the various parameter estimates due to the presence of outliers, especially for the parameter $\lambda$.\\ 

The analyses conducted here demostrate that experiments comparing instruments with different bins should be performed in a preliminary study on what methodologies are the most appropriate in accordance with the objectives of each experiment and, above all, with the characterisation of errors. \\

\chapterstyle{FancyUnnumberedChap}

\renewcommand{\chaptername}{}  

\part{Conclusions}
\chapter{Conclusions}

Throughout this dissertation, different issues have been broached relating to the properties of the drop size distribution of precipitation. The main questions posed are the following: 

\begin{itemize}
 \item At the pixel scale of the meteorological radar, does a spatial variation in the DSD exist? If so, how is it reflected? 
 \item There are different DSD models that nonetheless present difficulties in the case of high temporal resolutions or in which the underlying distribution is bimodal or multimodal. Does a more robust method exist for the modeling of DSD?

\end{itemize}

Answering the first question has been possible owing to the first key experiment, which was based on a dense and homogeneous network of disdrometers. The question was answered by posing the hypothesis of appreciable variation in the DSD at the kilometer scale. The hypothesis was validated by analysis of the different metrics (e.g., correlogram and analysis of the Z-R relationships).\\

In answering the second question, artificial DSDs were created, thereby allowing us to understand the processes of DSD modeling. The DSDs were used to validate the hypothesis that the method of maximum entropy offers adequate modeling for high temporal resolutions and multimodal DSDs and that it improves the existing models based on the gamma distribution.\\

As a complement, another source of bias in the precipitation measurements was analyzed. In this case, we offered an analysis of the relevant characteristics of the histogram creation process for the disdrometric measurements, as well as the effects of underestimation of small drops.\\

Next, we detail the results obtained in each of the issues analyzed. 

\section{Spatial variability in the drop-size distribution}

In light of the first question, the results indicate that:

\begin{itemize}
 \item The $v(D)$ relationships for different episodes and throughout the entire network show consistent results with that of prior studies, which used only one instrument. The values corresponding to the fits of the functional form $v(D)=\delta D^{\gamma}$ reveal certain systematic deviations with respect to the traditional relationship (\ref{eqn:AtlasVDequation}). The differences were quantified throughout the network via average values and typical deviations in the $(\delta,\gamma)$ coefficients.
 \item The set of precipitation-intensity time series shows a multiplicative bias in the rainfall-intensity estimate, specifically if it is determined using the standard deviation obtained throughout the instrument network with the precipitation-intensity value used as the average.
 \item The correlogram for the R and Z parameters exhibits a decrease in the correlations among time series throughout the network as the distance between instruments is increased.
 \item The previous assertion is independent of the adopted methodology in correlogram construction. This conclusion arises from the comparison of various methods, including the original time series, their logarithmic transformations or modeling via a bivariate mixed log-normal distribution. This last method shows a greater decay in the correlations with distance.
 \item The differences in the Z-R relationship estimates were determined throughout the instrument network. It was demonstrated that the variation within the same episode can surpass the variation among several episodes. This fact was analyzed via various methods of determination for the Z-R relationship. In particular, the issue was studied by means of the SIFT method with different hypotheses for its application. 
 \item The DSD was studied via the scaling method, reaching the conclusion that the variations observed throughout the network and resulting in the model can be as significant within an episode as among various episodes. 
 \item The scaled DSD was modeled via a gamma distribution, observing the relationships between the $\mu$ and $\lambda$ parameters that are of physical origin. This fact has been observed in experimental gamma distributions, but it had not been previously studied in scaled DSD modeling.
 \item This last aspect requires more investigation, in particular, verification of whether, after a scaling process, the sample problems persist and thus give rise to artificial relationships between $\mu$ and $\lambda$.
\end{itemize}

\section{Modeling via maximum entropy}

In the case of modeling via maximum entropy, the following conclusions were reached:

\begin{itemize}

\item The maximum-entropy model applied to the DSD improves other, more traditional methods for synthetic data and experimental data, mainly when the DSD temporal resolution is high, or multimodality exists. The improvement in MaxEnt was achieved both in terms of reproducing a given histogram, as well as in terms of the integral precipitation parameters.

\item In the cases in which there exists a sufficient number of drops and measurements of all sizes, the MLE method satisfactorily represents the sample. However, when either of these two conditions is not met, the method of moments improves the MLE method. In contrast to both, MaxEnt proves to be a more systematic and progressive approximation of the empirical information.

\item In the synthetic case, different methods were evaluated with an underlying hypothetical distribution. In the experimental case, errors appear in the model, decreasing the capacity to represent the experimental DSD via the usual methods. Compared with the method of maximum entropy, the method presented in this study permits finding the least biased distribution capable of fulfilling a given set of constraints imposed by the empirical information. Therefore, the spatial variability study is not based on the ability of a fixed functional form to represent a DSD in all the possible cases; instead, it is possible to select constraints or restrictions with physical and empirical sense. The analysis of the values of the $\lambda_{i}$ parameters provided by the maximum-entropy method gives the information necessary to develop a deeper understanding of the issues involving the DSD.\\

\item This finding raises the possibility of improving the precipitation predictions in two different aspects. The first aspect requires the incorporation of a parameterization with a more physical basis in light of the numerical prediction models. The second aspect requires a new method of analysis and prediction of the Z-R relationships that should be useful in both ground and orbital meteorological radar systems.

\end{itemize}

\section{Analysis of the relevance of the discretization process}

Similarly, the discretization process in disdrometric measurements was studied and characterized, with the main conclusions being given as the following:

\begin{itemize}
\item Utilizing the simulation of drop-size distributions, various aspects were investigated in which discretization processes can be relevant, and the size classifications given by real instruments were used directly. It was shown that discretization cannot be discounted in attempting to obtain precise results. 
\item Systematic deviations were observed in the estimation of moments due to the sensitivity of each moment to different parts of the drop-size spectrum. This indicates that a lower density of class intervals in drop diameters larger than 3 mm implies a systematic overestimation of reflectivity of approximately 5\%, which is also additive with respect to other potential sources of errors, such as insufficient sampling or uncertainties that arise due to errors in estimation of the parameters that define the DSD. 

\item It was observed that deviations in the moments possess dependencies both on the precipitation intensity (via the category classification used in this study), as well as on the order of the moment analyzed. Both factors were evaluated in the moment-error estimation for the DSD models. The relevance in the DSD-parameter estimation from the discretization processes also was evaluated, and it was found that small drop-measurement problems are the most relevant, affecting both the estimation via the moment method and the method based on maximum probability.

 \item It is possible to perform a better estimation utilizing a method of truncated moments (and its analog in MLE), but this implies that the distribution of the $\lambda$ parameter estimates will possess a certain skew.

 \item The possibility of having measurements from the 2DVD disdrometer that allow recording of the parameters of each drop, in turn, allows us to verify whether the previous assertions under the hypothesis of a gamma distribution are true for real measurements. To carry out this verification adequately, the different measurement areas for each instrument should be included as key information. It was shown that beyond the sampling differences implied by the existence of various capture areas, the discretization process continues to be present in real measurements. 
\item Similarly, an important result was obtained, specifically, the standard deviation that arises upon decreasing the capture area due to sampling problems in the drop distribution. 
 \item Based on the analyses carried out in instrument-comparison experiments, it is convenient to conduct a prior study that determines which methodologies are the most appropriate according to the objectives of each experiment, particularly with regard to error characterization.
\end{itemize}

	\renewcommand{\chaptername}{}
	
\chapter{Future research lines}

Las líneas de trabajo futuro se encuadran dentro de los tres ejes desarrollados en esta tesis, que son: variabilidad espacial, modelización y caracterización de errores instrumentales.

\section{Variabilidad espacial: escalas complementarias}

Hasta el momento se ha estudiado la variabilidad espacial en una escala de 3 $km$. Es interesante comparar dicho estudio con dos complementarios en las escalas anexas para realizar una interpretación global de los resultados:
\begin{itemize}
 \item  La escala de hasta 10 metros es interesante de estudiar para analizar la relevancia en posibles
caracterizaciones de la variabilidad espacial en distancias mayores \citep{Parsivel2011Loussane}.
 \item  Como se aprecia en las figuras obtenidas, disponer de una red similar, y situar dis\-dró\-me\-tros extra a distancias de entre 4 y 10 $km$ puede permitir una estimación más precisa de modelizaciones del correlograma. 
\end{itemize}

En el caso de la escala de hasta 10 metros se ha desarrollado una nueva campaña de recogida de datos que utiliza el mismo conjunto de instrumentos descrito en esta tesis. Esta nueva disposición de los instrumentos permitirá una caracterización adecuada del error de muestreo para instrumentos Parsivel OTT. Hasta el momento dicha caracterización solo se ha realizado con hasta tres disdrómetros idénticos, con lo que se espera que el análisis estadístico derivado de la comparación de 14 instrumentos sea mucho más fiable. Así, la nueva campaña posibilitará una descripción cuantitativa de la distribución los errores en los estimadores de la población subyacente de gotas. Además, el experimento se ha acompañado de un anemómetro para contrastar la hipótesis de si una variación mayor estas escalas (para determinados episodios) esta relacionada con fenómenos de turbulencia atmosférica.\\


\section{Máxima Entropía. Otras aplicaciones}

\subsection{Uso de MaxEnt para la variabilidad espacial}

Son posibles varias aplicaciones de métodos de máxima entropía al estudio de la variabilidad espacial. Una primera aproximación viene dada por el cálculo de la entropía relativa, de manera que dada una distribución tomada como referencia se puede estimar la separación de la distribución, en términos de información contenida en ella, en los diferentes disdrómetros de la red.

\begin{equation}
S_{rel}[N]=-\int N(D)\left[ log N(D)-log N_{ref}(D)\right] dD
 \label{eqn:entropyRelativa}
\end{equation}

Otra posible aplicación de máxima entropía viene dada por estudios de geoestadística construidos sobre dicho principio \citep{BMEChristakos}.

\subsection{Cuantificación de los factores de variabilidad}

El principal resultado relativo a la modelización mediante Máxima Entropía ha sido su viabilidad de aplicación y su capacidad descriptiva. La metodología permite modelizar DSD multimodales y con alta resolución temporal y codificar las características de estas en el conjunto de valores de los parámetros $\lambda_{i}$. Sin embargo, es posible realizar además un análisis posterior para identificar, de entre los valores de $\lambda_{i}$, aquellos que, para cada tipo de precipitación sean más responsables (en un sentido estadístico) de la variabilidad de la DSD. El método más tradicional para afrontar esta cuestión sería realizar un \textit{análisis de componentes principales}. Nótese que, añadido al análisis de componentes principales, se podría proponer una argumentación similar en su motivación a la de escalado de la DSD (véase \S\ref{sec:Scaling1moment}) pero basada en los parámetros $\lambda_{i}$ en lugar de los parámetros integrales de la precipitación.

\section{Caracterización de errores instrumentales}

En esta tesis se ha analizado la relevancia del proceso de discretización de la DSD que realizan los diferentes disdrómetros. Además es interesante, en el caso de disdrómetros ópticos, analizar los errores típicos debido a la no homogeneidad (y ruido intrínseco) de la señal que utiliza el disdrómetro óptico Parsivel OTT. Para ello se puede realizar un análisis de laboratorio con lluvia artificial que compare las estimaciones de cada instrumento con los valores nominales de las gotas artificiales tanto de tamaño como de velocidad de caída. A este respecto se dispone, en el experimento a escala de 10 metros en desarrollo, instrumentos Parsivel OTT que proceden de dos diseños de fabricación diferentes, tal y como se describió en el último capítulo de resultados. Esto permitirá hacer un estudio de que mejoras en la haz emitido por los disdrómetros ópticos posee más relevancia práctica. Un estudio similar para otro tipo de disdrómetros ha sido realizado recientemente por \citep{deMoraes2011}.\\

La mayoria de los experimentos describen la incertidumbre debida a muestreo mediante las diferencias entre dos disdrometros
anexos bajo la hipotesis de que su diferencia elimina fuentes de error debidas a la variabilidad natural. Este procedimiento se debe a
los trabajos realizados por Tokay (2005). Recientemente \cite{Parsivel2011Loussane} han realizado una estimacion de la incertidumbre en el
muestreo bajo hipotesis similares: misma DSD real para todos los instrumentos, modelado Gaussiano para el problema de muestreo y estimador de la desviacion tipica relativa mediante la diferenica entre los centiles 90 y 10 (propuesto como estimador robusto). Un estudio factible sería (1) Acotar la hipotesis de misma DSD real a casos de baja turbulencia (estimada mediante el modulo de la velocidad del viento) (2) Comparar diferentes estimadores tanto para parametros integrales de la DSD como para parametros que definan la DSD (3) Realizar un analisis mediante estimadores robustos de los parametros integrales aprovechando las posibilidades que ofrece la red densa situada en el Campus de la UCLM. Los datos de la red de la UCLM (a fecha de Mayo 2012) permiten el analisis para varios episodios estratiforme y un caso convectivo severo.

\chapterstyle{FancyChap}

\part{Appendices}

\appendix
\renewcommand{\chaptername}{}  
\renewcommand{\appendixname}{}  
\renewcommand\chapterillustration{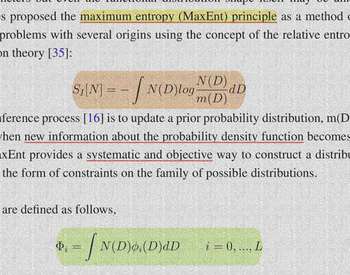}

\chapter{Extended analysis using Maximum Entropy modeling}
\label{chap:MaxEntAMPLIACION}

\vspace{0.5cm}

Results corresponding to the parametrization of the drop size distribution by using the MaxEnt method were presented for first time in this PhD dissertation.  It is also reasonable apply the same methodology to the modelization of the bi-variate joint distribution $f(D,v)$ and the experimental matrix $n(D,v)$, which in fact is the data retrieved by optical disdrometers. In this appendix is developed the parametrization of $n(D,v)$ by showing the method, preliminary results and further possibilities.


\section{Bi-dimensional estimation by MaxEnt}


The previously method exposed on \S\ref{sec:chapMAXENT} was concentrated on the one-dimensional case N(D). But the formal methodology may be applied to bi-dimensional distributions $f(x,y)$ (or in general N-dimensional distributions) with a direct extension of the formalism. Here is detailed the bi-dimensional case for $n(D,v)$ (see \S\ref{sec:defDSD}), 


\begin{equation}
\mathcal{S}_{I}[n]=-\int n(D,v)log \frac{n(D,v)}{m(D,v)} dDdv
 \label{eqn:entropy2dim}
\end{equation}

The constrains are defined as,
\begin{equation}
\Phi_{i}=\int n(D,v)\phi_{i}(D,v)dD dv\qquad i=0,...,L
\label{eqn:constrains2dim}
\end{equation}

The result of maximize the Entropy functional $\mathcal{S}$ constrained to (\ref{eqn:constrains2dim}) gives the following formal solution,
\begin{equation}
\hat{n}(D,v)=\frac{m(D,v)}{\mathbb{Z}}exp\left[-\sum_{i=0}\lambda_{i}\phi_{i}(D,v)\right],
\qquad \mathbb{Z}(\lambda_{1},...,\lambda_{k})=\int
m(D,v)e^{-\sum_{i=0}\lambda_{i}\phi_{i}(D,v)} \label{eqn:pdf}
\end{equation}

where the values $\lambda_{0},...,\lambda_{L}$ represents the free parameters that allows to $n(D,v)$ satisfy the relation (\ref{eqn:constrains2dim}) and maximize the uncertainty expressed by $\mathcal{S}$.



The algorithm presented on the section \S\ref{sec:metodoMaxEnt} is easily adapted to the bi-dimensional case, while the analogous constrains to include to parametrize the $n(D,v)$ are,

\begin{equation}
\Phi_{k,j}=\int n(D,v)D^{k}v^{j}dD dv\qquad i=0,...,L
\label{eqn:constrains2dimMOMENTS}
\end{equation}

Those constrains were previously proposed by  \citep{Dumouchel2009} to study the drop size distributions generated by sprays.\\
In the case in which the objective is describe $f(D,v)$, see section \S\ref{sec:defDSD}, the only change on the previous formalism is the constant $\lambda_{0}$ that allows a normalized solution of the distribution function if it is required.\\



With respect to the numerical solution it is feasible to solve the problem by the Newton-Raphson method (as described on \S\ref{sec:metodoMaxEnt} and successfully applied). However, it should be noted that when the number of constrains is large and, consequently the dimensionality of the system of equations to be solve, the Newton-Raphson method may be inefficient. An alternative is the Broyden–Fletcher–Goldfarb–Shanno (BFGS) method\footnote{Indeed the L-BFGS \citep{liu1989limited}, also have good convergence properties while the memory requirements on each iteration step is significantly lower and should be considered if computational limitations  were presented.} which usually implies more steps to achieve the convergence of the non-linear systems but the efficient of each step in term of computational cost is considerable larger.


\section{Examples of $n(D,v)$ parameterization}
\label{sec:2DIMentropyFUTURE}

\begin{center}
\begin{figure}[H] 
\vspace{1.75cm}
   \includegraphics[width=1.00\textwidth, angle=270]{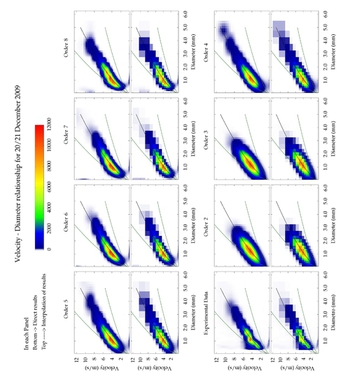}
\vspace{1.15cm}
   \caption[Experimental matrix $n(D,v)$ presented by a diagram $v(D)$. Event 20th December 2009. Liquid Rain event.  Modelization by MaxEnt]{\textbf{Experimental matrix $n(D,v)$ presented by a diagram $v(D)$. Event 20th December 2009. Liquid Rain event.  Modelization by MaxEnt.} The figure shows a comparison between experimental data (left-bottom panel) and progressive parametrization with more contains in the form of equation (\ref{eqn:constrains2dimMOMENTS}). They comparison is performed by introducing direct results and lineal interpolation as a second step. The usual $v(D)$ assumption (\ref{eqn:AtlasVDequation}) is also shown together with the Confidence Interval of 50\%. The experimental matrix $n(D,v)$ shown was not previously filtered or preprocessed.}
\label{fig:Entropy2dim20DIC}
\end{figure}
\end{center}

\begin{center}

\begin{figure}[H] 
\vspace{1.75cm}
   \includegraphics[width=1.00\textwidth, angle=270]{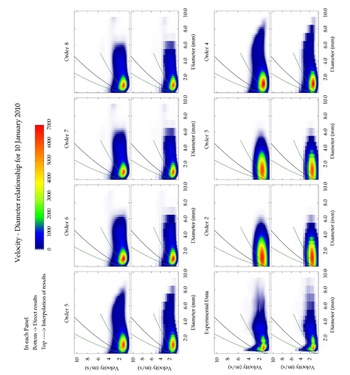}
   \caption[Experimental matrix $n(D,v)$ presented by a diagram $v(D)$. Event 10th January 2010. Snow event. Modelization by MaxEnt.]{\textbf{Experimental matrix $n(D,v)$ presented by a diagram $v(D)$. Event 10th January 2010. Snow event. Modelization by MaxEnt.} The figure shows a comparison between experimental data (left-bottom panel) and progressive parametrization with more constrains in the form of equation (\ref{eqn:constrains2dimMOMENTS}). They comparison is performed by introducing direct results and lineal interpolation as a second step. The usual $v(D)$ assumption (\ref{eqn:AtlasVDequation}) is also shown together with the Confidence Interval of 50\%. The case of $p=8$ exhibits a partial convergence in the terms described on \S\ref{sec:metodoMaxEnt} due to the inefficiency of the Newton-Raphson method. The experimental matrix $n(D,v)$ shown was not previously filtered or preprocessed.}

\vspace{1.15cm}
\label{fig:Entropy2dimNIEVE}
\end{figure}
\end{center}

As an example of the previously described methodology an study of the experimental $n(D,v)$ data of the events analyzed on this dissertation was carried on. In the figure (\ref{fig:Entropy2dim20DIC}), it is shown the results for the event of December 20th-2009, while the figure (\ref{fig:Entropy2dimNIEVE}) shows the case of the snow event measured the January 10th, 2010.
representa el caso de nieve registrado el 10 de enero del 2010.\\


The main objective, at this stage, was analyze the general viability with the proposed algorithm, and assert the general relevance of the several $\Phi_{k,j}$ in the $n(D,v)$ parametrization. With the goal of a preliminary systematic analysis was performed calculations for subsets of constrains $\Phi_{k,j}$ where $k+j=p$, while the value of p was increased progressively, The interpretation those results with a micro-physics point of view requires an evaluation of the relevance of the role of the several $(k,j)$ values\footnote{Generically investigate rigorously an interpretation within the micro-physics point of view would require a compressive comparative of a large number of events with similar micro-physical constrains asserted by external information, and perform the corresponding statistical interpretation in a second stage, situation that exceeds the information available and initial objectives more focuses on the viability of the numerical procedure.}.\\ 



\begin{center}
\begin{figure}[H] 
\vspace{1.15cm}
   \includegraphics[width=1.10\textwidth]{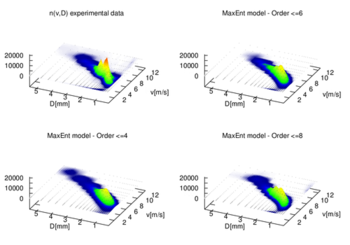}
\vspace{1.25cm}
   \caption[Experimental matrix $n(D,v)$ presented by a 3-dimensional plot. Event 20th January 2009. Liquid Rain event. Modelization by MaxEnt]{\textbf{Experimental matrix $n(D,v)$ presented by a 3-dimensional plot. Event 20th January 2009. Liquid Rain event. Modelization by MaxEnt.} The figure shows a comparison between experimental data (left-top panel) and progressive parametrization with more constrains in the form of equation (\ref{eqn:constrains2dimMOMENTS}). The sub-plots show the cases with p values until 4, until 6 and until 8. The case $p<=4$ allows main details of the matrix $n(D,v)$, the case $p<=6$ allows the analysis within the parametrization of the regions with small number of drops captured. The case $p<=8$ allows a representation of secondary peaks of the matrix $n(D,v)$. The experimental matrix $n(D,v)$ shown was not previously filtered or preprocessed. Slight differences of the the color scale between the sub-plots are present due to differences on the maximum value of the main peak. }
\vspace{0.95cm}
\label{fig:Entropy2dim3D}
\end{figure}
\end{center}

\vspace{2.5cm}

In the previously shown figures (\ref{fig:Entropy2dim20DIC}) and (\ref{fig:Entropy2dimNIEVE}) the total number of drops, $\Phi_{0,0}$, is conserved on all the calculations. The visualization is conditioned by the fact that the experimental peak is sharper than the parametrizations (even the case with maximum p value of 8), this fact is clear in the three-dimensional plots shown in the figure (\ref{fig:Entropy2dim3D}).\\



Evaluated positively the methodology there are several possible applications. Some of them are,

\begin{itemize}
    \item It is possible analyze the relationship $v(D)$ for several subsets of constrains $\Phi_{k,j}$. Assert the relevance of several physical amounts in different parametrization to understand the physical differences in terms of $v(D)$ of a number of various kinds of hydrometers.  
    \item It is feasible introduce physical modelizations or constrains on $n(D,v)$, that are not possible for N(D). Examples are the conservation of the total energy or the total surface of the ensemble of drops. After the analysis in terms of $n(D,v)$ we would retrieve the N(D) according with the premises. Consequently a new point of view in the analysis of the DSD would be available.   
\end{itemize}



Regarding with the meaning of the constrains $\Phi_{k,j}$, interesting cases of physical amounts conserved may be:

\begin{itemize}
 \item $\Phi_{0,0}\Rightarrow$ \textit{Total number of drops}.
 \item $\Phi_{3,0}\Rightarrow$ \textit{Mass} of the ensemble of drops.
 \item $\Phi_{3,1}\Rightarrow$ \textit{Total moment} of the ensemble of drops.
 \item $\Phi_{2,2}\Rightarrow$ \textit{Kinetic energy} of the ensemble of drops.
 \item $\Phi_{2,0}\Rightarrow$ \textit{Superficial energy} of the ensemble of drops.
\end{itemize}



\section{Correction of disdrometric measurements}


In the chapter \S\ref{sec:Instrumentacion} was explained how JWD disdrometers have difficulties to estimate with high accuracy very high intensity rain events because the response time inherent to this devides impose a practial limitation. The proposed methodology to correct the estimations of JWD rely on a multiplicative factor to the number of drops.\\


In an analogous situation it has been shown how disdrometric estimations by Parsivel OTT disdrometers are reliable but also presents difficulties to estimate faithfully the drops with diameters lower than $D\lessapprox 0.7 mm$. The comparison with other kinds of disdrometers like JWD, 2DVD, POSS \citep{thurai_petersen_ea_2011} is showing that in this interval of sizes,  $0<D\lessapprox 0.7 mm$, the uncertainty due to the different estimations of each instruments is higher than in others intervals of sizes\footnote{The relevance of this issue and therefore found a possible correction was proposed by Ali Tokay who also provide to the author of this dissertation with a data-set to compare 2DVD, Parsival OTT and JWD instruments. Several methodologies were proposed by Ali Tokay for the case of composite DSD for a events with a enough number of measured DSDs. Here in this section is discussed a methodology that relay on the same motivation, but applied to disdrometric measurements with a higher temporal resolutions that event by event composite DSDs.} \\

\begin{center}
\begin{figure}[H] 
   \includegraphics[width=1.00\textwidth]{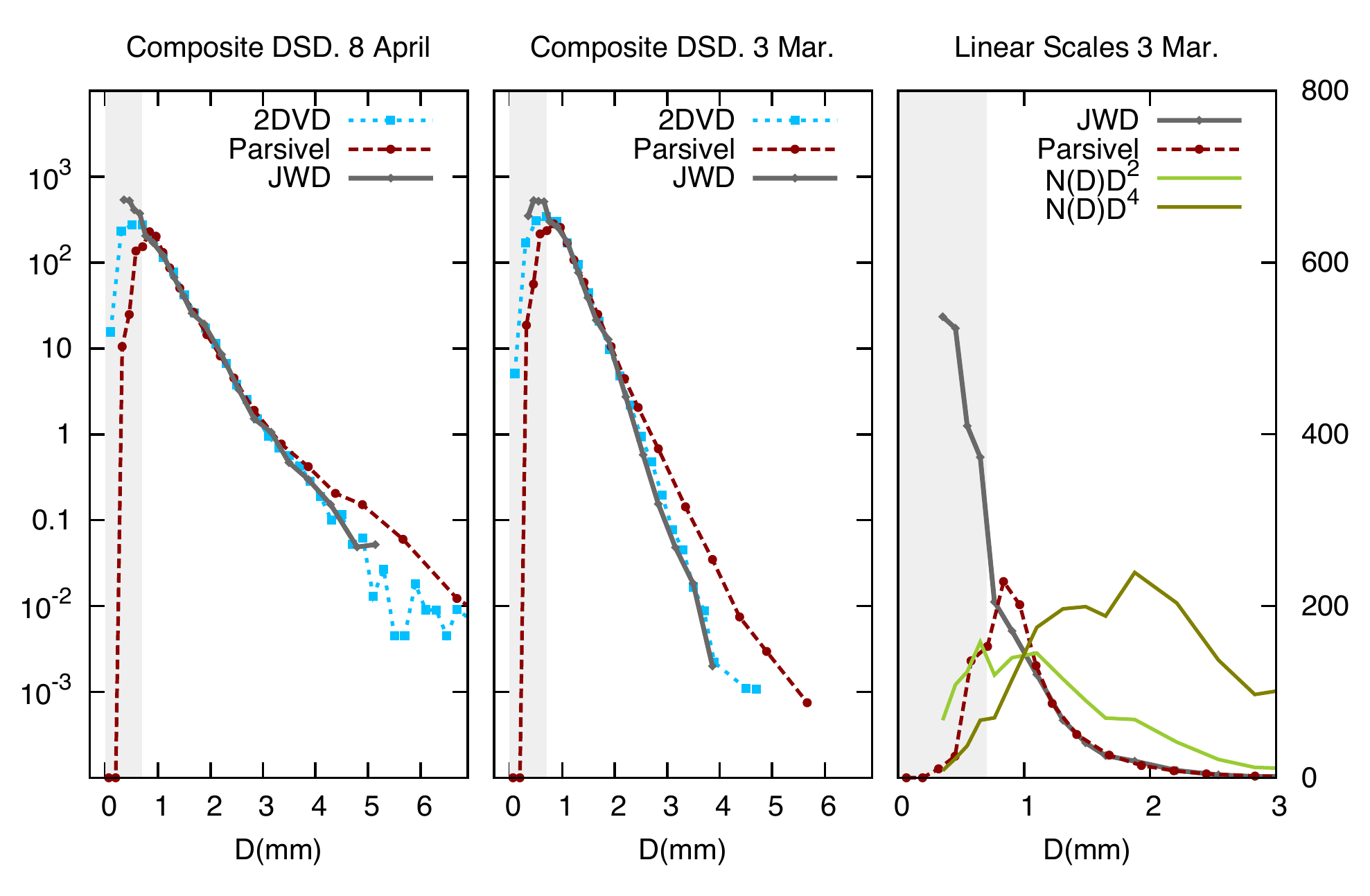}

   \caption[Comparison of disdrometric measurements for JWD, Parsivel OTT and 2DVD disdrometers. Source: Data-set provided by Ali Tokay]{\textbf{Comparison of disdrometric measurements for JWD, Parsivel OTT and 2DVD disdroemters. Source: Data-set provided by Ali Tokay.} A comparison of several DSD estimations for collocated disdrometers for 60 minutes of rainfall. The figure reports differences between JWD and two optical disdrometers. At the right: it is shown with a non-logarithmic scale. It is also shown the differences in the relevance of the several moments of the N(D) for the JWD disdrometer.} 

\label{fig:DisdroCOMPARATIVA}
\end{figure}
\end{center}

In this context should be useful define a filter to reduce the differences between instruments by selecting a confident device as a reference. In the chapter  \S\ref{sec:chapMAXENT} was proven how using a parametrization with MaxEnt it is possible to characterize the DSD with the moment of the distribution, while it was also shown that a reliable caracterization needs moments until 6th order. On the other side, the role of small drops on the DSD's moments is relevant only until 2th order, therfore it is relevant only for the moments $M_{0}$,  $M_{1}$ and $M_{2}$ (please refer to the figure \ref{fig:DisdroCOMPARATIVA}).\\


Here I define a procedure to \textit{reconstruct the DSD} with the MaxEnt method by the following steps:



\begin{itemize}
\item Estimation of correction multiplicative factors for $M_{0}$, $M_{1}$ and $M_{2}$ as,
\begin{equation}
M^{(ref)}_{i}=\alpha_{i} M^{(par)}_{i} \qquad i=0,1,2
\end{equation}
where $M^{(ref)}_{i}$ represents the reference disdrometer moment value of i-th order. The multiplicative factors $\alpha_{i}$, allow to retrieve the reference values for the specific disdrometer under analysis (this factors $\alpha_{i}$ are supposed to be characteristic of each disdrometer, with a second order dependence on the event type).
\item Given the experimental values of $M_{k}$ the MaxEnt methodology is applied to the original set of moments with orders k=3,...,6 together with the moments of orders k=0,1,2 corrected with the multiplicative factors $\alpha_{i}$. The results should be a faithful parametrization of the histogram for large and medium drops and a multiplicative correction for smallest drops. 
\end{itemize}

The previous procedure may be carried on within several rainfall conditions. In particular, it is appropriate compare convective events and stratiform events and also several rainfall intensity intervals.


\renewcommand\chapterillustration{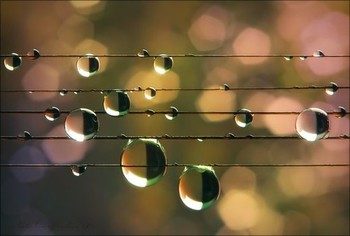}
\chapter{Extended analysis of spatial and temporal variability of DSD}

\label{sec:newVariability}

En el capítulo \S\ref{sec:chapVARSPACIAL1} se ha realizado un contraste de hipótesis sobre la existencia de variación espacial de la DSD a pequeña escala y se ha aplicado a comprobar su relevancia en las relaciones Z-R a esta escala. Uno de los objetivos siguientes es intentar dar una caracterización de dicha variabilidad. El estudiar cuantitativamente el problema involucra estudiar tanto la relevancia de posibles preprocesados de datos como analizar dicha variabilidad desde no solo parámetros integrales sino también modelizaciones de la DSD. En el capítulo \S\ref{chap:preprocesadoTOKAY} se especifica la metodología estándar de preprocesado de datos para comparar diferentes bases empíricas. En la siguiente sección se detalla la metodología elegida para la caracterización cuantitativa, para posteriormente aplicarla al conjunto de la base empírica disponible.


\section{Caracterización cuantitativa de la variabilidad espacial}

Para estudiar la variabilidad, se pueden realizar análisis tanto para el conjunto global de datos como evento por evento. Los análisis comprenden la comparación de histogramas a lo largo de la red, de histogramas de frecuencias y de las funciones de probabilidad acumuladas para cada parámetro. Desde el punto de vista de la comparación de series temporales y sus diferencias al incrementarse la distancia entre instrumentos realizamos dos análisis cuantitativos complementarios:

\subsection{Desviación estándar de la diferencia}
\begin{figure}[H] 
\vspace{0.05cm}
\begin{center}
   \includegraphics[width=0.80\textwidth]{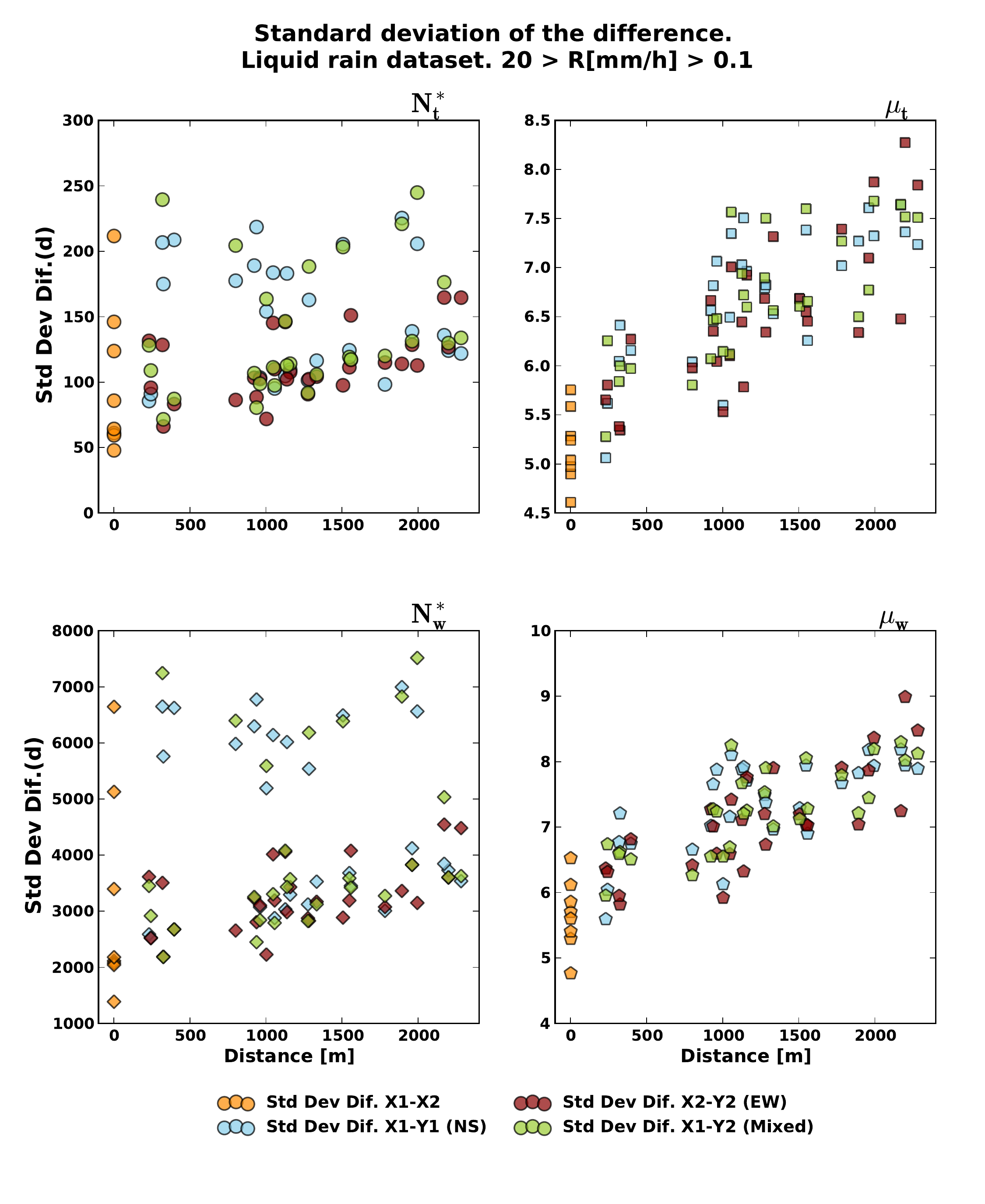}
\vspace{0.05cm}
   \caption[Desviación estándar de la diferencia de las series temporales para toda la base empírica de los parámetros de la DSD normalizada. Acotación en R entre 0.1 y 20 mm/h.]{ \textbf{[Desviación estándar de la diferencia de las series temporales para toda la base empírica de los parámetros de la DSD normalizada. Acotación en R entre 0.0 y 20 mm/h.}. Color \emph{azul}: Disdrómetros orientación NS. Color \emph{rojo}: Disdrómetros orientación EW. Color \emph{verde}: Desviación estándar de la diferencia entre disdrómetros con distinta orientación. Color \emph{naranja}: Desviación estándar de la diferencia entre instrumentos duales. Se indican las desviaciones estándar de la diferencia dada por (\ref{eqn:SDdiff}) para los parámetros de una distribución gamma normalizada minuto a minuto sobre toda la base empírica correspondiente a precipitación líquida.}
\label{fig:Var2stdevNORM_up20}
\end{center}
\end{figure}

\begin{center}
\begin{figure}[H] 
\vspace{0.15cm}
   \includegraphics[width=1.00\textwidth]{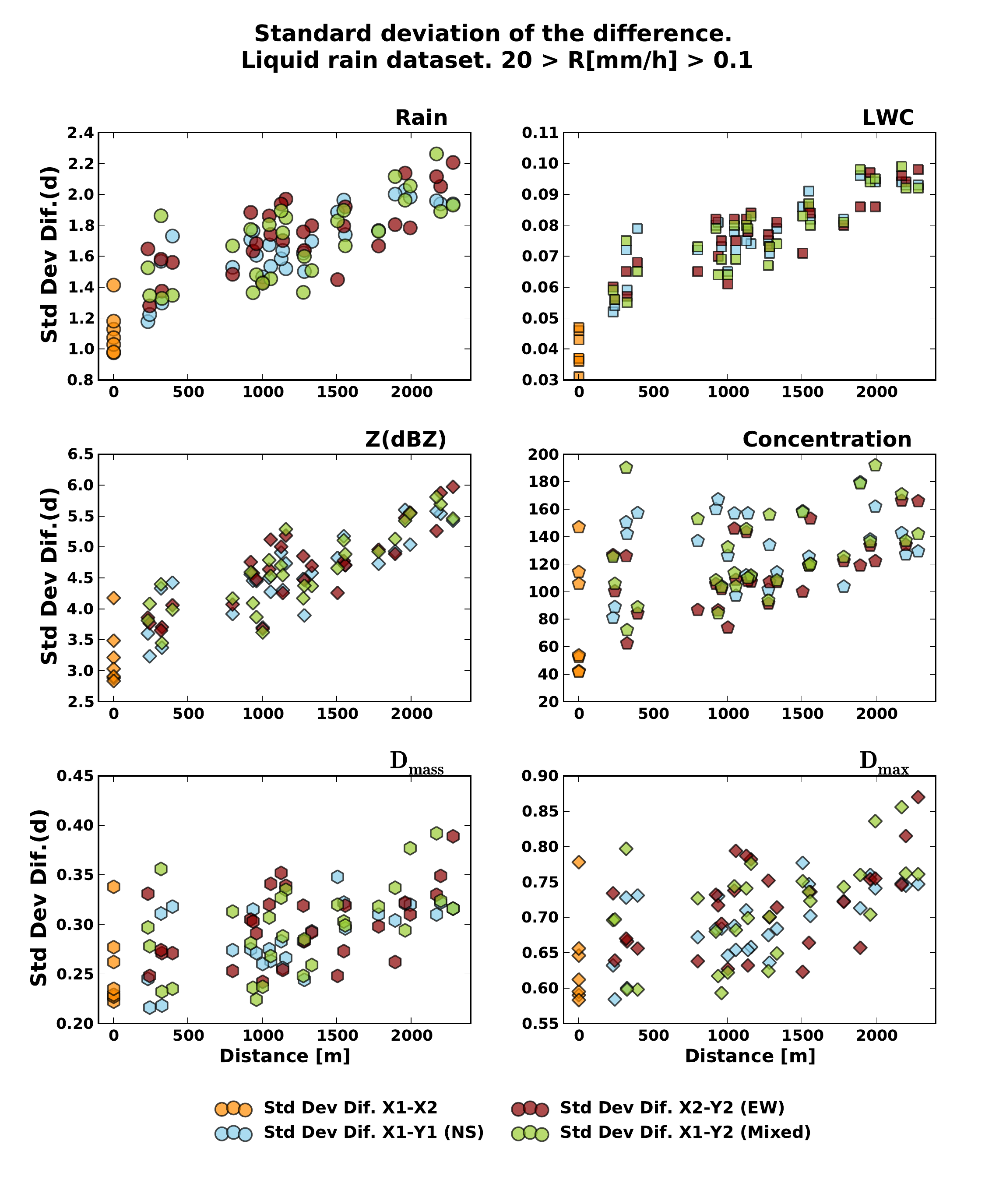}
\vspace{0.15cm}
   \caption[Desviación estándar de la diferencia de las series temporales para toda la base empírica de los parámetros integrales de la precipitación. Acotación en R entre 0.1 y 20 mm/h.]{ \textbf{Desviación estándar de la diferencia de las series temporales para toda la base empírica de los parámetros integrales de la precipitación. Acotación en R dada por $\mathbf{0.1< R[mm/h]<20}$}. Color \emph{azul}: Disdrómetros orientación NS. Color \emph{rojo}: Disdrómetros orientación EW. Color \emph{verde}: Desviación estándar de la diferencia entre disdrómetros con distinta orientación. Color \emph{naranja}: Desviación estándar de la diferencia entre instrumentos duales. Se muestran los resultados para la intensidad de precipitación (Rain), para el contenido en agua líquida (LWC), para la reflectividad, para la concentración y para el $D_{mass}$ y el diámetro máximo registrado en cada disdrómetro. Se indican las desviaciones estándar de la diferencia dada por (\ref{eqn:SDdiff}) sobre toda la base empírica correspondiente a precipitación líquida.}
\label{fig:Var2stdevRCWZ_up20}
\vspace{0.1cm}
\end{figure}
\end{center}

La desiviación estándar de la diferencia se define dadas las varianzas de cada serie de valores y la covarianza entre dos series de valores mediante la expresión:
\begin{equation}
SD(X_{i},X_{j})=\sqrt{Var(X_{i})+Var(X_{j})-2Cov(X_{i},X_{j})}
\label{eqn:SDdiff}
\end{equation}
donde es previsible un aumento de la desviación estándar de la diferencia conforme aumenta la distancia entre dos instrumentos dados. Cuestión que confirman las figuras 
(\ref{fig:Var2stdevNORM_up20}) y (\ref{fig:Var2stdevRCWZ_up20}).

\subsection{Modelización del correlograma $\rho(d)$}

\begin{figure}[H] 
\begin{center}
\vspace{1.3cm}
   \includegraphics[width=1.00\textwidth]{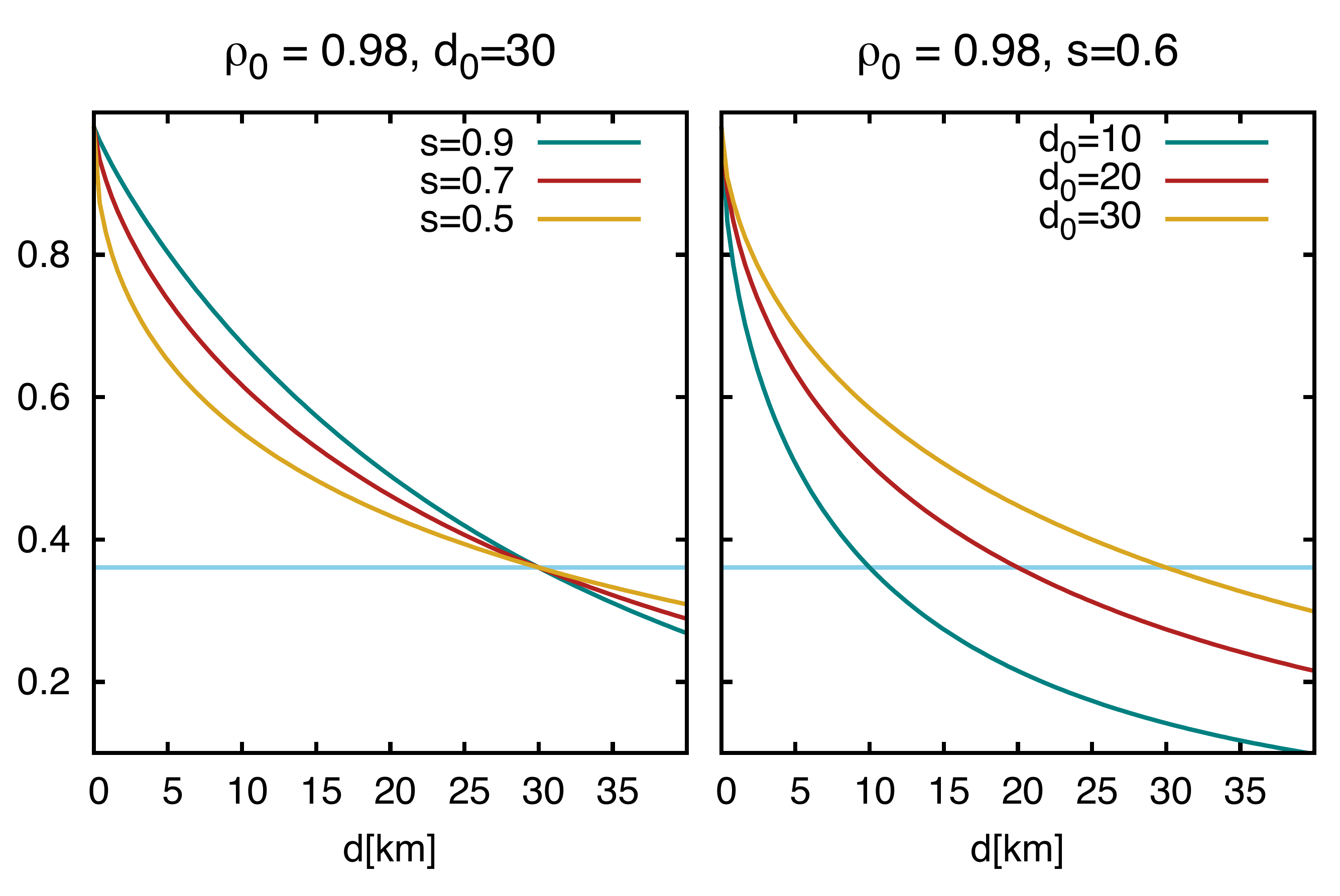}
\vspace{0.6cm}
   \caption[Figura representativa: modelización correlograma mediante un decaimiento exponencial modulado.]{\textbf{Figura representativa modelización correlograma mediante un decaimiento exponencial modulado.} Se representan diferentes valores de los parámetros libres de la ecuación (\ref{eqn:nuggeteq}) para apreciar su significado.}
\label{fig:var2NUGGET}
\end{center}
\vspace{1.4cm}
\end{figure}

Dados los valores de las correlaciones por la metodología mostrada en el capítulo \S\ref{sec:chapVARSPACIAL1}, véase la ecuación (\ref{eqn:PearsonTradicional}), es posible un estudio del correlograma mediante la relación:
\begin{equation}
\rho(d)=\rho_{0}e^{-(d/d_{0})^{s}}
\label{eqn:nuggeteq}
\end{equation}
El parámetro $\rho_{0}$ es llamado \emph{nugget parameter}. Suele ser determinado mediante dos instrumentos anexos, mientras que los otros parámetros $d_{0}$ y $s$ se determinan mediante ajustes al conjunto de datos. La figura (\ref{fig:var2NUGGET}) indica el papel jugado por los diferentes parámetros de la ecuación. La cantidad $d_{0}$ indica la longitud a la que se alcanza el valor de referencia $0.37\rho_{0}$. Menores valores de $s$ indican una disminución más rápida de la correlación a cortas distancias y más lenta a partir del valor de $d_{0}$.

Se ha procedido a calcular los parámetros libres de la ecuación (\ref{eqn:nuggeteq}) para diferentes parámetros integrales de la DSD, así como para estimaciones del modelo de la distribución gamma normalizada tal y como se describió en \S\ref{sec:AplicacionGammaNormalizada}. Se ha calculado el parámetro $\mu$ para las dos normalizaciones allí descritas.\\

Respecto del método de estudio se ha realizado del siguiente modo:
\begin{itemize}
 \item Se ha optado por fijar el parámetro de \emph{nugget}, $\rho_{0}$, como el valor de la correlación dada por las parejas de disdrómetros anexos, $\rho_{0}=\rho_{X1-X2}$. Sus valores pueden hacer depender parcialmente los otros valores del ajuste con la pareja elegida; estas diferencias pueden ser interpretadas como una incertidumbre en la estimación mediante la red experimental disponible. En las figuras siguientes se muestran los resultados tomando el parámetro de \emph{nugget}, $\rho_{0}$, como la mediana sobre el conjunto de valores de la red:
\begin{equation}
   \rho_{0}=M_{e}\left[\rho_{A1-A2}, \rho_{B1-B2},...,\rho_{H1-H2}  \right]
\end{equation}
 De esta manera se toma siempre el valor de la correlación correspondiente a una pareja anexa de instrumentos, y además la mediana es menos sensible (más robusta) que la media a valores anómalos que pudieran aparecer en algunos parámetros integrales \textemdash\,como en el caso que mostramos en la figura (\ref{fig:Var1CorreEpisodio12ENE}\textemdash\,. En el caso de toda la base empírica y para los parámetros integrales de la precipitación usuales los resultados de las correlaciones entre instrumentos duales son similares a lo largo de toda la red. Mientras que es posible una mayor dispersión en los valores de las correlaciones a pequeña distancia en el caso de los parámetros integrales correspondientes, bien a momentos de la DSD de orden alto, bien a momentos de la DSD de orden bajo (como es el caso de la concentración de gotas).\\

 Recalcar que el hecho de que $\rho_{0}<1$ implica tanto la existencia de errores aleatorios en la medida de la DSD como la posible existencia de una variabilidad intrínseca en la \textit{microescala}.
 \item Se han realizado cálculos de las correlaciones para disdrómetros con orientación NS y orientación EO, es decir, se ha aprovechado que la red utiliza disdrómetros duales para tener dos modelizaciones diferentes de la ecuación (\ref{fig:var2NUGGET}). Además es posible comparar las correlaciones entre instrumentos con orientaciones diferentes, que se ha representado también en las figuras siguientes. 
 \item Se ha realizado el estudio comparando los resultados al introducir acotaciones diferentes en la intensidad de lluvia a lo largo de toda la red (generado por tanto conjuntos consistentes de datos en cada caso).
\end{itemize}

 El tipo de interpretación del correlograma basado en la ecuación (\ref{eqn:nuggeteq}) ha sido utilizado en estudios pluviométricos, esencialmente en escalas de 10 a 100 km \citep{villarini_mandapaka_etal_2008_aa,2006CiachAWR}. Cabe notar que siguiendo a \citep{1974RodriguezIturbe} y \citep{1995Bacchi} el conjunto de datos que se poseen  de los experimentos de variabilidad no es suficiente para discriminar entre diferentes tipos de funciones. Lo usual es utilizar un modelo flexible que permita una interpretación sencilla de los parámetros que contenga.
\newpage

\begin{center}
\begin{figure}[H] 
\vspace{0.1cm}
   \includegraphics[width=0.95\textwidth]{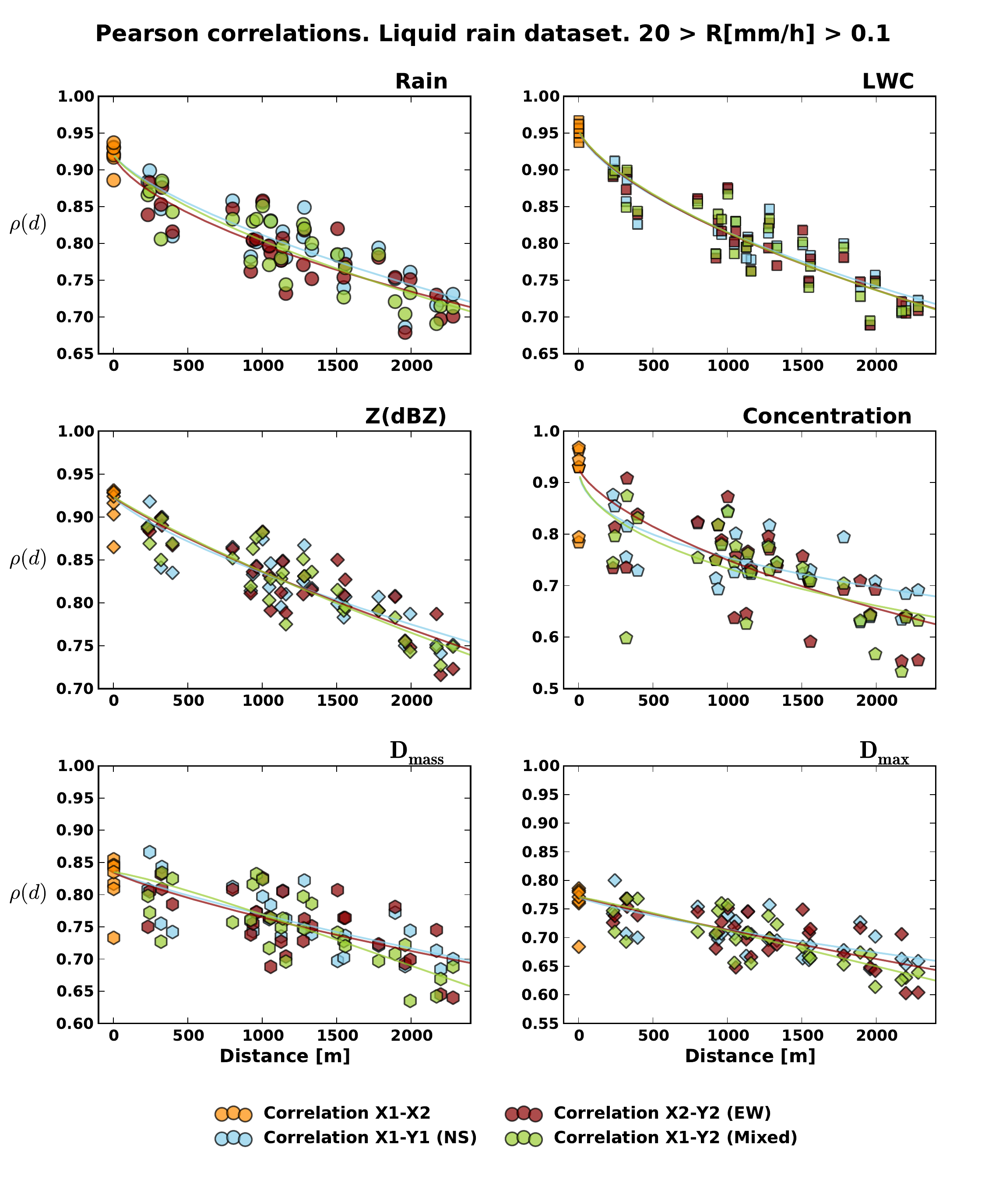}
\vspace{0.1cm}
   \caption[Correlograma modelizado de los parámetros integrales de la precipitación para toda la base empírica. Parámetro \textit{nugget} estimado desde correlaciones de instrumentos duales. Acotación en R entre 0.1 y 20 mm/h. ]{ \textbf{Correlograma modelizado de los parámetros integrales de la precipitación para toda la base empírica. Parámetro \textit{nugget} estimado desde correlaciones de instrumentos duales. Acotación en R dada por $\mathbf{0.0< R[mm/h]<20}$.} Color \emph{azul}: Disdrómetros orientación NS. Color \emph{rojo}: Disdrómetros orientación EW. Color \emph{verde}: Correlaciones entre disdrómetros con distinta orientación. Color \emph{naranja}: Correlación entre instrumentos duales. Se indican los correlogramas para la concentración total de gotas, el contenido en agua líquida, la intensidad de precipitación y la reflectividad. También se indica para $D_{mass}$ y el diámetro máximo registrado por cada disdrómetro. Las líneas son ajustes no lineales al correlograma según la ecuación (\ref{eqn:nuggeteq}). El ajuste no lineal hace mínima la diferencia cuadrática media entre los puntos y la curva.}
\label{fig:Var2CorrelogramaRCWZ_up10_A}
\vspace{0.1cm}
\end{figure}
\end{center}

\begin{center}
\begin{figure}[H] 
\vspace{0.75cm}
   \includegraphics[width=1.00\textwidth]{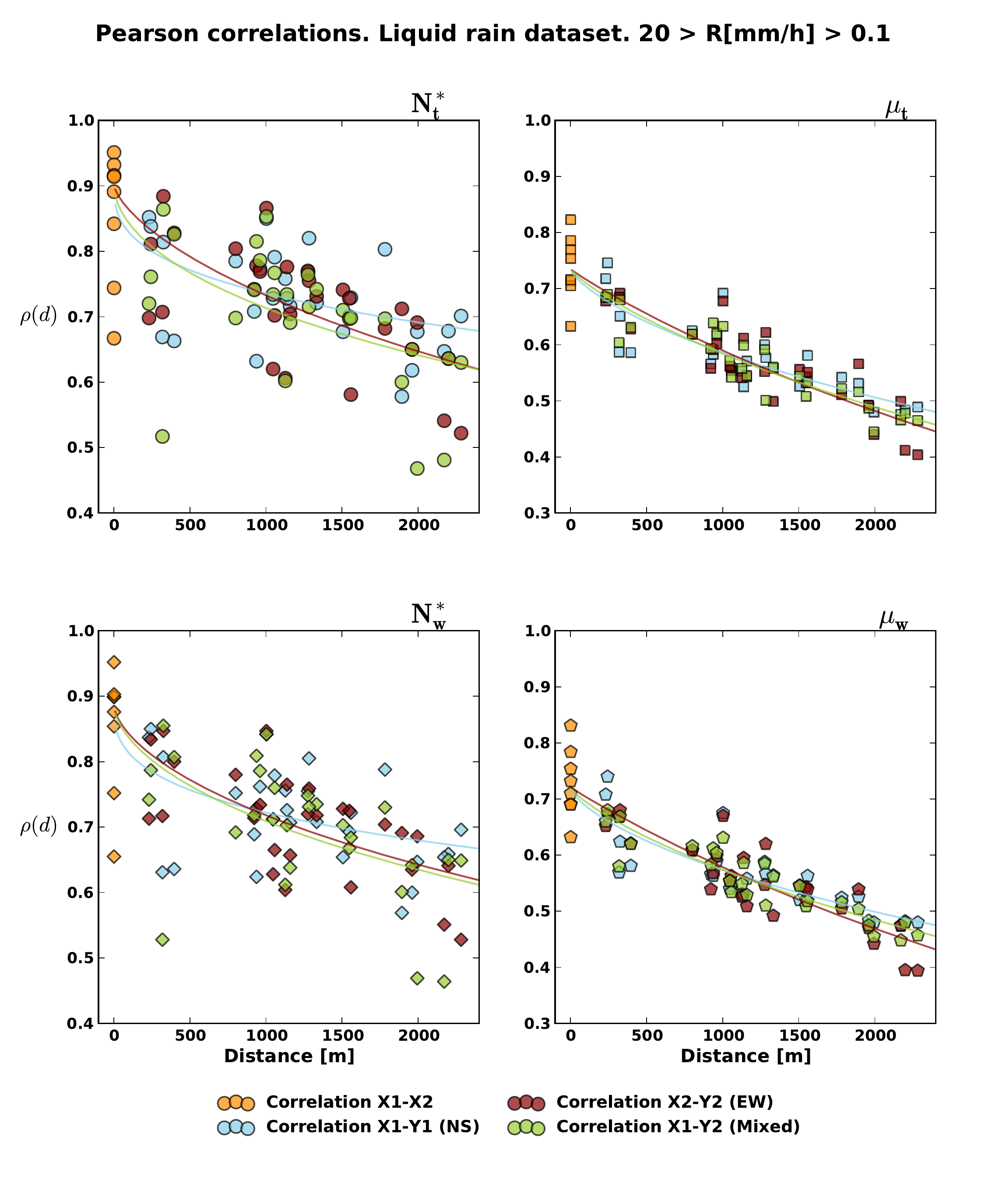}
\vspace{0.05cm}
   \caption[Correlograma modelizado de los parámetros de la DSD normalizada para toda la base empírica. Parámetro \textit{nugget} estimado desde correlaciones de instrumentos duales. Acotación en R entre 0.1 y 20 mm/h.]{ \textbf{Correlograma modelizado de los parámetros de la DSD normalizada para toda la base empírica. Parámetro \textit{nugget} estimado desde correlaciones de instrumentos duales. Acotación en R dada por $\mathbf{0.0< R[mm/h]<20}$}. Color \emph{azul}: Disdrómetros orientación NS. Color \emph{rojo}: Disdrómetros orientación EW. Color \emph{verde}: Correlaciones entre disdrómetros con distinta orientación. Color \emph{naranja}: Correlación entre instrumentos duales. Se indican los correlogramas para el parámetro de forma del modelo como distribución gamma de la DSD normalizada, así como la concentración utilizada en el normalizado de los espectros. Los detalles fueron indicados en \S\ref{sec:metodoNormalizada}. Las líneas son ajustes no-lineales al correlograma según la ecuación (\ref{eqn:nuggeteq}). El ajuste no-lineal hace mínima la diferencia cuadrática media entre los puntos y la curva.}
\label{fig:Var2CorrelogramaDSDparam_up20_F}
\vspace{0.25cm}
\end{figure}
\end{center}

\subsubsection{Tablas de estimaciones de los parámetros de la relación $\rho(d)$}

En esta sección se recogen los valores estimados de los parámetros que aparecen en la relación (\ref{eqn:nuggeteq}). Se muestran los resultados para diferentes acotaciones de precipitación mínima de 0.0 mm/h, 0.1 mm/h, 0.2 mm/h y 0.5 mm/h, debido a la relevancia que cada acotación puede tener en los varios algoritmos radar existentes, véase por ejemplo \citep{iguchi_kozu_ea_2000} y las referencias allí contenidas. 

\begin{table}[h]
\vspace{0.55cm}
\caption[Valores estimados para los parámetros de la ecuación (\ref{eqn:nuggeteq}). Parámetros integrales de la precipitación. Sin acotaciones en la intensidad de precipitación.]{\textbf{Valores estimados para los parámetros de la ecuación (\ref{eqn:nuggeteq}). Parámetros integrales de la precipitación. Sin acotaciones en la intensidad de precipitación.} Se muestran los valores estimados para varios parámetros integrales de la precipitación introducidos en la sección \S\ref{sec:parametrosintegralesDSD} y tres sub-redes consideradas según la orientación.  Se muestra la diferencia entre el modelo y los valores reales mediante la desviación cuadrática media (RMS).}\label{tabla:est_nugget_par_00_999}
\vspace{0.85cm}
\begin{center}
 \ra{1.30}
\begin{tabular}{l>{\columncolor[gray]{0.95}}rc>{\columncolor[gray]{0.95}}cc>{\columncolor[gray]{0.95}}cc}

\toprule
\textbf{Parámetro Integral} & \multicolumn{2}{c}{\textbf{Parámetros Libres}} & \multicolumn{1}{c}{\textbf{Ajuste}} &\multicolumn{2}{c}{\textbf{Sub-Red}}  \\

\cmidrule(r{.5em}){2-3}  \cmidrule(l{.5em}){5-6}

                         & $d_{0}[m]\,\,\,$   & $s$        &  RMS           & $\rho_{0}$  & Orientación \\
\midrule
Intensidad de lluvia     & 3\,739.10       & 0.9703      &  0.1194        & 0.9551      & EO \\
   & 3\,994.35       & 0.9662      &  0.171         & 0.9551      & NS \\
   & 3\,965.19       & 0.9448      &  0.1551        & 0.9551      & Mixto \\
\midrule
Reflectividad  (ZdBZ)          & 9\,161.31       & 1.1153      & 0.0165         & 0.9386      & EO\\
           & 12\,300.94      & 0.9682      & 0.0128         & 0.9386      & NS\\
           & 8\,862.97       & 1.1361      & 0.0142         &0.9386       & Mixto\\
\midrule
Contenido en agua líq.   & 4\,338.89       & 0.9392      & 0.0992         &0.9728       & EO\\
   & 4\,554.47       & 0.9456      & 0.1225         &0.9728       & NS\\
   & 4\,706.84       & 0.9086      & 0.1248         &0.9728       & Mixto\\
\midrule
Concentración            & 5\,355.75       & 1.1123      & 0.0893         &0.8938       & EO\\
          & 18\,070.10      & 0.5483      & 0.1008         &0.8938       & NS\\
         &  7\,107.58      & 0.8330      & 0.1397         &0.8938       & Mixto\\
\midrule
Diámetro Máximo          &  9\,005.17      & 1.2592      & 0.0265         &0.8188       & EO\\
        & 12\,417.94      & 1.0866      & 0.0123         &0.8188       & NS\\
       &  7\,062.93      & 1.4205      & 0.0189         &0.8188       & Mixto\\
\midrule
Diámetro $D_{mass}$      &  10\,243.28     & 1.1221      & 0.0343         &0.8586       & EO\\  
      &  10\,971.78     & 1.1211      & 0.0205         &0.8586       & NS\\  
     &  6\,143.62      & 1.5043      & 0.0297         &0.8586       & Mixto\\  
                                                                             
\bottomrule
\end{tabular}
\end{center}
\vspace{0.75cm}
\end{table}

\clearpage
\begin{table}[h]
\vspace{0.95cm}
\caption[Valores estimados para los parámetros de la ecuación (\ref{eqn:nuggeteq}). Parámetros integrales de la precipitación. Acotación en la intensidad de lluvia máxima de 20 mm/h y mínima de 0.1 mm/h.] {\textbf{Valores estimados para los parámetros de la ecuación (\ref{eqn:nuggeteq}). Parámetros integrales de la precipitación. Acotación en la intensidad de lluvia: $\mathbf{20 > R [mm/h] > 0.1}$}. Se muestran los valores estimados para varios parámetros integrales de la precipitación introducidos en la sección \S\ref{sec:parametrosintegralesDSD} y tres sub-redes consideradas según la orientación.  Se muestra la diferencia entre el modelo y los valores reales mediante la desviación cuadrática media (RMS).}\label{tabla:est_nugget_par_01_20}
\vspace{0.95cm}
\begin{center}
 \ra{1.30}
\begin{tabular}{l>{\columncolor[gray]{0.95}}rc>{\columncolor[gray]{0.95}}cc>{\columncolor[gray]{0.95}}cc}

\toprule
\textbf{Parámetro Integral} & \multicolumn{2}{c}{\textbf{Parámetros Libres}} & \multicolumn{1}{c}{\textbf{Ajuste}} &\multicolumn{2}{c}{\textbf{Sub-Red}}  \\

\cmidrule(r{.5em}){2-3}  \cmidrule(l{.5em}){5-6}

                         &$d_{0}[m]\,\,\,$       & $s$        &  RMS        & $\rho_{0}$  & Orientación \\
\midrule
Intensidad de lluvia     & 19\,508.63    &  0.6458     & 0.0246        & 0.9245      & EO \\
     & 16\,001.75    &  0.7339     & 0.0213        & 0.9245      & NS \\
    & 14\,464.64    &  0.7361     & 0.0233        & 0.9245      & Mixto \\
\midrule
Reflectividad  (ZdBZ)          & 11\,777.61    &  0.9832     & 0.0162        & 0.9149      & EO\\
        & 14\,856.44    &  0.8913     & 0.0152        & 0.9149      & NS\\
         & 10\,371.02    &  1.0399     & 0.0143        & 0.9149      & Mixto\\
\midrule
Contenido en agua líq.   & 13\,637.01    &  0.7014     & 0.0184        & 0.9554      & EO\\
   & 15\,359.72    &  0.6739     & 0.0220        & 0.9554      & NS\\
   & 13\,171.61    &  0.7137     & 0.0200        & 0.9554      & Mixto\\
\midrule
Concentración            & 10\,668.03    &  1.1688     & 0.0277        & 0.8994      & EO\\
         & 17\,330.35    &  0.9681     & 0.0213        & 0.8994      & NS\\
        & 10\,778.66    &  0.6824     & 0.1252        & 0.8994      & Mixto\\
\midrule
Diámetro Máximo           &  10\,667.82    &  1.1688      &  0.0277        &  0.7616      &  EO\\
         &  17\,330.36    &  0.9681   &  0.0213         &  0.7616      &  NS\\
	 &   8\,038.80    &  1.3103      &  0.0219           &  0.7616      &  Mixto\\

\midrule
Diámetro $D_{mass}$      &  11\,654.77   & 1.0930      & 0.0418        & 0.8215      & EO\\  
     &  11\,913.41   & 1.1088      & 0.0296        & 0.8215      & NS\\  
     &   6\,235.51   & 1.5187      & 0.0353        & 0.8215      & Mixto\\  
                                                                             
\bottomrule
\end{tabular}
\end{center}
\vspace{0.75cm}
\end{table}
\clearpage
\begin{table}[h]
\vspace{0.95cm}
\caption[Valores estimados para los parámetros de la ecuación (\ref{eqn:nuggeteq}). Parámetros integrales de la precipitación. Acotación en la intensidad de lluvia máxima de 20 mm/h y mínima de 0.2 mm/h.] {\textbf{Valores estimados para los parámetros de la ecuación (\ref{eqn:nuggeteq}). Parámetros integrales de la precipitación. Acotación en la intensidad de lluvia: $\mathbf{20 > R [mm/h] > 0.2}$.} Se muestran los valores estimados para varios parámetros integrales de la precipitación introducidos en la sección \S\ref{sec:parametrosintegralesDSD} y tres sub-redes consideradas según la orientación.  Se muestra la diferencia entre el modelo y los valores reales mediante la desviación cuadrática media (RMS).}\label{tabla:est_nugget_par_02_20}
\vspace{0.95cm}
\begin{center}
 \ra{1.30}
\begin{tabular}{l>{\columncolor[gray]{0.95}}rc>{\columncolor[gray]{0.95}}cc>{\columncolor[gray]{0.95}}cc}

\toprule
\textbf{Parámetro Integral} & \multicolumn{2}{c}{\textbf{Parámetros Libres}} & \multicolumn{1}{c}{\textbf{Ajuste}} &\multicolumn{2}{c}{\textbf{Sub-Red}}  \\

\cmidrule(r{.5em}){2-3}  \cmidrule(l{.5em}){5-6}

                         &$d_{0}[m]\,\,\,$    & $s$        &  RMS     & $\rho_{0}$  & Orientación \\
\midrule
Intensidad de lluvia     &  17\,592.38    &  0.6309    &  0.0303    &  0.9148     & EO \\
    &  13\,994.62    &  0.7397    &  0.0241    &  0.9148     & NS \\
    &  13\,057.13    &  0.7278    &  0.0264    &  0.9148     & Mixto \\
\midrule
Reflectividad   (ZdBZ)         &  11\,028.63    &  0.9731    &  0.0210    &  0.8916     & EO\\
         &  16\,232.28    &  0.8372    &  0.0178    &  0.8916     & NS\\
        &  10\,116.80    &  1.0068    &  0.0149    &  0.8916     & Mixto\\
\midrule
Contenido en agua líq.   &  11\,906.41    &  0.6967    &  0.0227    &  0.9496     & EO\\
   &  13\,305.68    &  0.6749    &  0.0253    &  0.9496     & NS\\
   &  11\,672.75    &  0.7067    &  0.0236    &  0.9496     & Mixto\\
\midrule
Concentración            &   6\,966.58    &  0.8979    &  0.1331    &  0.8999     & EO\\
           &  18\,964.20    &  0.6039    &  0.0552    &  0.8999     & NS\\
         &  11\,283.49    &  0.6653    &  0.1366    &  0.8999     & Mixto\\
\midrule
Diámetro Máximo          &  10\,027.60    &  1.1639    &  0.0336    &  0.7199     & EO\\
        &  18\,432.94    &  0.9355    &  0.0269    &  0.7199     & NS\\
        &   7\,919.42    &  1.2714    &  0.0245    &  0.7199     & Mixto\\
\midrule
Diámetro $D_{mass}$      &   11\,271.76   &  1.0971    &  0.0503    &  0.7967     & EO\\  
     &   10\,776.08   &  1.1829    &  0.0343    &  0.7967     & NS\\  
     &    5\,666.63   &  1.6172    &  0.0417    &  0.7967     & Mixto\\  
                                                                             
\bottomrule
\end{tabular}
\end{center}
\vspace{0.75cm}
\end{table}

\clearpage

\begin{table}[h]
\vspace{0.95cm}
\caption[Valores estimados para los parámetros de la ecuación (\ref{eqn:nuggeteq}). Parámetros integrales de la precipitación. Acotación en la intensidad de lluvia máxima de 20 mm/h y mínima de 0.5 mm/h.] {\textbf{Valores estimados para los parámetros de la ecuación (\ref{eqn:nuggeteq}). Parámetros integrales de la precipitación. Acotación en la intensidad de lluvia: $\mathbf{20 > R [mm/h] > 0.5}$}. Se muestran los valores estimados para varios parámetros integrales de la precipitación introducidos en la sección \S\ref{sec:parametrosintegralesDSD} y tres sub-redes consideradas según la orientación.  Se muestra la diferencia entre el modelo y los valores reales mediante la desviación cuadrática media (RMS).}\label{tabla:est_nugget_par_05_20}
\vspace{0.95cm}
\begin{center}
\ra{1.30}
\begin{tabular}{l>{\columncolor[gray]{0.95}}rc>{\columncolor[gray]{0.95}}cc>{\columncolor[gray]{0.95}}cc}

\toprule
\textbf{Parámetro Integral} & \multicolumn{2}{c}{\textbf{Parámetros Libres}} & \multicolumn{1}{c}{\textbf{Ajuste}} &\multicolumn{2}{c}{\textbf{Sub-Red}}  \\

\cmidrule(r{.5em}){2-3}  \cmidrule(l{.5em}){5-6}

                         & $d_{0}[m]\,\,\,$   & $s$        &  RMS      & $\rho_{0}$  & Orientación \\
\midrule
Intensidad de lluvia     &  19\,465.22    &  0.5688    &  0.0418   &   0.8926    & EO \\
     &  14\,102.04    &  0.6898    &  0.0316   &   0.8926    & NS \\
    &  12\,989.99    &  0.6784    &  0.0347   &   0.8926    & Mixto \\
\midrule
Reflectividad  (ZdBZ)          &  15\,775.97    &  0.8083    &  0.0251   &   0.8493    & EO\\
         &  16\,731.12    &  0.8007    &  0.0241   &   0.8493    & NS\\
         &   9\,859.84    &  0.9712    &  0.0198   &   0.8493    & Mixto\\
\midrule
Contenido en agua líq.   &  13\,168.15    &  0.6201    &  0.0311   &   0.9360    & EO\\
  &  14\,076.32    &  0.6118    &  0.0321   &   0.9360    & NS\\
  &  12\,453.71    &  0.6355    &  0.0306   &   0.9360    & Mixto\\
\midrule
Concentración            &   8\,707.07    &  0.7653    &  0.1979   &   0.9012    & EO\\
          &  18\,225.25    &  0.6677    &  0.0498   &   0.9012    & NS\\
           &  21\,475.75    &  0.5230    &  0.1850   &   0.9012    & Mixto\\
\midrule
 Diámetro Máximo          &  14\,824.06    &  0.9682    &  0.0325   &    0.6575    &  EO\\
			             &  18\,181.88    &  0.9369   &   0.0346    &    0.6575     &  NS\\
			             &   7\,507.13    &  1.2773     &  0.0289      &    0.6575     &  Mixto\\
\midrule
 Diámetro $D_{mass}$      &   17\,391.52    &  0.9560     &  0.0552   &   0.7656     &  EO\\  
		                     &    9\,848.13    &  1.2897     &  0.0414   &  0.7656     &  NS\\  
			             &    5\,289.56    &  1.7651     &  0.0533    &   0.7656    &  Mixto\\  
                                                                             
\bottomrule
\end{tabular}
\end{center}
\vspace{0.75cm}
\end{table}

\clearpage

\begin{table}[h]
\vspace{0.75cm}
\caption[Valores estimados para los parámetros de la ecuación (\ref{eqn:nuggeteq}). Parámetros de la DSD normalizada (asumida una distribución gamma). Acotación en la intensidad de lluvia.] {\textbf{Valores estimados para los parámetros de la ecuación (\ref{eqn:nuggeteq}). Parámetros de la DSD normalizada (asumida una distribución gamma). Acotación en la intensidad de lluvia.: $\mathbf{20 > R [mm/h] > 0.1}$}.  Se muestran los valores estimados para los parámetros de forma y concentración introducidos en la sección \S\ref{sec:AplicacionGammaNormalizada} y tres sub-redes consideradas según la orientación.  Se muestra la diferencia entre el modelo y los valores reales mediante la desviación cuadrática media (RMS). }\label{tabla:est_nugget_norm_01_20}
\vspace{0.95cm}
\begin{center}
 \ra{1.35}
\begin{tabular}{c>{\columncolor[gray]{0.95}}rc>{\columncolor[gray]{0.95}}cc>{\columncolor[gray]{0.95}}cc}

\toprule
\textbf{Parámetro DSD} & \multicolumn{2}{c}{\textbf{Parámetros Libres}} & \multicolumn{1}{c}{\textbf{Ajuste}} &\multicolumn{2}{c}{\textbf{Sub-Red}}  \\

\cmidrule(r{.5em}){2-3}  \cmidrule(l{.5em}){5-6}

 \textbf{normalizada}   & $d_{0}[m]\,\,\,$ & $s$        &  RMS       & $\rho_{0}$  & Orientación \\
\midrule
$N_{t}^{*}$     &   6\,508.48   &   1.0341   &   0.1291   &   0.8571   & EO \\
$N_{t}^{*}$     &  32\,157.79   &   0.5469   &   0.1109   &   0.8571   & NS \\
$N_{t}^{*}$     &   7\,287.58   &   0.9017   &   0.2051   &   0.8571   & Mixto \\
\midrule
$\mu_{t}$      &   5\,047.59   &   0.9239   &   0.0390   &   0.7375   & EO\\
$\mu_{t}$      &   7\,928.30   &   0.7094   &   0.0401   &   0.7375   & NS\\
$\mu_{t}$      &   5\,975.75   &   0.8132   &   0.0219   &   0.7375   & Mixto\\
\midrule
$N_{w}^{*}$     &   9\,111.68   &   0.8273   &   0.0993   &   0.8488   & EO\\
$N_{w}^{*}$     &  39\,249.68   &   0.5009   &   0.1248   &   0.8488   & NS\\
$N_{w}^{*}$     &   8\,108.95   &   0.8505   &   0.1935   &   0.8488   & Mixto\\
\midrule
$\mu_{w}$      &    4\,917.53  &    0.9182   &   0.0402  &    0.7280   & EO\\
$\mu_{w}$      &    8\,840.28  &   0.6559   &   0.0383  &    0.7280   & NS\\
$\mu_{w}$      &    6\,568.40  &   0.7590   &   0.0218  &    0.7280   & Mixto\\
                                                                      
\bottomrule
\end{tabular}
\end{center}
\vspace{0.15cm}
\end{table}

\clearpage

\begin{table}[h]
\vspace{0.75cm}
\caption[Valores estimados para los parámetros de la ecuación (\ref{eqn:nuggeteq}). Parámetros de la DSD normalizada (asumida una distribución gamma). Acotación en la intensidad de lluvia.] {\textbf{Valores estimados para los parámetros de la ecuación (\ref{eqn:nuggeteq}). Parámetros de la DSD normalizada (asumida una distribución gamma). Acotación en la intensidad de lluvia.: $\mathbf{20 > R [mm/h] > 0.2}$}. Se muestran los valores estimados para los parámetros de forma y concentración introducidos en la sección \S\ref{sec:AplicacionGammaNormalizada} y tres sub-redes consideradas según la orientación.  Se muestra la diferencia entre el modelo y los valores reales mediante la desviación cuadrática media (RMS).}\label{tabla:est_nugget_norm_02_20}
\vspace{0.95cm}
\begin{center}
 \ra{1.35}
\begin{tabular}{c>{\columncolor[gray]{0.95}}rc>{\columncolor[gray]{0.95}}cc>{\columncolor[gray]{0.95}}cc}

\toprule
\textbf{Parámetro DSD} & \multicolumn{2}{c}{\textbf{Parámetros Libres}} & \multicolumn{1}{c}{\textbf{Ajuste}} &\multicolumn{2}{c}{\textbf{Sub-Red}}  \\

\cmidrule(r{.5em}){2-3}  \cmidrule(l{.5em}){5-6}

 \textbf{normalizada}         & $d_{0}[m]\,\,\,$  & $s$       &  RMS    & $\rho_{0}$  & Orientación \\
\midrule
$N_{t}^{*}$     &   7\,099.23   &   1.0099  &  0.1430 &  0.8638  & EO \\
$N_{t}^{*}$     &  32\,944.65   &   0.5568  &  0.1077 &  0.8638  & NS \\
$N_{t}^{*}$     &   7\,432.72   &   0.9150  &  0.2131 &  0.8638  & Mixto \\
\midrule
$\mu_{t}$      &   5\,279.74   &   0.8794  &  0.0485 &  0.7614  & EO\\
$\mu_{t}$      &   9\,591.88   &   0.6427  &  0.0448 &  0.7614  & NS\\
$\mu_{t}$      &   7\,236.32   &   0.6873  &  0.0296 &  0.7614  & Mixto\\
\midrule
$N_{w}^{*}$     &   11\,609.01  &   0.7605  &  0.1002 &  0.8586  & EO\\
$N_{w}^{*}$     &   43\,378.29  &   0.5000  &  0.1202 &  0.8586  & NS\\
$N_{w}^{*}$     &    9\,008.04  &   0.8290  &  0.1893 &  0.8586  & Mixto\\
\midrule
$\mu_{w}$      &    5\,409.61  &   0.8332  &  0.0499 &  0.7554  & EO\\
$\mu_{w}$      &   11\,393.20  &   0.5875  &  0.0422 &  0.7554  & NS\\
$\mu_{w}$      &    8\,254.75  &   0.6291  &  0.0324 &  0.7554  & Mixto\\
                                                                      
\bottomrule
\end{tabular}
\end{center}
\vspace{0.55cm}
\end{table}

\clearpage

\begin{table}[h]
\vspace{0.75cm}
\caption[Valores estimados para los parámetros de la ecuación (\ref{eqn:nuggeteq}). Parámetros de la DSD normalizada (asumida una distribución gamma). Acotación en la intensidad de lluvia.] {\textbf{Valores estimados para los parámetros de la ecuación (\ref{eqn:nuggeteq}). Parámetros de la DSD normalizada (asumida una distribución gamma). Acotación en la intensidad de lluvia.: $\mathbf{20 > R [mm/h] > 0.5}$}. Se muestran los valores estimados para los parámetros de forma y concentración introducidos en la sección \S\ref{sec:AplicacionGammaNormalizada} y tres sub-redes consideradas según la orientación.  Se muestra la diferencia entre el modelo y los valores reales mediante la desviación cuadrática media (RMS).}\label{tabla:est_nugget_norm_05_20}
\vspace{0.95cm}
\begin{center}
 \ra{1.35}
\begin{tabular}{c>{\columncolor[gray]{0.95}}rc>{\columncolor[gray]{0.95}}cc>{\columncolor[gray]{0.95}}cc}

\toprule
\textbf{Parámetro DSD} & \multicolumn{2}{c}{\textbf{Parámetros Libres}} & \multicolumn{1}{c}{\textbf{Ajuste}} &\multicolumn{2}{c}{\textbf{Sub-Red}}  \\

\cmidrule(r{.5em}){2-3}  \cmidrule(l{.5em}){5-6}

 \textbf{normalizada}   & $d_{0}[m]\,\,\,$  & $s$        &  RMS      & $\rho_{0}$  & Orientación \\
\midrule
$N_{t}^{*}$     &   11\,201.86  &    0.7677  &    0.1814  &    0.8775 & EO \\
$N_{t}^{*}$     &   14\,519.86  &    0.8429  &    0.0631  &    0.8775 & NS \\
$N_{t}^{*}$     &    9\,937.30  &    0.8288  &    0.2113  &    0.8775 & Mixto \\
\midrule
$\mu_{t}$      &    3\,926.79  &    1.1843  &    0.0419  &    0.7650 & EO\\
$\mu_{t}$      &    5\,467.97  &    0.9912  &    0.0495  &    0.7650 & NS\\
$\mu_{t}$      &    4\,214.50  &    1.0977  &    0.0518  &    0.7650 & Mixto\\
\midrule
$N_{w}^{*}$     &   24\,078.82  &    0.5634  &    0.1220  &    0.8756 & EO\\
$N_{w}^{*}$     &   11\,220.59  &    0.9312  &    0.0608  &    0.8756 & NS\\
$N_{w}^{*}$     &   10\,186.21  &    0.8460  &    0.1664  &    0.8756 & Mixto\\
\midrule
$\mu_{w}$      &    3\,998.83  &    1.1087  &    0.0402  &    0.7685 & EO\\
$\mu_{w}$      &    6\,870.82  &    0.8076  &    0.0561  &    0.7685 & NS\\
$\mu_{w}$      &    4\,770.21  &    0.9513  &    0.0580  &    0.7685 & Mixto\\
                                                                      
\bottomrule
\end{tabular}
\end{center}
\vspace{0.55cm}
\end{table}

\clearpage

\begin{table}[h]
\vspace{0.95cm}
\caption[Valores estimados para los parámetros de la ecuación (\ref{eqn:nuggeteq}). Parámetros integrales de la precipitación. Acotación en la intensidad de lluvia máxima de 20 mm/h y mínima de 0.1 mm/h. Datos por episodios.] {\textbf{Valores estimados para los parámetros de la ecuación (\ref{eqn:nuggeteq}). Parámetros integrales de la precipitación. Acotación en la intensidad de lluvia: $\mathbf{20 > R [mm/h] > 0.1}$. Datos por episodios.} Se muestran los valores estimados para varios parámetros integrales de la precipitación introducidos en la sección \S\ref{sec:parametrosintegralesDSD} y los episodios con intensidades de lluvia acumulada mayores de 10 mm.  Se muestra la diferencia entre el modelo y los valores reales mediante la desviación cuadrática media (RMS). Los valores que se muestran corresponden a la sub-red X1-Y2. En este caso se ha determinado el parámetro \textit{nugget} usando la media en lugar de la mediana.}\label{tabla:est_nugget_par_01_20_episodios}
\vspace{0.95cm}
\begin{center}
\ra{1.30}
\begin{tabular}{l>{\columncolor[gray]{0.95}}rc>{\columncolor[gray]{0.95}}cc>{\columncolor[gray]{0.95}}cc}

\toprule
\textbf{Parámetro Integral} & \multicolumn{2}{c}{\textbf{Parámetros Libres}} & \multicolumn{1}{c}{\textbf{Ajuste}} &\multicolumn{2}{c}{\textbf{Sub-Red}}  \\

\cmidrule(r{.5em}){2-3}  \cmidrule(l{.5em}){5-6}

                         & $d_{0}[m]\,\,\,$   & $s$        &  RMS      & $\rho_{0}$  & Episodio \\
\midrule
 Intensidad de lluvia    &             &            &            &             &           \\ 

                         &  2\,314.10     &  0.8580    &  0.3054   &   0.9745    & 02 dic \\
                         &  10\,138.71    &  0.8990    &  0.0834   &   0.9432    & 20 dic \\
                         &  12\,540.76    &  0.5084    &  0.1324   &   0.9728    & 24 dic \\
                         &  72\,952.36    &  0.6025    &  0.0133   &   0.9706    & 03 ene \\
                         &  19\,785.75    &  0.8204    &  0.0235   &   0.8755    & 12 ene \\
\midrule
Reflectividad (ZdBZ)    &             &            &           &             &           \\

                         &   6\,851.20    &  0.6508    &  0.2079   &   0.9495    & 02 dic\\
                         &  11\,683.81    &  1.1520    &  0.0295   &   0.9141    & 20 dic\\
                         &  6\,206.89     &  0.5433    &  0.1109   &   0.9580    & 24 dic\\
                         &  15\,712.45    &  1.0290    &  0.0156   &   0.9566    & 03 ene\\
                         &  32\,202.88    &  1.0637    &  0.0202   &   0.9090    & 12 ene\\
\midrule
Contenido en agua líq.   &             &            &           &             &           \\ 
                         &  2\,054.85     &  0.9110    &  0.2835   &   0.9761    & 02 dic\\
                         &  9\,672.84     &  1.0468    &  0.0459   &   0.9646    & 20 dic\\
                         &  12\,040.00    &  0.4952    &  0.1344   &   0.9792    & 24 dic\\
                         &  201\,065.90   &  0.4783    &  0.0106   &   0.9779    & 03 ene\\
                         &  27\,247.64    &  0.7168    &  0.0123   &   0.9253    & 12 ene\\                                                                        
\bottomrule
\end{tabular}
\end{center}
\vspace{0.75cm}
\end{table}
\clearpage

\subsection{Análisis de la variabilidad espacial para diferentes acumulaciones temporales.}

Los estudios de correlación entre series temporales mostrados hasta el momento en esta memoria contienen series temporales a una resolución temporal de 1 minuto. Sin embargo, estudios pluviométricos \citep{2008VillariniJGR} y disdrométricos \citep{Tokay2010smallscaleDSD} muestran que una acumulación temporal previa al proceso de análisis de la correlación altera los valores de las correlaciones en bases empíricas amplias que involucren varios episodios de precipitación. En esta sección mostramos los resultados cuando se incluye progresivamente mayor acumulación temporal en cada serie temporal comparado entre disdrómetros\footnote{Notesé que el procedimiento es acumular y luego realizar la consistencia entre instrumentos para comparar exactamente los mismos intervalos de tiempo en cada instrumento, lo que lleva a las limitaciones lógicas para acumulaciones temporales suficientemente largas, por esta razón la comparación se ha limitado hasta el valor de 15 min.}.

\begin{table}[h]
\vspace{0.05cm}
\caption[Valores estimados para los parámetros de la ecuación (\ref{eqn:nuggeteq}). Parámetros integrales de la precipitación. Acotación en la intensidad de lluvia máxima de 20 mm/h y mínima de 0.1 mm/h. Acumulaciones temporales.] {\textbf{Valores estimados para los parámetros de la ecuación (\ref{eqn:nuggeteq}). Parámetros integrales de la precipitación. Acotación en la intensidad de lluvia: $\mathbf{20 > R [mm/h] > 0.1}$. Acumulaciones temporales.} Se muestran los valores estimados para varios parámetros integrales de la precipitación introducidos en la sección \S\ref{sec:parametrosintegralesDSD} y los episodios con intensidades de lluvia acumulada mayores de 10 mm.  Se muestra la diferencia entre el modelo y los valores reales mediante la desviación cuadrática media (RMS). Los valores que se muestran corresponden a la sub-red X1-Y2.}\label{tabla:est_nugget_par_01_20_acctem}
\vspace{0.10cm}
\begin{center}
\ra{1.10}
\begin{tabular}{l>{\columncolor[gray]{0.95}}rc>{\columncolor[gray]{0.95}}cc>{\columncolor[gray]{0.95}}cc}

\toprule
\textbf{Parámetro Integral} & \multicolumn{2}{c}{\textbf{Parámetros Libres}} & \multicolumn{1}{c}{\textbf{Ajuste}} &\multicolumn{2}{c}{\textbf{Sub-Red}}  \\

\cmidrule(r{.5em}){2-3}  \cmidrule(l{.5em}){5-6}

                         & $d_{0}[m]\,\,\,$   & $s$        &  RMS      & $\rho_{0}$  & Acumulación \\
\midrule
  Intensidad de lluvia   &  8\,366.55     &  1.1143    &  0.0327   &   0.9420    & 03 min \\
                         &  8\,323.96     &  1.1671    &  0.0305   &   0.9435    & 04 min \\
                         &  6\,993.54     &  1.3838    &  0.0296   &   0.9450    & 06 min \\
                         &   7\,075.56    &  1.6097    &  0.0238   &   0.9464    & 10 min \\
                         &  7\,699.69     &  1.4796    &  0.0196   &   0.9505    & 12 min \\
                         &  7\,856.93     &  1.5725    &  0.0228   &   0.9535    & 15 min \\
\midrule
 Reflectividad (ZdBZ)    &   9\,297.64    &  1.1392    &  0.0194   &   0.9395  & 03 min\\
                         &   9\,515.24    &  1.1812    &  0.0214   &   0.9420  & 04 min\\
                         &   7\,504.81    &  1.4426    &  0.0195   &   0.9395  & 06 min\\
                         &   6\,120.62    &  1.8569    &  0.0191   &   0.9327  & 10 min\\                  
                         &   6\,156.46    &  1.7812    &  0.0200   &   0.9395  & 12 min\\
                         &   6\,283.46    &  1.7793    &  0.0221   &   0.9400  & 15 min\\
\midrule
 Contenido en agua líq.  &  8\,533.04     &  1.0143    &  0.0290   &   0.9685    & 03 min\\
                         &  8\,992.65     &  1.0242    &  0.0294   &   0.9715   & 04 min\\
                         &  7\,227.15     &  1.2336    &  0.0283   &   0.9730    & 06 min\\
                         &  8\,790.78     &  1.3291    &  0.0142   &   0.9710    & 10 min\\
                         &  8\,316.46     &  1.3403    &  0.0151   &   0.9745  & 12 min\\
                         &  9\,094.55     &  1.3616    &  0.0161   &   0.9750   & 15 min\\                                                                        
\bottomrule
\end{tabular}
\end{center}
\vspace{0.05cm}
\end{table}

\clearpage

\begin{table}[h]
\vspace{0.95cm}
\caption[Valores estimados para los parámetros de la ecuación (\ref{eqn:nuggeteq}). Parámetros integrales de la precipitación. Acotación en la intensidad de lluvia máxima de 20 mm/h y mínima de 0.5 mm/h. Acumulaciones temporales.] {\textbf{Valores estimados para los parámetros de la ecuación (\ref{eqn:nuggeteq}). Parámetros integrales de la precipitación. Acotación en la intensidad de lluvia: $\mathbf{20 > R [mm/h] > 0.5}$. Acumulaciones temporales.} Se muestran los valores estimados para varios parámetros integrales de la precipitación introducidos en la sección \S\ref{sec:parametrosintegralesDSD} y los episodios con intensidades de lluvia acumulada mayores de 10 mm.  Se muestra la diferencia entre el modelo y los valores reales mediante la desviación cuadrática media (RMS). Los valores que se muestran corresponden a la sub-red X1-Y2.}\label{tabla:est_nugget_par_05_20_acctem}
\vspace{0.95cm}
\begin{center}
\ra{1.30}
\begin{tabular}{l>{\columncolor[gray]{0.95}}rc>{\columncolor[gray]{0.95}}cc>{\columncolor[gray]{0.95}}cc}

\toprule
\textbf{Parámetro Integral} & \multicolumn{2}{c}{\textbf{Parámetros Libres}} & \multicolumn{1}{c}{\textbf{Ajuste}} &\multicolumn{2}{c}{\textbf{Sub-Red}}  \\

\cmidrule(r{.5em}){2-3}  \cmidrule(l{.5em}){5-6}

                         & $d_{0}[m]\,\,\,$   & $s$        &  RMS      & $\rho_{0}$  & Acumulación \\
\midrule
 Intensidad de lluvia    &  6\,877.55     &  1.0754    &  0.0531   &   0.9170    & 03 min \\
                         &  7\,332.73     &  1.1185    &  0.0428   &   0.9175    & 04 min \\
                         &  5\,673.54     &  1.3654    &  0.0504   &   0.9220    & 06 min \\
                         &  5\,926.67     &  1.6143    &  0.0458   &   0.9224    & 10 min \\
                         &  6\,704.78     &  1.4409    &  0.0373   &   0.9295    & 12 min \\
                         &  7\,572.86     &  1.5513    &  0.0326   &   0.9330    & 15 min \\
\midrule
Reflectividad (ZdBZ)     &  6\,180.00     &  1.3836    &  0.0296   &   0.8745    & 03 min\\
                         &  6\,569.87     &  1.4375    &  0.0288   &   0.8810    & 04 min\\
                         &  5\,316.19     &  1.7555    &  0.0381   &   0.8815    & 06 min\\
                         &  5\,119.29     &  2.1428    &  0.0488   &   0.8760    & 10 min\\                  
                         &  5\,053.88     &  2.0511    &  0.0391   &   0.8895    & 12 min\\
                         &  5\,836.31     &  1.9426    &  0.0449   &   0.8965    & 15 min\\
\midrule
 Contenido en agua líq.  &  6\,775.66     &  0.9802    &  0.0493   &   0.9545   & 03 min\\
                         &  7\,906.61     &  0.9473    &  0.0427   &   0.9595   & 04 min\\
                         &  5\,595.67     &  1.2241    &  0.0514   &   0.9620   & 06 min\\
                         &  7\,245.95     &  1.2959    &  0.0276   &   0.9587   & 10 min\\
                         &  7\,167.04     &  1.2917    &  0.0278   &   0.9640   & 12 min\\
                         &  9\,578.29     &  1.2455    &  0.0201   &   0.9655   & 15 min\\                                                                        
\bottomrule
\end{tabular}
\end{center}
\vspace{0.75cm}
\end{table}

\subsubsection{Parámetros de la ecuación de nugget para acumulaciones temporales de hasta 30 min}

Como complemento a las tablas anteriores se incluyen figuras en que se muestran resultados para acumulaciones temporales de hasta 30 min y varias acotaciones de intensidades máxima y mínima.

\begin{figure}[H] 
\begin{center}
   \includegraphics[width=1.00\textwidth]{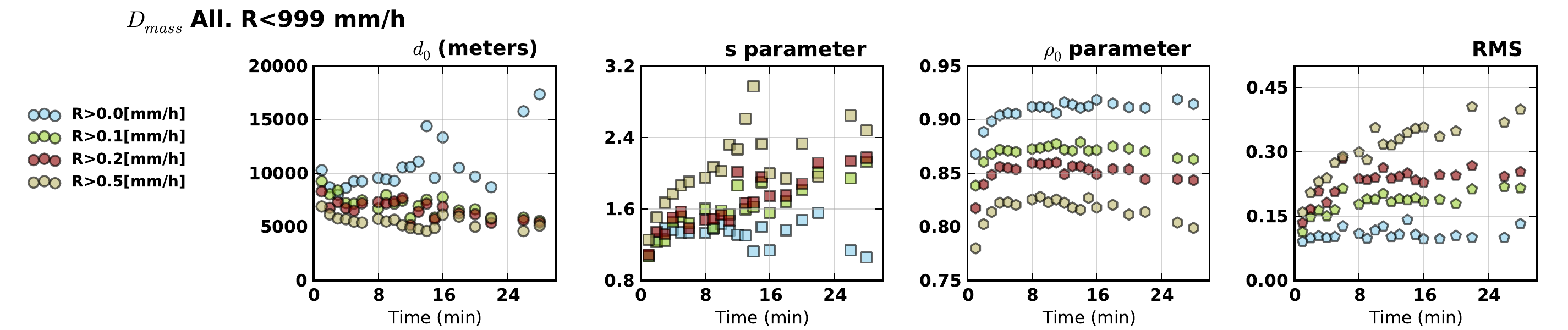}
   \includegraphics[width=1.00\textwidth]{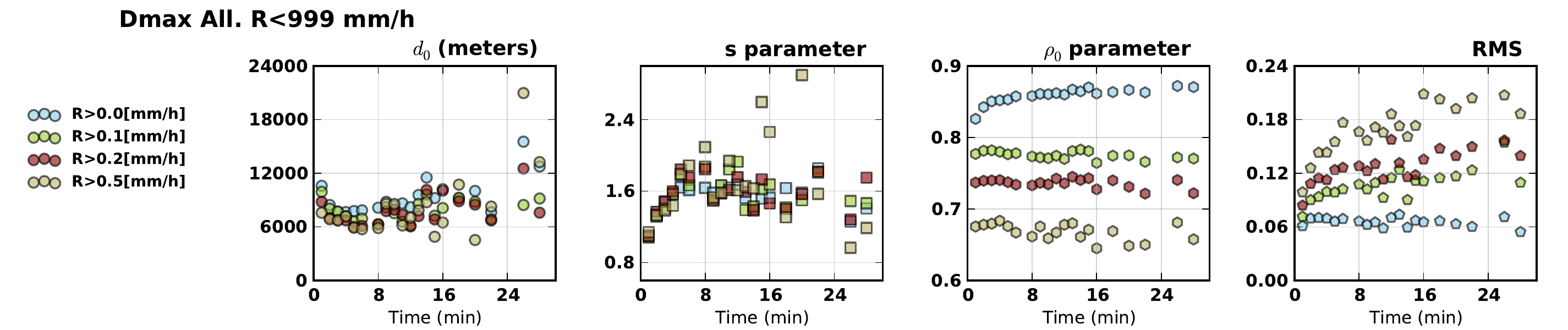}
   \includegraphics[width=1.00\textwidth]{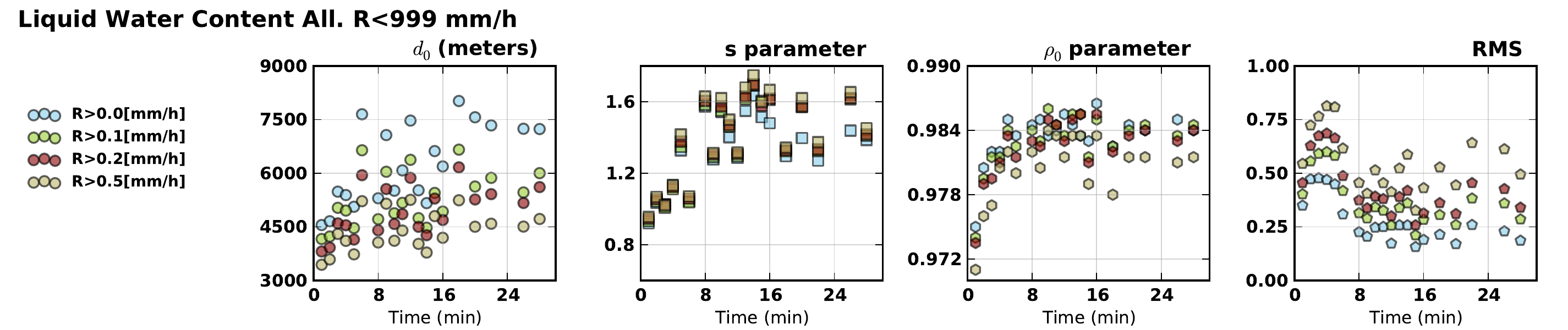}
   \includegraphics[width=1.00\textwidth]{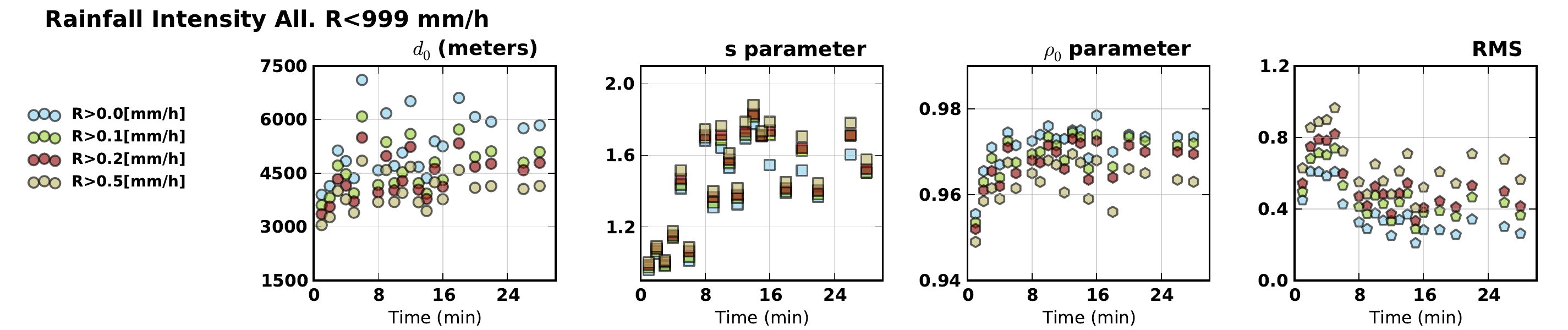}
   \includegraphics[width=1.00\textwidth]{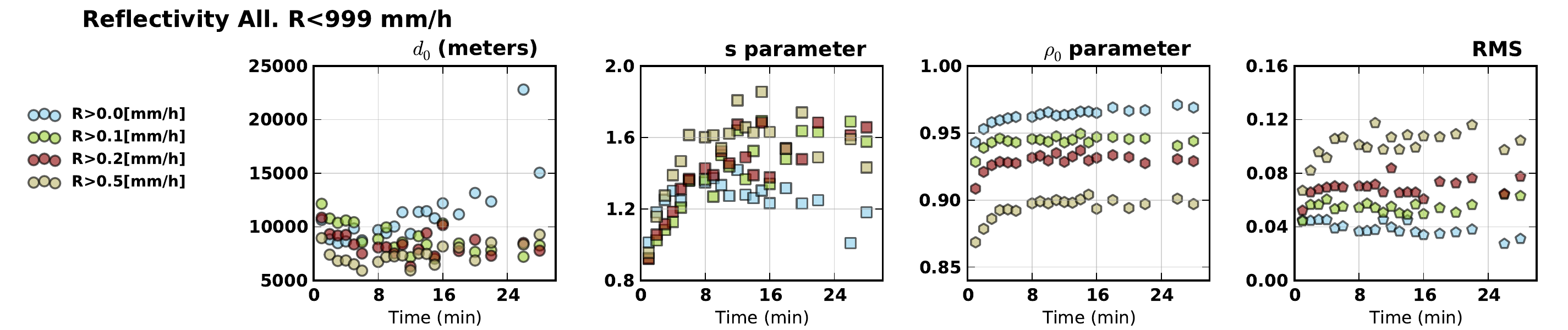}

   \caption[Parámetros estimados de la ecuación de nugget para acumulaciones temporales de hasta 30 min. Sin acotación de intensidad de precipitación máxima.]{\textbf{Parámetros estimados de la ecuación de nugget para acumulaciones temporales de hasta 30 min. Sin acotación de intensidad de precipitación máxima.} Se muestran los valores de los parámetros de la ecuación de nugget estimados para acumulaciones temporales de hasta 30 min para varios parámetros integrales. Se incluyen cuatro casos diferentes de intensidad mínima.}
\label{fig:nugget999_30min}
\end{center}
\vspace{0.5cm}
\end{figure}

\begin{figure}[H] 
\begin{center}
   \includegraphics[width=1.00\textwidth]{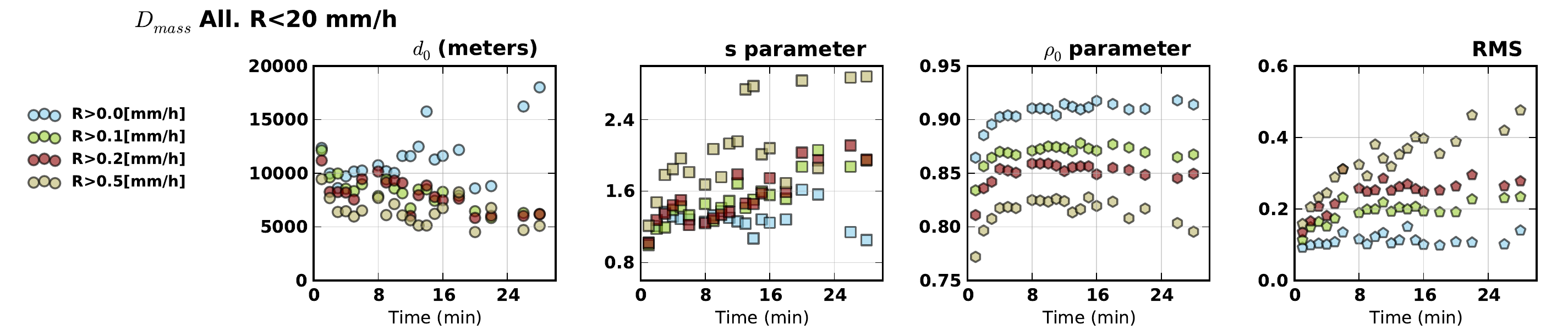}
   \includegraphics[width=1.00\textwidth]{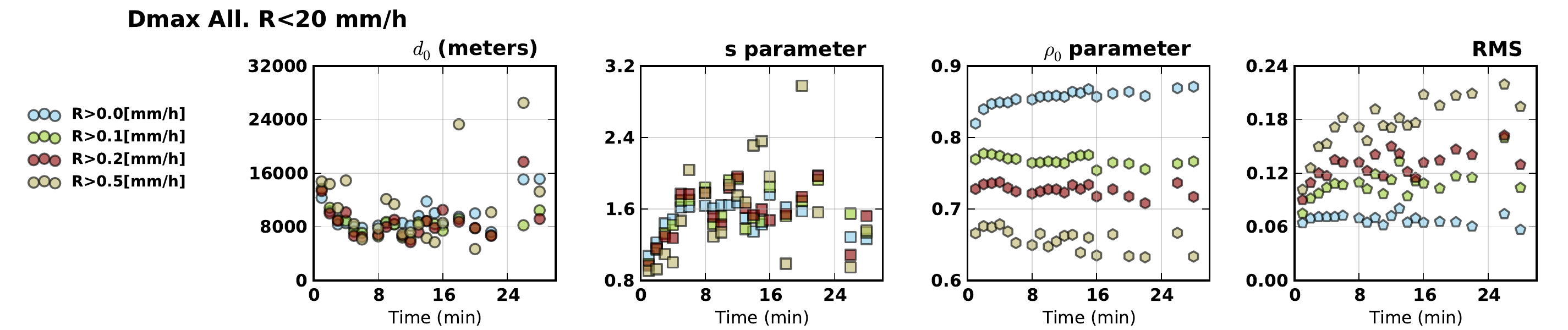}
   \includegraphics[width=1.00\textwidth]{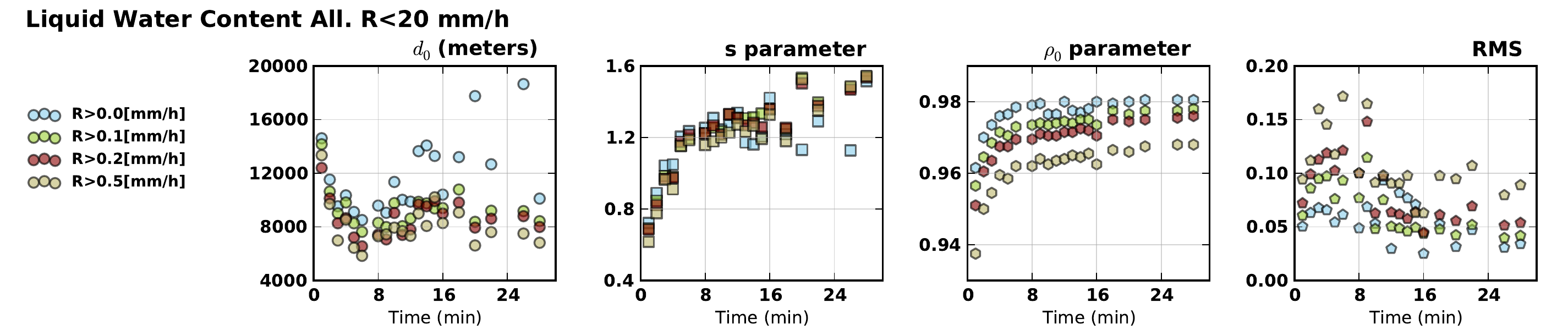}
   \includegraphics[width=1.00\textwidth]{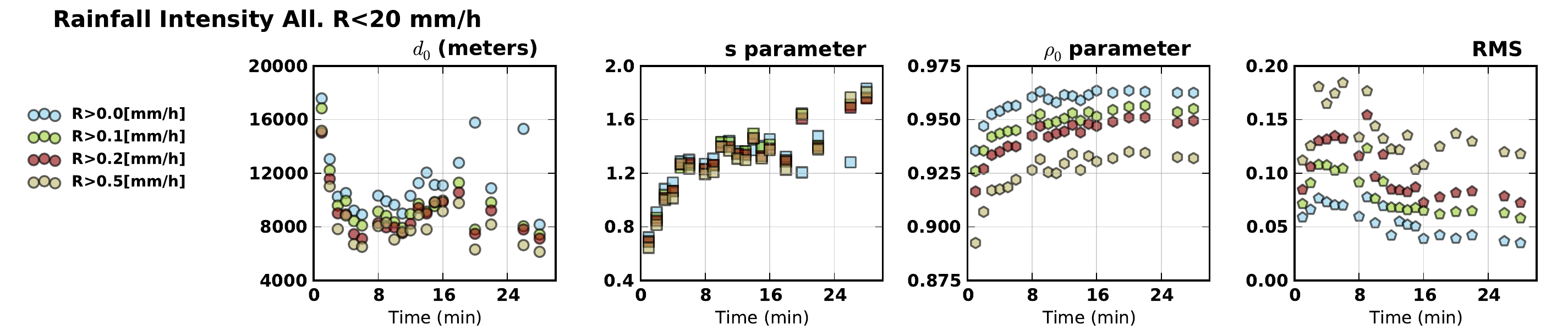}
   \includegraphics[width=1.00\textwidth]{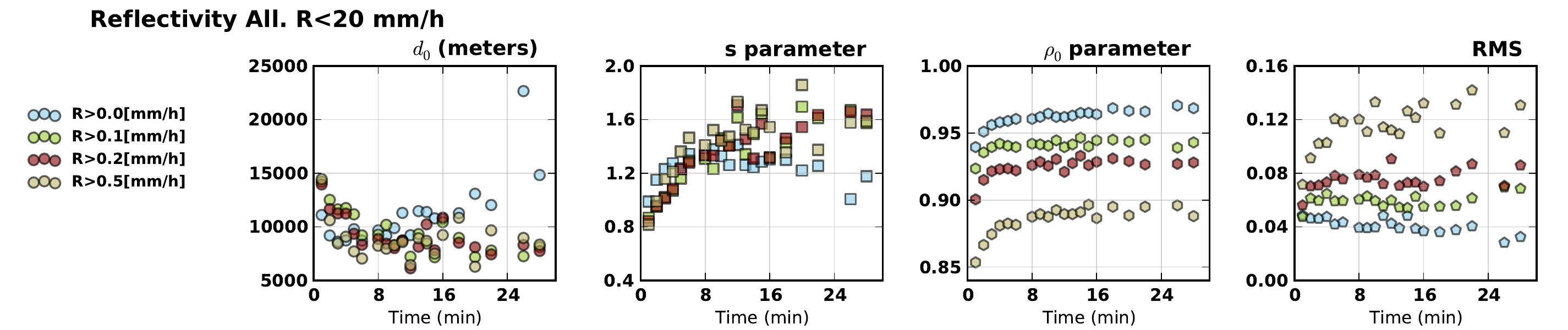}

   \caption[Parámetros estimados de la ecuación de nugget para acumulaciones temporales de hasta 30 min. Intensidad de precipitación máxima de 20 mm/h]{\textbf{Parámetros estimados de la ecuación de nugget para acumulaciones temporales de hasta 30 min. Intensidad de precipitación máxima de 20 mm/h.} Se muestran los valores de los parámetros de la ecuación de nugget estimados para acumulaciones temporales de hasta 30 min para varios parámetros integrales. Se incluyen cuatro casos diferentes de intensidad mínima.}
\label{fig:nugget20_30min}
\end{center}
\vspace{0.5cm}
\end{figure}

\subsubsection{Parámetros de la ecuación de nugget por episodios}

En la tabla (\ref{tabla:est_nugget_par_01_20_episodios}) se da información para los episodios analizados en el texto principal de los diferentes parámetros de la ecuación de nugget. Aquí se detallan figuras en que aparecen todos los episodios que poseen más de 100 min y más de 4 mm de precipitación acumulada. Se muestran los valores de los parámetros estimados en cada episodio para toda la red completa de instrumentos (por tanto difieren levemente de los datos suministrados en la tabla  (\ref{tabla:est_nugget_par_01_20_episodios}) que incluian solo la sub-red X1-Y2.

\begin{figure}[H] 
\begin{center}
   \includegraphics[width=0.97\textwidth]{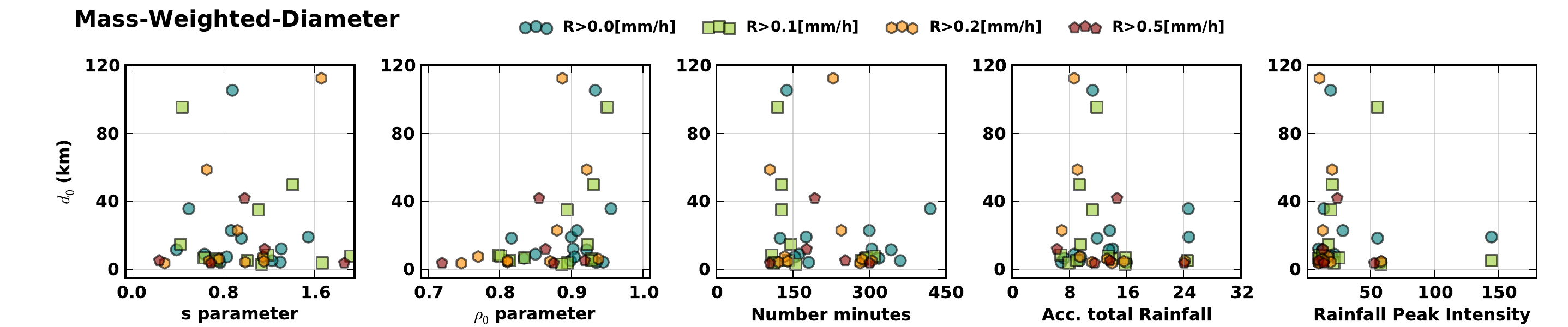}
   \includegraphics[width=0.97\textwidth]{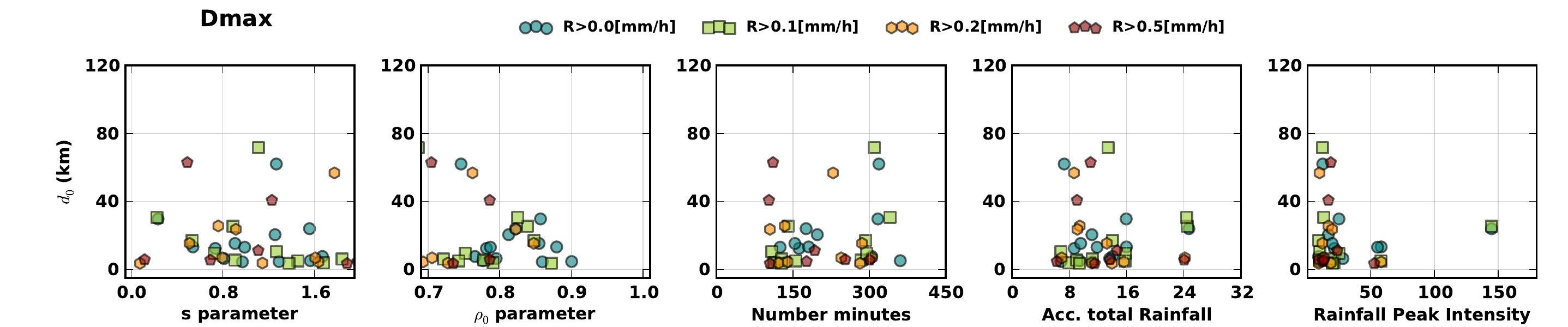}
   \includegraphics[width=0.97\textwidth]{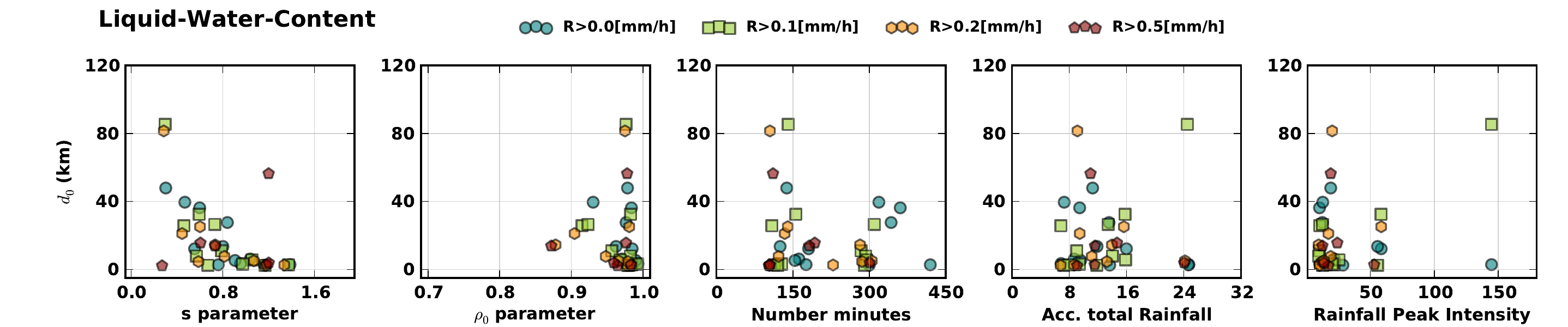}
   \includegraphics[width=0.97\textwidth]{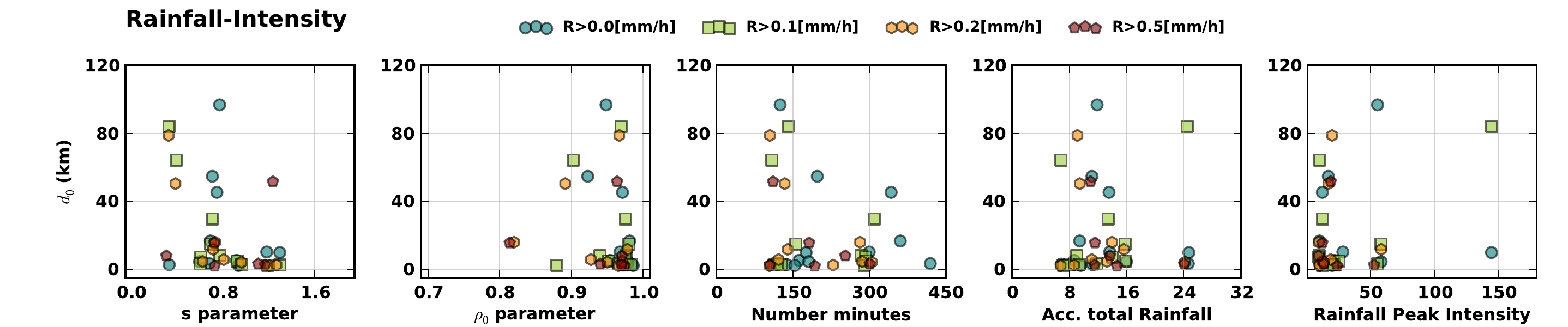}
   \includegraphics[width=0.97\textwidth]{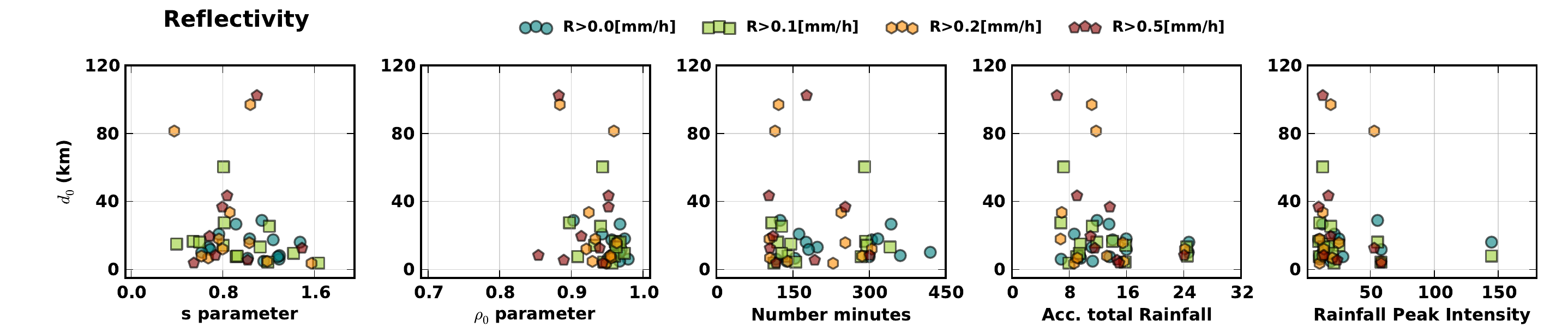}

   \caption[Parámetros estimados de la ecuación de nugget para los episodios con al menos 100 min de lluvia significativa y 4 mm de precipitación acumulada total. Sin acotación de intensidad de precipitación máxima.]{\textbf{Parámetros estimados de la ecuación de nugget para los episodios con al menos 100 min de lluvia significativa y 4 mm de precipitación acumulada total. Sin acotación de intensidad de precipitación máxima.} Se muestran los valores de los parámetros de la ecuación de nugget estimados para varios parámetros integrales. Se incluyen cuatro casos diferentes de intensidad mínima.}
\label{fig:nugget999_episodios}
\end{center}
\vspace{0.5cm}
\end{figure}

\begin{figure}[H] 
\begin{center}
   \includegraphics[width=0.97\textwidth]{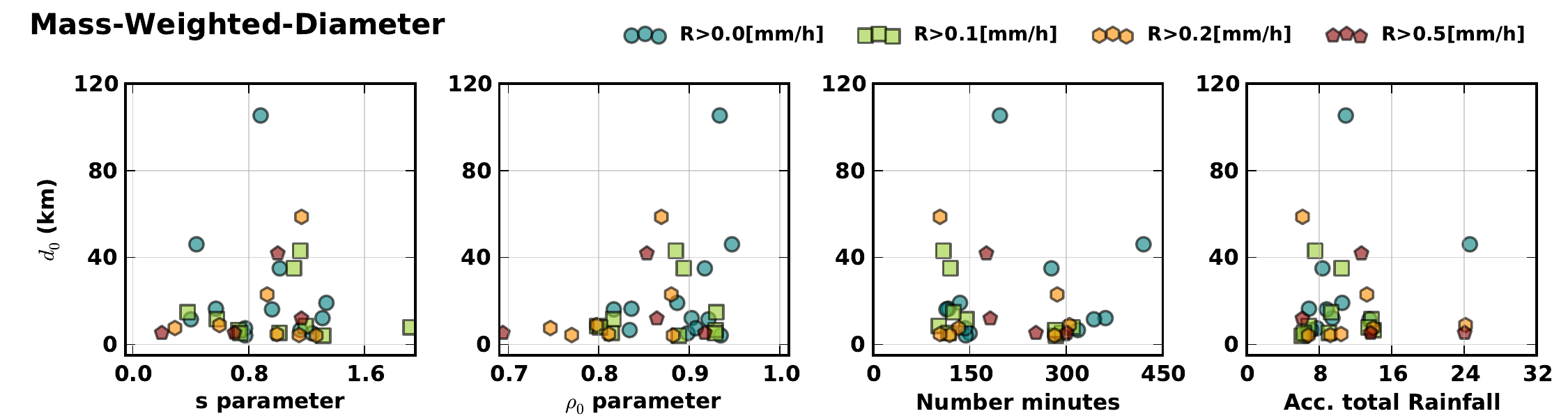}
   \includegraphics[width=0.97\textwidth]{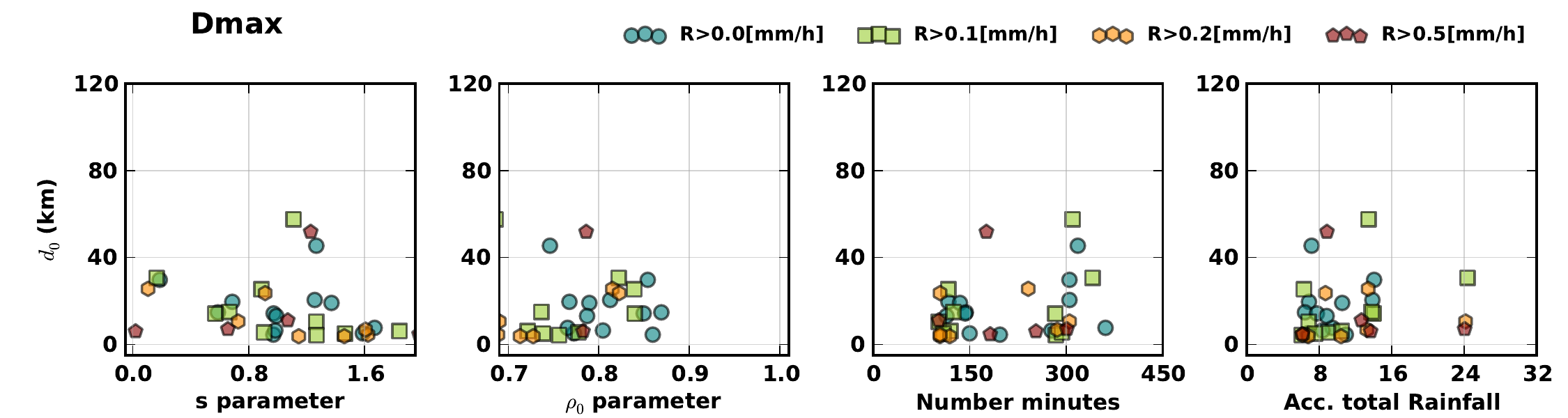}
   \includegraphics[width=0.97\textwidth]{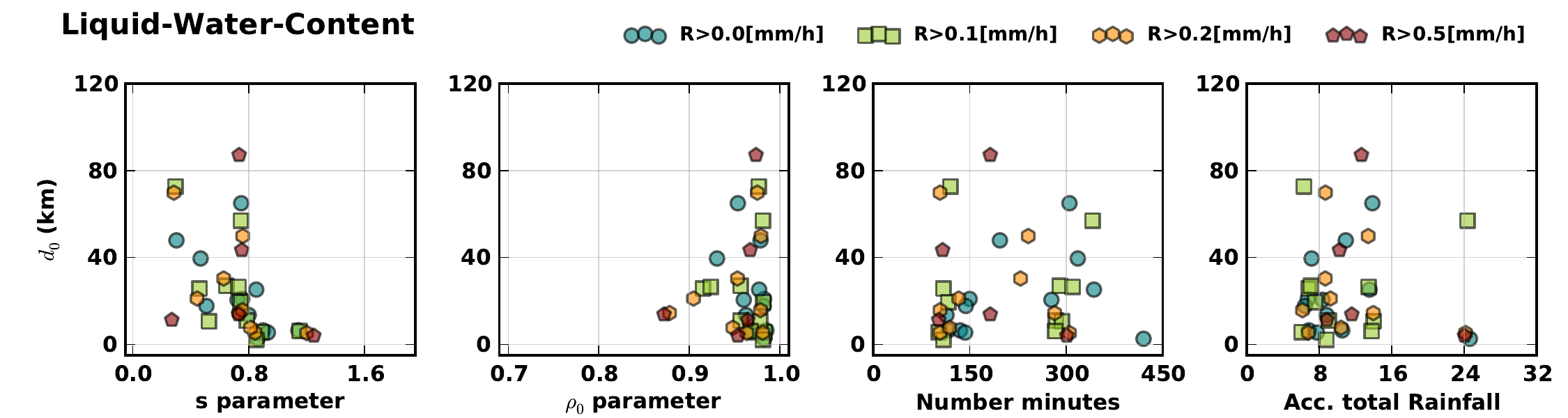}
   \includegraphics[width=0.97\textwidth]{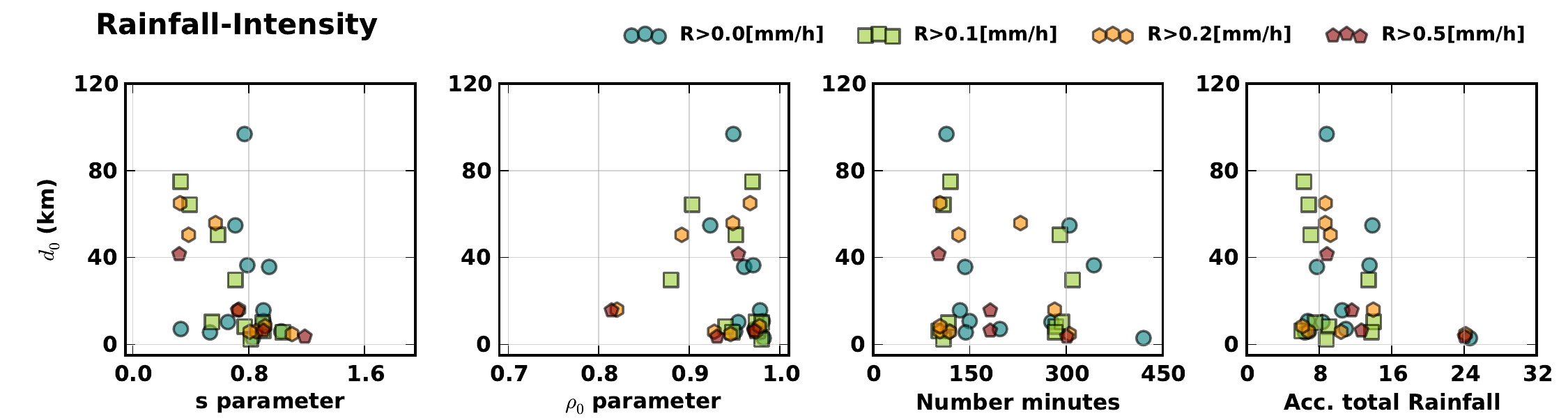}
   \includegraphics[width=0.97\textwidth]{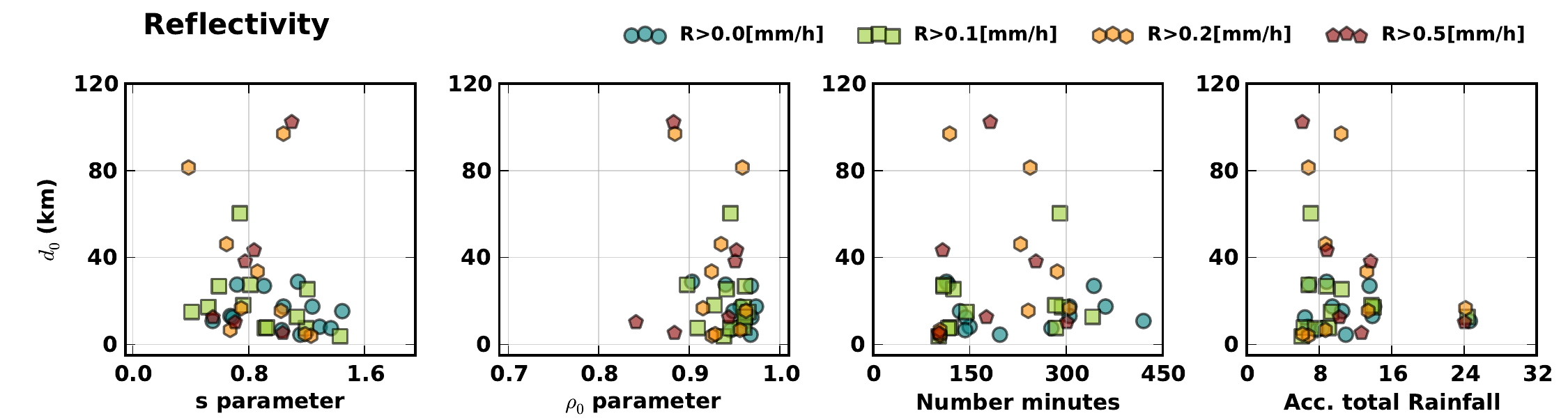}

   \caption[Parámetros estimados de la ecuación de nugget para los episodios con al menos 100 min de lluvia significativa y 4 mm de precipitación acumulada total. Intensidad de precipitación máxima de 20 mm/h]{\textbf{Parámetros estimados de la ecuación de nugget para los episodios con al menos 100 min de lluvia significativa y 4 mm de precipitación acumulada total. Intensidad de precipitación máxima de 20 mm/h.} Se muestran los valores de los parámetros de la ecuación de nugget estimados para varios parámetros integrales. Se incluyen cuatro casos diferentes de intensidad mínima.}
\label{fig:nugget20_episodios}
\end{center}
\vspace{0.5cm}
\end{figure}

\clearpage
\subsubsection{Papel de la acumulación temporal en las relaciones Z-R.}

La posibilidad de que las acumulaciones temporales puedan reducir los errores de muestreo parcialmente motiva las siguientes figuras. En ellas se muestran los resultados obtenidos para las relaciones Z-R a lo largo de la red y para toda la base empírica cuando se realiza una acumulación temporal (promedio) con una resolución de 10 min.

\begin{figure}[H] 
\begin{center}
   \includegraphics[width=0.95\textwidth]{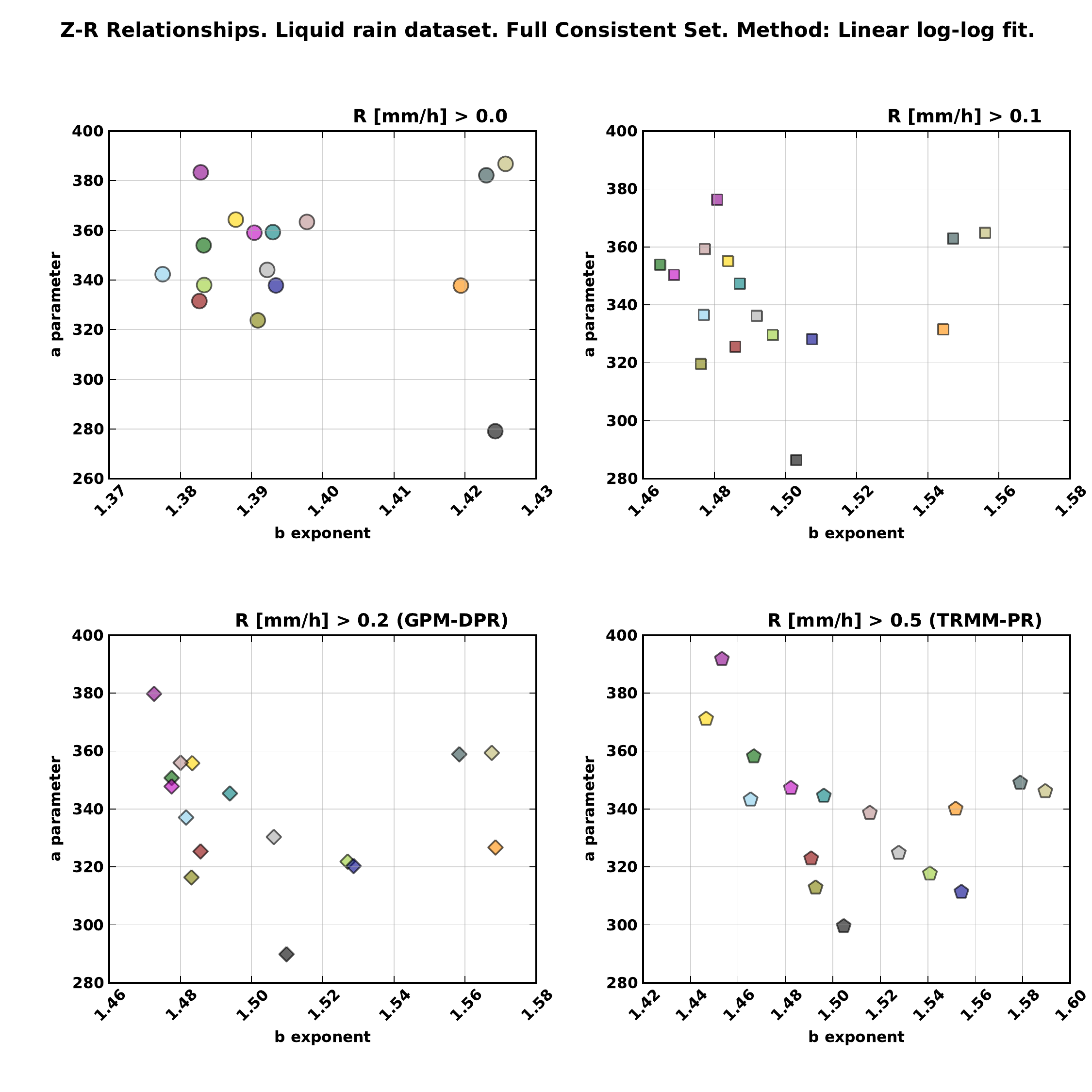}
   \caption[Relaciones Z-R para la base empírica completa y diferentes intensidades de lluvia mínimas. Datos provienen de acumulaciones de 10 min.]{\textbf{Relaciones Z-R para la base empírica completa y diferentes intensidades de lluvia mínimas. Datos provienen de acumulaciones de 10 min.} Se muestran las relaciones Z-R para la base empírica completa (restringida a todos los episodios de precipitación líquida con intensidades de precipitación máximas de 20 mm/h ) calculada para todos los disdrómetros de la red y bajo cuatro intensidades de precipitación mínima en todos los disdrómetros a un tiempo. Los datos provienen de acumulaciones de 10 min. Simula las sensibilidades estimadas para varios sensores radar en términos de intensidad de precipitación mínima detectable. Los coeficientes de la relación Z-R han sido calculados mediante regresión lineal simple en escala logarítmica.}
\label{fig:ZRwhole_10min}
\end{center}
\vspace{0.5cm}
\end{figure}

\begin{figure}[H] 
\begin{center}
   \includegraphics[width=0.95\textwidth]{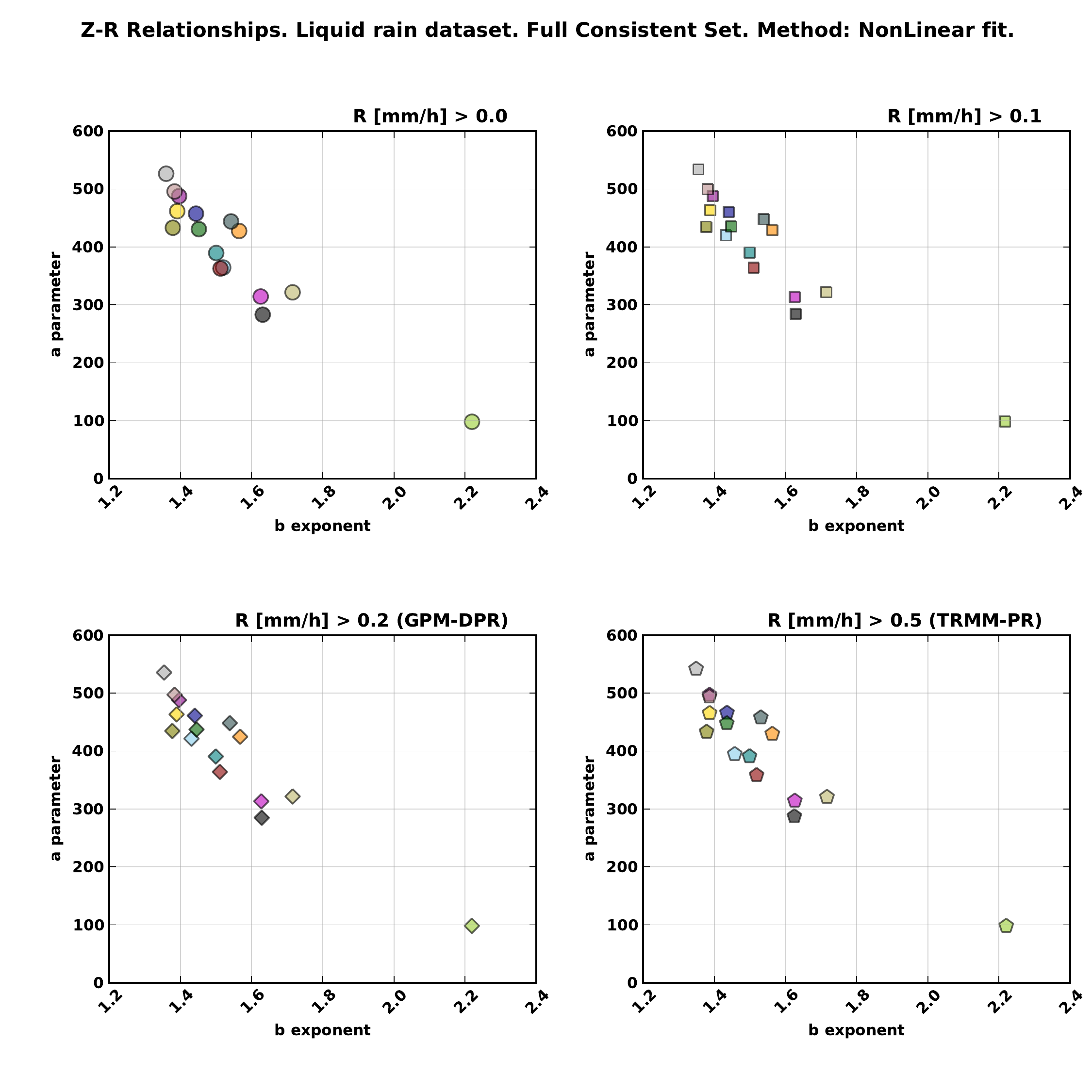}
   \caption[Relaciones Z-R para la base empírica completa y diferentes intensidades de lluvia mínimas. Datos provienen de acumulaciones de 10 min.]{\textbf{Relaciones Z-R para la base empírica completa y diferentes intensidades de lluvia mínimas. Datos provienen de acumulaciones de 10 min.} Se muestran las relaciones Z-R para la base empírica completa (restringida a todos los episodios de precipitación líquida con intensidades de precipitación máximas de 20 mm/h ) calculada para todos los disdrómetros de la red y bajo cuatro intensidades de precipitación mínima en todos los disdrómetros a un tiempo. Los datos provienen de acumulaciones de 10 min. Simula las sensibilidades estimadas para varios sensores radar en términos de intensidad de precipitación mínima detectable.Los coeficientes de la relación Z-R han sido calculados mediante un ajuste no-lineal de mínimos cuadrados.}
\label{fig:ZRwhole_10min_nonlineal}
\end{center}
\vspace{0.5cm}
\end{figure}

\clearpage
\subsection{Análisis de un episodio de precipitación en forma de nieve. Método de escalado en un parámetro integral.}

Como se comentó en \S\ref{sec:escaladado_variab_esp} es posible aplicar formalmente el procedimiento de escalado de la distribución de tamaños de gota mediante un parámetro integral a distribuciones que provengan de otros agregados no necesariamente líquidos, como nieve o granizo. La interpretación microfísica de los resultados será en parte diferente, sobretodo en lo que respecta al significado de los parámetros $\alpha$ y $\beta$ que se sugería en \S\ref{sec:microfisica_escalado_esp}. Resulta por otra parte sugerente realizar el mismo análisis para los mismos parámetros de escalado que utilizábamos en la precipitación líquida para comprobar si observamos alguna característica peculiar.\\

La primera parte del análisis comprendía las relaciones entre $\alpha$ y $\beta$ para comprobar la consistencia general del método y observar el rango de valores que tenemos para ellos. Llaman la atención tres hechos:
\begin{itemize}
   \item La relación de consistencia se mantiene, compartiendo el patrón por el cual cada uno de los diferentes subconjuntos de momentos utilizados en las relaciones de leyes de potencias entre parámetros condicionan posibles desviaciones sistemáticas en la ley de consistencia. Tal y como sucedía en la precipitación líquida evitar momentos de orden bajo y alto incrementa la consistencia (salvo en el caso de escalar respecto de $D_{mass}$ posiblemente debido al mayor rango de tamaños diferentes que posee la nieve y los valores generalmente más altos de $D_{mass}$).
   \item Los valores de $\alpha$ no presentan valores negativos, aspecto que, para determinados episodios de precipitación líquida, sucedía en el caso de lluvia estratiforme. 
   \item El rango de valores en $\alpha$ y $\beta$ es amplio, los episodios de precipitación líquida no suelen poseer tanta amplitud de variación.
\end{itemize}

La principal característica de las modelizaciones de la función g(x), tal y como apreciamos en las figuras (\ref{fig:Var1ScalingFUNCIONgDISDROA1_nieve}) y (\ref{fig:Var1Scalingmulambda_nieve}) son notablemente diferentes del caso líquido ya que implican valores ne\-ga\-tivos de $\mu$ (en el caso más consistente en que se realiza una transformación logarítmica antes del ajuste). Observando la figura (\ref{fig:Var1Scalingmulambda_nieve}) vemos además que este hecho lo comparten todos los escalados cualesquiera que sean los parámetros de referencia. Solo el disdrómetro F1 sistemáticamente da valores de $\mu$ cercanos a 1.5 desviación sistemática que puede ser explicada por alguna particularidad instrumental en un episodio de nieve\footnote{Los disdrómetros Parsivel OTT poseen una resistencia para mantener la temperatura homogénea dentro del sistema de medida, en episodios de baja temperatura algún instrumento puede sufrir una bajada de temperatura en unos minutos mientras se activa la resistencia, esto puede ser la causa de la desviación del instrumento F1 ya que implica unos minutos con diferente concentración de gotas.}.\\

Tal y como sucedía en el caso de precipitación líquida parece una relación aproximadamente lineal entre las estimaciones de $\mu$ y $\lambda$ que dados los estudios previos sobre muestreo \S\ref{teo:relacionmulb}, pudiera se debida al problema de muestreo. Un análisis de otros episodios de nieve a diferentes escalas permitiría despejar la incógnita de si la variación es natural o se debe a muestreo insuficiente de la población de copos de nieve.

\begin{center}
\begin{figure}[H] 
\vspace{0.1cm}
   \includegraphics[width=0.95\textwidth]{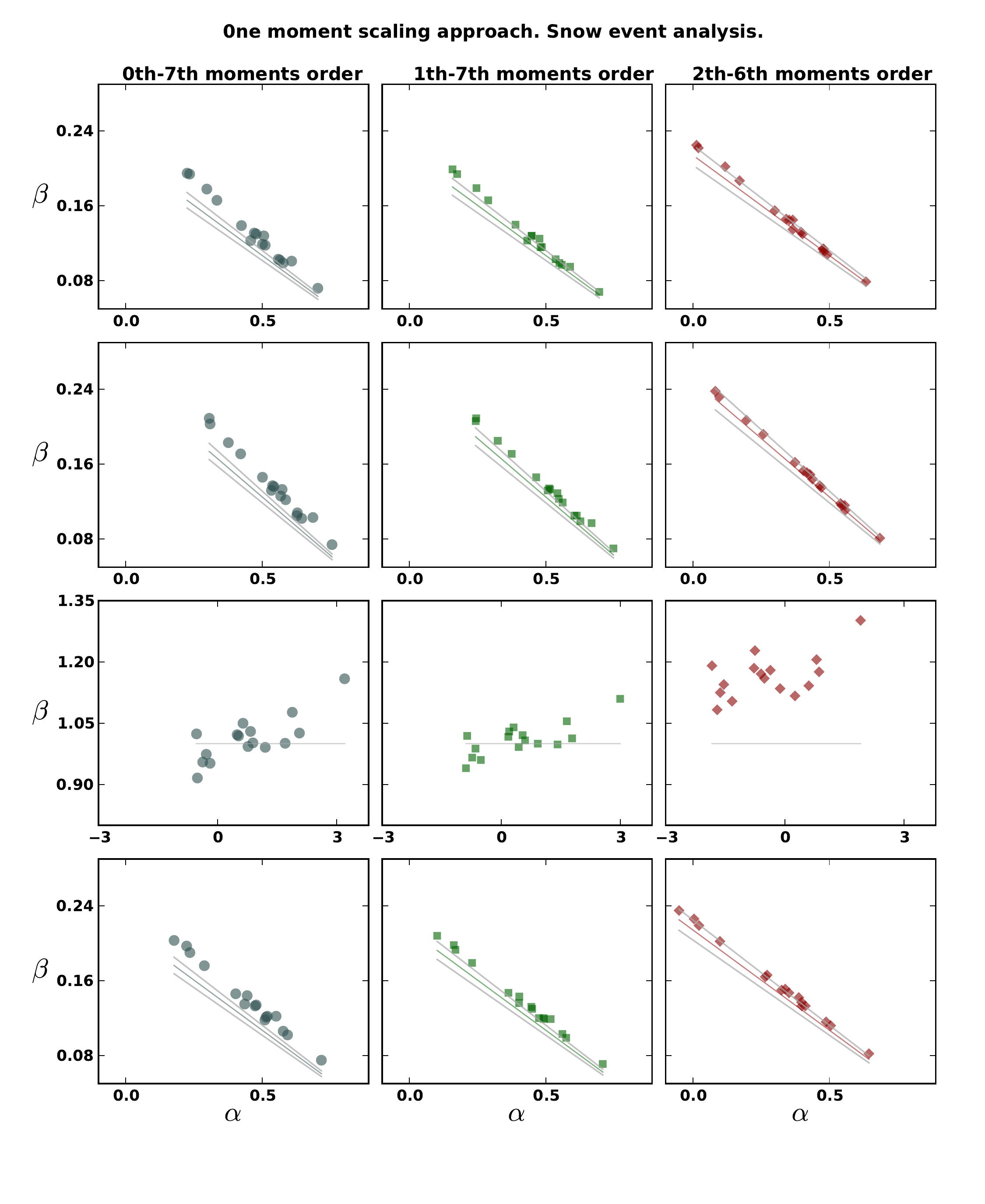}
\vspace{0.1cm}
   \caption[Representación de las estimaciones de $\alpha$ y $\beta$ a lo largo de la red de disdrómetros. Episodio de nieve del 10 de enero del 2010.]{\textbf{Representación de las estimaciones de $\alpha$ y $\beta$ a lo largo de la red de disdrómetros. Episodio de nieve del 10 de enero del 2010.}. Cada punto representa la modelización mediante el escalado en un parámetro de referencia y para cada disdrómetro. La primera línea representa los resultados cuando el parámetro de escalado es la intensidad de precipitación. La segunda línea representa el caso de escalar respecto del contenido en agua líquida. La tercera contiene muestra los resultados de escalar respecto de $D_{mass}$, finalmente la última fila representa el escalado respecto de $R^{*}$ que es el momento de orden 3.67 de la DSD.}
\label{fig:Var1ScalingAlfa1_nieve}
\vspace{0.1cm}
\end{figure}
\end{center}

\begin{center}
\begin{figure}[H] 
\vspace{0.75cm}
   \includegraphics[width=0.95\textwidth]{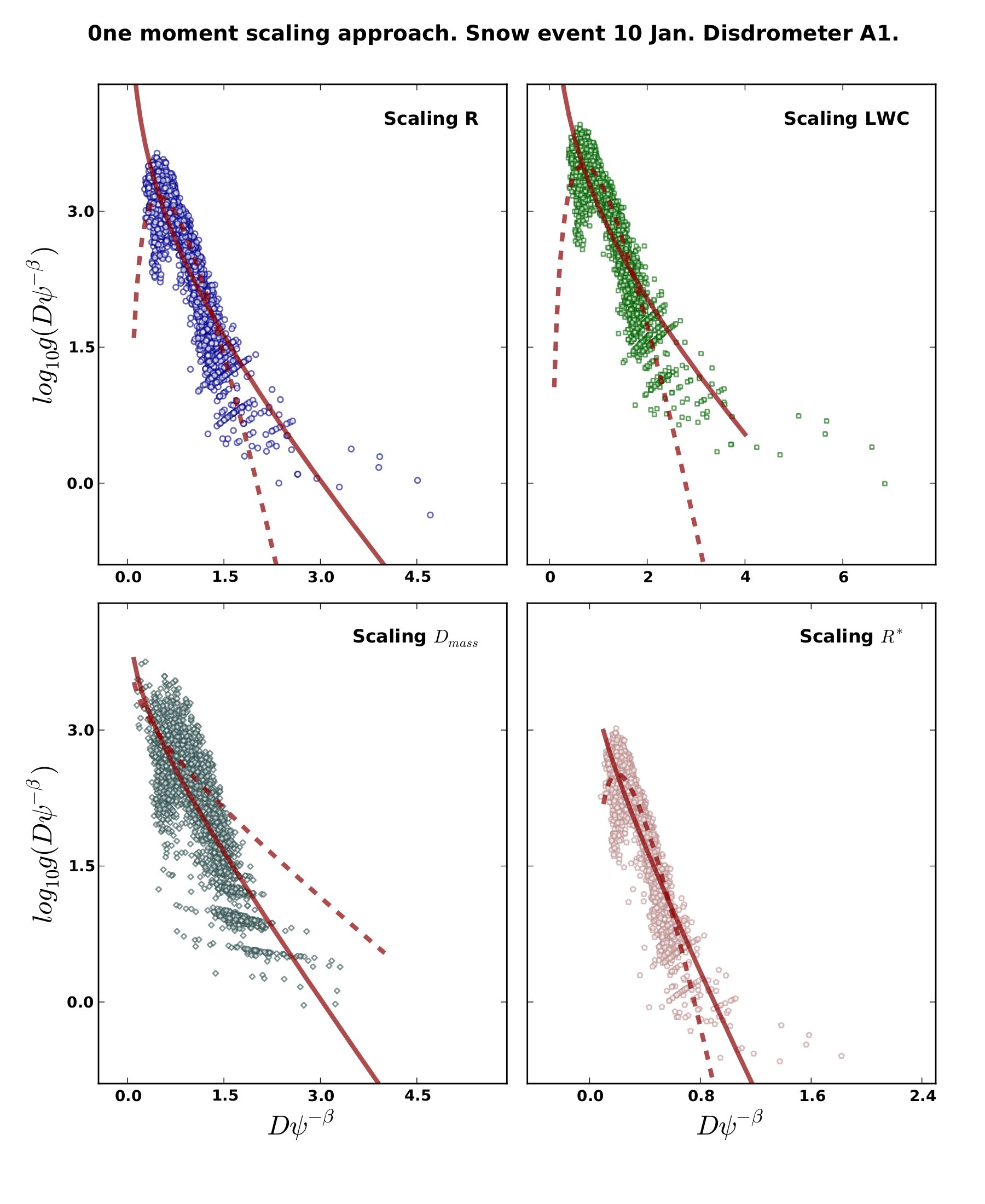}
\vspace{0.05cm}
   \caption[Representación de las funciones N(D) escaladas usando como parámetro la intensidad de precipitación R, el contenido en agua líquida, $D_{mass}$ y el momento de orden 3.67 ($R^{*}$). Junto con modelizaciones que utilizan la distribución gamma, para el episodio de nieve del 10 de enero del 2010.]{\textbf{Representación de las funciones N(D) escaladas usando como parámetro la intensidad de precipitación R, el contenido en agua líquida, $\mathbf{D_{mass}}$ y el momento de orden 3.67 ($R^{*}$). Junto con modelizaciones que utilizan la distribución gamma, para el episodio de nieve del 10 de enero del 2010.} Símbolos constituyen los datos obtenidos para $g(D\,\psi^{-\beta})$ del experimento. Las líneas en rojo representan las dos metodologías de ajuste se basan en ajustes no lineales de Levenberg\textendash Marquardt, las líneas discontinuas son ajustes directos, las continuas bajo una transformación logarítmica. Esta transformación permite tanto ajustar mejor todo el espectro de tamaños como respetar la condición de consistencia (\ref{eqn:consistenciaGammaK-1moment}) aunque en el caso de episodios de nieve esta última presente más diferencias relativas que en la tabla (\ref{TablaKAPPA}) .}
\label{fig:Var1ScalingFUNCIONgDISDROA1_nieve}
\vspace{0.25cm}
\end{figure}
\end{center}

\begin{center}
\begin{figure}[H] 
\vspace{0.75cm}
   \includegraphics[width=0.95\textwidth]{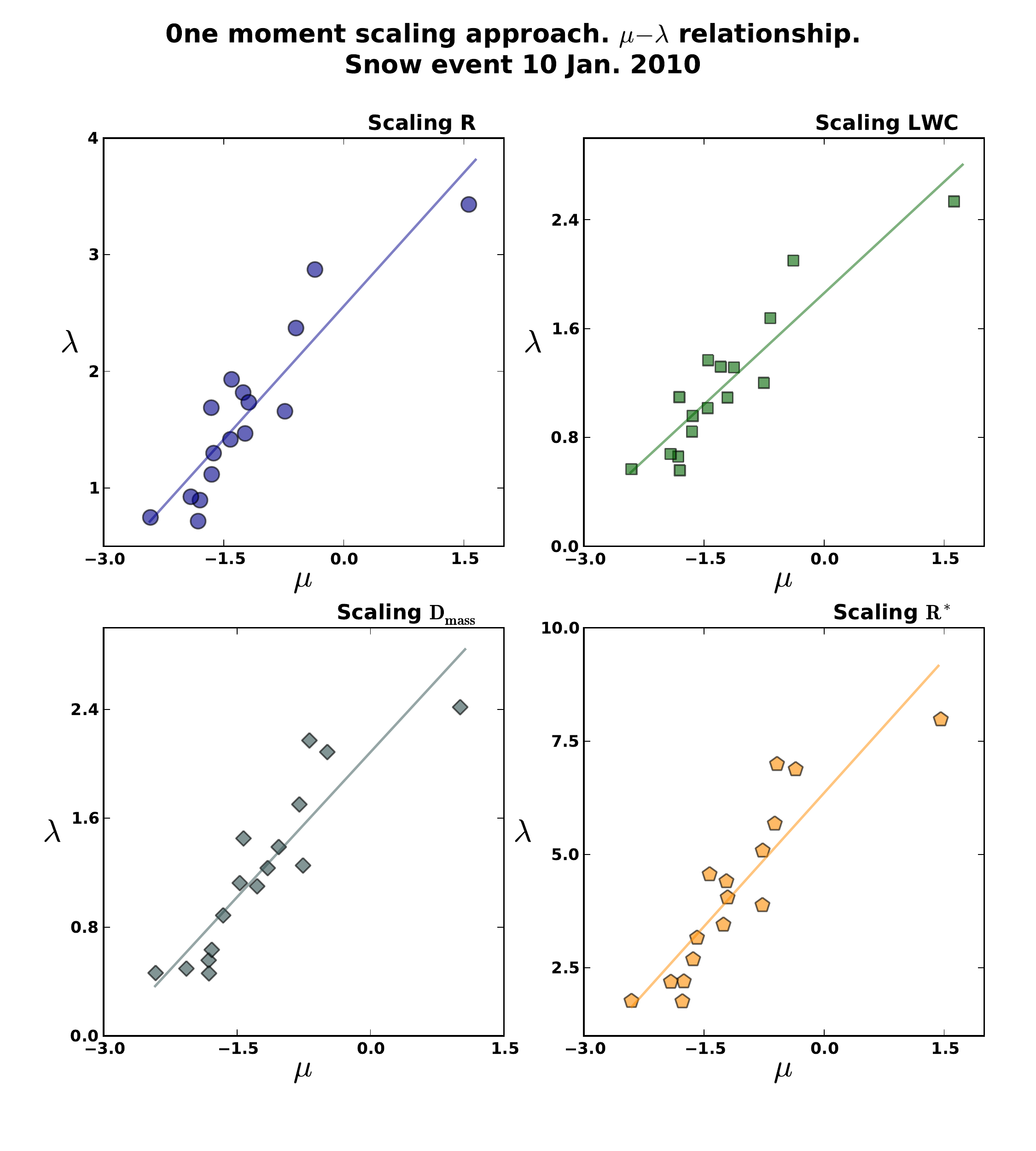}
\vspace{0.05cm}
   \caption[Representación de las estimaciones de $\mu$ y $\lambda$ correspondientes a una distribución gamma, dada por ec. (\ref{eqn:gammaScaling1moment}), bajo los escalados en R, W, $D_{mass}$ y $R^{*}$.]{\textbf{Representación de las estimaciones de $\mu$ y $\lambda$ correspondientes a una distribución gamma, dada por ec. (\ref{eqn:gammaScaling1moment}), bajo los escalados en R, W, $D_{mass}$ y $R^{*}$.} Cada punto representa la modelización para cada disdrómetro de la red. Los valores provienen de ajustes no lineales como los dados en la figura (\ref{fig:Var1ScalingFUNCIONgDISDROA1_rain}). En la metodología usada en las dos primeras líneas se han incluido los errores cuadráticos medios que resultan del método de ajuste no lineal.}
\label{fig:Var1Scalingmulambda_nieve}
\vspace{0.25cm}
\end{figure}
\end{center}



\renewcommand\chapterillustration{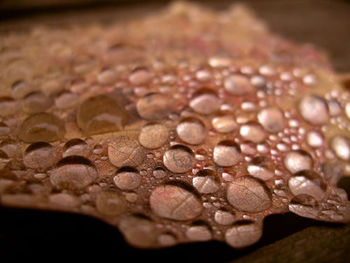}
\chapter{Extended analysis of discretization of disdrometric measurements}

\label{sec:newBINS}

In the chapter \S\ref{chap:BINNING} were shown results asserting the relevance of discretisation processes typical of disdrometric measurements. In this appendix that analyzes is extended by a methodological analysis of the errors that appear on integral rainfall parameters of precipitation (given an uncertainty on the values of the parameters of the DSD when a gamma model is assumed). Also an analysis of the errors of the DSD parameters of a gamma distribution are estimated when uncertainties on integral rainfall parameters exists. This study is completed by evaluating the consequences of discretisation in the relations $\mu-\lambda$ (see section \S\ref{teo:relacionmulb}).\\

Along this appendix the author has chosen another set of possible discretisation processes (which complements the \S\ref{chap:BINNING} analyzes). The main motivation is that disdrometer JWD classifies internally the drops sizes within 127 class intervals, that in a second step, are re-binned within 20 interval class (bins). The election of that second set of bins depends slightly with the experiment. In the case of \citep{caracciolo_prodi_etal_2006_aa} that was shown on \S\ref{chap:BINNING} the interval class distance $\Delta D_{i}$ decrease just in the same diameter values in which the disdrometer Parsivel OTT increase the interval class diameter. For this reason for the JWD disdrometers three configurations are included in this appendix: JWD-Alabama, JWD-Campos y JWD-Sheppard obtained from the study\citep{2000Campos}, the experiment carried on at Alabama by NASA-GPM, and the study \citep{sheppard_joe_1994_aa}, respectively. The last two cases are similar between them therefore the results are similar to 
the previously shown on \S\ref{chap:BINNING} for a generic JWD disdrometer. Finally to compare the results for these sets of bins, two additional artificial discretisations are defined, both with bins between 0.1 to 8 mm, the first is a lineal scale with 50 class intervals, the second is a logaritmic scale with 30 intervals.\\

 \begin{figure}[t]
 \vspace{0.10cm}
 \begin{center}
 \includegraphics[width=12.0cm]{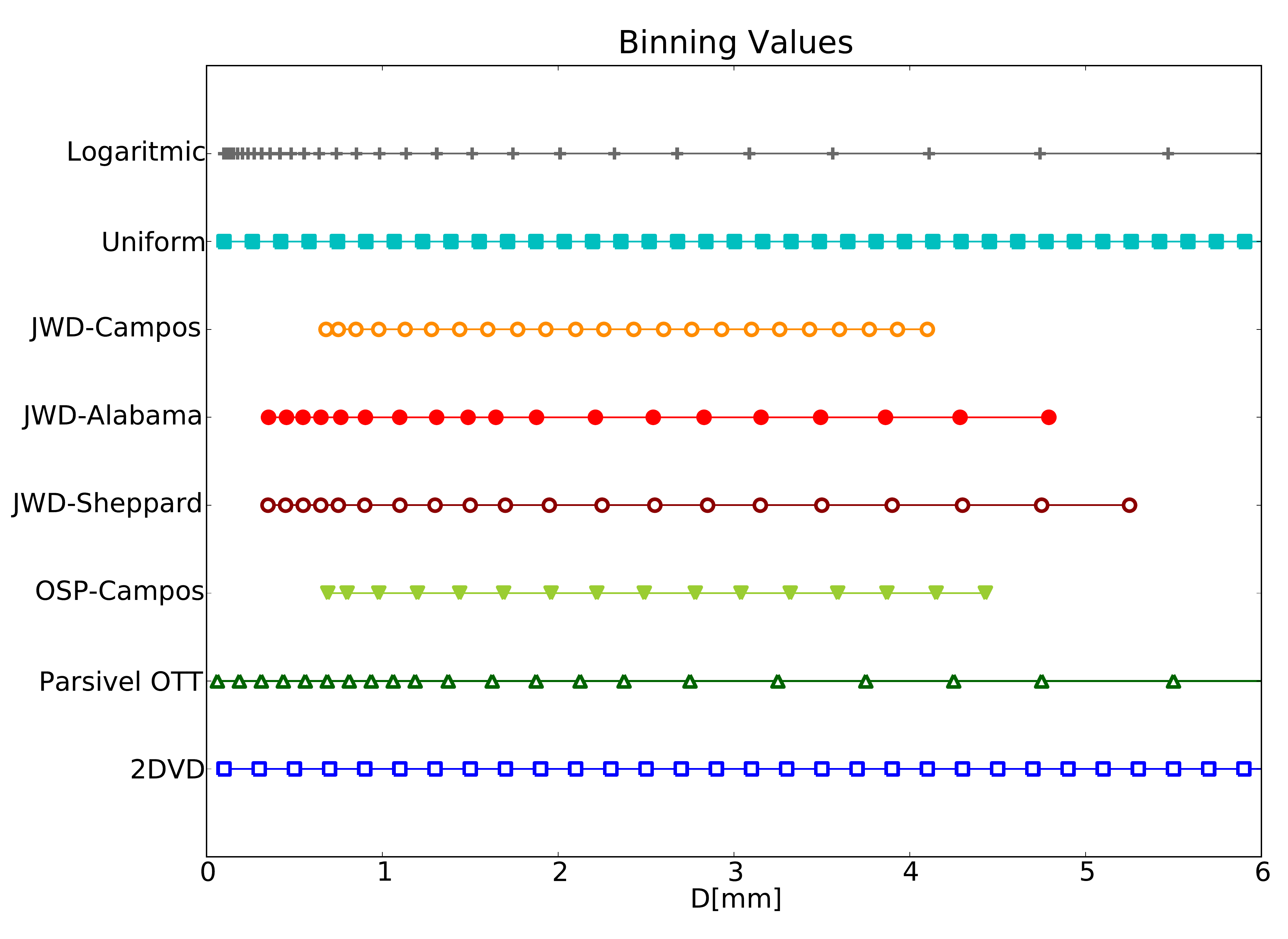}
 \vspace{0.10cm}
\caption[Binnings analysed along the appendix corresponding to several JWD devices, Parsivel OTT, OSP and 2DVD]{\textbf{Binnings analysed along the appendix corresponding to several JWD devices, Parsivel OTT, OSP and 2DVD}. Bins analysed in this study showing the central size classification values used by each instrument, as extracted from \citep{2000Campos,sheppard_joe_1994_aa,loffler-mang_joss_2000_aa} and the NASA-GPM-\textit{Ground Validation} project in the case of JWD-Alabama.}\label{fig:BINNINGS_appx}
 \end{center}
\vspace{0.05cm}
 \end{figure}
 
\begin{table}[t]
 \vspace{0.05cm}
\caption[Precipitation categories from \citep{TokayShort1996} and minimum values for the function g(k) as defined by equation (\ref{eqn:gdek})]{\textbf{Precipitation categories from \citep{TokayShort1996} and minimum values for the function g(k) as defined by equation (\ref{eqn:gdek})}.  For the precipitation categories from \citep{TokayShort1996} previously introduced in the section \S\ref{sec:defsigmas}, see table (\ref{t1}), the value of the Gaussian width $\sigma$ value equal to 20\% of mean value. The $k_{0}$ values, that for each category produce a minimum for the function g(k) as defined by equation (\ref{eqn:gdek}), are also shown.}\label{t1_appx}
\vspace{0.25cm}
\ra{1.20}
 \begin{center}
 \begin{tabular}{lccc}
 \toprule
 \textbf{Category}$^{*}$   & $\mathbf{\sigma(N_{0})}$   & $\mathbf{\sigma(\mu)}$  & \textbf{min g(k)}\\
 \midrule

 very light (vl)   &   1058   & 0.34 & 2.80\\
 light (l)         &   2620   & 0.46 & 2.21\\
 moderate (m)      &   4820   & 0.58 & 1.61\\
 heavy (h)         &   16020  & 0.78 & 1.12\\
 very heavy (vh)   &   66400  & 1.22 & 0.05\\
 extreme (e)       &   85200  & 1.78 & -2.70\\
 
 \bottomrule
 \end{tabular}
\end{center}
\vspace{0.7cm}
 \end{table}

\section{Sensitivity in the estimates of integral parameters due to errors in the DSD parameters}

The main objective of this section is understand the relationship between uncertainties of integral rainfall parameters and the uncertainties of DSD parameters when a Gamma distribution is used to model the raindrop size distribution. To achieve this goal it is performed an analysis of the stability of the estimations of integral rainfall parameters when we assume a specific error (that we will model as Gaussian noise) around the values that define each precipitation category. This problem can be understood from two different points of view:

\begin{itemize}
   \item The first supposes that a gamma distribution exists with fixed values $(\mu_{0},\lambda_{0},N^{(g)}_{0})$, but due to sampling and binning issues, we have an estimate $(\mu,\lambda,N^{(g)})$ which differ from the real amounts in the quantities $\delta \mu$, $\delta \lambda$ and $\delta N^{(g)}$. Given this situation, we ask what the expected errors are for the $M_{k}$ integral parameters for different error rates in the DSD parameters. 
   \item The second point of view supposes that the real distribution has some values $(\mu,\lambda,N^{(g)})$ that fluctuate due to physical or stochastic variations around the model with parameters for $(\mu_{0},\lambda_{0},N^{(g)}_{0})$. In this case, we ask what differences exist between the real and modeled moments\footnote{In the following sections to avoid a too heavy notation the super-index of $N^{(g)}$ it is omitted.}.
\end{itemize}

Generically, given a parameter vector of $\vec{p}_{0}=(\mu_{0},\lambda_{0},N^{(g)}_{0})$ and a fluctuation of $\delta \vec{p}=(\delta \mu$, $\delta \lambda$, $\delta N^{(g)}$), the following is true:
\begin{equation}
M_{k}(\vec{p}_{0}+\delta \vec{p})\simeq M_{k}(\vec{p}_{0})+\vec{f}_{M}^{T}(\vec{p}_{0})\cdot\delta\vec{p}+\frac{1}{2}\delta  \vec{p}^{T} \cdot\hat{\mathcal{H}}_{M}(\vec{p}_{0}) \cdot\delta \vec{p}
\label{eqn:desarrolloM}
\end{equation}
where $\vec{f}_{M}$ is the gradient vector of $M_{k}$ and $\hat{\mathcal{H}}_{M}$ is the Hessian matrix of $M_{k}$ defined as
\begin{equation}
\left(\vec{f}_{M}^{T}(\vec{p})\right)_{i}=\frac{\partial M_{k}}{\partial p_{i} }
\end{equation}
\begin{equation}
\left(\hat{\mathcal{H}}_{M}(\vec{p})\right)_{i,j}=\frac{\partial M_{k}}{\partial p_{i} \partial p_{j}}
\end{equation}

 \begin{figure}[t] 
\vspace{0.3cm}
 \begin{center}
 \includegraphics[width=11.8cm]{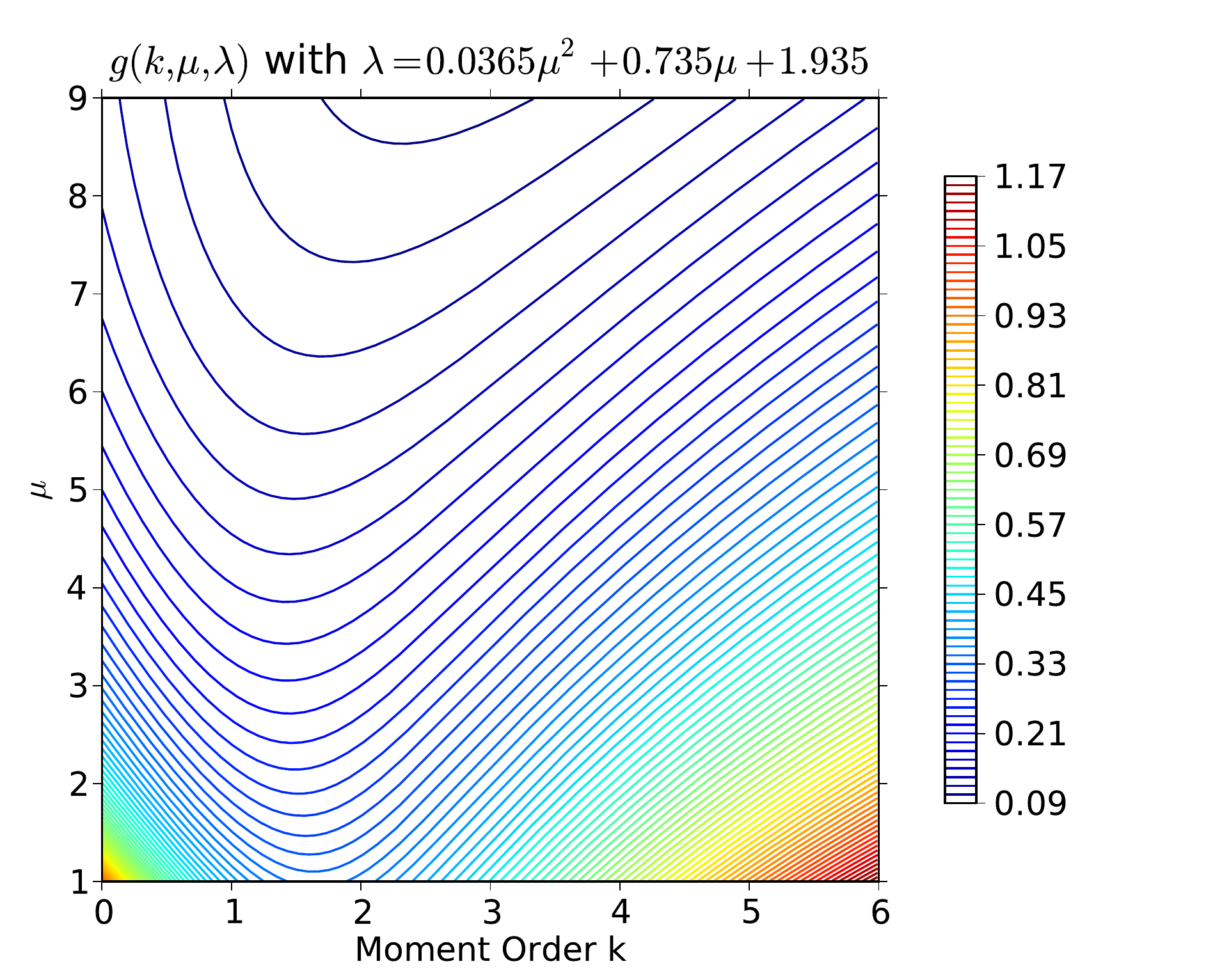}
 \end{center}
\vspace{0.3cm}
  \caption[Study of second order correction of $M_{k}$ error]{\textbf{Study of second order correction of $M_{k}$ error}. Function $g(\mu,l,k)$ defined by (\ref{eqn:gdek}) for k values between $[0,6]$ and $\mu$ between $[1,9]$. The value of $\lambda$ was determined by the relationship $\lambda=0.0365\mu^{2}+0.735\mu+1.935$. The presence of a minimum around k of 1.5 was observed with slight variations when $\mu$ is increased. }\label{fig6a_appx}
\vspace{0.3cm}
 \end{figure}

The sensitivity analysis involve taking a broad set of samples $S_{i}$ where each sample has a total number of drops conditioned by the values of $N^{(g)}$ and a Poisson process that governs the number of drops that reach the simulated instrument. In general, we refer to $\mathcal{S}=(S_{1},...,S_{N})$ as the set of samples over which we average the values of (\ref{eqn:desarrolloM}). That is,\\
\begin{equation}
\left<M_{k}(\vec{p}_{0}+\delta \vec{p})\right>_{\mathcal{S}}=\frac{1}{N}\sum_{S_{i}=S_{1}}^{S_{N}}M_{k}^{(S_{i})}(\vec{p}_{0}+\delta \vec{p})
\label{eqn:mediaMkfull}
\end{equation}
where $M_{k}^{(S_{i})}$ is calculated based on the estimate of the $(\mu,\lambda,N^{(g)})$ values for sample $S_{i}$. The value of N is supposed to be sufficient to reach stable values and arrive at a situation of asymptotic behavior in (\ref{eqn:mediaMkfull}). The Taylor series structure (\ref{eqn:desarrolloM}) indicates that if the distribution of errors $\delta\vec{p}$ in $\mathcal{S}$ is symmetric around zero; we then have
\begin{equation}
\left<M_{k}(\vec{p}_{0}+\delta \vec{p})\right>_{\mathcal{S}} \simeq  \left<M_{k}(\vec{p}_{0})\right>_{\mathcal{S}}=M_{k}(\vec{p}_{0})
\label{eqn:mediaMk}
\end{equation}

as long as $\delta \vec{p}$ is small. Significant relative deviations in $(\delta\vec{p})_{i}$ can imply the need to include a second term based on $\hat{\mathcal{H}}_{M}(\vec{p})$ even if the distribution of errors is symmetric. For the case in which the $(\delta\vec{p})_{i}$ fluctuations in $\mathcal{S}$ are not identically distributed around zero, we also include a term that corresponds to the gradient in the average.\\

For the case in which we wish to determine the variance over the set $\mathcal{S}$ of $M_{k}(\vec{p}_{0}+\delta \vec{p})$, we begin from
\begin{equation}
\sigma^{2}_{\mathcal{S}}(M_{k}(\vec{p}_{0}+\delta \vec{p}))=\frac{1}{N}\sum_{S_{i}=S_{1}}^{S_{N}}\left [M_{k}^{(S_{i})}(\vec{p}_{0}+\delta \vec{p})- \overline{M}_{k}^{(S_{i})} \right]^{2}
\label{eqn:sigmaMk}
\end{equation}
and if the values of $(\delta \vec{p})_{i}$ are sufficiently small so that the equation (\ref{eqn:mediaMk}) is adequate, then it is possible to express the variance of $M_{k}$ in $\mathcal{S}$ as
\begin{equation}
\sigma^{2}_{\mathcal{S}}(M_{k}(\vec{p}_{0}+\delta \vec{p}))=\frac{1}{N}\sum_{S_{i}=S_{1}}^{S_{n}}\left [\vec{f}_{M}^{T}(\vec{p}) \cdot \delta\vec{p} \right]^{2}
\label{eqn:sigmaMk_aprox}
\end{equation}
This expression gives rise to an equation in terms of variances and covariances. The variances arise from the terms $(\delta \vec{p})_{i}^{2}$ and the covariance from the terms $(\delta \vec{p})_{i}(\delta \vec{p})_{j}$. For the case in which the covariance is the product of the variances, we have the usual equation for the propagation of errors as the quadratic sum of errors for each variable.\\

For the case of disdrometric measurements, it is normal for some of the hypotheses to fail (either because the relative errors of some component of $\delta \vec{p}$ are moderated or because of the presence of errors between components), which is why the previous considerations must be performed carefully.\\

This work analyzes the average values of the moments. To answer the question of the sensitivity of these integral rainfall parameters to errors in DSD parameters estimations, the most convenient method is to analyze each of the contributions $\mu$, $\lambda$ and $N^{(g)}$ individually. Given the possibility of coupling between the first two contributions (either a statistical or implicit physical coupling between the parameters), the role of $\mu$ and $N^{(g)}$ has been analyzed (the expressions corresponding to $\lambda$ are given under appendix \ref{app:mmethod}). The supposition is made that around the values given in Table (\ref{t1}) there is a Gaussian distribution of errors of $\delta\mu$ and $\delta N^{(g)}$. As we have demonstrated previously, it is possible to analytically determine which errors are expected for a specific value of these fluctuations, although the results have to be averaged, in our case, over the Gaussian distributions (choose to include or not include the restrictions under the 
sign -positive or negative- for those distributions).
We can estimate the relevance of the following moments under the hypothesis of small errors in $N^{(g)}$,\\
\begin{equation}
M_{k}(N^{(g)}_{0}+\delta N^{(g)})\simeq M_{k}(N^{(g)}_{0})\left[1+\frac{\delta N^{(g)}}{N^{(g)}_{0}}\right]+O(\delta N^{(g)})^{2}
\end{equation}
The calculations made here corroborate a significant property of the previous expression in which the corrections under Gaussian noise are small in the first order and tend to be eliminated with symmetrical values distributions for $\delta N^{(g)}_{0}$. Restrictions in the fluctuation symbol imply that systematic biases in the estimates of the moments are proportionate to $\delta N^{(g)}_{0}$ without any significant dependencies in k.\\ 

For the case of the $\mu$ parameter up to the second order,\\
\begin{equation}
M_{k}(\mu_{0}+\delta \mu)\simeq M_{k}(\mu_{0})\left [1+f(\mu_{0})\delta\mu+\frac{1}{2}g(\mu_{0})(\delta \mu)^{2}\right]
\end{equation}
where 
\begin{equation}
f(\mu_{0},k,\lambda)=\psi_{0}(k+\mu_{0}+1)-ln(\lambda)
\end{equation}
if the gamma function is $\psi_{0}$. The correction of the first order under a sufficient average (large value of N in the expression \ref{eqn:mediaMkfull}) on the fluctuation $\delta \mu$ that follows a Gaussian distribution must be annulled; therefore, the deviations in the moments should be insignificant. However, the role of the following term, which corresponds to one of the elements of $\hat{\mathcal{H}}_{m}$, is noteworthy and appears even for errors in $\mu$ of 10-15\%. This correction is written as\\
\begin{equation}
g(\mu_{0},k,\lambda)=\psi_{1}(z)+\psi_{0}(z)^{2}-2ln(\lambda)\psi_{0}(z)+ln(\lambda)^{2}
\label{eqn:gdek}
\end{equation}
where $z=\mu_{0}+k+1$ y $\psi_{1}$ is the first derivative of the digamma function. The function $g(\mu,\lambda,k)$ is a positive function for the usual values of $(\mu,\lambda)$ that present a unique minimum in k, which depends on the value pair $(\mu,\lambda)$. Analysis of these categories can be observed in Table (\ref{t1}). These analysis explain the behavior in the predictions of $M_{k}$ when errors in $\mu$ are $\geq 10\%$, as in Tables (\ref{t2}) and (\ref{t3}). For the case of weak rain, the error implies an overestimation of the concentration and reflectivity but not of the content of liquid water, a fact that is explained by the position of the minimum of g(k) near a k of 3. For the case of intense rains, this property is lacking as the minimum is located closer to a k of 0, and in this way all of the moments are overestimated by progressively greater amounts as k increases.\\

Regarding the position of this minimum, under the hypothesis of the $\mu-\lambda$ relationship suggested by \citep{zhang_vivekanandan_etal_2003_aa}, the minimum is stabilised at an approximate k of 1.5, and significant differences appear in this second term for small values of $\mu$, as can be appreciated in Figure \ref{fig6a_appx}).\\

Studying the deviations assuming that errors in $\delta \mu$ include errors introduced by the binning process, all of the disdrometers that do not measure small drops exhibit a similar pattern to the previous results. The binning process continues to affect the error when additive errors are implied over the sampling problems that constitute that the differences between the Parsivel OTT and the 2DVD disdrometer, which continue to be approximately 5\%, for the reflectivity (k=6).\\

\begin{table}[t]
 \vspace{0.10cm}

\caption{Gaussian errors $\delta \mu$ are generated by the normal distribution $\mathcal{N}(\mu_{0},(\delta \mu)^{2})$ and added to $\mu_{0}$, and the value $\mu=\mu_{0}+\delta \mu$ is obtained. The amount $M_{k}(\mu_{0}+\delta \mu)$ is shown relative to the analytical value $M_{k}(\mu_{0})$. Then  $m_{k}=M_{k}(\mu_{0}+\delta\mu)/M_{k}(\mu_{0})$ is presented. For the cases $M_{0}$ (Concentration), $M_{3}$ (Liquid Water Content) and $M_{6}$ (Reflectivity). The calculations arise from the average of 1000 different samples. Category \textit{Heavy Rain}. To understand the relevance of the different orders that contribute to $M_{k}(\mu_{0}+\delta \mu)$, restrictions are included in the sign of $\delta \mu$. In a similar way, $m_{k}=M_{k}(N^{(g)}_{0}+\delta N^{(g)})/M_{k}(N^{(g)}_{0})$ is presented in the case where $N^{(g)}$ is varied.}\label{t2_appx}

 \vskip10mm
\ra{1.12}
 \centering
 \begin{tabular}{ccccc}
 \toprule
Restriction   & $\sigma$     & $m_{0}$   & $m_{3}$  &  $m_{6}$   \\
\midrule
              & $0.05\mu$    & 1.002 & 1.003 & 1.005 \\
              & $0.10\mu$    & 1.02  & 1.02  & 1.03 \\
None          & $0.20\mu$    & 1.10  & 1.07  & 1.16 \\
              & $0.25\mu$    & 1.16  & 1.12  & 1.29 \\
              & $0.30\mu$    & 1.36  & 1.20  & 1.52 \\
\midrule
$\delta \mu >0$ & $0.15\mu$    & 0.98  & 1.12 & 1.25 \\
$\delta \mu >0$ & $0.20\mu$    & 0.98  & 1.16 & 1.34 \\
$\delta \mu <0$ & $0.15\mu$    & 1.07  & 0.93 & 0.87 \\
$\delta \mu <0$ & $0.20\mu$    & 1.12  & 0.97 & 0.85 \\

\midrule
      & $0.05N_{0}$    & 1.002  & 1.003 & 1.006 \\
None  & $0.20N_{0}$    & 1.002  & 1.001 & 0.99 \\
      & $0.25N_{0}$    & 1.010  & 1.002 & 0.99 \\
\midrule
$\delta N_{0} <0$   & $0.20N_{0}$    & 0.92   & 0.92  &  0.92 \\
$\delta N_{0} >0$   & $0.20N_{0}$    & 1.07  & 1.06 & 1.05 \\
 \bottomrule
 \end{tabular}
 \vspace{10mm}
 \end{table}

 \begin{table}[t]
  \vspace{0.10cm}

\caption{Gaussian errors $\delta \mu$ are generated by the normal distribution $\mathcal{N}(\mu_{0},(\delta \mu)^{2})$ and added to $\mu_{0}$, and the value $\mu=\mu_{0}+\delta \mu$ is obtained. The amount $M_{k}(\mu_{0}+\delta \mu)$ is shown relative to the analytical value $M_{k}(\mu_{0})$. Then  $m_{k}=M_{k}(\mu_{0}+\delta\mu)/M_{k}(\mu_{0})$ is presented. For the cases $M_{0}$ (Concentration), $M_{3}$ (Liquid Water Content) and $M_{6}$ (Reflectivity). The calculations arise from the average of 1000 different samples. Category \textit{Light Rain}. To understand the relevance of the different orders that contribute to $M_{k}(\mu_{0}+\delta \mu)$, restrictions are included in the sign of $\delta \mu$. In a similar way, $m_{k}=M_{k}(N^{(g)}_{0}+\delta N^{(g)})/M_{k}(N^{(g)}_{0})$ is presented in the case where $N^{(g)}$ is varied.}\label{t2b_appx}

 \vskip8mm
\ra{1.12}

 \centering
 \begin{tabular}{ccccc}
 \toprule
Restriction   & $\sigma$     & $m_{0}$   & $m_{3}$  &  $m_{6}$   \\
\midrule
                & $0.05\mu$    & 1.04  & 1.01 & 1.05 \\
                & $0.10\mu$    & 1.01  & 0.99 & 0.97 \\
 None           & $0.20\mu$    & 1.08  & 1.02 & 1.05 \\
                & $0.25\mu$    & 1.13  & 1.04 & 1.10 \\
                & $0.30\mu$    & 1.20  & 1.05 & 1.10 \\
\midrule
$\delta \mu >0$ & $0.10\mu$    & 0.96  & 1.03 & 1.07 \\
$\delta \mu >0$ & $0.30\mu$    & 0.90  & 1.09 & 1.32 \\
$\delta \mu <0$ & $0.10\mu$    & 1.05  & 0.98 & 0.92 \\
$\delta \mu <0$ & $0.30\mu$    & 1.33  & 0.97 & 0.88 \\

\midrule    
      & $0.05N_{0}$    & 1.00  & 0.99 & 0.97 \\
None  & $0.20N_{0}$    & 1.01  & 1.01 & 0.99 \\
      & $0.30N_{0}$    & 1.003  & 1.003 & 1.009 \\

\midrule  
$\delta N_{0} <0$   & $0.20N_{0}$    & 0.916   & 0.91  &  0.91 \\
$\delta N_{0} >0$   & $0.20N_{0}$    & 0.96    & 1.03 & 1.07 \\
 \bottomrule
 \end{tabular}
 \vspace{0.2cm}
 \end{table}

\section{Moments Method. Errors in the estimation of DSD parameters}

In a similar way to the sensitivity or stability previoulsy explained for the case of integral rainfall parameters (as DSD moments) is natural estimate the stability of the DSD parameters when the are estimated with the moments method by supposing that the set  $(M_{l},M_{k},M_{m})$ has errors  $(\delta M_{l},\delta M_{k},\delta M_{m})$. The sensivisity analysis in this case implies estimate the error $\delta \mu$ as a first step.\\

By inspecting the expresion (\ref{eqn:desarrolloM}) is clear that given the known integral rainfall parameters, expresed by the vector, $\vec{v}_{0}=(M_{l},M_{k},M_{m})$ and their errors given by $\delta \vec{v}=(\delta M_{l},\delta M_{k},\delta M_{m})$ then, 

\begin{equation}
G(\vec{v}_{0}+\delta \vec{v}_{0})\simeq G(\vec{v}_{0})+\vec{\nabla}_{G}^{T}(\vec{v}_{0})\delta\vec{v}+\frac{1}{2}\delta \vec{v}^{T}\hat{\mathcal{H}}_{G}(\vec{v}_{0})\delta \vec{v}
\end{equation}

In the generic case of the method MM-lkm, the expresions for the gradient $\vec{\nabla}_{G}^{T}$ were given in the section (\ref{app:mmethod}). \\

Given  $\delta G$ we can write,

\begin{equation}
\delta \mu \simeq \frac{d \mu} {dG}\delta G
\end{equation}
equation that allow us to calculate $\delta \mu$. \\

With the value of  $\delta \mu$ it is possible to calculate $\delta \lambda$. Just we apply that $\lambda$ may be written as,
\begin{equation}
 \lambda=\left[\frac{M_{k}}{M_{l}}\frac{\Gamma(\mu+l+1)}{\Gamma(\mu+k+1)}\right]^{\frac{1}{l-k}}
 \label{eqn:lambda1}
 \end{equation}
 while for k and l natural numbers ($l-k>0$) it is written as:
 \begin{equation}
 \lambda=\left[\frac{M_{k}}{M_{l}}(\mu+l)\cdots(\mu+k+1)\right]^{\frac{1}{l-k}}
 \label{eqn:lambda2}
 \end{equation}
Therefore an expression for $\delta \lambda$ is,
\begin{equation}
\lambda(\vec{u}_{0}+\delta \vec{u}_{0})\simeq \lambda(\vec{u}_{0})+\vec{\nabla}_{\lambda}^{T}(\vec{u}_{0})\delta\vec{u}+\frac{1}{2}\delta \vec{u}^{T}\hat{\mathcal{H}}_{\lambda}(\vec{u}_{0})\delta \vec{u}
\end{equation}
where $\vec{u}_{0}=(\mu,M_{l},M_{k})$ and their errors are $\delta \vec{u}=(\delta \mu,\delta M_{l},\delta M_{k})$. The expressions at first order are given in section \S(\ref{app:mmethod}).\\

This is the first time that those general expressions were written, while in the bibliography the only analysed case was the particular case MM-246.

\begin{figure}[H]
 \vspace{1cm}
 \begin{center}
 \includegraphics[width=15.0cm]{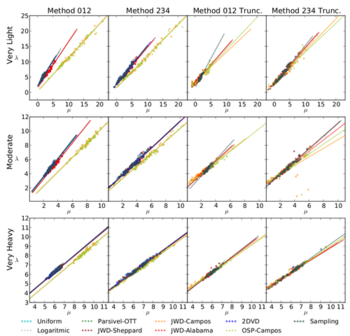}
 \end{center}
 \vspace{1.72cm}
\caption[Relationships $\hat{\mu}-\hat{\lambda}$ for several estimation methods and instruments]{\textbf{Relationships $\hat{\mu}-\hat{\lambda}$ for several estimation methods and instruments}. The artificial relationships $\mu-\lambda$ are compared for the different disdrometers, three precipitation categories and the methods of moments (usual and truncated). They were obtained with 50 DSDs within each category. The lines represents linear fits performed by regression method. Just in two case for moderated category by MM234 truncated the correlation coefficient was around 0.80, in all other cases it was larger than 0.94}\label{fig10a_appx}
 \vspace{1.92cm}
 \end{figure}

  \begin{figure}[H]
 \vspace{0.35cm}
 \begin{center}
 \includegraphics[width=10.5cm]{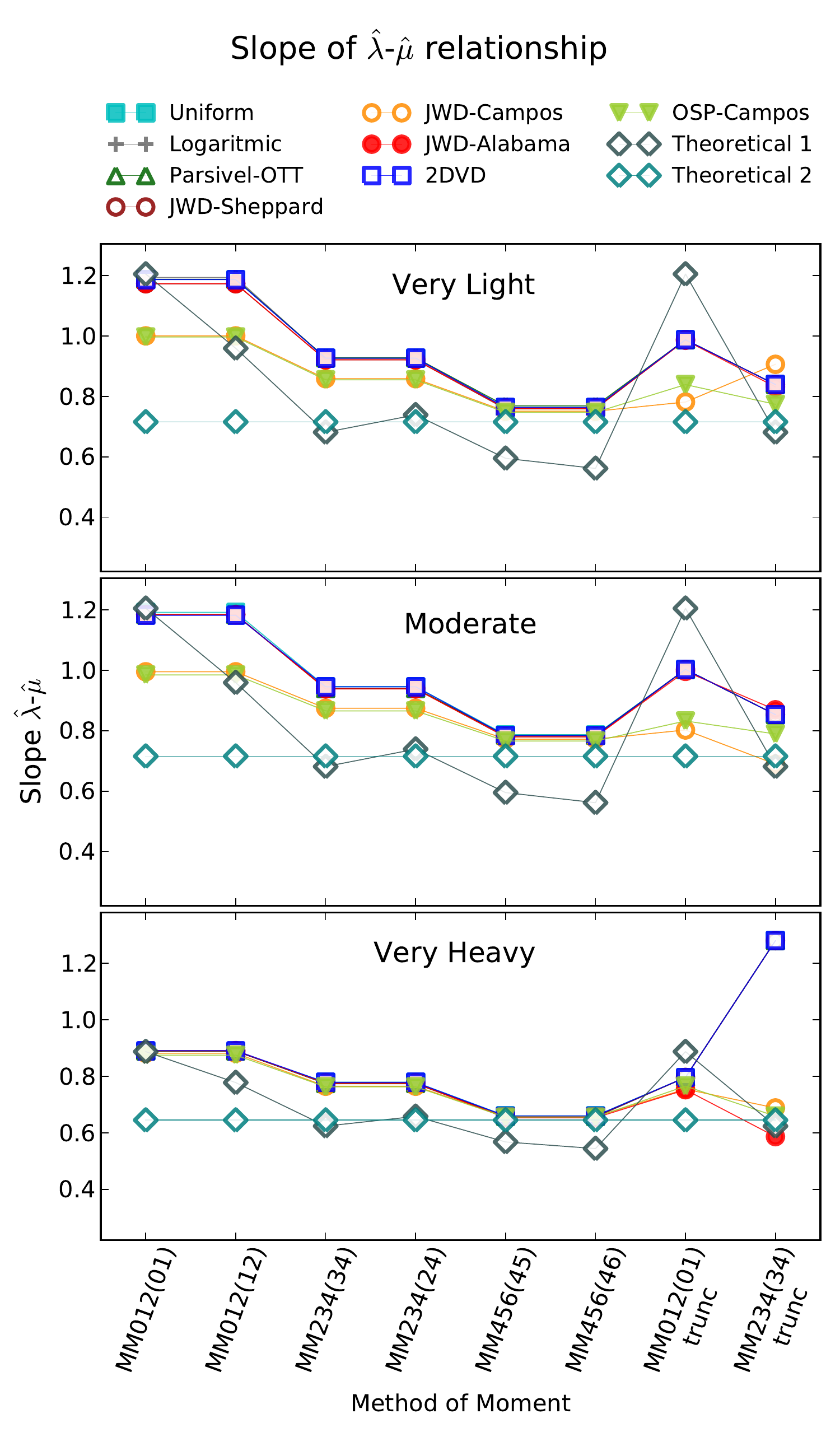}
 \end{center}
\vspace{0.95cm}
  \caption[Compartive of slope values of the relationship $\hat{\lambda}=a\hat{\mu}+b$]{\textbf{Compartive of slope values of the relationship $\hat{\lambda}=a\hat{\mu}+b$} Slopes of relationship  $\hat{\mu}-\hat{\lambda}$ that corresponds to figure (\ref{fig10a_appx}). Two analytical methodologies are shown to estimate the potential relationship, as defined in \S(\ref{sec:coefMULAMB})}\label{fig12a_appx}
 \end{figure}
\vspace{1.85cm}

\section{Analyses of $\mu-\lambda$ relationships}

A common question regarding the predictions of a gamma distribution function is the possible relationship between the $(\mu)$ form parameters and those of a ($\lambda$) scale for the experimental distributions of drop sizes. Several studies have proposed \citep{2008ChuandSu,zhang_vivekanandan_etal_2003_aa} that a relationship exists between these two parameters. This relationship would allow for the distribution function to be estimated from two parameters measured by means of remote sensing, which has led to interest in the basic physics of these relationships. The usual form in which these relationships are parameterized by means of the three coefficients, $\alpha$, $\beta$ and $\gamma$,

\begin{equation}
\lambda=\alpha \mu^{2}+\beta \mu+\gamma
\end{equation}

The primary problem is that artificial relationships appear due to the sampling problem between $\mu$ and $\lambda$, which mask the possible determination of a physical relationship \citep{moisseev_chandrasekar_2007_aa}. What usually occurs is that given some fixed values, $\mu$ and $\lambda$, the artificial relationships due to insufficient sampling correspond to a linear relationship, 

\begin{equation}
\hat{\lambda}=a \hat{\mu}+b
\label{eqn:relacARTIFICIAL}
\end{equation}

where the relationship does not appear between the real values but between their estimators $\hat{\mu}$ and $\hat{\lambda}$.

In the studies by \citep{moisseev_chandrasekar_2007_aa} and \citep{1987Chandra}, the relevance of this artificial relationship prediction was indicated by the measurement issues of small drops. Therefore, we analysed the relationships that arise due to the sampling and binning processes of the different disdrometers. To this effect, we studied which possible artificial relationships will appear for different precipitation categories. We adopted two different methodological schemes:
\begin{itemize}
\item The first set ($\mu$,$\lambda$) of each category can determine the artificial relationship among its estimators $\hat{\mu}$ and $\hat{\lambda}$ due to the sampling and discretisation problems. In the study by \citep{zhang_vivekanandan_etal_2003_aa}, possible analytical relationships were established for the estimators in the MM246 method. In our study, we compared the relationship between $\hat{\mu}$ and $\hat{\lambda}$ with general analytical expressions for MMlkm, as shown in \ref{sec:coefMULAMB}. 

\item The other approach \citep{2008ChuandSu} is based on establishing a general analytical relationship between $\lambda$ and $\mu$ based on the quotient between the number of sample drops with a diameter equal to the average diameter and the total number of drops per volume unit. In this way, the quotient, $N(\overline{D})/M_{0}$, is adjusted to a polynomial of order 2 in $\lambda$, and from this relationship, we find the estimate for $\mu-\lambda$ according to equation (\ref{eqn:relacARTIFICIAL}). In our work, we adopted a more general comparison expression with $N(\overline{D}_{k})/M_{k}$, as derived in the appendix \ref{sec:coefMULAMB2}. 
\end{itemize}

One of the striking results in our analyses is the dependence of artificial relationships on the disdrometer used, especially in the case of the biased estimates of $\mu$ and $\lambda$ when a deficit is present in the measurement of small drops. Even disdrometers, such as the JWD-Sheppard instrument, present different predictions of the $\mu-\lambda$ relationship. Aside from the different methods, apparent different relationships can be presented. Figure (\ref{fig11}) presents the artificial relationships for each method and each disdrometer combined with linear adjustments, $\hat{\lambda}=a\hat{\mu}+b$, noting that the correlation coefficients obtained are always at least 0.94 (except for the case of MM234, which has a value of 0.8).\\

\begin{figure}[H]
 \vspace*{2mm}
 \begin{center}
 \includegraphics[width=15.0cm]{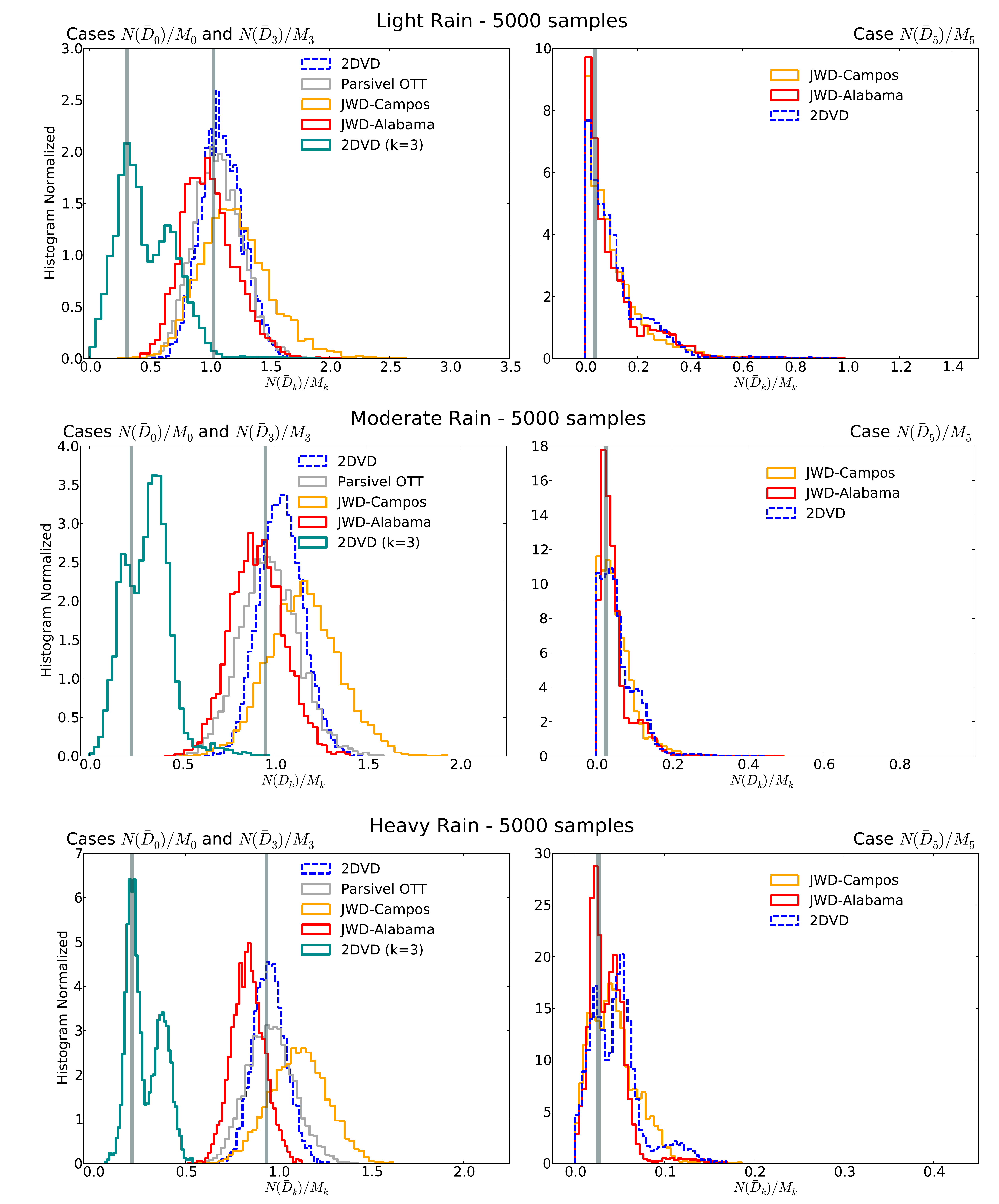}
 \vspace{1.50 cm}
 \end{center}
\caption[Histograms for the relationship given by equation (\ref{eqn:relationSHULU}) for several instruments]{\textbf{Histograms for the relationship given by equation (\ref{eqn:relationSHULU}) for several instruments}. The histograms of $N(\bar{D}_{k})/M_{k}$ are shown for 5\,000 different samples and several binning processes. Two extremes cases were indicated: k=0 y k=5. The case  k=3 in which $\bar{D}_{3}=D_{mass}$ its shown for the 2DVD. The vertical lines corresponds to the analytic estimation given by equation (\ref{eqn:relationSHULU}) for the values of $(\mu,\lambda)$ related with the 4 precipitation categories.}\label{fig11a_appx}
 \vspace{1.50 cm}
 \end{figure}
 
The expressions of the coefficients a and b can be contrasted with the analytical predictions shown in Figure (\ref{fig12a_appx}). The slope follows the pattern given by the (\ref{eqn:mulambdaT1}) relationship, while the relationship (\ref{eqn:mulambdaT2}) is only feasible for case MM456. Similar to the findings of \citep{KumarIEEE2011}, the difference between a truncated and non-truncated method is not critical to the slope estimation for cases of high precipitation intensity.\\

As shown by \citep{2008ChuandSu}, the uncertainties in the approximation are expressed in the probability distributions in Figure (\ref{fig11a_appx}) as histograms. These histograms show that dispersion introduces a bias between the analytical prediction and results of different disdrometers, indicating the relevance of the binning process in the $\hat{\mu}-\hat{\lambda}$ relationships that arise from this procedure. Different values of k produce distributions with different degrees of dispersion and with different deviations between the maximum distribution value and the theoretical relationship represented in the figures by the vertical line. We also observe cases with bimodal distributions (in some cases multimodal) as a consequence of the location of $\overline{D}_{k}$ in the discretised spectrum that can oscillate depending on the sample, according to the $\mu,\lambda$ values and disdrometer, between two or more values. 

\section{Equations of the moments method and error analysis}    
\label{app:mmethod}

In the table (\ref{tablaMMmetodosGAMMA_errores}) the expresions of $\mu(G)$ and $d\mu/dG$ are shown for several moments sets MM-lkm. The expressions in the case of MM012 reproduce (after algebraic operations) the usual equations which define the parameters  $\mu$ and $\lambda$ by using the mean and the variance in the size of the DSD:
\begin{equation}
\lambda=\frac{\overline{D}}{\sigma^{2}}
\end{equation}
\begin{equation}
\mu=\frac{\overline{D}^{2}}{\sigma^{2}}
\end{equation}
where $\overline{D}=M_{1}/M_{0}$ is the mean value of the diameter under the gamma distribution and $\sigma^{2}$ is the variance:
\begin{equation}
\sigma^{2}M_{0}=\int(D-\overline{D})N(D)dD
\end{equation}

\subsection{Estimation Error of the parameter $\mu$} 

In the case of estimating the error in $\mu$, aside from the following equation
(\ref{eqn:errorMUdesdeG}), we also approximate by the development of the first
order (together with the application of the chain rule):
\begin{equation}
\delta\mu=\frac{d\mu}{dG}\delta G
\label{eqn:errorMUdesdeG}
\end{equation}
and if we approximate also  $\delta G \simeq \vec{\nabla}_{G}^{T}(\vec{v}_{0})\delta \vec{v}$, which is the expansion until first order, we have:
\begin{equation}
\delta G=\frac{\partial G}{\partial M_{l}}\delta M_{l}+\frac{\partial G}{\partial M_{k}}\delta M_{k}+\frac{\partial G}{\partial M_{m}}\delta M_{m}
\end{equation}
As
\begin{equation}
G=\frac{M^{a}_{l}}{M^{b}_{k}M^{c}_{m}}
\end{equation}
the three components of $\vec{\nabla}_{G}^{T}(\vec{v}_{0})$ are:
\begin{equation}
\frac{\partial G}{\partial M_{l}}=a\frac{G}{M_{l}}
\end{equation}
\begin{equation}
\frac{\partial G}{\partial M_{k}}=-b\frac{G}{M_{k}}
\end{equation}
\begin{equation}
\frac{\partial G}{\partial M_{m}}=-c\frac{G}{M_{m}}
\end{equation}

This expansion is an extended versions of the  \citep{zhang_vivekanandan_etal_2003_aa}  and it allows the calculation of the errors on the gamma distribution parameters for any method MM-lkm.

Given that $a=b+c$ and if $\delta M_{n}/M_{n}$ is the same order for n =  k,l and m, then the prediction would be small errors on  $\mu$; but also it was shown that differences on $\delta M_{n}/M_{n}$ may be amplificated. Therefore the errors on $\mu$ would be significant.

\vskip10mm
\begin{table}[t]
\caption[Estimation of gamma distribution parameters using the moment method and its errors.]{\textbf{ Estimation of gamma distribution parameters and its errors using the moment method.} Four broadly used methods are shown. In regards to the methodology used to obtain the expressions, the generic method is introduced in the text, see  \S\ref{sec:metodoMomentos}. Also $d\mu/dG$ necessary for the estimation of $\mu$ and $\lambda$ errors is shown.} \label{tablaMMmetodosGAMMA_errores}

 \vskip10mm
\ra{1.30}
 \centering
 \begin{tabular}{lccc}
 \toprule
 \textbf{Method}         &  \textbf{Function G}                      &  $\mathbf{\mu(G)}$ &   $\mathbf{d\mu / dG}$ \\
 \midrule
MM012(01)             &  $\displaystyle\frac{M_{1}^{2}}{M_{0}M_{2}}$   &$\displaystyle\frac{1}{1-G}-2$  & $\displaystyle\frac{1}{(1-G)^2}$ \\[1.0 ex]
MM246(24)             &  $\displaystyle\frac{M_{4}^{2}}{M_{2}M_{6}}$   & $\displaystyle\frac{7-11G-\sqrt{14G^{2}+G+1}}{2(G-1)}$    & $\displaystyle\frac{8\sqrt{14G^{2}+G+1}+29G+3}{(4G^{2}-8G+4)\sqrt{14G^{2}+G+1}}$\\[2.0 ex]

MM346(34)             &  $\displaystyle\frac{M_{4}^{3}}{M^{2}_{3}M_{6}}$  &   $\displaystyle\frac{-8+11G+\sqrt{G^{2}+8G}}{2(1-G)}$   &  $\displaystyle\frac{3\sqrt{G^{2}+8G}+5G+4}{(2G^{2}-3G+2)\sqrt{G^{2}+8G}}$             \\[2.0 ex]

MM234(23)             &  $\displaystyle\frac{M_{3}^{2}}{M_{2}M_{4}}$      &  $\displaystyle\frac{1}{1-G}-4$     & $\displaystyle\frac{1}{(1-G)^2}$\\[2.0 ex]

MM456(45)             &  $\displaystyle\frac{M_{5}^{2}}{M_{4}M_{6}}$            &  $\displaystyle\frac{1}{1-G}-6$       & $\displaystyle\frac{1}{(1-G)^2}$              
\\[1.0 ex]

 \bottomrule
 \end{tabular}
 \vskip14mm
 \end{table}

\subsection{Estimating the error in parameter $\lambda$} 

Generally, the contributions of the first order use $\vec{\nabla}_{\lambda}=(\frac{\delta \lambda}{\delta \mu},\frac{\delta \lambda}{\delta M_{l}},\frac{\delta \lambda}{\delta M_{k}})$ and are given by,
\begin{equation}
\delta\lambda = \vec{\nabla}_{\lambda}^{T}\cdot (\delta \mu,\delta M_{l},\delta M_{k})
\end{equation}
\begin{equation}
\delta\lambda \simeq \frac{\partial\lambda}{\partial\mu}\delta \mu+\frac{\partial\lambda}{\partial M_{k}}\delta M_{k}+\frac{\partial\lambda}{\partial M_{l}}\delta M_{l}
\end{equation}

Given the value of $\lambda$, which can generally be written as
\begin{equation}
\lambda=\left[\frac{M_{k}}{M_{l}}\frac{\Gamma(\mu+l+1)}{\Gamma(\mu+k+1)}\right]^{\frac{1}{l-k}}
\end{equation}
the primary contribution to error of $\lambda$ supposes an error in $\mu$ and is given by
\begin{equation}
\delta\lambda\simeq \lambda\frac{1}{l-k}\left[\left( \psi_{0}(\mu+l+1)- \psi_{0}(\mu+k+1)\right)\delta \mu +\frac{1}{M_{k}}\delta M_{k}+\frac{-1}{M_{l}}\delta M_{l}\right]
\end{equation}
Making use of the recurrent relationship of the gamma function, $\psi_{0}(z+1)=\psi_{0}(z)+1/z$, different simple expressions are reached for each pair of values of k and l (an expression that will only depend on $\mu$ and m=l-k). 
\begin{equation}
\delta \lambda\simeq \frac{\lambda}{l-k}\left[\left(\sum_{i=0}^{l-k}\frac{1}{\mu+l-i}\right)\delta\mu+\frac{1}{M_{k}}\delta M_{k}+\frac{-1}{M_{l}}\delta M_{l}\right]
\label{eqn:errorLB}
\end{equation}

The error in $\lambda$ therefore appears to depend not only on the error in $\mu$ and on the value of $\mu$ but also on the method chosen to estimate $\lambda$ through the moments of order k and l. \\

\section{Coefficient estimations in relationship $\hat{\mu}-\hat{\lambda}$} 
\label{sec:coefMULAMB}

Based on the expression $\delta \lambda\simeq (\partial \lambda/\partial \mu)\delta \mu$, where, following the suggestion of \citep{zhang_vivekanandan_etal_2003_aa}, the quantity $\delta \lambda$, and hence the equation (\ref{eqn:errorLB}), is conditioned above all by $\delta \mu$,

\begin{equation}
\delta \lambda\simeq\frac{\lambda}{l-k}\left[\left(\sum_{i=0}^{l-k}\frac{1}{\mu+l-i}\right)\delta\mu\right]
\end{equation}

If $\delta \lambda=\lambda-\hat{\lambda}$ and $\delta \mu=\mu-\hat{\mu}$, the following can be written: 

\begin{equation}
\hat{\lambda}=\frac{\lambda}{l-k}\left[\left(\sum_{i=0}^{l-k}\frac{1}{\mu+l-i}\right)\right](\hat{\mu}-\mu)+\lambda
\label{eqn:mulambdaT1}
\end{equation}

from which we can estimate the ordinate of the origin and the slope given in
Figure (\ref{fig11}). This approximation is termed \textit{Theoretical 1} in
Figure (\label{fig12a_appx}). Conducting a stricter approximation is possible if
we know that,

\begin{equation}
\delta \lambda=(1/\overline{D})\delta\mu
\label{eqn:mulambdaT2}
\end{equation}
and use the following analytical prediction: $\overline{D}=(\mu+3.67)/\lambda$. We named this approximation \textit{Theoretical 2} in figure (\ref{fig12a_appx}).

\section{Estimation by means of $N(\overline{D}_{k})$ y $\overline{D}_{k}$ of the relationship $\mu-\lambda$} 
\label{sec:coefMULAMB2}

We define $\overline{D}_{k}$ and use the values of the $M_{k}$ moments given by (\ref{eqn:Mk}) to express:
\begin{equation}
\overline{D}_{k}=\frac{M_{k+1}}{M_{k}}=\frac{\mu+k+1}{\lambda}
\end{equation}
Given the same expression (\ref{eqn:Mk}) and definition (\ref{eqn:gamma}), 
\begin{equation}
\frac{M_{n}\lambda^{\mu+n+1}}{\Gamma(\mu+n+1)}=N(\overline{D}_{k})\overline{D}_{k}^{-\mu}e^{\lambda\overline{D}_{k}}
\end{equation}
which allows the analytical expression for $\overline{D}_{k}$ to be written as,
\begin{equation}
\lambda^{n+1}=\frac{N(\overline{D}_{k})}{M_{n}}\frac{e^{\mu+k}}{(\mu+k)^{\mu}}\Gamma(\mu+k+1)
\label{eqn:lambdaN1}
\end{equation}
As \citep{2008ChuandSu} indicated, it is possible to introduce a development of $\Gamma(z)$ as
\begin{equation}
\Gamma(z)\simeq z^{z-0.5}e^{-z}\sqrt{2\pi}\left[1+\frac{1}{12z}+\cdots \right]
\label{eqn:desarrolloGamma}
\end{equation}
This asymptotic development permits the adequate representation of the gamma function given that $\Gamma(1.5)=0.886$, and by means of its development, we obtain 0.83. The relative differences diminish when the $z$ module is increased in Equation (\ref{eqn:desarrolloGamma}). If we introduce this development and use the expression (\ref{eqn:lambdaN1}), which is valid for all of n, with n equals k then,
\begin{equation}
\lambda^{k+1}\simeq \frac{N(\overline{D}_{k})}{M_{k}}(\mu+k+1)^{k+0.5}\sqrt{2\pi}
\label{eqn:relationSHULU}
\end{equation}
que es general y resulta tanto más aproximada cuando mayor sea el valor de $\mu+k+1$ (desde el punto de vista de la aproximación bastante adecuada que supone la ecuación (\ref{eqn:desarrolloGamma})).\\

which is general and it is a better approximation for large values of $\mu+k+1$ (from the point of view of Equation (\ref{eqn:desarrolloGamma})). Given the size classification processes used in this study, an n of 4 and k of 3 can be quite adequate for studying disdrometric measurements. Thus, the procedure is to adjust $N(\overline{D}_{k})/M_{k}$ to a polynomial in $\lambda$ and use the above expression to reach the required $\mu-\lambda$ relationship.

\section{Comentarios}

En este apéndice, usando la misma metodología de generación de distribuciones simuladas de tamaños de gota que en el capítulo \S\ref{chap:BINNING}, se ha comprobado los efectos de los procesos de discretización en las  relaciones $\mu-\lambda$. En el capítulo \S\ref{chap:BINNING} ya comprobamos la presencia de valores anómalos de $\lambda$ cuando utilizamos métodos truncados para la estimación de los parámetros de la distribución gamma. La presencia de estos valores marginales puede dificultar la estimación de relaciones $\mu-\lambda$. Por tanto, ambos factores, la discretización y el muestreo insuficiente, condicionan las interpretaciones de dichas relaciones. Respecto de estas últimas, los procesos de clasificación en intervalos de clase discretos pueden provocar una relación artificial diferente de la que originan los procesos de muestreo.\\

Las metodologías introducidas en este apéndice acerca de la propagación de errores entre los parámetros integrales y los parámetros de la distribución gamma permiten estimar la relevancia, tanto de variaciones físicas reales de la DSD en los parámetros integrales como de los errores de estos en la estimación de los parámetros de forma y escala de la distribución gamma. Las metodologías desarrolladas en esta tesis permiten además su aplicación a cualquier método de los momentos cuestión no aparecida aun en la bibliografía.

\selectlanguage{english}

\chapterstyle{FancyUnnumberedChap}

\bibliographystyle{Biblio/ametsoc_RAM_Thesis}
\bibliography{Biblio/atmospheric_bib}

\begin{thebibliography}{138}
\expandafter\ifx\csname natexlab\endcsname\relax\def\natexlab#1{#1}\fi
\expandafter\ifx\csname url\endcsname\relax
  \def\url#1{{\tt #1}}\fi
\expandafter\ifx\csname urlprefix\endcsname\relax\def\urlprefix{URL }\fi
\expandafter\ifx\csname doiprefix\endcsname\relax\def\doiprefix{doi:}\fi

\bibitem[{{Abramov}(2006)}]{Abramov2006}
{Abramov}, R., 2006: A practical computational framework for the
  multidimensional moment-constrained maximum entropy principle. {\textit{
  Journal of Computational Physics}\/}, {\textbf{ 211}\/}, 198--209.

\bibitem[{{Abramov}(2009)}]{Abramov2009}
--- 2009: The multidimensional moment-constrained maximum entropy problem: A
  bfgs algorithm with constraint scaling. {\textit{ Journal of Computational
  Physics}\/}, {\textbf{ 228}\/}, 96--108.

\bibitem[{{Atlas} et al.(1973){Atlas}, {Srivastava}, and {Sekhon}}]{1973Atlas}
{Atlas}, D., R.~C. {Srivastava}, and R.~S. {Sekhon}, 1973: Doppler radar
  characteristics of precipitation at vertical incidence. {\textit{ Reviews of
  Geophysics and Space Physics}\/}, {\textbf{ 11}\/}, 1--+.

\bibitem[{Atlas and Ulbrich(2000)}]{AtlasUlbrich200microDSD}
Atlas, D. and C.~W. Ulbrich, 2000: An observationally based conceptual model of
  warm oceanic convective rain in the tropics. {\textit{ Journal of Applied
  Meteorology}\/}, {\textbf{ 39}\/}, 2165--2181.

\bibitem[{Babinsky and Sojka(2002)}]{Babinsky2002}
Babinsky, E. and P.~E. Sojka, 2002: Modeling drop size distributions. {\textit{
  Progress in Energy and Combustion Science}\/}, {\textbf{ 28}\/}, 303 -- 329.

\bibitem[{{Bacchi}(1995)}]{1995Bacchi}
{Bacchi}, B., 1995: Identification and calibration of spatial correlation
  patterns of rainfall. {\textit{ Journal of Hydrology}\/}, {\textbf{ 165}\/},
  311--348.

\bibitem[{Battaglia et al.(2010)Battaglia, Rustemeier, Tokay, Blahak, and
  Simmer}]{ParsivelSNOW}
Battaglia, A., E.~Rustemeier, A.~Tokay, U.~Blahak, and C.~Simmer, 2010:
  Parsivel snow observations: A critical assessment. {\textit{ Journal of
  Atmospheric and Oceanic Technology}\/}, {\textbf{ 27}\/}, 333--344.

\bibitem[{Beard(1976)}]{Beard1976vD}
Beard, K., 1976: Terminal velocity and shape of cloud and precipitation drops
  aloft. {\textit{ Journal of Atmospheric Sciences}\/}, {\textbf{ 33}\/},
  851--864.

\bibitem[{Berne et al.(2009)Berne, Delrieu, and Boudevillain}]{Berne2009flood}
Berne, A., G.~Delrieu, and B.~Boudevillain, 2009: Variability of the spatial
  structure of intense mediterranean precipitation. {\textit{ Advances in Water
  Resources}\/}, {\textbf{ 1}\/}, 1031 -- 1042.

\bibitem[{Brandes et al.(2007)Brandes, Ikeda, Zhang, Sch{\"o}nhuber, and
  Rasmussen}]{brandes_ikeda_etal_2007_aa}
Brandes, E.~A., K.~Ikeda, G.~Zhang, M.~Sch{\"o}nhuber, and R.~M. Rasmussen,
  2007: A statistical and physical description of hydrometeor distributions in
  colorado snowstorms using a video disdrometer. {\textit{ Journal of Applied
  Meteorology and Climatology}\/}, {\textbf{ 46}\/}, 634--650.

\bibitem[{Brawn and Upton(2008)}]{brawn_upton_2008_aa}
Brawn, D. and G.~Upton, 2008: Estimation of an atmospheric gamma drop size
  distribution using disdrometer data. {\textit{ Atmospheric Research}\/},
  {\textbf{ 87}\/}, 66 -- 79.

\bibitem[{{Bringi} et al.(2003){Bringi}, {Chandrasekar}, {Hubbert}, {Gorgucci},
  {Randeu}, and {Schoenhuber}}]{bringi_chandrasekar_etal_2003_aa}
{Bringi}, V.~N., V.~{Chandrasekar}, J.~{Hubbert}, E.~{Gorgucci}, W.~L.
  {Randeu}, and M.~{Schoenhuber}, 2003: {Raindrop Size Distribution in
  Different Climatic Regimes from Disdrometer and Dual-Polarized Radar
  Analysis.} {\textit{ Journal of Atmospheric Sciences}\/}, {\textbf{ 60}\/},
  354--365.

\bibitem[{{Bringi} et al.(2002){Bringi}, {Huang}, {Chandrasekar}, and
  {Gorgucci}}]{bringi_huang_etal_2002_aa}
{Bringi}, V.~N., G.~{Huang}, V.~{Chandrasekar}, and E.~{Gorgucci}, 2002: A
  methodology for estimating the parameters of a gamma raindrop size
  distribution model from polarimetric radar data: Application to a squall-line
  event from the trmm/brazil campaign. {\textit{ Journal of Atmospheric and
  Oceanic Technology}\/}, {\textbf{ 19}\/}, 633--645.

\bibitem[{Brommundt and Bardossy(2007)}]{Brommundt2007spatialcorre}
Brommundt, J. and A.~Bardossy, 2007: Spatial correlation of radar and gauge
  precipitation data in high temporal resolution. {\textit{ Advances in
  Geosciences}\/}, {\textbf{ 10}\/}, 103--109.

\bibitem[{{Brown}(1989)}]{MicroDSDBROWN}
{Brown}, P.~S., Jr., 1989: Coalescence and breakup-induced oscillations in the
  evolution of the raindrop size distribution. {\textit{ Journal of Atmospheric
  Sciences}\/}, {\textbf{ 46}\/}, 1186--1192.

\bibitem[{{Campos} and {Zawadzki}(2000)}]{2000Campos}
{Campos}, E. and I.~{Zawadzki}, 2000: {Instrumental Uncertainties in Z-R
  Relations.} {\textit{ Journal of Applied Meteorology}\/}, {\textbf{ 39}\/},
  1088--1102.

\bibitem[{Cao and Zhang(2009)}]{cao_zhang_2009_aa}
Cao, Q. and G.~Zhang, 2009: Errors in estimating raindrop size distribution
  parameters employing disdrometer and simulated raindrop spectra. {\textit{
  Journal of Applied Meteorology and Climatology}\/}, {\textbf{ 48}\/},
  406--425.

\bibitem[{{Caracciolo} et al.(2006){Caracciolo}, {Prodi}, and
  {Uijlenhoet}}]{caracciolo_prodi_etal_2006_aa}
{Caracciolo}, C., F.~{Prodi}, and R.~{Uijlenhoet}, 2006: {Comparison between
  Pludix and impact/optical disdrometers during rainfall measurement
  campaigns}. {\textit{ Atmospheric Research}\/}, {\textbf{ 82}\/}, 137--163.

\bibitem[{{Caticha}(2001)}]{CatichaEntropyFluctuations2001}
{Caticha}, A., 2001: Maximum entropy, fluctuations and priors. {\textit{
  Bayesian Inference and Maximum Entropy Methods in Science and
  Engineering}\/}, {A.~Mohammad-Djafari}, ed., volume 568 of {\textit{ American
  Institute of Physics Conference Series}\/}, 94--105.

\bibitem[{{Chandrasekar} and {Bringi}(1987)}]{1987Chandra}
{Chandrasekar}, V. and V.~N. {Bringi}, 1987: {Simulation of Radar Reflectivity
  and Surface Measurements of Rainfall}. {\textit{ Journal of Atmospheric and
  Oceanic Technology}\/}, {\textbf{ 4}\/}, 464--+.

\bibitem[{Chandrasekar and Gori(1991)}]{chandrasekar_gori_1991_aa}
Chandrasekar, V. and E.~G. Gori, 1991: Multiple disdrometer observations of
  rainfall. {\textit{ Journal of Applied Meteorology}\/}, {\textbf{ 30}\/},
  1514--1520.

\bibitem[{{Chandrasekar} et al.(2008){Chandrasekar}, {Hou}, {Smith}, {Bringi},
  {Rutledge}, {Gorgucci}, {Petersen}, and
  {Jackson}}]{2008BAMSpolarizationRADARforGPM}
{Chandrasekar}, V., A.~{Hou}, E.~{Smith}, V.~N. {Bringi}, S.~A. {Rutledge},
  E.~{Gorgucci}, W.~A. {Petersen}, and G.~S. {Jackson}, 2008: Potential role of
  dual- polarization radar in the validation of satellite precipitation
  measurements: Rationale and opportunities. {\textit{ Bulletin of the American
  Meteorological Society}\/}, {\textbf{ 89}\/}, 1127--+.

\bibitem[{Chapon et al.(2008)Chapon, Delrieu, Gosset, and
  Boudevillain}]{Chapon200852}
Chapon, B., G.~Delrieu, M.~Gosset, and B.~Boudevillain, 2008: Variability of
  rain drop size distribution and its effect on the z-r relationship: A case
  study for intense mediterranean rainfall. {\textit{ Atmospheric Research}\/},
  {\textbf{ 87}\/}, 52 -- 65.

\bibitem[{Checa and Tapiador(2011)}]{checa_tapiador_2011_aa}
Checa, R. and F.~J. Tapiador, 2011: A maximum entropy modelling of the rain
  drop size distribution. {\textit{ Entropy}\/}, {\textbf{ 13}\/}, 293--315.

\bibitem[{Christakos(1990)}]{BMEChristakos}
Christakos, G., 1990: A bayesian/maximum-entropy view to the spatial estimation
  problem. {\textit{ Mathematical Geology}\/}, {\textbf{ 22}\/}, 763--777,
  10.1007/BF00890661.

\bibitem[{{Chu} and {Su}(2008)}]{2008ChuandSu}
{Chu}, Y.-H. and C.-L. {Su}, 2008: {An Investigation of the Slope-Shape
  Relation for Gamma Raindrop Size Distribution}. {\textit{ Journal of Applied
  Meteorology and Climatology}\/}, {\textbf{ 47}\/}, 2531--+.

\bibitem[{Ciach and Krajewski(1999)}]{Ciach1999585}
Ciach, G.~J. and W.~F. Krajewski, 1999: On the estimation of radar rainfall
  error variance. {\textit{ Advances in Water Resources}\/}, {\textbf{ 22}\/},
  585 -- 595.

\bibitem[{{Ciach} and {Krajewski}(2006)}]{2006CiachAWR}
{Ciach}, G.~J. and W.~F. {Krajewski}, 2006: Analysis and modeling of spatial
  correlation structure in small-scale rainfall in central oklahoma. {\textit{
  Advances in Water Resources}\/}, {\textbf{ 29}\/}, 1450--1463.

\bibitem[{Ciach et al.(1997)Ciach, Krajewski, and
  Smith}]{ciach_krajewski_ea_1997}
Ciach, G.~J., W.~F. Krajewski, and J.~A. Smith, 1997: Comments on the window
  probability matching method for rainfall measurements with radar. {\textit{
  J. Appl. Meteor.}\/}, {\textbf{ 36}\/}, 243--246.

\bibitem[{de~Moraes et al.(2011)de~Moraes, de~Cunha, and
  Krajewski}]{deMoraes2011}
de~Moraes, R.~P., L.~K. de~Cunha, and W.~F. Krajewski, 2011: Assessment of the
  thies optical disdrometer performance. {\textit{ Atmospheric Research}\/},
  {\textbf{ 101}\/}, 237--255.

\bibitem[{Delrieu et al.(1991)Delrieu, Creutin, and
  Saint-Andre}]{AttenuationRelevance}
Delrieu, G., J.~D. Creutin, and I.~Saint-Andre, 1991: Mean k-r relationships:
  Practical results for typical weather radar wavelengths. {\textit{ Journal of
  Atmospheric and Oceanic Technology}\/}, {\textbf{ 8}\/}, 467--476.

\bibitem[{Dumouchel(2009)}]{Dumouchel2009}
Dumouchel, C., 2009: The maximum entropy formalism and the prediction of liquid
  spray drop-size distribution. {\textit{ Entropy}\/}, {\textbf{ 11}\/},
  713--747.

\bibitem[{Efron(1979)}]{Efron1979}
Efron, B., 1979: Bootstrap methods: another look at the jackknife. {\textit{
  The annals of Statistics}\/}, {\textbf{ 7}\/}, 1--26.

\bibitem[{{Feingold} and {Levin}(1986)}]{FeingoldLevin1986}
{Feingold}, G. and Z.~{Levin}, 1986: The lognormal fit to raindrop spectra from
  frontal convective clouds in israel. {\textit{ Journal of Applied
  Meteorology}\/}, {\textbf{ 25}\/}, 1346--1364.

\bibitem[{Feingold and Levin(1987)}]{FeingoldLevin1987}
Feingold, G. and Z.~Levin, 1987: Application of the lognormal raindrop
  distribution to differential reflectivity radar measurement (zdr). {\textit{
  Journal of Atmospheric and Oceanic Technology}\/}, {\textbf{ 4}\/}, 377--382.

\bibitem[{Fujiwara(1965)}]{Fujiwara1965ZRclasification}
Fujiwara, M., 1965: Raindrop-size distribution from individual storms.
  {\textit{ Journal of the Atmospheric Sciences}\/}, {\textbf{ 22}\/},
  585--591.

\bibitem[{{Gebremichael} and {Krajewski}(2004)}]{2004Gebremichael}
{Gebremichael}, M. and W.~F. {Krajewski}, 2004: Assessment of the statistical
  characterization of small-scale rainfall variability from radar: Analysis of
  trmm ground validation datasets. {\textit{ Journal of Applied
  Meteorology}\/}, {\textbf{ 43}\/}, 1180--1199.

\bibitem[{Gosset and Zawadzki(2001)}]{MGossetNUBF2001}
Gosset, M. and I.~Zawadzki, 2001: Effect of nonuniform beam filling on the
  propagation of the radar signal at x-band frequencies. part (i): Changes in
  the k(z) relationship. {\textit{ Journal of Atmospheric and Oceanic
  Technology}\/}, {\textbf{ 18}\/}, 1113--1126.

\bibitem[{Gumm and Kinzer(1949)}]{GunnKinzer1948}
Gumm, R. and G.~Kinzer, 1949: The terminal velocity of fall for water droplets
  in stagnant air. {\textit{ Journal of Meteorology}\/}, {\textbf{ 6}\/},
  243--248.

\bibitem[{{Ha} and {North}(1999)}]{1999HaNorth}
{Ha}, E. and G.~R. {North}, 1999: Error analysis for some ground validation
  designs for satellite observations of precipitation. {\textit{ Journal of
  Atmospheric and Oceanic Technology}\/}, {\textbf{ 16}\/}, 1949--+.

\bibitem[{{Habib} and {Krajewski}(2002)}]{2002TeflunBcampaingHabib}
{Habib}, E. and W.~F. {Krajewski}, 2002: Uncertainty analysis of the trmm
  ground-validation radar-rainfall products: Application to the teflun-b field
  campaign. {\textit{ Journal of Applied Meteorology}\/}, {\textbf{ 41}\/},
  558--572.

\bibitem[{{Habib} et al.(2001){Habib}, {Krajewski}, and
  {Ciach}}]{2001HabibKrajewski}
{Habib}, E., W.~F. {Krajewski}, and G.~J. {Ciach}, 2001: Estimation of rainfall
  interstation correlation. {\textit{ Journal of Hydrometeorology}\/},
  {\textbf{ 2}\/}, 621--+.

\bibitem[{{Haddad} et al.(2006){Haddad}, {Meagher}, {Durden}, {Smith}, and
  {Im}}]{haddad_meagher_etal_2006_aa}
{Haddad}, Z.~S., J.~P. {Meagher}, S.~L. {Durden}, E.~A. {Smith}, and E.~{Im},
  2006: Drop size ambiguities in the retrieval of precipitation profiles from
  dual-frequency radar measurements. {\textit{ Journal of Atmospheric
  Sciences}\/}, {\textbf{ 63}\/}, 204--217.

\bibitem[{Hartmann(2007)}]{PeterHartmann2007}
Hartmann, P., 2007: {\textit{ Analysis of Microphysical Processes from
  Profiling Radar Measurements}\/}. Master's thesis, Meteorologisches Institut
  der Rheinischen Friedrich-Wilhelms-Universität Bonn, master Thesis. Data set
  of Wallops Facilities Center - NASA (Virginia).

\bibitem[{Hauser et al.(1984)Hauser, Amayenc, Nutten, and
  Waldteufel}]{hauser_amayenc_etal_1984_aa}
Hauser, D., P.~Amayenc, B.~Nutten, and P.~Waldteufel, 1984: A new optical
  instrument for simultaneous measurement of raindrop diameter and fallspeed
  distributions. {\textit{ Journal of Atmospheric and Oceanic Technology}\/},
  {\textbf{ 1}\/}, 256--296.

\bibitem[{Hou et al.(2008)Hou, Skofronick-Jackson, Kummerow, and
  Shepherd}]{BookPrecipitation}
Hou, A.~Y., G.~Skofronick-Jackson, C.~D. Kummerow, and J.~M. Shepherd, 2008:
  {\textit{ Precipitation: Advances in Measurement, Estimation and Prediction.
  Chapter 6: Global Precipitation Measurement}\/}. Springer (Editor:
  Michaelides, Silas C.).

\bibitem[{{Hu} and {Srivastava}(1995)}]{MicroDSDSrivastava1995}
{Hu}, Z. and R.~C. {Srivastava}, 1995: Evolution of raindrop size distribution
  by coalescence, breakup, and evaporation: Theory and observations. {\textit{
  Journal of Atmospheric Sciences}\/}, {\textbf{ 52}\/}, 1761--1783.

\bibitem[{Iguchi et al.(2000)Iguchi, Kozu, Meneghini, Awaka, and
  Okamoto}]{iguchi_kozu_ea_2000}
Iguchi, T., T.~Kozu, R.~Meneghini, J.~Awaka, and K.~Okamoto, 2000:
  Rain-profiling algorithm for the trmm precipitation radar. {\textit{ J. Appl.
  Meteor.}\/}, {\textbf{ 39}\/}, 2038--2052.

\bibitem[{{Jaffrain} and {Berne}(2011)}]{Jafrain_Berne_2011}
{Jaffrain}, J. and A.~{Berne}, 2011: {Experimental Quantification of the
  Sampling Uncertainty Associated with Measurements from PARSIVEL
  Disdrometers}. {\textit{ Journal of Hydrometeorology}\/}, {\textbf{ 12}\/},
  352--370.

\bibitem[{Jaffrain and Berne(2011)}]{Parsivel2011Loussane}
Jaffrain, J. and A.~Berne, 2011: Experimental quantification of the sampling
  uncertainty associated with measurements from parsivel disdrometers.
  {\textit{ Journal of Hydrometeorology}\/}, {\textbf{ In press}\/}.

\bibitem[{{Jameson} and {Kostinski}(2002)}]{jameson_kostinski_2002_aa}
{Jameson}, A.~R. and A.~B. {Kostinski}, 2002: {When is Rain Steady?.} {\textit{
  Journal of Applied Meteorology}\/}, {\textbf{ 41}\/}, 83--90.

\bibitem[{Jaynes(1957)}]{Jaynes1957a}
Jaynes, E., 1957: Information theory and statistical mechanics. {\textit{
  Physical Review}\/}, {\textbf{ 106}\/}, 620--630.

\bibitem[{{Joss} and {Waldvogel}(1969)}]{joss_waldvogel_1969_aa}
{Joss}, J. and A.~{Waldvogel}, 1969: {Raindrop Size Distribution and Sampling
  Size Errors.} {\textit{ Journal of Atmospheric Sciences}\/}, {\textbf{
  26}\/}, 566--569.

\bibitem[{Kliche et al.(2008)Kliche, Smith, and
  Johnson}]{kliche_smith_etal_2008_aa}
Kliche, D.~V., P.~L. Smith, and R.~W. Johnson, 2008: L-moment estimators as
  applied to gamma drop size distributions. {\textit{ Journal of Applied
  Meteorology and Climatology}\/}, {\textbf{ 47}\/}, 3117--3130.

\bibitem[{Kostinski et al.(2006)Kostinski, Larsen, and
  Jameson}]{Kostinski200638}
Kostinski, A., M.~Larsen, and A.~Jameson, 2006: The texture of rain: Exploring
  stochastic micro-structure at small scales. {\textit{ Journal of
  Hydrology}\/}, {\textbf{ 328}\/}, 38 -- 45, measurement and Parameterization
  of Rainfall Microstructure.

\bibitem[{{Kozu} and {Nakamura}(1991)}]{kozu_nakamura_1991_aa}
{Kozu}, T. and K.~{Nakamura}, 1991: {Rainfall Parameter Estimation from
  Dual-Radar Measurements Combining Reflectivity Profile and Path-integrated
  Attenuation}. {\textit{ Journal of Atmospheric and Oceanic Technology}\/},
  {\textbf{ 8}\/}, 259--270.

\bibitem[{Krajewski et al.(2006)Krajewski, Kruger, Caraicciolo, Gol{\'e},
  Barthes, Creutin, Delahaye, Nikolopoulus, Odgen, and
  Visoni}]{krajewski_kruger_etal_2006_aa}
Krajewski, W., A.~Kruger, C.~Caraicciolo, P.~Gol{\'e}, L.~Barthes, J.~Creutin,
  J.~Delahaye, E.~Nikolopoulus, F.~Odgen, and J.~Visoni, 2006:
  Devex-disdrometer evaluation experiment: Basic results and implications for
  hydrologic studies. {\textit{ Advances in Water Resources}\/}, {\textbf{
  1}\/}, 311 -- 325.

\bibitem[{Krajewski and Smith(2002)}]{krajewski_smith_2002_aa}
Krajewski, W. and J.~Smith, 2002: Radar hydrology: rainfall estimation.
  {\textit{ Advances in Water Resources}\/}, {\textbf{ 25}\/}, 1387 -- 1394.

\bibitem[{Kumar et al.(2010)Kumar, Lee, and Ong}]{KumarIEEE2010}
Kumar, L., Y.~H. Lee, and J.~T. Ong, 2010: Truncated gamma drop size
  distribution models for rain attenuation in singapore. {\textit{ Antennas and
  Propagation, IEEE Transactions on}\/}, {\textbf{ 58}\/}, 1325 --1335.

\bibitem[{Kumar et al.(2011)Kumar, Lee, and Ong}]{KumarIEEE2011}
--- 2011: Two-parameter gamma drop size distribution models for singapore.
  {\textit{ Geoscience and Remote Sensing, IEEE Transactions on}\/}, {\textbf{
  49}\/}, 3371 --3380.

\bibitem[{{Larsen} et al.(2005){Larsen}, {Kostinski}, and
  {Tokay}}]{larsen_kostinski_etal_2005_aa}
{Larsen}, M.~L., A.~B. {Kostinski}, and A.~{Tokay}, 2005: {Observations and
  Analysis of Uncorrelated Rain.} {\textit{ Journal of Atmospheric
  Sciences}\/}, {\textbf{ 62}\/}, 4071--4083.

\bibitem[{Laws and Parsons(1943)}]{LawsParsons1943}
Laws, J.~O. and D.~Parsons, 1943: The relation of raindrop-size to intensity.
  {\textit{ Transactions American Geophysical Union}\/}, {\textbf{ 24}\/},
  452--460.

\bibitem[{{Lee} et al.(2009){Lee}, {Lee}, {Zawadzki}, and
  {Kim}}]{2009ChoongLEE}
{Lee}, C.~K., G.~{Lee}, I.~{Zawadzki}, and K.~{Kim}, 2009: A preliminary
  analysis of spatial variability of raindrop size distributions during
  stratiform rain events. {\textit{ Journal of Applied Meteorology and
  Climatology}\/}, {\textbf{ 48}\/}, 270--+.

\bibitem[{{Lee} and {Zawadzki}(2005)}]{LeeSIFTmethod2005}
{Lee}, G. and I.~{Zawadzki}, 2005: Variability of drop size distributions:
  Noise and noise filtering in disdrometric data. {\textit{ Journal of Applied
  Meteorology}\/}, {\textbf{ 44}\/}, 634--652.

\bibitem[{{Lee} et al.(2004){Lee}, {Zawadzki}, {Szyrmer}, {Sempere-Torres}, and
  {Uijlenhoet}}]{GyuWonLee2004}
{Lee}, G., I.~{Zawadzki}, W.~{Szyrmer}, D.~{Sempere-Torres}, and
  R.~{Uijlenhoet}, 2004: A general approach to double-moment normalization of
  drop size distributions. {\textit{ Journal of Applied Meteorology}\/},
  {\textbf{ 43}\/}, 264--281.

\bibitem[{Leijnse and Uijlenhoet(2010)}]{Leijnse2010}
Leijnse, H. and R.~Uijlenhoet, 2010: The effect of reported high-velocity small
  raindrops on inferred drop size distributions and derived power laws.
  {\textit{ Atmospheric Chemistry and Physics}\/}, {\textbf{ 10}\/},
  6807--6818.

\bibitem[{{Lilley} et al.(2006){Lilley}, {Lovejoy}, {Desaulniers-Soucy}, and
  {Schertzer}}]{2006JHydropExperiment}
{Lilley}, M., S.~{Lovejoy}, N.~{Desaulniers-Soucy}, and D.~{Schertzer}, 2006:
  Multifractal large number of drops limit in rain. {\textit{ Journal of
  Hydrology}\/}, {\textbf{ 328}\/}, 20--37.

\bibitem[{{List}(1988)}]{1988List}
{List}, R., 1988: A linear radar reflectivity-rainrate relationship for steady
  tropical rain. {\textit{ Journal of Atmospheric Sciences}\/}, {\textbf{
  45}\/}, 3564--3572.

\bibitem[{Liu and Nocedal(1989)}]{liu1989limited}
Liu, D. and J.~Nocedal, 1989: On the limited memory bfgs method for large scale
  optimization. {\textit{ Mathematical programming}\/}, {\textbf{ 45}\/},
  503--528.

\bibitem[{Liu et al.(2002)Liu, Daum, and Hallett}]{LiuYangang2002}
Liu, Y., P.~H. Daum, and J.~Hallett, 2002: A generalized systems theory for the
  effect of varying fluctuations on cloud droplet size distributions. {\textit{
  Journal of the Atmospheric Sciences}\/}, {\textbf{ 59}\/}, 2279--2290.

\bibitem[{Liu and Hallett(1998)}]{LiuYangang1998}
Liu, Y. and J.~Hallett, 1998: On size distributions of cloud droplets growing
  by condensation: A new conceptual model. {\textit{ Journal of the Atmospheric
  Sciences}\/}, {\textbf{ 55}\/}, 527--536.

\bibitem[{Liu et al.(1995)Liu, Laiguang, Weinong, and Feng}]{LiuYangang1995}
Liu, Y., Y.~Laiguang, Y.~Weinong, and L.~Feng, 1995: On the size distribution
  of cloud droplets. {\textit{ Atmospheric Research}\/}, {\textbf{ 35}\/}, 201
  -- 216.

\bibitem[{Loffler-Mang and Joss(2000)}]{loffler-mang_joss_2000_aa}
Loffler-Mang, M. and J.~Joss, 2000: An optical disdrometer for measuring size
  and velocity of hydrometeors. {\textit{ Journal of Atmospheric and Oceanic
  Technology}\/}, {\textbf{ 17}\/}, 130--139.

\bibitem[{Loucks et al.(2005)Loucks, Van~Beek, Stedinger, Dijkman, and
  Villars}]{loucks2005water}
Loucks, D., E.~Van~Beek, J.~Stedinger, J.~Dijkman, and M.~Villars, 2005:
  {\textit{ Water resources systems planning and management: an introduction to
  methods, models and applications}\/}. Paris: UNESCO.

\bibitem[{{Lovejoy} and {Schertzer}(1990)}]{1990JApMe291167L}
{Lovejoy}, S. and D.~{Schertzer}, 1990: {Fractals, Raindrops and Resolution
  Dependence of Rain Measurements.} {\textit{ Journal of Applied
  Meteorology}\/}, {\textbf{ 29}\/}, 1167--1170.

\bibitem[{{Mallet} and {Barthes}(2009)}]{mallet_barthes_2009_aa}
{Mallet}, C. and L.~{Barthes}, 2009: {Estimation of Gamma Raindrop Size
  Distribution Parameters: Statistical Fluctuations and Estimation Errors}.
  {\textit{ Journal of Atmospheric and Oceanic Technology}\/}, {\textbf{
  26}\/}, 1572--+.

\bibitem[{Marshall and Palmer(1948)}]{MarshallPalmer1948}
Marshall, J. and W.~Palmer, 1948: The distribution of raindrops with size.
  {\textit{ Journal of Atmospheric Sciences}\/}, {\textbf{ 5}\/}, 165--166.

\bibitem[{{Marzano} et al.(2004){Marzano}, {Tapiador}, {Kidd}, and
  {Levizzani}}]{Tapiador2004}
{Marzano}, F.~S., F.~J. {Tapiador}, C.~{Kidd}, and V.~{Levizzani}, 2004: A
  maximum entropy approach to satellite quantitative precipitation estimation
  (qpe). {\textit{ International Journal of Remote Sensing}\/}, {\textbf{
  25}\/}, 4629--4639.

\bibitem[{Marzuki et al.(2010)Marzuki, Randeu, Schoandnhuber, Bringi, Kozu, and
  Shimomai}]{Marzuki2010}
Marzuki, M., W.~Randeu, M.~Schoandnhuber, V.~Bringi, T.~Kozu, and T.~Shimomai,
  2010: Raindrop size distribution parameters of distrometer data with
  different bin sizes. {\textit{ Geoscience and Remote Sensing, IEEE
  Transactions on}\/}, {\textbf{ 48}\/}, 3075 --3080.

\bibitem[{Matsumoto and Nishimura(1998)}]{MersenneTwister}
Matsumoto, M. and T.~Nishimura, 1998: Mersenne twister: a 623-dimensionally
  equidistributed uniform pseudo-random number generator. {\textit{ ACM Trans.
  Model. Comput. Simul.}\/}, {\textbf{ 8}\/}, 3--30.

\bibitem[{{Michaelides} et al.(2010){Michaelides}, {Levizzani}, {Anagnostou},
  {Bauer}, {Kasparis}, and {Lane}}]{michaelides_levizzani_etal_2010_aa}
{Michaelides}, S., V.~{Levizzani}, E.~{Anagnostou}, P.~{Bauer}, T.~{Kasparis},
  and J.~E. {Lane}, 2010: {Precipitation: Measurement, remote sensing,
  climatology and modeling}. {\textit{ Atmospheric Research}\/}, {\textbf{
  95}\/}, 512--533.

\bibitem[{Miriovsky et al.(2004)Miriovsky, Bradley, Eichinger, Krajewski,
  Kruger, and Nelson}]{miriovsky_bradley_etal_2004_aa}
Miriovsky, B., A.~Bradley, W.~Eichinger, W.~F. Krajewski, A.~Kruger, and
  B.~Nelson, 2004: An experimental study od small-scale variability of radar
  reflectivity using disdrometer observations. {\textit{ Journal of Applied
  Meteorology}\/}, {\textbf{ 43}\/}, 106--118.

\bibitem[{{Mohammad-Djafari}(2001)}]{Djafari1991}
{Mohammad-Djafari}, A., 2001: A matlab program to calculate the maximum entropy
  distributions. {\textit{ ArXiv Physics e-prints}\/}.

\bibitem[{Moisseev and Chandrasekar(2007)}]{moisseev_chandrasekar_2007_aa}
Moisseev, D.~N. and V.~Chandrasekar, 2007: Examination of the shape--slope
  relation suggested for drop size distribution parameters. {\textit{ Journal
  of Atmospheric and Oceanic Technology}\/}, {\textbf{ 24}\/}, 847--855.

\bibitem[{Mondal et al.(2004)Mondal, Datta, and Sarkar}]{SpraysMaxEntMODAL2004}
Mondal, D., A.~Datta, and A.~Sarkar, 2004: Droplet size and velocity
  distributions in a spray from a pressure swirl atomizer: application of
  maximum entropy formalism. {\textit{ Proceedings of the Institution of
  Mechanical Engineers, Part C: Journal of Mechanical Engineering Science}\/},
  {\textbf{ 218}\/}, 737--749.

\bibitem[{Montero-Martinez et al.(2009)Montero-Martinez, Kostinski, Shaw, and
  Garcia-Garcia}]{Montero-Martinez:2009kx}
Montero-Martinez, G., A.~Kostinski, R.~Shaw, and F.~Garcia-Garcia, 2009: Do all
  raindrops fall at terminal speed? {\textit{ Geophysical Research Letters}\/},
  {\textbf{ 36}\/}, L11818.

\bibitem[{Moumouni et al.(2009)Moumouni, Gosset, and
  Houngninou}]{moumouni_gosset_etal_2009_aa}
Moumouni, S., S.~Gosset, and E.~Houngninou, 2009: Main features of rain drop
  size distributions observed in benin, west africa, with optical disdrometers.
  {\textit{ Geophysical Research Letters}\/}, {\textbf{ 35}\/}, 23807.

\bibitem[{Niu et al.(2010)Niu, Jia, Sang, Liu, Lu, and
  Liu}]{ChinosPARSIVELestudio}
Niu, S., X.~Jia, J.~Sang, X.~Liu, C.~Lu, and Y.~Liu, 2010: Distributions of
  raindrop sizes and fall velocities in a semiarid plateau climate: Convective
  versus stratiform rains. {\textit{ Journal of Applied Meteorology and
  Climatology}\/}, {\textbf{ 49}\/}, 632--645.

\bibitem[{{Niven}(2009)}]{Niven2009}
{Niven}, R.~K., 2009: Steady state of a dissipative flow-controlled system and
  the maximum entropy production principle. {\textit{ Physical Review E}\/},
  {\textbf{ 80}\/}, 021113--+.

\bibitem[{{Nzeukou} et al.(2004){Nzeukou}, {Sauvageot}, {Delfin Ochou}, and
  {Mouhamed Fadel Kebe}}]{2004Nzeukou}
{Nzeukou}, A., H.~{Sauvageot}, A.~{Delfin Ochou}, and C.~{Mouhamed Fadel Kebe},
  2004: {Raindrop Size Distribution and Radar Parameters at Cape Verde.}
  {\textit{ Journal of Applied Meteorology}\/}, {\textbf{ 43}\/}, 90--105.

\bibitem[{Radhakrishna and Narayana~Rao(2009)}]{Multimodal2}
Radhakrishna, B. and T.~Narayana~Rao, 2009: Statistical characteristics of
  multipeak raindrop size distributions at the surface and aloft in different
  rain regimes. {\textit{ Monthly Weather Review}\/}, {\textbf{ 137}\/},
  3501--3518.

\bibitem[{{Rodriguez-Iturbe} and {Mej{\'{\i}}a}(1974)}]{1974RodriguezIturbe}
{Rodriguez-Iturbe}, I. and J.~M. {Mej{\'{\i}}a}, 1974: On the transformation of
  point rainfall to areal rainfall. {\textit{ Water Resources Research}\/},
  {\textbf{ 10}\/}, 729--735.

\bibitem[{Roger(1976)}]{RogersBOOK}
Roger, R.~R., 1976: {\textit{ A short Course in Cloud Physics}\/}. Pergamon
  Press, Oxford.

\bibitem[{{Sauvageot} and {Koffi}(2000)}]{Multimodal1}
{Sauvageot}, H. and M.~{Koffi}, 2000: Multimodal raindrop size distributions.
  {\textit{ Journal of Atmospheric Sciences}\/}, {\textbf{ 57}\/}, 2480--2492.

\bibitem[{Sekhon and Srivastava(1971)}]{SekhonSrivastava1971}
Sekhon, R.~S. and R.~C. Srivastava, 1971: Doppler radar observations of
  drop-size distributions in a thunderstorm. {\textit{ Journal of the
  Atmospheric Sciences}\/}, {\textbf{ 28}\/}, 983--994.

\bibitem[{Sempere-Torres et al.(1998)Sempere-Torres, Porra, and
  Creutin}]{SempereTorres1998}
Sempere-Torres, D., J.~Porra, and J.~Creutin, 1998: Experimental evidence of a
  general description for raindrop size distribution properties. {\textit{
  Journal of Geophysical Research}\/}, {\textbf{ 103}\/}, 1785--1798.

\bibitem[{{Sempere Torres} et al.(1994){Sempere Torres}, {Porra}, and
  {Creutin}}]{semperetorres_porra_ea_1994}
{Sempere Torres}, D., J.~M. {Porra}, and J.~{Creutin}, 1994: A general
  formulation for raindrop size distribution. {\textit{ Journal Of Applied
  Meteorology}\/}, {\textbf{ 33}\/}, 1494--1502.

\bibitem[{Shannon(2001)}]{Shannon1948}
Shannon, C.~E., 2001: A mathematical theory of communication. {\textit{
  SIGMOBILE Mob. Comput. Commun. Rev.}\/}, {\textbf{ 5}\/}, 3--55.

\bibitem[{{Sheppard} and {Joe}(1994)}]{sheppard_joe_1994_aa}
{Sheppard}, B. and P.~{Joe}, 1994: {Comparison of raindrop size distribution
  measurements by a Joss-Wadvogel disdrometer, a PMS 2DG spectrometer and a
  POSS Doppler radar.} {\textit{ Journal of Atmospheric and Oceanic
  Technology}\/}, {\textbf{ 11}\/}, 874--887.

\bibitem[{{Shimizu}(1993)}]{1993Shimizu}
{Shimizu}, K., 1993: A bivariate mixed lognormal distribution with an analysis
  of rainfall data. {\textit{ Journal of Applied Meteorology}\/}, {\textbf{
  32}\/}, 161--171.

\bibitem[{Singh et al.(1986)Singh, Rajagopal, and Singh}]{POME1986}
Singh, V.~P., A.~K. Rajagopal, and K.~Singh, 1986: Derivation of some frequency
  distributions using the principle of maximum entropy (pome). {\textit{
  Advances in Water Resources}\/}, {\textbf{ 9}\/}, 91 -- 106.

\bibitem[{{Smith} and {Krajewski}(1993)}]{SmithKrajewski1993}
{Smith}, J.~A. and W.~F. {Krajewski}, 1993: A modeling study of rainfall
  rate-reflectivity relationships. {\textit{ Water Resources Research}\/},
  {\textbf{ 29}\/}, 2505--2514.

\bibitem[{{Smith}(2003)}]{Smith2003}
{Smith}, P.~L., 2003: Raindrop size distributions: Exponential or gamma--does
  the difference matter?. {\textit{ Journal of Applied Meteorology}\/},
  {\textbf{ 42}\/}, 1031--1034.

\bibitem[{{Smith} and {Kliche}(2005)}]{smith_kliche_2005_aa}
{Smith}, P.~L. and D.~V. {Kliche}, 2005: {The Bias in Moment Estimators for
  Parameters of Drop Size Distribution Functions: Sampling from Exponential
  Distributions.} {\textit{ Journal of Applied Meteorology}\/}, {\textbf{
  44}\/}, 1195--1205.

\bibitem[{{Steiner} and {Smith}(2000)}]{SteinerKINETIC2000}
{Steiner}, M. and J.~A. {Smith}, 2000: Reflectivity, rain rate, and kinetic
  energy flux relationships based on raindrop spectra. {\textit{ Journal of
  Applied Meteorology}\/}, {\textbf{ 39}\/}, 1923--1940.

\bibitem[{Steiner et al.(2004)Steiner, Smith, and
  Uijlenhoet}]{Steiner2004microphysisofZRgamma}
Steiner, M., J.~A. Smith, and R.~Uijlenhoet, 2004: A microphysical
  interpretation of radar reflectivity–rain rate relationships. {\textit{
  Journal of the Atmospheric Sciences}\/}, {\textbf{ 61}\/}, 1114--1131.

\bibitem[{Straka(2009)}]{StrakaBOOK}
Straka, J., 2009: {\textit{ Cloud and Precipitation MicroPhysics (Principles
  and Parameterizations)}\/}. Cambride UNiversity Press.

\bibitem[{Tapiador et al.(2011{\natexlab{a}})Tapiador, Turk, Petersen, Hou,
  Garcia-Ortega, Machado, Angelis, Salio, Kidd, Huffman, and
  de~Castro}]{TapiadorGPM2011}
Tapiador, F., J.~Turk, W.~Petersen, A.~Hou, E.~Garcia-Ortega, L.~Machado,
  C.~Angelis, P.~Salio, C.~Kidd, G.~Huffman, and M.~de~Castro,
  2011{\natexlab{a}}: Global precipitation measurement: Methods, datasets and
  applications. {\textit{ Atmospheric Research}\/}, --.

\bibitem[{Tapiador(2007)}]{Tapiador2007}
Tapiador, F.~J., 2007: A maximum entropy analysis of global monthly series of
  rainfall from merged satellite data. {\textit{ Int. J. Remote Sens.}\/},
  {\textbf{ 28}\/}, 1113--1121.

\bibitem[{{Tapiador}(2008)}]{Tapiador2008}
{Tapiador}, F.~J., 2008: Hurricane footprints in global climate models.
  {\textit{ Entropy}\/}, {\textbf{ 10}\/}, 613--620.

\bibitem[{{Tapiador} and {Casanova}(2002)}]{Tapiador2002}
{Tapiador}, F.~J. and J.~L. {Casanova}, 2002: An algorithm for the fusion of
  images based on jaynes' maximum entropy method. {\textit{ International
  Journal of Remote Sensing}\/}, {\textbf{ 23}\/}, 777--785.

\bibitem[{{Tapiador} et al.(2010){Tapiador}, {Checa}, and {de
  Castro}}]{Tapiador2010}
{Tapiador}, F.~J., R.~{Checa}, and M.~{de Castro}, 2010: An experiment to
  measure the spatial variability of rain drop size distribution using sixteen
  laser disdrometers. {\textit{ Geophysical Research Letters}\/}, {\textbf{
  37}\/}, 16803--+.

\bibitem[{Tapiador et al.(2011{\natexlab{b}})Tapiador, Hou, de~Castro, Checa,
  Cuartero, and Barros}]{HydroPower2011}
Tapiador, F.~J., A.~Y. Hou, M.~de~Castro, R.~Checa, F.~Cuartero, and A.~P.
  Barros, 2011{\natexlab{b}}: Precipitation estimates for hydroelectricity.
  {\textit{ Energy Environ. Sci.}\/}, {\textbf{ 4}\/}, 4435--4448.

\bibitem[{{Testud} et al.(2001){Testud}, {Oury}, Black, Amayenc, and
  Dou}]{NormalizadaTestud2001}
{Testud}, J., S.~{Oury}, R.~Black, P.~Amayenc, and X.~Dou, 2001: The concept of
  normalized distribution to describe raindrop spectra: A tool for cloud
  physics and cloud remote sensing. {\textit{ Journal of Applied
  Meteorology}\/}, {\textbf{ 40}\/}, 1118--1140.

\bibitem[{Testud et al.(2001)Testud, Oury, Black, Amayenc, and
  Dou}]{Testud2001}
Testud, J., S.~Oury, R.~A. Black, P.~Amayenc, and X.~Dou, 2001: The concept of
  “normalized” distribution to describe raindrop spectra: A tool for cloud
  physics and cloud remote sensing. {\textit{ Journal of Applied
  Meteorology}\/}, {\textbf{ 40}\/}, 1118--1140.

\bibitem[{Thurai et al.(2011)Thurai, Petersen, Tokay, Schultz, and
  Gatlin}]{thurai_petersen_ea_2011}
Thurai, M., W.~A. Petersen, A.~Tokay, C.~Schultz, and P.~Gatlin, 2011: Drop
  size distribution comparisons between parsivel and 2-d video disdrometers.
  {\textit{ Advances in Geosciences}\/}, {\textbf{ 30}\/}, 3--9.

\bibitem[{Tokay et al.(2005)Tokay, Bashor, and Wolff}]{TokayCOLLOCATEDjwd2005}
Tokay, A., P.~Bashor, and K.~Wolff, 2005: Error characteristic of rainfall
  measurements by collocated joss-waldvogel disdrometers. {\textit{ Journal of
  Atmospheric and Oceanic Technology}\/}, {\textbf{ 22}\/}, 513--527.

\bibitem[{Tokay and Bashor(2010)}]{Tokay2010smallscaleDSD}
Tokay, A. and P.~G. Bashor, 2010: An experimental study of small-scale
  variability of raindrop size distribution. {\textit{ Journal of Applied
  Meteorology and Climatology}\/}, {\textbf{ 49}\/}, 2348--2365.

\bibitem[{{Tokay} et al.(2005){Tokay}, {Bashor}, and
  {Wolff}}]{tokay_bashor_etal_2005_aa}
{Tokay}, A., P.~G. {Bashor}, and K.~R. {Wolff}, 2005: {Error Characteristics of
  Rainfall Measurements by Collocated Joss Waldvogel Disdrometers}. {\textit{
  Journal of Atmospheric and Oceanic Technology}\/}, {\textbf{ 22}\/}, 513--+.

\bibitem[{Tokay and Beard(1996)}]{TokayBeard1996drops}
Tokay, A. and K.~V. Beard, 1996: A field study of raindrop oscillations. part
  i: Observation of size spectra and evaluation of oscillation causes.
  {\textit{ Journal of Applied Meteorology}\/}, {\textbf{ 35}\/}, 1671--1687.

\bibitem[{{Tokay} et al.(2001){Tokay}, {Kruger}, and
  {Krajewski}}]{tokay_kruger_etal_2001_aa}
{Tokay}, A., A.~{Kruger}, and W.~{Krajewski}, 2001: {Comparison of Drop Size
  Distribution Measurements by Impact and Optical Disdrometers.} {\textit{
  Journal of Applied Meteorology}\/}, {\textbf{ 40}\/}, 2083--2097.

\bibitem[{{Tokay} and {Short}(1996)}]{TokayShort1996}
{Tokay}, A. and D.~A. {Short}, 1996: Evidence from tropical raindrop spectra of
  the origin of rain from stratiform versus convective clouds. {\textit{
  Journal of Applied Meteorology}\/}, {\textbf{ 35}\/}, 355--371.

\bibitem[{{Tokay} et al.(1999){Tokay}, {Short}, {Williams}, {Ecklund}, and
  {Gage}}]{tokay_short_etal_1999_aa}
{Tokay}, A., D.~A. {Short}, C.~R. {Williams}, W.~L. {Ecklund}, and K.~S.
  {Gage}, 1999: {Tropical Rainfall Associated with Convective and Stratiform
  Clouds: Intercomparison of Disdrometer and Profiler Measurements.} {\textit{
  Journal of Applied Meteorology}\/}, {\textbf{ 38}\/}, 302--320.

\bibitem[{Tokay et al.(2003)Tokay, Wolff, Wolff, and
  Bashor}]{Tokay2003disdrometerKAMP}
Tokay, A., D.~B. Wolff, K.~R. Wolff, and P.~Bashor, 2003: Rain gauge and
  disdrometer measurements during the keys area microphysics project (kamp).
  {\textit{ Journal of Atmospheric and Oceanic Technology}\/}, {\textbf{
  20}\/}, 1460--1477.

\bibitem[{Uffink(1995)}]{Uffink1995223}
Uffink, J., 1995: Can the maximum entropy principle be explained as a
  consistency requirement? {\textit{ Studies In History and Philosophy of
  Science Part B: Studies In History and Philosophy of Modern Physics}\/},
  {\textbf{ 26}\/}, 223 -- 261.

\bibitem[{Uijlenhoet and Pomeroy(2001)}]{uijlenhoet_pomeroy_2001_aa}
Uijlenhoet, R. and J.~H. Pomeroy, 2001: Raindrop size distributions and radar
  reflectivity--rain rate relationships for radar hydrology. {\textit{
  Hydrology and Earth System Sciences}\/}, {\textbf{ 5}\/}, 615--628.

\bibitem[{{Uijlenhoet} et al.(2009){Uijlenhoet}, {Porra}, {Sempere Torres}, and
  {Creutin}}]{2009NPGeo16287U}
{Uijlenhoet}, R., J.~{Porra}, D.~{Sempere Torres}, and J.~{Creutin}, 2009: Edge
  effect causes apparent fractal correlation dimension of uniform spatial
  raindrop distribution. {\textit{ Nonlinear Processes in Geophysics}\/},
  {\textbf{ 16}\/}, 287--297.

\bibitem[{Uijlenhoet et al.(2006)Uijlenhoet, Porra, Torres, and
  Creutin}]{uijlenhoet2006analytical}
Uijlenhoet, R., J.~Porra, D.~Torres, and J.~Creutin, 2006: Analytical solutions
  to sampling effects in drop size distribution measurements during stationary
  rainfall: Estimation of bulk rainfall variables. {\textit{ Journal of
  Hydrology}\/}, {\textbf{ 328}\/}, 65--82.

\bibitem[{{Uijlenhoet} et al.(2003){Uijlenhoet}, {Steiner}, and
  {Smith}}]{uijlenhoet_steiner_etal_2003_aa}
{Uijlenhoet}, R., M.~{Steiner}, and J.~A. {Smith}, 2003: {Variability of
  Raindrop Size Distributions in a Squall Line and Implications for Radar
  Rainfall Estimation}. {\textit{ Journal of Hydrometeorology}\/}, {\textbf{
  4}\/}, 43--+.

\bibitem[{{Ulbrich}(1983)}]{ulbrich_1983_aa}
{Ulbrich}, C.~W., 1983: {Natural Variations in the Analytical Form of the
  Raindrop Size Distribution.} {\textit{ Journal of Applied Meteorology}\/},
  {\textbf{ 22}\/}, 1764--1775.

\bibitem[{{Ulbrich} and {Atlas}(1998)}]{UlbrichAtlas1998}
{Ulbrich}, C.~W. and D.~{Atlas}, 1998: Rainfall microphysics and radar
  properties: Analysis methods for drop size spectra. {\textit{ Journal of
  Applied Meteorology}\/}, {\textbf{ 37}\/}, 912--923.

\bibitem[{{Villarini} et al.(2008{\natexlab{a}}){Villarini}, {Mandapaka},
  {Krajewski}, and {Moore}}]{villarini_mandapaka_etal_2008_aa}
{Villarini}, G., P.~V. {Mandapaka}, W.~F. {Krajewski}, and R.~J. {Moore},
  2008{\natexlab{a}}: {Rainfall and sampling uncertainties: A rain gauge
  perspective}. {\textit{ Journal of Geophysical Research (Atmospheres)}\/},
  {\textbf{ 113}\/}, 11102--+.

\bibitem[{{Villarini} et al.(2008{\natexlab{b}}){Villarini}, {Mandapaka},
  {Krajewski}, and {Moore}}]{2008VillariniJGR}
--- 2008{\natexlab{b}}: Rainfall and sampling uncertainties: A rain gauge
  perspective. {\textit{ Journal of Geophysical Research (Atmospheres)}\/},
  {\textbf{ 113}\/}, 11102--+.

\bibitem[{{Villermaux} and {Bossa}(2009)}]{Villermaux2009Nature}
{Villermaux}, E. and B.~{Bossa}, 2009: Single-drop fragmentation determines
  size distribution of raindrops. {\textit{ Nature Physics}\/}, {\textbf{
  5}\/}, 697--702.

\bibitem[{Vivekanandam et al.(2004)Vivekanandam, Zhang, and
  Brandes}]{Vivekanandam2004}
Vivekanandam, J., G.~Zhang, and E.~Brandes, 2004: Polarimetric radar estimators
  based on a constrained gamma drop size distribution model. {\textit{ Journal
  of Applied Meteorology}\/}, {\textbf{ 43}\/}, 217--230.

\bibitem[{Williams et al.(2007)Williams, White, Gage, and
  Ralph}]{MicrofisicaNOAAprofiler}
Williams, C.~R., A.~B. White, K.~S. Gage, and F.~M. Ralph, 2007: Vertical
  structure of precipitation and related microphysics observed by noaa
  profilers and trmm during name 2004. {\textit{ Journal of Climate}\/},
  {\textbf{ 20}\/}, 1693--1712.

\bibitem[{Yuter et al.(2006)Yuter, Kingsmill, Nance, and
  Löffler-Mang}]{RainWetSnow2006parsivel}
Yuter, S.~E., D.~E. Kingsmill, L.~B. Nance, and M.~Löffler-Mang, 2006:
  Observations of precipitation size and fall speed characteristics within
  coexisting rain and wet snow. {\textit{ Journal of Applied Meteorology and
  Climatology}\/}, {\textbf{ 45}\/}, 1450--1464.

\bibitem[{Zhang et al.(2003)Zhang, Vivekanandan, Brandes, Meneghini, and
  Kozu}]{zhang_vivekanandan_etal_2003_aa}
Zhang, G., J.~Vivekanandan, E.~A. Brandes, R.~Meneghini, and T.~Kozu, 2003: The
  shape--slope relation in observed gamma raindrop size distributions:
  Statistical error or useful information? {\textit{ Journal of Atmospheric and
  Oceanic Technology}\/}, {\textbf{ 20}\/}, 1106--1119.

\end{thebibliography}
\newpage

\setlength{\cftfigurenumwidth}{4em}

\listoffigures
\newpage
\listoftables
\newpage

\end{document}